\documentclass{SciPost}
\hypersetup{
    colorlinks,
    linkcolor={red!50!black},
    citecolor={blue!50!black},
    urlcolor={blue!80!black}
}

\usepackage[bitstream-charter]{mathdesign}
\usepackage[most]{tcolorbox}
\tcbuselibrary{breakable,skins}
\usepackage{xparse}
\usepackage{etoolbox}
\usepackage{enumitem}
\usepackage{tikz}
\usetikzlibrary{arrows.meta,positioning,calc,fit}
\usepackage{pgfplots}
\pgfplotsset{compat=1.18}

\colorlet{SMRMExampleFrame}{red!50!black}
\colorlet{SMRMExampleBack}{red!3!white}
\colorlet{SMRMExampleTitleBack}{red!8!white}

\colorlet{SMRMExerciseFrame}{blue!50!black}
\colorlet{SMRMExerciseBack}{blue!3!white}
\colorlet{SMRMExerciseTitleBack}{blue!8!white}

\colorlet{SMRMNeutralFrame}{black!55}
\colorlet{SMRMNeutralBack}{black!3}

\newcounter{smrmexample}[section]
\renewcommand{\thesmrmexample}{\thesection.\arabic{smrmexample}}

\newcounter{smrmexercise}[section]
\renewcommand{\thesmrmexercise}{\thesection.\arabic{smrmexercise}}

\newcommand{\SMRMoptionalcolon}[1]{%
  \ifstrempty{#1}{}{\,: #1}%
}

\NewDocumentEnvironment{examplebox}{O{}}
{%
  \refstepcounter{smrmexample}%
  \begin{tcolorbox}[
    enhanced,
    breakable,
    sharp corners=downhill,
    boxrule=0.6pt,
    arc=1.2mm,
    colframe=SMRMExampleFrame,
    colback=SMRMExampleBack,
    coltitle=SMRMExampleFrame,
    colbacktitle=SMRMExampleTitleBack,
    fonttitle=\bfseries,
    title={Example~\thesmrmexample\SMRMoptionalcolon{#1}},
    left=1.5mm,
    right=1.5mm,
    top=1.2mm,
    bottom=1.2mm,
    before skip=10pt,
    after skip=10pt
  ]%
}
{%
  \end{tcolorbox}%
}

\NewDocumentEnvironment{exercisebox}{O{}}
{%
  \refstepcounter{smrmexercise}%
  \begin{tcolorbox}[
    enhanced,
    breakable,
    sharp corners=downhill,
    boxrule=0.6pt,
    arc=1.2mm,
    colframe=SMRMExerciseFrame,
    colback=SMRMExerciseBack,
    coltitle=SMRMExerciseFrame,
    colbacktitle=SMRMExerciseTitleBack,
    fonttitle=\bfseries,
    title={Exercise~\thesmrmexercise\SMRMoptionalcolon{#1}},
    left=1.5mm,
    right=1.5mm,
    top=1.2mm,
    bottom=1.2mm,
    before skip=10pt,
    after skip=10pt
  ]%
}
{%
  \end{tcolorbox}%
}

\NewDocumentEnvironment{exerciseblock}{O{Exercises for this section}}
{%
  \begin{tcolorbox}[
    enhanced,
    breakable,
    sharp corners=downhill,
    boxrule=0.6pt,
    arc=1.2mm,
    colframe=SMRMExerciseFrame,
    colback=SMRMExerciseBack,
    coltitle=SMRMExerciseFrame,
    colbacktitle=SMRMExerciseTitleBack,
    fonttitle=\bfseries,
    title={#1},
    left=1.5mm,
    right=1.5mm,
    top=1.2mm,
    bottom=1.2mm,
    before skip=10pt,
    after skip=10pt
  ]%
  \begin{list}{}{%
    \setlength{\leftmargin}{0pt}%
    \setlength{\labelwidth}{0pt}%
    \setlength{\labelsep}{0pt}%
    \setlength{\itemindent}{0pt}%
    \setlength{\itemsep}{8pt}%
    \setlength{\parsep}{4pt}%
    \setlength{\topsep}{0pt}%
  }%
}
{%
  \end{list}%
  \end{tcolorbox}%
}

\NewDocumentCommand{\exitem}{O{}}
{%
  \refstepcounter{smrmexercise}%
  \item
  \noindent\textbf{Exercise~\thesmrmexercise\SMRMoptionalcolon{#1}.}\quad
}

\NewDocumentEnvironment{remarkbox}{O{Remark}}
{%
  \begin{tcolorbox}[
    enhanced,
    breakable,
    sharp corners=downhill,
    boxrule=0.5pt,
    arc=1.2mm,
    colframe=SMRMNeutralFrame,
    colback=SMRMNeutralBack,
    coltitle=SMRMNeutralFrame,
    colbacktitle=black!6,
    fonttitle=\bfseries,
    title={#1},
    left=1.5mm,
    right=1.5mm,
    top=1.2mm,
    bottom=1.2mm,
    before skip=10pt,
    after skip=10pt
  ]%
}
{%
  \end{tcolorbox}%
}

% ...
% ...
% ...
% ...
% ...

\usepackage{bm}  
\DeclareSymbolFont{usualmathcal}{OMS}{cmsy}{m}{n}
\DeclareSymbolFontAlphabet{\mathcal}{usualmathcal}

\fancypagestyle{SPstyle}{
\fancyhf{}
\lhead{\colorbox{scipostblue}{\bf \color{white} ~SciPost Physics Lecture Notes }}
\rhead{{\bf \color{scipostdeepblue} ~Submission }}

\fancyfoot[C]{\textbf{\thepage}}
}

\begin{document}

\pagestyle{SPstyle}

\begin{center}{\Large \textbf{\color{scipostdeepblue}{
Statistical Mechanics of Random Matrices\\
}}}\end{center}

\begin{center}\textbf{
Isaac P\'erez Castillo
}\end{center}

\begin{center}
Departamento de Física, Universidad Autónoma Metropolitana-Iztapalapa,
 San Rafael Atlixco 186, Ciudad de México 09340, México\\
 and\\
Instituto de Ciencias F\'isicas, Universidad Nacional Aut\'onoma de M\'exico (UNAM), Av. Universidad s/n, Col. Chamilpa, CP 62210 Cuernavaca, Mor., Mexico
\\[\baselineskip]
\href{mailto:iperez@izt.uam.mx}{\small iperez@izt.uam.mx}
\end{center}

\section*{\color{scipostdeepblue}{Abstract}}
\boldmath\textbf{
These lecture notes are based on the lectures on \emph{Statistical Mechanics of Random Matrices} delivered at the Spring College on the Physics of Complex Systems, held at the Abdus Salam International Centre for Theoretical Physics, Trieste, Italy, from 19 February to 15 March 2024. Their aim is to present a statistical-mechanics route to the spectral theory of sparse and diluted random matrices, with emphasis on cavity and replica methods, resolvent techniques, population dynamics, typical spectral densities, spectral-count fluctuations and large deviations, conditioned spectra, and non-Hermitian extensions.\\
The written form of the notes has been deliberately expanded beyond the material actually covered during the lectures. This is partly because a set of lecture notes can afford a more systematic development than a sequence of blackboard lectures, and partly because several natural continuations of the material become clearer once the central methods have been introduced. Consequently, not every topic discussed here was presented during the College. The additional material is included to give a more coherent account of the subject and to indicate directions that, hopefully, can be covered in greater detail in future lectures or schools.
Since these are lecture notes rather than a state-of-the-art review, the choice of topics is necessarily selective and is naturally tilted towards the author's own work and collaborations on this subject. I have nevertheless tried, within the limits of this format, to place the material in contact with the broader literature and to represent the surrounding state of the art as fairly as possible. Inevitably, some relevant contributions may be missing or treated too briefly; such omissions are unintentional and reflect the pedagogical scope of the notes rather than a judgement on their importance.
}

\vspace{\baselineskip}
\noindent\textcolor{white!90!black}{%
\fbox{\parbox{0.975\linewidth}{%
\textcolor{white!40!black}{\begin{tabular}{lr}%
  \begin{minipage}{0.6\textwidth}%
    {\small Copyright attribution to authors. \newline
    This work is a submission to SciPost Physics Lecture Notes. \newline
    License information to appear upon publication. \newline
    Publication information to appear upon publication.}
  \end{minipage} & \begin{minipage}{0.4\textwidth}
    {\small Received Date \newline Accepted Date \newline Published Date}%
  \end{minipage}
\end{tabular}}
}}
}

\vspace{10pt}
\noindent\rule{\textwidth}{1pt}
\tableofcontents
\noindent\rule{\textwidth}{1pt}
\vspace{10pt}

\section{Introduction: sparse spectra as a statistical-mechanics problem}
\label{sec:introduction}
Random matrix theory began with the rather simple idea of replacing the intricacies of microscopic complexity by probabilistic structure. In multivariate statistics, Wishart introduced the distribution of random covariance matrices as early as 1928 \cite{Wishart1928}. In numerical analysis, von Neumann and Goldstine used random matrices in the study of large systems of linear equations and matrix inversion, thereby providing another early route through which randomness entered matrix theory \cite{vonNeumannGoldstine1947}. In nuclear physics, Wigner proposed random Hamiltonians as effective models for the energy levels of complex quantum systems \cite{Wigner1955}; and Dyson developed the symmetry classification and the Brownian-motion viewpoint that made the analogy with a gas of interacting eigenvalues explicit \cite{Dyson1962}. These ideas led to a mature theory of dense random matrices, in which global laws, local correlations, eigenvalue spacings, fluctuations of linear statistics, extreme eigenvalues, eigenvectors, and singular-value observables are studied through algebraic, analytic, probabilistic, and Coulomb-gas methods \cite{Mehta2004,Forrester2010,AndersonGuionnetZeitouni2010}. Two limiting laws became especially emblematic at the level of global spectral densities: the Wigner semicircle law for large Hermitian matrices with independent entries and the Mar\v{c}enko--Pastur law for empirical covariance matrices \cite{MarchenkoPastur1967}. In the non-Hermitian setting, the Ginibre ensembles, Girko's circular law, and the method of Hermitization established a parallel tradition for spectra in the complex plane \cite{Ginibre1965,Girko1984,FeinbergZee1997,TaoVuKrishnapur2010}. 

The present notes concern a different, and in many respects less universal, corner of random matrix theory. The matrices considered here are sparse or diluted: only a number of entries proportional to the matrix size is nonzero. Equivalently, the matrix is naturally associated with a graph of finite average degree, or with a sparse bipartite graph in the covariance and Wishart cases. This apparently modest change has major consequences. A dense random matrix averages over many weak interactions at every site while a diluted matrix retains strong local fluctuations in its underlying graph. The spectrum is therefore sensitive not only to the distribution of matrix entries, but also to local degrees, finite motifs, dangling trees, degree--degree correlations, community structure, and directed or asymmetric connectivity. In this regime the standard invariant-ensemble picture is no longer sufficient, because the joint law of eigenvalues is either unavailable or not the most natural object. The central problem then becomes that of finding ways to compute spectral observables directly from the graphical and probabilistic structure of the ensemble.

The first basic object that will concern us is the empirical spectral density, which, without entering yet into detailed definitions, has the form
\begin{equation}
\rho_N(\lambda)=\frac{1}{N}\sum_{i=1}^N\delta(\lambda-\lambda_i)\,,
\end{equation}
for Hermitian or real symmetric matrices, or its non-Hermitian analogue in the complex plane. Here $\delta(x)$ denotes the Dirac delta function. Closely related quantities are the resolvent, local Green functions, eigenvector observables, the number of eigenvalues in an interval or domain, and large-deviation probabilities for atypical spectral samples. In dense invariant ensembles, such questions often admit formulations in terms of orthogonal polynomials, loop equations, determinantal processes, or Coulomb gases. In sparse ensembles, however, the more natural language is that of disordered systems on locally tree-like graphs: Gaussian integral representations, replicas, cavity equations, belief propagation, and population dynamics.

Random graphs provide the simplest geometric support for diluted matrices. In particular, Erd\H{o}s--R\'enyi graphs already contain the essential feature that the local neighborhood of a typical vertex converges to a random tree \cite{ErdosRenyi1959}. More structured network ensembles may be used to introduce fixed degrees, heterogeneous degree distributions, degree correlations, and modularity \cite{Newman2010}. From the spectral point of view, these graph-theoretic distinctions are not cosmetic. As we will see, they directly alter the self-consistency equations for local resolvents and can create spectral tails, isolated eigenvalues, high-density regions, or singular components associated with localized states. This is one reason why the spectral theory of sparse matrices has developed more slowly than the theory of dense invariant ensembles: there is no single universal law that replaces the dense-ensemble benchmarks while also capturing the graph-local information. In my opinion, this is precisely what makes the problem technically demanding and conceptually interesting at the same time.

A statistical-mechanics route to spectra was opened by Edwards and Jones, who represented the spectral density of a large symmetric random matrix through a Gaussian integral and a disorder-averaged logarithm \cite{EdwardsJones1976}. This approach is natural because the same structure appears in the computation of quenched free energies in spin glasses. The logarithm is the source of the main difficulty, and it naturally invites either the replica method or a cavity formulation. The early replica treatment of sparse random matrices by Rodgers and Bray was a seminal step in this direction \cite{RodgersBray1988,BrayRodgers1988}. Subsequent work clarified several aspects of the sparse problem, including localized states, finite-connectivity corrections, and approximate or exact recursive formulations on locally tree-like structures \cite{BiroliMonasson1999,SemerjianCugliandolo2002,Kuhn2008}. The modern pedagogical review of Susca, Vivo, and K\"uhn gives a detailed account of this line for sparse symmetric matrices and makes explicit the equivalence between cavity and replica viewpoints in that setting \cite{SuscaVivoKuhn2021}.

There is also an important rigorous literature on spectra of sparse random graphs and their resolvents. Local weak convergence and resolvent methods have made it possible to connect finite random graphs with limiting infinite random trees \cite{BordenaveLelarge2010}; for random regular graphs, refined results describe spectral densities, eigenvectors, and local laws in regimes where the degree is fixed or grows with the system size \cite{DumitriuPal2012,BauerschmidtHuangYau2019}. These developments are not the main technical route followed in these notes, which are written from a theoretical-physics viewpoint, but they are part of the same broad attempt to understand what survives, and what fails, when random matrix theory is placed on a sparse graph.

The methods used here are inherited from the statistical mechanics of disordered systems, where probabilistic ideas are combined with variational, graphical, and asymptotic methods to study systems with quenched disorder. The replica method, the cavity method, and the Bethe approximation were developed to describe systems with quenched disorder and many interacting degrees of freedom \cite{MezardParisiVirasoro1987,MezardMontanari2009}. In the present context, as will be shown below, the ``spins'' are continuous Gaussian variables introduced by the determinant or resolvent representation of the spectral observable. For locally tree-like matrices, the cavity method gives recursive equations for single-site or single-edge Gaussian marginals, the so-called cavity equations. At the level of a single large matrix, these equations are belief-propagation equations for local resolvents. At the ensemble level, they become distributional self-consistency equations, which are typically solved by population dynamics. This correspondence between random matrices, Gaussian graphical models, and disordered systems is the organizing principle of the notes.

A first objective is to understand typical spectral densities. For sparse symmetric random matrices and sparse covariance matrices, the cavity method gives closed equations for the spectral density and recovers the Wigner and Mar\v{c}enko--Pastur laws in appropriate dense limits \cite{RogersTakedaPerezCastilloKuhn2008}. The same formalism can be adapted to graph ensembles with topological constraints, such as degree--degree correlations and community structure, where the spectral density reflects the imposed graph statistics \cite{RogersPerezVicenteTakedaPerezCastillo2010}. Sparse covariance and Wishart-type ensembles require a bipartite formulation, since the matrix of interest is built from rectangular data matrices. Earlier work on sparse sample covariance matrices and later work on products of diluted Wishart matrices show how the covariance structure connects the finite-connectivity problem with classical dense limits \cite{NagaoTanaka2007,DupicPerezCastillo2014}.

A second objective is to treat non-Hermitian sparse matrices. Non-Hermiticity appears whenever interactions are directed, asymmetric, dissipative, or out of equilibrium. In dense non-Hermitian random matrix theory, Hermitization and logarithmic-potential methods provide the standard route to the complex spectral density \cite{Ginibre1965,Girko1984,FeinbergZee1997}. In the sparse case, one must combine Hermitization with graph-local recursion. The cavity approach to non-Hermitian sparse matrices gives equations for the spectral density in the complex plane and recovers generalized dense laws in suitable limits \cite{RogersPerezCastillo2009}. Subsequent analytical work and reviews developed this sparse non-Hermitian direction further, including exact solutions and outlier theory for locally tree-like non-Hermitian ensembles \cite{NeriMetz2012,MetzNeriRogers2019}.

A third objective is to go beyond the average spectral density. Spectral counting observables, such as the number of eigenvalues in an interval, probe fluctuations of the spectrum at a level not captured by the typical density. For invariant ensembles, such observables can be naturally related to Coulomb-gas large deviations and index distributions \cite{DeanMajumdar2006,MajumdarNadalScardicchioVivo2009}, a viewpoint that is especially useful when the joint eigenvalue density is available. For sparse matrices, however, the absence or weakness of eigenvalue repulsion changes the structure of the large-deviation problem. A replica method for the number of eigenvalues of sparse random graphs inside an interval yields the corresponding rate function and reveals features that differ from the rotationally invariant case \cite{MetzPerezCastillo2016}. Applied to the Anderson model on random regular graphs, the same framework connects spectral counting to level compressibility and localization questions \cite{Anderson1958,MetzPerezCastillo2017}. For diluted Wishart matrices, the analogous theory gives the cumulant-generating function, cumulants, and rate function for the number of eigenvalues below a threshold \cite{PerezCastilloMetz2018Wishart}.

A fourth objective is to understand conditioned spectra. Once the number of eigenvalues in a region is fixed to be atypical, one may ask what the entire spectral density looks like under that constraint. For invariant ensembles this problem again has a natural Coulomb-gas interpretation, because the eigenvalues behave as strongly repelling charges. For sparse and, more generally, non-invariant ensembles the situation is qualitatively different: the conditioned density need not display the same accumulation phenomena familiar from invariant models. The theory for conditioned spectral densities of non-invariant random matrices provides a way to compute such atypical spectra directly, and shows explicitly how sparse ensembles depart from the classical picture \cite{PerezCastilloMetz2018Conditioned}. In the non-Hermitian setting, the analogous counting problem concerns the number of eigenvalues inside a contour in the complex plane; an analytic path-integral approach gives the cumulant-generating function and large-deviation rate function for this observable even when the joint eigenvalue density is not known \cite{RamosSanchezGuzmanGonzalezPerezCastilloMetz2021}.

A further part of this program concerns generalized diluted Wishart and cross-correlation ensembles. These ensembles are motivated by the fact that empirical covariance matrices in data analysis are often both high-dimensional and sparse: the observed rectangular data matrix may itself be supported on a sparse bipartite graph. The generalized diluted Wishart ensemble introduces two diluted rectangular matrices with correlated entries and studies the symmetric cross-correlation matrix built from them. This model interpolates between several known limits by varying the dilution, rectangularity, and entry correlations, and it provides a compact setting in which sparse graph structure and covariance structure coexist \cite{PerezCastillo2022Generalized}. A non-Hermitian extension adds an asymmetry parameter and brings the diluted Wishart program into contact with the non-Hermitian spectral theory of sparse matrices \cite{GuzmanGonzalezPerezCastillo2025}.

These notes are based on lectures delivered at the Spring College on the Physics of Complex Systems, held at the Abdus Salam International Centre for Theoretical Physics in Trieste from 19 February to 15 March 2024. The written version is broader than the material covered during the lectures. This expansion is intentional. The lectures were constrained by time and by the pedagogical need to introduce the replica and cavity methods before applying them to spectral problems. The present text uses the lecture format as a starting point, but develops the material into a more systematic pedagogical account of statistical-mechanics methods for diluted random matrices. Some of the topics included here were only mentioned briefly, or not presented at all, during the College; they are included because they complete the conceptual line from typical densities to fluctuations, conditioned spectra, and non-Hermitian extensions.

The viewpoint adopted here is therefore deliberately selective. This is not a general monograph on random matrix theory, nor a complete mathematical review of sparse random graphs. It is a set of lecture notes on how methods from the statistical mechanics of disordered systems can be used to compute spectral observables of sparse and diluted random matrices. The guiding idea is that a sparse random matrix may be treated as a graphical model. Its resolvent is a local observable of an associated Gaussian field; its spectral density follows from cavity marginals; its fluctuations are encoded in replicated or biased ensembles; and its non-Hermitian extensions can be handled by combining Hermitization with the same graphical machinery.

The preceding paragraphs should also be read as a guide to the structure of the notes. The path is methodological rather than encyclopedic. We begin with the graph and matrix ensembles, then introduce spectral observables and rewrite them as Gaussian disordered systems. We then derive local equations on sparse graphs and discuss how to solve the corresponding distributional equations by population dynamics. The same circle of ideas is then revisited in progressively richer settings: covariance matrices, topologically constrained graphs, non-Hermitian matrices, spectral-count fluctuations, conditioned ensembles, and diluted Wishart-type extensions. Each revisit is not merely a reference back to the first derivation, but a controlled reintroduction of the method in a new setting, emphasizing which part of the construction is unchanged and which part is modified by the new ensemble or observable. In this sense the later sections are not independent review chapters, but pedagogical variations on a common statistical-mechanics route to sparse spectra.

Several topics discussed below are closely connected to work by the author and collaborators on sparse spectra, diluted Wishart matrices, spectral-count large deviations, conditioned densities, generalized cross-correlation ensembles, and non-Hermitian extensions. This personal emphasis is natural in lecture notes: the selection is tilted towards problems and methods that I know well, rather than towards an exhaustive survey of the field. I have nevertheless tried, within this format, to place the material in contact with the surrounding literature and to represent the state of the art as fairly as possible. Inevitably, some relevant contributions will be missing or treated too briefly; such omissions are unintentional and should be understood as limitations of the pedagogical scope of the notes.

At the same time, the presentation does not attempt to reproduce the original papers verbatim. The lecture-note format gives an opportunity to rederive familiar results from a unified point of view, sometimes with different intermediate steps, notation, or emphasis. Part of the motivation was also personal: rederiving familiar results by different routes is one of the most useful ways of understanding what the methods are really doing. The aim is therefore not only to summarize individual results, but to expose the repeated structure behind them: graphical disorder, Gaussian integral representations, cavity marginals, population dynamics, and biased ensembles for fluctuations or conditioning.

These notes intentionally retain a certain amount of repetition. The same objects---resolvents, local Green functions, Gaussian integral representations, Hermitization, cavity messages, belief propagation, and population dynamics---are reintroduced in several ensembles, rather than only recalled by referring back to earlier formulas. This is not only a stylistic choice but a pedagogical one. Repetition helps the reader become familiar with the mathematical objects, while each recurrence also makes clear what changes, and what remains invariant, when one passes from sparse symmetric matrices to covariance matrices, non-Hermitian matrices, large deviations, and conditioned spectra. The repetition is therefore meant to be controlled rather than mechanical: full derivations are given when a new mechanism appears, while later appearances are treated as reminders, adaptations, or consistency checks.

The notes are organized accordingly. We first introduce random graphs and diluted matrix ensembles, then spectral observables and their Gaussian integral representations. We next review the replica and cavity methods in the minimal form needed for spectral calculations, together with belief propagation and population dynamics. The central sections treat sparse symmetric matrices, graph ensembles with topological constraints, sparse covariance and diluted Wishart matrices, products of Wishart matrices, non-Hermitian sparse matrices, and generalized diluted Wishart and cross-correlation ensembles. The later sections address spectral-count large deviations, diluted Wishart large deviations, conditioned spectral densities, and non-Hermitian number statistics. The appendices collect technical material on Gaussian integrals and resolvent conventions, dense-limit reductions, population-dynamics algorithms, and replica-symmetric saddle-point derivations. Examples and exercises are used throughout as part of the pedagogical structure, both to test the formal manipulations and to make the recurring construction explicit in concrete cases.

\begin{exerciseblock}
\exitem[Historical reading: how randomness entered matrix theory]
Choose three or four historical references among
\cite{Wishart1928,vonNeumannGoldstine1947,Wigner1955,Dyson1962,Ginibre1965,Girko1984,EdwardsJones1976,RodgersBray1988}. Read them with the following question in mind: what problem was randomness introduced to solve? For each reference, identify the matrix ensemble or class of matrices being considered, the main quantity of interest, and the method or physical picture used by the authors.

Write a short essay, of about one or two pages, comparing the role of randomness in these works. Your discussion may be organized around questions such as: Is the matrix used as a statistical model for data, as an effective Hamiltonian, as a numerical test object, as a gas of interacting eigenvalues, as a non-Hermitian ensemble, or as a disordered system? What is the spectral question being asked? Is the emphasis on global densities, level statistics, stability of linear systems, eigenvalue correlations, or a field-theoretic representation of the resolvent?

As a final part of the exercise, draw a concept map summarizing your essay. The map should not try to be exhaustive; it should simply show how different meanings of randomness in matrix theory emerged from different problems: statistics, numerical linear algebra, nuclear spectra, eigenvalue gases, non-Hermitian spectra, and disordered systems.

\exitem[Dense versus sparse scaling]
A diluted matrix has only a number of nonzero entries proportional to the matrix size. Let $\pmb C$ be the $N\times N$ adjacency matrix of a simple undirected random graph with vertex set $\{1,\dots,N\}$, so that $C_{ij}=1$ if vertices $i$ and $j$ are connected, $C_{ij}=0$ otherwise, $C_{ij}=C_{ji}$, and $C_{ii}=0$. Suppose that, for $i<j$,
\begin{equation}
{\rm Prob}(C_{ij}=1)=\frac{c}{N}\,,\qquad{\rm Prob}(C_{ij}=0)=1-\frac{c}{N}\,,
\label{eq:intro-ex-er-probability}
\end{equation}
with fixed $c=O(1)$ as $N\to\infty$. Show that the expected number of nonzero ordered off-diagonal matrix entries is
\begin{equation}
\mathbb E\left[\sum_{i\neq j} C_{ij}\right]=c(N-1)\,,
\label{eq:intro-ex-nonzero-entries}
\end{equation}
and that the expected degree of a fixed vertex is $c+O(N^{-1})$. Explain why this scaling is fundamentally different from a dense matrix whose entries are generically all nonzero.

\exitem[Why local graph structure survives in the spectrum]
One of the key claims of the introduction is that sparse spectra remain sensitive to local degrees and motifs. Let $\pmb C$ be the $N\times N$ adjacency matrix of a simple undirected graph with vertex set $\{1,\dots,N\}$, so that $C_{ij}=1$ if vertices $i$ and $j$ are connected, $C_{ij}=0$ otherwise, $C_{ij}=C_{ji}$, and $C_{ii}=0$. Show that
\begin{equation}
{\rm Tr}\pmb C^2=\sum_{i=1}^{N} k_i\,,
\label{eq:intro-ex-second-moment}
\end{equation}
where $k_i$ is the degree of vertex $i$. Then show that
\begin{equation}
{\rm Tr}\pmb C^3=6T\,,
\label{eq:intro-ex-third-moment}
\end{equation}
where $T$ is the number of triangles in the graph. Explain how these two identities illustrate the statement in the introduction that sparse spectra are sensitive not only to average connectivity but also to finite motifs.

\exitem[Why locally tree-like graphs are natural]
Let $\pmb C$ be the $N\times N$ adjacency matrix of an Erd\H{o}s--R\'enyi random graph on the vertex set $\{1,\dots,N\}$, where each unordered edge $\{i,j\}$, with $i<j$, is present independently with probability $c/N$, and where $c=O(1)$ as $N\to\infty$. Consider a fixed vertex $i$. Show that the expected number of triangles containing $i$ is
\begin{equation}
\binom{N-1}{2}\left(\frac{c}{N}\right)^3\,.
\label{eq:intro-ex-triangle-count}
\end{equation}
Deduce that this quantity vanishes as $N\to\infty$ for fixed $c$. Explain why this supports, at least heuristically, the use of cavity recursions on locally tree-like neighborhoods.

\exitem[The observables introduced in the section]
Let $\pmb A$ be an $N\times N$ Hermitian or real symmetric matrix with real eigenvalues $\lambda_1,\dots,\lambda_N$. Starting from the empirical spectral density
\begin{equation}
\rho_N(\lambda)=\frac{1}{N}\sum_{i=1}^{N}\delta(\lambda-\lambda_i)\,,
\label{eq:intro-ex-empirical-density}
\end{equation}
show that
\begin{equation}
\frac{1}{N}\mathcal N_{[a,b]}=\int_a^b d\lambda\, \rho_N(\lambda)\,,
\label{eq:intro-ex-interval-count}
\end{equation}
where $\mathcal N_{[a,b]}$ is the number of eigenvalues in the interval $[a,b]$. Assume for simplicity that no eigenvalue lies exactly at the endpoints $a$ and $b$. Then define the non-Hermitian empirical density
\begin{equation}
\rho_N^{(2)}(z)=\frac{1}{N}\sum_{i=1}^{N}\delta^{(2)}(z-z_i)
\end{equation}
for an $N\times N$ non-Hermitian matrix with complex eigenvalues $z_1,\dots,z_N$, and show that
\begin{equation}
\frac{1}{N}\mathcal N_{\mathcal D}=\int_{\mathcal D} d^2z \rho_N^{(2)}(z)
\end{equation}
counts the fraction of eigenvalues in a domain $\mathcal D\subset\mathbb C$, up to the usual boundary convention. The purpose of the exercise is to recognize that all these objects are related from the very beginning.

\exitem[Why the logarithm is the real difficulty]
Let $\pmb A$ be an $N\times N$ real symmetric matrix, let $\pmb I$ be the $N\times N$ identity matrix, and let $z=\lambda-i\epsilon$, with $\epsilon>0$, or more generally $\operatorname{Im}z<0$, so that the Gaussian integral below is understood with the convergence prescription used throughout these notes. Define
\begin{equation}
Z_{\pmb A}(z)=\int\left[\prod_{i=1}^{N}\frac{du_i}{\sqrt{2\pi}}\right]\exp\left[-\frac{i}{2}\pmb u^{\rm T}(z\pmb I-\pmb A)\pmb u\right]\,,
\label{eq:intro-ex-ej-partition}
\end{equation}
where $\pmb u=(u_1,\dots,u_N)^{\rm T}$. Show formally that
\begin{equation}
\frac{\partial}{\partial z}\log\det(z\pmb I-\pmb A)={\rm Tr} (z\pmb I-\pmb A)^{-1}\,.
\label{eq:intro-ex-logdet-derivative}
\end{equation}
Then explain why the average spectral density involves $\overline{\log Z_{\pmb A}(z)}$ rather than $\log \overline{Z_{\pmb A}(z)}$, and why this is the point at which replica or cavity methods enter naturally.

\exitem[Invariant versus non-invariant conditioned spectra]
Conditioned spectral densities behave differently in sparse non-invariant ensembles than in invariant ensembles. Explain why the invariant case naturally suggests a Coulomb-gas picture, while the sparse case naturally suggests a graphical or cavity-based picture. Your answer should use only ideas already stated in the introduction and should make clear which feature of the ensemble is responsible for the difference.

\exitem[Why sparse covariance matrices are bipartite objects]
Let $\pmb X$ be an $N\times P$ rectangular matrix with entries $X_i^\mu$, where $i=1,\dots,N$ and $\mu=1,\dots,P$. Define the $N\times N$ matrix
\begin{equation}
\pmb W=\frac{1}{P}\pmb X\pmb X^{\rm T}\,.
\label{eq:intro-ex-wishart}
\end{equation}
Here the factor $1/P$ is the conventional covariance normalization used for this introductory exercise; later diluted Wishart sections will use the finite-connectivity normalization $1/d$. Show that for every vector $\pmb v\in\mathbb R^N$,
\begin{equation}
\pmb v^{\rm T}\pmb W\pmb v=\frac{1}{P}\sum_{\mu=1}^{P}\left(\sum_{i=1}^{N}X_i^\mu v_i\right)^2\geq 0\,.
\label{eq:intro-ex-psd}
\end{equation}
Then show that
\begin{equation}
{\rm rank}\pmb W\leq P\,.
\label{eq:intro-ex-rank-bound}
\end{equation}
Explain why these two elementary facts already suggest that sparse covariance matrices are better thought of in terms of a rectangular data matrix and a bipartite graph than in terms of an ordinary square sparse matrix.

\exitem[Programming exercise: reading the introduction numerically]
Fix a mean degree $c=O(1)$, a list of matrix sizes $N$, and a number $S$ of independent samples for each $N$. For each value of $N$, generate numerically two ensembles. For the dense ensemble, let $\pmb X$ be an $N\times N$ real symmetric random matrix whose independent entries on and above the diagonal are distributed as
\begin{equation}
X_{ij}=X_{ji}\sim\mathcal N(0,1)\quad (i<j)\,,\qquad
X_{ii}\sim\mathcal N(0,2)\quad (i=1,\dots,N)\,,
\label{eq:intro-ex-dense-random-entries}
\end{equation}
and set
\begin{equation}
\pmb H_{\rm dense}=\frac{1}{\sqrt{N}}\pmb X\,.
\label{eq:intro-ex-dense-program}
\end{equation}
For the sparse ensemble, let $\pmb C$ be the $N\times N$ adjacency matrix of an Erd\H{o}s--R\'enyi random graph on the vertex set $\{1,\dots,N\}$, with $C_{ii}=0$, $C_{ij}=C_{ji}$, and independent upper-triangular entries satisfying
\begin{equation}
{\rm Prob}(C_{ij}=1)=\frac{c}{N}\,,\qquad
{\rm Prob}(C_{ij}=0)=1-\frac{c}{N}
\quad (i<j)\,,
\label{eq:intro-ex-sparse-random-entries}
\end{equation}
where $c=O(1)$ as $N\to\infty$. Set
\begin{equation}
\pmb H_{\rm sparse}=\frac{1}{\sqrt{c}}\pmb C\,.
\label{eq:intro-ex-sparse-program}
\end{equation}
Use the same eigenvalue grid or histogram bins for the two ensembles at fixed $N$, and report the values of $c$, $N$, $S$, and the bin width or smoothing parameter used in the comparison. Compare their eigenvalue histograms for increasing $N$. Then compute numerically:
\begin{equation}
\frac{1}{N}{\rm Tr}\pmb H_{\rm sparse}^2\,,\qquad \frac{1}{N}{\rm Tr}\pmb H_{\rm sparse}^3\,,
\label{eq:intro-ex-program-moments}
\end{equation}
and relate your observations to the dense-versus-sparse contrast stated in the introduction.
\end{exerciseblock}

\section{Random graphs and diluted matrix ensembles}
\label{sec:random-graphs-diluted-matrices}
The matrices studied in these notes are sparse in the strong sense that the number of nonzero entries grows linearly, rather than quadratically, with the matrix size. This is the defining feature of a diluted random matrix. It is useful to separate two pieces of randomness. The first is a graph, or in some cases a hypergraph or bipartite graph, which specifies which entries are present. The second is a collection of random weights placed on the occupied entries. In dense random matrix theory the graph is essentially complete and therefore invisible. In diluted ensembles the graph remains visible in the thermodynamic limit, and its local structure becomes part of the spectral problem.

A graph $\mathcal{G}=(V,E)$ is specified by a vertex set $V=\{1,\ldots,N\}$ and an edge set $E$. For a simple undirected graph, the edges are unordered pairs $\{i,j\}$ with $i\neq j$; equivalently, one may specify only the pairs with $i<j$. For a directed graph, the edges are ordered pairs $(i,j)$, and the presence of $(i,j)$ does not imply the presence of $(j,i)$. The connectivity information is encoded in the $N\times N$ adjacency matrix $\pmb C$. For a simple unweighted undirected graph without self-edges, its entries are
\begin{equation}
 C_{ij}=
\begin{cases}
1, & \{i,j\}\in E\,,\\
0, & \{i,j\}\notin E\,,
\end{cases}
\qquad i\neq j,
\end{equation}
together with $C_{ii}=0$ and $C_{ij}=C_{ji}$. For a directed graph one instead uses
\begin{equation}
 C_{ij}=
\begin{cases}
1, & (i,j)\in E\,,\\
0, & (i,j)\notin E\,,
\end{cases}
\end{equation}
so that, in general, $C_{ij}$ and $C_{ji}$ are separate pieces of information. We shall mostly exclude self-edges and impose $C_{ii}=0$. Diagonal terms will be introduced separately when needed, so that self-loops do not have to be treated as part of the graph support.

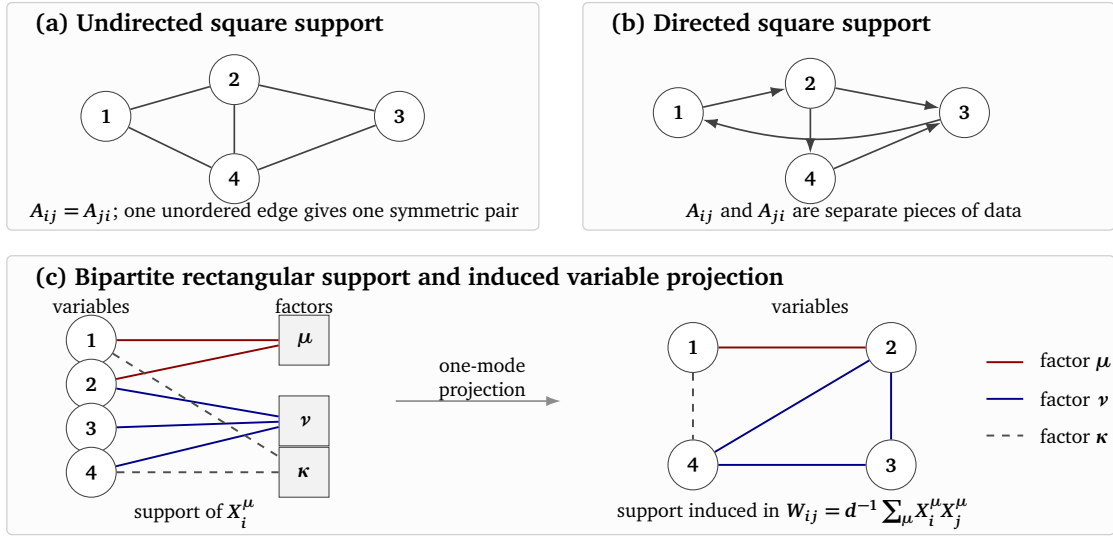
\begin{figure}[t]
\centering
\resizebox{0.98\textwidth}{!}{%
\begin{tikzpicture}[
    x=1cm,
    y=1cm,
    >=Latex,
    panel/.style={draw=black!18, fill=black!1, rounded corners=2pt, line width=0.5pt},
    vnode/.style={circle, draw=black!75, fill=white, minimum size=6.8mm, inner sep=0pt, font=\scriptsize},
    fnode/.style={rectangle, draw=black!75, fill=black!5, minimum size=6.8mm, inner sep=0pt, font=\scriptsize},
    edge/.style={draw=black!75, line width=0.65pt},
    arrowedge/.style={edge, -{Latex[length=2.0mm,width=1.4mm]}},
    muedge/.style={draw=red!55!black, line width=0.75pt},
    nuedge/.style={draw=blue!55!black, line width=0.75pt},
    kappaedge/.style={draw=black!70, line width=0.75pt, dashed},
    ptitle/.style={font=\bfseries\small, anchor=west},
    paneltext/.style={font=\scriptsize, align=center},
    legend/.style={font=\scriptsize, anchor=west}
]
\draw[panel] (0,4.15) rectangle (7.25,7.25);
\node[ptitle] at (0.25,6.95) {(a) Undirected square support};

\node[vnode] (u1) at (1.35,5.70) {$1$};
\node[vnode] (u2) at (3.10,6.20) {$2$};
\node[vnode] (u3) at (5.35,5.70) {$3$};
\node[vnode] (u4) at (3.10,4.85) {$4$};

\draw[edge] (u1) -- (u2);
\draw[edge] (u2) -- (u3);
\draw[edge] (u3) -- (u4);
\draw[edge] (u4) -- (u1);
\draw[edge] (u2) -- (u4);

\node[paneltext] at (3.65,4.38)
{$A_{ij}=A_{ji}$; one unordered edge gives one symmetric pair};

\draw[panel] (7.85,4.15) rectangle (15.10,7.25);
\node[ptitle] at (8.10,6.95) {(b) Directed square support};

\node[vnode] (d1) at (9.15,5.75) {$1$};
\node[vnode] (d2) at (10.95,6.15) {$2$};
\node[vnode] (d3) at (13.05,5.75) {$3$};
\node[vnode] (d4) at (10.95,4.85) {$4$};

\draw[arrowedge] (d1) -- (d2);
\draw[arrowedge] (d2) -- (d3);
\draw[arrowedge] (d3) to[bend left=16] (d1);
\draw[arrowedge] (d2) -- (d4);
\draw[arrowedge] (d4) -- (d3);

\node[paneltext] at (11.55,4.38)
{$A_{ij}$ and $A_{ji}$ are separate pieces of data};

\draw[panel] (0,0.00) rectangle (15.10,3.80);
\node[ptitle] at (0.25,3.50)
{(c) Bipartite rectangular support and induced variable projection};

\node[paneltext] at (1.15,3.12) {variables};
\node[paneltext] at (4.05,3.12) {factors};
\node[paneltext] at (10.95,3.12) {variables};

\node[vnode] (v1) at (1.15,2.65) {$1$};
\node[vnode] (v2) at (1.15,2.05) {$2$};
\node[vnode] (v3) at (1.15,1.45) {$3$};
\node[vnode] (v4) at (1.15,0.85) {$4$};

\node[fnode] (muf) at (4.05,2.65) {$\mu$};
\node[fnode] (nuf) at (4.05,1.55) {$\nu$};
\node[fnode] (kapf) at (4.05,0.85) {$\kappa$};

\draw[muedge] (v1) -- (muf);
\draw[muedge] (v2) -- (muf);

\draw[nuedge] (v2) -- (nuf);
\draw[nuedge] (v3) -- (nuf);
\draw[nuedge] (v4) -- (nuf);

\draw[kappaedge] (v1) -- (kapf);
\draw[kappaedge] (v4) -- (kapf);

\draw[arrowedge, draw=black!45] (5.30,1.82) -- (7.55,1.82);
\node[paneltext] at (6.48,2.12) {one-mode\\ projection};

\node[vnode] (p1) at (9.35,2.55) {$1$};
\node[vnode] (p2) at (12.05,2.55) {$2$};
\node[vnode] (p3) at (12.05,0.95) {$3$};
\node[vnode] (p4) at (9.35,0.95) {$4$};

\draw[muedge] (p1) -- (p2);

\draw[nuedge] (p2) -- (p3);
\draw[nuedge] (p3) -- (p4);
\draw[nuedge] (p2) -- (p4);

\draw[kappaedge] (p1) -- (p4);

\node[paneltext] at (2.55,0.28)
{support of $X_i^\mu$};

\node[paneltext] at (10.70,0.28)
{support induced in $W_{ij}=d^{-1}\sum_\mu X_i^\mu X_j^\mu$};

\draw[muedge] (13.35,2.35) -- (13.85,2.35);
\node[legend] at (13.95,2.35) {factor $\mu$};

\draw[nuedge] (13.35,1.85) -- (13.85,1.85);
\node[legend] at (13.95,1.85) {factor $\nu$};

\draw[kappaedge] (13.35,1.35) -- (13.85,1.35);
\node[legend] at (13.95,1.35) {factor $\kappa$};
\end{tikzpicture}%
}
\caption{Schematic support conventions for diluted matrix ensembles. In the bipartite case, the one-mode projection shows how factor nodes generate induced variable--variable couplings.}
\label{fig:rgdme-support-conventions}
\end{figure}

For an undirected graph, the degree of vertex $i$, namely the number of neighbours of $i$, is
\begin{equation}
k_i=\sum_{j=1}^N C_{ij}\,,
\end{equation}
and the empirical degree distribution is
\begin{equation}
p_N(k)=\frac{1}{N}\sum_{i=1}^N\delta_{k,k_i}\,.
\end{equation}
For directed graphs one must distinguish the in-degree and out-degree,
\begin{equation}
k_i^{\rm in}=\sum_{j=1}^N C_{ji}\,,\qquad k_i^{\rm out}=\sum_{j=1}^N C_{ij}\,.
\end{equation}
These elementary definitions already indicate why sparse random matrices are graph-theoretic objects: in a square graph-supported matrix, a row contains only vertices connected to the corresponding vertex, with outgoing neighbours playing this role in the directed case. Figure~\ref{fig:rgdme-support-conventions} fixes the support conventions used in the examples and ensembles below.

\begin{examplebox}[One sparse support, three matrix problems]
Consider first the undirected graph with vertex set $V=\{1,2,3\}$ and unordered edge set
\begin{equation}
E=\{\{1,2\},\{2,3\}\}\,.
\label{eq:rgdme-example-undirected-edges}
\end{equation}
Its adjacency matrix is
\begin{equation}
\pmb C_{\rm und}=
\begin{pmatrix}
0 & 1 & 0\\
1 & 0 & 1\\
0 & 1 & 0
\end{pmatrix}\,,
\label{eq:rgdme-example-undirected-adjacency}
\end{equation}
and the degrees are
\begin{equation}
k_1=1\,,\qquad k_2=2\,, \qquad k_3=1\,.
\label{eq:rgdme-example-undirected-degrees}
\end{equation}
If we attach weights $J_{12}=a$ and $J_{23}=b$, and diagonal terms $D_1,D_2,D_3$, then the associated symmetric diluted matrix is
\begin{equation}
\pmb A_{\rm sym}=
\begin{pmatrix}
D_1 & a & 0\\
a & D_2 & b\\
0 & b & D_3
\end{pmatrix}\,.
\label{eq:rgdme-example-symmetric-matrix}
\end{equation}
Thus an undirected graph naturally produces a real symmetric matrix problem.

Now orient the same support by keeping the arrows
\begin{equation}
1\to 2\,, \qquad 2\to 3\,,
\label{eq:rgdme-example-directed-edges}
\end{equation}
with no reverse links. The directed adjacency matrix becomes
\begin{equation}
\pmb C_{\rm dir}=
\begin{pmatrix}
0 & 1 & 0\\
0 & 0 & 1\\
0 & 0 & 0
\end{pmatrix}\,,
\label{eq:rgdme-example-directed-adjacency}
\end{equation}
so that
\begin{equation}
k_1^{\rm out}=1,\quad k_2^{\rm out}=1,\quad k_3^{\rm out}=0,\qquad k_1^{\rm in}=0,\quad k_2^{\rm in}=1,\quad k_3^{\rm in}=1\,.
\label{eq:rgdme-example-in-out-degrees}
\end{equation}
If the nonzero weights are $A_{12}=a$ and $A_{23}=b$, then
\begin{equation}
\pmb A_{\rm nh}=
\begin{pmatrix}
0 & a & 0\\
0 & 0 & b\\
0 & 0 & 0
\end{pmatrix}\,,
\label{eq:rgdme-example-nonhermitian-matrix}
\end{equation}
which is already non-Hermitian.

Finally, consider the sparse rectangular matrix
\begin{equation}
\pmb X=
\begin{pmatrix}
1 & 0\\
0 & 1\\
1 & 1
\end{pmatrix}\,.
\label{eq:rgdme-example-rectangular-matrix}
\end{equation}
It defines a bipartite graph with variable nodes $i=1,2,3$ and factor nodes $\mu=1,2$, where $X_i^\mu\neq 0$ means that variable $i$ is connected to factor $\mu$. Without any normalization, the associated covariance-type matrix is
\begin{equation}
\pmb W=\pmb X\pmb X^{\rm T}=
\begin{pmatrix}
1 & 0 & 1\\
0 & 1 & 1\\
1 & 1 & 2
\end{pmatrix}\,.
\label{eq:rgdme-example-wishart-matrix}
\end{equation}
This matrix is symmetric and positive semidefinite, but it is not the adjacency matrix of the original bipartite graph. Instead, two variables interact in $\pmb W$ when they share a common factor node. This is the basic mechanism behind sparse covariance and diluted Wishart ensembles.

The lesson of the example is simple. The same sparse-support language leads naturally to three different matrix problems: symmetric matrices on undirected graphs, non-Hermitian matrices on directed graphs, and covariance-type matrices on bipartite graphs.
\end{examplebox}

The simplest ensemble is the Erd\H{o}s--R\'enyi graph. In the sparse regime one connects each unordered pair of vertices independently with probability $c/N$, where $c=O(1)$ as $N\to\infty$ \cite{ErdosRenyi1959,Bollobas2001}\footnote{A brief remark on terminology is in order. In some older literature, especially on diluted neural networks, the word ‘dilution’ was also used for connectivities that still diverge with system size, for instance when the expected number of neighbors grows like $\log N$. Equivalently, the connection probability may vanish as $(\log N)/N$ while the mean degree still diverges. In these notes, by contrast, ‘sparse’ or ‘diluted’ will always mean finite connectivity in the thermodynamic limit, that is, an expected degree of order $O(1)$.}. Its probability measure can be written as
\begin{equation}
P_{\rm ER}(\pmb{C})=\prod_{i<j}\left\{\left[\frac{c}{N}\delta_{C_{ij},1}+\left(1-\frac{c}{N}\right)\delta_{C_{ij},0}\right]\delta_{C_{ji},C_{ij}}\right\}\prod_{i=1}^N\delta_{C_{ii},0}\,.
\label{eq:ER-graph-measure}
\end{equation}
With this scaling the expected degree remains finite, so the graph has $O(N)$ edges rather than $O(N^2)$. Notice that this ensemble defines the entries of the adjacency matrix--sometimes also referred to as the connectivity matrix--as Bernoulli random variables. For this reason, the probability of a vertex $i$ having a degree $k$ follows the binomial distribution,
\begin{equation}
{\rm Prob}(k_i=k)=\binom{N-1}{k}\left(\frac{c}{N}\right)^k\left(1-\frac{c}{N}\right)^{N-1-k}\,,
\end{equation}
and therefore in the limit of an infinitely large graph ($N\to\infty$, the thermodynamic limit in statistical mechanics) we obtain
\begin{equation}
{\rm Prob}(k_i=k)\longrightarrow e^{-c}\frac{c^k}{k!}\,.
\label{eq:ER-degree-poisson}
\end{equation}
Thus the sparse Erd\H{o}s--R\'enyi graph has a Poisson degree distribution of mean $c$. This is why sparse Erd\H{o}s--R\'enyi graphs are often called Poissonian graphs. They are also locally tree-like: if one chooses a vertex uniformly at random and inspects only its neighborhood up to some fixed graph distance $r$, keeping $r$ fixed as $N\to\infty$, then this neighborhood typically looks like a finite random tree. The root has approximately ${\rm Poisson}(c)$ neighbours, each vertex reached from the root has independently approximately ${\rm Poisson}(c)$ further neighbours beyond the edge through which it was reached, and the probability of seeing a cycle within distance $r$ tends to zero. Formally, one says that the neighborhood of a uniformly chosen vertex converges in distribution to a Galton--Watson tree with Poisson offspring distribution, that is, a random rooted tree generated by letting each vertex independently produce a ${\rm Poisson}(c)$ number of children. This local tree property is the reason why cavity and message-passing recursions provide asymptotically exact descriptions of many typical local observables in the thermodynamic limit. It is also the reason why the Erd\H{o}s--R\'enyi ensemble will serve throughout these notes---as it has done throughout much of the literature---as the canonical reference point of a diluted matrix support.

A second important ensemble is the ensemble of random regular graphs. In a $c$-regular graph every vertex has degree exactly $c$. Formally, the uniform measure over simple $c$-regular graphs can be written as
\begin{equation}
P_{\rm RRG}(\pmb C)=\frac{1}{Z_{\rm RRG}}\prod_{i=1}^N\delta_{\sum_{j=1}^N C_{ij},c}\prod_{i<j}\delta_{C_{ij},C_{ji}}\prod_{i=1}^N\delta_{C_{ii},0}\,,
\label{eq:RRG-measure}
\end{equation}
where $Z_{\rm RRG}$ is the normalization factor, or, equivalently, the number of simple $c$-regular graphs on $N$ labelled vertices. Random regular graphs are locally tree-like as well, but their local weak limit is now the infinite $c$-regular tree. Their adjacency spectrum has a classical limiting density, known as the Kesten--McKay law \cite{Kesten1959,McKay1981},
\begin{equation}
\rho_{\rm KM}(\lambda) =\frac{c}{2\pi}\frac{\sqrt{4(c-1)-\lambda^2}}{c^2-\lambda^2}
\mathbf{1}_{|\lambda|\leq 2\sqrt{c-1}}\,,
\label{eq:kesten-mckay-density}
\end{equation}
where $\mathbf{1}_{{|\lambda|\leq 2\sqrt{c-1}}}$ denotes the indicator of the support. For a connected $c$-regular graph, the uniform vector is an eigenvector with eigenvalue $\lambda=c$. This eigenvalue has spectral weight $1/N$ and disappears from the empirical spectral density as $N\to\infty$. More generally, the multiplicity of the eigenvalue $c$ is the number of connected components. The Kesten--McKay law is an important benchmark. It shows that even for an unweighted adjacency matrix, the sparse spectral density is not generically the Wigner semicircle law. The semicircle appears only after an additional dense-connectivity limit, with the appropriate treatment of the trivial mean mode and the usual bulk scaling.

The Erd\H{o}s--R\'enyi and random regular ensembles are analytically convenient, but many applications require heterogeneous degrees. A standard way of constructing graphs with a prescribed degree distribution is the configuration model \cite{MolloyReed1995,NewmanStrogatzWatts2001}. This works as follows: given a degree sequence $\pmb{k}^{(N)}=(k_1,\ldots,k_N)$ with even total degree $\sum_{i}k_i$, one attaches $k_i$ half-edges to vertex $i$ and pairs all half-edges uniformly at random. The pairing construction naturally produces a random multigraph, since self-edges and multiple edges may occur. Conditioning on the event that no such defects occur gives a random simple graph with the prescribed degree sequence. For local weak-limit calculations, one often obtains the same limiting rooted-neighborhood statistics from the multigraph construction, provided the degree sequence is in a regime where self-edges and multiple edges have negligible local effect. Equivalently, one may work directly with the following formal measure over simple graphs:
\begin{equation}
P_{\rm CM}(\pmb{C}\mid \pmb{k}^{(N)})=\frac{1}{Z_{\rm CM}(\pmb{k}^{(N)})}\prod_{i=1}^N\delta_{\sum_{j=1}^N C_{ij},k_i}\prod_{i<j}\delta_{C_{ij},C_{ji}}\prod_{i=1}^N\delta_{C_{ii},0}\,.
\label{eq:configuration-model-measure}
\end{equation}
Here $Z_{\rm CM}(\pmb{k}^{(N)})$ is the number of simple graphs with degree sequence $\pmb{k}^{(N)}$. To describe the large-$N$ limit, it is convenient to introduce the empirical degree distribution
\begin{equation}
p_{N}(k)=\frac{1}{N}\sum_{i=1}^N\delta_{k,k_i}\,.    
\end{equation}
Suppose that, along a sequence of degree sequences $\pmb{k}^{(N)}$, one has
\begin{equation}
p_{N}(k)\to p(k)    
\end{equation}
for each fixed $k$, and that the limiting mean degree is finite
\begin{equation}
\langle k\rangle=\lim_{N\to\infty}\frac{1}{N}\sum_{i=1}^N k_i=\sum_{k\geq0} kp(k)\,.
\end{equation}
Then the degree of a vertex seen by following a uniformly chosen edge is not distributed according to $p(k)$ but biased by $k$, according to the size-biased law
\begin{equation}
\frac{kp(k)}{\langle k\rangle}\,.
\end{equation}
Equivalently, the number of additional neighbors beyond the arrival edge, the so-called excess-degree distribution, is
\begin{equation}
q(\ell)=\frac{(\ell+1)p(\ell+1)}{\langle k\rangle}\,,\qquad \ell=0,1,2,\ldots\,.
\label{eq:excess-degree-distribution}
\end{equation}

\begin{examplebox}[Why the excess-degree distribution is different from the degree distribution.]
Suppose that a random graph has degree distribution
\begin{equation}
p(k)=\frac{1}{2}\delta_{k,1}+\frac{1}{2}\delta_{k,3}\,.
\label{eq:rgdme-example-degree-distribution}
\end{equation}
Then the mean degree is
\begin{equation}
\langle k\rangle=\sum_{k\geq0}kp(k)=\frac{1}{2}\cdot 1+\frac{1}{2}\cdot 3=2\,.
\label{eq:rgdme-example-mean-degree}
\end{equation}
The excess-degree distribution is
\begin{equation}
q(\ell)=\frac{(\ell+1)p(\ell+1)}{\langle k\rangle}\,.
\label{eq:rgdme-example-excess-definition}
\end{equation}
Therefore
\begin{equation}
q(0)=\frac{1\cdot p(1)}{2}=\frac{1}{4}\,,\qquad q(2)=\frac{3\cdot p(3)}{2}=\frac{3}{4}\,,
\label{eq:rgdme-example-excess-values}
\end{equation}
and
\begin{equation}
q(\ell)=0\qquad\text{for all other }\ell\,.
\label{eq:rgdme-example-excess-zero}
\end{equation}
Thus, although half the vertices have degree one and half have degree three, a vertex reached by following a uniformly chosen edge is \emph{not} equally likely to have degree one or degree three. It is more likely to have degree three, simply because degree-three vertices carry more edges. This is why cavity recursions are governed by the excess-degree law rather than by the original degree law.

Here $q(\ell)$ is a distribution over excess degrees $\ell$, not over the original degrees $k$.
\end{examplebox}

This size biasing is one of the basic mechanisms by which degree heterogeneity enters cavity equations. Under the usual assumptions of the configuration model, it is also the origin of the Molloy--Reed criterion for the emergence of a giant connected component,
\begin{equation}
\sum_{k\geq 0} k(k-2)p(k)>0\,.
\label{eq:molloy-reed-criterion}
\end{equation}
Although this criterion concerns connectivity rather than spectra, it illustrates a general point that will recur repeatedly: sparse graph ensembles are controlled by local branching statistics, and spectral observables inherit this dependence.

A related analytically tractable family consists of expected-degree or Chung--Lu graphs \cite{ChungLu2002}. Given positive weights $w_1,\ldots,w_N$, one connects vertices independently with probabilities approximately proportional to $w_iw_j$. In its simplest form,
\begin{equation}
{\rm Prob}(C_{ij}=1)=\frac{w_iw_j}{\sum_{\ell=1}^N w_\ell}\,,\qquad i<j\,,
\label{eq:chung-lu-probability}
\end{equation}
provided the right-hand side is at most one. If self-edges are excluded, then
\begin{equation}
\mathbb E[k_i]=\sum_{j\neq i}\frac{w_iw_j}{\sum_{\ell=1}^N w_\ell}
=w_i\left(1-\frac{w_i}{\sum_{\ell=1}^N w_\ell}\right)\,,
\end{equation}
and hence $\mathbb E[k_i]\simeq w_i$ when no single weight dominates the total weight. Such ensembles are useful because they interpolate between homogeneous Erd\H{o}s--R\'enyi graphs and highly heterogeneous networks, while preserving enough independence to remain analytically manageable. More generally, random graph theory and network science contain many ensembles designed to incorporate clustering, short paths, heavy-tailed degrees, or growth mechanisms, such as small-world and preferential-attachment networks \cite{WattsStrogatz1998,BarabasiAlbert1999,Newman2010}. The present notes will not attempt a complete review of such models. We shall focus on ensembles whose local weak structure can be described by tree-like recursions, since these are the natural domain of the cavity method, at least in its most basic implementation.

Graph ensembles can also include correlations and mesoscopic organization. For instance, vertices may carry types or community labels $g_i\in\{1,\ldots,B\}$, and the edge probability may depend on the pair of groups,
\begin{equation}
{\rm Prob}(C_{ij}=1|g_i=a,g_j=b)=\frac{c_{ab}}{N}\,.
\label{eq:block-edge-probability}
\end{equation}
For undirected graphs one imposes $c_{ab}=c_{ba}$. This is the sparse version of a block-structured random graph. One may also prescribe degree--degree correlations, so that the probability of an edge depends on the degrees or types of its endpoints. Such constraints modify the local branching process and therefore the spectral self-consistency equations. The spectral density of random graphs with topological constraints, including generalized degree--degree correlations and community structure, was analyzed using replica methods in \cite{RogersPerezVicenteTakedaPerezCastillo2010}. This class of examples is important for the present notes because it makes clear that sparse spectra are not determined only by a mean degree. They depend on the full statistical law of the local environment.

We now turn from graphs to matrices. The adjacency matrix itself, in the context of sparse ensembles, is the simplest sparse random matrix, but most ensembles of interest introduce weights on the edges. A general real symmetric diluted matrix can be written as
\begin{equation}
A_{ij}=C_{ij}J_{ij}+D_i\delta_{ij}\,,\qquad A_{ji}=A_{ij}\,,
\label{eq:symmetric-diluted-matrix}
\end{equation}
where $\pmb{C}$ is the adjacency matrix of an undirected random graph, $J_{ij}=J_{ji}$ are random edge weights, and $D_i$ are random diagonal terms. For the sparse Erd\H{o}s--R\'enyi support, the off-diagonal entries are equivalently encoded by the single-entry distribution
\begin{equation}
P(A_{ij})=\left(1-\frac{c}{N}\right)\delta(A_{ij})+\frac{c}{N}p_J(A_{ij})\,,\qquad i<j\,,
\label{eq:sparse-entry-distribution}
\end{equation}
together with the symmetry condition $A_{ji}=A_{ij}$. Here $p_J$ is the distribution of nonzero edge weights. The diagonal terms are specified separately; their distribution may be independent of the off-diagonal weights, or it may be graph-dependent. For example, the graph Laplacian is obtained from
\begin{equation}
L_{ij}=\left(\sum_{\ell=1}^N C_{i\ell}\right)\delta_{ij}-C_{ij}\,,
\label{eq:graph-laplacian}
\end{equation}
while Anderson-type operators on graphs include random diagonal potentials in addition to sparse hopping terms. These examples all belong to the same broad class of sparse Hermitian operators on random graphs.

The spectral density of sparse symmetric matrices was studied by replica and cavity methods in a sequence of works beginning with the analysis of Rodgers and Bray \cite{RodgersBray1988}. Subsequent developments clarified the role of local graph structure, localized states, and the relation between resolvent recursions and cavity equations \cite{Kuhn2008}. The cavity approach developed in \cite{RogersTakedaPerezCastilloKuhn2008} gives a particularly direct formulation for locally tree-like sparse matrices and sparse covariance matrices. In that framework the resolvent is represented through Gaussian variables on a graph, and the cavity messages are local variances or inverse variances. The resulting equations are not universal in the sense of dense random matrix theory, because finite-connectivity spectra retain the local graph-and-weight statistics: changing the degree distribution, edge-weight law, or topological constraints changes the corresponding ensemble self-consistency equations.

This non-universality is not a defect of the method but rather a feature of the sparse problem. Sparse graph spectra are sensitive to local structures that survive in the thermodynamic limit. Leaves, finite trees, short dangling components, and degree fluctuations may generate localized eigenvectors, spectral tails, and singular contributions. Numerical and analytical studies of graph spectra in network models showed early on that sparse and correlated graphs can deviate strongly from the semicircle law \cite{FarkasDerenyiBarabasiVicsek2001}. From the perspective of the present notes, this is precisely why a statistical-mechanics description is useful. The matrix is not merely a random array of entries; it is an operator on a random locally tree-like environment.

Dense random matrix laws nevertheless remain important as limiting checks. If the average degree $c$ is sent to infinity after the thermodynamic limit, and if the nonzero entries are scaled appropriately, sparse symmetric ensembles recover Wigner-type behavior\footnote{For typical spectral densities, the dense-connectivity limit of sparse ensembles provides a useful check against the Wigner or Mar\v{c}enko--Pastur laws. For large-deviation rate functions, however, the comparison is more subtle: the dense-ensemble rate functions are not recovered by a naive dense-connectivity limit unless the relevant finite-size corrections are retained before taking the limit. This issue lies beyond the scope of the present lecture notes.}. Similarly, sparse covariance ensembles recover the Mar\v{c}enko--Pastur law in an appropriate dense-connectivity limit \cite{MarchenkoPastur1967,RogersTakedaPerezCastilloKuhn2008}. These dense limits are useful sanity checks for the cavity equations, but they should not obscure the main finite-connectivity problem. Throughout most of these notes $c$ will remain finite as $N\to\infty$.

Sparse covariance matrices require a bipartite construction. Let $\pmb{X}$ be an $N\times P$ rectangular matrix. In the usual dense Wishart setting one studies
\begin{equation}
\pmb{W}=\frac{1}{P}\pmb X\pmb X^{\rm T}\,,
\label{eq:dense-wishart-form}
\end{equation}
and takes the limit $N,P\to\infty$ with $N/P$ fixed. In a diluted Wishart or sparse covariance ensemble, the entries of $\pmb X$ are themselves sparse. A convenient representation is
\begin{equation}
X_i^\mu=B_i^\mu \xi_i^\mu\,,\qquad {\rm Prob}(B_i^\mu=1)=\frac{d}{N}\,,\qquad {\rm Prob}(B_i^\mu=0)=1-\frac{d}{N}\,,
\label{eq:sparse-rectangular-matrix}
\end{equation}
where $i=1,\ldots,N$, $\mu=1,\ldots,P$, and $\xi_i^\mu$ are nonzero weights drawn from a prescribed distribution. With the convention $\alpha=N/P$, a column of $\pmb{X}$ has expected degree $d$, while a row has expected degree $d/\alpha$. The associated covariance matrix may be normalized as
\begin{equation}
\pmb{W}=\frac{1}{d}\pmb X\pmb X^{\rm T}\,,\qquad W_{ij}=\frac{1}{d}\sum_{\mu=1}^P B_i^\mu B_j^\mu\xi_i^\mu \xi_j^\mu\,.
\label{eq:diluted-wishart}
\end{equation}
For $\mathbb E[(\xi_i^\mu)^2]=1$, the normalization by $d$ gives $\mathbb E[W_{ii}]=1/\alpha$, so the typical diagonal entries of $\pmb W$ remain of order one at fixed $\alpha$. The graph underlying $\pmb{X}$ is bipartite: variable nodes $i$ are connected to factor or sample nodes $\mu$. Two variables $i$ and $j$ interact in $\pmb W$ when they share a common neighboring sample node. Thus the covariance matrix is sparse because overlaps between rows of a sparse rectangular matrix are rare. Sparse sample covariance matrices and diluted Wishart matrices have been studied with methods closely related to those used for sparse symmetric matrices \cite{NagaoTanaka2007,RogersTakedaPerezCastilloKuhn2008}. They will play a central role later because covariance-type matrices are the natural sparse analogues of Wishart ensembles.

The bipartite viewpoint is more than a convenient construction. It identifies the correct graphical model for the statistical-mechanics calculation. The Gaussian integral associated with the resolvent of $\pmb{W}$ may be written either directly in terms of the $N$ variables of the covariance matrix or in an enlarged representation involving the bipartite structure. The latter representation often leads to simpler cavity equations, because the sparsity is elementary at the level of $\pmb X$ even when the induced covariance matrix contains effective interactions generated by shared sample nodes. This is the same conceptual move used in spin-glass and constraint-satisfaction problems: one works on the factor graph on which the disorder is locally defined.

Non-Hermitian sparse matrices arise when the underlying graph is directed or when the weights are asymmetric. A basic directed Erd\H{o}s--R\'enyi support is obtained from
\begin{equation}
P_{\rm dir}(\pmb C)=\prod_{i\neq j}\left[\frac{c}{N}\delta_{C_{ij},1}+\left(1-\frac{c}{N}\right)\delta_{C_{ij},0}\right]\prod_{i=1}^N\delta_{C_{ii},0}\,,
\label{eq:directed-ER-measure}
\end{equation}
and a general sparse non-Hermitian matrix may be written as
\begin{equation}
A_{ij}=C_{ij}J_{ij}\,,
\label{eq:nonhermitian-diluted-matrix}
\end{equation}
where, in general, $A_{ij}\neq A_{ji}$. Diagonal terms can again be added separately when needed. The spectral density is then a two-dimensional density in the complex plane rather than a density on the real line. The dense non-Hermitian problem can often be approached through Hermitization and logarithmic potentials, but in the sparse case these tools must be combined with graph-local recursions. The cavity approach to the spectral density of non-Hermitian sparse matrices was developed in \cite{RogersPerezCastillo2009}. In comparison with the Hermitian case, the messages become matrix-valued objects after Hermitization, but the underlying reason for the method is the same: the graph is locally tree-like and the resolvent can be represented through Gaussian variables.

A further class of ensembles, important for these notes, consists of generalized diluted Wishart or cross-correlation matrices. Let $\pmb{X}$ and $\pmb{Y}$ be two diluted $N\times P$ rectangular matrices whose entries are jointly distributed according to
\begin{equation}
P(x_i^\mu,y_i^\mu)=\left(1-\frac{d}{N}\right)\delta(x_i^\mu)\delta(y_i^\mu)+\frac{d}{N}\varrho(x_i^\mu,y_i^\mu)\,,
\label{eq:joint-diluted-rectangular-distribution}
\end{equation}
where $\varrho(x,y)$ controls the joint distribution of nonzero entries. The generalized diluted Wishart matrix is
\begin{equation}
\pmb{F}=\frac{1}{2d}\left(\pmb X\pmb Y^{\rm T}+\pmb Y\pmb X^{\rm T}\right)\,.
\label{eq:generalized-diluted-wishart}
\end{equation}
This ensemble interpolates between several cases by tuning the dilution $d$, the rectangularity $\alpha=N/P$, and the correlation structure encoded in $\varrho(x,y)$ \cite{PerezCastillo2022Generalized}. When $\pmb X=\pmb Y$, or equivalently when the joint distribution forces $x=y$, one recovers a diluted Wishart-type structure. More general choices of $\varrho$ generate sparse cross-correlation matrices. This class of models is useful because it combines the two main sources of complexity emphasized in these notes: sparse graph support and covariance-type matrix structure.

\begin{examplebox}[The generalized diluted Wishart ensemble need not be positive semidefinite.]
Take the smallest possible nontrivial example with $N=2$, $P=1$, and $d=1$. Then $\pmb X$ and $\pmb Y$ are two column vectors,
\begin{equation}
\pmb X=\begin{pmatrix}
x_1\\
x_2
\end{pmatrix}\,,\qquad
\pmb Y=\begin{pmatrix}
y_1\\
y_2
\end{pmatrix}\,.
\label{eq:rgdme-example-generalized-vectors}
\end{equation}
The generalized diluted Wishart matrix is
\begin{equation}
\pmb F=\frac{1}{2}\left(\pmb X\pmb Y^{\rm T}+\pmb Y\pmb X^{\rm T}\right)=
\begin{pmatrix}
x_1y_1 & \frac{x_1y_2+y_1x_2}{2}\\
\frac{x_1y_2+y_1x_2}{2} & x_2y_2
\end{pmatrix}\,.
\label{eq:rgdme-example-generalized-matrix}
\end{equation}

If $\pmb Y=\pmb X$, then
\begin{equation}
\pmb F=\pmb X\pmb X^{\rm T}\,,
\label{eq:rgdme-example-generalized-wishart-limit}
\end{equation}
which is positive semidefinite. For example, with
\begin{equation}
\pmb X=
\begin{pmatrix}
1\\
1
\end{pmatrix}\,\qquad
\pmb Y=
\begin{pmatrix}
1\\
1
\end{pmatrix}\,,
\label{eq:rgdme-example-positive-choice}
\end{equation}
we obtain
\begin{equation}
\pmb F=
\begin{pmatrix}
1 & 1\\
1 & 1
\end{pmatrix}\,,
\label{eq:rgdme-example-positive-matrix}
\end{equation}
whose eigenvalues are $2$ and $0$.

If instead
\begin{equation}
\pmb X=
\begin{pmatrix}
1\\
1
\end{pmatrix}\,,
\qquad\pmb Y=
\begin{pmatrix}
1\\
-1
\end{pmatrix}\,,
\label{eq:rgdme-example-indefinite-choice}
\end{equation}
then
\begin{equation}
\pmb F=\frac{1}{2}\left[
\begin{pmatrix}
1 & -1\\
1 & -1
\end{pmatrix}+
\begin{pmatrix}
1 & 1\\
-1 & -1
\end{pmatrix}\right]=
\begin{pmatrix}
1 & 0\\
0 & -1
\end{pmatrix}\,,
    \label{eq:rgdme-example-indefinite-matrix}
\end{equation}
whose eigenvalues are $1$ and $-1$. Thus the generalized diluted Wishart ensemble interpolates between an ordinary positive covariance-type matrix and a genuinely sign-indefinite cross-correlation matrix.
\end{examplebox}

The ensembles introduced above share a common feature. Their randomness is quenched: one first samples a graph and its weights, then studies the spectrum of the resulting matrix. Disorder averages of spectral observables are taken only after the spectral quantity has been defined for the individual realization. We shall denote such averages by an overbar $\overline{(\cdots)}$. For example, if $\rho_{\pmb A}(\lambda)$ is the empirical spectral density of a matrix $\pmb{A}$, then the ensemble-averaged spectral density is
\begin{equation}
\overline{\rho_{\pmb A}(\lambda)}=\int d\pmb A P(\pmb A)\rho_{\pmb A}(\lambda)\,.
\label{eq:quenched-spectral-average}
\end{equation}
Here the notation $\int d\pmb A P(\pmb A)$ denotes the appropriate combination of sums over graph variables and integrals over random weights. When the empirical spectral density is self-averaging, this ensemble average is also the typical spectral density. It is the main object in the first part of the notes, while later sections will study fluctuations of spectral counting observables and conditioned spectral densities.

The distinction between annealed and quenched quantities is central. In many graph problems, averaging the partition function or determinant is much easier than averaging its logarithm. In the Edwards--Jones route, spectral observables are expressed through logarithms of determinants, or through resolvents obtained by differentiating such logarithms, and therefore lead to quenched averages. This is the point at which the statistical mechanics of disordered systems enters the theory. Replica and cavity methods are designed precisely to handle such quenched quantities in locally tree-like random environments.

To summarize, random graphs provide the support of diluted matrices, but the matrix ensemble is specified only after one also chooses weights, symmetry constraints, diagonal terms, and possible rectangular or directed structure. Erd\H{o}s--R\'enyi graphs give the simplest Poissonian support; random regular graphs give a homogeneous finite-connectivity benchmark; configuration and expected-degree models introduce degree heterogeneity; block and correlated graph ensembles introduce topological constraints; bipartite graphs generate sparse covariance and Wishart matrices; and directed supports lead to non-Hermitian spectra. The rest of the notes develop a common statistical-mechanics language for these ensembles. The next step is to express spectral observables in a form suitable for the replica and cavity methods, and then to discuss their numerical solution by population dynamics.

\begin{exerciseblock}
\exitem[From graph to matrix] For the undirected graph with vertex set $V=\{1,2,3,4\}$ and edge set
\begin{equation}
E=\{\{1,2\},\{1,3\},\{2,4\}\}\,,
\label{eq:rgdme-ex1-edges}
\end{equation}
write its adjacency matrix $\pmb C$, compute the degrees $k_i$, and compute the empirical degree distribution $p_N(k)$.

\exitem[Directed support] For the directed graph with vertex set $V=\{1,2,3\}$ and arrows
\begin{equation}
1\to 2\,,\qquad 2\to 1\,,\qquad 2\to 3\,,\qquad 3\to 1\,,
\label{eq:rgdme-ex2-directed-edges}
\end{equation}
write the directed adjacency matrix $\pmb C$ and compute $(k_i^{\rm in},k_i^{\rm out})$ for all vertices.

\exitem[Poisson degree law in the sparse Erd\H{o}s--R\'enyi regime] Starting from
\begin{equation}
{\rm Prob}(k_i=k)=\binom{N-1}{k}\left(\frac{c}{N}\right)^k\left(1-\frac{c}{N}\right)^{N-1-k}\,,
\label{eq:rgdme-ex3-binomial}
\end{equation}
show explicitly that, for fixed $k$ and fixed $c$,
\begin{equation}
{\rm Prob}(k_i=k)\longrightarrow e^{-c}\frac{c^k}{k!}\qquad (N\to\infty)\,.
\label{eq:rgdme-ex3-poisson}
\end{equation}

\exitem[A first local-tree check]
In the sparse Erd\H{o}s--R\'enyi ensemble on $N$ vertices, with each unordered edge present independently with probability $c/N$, compute the expected number of triangles containing a fixed vertex $i$. Show that this expectation vanishes as $N\to\infty$ for fixed $c$. Explain why this supports the statement that the local neighborhood is tree-like at finite graph distance.

\exitem[Random regular graphs]
Let $\pmb C$ be the adjacency matrix of a connected $c$-regular graph on $N$ vertices. Show directly from the definition of a $c$-regular graph that the vector
\begin{equation}
\pmb 1=(1,\ldots,1)^{\rm T}
\label{eq:rgdme-ex5-uniform-vector}
\end{equation}
is an eigenvector of $\pmb C$ with eigenvalue $c$. Then explain why this eigenvalue has weight $1/N$ in the empirical spectral density and therefore vanishes from the limiting spectral density as $N\to\infty$.

\exitem[Excess-degree distribution]
For the degree distribution
\begin{equation}
p(k)=\frac{1}{3}\delta_{k,1}+\frac{2}{3}\delta_{k,4}\,,
\label{eq:rgdme-ex6-degree-law}
\end{equation}
compute $\langle k\rangle$ and the excess-degree distribution $q(\ell)$. Compare the original degree distribution $p(k)$, the size-biased degree distribution $kp(k)/\langle k\rangle$, and the excess-degree distribution $q(\ell)$. Explain in words why these distributions are different.

\exitem[Molloy--Reed criterion]
Verify that for a Poisson degree distribution with mean $c$,
\begin{equation}
\sum_{k\geq0}k(k-2)p(k)=c(c-1)\,.
\label{eq:rgdme-ex7-poisson-molloy-reed}
\end{equation}
Deduce that the Molloy--Reed criterion predicts the appearance of a giant component at $c=1$.

\exitem[Expected-degree ensembles]
In the Chung--Lu model, define
\begin{equation}
S=\sum_{\ell=1}^{N}w_\ell
\label{eq:rgdme-ex8-total-weight}
\end{equation}
and suppose that
\begin{equation}
{\rm Prob}(C_{ij}=1)=\frac{w_iw_j}{S}\,,\qquad i<j\,,
\label{eq:rgdme-ex8-chung-lu}
\end{equation}
assuming that $w_iw_j/S\leq 1$ for all pairs considered. Show that, when self-edges are excluded,
\begin{equation}
\mathbb E[k_i]=\sum_{j\neq i}\frac{w_iw_j}{S}
=w_i\left(1-\frac{w_i}{S}\right)\,.
\label{eq:rgdme-ex8-expected-degree}
\end{equation}
Hence explain in what sense $w_i$ is the expected degree when no single weight dominates the sum.

\exitem[Block-structured graphs]
For the sparse block model with $N$ vertices and group labels $g_i\in\{1,\ldots,B\}$, suppose that a fraction $p_b$ of vertices belongs to group $b$ and that
\begin{equation}
{\rm Prob}(C_{ij}=1|g_i=a,g_j=b)=\frac{c_{ab}}{N}\,.
\label{eq:rgdme-ex9-block}
\end{equation}
Show that, up to corrections that vanish as $N\to\infty$, the expected degree of a vertex in group $a$ is
\begin{equation}
c_a=\sum_{b=1}^{B}p_b c_{ab}\,.
\label{eq:rgdme-ex9-expected-block-degree}
\end{equation}
Under what condition does this reduce to the ordinary sparse Erd\H{o}s--R\'enyi model?

\exitem[From sparse rectangular matrices to covariance matrices]
Let
\begin{equation}
\pmb X=
\begin{pmatrix}
1 & 0\\
0 & 1\\
1 & 1\\
1 & -1
\end{pmatrix}\,,\qquad\pmb W=\pmb X\pmb X^{\rm T}\,.
\label{eq:rgdme-ex10-bipartite}
\end{equation}
Compute $\pmb W$ explicitly. Then show that
\begin{equation}
{\rm rank}\pmb W \leq {\rm rank}\pmb X\leq 2\,.
\label{eq:rgdme-ex10-rank}
\end{equation}
What does this imply about the number of zero eigenvalues of $\pmb W$?

\exitem[Positivity of diluted Wishart matrices]
Assume $d>0$. Starting from
\begin{equation}
\pmb W=\frac{1}{d}\pmb X\pmb X^{\rm T}\,,
\label{eq:rgdme-ex11-wishart}
\end{equation}
show that for every $\pmb v\in\mathbb R^N$,
\begin{equation}
\pmb v^{\rm T}\pmb W\pmb v=\frac{1}{d}\sum_{\mu=1}^{P}\left(\sum_{i=1}^{N}X_i^\mu v_i\right)^2\geq 0\,.
\label{eq:rgdme-ex11-positivity}
\end{equation}
Explain why this argument fails for the generalized diluted Wishart ensemble.

\exitem[Generalized diluted Wishart reduction]
Starting from
\begin{equation}
\pmb F=\frac{1}{2d}\left(\pmb X\pmb Y^{\rm T}+\pmb Y\pmb X^{\rm T}\right)\,,
\label{eq:rgdme-ex12-generalized}
\end{equation}
show that if $\pmb X=\pmb Y$ then
\begin{equation}
\pmb F=\frac{1}{d}\pmb X\pmb X^{\rm T}.
\label{eq:rgdme-ex12-reduction}
\end{equation}
Then, taking $N=2$, $P=1$, and $d=1$, construct a $2\times2$ example, different from the one in the worked example above, showing that $\pmb F$ can have a negative eigenvalue when $\pmb X\neq\pmb Y$.

\exitem[Quenched versus annealed averages]
Let $Z(\omega)>0$ be a positive random quantity depending on disorder $\omega$. Show using Jensen's inequality that
\begin{equation}
\overline{\log Z}\leq\log \overline{Z}\,.
\label{eq:rgdme-ex13-jensen}
\end{equation}
Explain why the section emphasizes quenched rather than annealed averages for spectral observables.

\exitem[Programming exercise: graph ensembles]
Fix a list of system sizes $N$, an integer degree parameter $c\geq 2$ such that $cN$ is even for the random regular graph, and a number $S$ of independent samples for each $N$. Generate numerically:
\begin{itemize}
    \item  a sparse Erd\H{o}s--R\'enyi graph with mean degree $c$,
    \item  a random regular graph with degree $c$,
    \item  a configuration-model graph with a prescribed degree distribution $p(k)$, for example $p(k)=\frac{1}{2}\delta_{k,1}+\frac{1}{2}\delta_{k,3}$, using degree sequences with even total degree.
\end{itemize}
For each ensemble and each value of $N$, compute the empirical degree distribution, average it over the $S$ samples, and compare it with the theoretical one. Then estimate the size-biased degree distribution by sampling uniformly random oriented edges, equivalently by choosing an undirected edge uniformly and then choosing one of its two endpoints uniformly, and recording the degree of the reached vertex. Finally, estimate the excess-degree distribution by recording the degree of the reached vertex minus one, and compare it with
\begin{equation}
q(\ell)=\frac{(\ell+1)p(\ell+1)}{\langle k\rangle}\,.
\label{eq:rgdme-ex14-excess-degree-law}
\end{equation}
Explain why, for a Poisson degree distribution, the excess-degree distribution is again Poisson with the same mean, while for a regular graph it is concentrated at $c-1$.

\exitem[Programming exercise: sparse covariance matrices]
Fix a list of system sizes $N$, a dilution $d>0$, an aspect ratio $\alpha>0$ such that $P=N/\alpha$ is an integer, a number $S$ of independent samples for each $N$, and a choice of nonzero-weight distribution $p_\xi$.
Generate a sparse rectangular matrix $\pmb X$ with Bernoulli support,
\begin{equation}
X_i^\mu=B_i^\mu\xi_i^\mu\,,\qquad {\rm Prob}(B_i^\mu=1)=\frac{d}{N}\,,\qquad {\rm Prob}(B_i^\mu=0)=1-\frac{d}{N}\,,
\label{eq:rgdme-ex15-program-wishart}
\end{equation}
where the nonzero weights $\xi_i^\mu$ may be taken either equal to one or drawn independently from a standard Gaussian distribution. Form
\begin{equation}
\pmb W=\frac{1}{d}\pmb X\pmb X^{\rm T}\,,
\label{eq:rgdme-ex15-program-W}
\end{equation}
and compare numerically the ranks of $\pmb X$ and $\pmb W$ using the same numerical tolerance for singular values or eigenvalues. For each sample, compare the multiplicity of zero eigenvalues of $\pmb W$ with $N-{\rm rank}\pmb X$, and then examine how this quantity changes with $\alpha=N/P$.
\end{exerciseblock}

\section{Spectral observables, resolvents, and Gaussian integral representations}
\label{sec:spectral-observables-resolvents}
As we have seen in the previous section, the random graph or diluted matrix ensemble specifies the disorder. The next step is to decide which spectral observables are to be computed and how they can be represented in a form suitable for statistical-mechanics methods. For dense invariant ensembles, one often starts from the joint probability density of eigenvalues. For the sparse and diluted ensembles considered in these notes, such a description is usually unavailable, or at least not the most useful starting point. The natural objects are instead resolvents, determinants, local Green functions, and Gaussian integral representations. These objects retain the graph structure of the matrix and lead directly to cavity and replica methods. We now introduce these objects one by one.

Let $\pmb A$ be a real symmetric or Hermitian $N\times N$ matrix with eigenvalues $\lambda_1,\ldots,\lambda_N$. Its empirical spectral density is
\begin{equation}
\rho_{\pmb A}(\lambda)=\frac{1}{N}\sum_{i=1}^N\delta(\lambda-\lambda_i)\,.
\label{eq:empirical-spectral-density}
\end{equation}
More formally, one may regard
\begin{equation}
\mu_{\pmb A}=\frac{1}{N}\sum_{i=1}^N\delta_{\lambda_i}
    \label{eq:empirical-spectral-measure}
\end{equation}
as the empirical spectral measure of $\pmb A$. The standard analytic object associated with this measure is its Cauchy--Stieltjes transform, or normalized trace resolvent. With the convention
\begin{equation}
z=\lambda_{\epsilon}\equiv \lambda-i\epsilon\,,\qquad \epsilon>0\,,
\label{eq:lambda-epsilon-convention}
\end{equation}
we define the matrix resolvent, or Green function, and its normalized trace by
\begin{equation}
\pmb G_{\pmb A}(z)=(z\pmb I-\pmb A)^{-1}\,,\qquad g_{\pmb A}(z)=\frac{1}{N}{\rm Tr}\pmb G_{\pmb A}(z)\,.
\label{eq:resolvent-definition}
\end{equation}
Since
\begin{equation}
g_{\pmb A}(z)=\frac{1}{N}\sum_{i=1}^N\frac{1}{z-\lambda_i}\,,
\label{eq:resolvent-spectral-representation}
\end{equation}
one can equivalently write
\begin{equation}
g_{\pmb A}(z)=\int_{\mathbb{R}}\frac{\mu_{\pmb A}(d\lambda')}{z-\lambda'}
=\int_{\mathbb{R}} d\lambda'\frac{\rho_{\pmb A}(\lambda')}{z-\lambda'}\,.
\end{equation}
Thus $g_{\pmb A}(z)$ is precisely the Cauchy--Stieltjes transform of the empirical spectral measure. In this way, the spectral density is recovered by the inversion formula
\begin{equation}
\rho_{\pmb{A}}(\lambda)=\frac{1}{\pi}\lim_{\epsilon\downarrow 0}{\rm Im} g_{\pmb{A}}(\lambda-i\epsilon)\,.
\label{eq:stieltjes-inversion}
\end{equation}
This is the basic relation between spectra and resolvents used throughout random matrix theory \cite{Mehta2004,Forrester2010,AndersonGuionnetZeitouni2010,BaiSilverstein2010,PasturShcherbina2011}. Note that the sign in the $\epsilon$-limit of Eq.~\eqref{eq:stieltjes-inversion} follows from the lower-half-plane convention \eqref{eq:lambda-epsilon-convention}, for which
\begin{equation}
{\rm Im}\frac{1}{\lambda-i\epsilon-\lambda_i}= \frac{\epsilon}{(\lambda-\lambda_i)^2+\epsilon^2}\,.
    \label{eq:lorentzian-broadening}
\end{equation}
Thus the finite-$\epsilon$ object
\begin{equation}
\rho_{\pmb A,\epsilon}(\lambda)=\frac{1}{\pi N}\sum_{i=1}^N\frac{\epsilon}{(\lambda-\lambda_i)^2+\epsilon^2}
    \label{eq:regularized-density}
\end{equation}
is a Lorentzian regularization of the empirical spectral density. In sparse random matrices the regulator $\epsilon$ is not merely a harmless technical parameter. It controls the resolution with which localized states, isolated peaks, and singular spectral components are seen. For this reason, in some cases, the order of the limits $N\to\infty$ and $\epsilon\downarrow 0$ has to be handled with care.

The diagonal entries of the resolvent define local spectral densities. Choosing an orthonormal eigenbasis $\{\pmb u^{(i)}\}_{i=1}^N$, with $\pmb u^{(i)}$ associated with $\lambda_i$, one has
\begin{equation}
G_{jj}(z)= \sum_{i=1}^N\frac{|u_j^{(i)}|^2}{z-\lambda_i}\,.
\label{eq:diagonal-resolvent-eigenvectors}
\end{equation}
Hence
\begin{equation}
\rho_j(\lambda)=\sum_{i=1}^N|u_j^{(i)}|^2\delta(\lambda-\lambda_i)=\frac{1}{\pi}\lim_{\epsilon\downarrow 0}{\rm Im} G_{jj}(\lambda-i\epsilon)\,,
\label{eq:local-density-of-states}
\end{equation}
and
\begin{equation}
\rho_{\pmb A}(\lambda)=\frac{1}{N} \sum_{j=1}^N\rho_j(\lambda)\,.
\label{eq:global-density-from-local-densities}
\end{equation}
This local formulation is essential for sparse matrices. On a fixed locally tree-like graph, cavity equations describe diagonal resolvent entries through local messages. In an ensemble formulation, the statistics of these messages give the distribution of local densities of states. This is one of the first indications that, in contrast with dense invariant ensembles, the local environment of a vertex remains relevant in the thermodynamic limit.

\begin{examplebox}[Resolvent, global density, and local density of states for a $2\times2$ matrix]
Consider
\begin{equation}
\pmb A=\begin{pmatrix}
0 & 1\\
1 & 0
\end{pmatrix}\,.
\label{eq:sor-example-two-by-two}
\end{equation}
Its eigenvalues are $\lambda_{\pm}=\pm1$, with normalized eigenvectors
\begin{equation}
\pmb u^{(+)}=\frac{1}{\sqrt2}\begin{pmatrix}
1\\
1
\end{pmatrix}\,,\qquad
\pmb u^{(-)}=\frac{1}{\sqrt2}
\begin{pmatrix}
1\\
-1
\end{pmatrix}\,.
\label{eq:sor-example-eigenvectors}
\end{equation}
The resolvent is
\begin{equation}
\pmb G(z)=(z\pmb I-\pmb A)^{-1}=\frac{1}{z^2-1}
\begin{pmatrix}
z & 1\\
1 & z
\end{pmatrix}\,,
\label{eq:sor-example-resolvent}
\end{equation}
since
\begin{equation}
(z\pmb I-\pmb A)=\begin{pmatrix}
z & -1\\
-1 & z
\end{pmatrix}\,,\qquad\det(z\pmb I-\pmb A)=z^2-1\,.
\label{eq:sor-example-determinant}
\end{equation}
Hence the normalized trace resolvent is
\begin{equation}
g(z)=\frac{1}{2}{\rm Tr}\pmb G(z)=\frac{z}{z^2-1}=\frac{1}{2}\left(\frac{1}{z-1}+\frac{1}{z+1}\right)\,.
\label{eq:sor-example-trace-resolvent}
\end{equation}
Using the convention $z=\lambda-i\epsilon$ with $\epsilon>0$, the regularized density is therefore
\begin{equation}
\rho_{\pmb A,\epsilon}(\lambda)=\frac{1}{2\pi}\left[\frac{\epsilon}{(\lambda-1)^2+\epsilon^2}+\frac{\epsilon}{(\lambda+1)^2+\epsilon^2}\right]\,.
\label{eq:sor-example-regularized-density}
\end{equation}
Taking $\epsilon\downarrow0$ gives
\begin{equation}
\rho_{\pmb A}(\lambda)=\frac{1}{2}\delta(\lambda-1)+\frac{1}{2}\delta(\lambda+1)\,,
\label{eq:sor-example-density}
\end{equation}
as expected.

Now let us compute the local density of states. From \eqref{eq:sor-example-resolvent},
\begin{equation}
G_{11}(z)=G_{22}(z)=\frac{z}{z^2-1}\,.
\label{eq:sor-example-local-resolvent}
\end{equation}
Hence
\begin{equation}
\rho_1(\lambda)=\rho_2(\lambda)=\frac{1}{2}\delta(\lambda-1)+\frac{1}{2}\delta(\lambda+1)\,.
\label{eq:sor-example-local-density}
\end{equation}
This also follows directly from the eigenvectors, since
\begin{equation}
|u_1^{(+)}|^2=|u_1^{(-)}|^2=|u_2^{(+)}|^2=|u_2^{(-)}|^2=\frac{1}{2}\,.
\label{eq:sor-example-local-weights}
\end{equation}
Thus this simple example shows explicitly how the global density, the diagonal resolvent, and the local density of states fit together.
\end{examplebox}

The resolvent can also be written as a derivative of a determinant:
\begin{equation}
g_{\pmb A}(z)=\frac{1}{N}\frac{\partial}{\partial z}\log\det(z\pmb I-\pmb A)\,.
\label{eq:resolvent-log-det}
\end{equation}
This harmless-looking identity is the bridge to statistical mechanics. In the real symmetric case, Edwards and Jones observed that the determinant in \eqref{eq:resolvent-log-det} can be represented by a Gaussian integral, turning the computation of the spectral density into the computation of a free energy \cite{EdwardsJones1976}. With the convention \eqref{eq:lambda-epsilon-convention}, define
\begin{equation}
Z_{\pmb A}(z)=\int_{\mathbb{R}^N}\left[\prod_{i=1}^N\frac{dx_i}{\sqrt{2\pi}}\right]\exp\left[-\frac{i}{2}\pmb x^{\rm T}(z\pmb I-\pmb A)\pmb x\right]\,,
\label{eq:edwards-jones-partition-function}
\end{equation}
with $\pmb{x}^{\rm T}=(x_1,\dots,x_N)$. The convergence of this integral follows from ${\rm Im} z<0$, because the exponent contains the damping factor $-\epsilon \sum_i x_i^2/2$. Evaluating the Gaussian integral gives
\begin{equation}
Z_{\pmb A}(z) =\det\left[i(z\pmb I-\pmb A)\right]^{-1/2}\,,
\label{eq:edwards-jones-determinant}
\end{equation}
up to a $z$-independent normalization convention. Consequently,
\begin{equation}
\frac{\partial}{\partial \lambda}\log Z_{\pmb A}(\lambda-i\epsilon)=-\frac{1}{2}{\rm Tr}[(\lambda-i\epsilon)\pmb I-\pmb A]^{-1}\,,
\label{eq:derivative-partition-function}
\end{equation}
and the spectral density may be written as
\begin{equation}
\rho_{\pmb A}(\lambda)=-\frac{2}{\pi N} \lim_{\epsilon\downarrow 0}{\rm Im}\frac{\partial}{\partial \lambda}\log Z_{\pmb A}(\lambda-i\epsilon)\,.
\label{eq:density-from-free-energy}
\end{equation}
This is the Edwards--Jones representation of the density of states. The Gaussian identities, resolvent sign convention, and determinant conventions used here are collected in Appendix~\ref{app:gaussian-identities-resolvents}.

\begin{examplebox}[Checking the Edwards--Jones representation on one site]
Take the $1\times1$ matrix
\begin{equation}
\pmb A=(a)\,,
\label{eq:sor-example-one-site-matrix}
\end{equation}
so its only eigenvalue is $a$. The partition function \eqref{eq:edwards-jones-partition-function} becomes
\begin{equation}
Z_{\pmb A}(z)=\int_{-\infty}^{\infty}\frac{dx}{\sqrt{2\pi}}\exp\left[-\frac{i}{2}(z-a)x^2\right]\,.
\label{eq:sor-example-one-site-partition}
\end{equation}
Since ${\rm Im} z<0$, the real part of $i(z-a)$ is positive, so the Gaussian integral converges. Using the one-dimensional Gaussian formula,
\begin{equation}
Z_{\pmb A}(z)=\left[i(z-a)\right]^{-1/2}\,.
\label{eq:sor-example-one-site-gaussian}
\end{equation}
Hence
\begin{equation}
\log Z_{\pmb A}(z)=-\frac{1}{2}\log\!\left[i(z-a)\right]\,,
\label{eq:sor-example-one-site-logZ}
\end{equation}
and therefore
\begin{equation}
-2\frac{\partial}{\partial z}\log Z_{\pmb A}(z)=\frac{1}{z-a}\,.
\label{eq:sor-example-one-site-resolvent}
\end{equation}
This is exactly the resolvent of the one-site matrix. Substituting $z=\lambda-i\epsilon$, we get
\begin{equation}
\rho_{\pmb A,\epsilon}(\lambda)=\frac{1}{\pi}{\rm Im}\frac{1}{\lambda-i\epsilon-a}=\frac{1}{\pi}\frac{\epsilon}{(\lambda-a)^2+\epsilon^2}\,,
\label{eq:sor-example-one-site-regularized-density}
\end{equation}
and the limit $\epsilon\downarrow0$ gives
\begin{equation}
\rho_{\pmb A}(\lambda)=\delta(\lambda-a)\,.
\label{eq:sor-example-one-site-density}
\end{equation}
Thus the Edwards--Jones representation already works exactly in the simplest possible example.
\end{examplebox}

Equation \eqref{eq:edwards-jones-partition-function} has the form of a partition function for a system of $N$ continuous variables $x_i$. If $\pmb A$ is sparse, the quadratic form has the graphical structure
\begin{equation}
\pmb x^{\rm T}(z\pmb I-\pmb A)\pmb x=z\sum_{i=1}^N x_i^2-\sum_{i,j=1}^N A_{ij}x_i x_j\,.
\label{eq:quadratic-form-graphical}
\end{equation}
For a symmetric diluted matrix with $A_{ij}=C_{ij}J_{ij}$, the interactions $x_i x_j$ occur only along the edges of the underlying graph, with $J_{ij}$ setting the strength of the interaction. The spectral problem has therefore been transformed into a Gaussian graphical model with complex couplings. This is the point at which the methods of disordered systems enter. The disorder average of the spectral density involves the disorder average of $\log Z_{\pmb A}$, not of $Z_{\pmb A}$ itself:
\begin{equation}
\overline{\rho_{\pmb A}(\lambda)}=-\frac{2}{\pi N}\lim_{\epsilon\downarrow 0}{\rm Im}\frac{\partial}{\partial \lambda}\overline{\log Z_{\pmb A}(\lambda-i\epsilon)}\,.
\label{eq:quenched-density-from-free-energy}
\end{equation}
The average is therefore quenched. Replica methods address the appearance of the logarithm by computing integer moments of the partition function and then using, at the level of the thermodynamic free-energy density, the formal relation
\begin{equation}
\lim_{N\to\infty}\frac{1}{N}\overline{\log Z_{\pmb A}}
=
\lim_{n\to 0}\lim_{N\to\infty}\frac{1}{nN}\log\overline{Z_{\pmb A}^{n}}\,.
\label{eq:replica-identity-spectral}
\end{equation}
The calculation is first performed for positive integer $n$, rewritten in a form that can be analytically continued, and then evaluated at $n=0$. To carry it out explicitly, one must make an ansatz about the symmetry of the replicated order parameter under permutations of replicas. The same Gaussian representation also motivates the cavity method, which avoids the explicit analytic continuation by exploiting directly the local tree structure of sparse graphs.

Visually, the cavity method is based on a simple operation: remove a vertex or cut an edge so that the graph separates into independent branches. On a tree, removing a vertex $i$ disconnects the branches rooted at the neighbors of $i$; equivalently, cutting the edge between $j$ and $i$ isolates the branch rooted at $j$ from the rest of the graph through $i$. The statistical effect of this graphical separation is that the variables living in different branches become independent in the cavity graph. This suggests introducing, for every directed edge $j\to i$, the \emph{cavity marginal} $P_{j\to i}(x_j)$, namely the marginal distribution of the variable $x_j$ in the branch rooted at $j$ after the edge between $j$ and $i$ has been removed. Equivalently, it is the marginal in the branch rooted at $j$ after cutting its connection to $i$. In the Gaussian models considered here, this cavity marginal remains Gaussian, so it can be parametrized as
\begin{equation}
P_{j\to i}(x_j)\propto\exp\left(-\frac{1}{2}\omega_{j\to i}x_j^2\right)\,.
    \label{eq:gaussian-cavity-message}
\end{equation}
The quantity $\omega_{j\to i}$ is the cavity inverse variance. Integrating over neighboring cavity variables then leads to recursions of the form
\begin{equation}
\omega_{j\to i} =i(z-A_{jj})+\sum_{\ell\in\partial j\setminus i}\frac{A_{j\ell}^2}{\omega_{\ell\to j}}\,.
\label{eq:preview-cavity-recursion}
\end{equation}
In later sections, we will see that these cavity marginals may be interpreted as messages passed along directed edges.

The corresponding full single-site parameter is
\begin{equation}
\omega_j=i(z-A_{jj})+\sum_{\ell\in\partial j}\frac{A_{j\ell}^2}{\omega_{\ell\to j}}\,,
\label{eq:preview-full-cavity-parameter}
\end{equation}
and the local resolvent follows from
\begin{equation}
G_{jj}(z)=\frac{i}{\omega_j}\,.
\label{eq:local-resolvent-from-cavity-parameter}
\end{equation}
These equations are written here only to show why Gaussian integral representations are so useful. The detailed derivation of the local recursions and their ensemble-level self-consistency equations will be developed later. Their structure is already clear: sparse spectral theory becomes a theory of local messages on a locally tree-like graph, with message statistics determined by the graph ensemble. This observation underlies the cavity approach to sparse symmetric matrices and its extensions \cite{RodgersBray1988,Kuhn2008,RogersTakedaPerezCastilloKuhn2008,SuscaVivoKuhn2021}.

For covariance and Wishart-type matrices there is a closely related representation. Let $\pmb X$ be a real $N\times P$ rectangular matrix and let
\begin{equation}
\pmb W=\frac{1}{d}\pmb X\pmb X^{\rm T}\,.
\label{eq:wishart-observable-matrix}
\end{equation}
The spectral density of $\pmb W$ can again be obtained from the resolvent
\begin{equation}
\pmb G_{\pmb W}(z)=(z\pmb I_N-\pmb W)^{-1}\,.
\label{eq:wishart-resolvent}
\end{equation}
However, when $\pmb X$ is sparse it is often advantageous to use the bipartite structure directly. A standard linearization is the Hermitian matrix
\begin{equation}
\pmb{\mathcal L} =\frac{1}{\sqrt{d}}\begin{pmatrix}
\pmb 0_{N\times N} & \pmb X\\
\pmb X^{\rm T} & \pmb 0_{P\times P}
\end{pmatrix}\,.
\label{eq:bipartite-linearization}
\end{equation}
The nonzero eigenvalues of $\pmb{\mathcal L}$ occur in pairs $\pm\sqrt{\nu}$, where $\nu$ runs over the nonzero eigenvalues of $\pmb W$. Thus the covariance problem can be reformulated as a spectral problem for a sparse Hermitian operator on a bipartite graph. This reformulation is particularly natural for diluted Wishart ensembles, because the elementary disorder variables live on the bipartite graph rather than directly on the induced covariance matrix \cite{NagaoTanaka2007,RogersTakedaPerezCastilloKuhn2008,PerezCastilloMetz2018Wishart,PerezCastillo2022Generalized}.

\begin{examplebox}[The bipartite linearization of a rank-one covariance matrix]
Take
\begin{equation}
\pmb X=
\begin{pmatrix}
x_1\\
x_2
\end{pmatrix}\,,\qquad N=2\,,\qquad P=1\,, \qquad d=1\,.
\label{eq:sor-example-rectangular-vector}
\end{equation}
Then
\begin{equation}
\pmb W=\pmb X\pmb X^{\rm T}=\begin{pmatrix}
x_1^2 & x_1x_2\\
x_1x_2 & x_2^2
\end{pmatrix}\,.
\label{eq:sor-example-rank-one-wishart}
\end{equation}
Since $\pmb W$ has rank one, its eigenvalues are
\begin{equation}
\nu_1=x_1^2+x_2^2\,,\qquad\nu_2=0\,.
\label{eq:sor-example-rank-one-eigenvalues}
\end{equation}

Now consider the bipartite linearization
\begin{equation}
\pmb{\mathcal L}=\begin{pmatrix}
0 & 0 & x_1\\
0 & 0 & x_2\\
x_1 & x_2 & 0
\end{pmatrix}\,.
\label{eq:sor-example-linearization}
\end{equation}
A direct multiplication gives
\begin{equation}
\pmb{\mathcal L}^2=\begin{pmatrix}
x_1^2 & x_1x_2 & 0\\
x_1x_2 & x_2^2 & 0\\
0 & 0 & x_1^2+x_2^2
\end{pmatrix}=
\begin{pmatrix}
\pmb W & 0\\
0 & x_1^2+x_2^2
\end{pmatrix}\,.
\label{eq:sor-example-linearization-square}
\end{equation}
Hence the eigenvalues of $\pmb{\mathcal L}$ are
\begin{equation}
+\sqrt{x_1^2+x_2^2}\,,\qquad-\sqrt{x_1^2+x_2^2}\,,\qquad 0\,.
\label{eq:sor-example-linearization-eigenvalues}
\end{equation}
This simple calculation makes explicit the general statement in the text: the nonzero eigenvalues of the bipartite linearization come in opposite pairs, and their squares reproduce the nonzero eigenvalues of the covariance matrix.
\end{examplebox} 

Spectral counting observables, which are an integral part of these lecture notes in later sections, are also expressible in terms of the resolvent. For an interval $I=[a,b]\subset\mathbb{R}$, define
\begin{equation}
\mathcal{N}_{\pmb A}(I)=\sum_{i=1}^N\mathbf{1}_{\lambda_i\in I}\,,
\label{eq:number-eigenvalues-interval}
\end{equation}
as the number of eigenvalues of $\pmb A$ inside the interval $I$. Assume, for this formula, that no eigenvalue lies exactly at $a$ or $b$, or adopt the usual boundary convention for endpoint eigenvalues. Using Eq.~\eqref{eq:stieltjes-inversion}, we can express it as
\begin{equation}
\mathcal{N}_{\pmb A}(I)=\frac{1}{\pi}\lim_{\epsilon\downarrow 0}{\rm Im}\int_a^b d\lambda\,{\rm Tr}[(\lambda-i\epsilon)\pmb I-\pmb A]^{-1}\,.
\label{eq:index-from-resolvent}
\end{equation}
This observable is the starting point for index statistics and large-deviation theory. In invariant ensembles, the fluctuations of such counting functions can be studied using Coulomb-gas methods and the joint eigenvalue density \cite{DeanMajumdar2006,MajumdarNadalScardicchioVivo2009}. For sparse and non-invariant matrices, one instead returns to determinant and resolvent representations, introducing biased replicated partition functions that count eigenvalues in a prescribed interval. This strategy leads to large-deviation functions for sparse random graphs and diluted Wishart matrices, and later to conditioned spectral densities \cite{MetzPerezCastillo2016,MetzPerezCastillo2017,PerezCastilloMetz2018Wishart,PerezCastilloMetz2018Conditioned}.

For non-Hermitian matrices the eigenvalues are complex, and the resolvent is not sufficient in the same way. Let $\pmb A$ be a general real or complex $N\times N$ matrix with eigenvalues $z_1,\ldots,z_N\in\mathbb{C}$. The empirical spectral density in the complex plane is
\begin{equation}
\rho_{\pmb A}(z)=\frac{1}{N}\sum_{i=1}^N\delta^{(2)}(z-z_i)\,,\qquad z=x+iy\,.
\label{eq:nonhermitian-spectral-density}
\end{equation}
The non-Hermitian analogue of the logarithmic potential is
\begin{equation}
\Phi_{\pmb A,\eta}(z,z^*)=\frac{1}{N}\log\det\left[(z\pmb I-\pmb A)(z^*\pmb I-\pmb A^\dagger) +\eta^2\pmb I\right]\,,\qquad \eta>0\,.
\label{eq:nonhermitian-log-potential}
\end{equation}
Using the identity
\begin{equation}
\partial_{z^*}\frac{1}{z-z_i} =\pi\delta^{(2)}(z-z_i),
\label{eq:complex-delta-identity}
\end{equation}
one obtains the expression
\begin{equation}
\rho_{\pmb A}(z)=\frac{1}{\pi}\lim_{\eta\downarrow 0}\partial_{z^*}\partial_z\Phi_{\pmb A,\eta}(z,z^*)\,,
\label{eq:nonhermitian-density-log-potential}
\end{equation}
or equivalently
\begin{equation}
\rho_{\pmb A}(z)=\frac{1}{4\pi}\lim_{\eta\downarrow 0}\Delta_z\Phi_{\pmb A,\eta}(z,z^*)\,,\qquad\Delta_z=\partial_x^2+\partial_y^2\,.
\label{eq:nonhermitian-density-laplacian}
\end{equation}
This is Girko's Hermitization principle in the language used in physics and random matrix theory \cite{Girko1984,FeinbergZee1997,TaoVuKrishnapur2010}.

\begin{examplebox}[Hermitization for a $1\times1$ non-Hermitian matrix]
Take the $1\times1$ matrix
\begin{equation}
\pmb A=(a)\,,\qquad a\in\mathbb C\,.
\label{eq:sor-example-one-by-one-nh}
\end{equation}
Its spectral density should be
\begin{equation}
\rho_{\pmb A}(z)=\delta^{(2)}(z-a)\,.
\label{eq:sor-example-one-by-one-target}
\end{equation}
Let us recover this from the Hermitized logarithmic potential. Since
\begin{equation}
(z\pmb I-\pmb A)(z^*\pmb I-\pmb A^\dagger)+\eta^2\pmb I=|z-a|^2+\eta^2\,,
\label{eq:sor-example-hermitized-scalar}
\end{equation}
we have
\begin{equation}
\Phi_{\pmb A,\eta}(z,z^*)=\log\left(|z-a|^2+\eta^2\right)\,.
\label{eq:sor-example-log-potential}
\end{equation}
Now
\begin{equation}
\partial_z\Phi_{\pmb A,\eta}(z,z^*)=\frac{z^*-a^*}{|z-a|^2+\eta^2}\,,
\label{eq:sor-example-dz}
\end{equation}
and differentiating once more,
\begin{equation}
\partial_{z^*}\partial_z\Phi_{\pmb A,\eta}(z,z^*)=\frac{\eta^2}{\left(|z-a|^2+\eta^2\right)^2}\,.
\label{eq:sor-example-dzdzbar}
\end{equation}
Therefore
\begin{equation}
\rho_{\pmb A,\eta}(z)=\frac{1}{\pi}\frac{\eta^2}{\left(|z-a|^2+\eta^2\right)^2}\,.
\label{eq:sor-example-regularized-two-dimensional-density}
\end{equation}
This is a normalized two-dimensional mollifier centered at $a$, and in the limit $\eta\downarrow0$ it converges in the sense of distributions to
\begin{equation}
\rho_{\pmb A}(z)=\delta^{(2)}(z-a)\,.
\label{eq:sor-example-two-dimensional-delta}
\end{equation}
Thus Hermitization reproduces the correct point spectrum in the simplest non-Hermitian example.
\end{examplebox}

The determinant in \eqref{eq:nonhermitian-log-potential} can again be represented by a Gaussian integral, now after doubling the degrees of freedom. Introduce the $2N\times 2N$ Hermitized block matrix
\begin{equation}
\pmb{\mathcal B}_{\pmb A}(z,\eta)=
\begin{pmatrix}
\eta\pmb I & i(z\pmb I-\pmb A)\\
i(z^*\pmb I-\pmb A^\dagger) & \eta\pmb I
\end{pmatrix}\,.
\label{eq:hermitized-block-matrix}
\end{equation}
A block-determinant identity gives
\[
\det \pmb{\mathcal B}_{\pmb A}(z,\eta)=\det\left[(z\pmb I-\pmb A)(z^*\pmb I-\pmb A^\dagger)+\eta^2\pmb I\right]\,.
\]
Thus the non-Hermitian spectral density is again obtained from the free energy of a Gaussian theory, but with two coupled fields per vertex. In sparse non-Hermitian ensembles the resulting cavity inverse variances are therefore matrix-valued rather than scalar. This is the basis of the cavity approach to non-Hermitian sparse matrices and of later developments on non-Hermitian number statistics \cite{RogersPerezCastillo2009,MetzNeriRogers2019,RamosSanchezGuzmanGonzalezPerezCastilloMetz2021}.

The observables introduced in this section form the technical backbone of the notes. The empirical density is recovered from the imaginary part of the trace resolvent; local densities are recovered from diagonal resolvent entries; covariance matrices can be linearized on bipartite graphs; counting observables are integrals of resolvents; and non-Hermitian spectral densities are obtained by Hermitization. The common feature is that all these objects can be written in terms of logarithms of determinants or Gaussian partition functions. Once this translation has been made, sparse random matrix theory becomes a problem in the statistical mechanics of disordered Gaussian models on sparse graphs.

\begin{exerciseblock}
 \exitem[Stieltjes inversion in the sign convention of the notes]
Let $\pmb A$ be an $N\times N$ real symmetric or Hermitian matrix with real eigenvalues $\lambda_1,\ldots,\lambda_N$. Starting from
\begin{equation}
g_{\pmb A}(z)=\frac{1}{N}\sum_{i=1}^{N}\frac{1}{z-\lambda_i}\,,\qquad z=\lambda-i\epsilon\,,\qquad\epsilon>0\,,
\label{eq:sor-ex1-resolvent}
\end{equation}
show that
\begin{equation}
{\rm Im}\frac{1}{\lambda-i\epsilon-\lambda_i}=\frac{\epsilon}{(\lambda-\lambda_i)^2+\epsilon^2}\,,
\label{eq:sor-ex1-lorentzian}
\end{equation}
and deduce
\begin{equation}
\rho_{\pmb A}(\lambda)=\frac{1}{\pi}\lim_{\epsilon\downarrow0}{\rm Im} g_{\pmb A}(\lambda-i\epsilon)\,.
\label{eq:sor-ex1-density}
\end{equation}

\exitem[Normalization of the regularized density]
Let $\pmb A$ be an $N\times N$ real symmetric or Hermitian matrix with eigenvalues $\lambda_1,\ldots,\lambda_N$, and define
\begin{equation}
\rho_{\pmb A,\epsilon}(\lambda)=\frac{1}{\pi N}\sum_{i=1}^N\frac{\epsilon}{(\lambda-\lambda_i)^2+\epsilon^2}\,.
\label{eq:sor-ex2-regularized-density}
\end{equation}
Show that for every fixed $\epsilon>0$,
\begin{equation}
\int_{-\infty}^{\infty}d\lambda \rho_{\pmb A,\epsilon}(\lambda)=1\,.
\label{eq:sor-ex2-normalization}
\end{equation}
Why is this a useful numerical check in later sections?

\exitem[Local density of states from eigenvectors]
Let $\pmb A$ be an $N\times N$ real symmetric or Hermitian matrix with normalized eigenvectors $\pmb u^{(j)}$ and eigenvalues $\lambda_j$. Starting from the spectral decomposition of the resolvent,
\begin{equation}
G_{ii}(z)=\sum_{j=1}^{N}\frac{|u_i^{(j)}|^2}{z-\lambda_j}\,,
\label{eq:sor-ex3-local-resolvent}
\end{equation}
derive
\begin{equation}
\rho_i(\lambda)=\sum_{j=1}^{N}|u_i^{(j)}|^2\delta(\lambda-\lambda_j)\,.
\label{eq:sor-ex3-local-density}
\end{equation}
Then verify explicitly that
\begin{equation}
\rho_{\pmb A}(\lambda)=\frac{1}{N}\sum_{i=1}^{N}\rho_i(\lambda)\,.
\label{eq:sor-ex3-global-local-relation}
\end{equation}

\exitem[Derivative of the logarithmic determinant]
Let $\pmb A$ be an $N\times N$ matrix, let $\pmb I$ be the $N\times N$ identity matrix, and let $z$ be such that $z\pmb I-\pmb A$ is invertible. Prove the matrix identity
\begin{equation}
\frac{\partial}{\partial z}\log\det(z\pmb I-\pmb A)={\rm Tr} (z\pmb I-\pmb A)^{-1}\,.
\label{eq:sor-ex4-logdet}
\end{equation}
Then use it to recover equation \eqref{eq:resolvent-log-det} from the text.

\exitem[The Edwards--Jones Gaussian integral]
Let $z\in\mathbb C$ with ${\rm Im}\,z<0$. Diagonalize an $N\times N$ real symmetric matrix $\pmb A$ and use the one-dimensional Gaussian integral
\begin{equation}
\int_{-\infty}^{\infty}\frac{dx}{\sqrt{2\pi}}e^{-\frac{a}{2}x^2}=a^{-1/2}\,,\qquad{\rm Re} a>0\,,
\label{eq:sor-ex5-one-dimensional-gaussian}
\end{equation}
to derive
\begin{equation}
Z_{\pmb A}(z)=\det\left[i(z\pmb I-\pmb A)\right]^{-1/2}\,.
\label{eq:sor-ex5-edwards-jones}
\end{equation}
Why is the choice ${\rm Im}\,z<0$ convenient for convergence?

\exitem[Schur complement and the preview cavity recursion]
Consider a real symmetric matrix $\pmb A$ supported on a tree, and a vertex $j$ connected to neighbors in the set $\partial j$. Let $G_{\ell\to j}(z)$ denote the diagonal resolvent entry at vertex $\ell$ in the branch obtained by removing the edge between $\ell$ and $j$. Write the matrix $z\pmb I-\pmb A$ in block form with the row and column corresponding to vertex $j$ separated from the remaining vertices, and use the Schur complement formula to derive
\begin{equation}
G_{jj}(z)=\frac{1}{z-A_{jj}-\displaystyle\sum_{\ell\in\partial j}A_{j\ell}^2G_{\ell\to j}(z)}\,.
\label{eq:sor-ex6-schur}
\end{equation}
This reproduces the cavity structure previewed in the section.

\exitem[A small exact cavity check]
Take the three-site chain
\begin{equation}
\pmb A=\begin{pmatrix}
0 & 1 & 0\\
1 & 0 & 1\\
0 & 1 & 0
\end{pmatrix}\,.
\label{eq:sor-ex7-three-site-chain}
\end{equation}
Compute the full resolvent entry $G_{22}(z)$ directly from $(z\pmb I-\pmb A)^{-1}$. Then compute it again using the Schur-complement or cavity recursion of Exercise 3.6. Verify that the two answers agree exactly.

\exitem[Bipartite linearization]
Let $\pmb X$ be a real $N\times P$ matrix, let $d>0$, and define
\begin{equation}
\pmb W=\frac{1}{d}\pmb X\pmb X^{\rm T}\,,\qquad\pmb{\mathcal L}=\frac{1}{\sqrt d}
\begin{pmatrix}
\pmb 0_{N\times N} & \pmb X\\
\pmb X^{\rm T} & \pmb 0_{P\times P}
\end{pmatrix}\,.
\label{eq:sor-ex8-linearization}
\end{equation}
Show that
\begin{equation}
\pmb{\mathcal L}^2=\frac{1}{d}\begin{pmatrix}
\pmb X\pmb X^{\rm T} & 0\\
0 & \pmb X^{\rm T}\pmb X
\end{pmatrix}\,.
\label{eq:sor-ex8-square}
\end{equation}
Deduce that the nonzero eigenvalues of $\pmb{\mathcal L}$ are $\pm\sqrt{\nu}$, where $\nu$ runs over the nonzero eigenvalues of $\pmb W$.

\exitem[Interval counts from the resolvent]
Let $\pmb A$ be an $N\times N$ real symmetric or Hermitian matrix with eigenvalues $\lambda_1,\ldots,\lambda_N$, and assume that no eigenvalue lies exactly at $a$ or $b$. Starting from
\begin{equation}
{\rm Tr}(z\pmb I-\pmb A)^{-1}=\sum_{i=1}^{N}\frac{1}{z-\lambda_i}\,,
\label{eq:sor-ex9-trace-resolvent}
\end{equation}
show that
\begin{equation}
\mathcal N_{\pmb A}([a,b])=\frac{1}{\pi}\lim_{\epsilon\downarrow0}{\rm Im}\int_a^b d\lambda{\rm Tr}\left[(\lambda-i\epsilon)\pmb I-\pmb A\right]^{-1}\,.
    \label{eq:sor-ex9-interval-count}
\end{equation}
Explain why this formula is the natural precursor of the index-number large-deviation theory.

\exitem[A direct check of Hermitization]
For the scalar non-Hermitian matrix $\pmb A=(a)$, with $a\in\mathbb C$, derive explicitly
\begin{equation}
\rho_{\pmb A,\eta}(z)=\frac{1}{\pi}\frac{\eta^2}{\left(|z-a|^2+\eta^2\right)^2}\,,
\label{eq:sor-ex10-hermitized-density}
\end{equation}
starting from the logarithmic potential. Then verify directly that
\begin{equation}
\int_{\mathbb C} d^2z\rho_{\pmb A,\eta}(z)=1\,.
\label{eq:sor-ex10-normalization}
\end{equation}

\exitem[Programming exercise: diagonalization versus resolvent]
Fix a mean degree $c=O(1)$, a list of system sizes $N$, a number $S$ of independent samples for each $N$, a regulator $\epsilon>0$, and a grid of real values of $\lambda$. For each value of $N$ and each sample, generate an $N\times N$ real symmetric sparse Erd\H{o}s--R\'enyi adjacency matrix $\pmb A$ with $A_{ii}=0$, $A_{ij}=A_{ji}$, and
\begin{equation}
{\rm Prob}(A_{ij}=1)=\frac{c}{N}\,,\qquad {\rm Prob}(A_{ij}=0)=1-\frac{c}{N}\,,\qquad i<j\,.
\label{eq:sor-ex11-er-program}
\end{equation}
For that sample:
\begin{itemize}
\item diagonalize the matrix and build the broadened density from the eigenvalues,
\begin{equation}
\rho_{\epsilon}^{\rm diag}(\lambda)=\frac{1}{\pi N}\sum_{i=1}^{N}\frac{\epsilon}{(\lambda-\lambda_i)^2+\epsilon^2}\,,
\label{eq:sor-ex11-diag-density}
\end{equation}
and
\item compute the trace of the resolvent on a grid of $\lambda$ values and build
\begin{equation}
\rho_{\epsilon}^{\rm res}(\lambda)=\frac{1}{\pi}{\rm Im}\frac{1}{N}{\rm Tr}\left[(\lambda-i\epsilon)\pmb I-\pmb A\right]^{-1}\,.
\label{eq:sor-ex11-resolvent-density}
\end{equation}
\end{itemize}
Verify numerically, on the same $\lambda$ grid and at the same value of $\epsilon$, that the two procedures agree. Report the values of $c$, $N$, $S$, $\epsilon$, the grid spacing, and a simple discrepancy measure such as the maximum absolute difference between $\rho_{\epsilon}^{\rm diag}$ and $\rho_{\epsilon}^{\rm res}$ on the grid.
\end{exerciseblock}

\section{Replica and cavity methods for disordered systems}
\label{sec:replica-cavity-methods}
The previous section showed that spectral observables of sparse matrices can be represented through logarithms of Gaussian partition functions. This places the spectral problem in the same formal class as the computation of quenched free energies in disordered statistical mechanics. The purpose of this section is to recall the two main tools that will be used throughout these notes: the replica method and the cavity method. For the locally tree-like finite-connectivity systems that will concern us, these two approaches are closely related and, at the replica-symmetric level, often lead to the same self-consistency equations. In that sense, choosing one or the other is sometimes a matter of taste. Their practical use, however, is not equally transparent in all problems: the cavity method is often more natural when the graph structure directly suggests recursive local equations and message-passing algorithms, whereas an equivalent replica derivation may be considerably more cumbersome. Conversely, replica methods can be more convenient when one wants direct access to disorder-averaged free energies or generating functions. We therefore introduce both viewpoints, first in the more familiar setting of spin systems on sparse random graphs, and then explain how the same logic applies to Gaussian models generated by resolvent and determinant representations.

The historical origin of these ideas lies in the statistical mechanics of disordered systems. In a spin glass, the interactions are random and remain fixed during the thermal equilibration of the spins. The disorder is therefore quenched. The Edwards--Anderson model introduced this viewpoint for finite-dimensional spin glasses \cite{EdwardsAnderson1975}, while the Sherrington--Kirkpatrick model provided a model with infinite-range random couplings for which mean field theory became exact \cite{SherringtonKirkpatrick1975}. The replica method became the central analytic tool for this class of problems, and a systematic and historical development can be reviewed in \cite{MezardParisiVirasoro1987}. For finite-connectivity systems, such as diluted spin glasses and spin models on random graphs, the relevant mean-field geometry is not the complete graph but a locally tree-like sparse graph. The Viana--Bray model is the canonical diluted spin-glass example \cite{VianaBray1985}, and the cavity method gives the natural description of such finite-connectivity systems \cite{MezardParisi2001,MezardMontanari2009}.

To introduce these ideas, we will use one of the most important models in physics, the Ising model, but now placed on a simple undirected graph $\mathcal{G}=(V,E)$, where $V=\{1,\ldots,N\}$ is the set of vertices and $E$ is a set of unordered edges. The spin variables are $\sigma_i\in\{-1,1\}$, and the Hamiltonian is
\begin{equation}
 H_{\mathcal{G},\pmb{J},\pmb{B}}(\pmb{\sigma})=-\sum_{(i,j)\in E}J_{ij}\sigma_i\sigma_j-\sum_{i=1}^N B_i\sigma_i\,.
\label{eq:ising-hamiltonian-random-graph}
\end{equation}
Here the graph $\mathcal{G}$, the couplings $\pmb{J}=\{J_{ij}\}$, and possibly the fields $\pmb{B}=\{B_i\}$ are all considered quenched random variables. For this fixed disorder, the partition function is
\begin{equation}
\mathcal{Z}_{\mathcal{G},\pmb{J},\pmb{B}}=\sum_{\pmb{\sigma}\in\{-1,1\}^N}\exp\left[-\beta H_{\mathcal{G},\pmb{J},\pmb{B}}(\pmb{\sigma})\right]\,,
\label{eq:ising-partition-function-random-graph}
\end{equation}
and the corresponding free energy is
\begin{equation}
-\beta F_{\mathcal{G},\pmb{J},\pmb{B}}=\log\mathcal{Z}_{\mathcal{G},\pmb{J},\pmb{B}}\,.
\label{eq:sample-free-energy}
\end{equation}
The thermodynamic quantity of interest is usually not the free energy of a particular realization, but its typical value. In the thermodynamic limit this is encoded by the quenched free-energy density
\begin{equation}
-\beta f=\lim_{N\to\infty}\frac{1}{N}\overline{\log\mathcal{Z}_{\mathcal{G},\pmb{J},\pmb{B}}}\,,
\label{eq:quenched-free-energy-density}
\end{equation}
where the overbar denotes the average over the graph and coupling disorder, and $f$ is the limiting disorder-averaged free-energy density. The central difficulty is the logarithm. An annealed calculation would average $\mathcal{Z}$ before taking the logarithm, but the quenched problem requires the average of $\log\mathcal{Z}$. This distinction is the statistical-mechanics analogue of the distinction, in sparse random matrix theory, between averaging a determinant and averaging the logarithm of a determinant.

\begin{examplebox}[Annealed versus quenched Edwards--Jones at small connectivity]
It is useful to see explicitly why replacing a quenched average by an annealed one is not an innocent operation in sparse spectral problems. We illustrate this in the simplest possible setting: the adjacency matrix of an Erd\H{o}s--R\'enyi graph in the low-connectivity expansion.

Let $\pmb A$ be the adjacency matrix of an Erd\H{o}s--R\'enyi graph with edge probability $c/N$, and consider the Edwards--Jones partition function
\begin{equation}
Z_{\pmb A}(z)=\int\left[\prod_{i=1}^{N}\frac{du_i}{\sqrt{2\pi}}\right]\exp\left[-\frac{i}{2}\pmb u^{\rm T}(z\pmb I-\pmb A)\pmb u\right]\,,\qquad z=\lambda-i\epsilon\,,\qquad\epsilon>0\,.
\label{eq:rcmds-ann-ej-partition}
\end{equation}
For the empty graph, $\pmb A=\pmb 0$, the partition function is
\begin{equation}
Z_0(z)=(iz)^{-N/2}\,.
\label{eq:rcmds-ann-empty-partition}
\end{equation}
Now add a single edge between two vertices. The only nontrivial block is
\begin{equation}
\pmb A_{\rm edge}=\begin{pmatrix}
0 & 1\\
1 & 0
\end{pmatrix}\,,
\label{eq:rcmds-ann-edge-block}
\end{equation}
and the ratio between the two-site edge partition function and two isolated one-site partition functions is
\begin{equation}
r(z)=\frac{\det[i(z\pmb I_2-\pmb A_{\rm edge})]^{-1/2}}{(iz)^{-1}}=\left(1-\frac{1}{z^2}\right)^{-1/2}\,,
\label{eq:rcmds-ann-edge-ratio}
\end{equation}
up to a $z$-independent phase, which does not affect the spectral density.

At small $c$, the graph is, to first order in $c$, a collection of isolated vertices and isolated edges. Components with two or more adjacent edges contribute only at order $c^2$. Hence the quenched free energy per site is
\begin{equation}
\frac{1}{N}\overline{\log Z_{\pmb A}(z)}=-\frac{1}{2}\log(iz)+\frac{c}{2}\log r(z)+O(c^2)\,.
\label{eq:rcmds-ann-quenched-free-energy}
\end{equation}
Using \eqref{eq:rcmds-ann-edge-ratio}, this becomes
\begin{equation}
\frac{1}{N}\overline{\log Z_{\pmb A}(z)}=-\frac{1}{2}\log(iz)-\frac{c}{4}\log\left(1-\frac{1}{z^2}\right)+O(c^2)\,.
\label{eq:rcmds-ann-quenched-free-energy-explicit}
\end{equation}
The corresponding quenched resolvent is
\begin{equation}
g_{\rm q}(z)=-2\frac{\partial}{\partial z}\left[\frac{1}{N}\overline{\log Z_{\pmb A}(z)}\right]\,,
\label{eq:rcmds-ann-quenched-resolvent-def}
\end{equation}
and therefore
\begin{equation}
g_{\rm q}(z)=\frac{1}{z}+\frac{c}{z(z^2-1)}+O(c^2)\,.
\label{eq:rcmds-ann-quenched-resolvent}
\end{equation}
Equivalently,
\begin{equation}
g_{\rm q}(z)=\frac{1-c}{z}+\frac{c}{2}\left(\frac{1}{z-1}+\frac{1}{z+1}\right)+O(c^2)\,.
    \label{eq:rcmds-ann-quenched-physical-form}
\end{equation}
This is exactly the physical low-connectivity picture: with probability $1-c+O(c^2)$ a uniformly chosen vertex is isolated and contributes an eigenvalue at zero, while with probability $c+O(c^2)$ it belongs to an isolated edge, whose two eigenvalues are $+1$ and $-1$. Thus
\begin{equation}
\rho_{\rm q}(\lambda)=(1-c)\delta(\lambda)+\frac{c}{2}\delta(\lambda-1)+\frac{c}{2}\delta(\lambda+1)+O(c^2)\,.
\label{eq:rcmds-ann-quenched-density}
\end{equation}

Now repeat the calculation in the annealed approximation. Instead of computing
$\overline{\log Z_{\pmb A}(z)}$, one computes $\log\overline{Z_{\pmb A}(z)}$. To first order in $c$, each possible edge contributes the factor $r(z)-1$, and there are asymptotically $N^2/2$ possible unordered pairs, each present with probability $c/N$. Hence
\begin{equation}
\overline{Z_{\pmb A}(z)}=Z_0(z)\exp\left[\frac{cN}{2}\big(r(z)-1\big)+O(c^2N)\right]\,,
\label{eq:rcmds-ann-annealed-partition}
\end{equation}
so that
\begin{equation}
\frac{1}{N}\log\overline{Z_{\pmb A}(z)}=-\frac{1}{2}\log(iz)+\frac{c}{2}\left[\left(1-\frac{1}{z^2}\right)^{-1/2}-1\right]+O(c^2)\,.
    \label{eq:rcmds-ann-annealed-free-energy}
\end{equation}
The corresponding annealed resolvent is
\begin{equation}
g_{\rm a}(z)=-2\frac{\partial}{\partial z}\left[\frac{1}{N}\log\overline{Z_{\pmb A}(z)}\right]\,,
\label{eq:rcmds-ann-annealed-resolvent-def}
\end{equation}
and therefore
\begin{equation}
g_{\rm a}(z)=\frac{1}{z}+\frac{c}{z^3}\left(1-\frac{1}{z^2}\right)^{-3/2}+O(c^2)\,.
    \label{eq:rcmds-ann-annealed-resolvent}
\end{equation}
This differs from \eqref{eq:rcmds-ann-quenched-resolvent}. In particular, the quenched expression has simple poles at $z=0,\pm1$, reflecting isolated vertices and isolated edges, whereas the annealed expression has a branch singularity inherited from the averaged partition function. Thus, already at order $c$, the annealed Edwards--Jones calculation does not reproduce the finite-connectivity spectral density.

The lesson is not that annealed calculations are never useful. In dense Gaussian ensembles, and in some high-temperature disordered systems, annealed and quenched free energies may agree at leading order. The lesson is more precise: in sparse spectral problems, annealing may wash out the local component structure that is visible to the quenched logarithm. The quenched average sees the logarithm of the component contribution; the annealed average sees the component contribution before taking the logarithm. These two operations are not equivalent.
\end{examplebox}

The replica method addresses the logarithm through the identity
\begin{equation}
\log \mathcal{Z}=\lim_{n\to 0}\frac{\mathcal{Z}^n-1}{n}\,.
\label{eq:basic-replica-identity}
\end{equation}
In thermodynamic calculations one usually implements this in the form
\begin{equation}
-\beta f=\lim_{n\to 0}\lim_{N\to\infty}\frac{1}{nN}\log\overline{\mathcal{Z}_{\mathcal{G},\pmb{J},\pmb{B}}^{n}}\,,
    \label{eq:replicated-free-energy-density}
\end{equation}
where the interchange of the replica and thermodynamic limits is a formal step of the method. In these notes we shall use this formula as an organizing device, without entering into the mathematical problem of justifying this interchange.

For positive integer $n$, the quantity $\mathcal{Z}^n$ is the partition function of $n$ identical copies, or replicas, of the original system. The replicated spins are
\begin{equation}
\underline{\sigma}_i=(\sigma_i^1,\ldots,\sigma_i^n)\,,\qquad\sigma_i^a\in\{-1,1\}\,,\qquad a=1,\ldots,n\,.
\label{eq:replicated-spin-vector}
\end{equation}
The disorder average couples the replicas, even though the replicas were independent before the average is performed. For an Erd\H{o}s--R\'enyi graph with mean degree $c$, independent couplings drawn from a distribution $p_J(J)$, and fixed external fields $B_i$, one obtains schematically
\begin{equation}
\overline{\mathcal{Z}_{\mathcal{G},\pmb{J},\pmb{B}}^{n}}=\sum_{\{\sigma_i^a\}}\exp\left[\frac{c}{2N}\sum_{i,j=1}^N\left(\int dJ\,p_J(J)\exp\left[\beta J\sum_{a=1}^n\sigma_i^a\sigma_j^a\right]-1\right)+\beta\sum_{i=1}^N\sum_{a=1}^nB_i\sigma_i^a\right]\,,
\label{eq:replicated-ising-er-average}
\end{equation}
where terms subleading in $N$ have been ignored as they are not relevant for the typical properties of the system in the thermodynamic limit. This expression illustrates the characteristic structure of finite-connectivity replica calculations. The effective interaction after disorder averaging depends on the whole replicated spin vector at each site, not only on a small number of scalar order parameters.

The natural order parameter is therefore the empirical distribution of replicated spin vectors,
\begin{equation}
P(\underline{\sigma})=\frac{1}{N}\sum_{i=1}^N\delta_{\underline{\sigma},\underline{\sigma}_i}\,,\qquad\sum_{\underline{\sigma}\in\{-1,1\}^n}P(\underline{\sigma})=1\,.
\label{eq:replicated-spin-order-parameter}
\end{equation}

\begin{examplebox}[Why the finite-connectivity order parameter is a distribution]
For $n=2$ replicas, the replicated spin at one site can take four values,
\begin{equation}
(\sigma^1,\sigma^2)\in\{(++),(+-),(-+),(--)\}\,.\label{eq:rcmds-example-replicated-states}
\end{equation}
Let us write
\begin{equation}
P_{++}=P(+,+)\,,\qquad P_{+-}=P(+,-)\,, \qquad P_{-+}=P(-,+)\,, \qquad P_{--}=P(-,-)\,,
\label{eq:rcmds-example-four-probabilities}
\end{equation}
with
\begin{equation}
P_{++}+P_{+-}+P_{-+}+P_{--}=1\,.
\label{eq:rcmds-example-normalization}
\end{equation}
From this distribution one can reconstruct the replica magnetizations and the overlap:
\begin{align}
m_1&=\sum_{\sigma^1,\sigma^2}P(\sigma^1,\sigma^2)\sigma^1=P_{++}+P_{+-}-P_{-+}-P_{--}\,,\label{eq:rcmds-example-m1}\\
m_2&=\sum_{\sigma^1,\sigma^2}P(\sigma^1,\sigma^2)\sigma^2=P_{++}-P_{+-}+P_{-+}-P_{--}\,,
\label{eq:rcmds-example-m2}\\
q_{12}&=\sum_{\sigma^1,\sigma^2}P(\sigma^1,\sigma^2)\sigma^1\sigma^2=P_{++}+P_{--}-P_{+-}-P_{-+}\,.
\label{eq:rcmds-example-q12}
\end{align}
Thus the overlap is only one moment of the full distribution $P(\sigma^1,\sigma^2)$.

To see that $P$ contains more information than $q_{12}$ alone, consider the two distributions
\begin{equation}
P^{(A)}_{++}=\frac{1}{2}\,,\qquad P^{(A)}_{--}=\frac{1}{2}\,,\qquad P^{(A)}_{+-}=P^{(A)}_{-+}=0\,,
\label{eq:rcmds-example-distribution-A}
\end{equation}
and
\begin{equation}
P^{(B)}_{++}=1\,, \qquad P^{(B)}_{+-}=P^{(B)}_{-+}=P^{(B)}_{--}=0\,.
\label{eq:rcmds-example-distribution-B}
\end{equation}
In both cases
\begin{equation}
q_{12}=1\,,
\label{eq:rcmds-example-same-overlap}
\end{equation}
but the magnetizations are different:
\begin{equation}
m_1^{(A)}=m_2^{(A)}=0\,,\qquad m_1^{(B)}=m_2^{(B)}=1\,.
\label{eq:rcmds-example-different-magnetizations}
\end{equation}
This simple calculation shows why, in finite-connectivity systems, the natural order parameter is the whole empirical distribution of replicated local configurations rather than only the overlap matrix.
\end{examplebox}

In fully connected spin glasses the standard replica order parameter is the overlap matrix $q_{ab}=N^{-1}\sum_i \sigma_i^a\sigma_i^b$. In finite-connectivity systems the correct order parameter is, however, richer: it is a probability distribution over replicated local configurations $P(\underline{\sigma})$. This is the replica signature of the fact that vertices have different local neighborhoods and therefore experience different effective fields. After introducing the order parameter function \eqref{eq:replicated-spin-order-parameter}, the replicated partition function takes a large-deviation form
\begin{equation}
\overline{\mathcal{Z}_{\mathcal{G},\pmb{J},\pmb{B}}^{n}}=\int \mathcal{D}P\exp\left[ N\Psi_n[P]\right]\,,
\label{eq:replicated-functional-integral}
\end{equation}
and the thermodynamic limit is evaluated by the saddle-point method,
\begin{equation}
\lim_{N\to\infty}\frac{1}{N}\log\overline{\mathcal{Z}_{\mathcal{G},\pmb{J},\pmb{B}}^{n}}=\operatorname*{extr}_{P}\Psi_n[P]\,.
\label{eq:replicated-saddle-point}
\end{equation}
The analytic continuation $n\to 0$ is the formal step of the replica method. It is powerful, but it is also the step at which care is needed: one makes an ansatz about how the system of replicas behaves under permutations of the replicas, computes $\Psi_n[P]$ for integer $n$ under this ansatz, rewrites the result in a form suitable for analytic continuation, and finally evaluates the $n\to 0$ limit.

Within the replica-symmetric (RS) ansatz, the order parameter \eqref{eq:replicated-spin-order-parameter} is assumed to be invariant under permutations of replicas. For finite-connectivity Ising systems the order parameter function $P(\underline{\sigma})$ then takes the form
\begin{equation}
P_{\rm RS}(\sigma^1,\ldots,\sigma^n)=\int dh\,W(h)\prod_{a=1}^n\frac{e^{\beta h\sigma^a}}{2\cosh(\beta h)}\,,
\label{eq:rs-field-distribution-ansatz}
\end{equation}
where $W(h)$ is the mixing distribution of effective fields in the replica-symmetric order parameter. Thus the replica-symmetric order parameter is not a single magnetization or overlap, but a full distribution of fields. In the cavity interpretation below, this distribution is identified with the ensemble law of a typical directed-edge cavity field. This is the point where the replica and cavity descriptions meet. Appendix~\ref{app:replica-symmetric-saddle-points} gives an explicit replica-symmetric saddle-point derivation showing how the corresponding cavity-message distribution appears from the replicated disorder average.

The cavity method starts from the observation that sparse random graphs are locally tree-like. If the edge $(i,j)$ is removed, the branches attached to $i$ through the neighbors $k\in\partial i\setminus j$ are disconnected from the rest of the graph through $j$; on a tree, these branches are independent once $\sigma_i$ is fixed. One therefore introduces a cavity marginal $\psi_{i\to j}(\sigma_i)$ which corresponds to the marginal distribution of $\sigma_i$ in the graph where the edge between $i$ and $j$ has been deleted. On a tree these cavity marginals obey the exact recursion
\begin{equation}
\psi_{i\to j}(\sigma_i)=\frac{1}{Z_{i\to j}}e^{\beta B_i\sigma_i}\prod_{k\in\partial i\setminus j}\left[\sum_{\sigma_k=\pm1}e^{\beta J_{ik}\sigma_i\sigma_k}\psi_{k\to i}(\sigma_k)\right]\,,
\label{eq:bp-ising-message}
\end{equation}
where $\partial i$ is the set of neighbors of $i$, and $Z_{i\to j}$ is a normalization factor. This is the belief-propagation equation for the Ising model on a graph. Belief propagation was developed in probability and information theory as a message-passing algorithm on graphical models \cite{Pearl1988,KschischangFreyLoeliger2001}, while in statistical mechanics its fixed points are the stationary points of the Bethe free energy \cite{Bethe1935,YedidiaFreemanWeiss2005}. On a tree the equations are exact. On a sparse random graph they are asymptotically exact for local observables in regimes where the locally tree-like approximation and the assumed pure-state structure are valid.

It is often convenient to parametrize the binary cavity marginals in terms of a cavity field,
\begin{equation}
\psi_{i\to j}(\sigma_i)=\frac{e^{\beta h_{i\to j}\sigma_i}}{2\cosh(\beta h_{i\to j})}\,.
\label{eq:ising-message-field-parametrization}
\end{equation}
Substituting \eqref{eq:ising-message-field-parametrization} into \eqref{eq:bp-ising-message} gives the following set of self-consistency equations known as cavity equations
\begin{equation}
h_{i\to j}=B_i+\sum_{k\in\partial i\setminus j}u(J_{ik},h_{k\to i})\,,
\label{eq:ising-cavity-field-recursion}
\end{equation}
with $ u(J,h)$ being the propagating field
\begin{equation}
u(J,h) =\frac{1}{\beta}\operatorname{arctanh}\left[\tanh(\beta J)\tanh(\beta h)\right]\,.
\label{eq:ising-message-function}
\end{equation}
From the cavity marginals one can reconstruct the physical single-site marginal $P_i(\sigma_i)$. It can be parametrized in the same way by a full effective, or physical, field $h_i$, given by
\begin{equation}
h_i=B_i+\sum_{k\in\partial i}u(J_{ik},h_{k\to i})\,.
\label{eq:ising-full-field}
\end{equation}
Local observables then follow from the corresponding single-site marginalization. For example,
\begin{equation}
\langle \sigma_i\rangle=\tanh(\beta h_i)\,.
\label{eq:ising-local-magnetization}
\end{equation}

\begin{examplebox}[Cavity fields on a three-spin chain]
Consider the Ising chain
\begin{equation}
1\!-\!2\!-\!3
\label{eq:rcmds-example-chain-graph}
\end{equation}
with Hamiltonian
\begin{equation}
H(\sigma_1,\sigma_2,\sigma_3)=-J_{12}\sigma_1\sigma_2-J_{23}\sigma_2\sigma_3-B_1\sigma_1-B_2\sigma_2-B_3\sigma_3\,.
\label{eq:rcmds-example-chain-hamiltonian}
\end{equation}
Since vertex $3$ is a leaf, the cavity marginal sent from $3$ to $2$ is simply
\begin{equation}
\psi_{3\to 2}(\sigma_3)\propto e^{\beta B_3\sigma_3}\,.
\label{eq:rcmds-example-leaf-message}
\end{equation}
Comparing with \eqref{eq:ising-message-field-parametrization}, we read off
\begin{equation}
h_{3\to 2}=B_3\,.
\label{eq:rcmds-example-leaf-field}
\end{equation}

Now let us compute the cavity marginal sent from $2$ to $1$. By definition,
\begin{equation}
\psi_{2\to 1}(\sigma_2)\propto e^{\beta B_2\sigma_2} \sum_{\sigma_3=\pm1} e^{\beta J_{23}\sigma_2\sigma_3} \psi_{3\to 2}(\sigma_3)\,.
\label{eq:rcmds-example-middle-message-start}
\end{equation}
Using \eqref{eq:rcmds-example-leaf-message},
\begin{equation}
\psi_{2\to 1}(\sigma_2) \propto e^{\beta B_2\sigma_2} \sum_{\sigma_3=\pm1} e^{\beta (J_{23}\sigma_2+B_3)\sigma_3} = e^{\beta B_2\sigma_2} 2\cosh\left[\beta(J_{23}\sigma_2+B_3)\right]\,.
\label{eq:rcmds-example-middle-message-cosh}
\end{equation}
To identify the corresponding cavity field, take the ratio
\begin{equation}
\frac{\psi_{2\to 1}(+1)}{\psi_{2\to 1}(-1)}=\exp\left(2\beta h_{2\to 1}\right)\,.
\label{eq:rcmds-example-ratio-definition}
\end{equation}
From \eqref{eq:rcmds-example-middle-message-cosh},
\begin{equation}
\exp\left(2\beta h_{2\to 1}\right)=e^{2\beta B_2}\frac{\cosh[\beta(B_3+J_{23})]}{\cosh[\beta(B_3-J_{23})]}\,.
\label{eq:rcmds-example-ratio-explicit}
\end{equation}
Hence
\begin{equation}
h_{2\to 1}=B_2+\frac{1}{2\beta}\log\frac{\cosh[\beta(B_3+J_{23})]}{\cosh[\beta(B_3-J_{23})]}\,.
\label{eq:rcmds-example-middle-field-log}
\end{equation}
Using the identity
\begin{equation}
\frac{1}{2}\log\frac{\cosh(x+y)}{\cosh(x-y)}=\operatorname{arctanh}\!\big(\tanh x\,\tanh y\big)\,,
\label{eq:rcmds-example-cosh-identity}
\end{equation}
we recover the cavity-field recursion
\begin{equation}
h_{2\to 1}=B_2+u(J_{23},B_3)\,,\qquad u(J,h) =\frac{1}{\beta}\operatorname{arctanh}\!\big[\tanh(\beta J)\tanh(\beta h)\big]\,.
\label{eq:rcmds-example-middle-field-u}
\end{equation}

The full effective field at the middle spin is obtained by including both neighbors:
\begin{equation}
h_2=B_2+u(J_{12},B_1)+u(J_{23},B_3)\,,
\label{eq:rcmds-example-full-middle-field}
\end{equation}
and therefore
\begin{equation}
\langle \sigma_2\rangle=\tanh(\beta h_2)\,.
\label{eq:rcmds-example-middle-magnetization}
\end{equation}
This is the simplest explicit derivation of the belief-propagation/cavity recursion on a tree.
\end{examplebox}

At the ensemble level one describes the statistics of a typical cavity field. For a finite undirected graph $\mathcal{G}=(V,E)$, let $\overrightarrow E=\{(i,j),(j,i):\{i,j\}\in E\}$ be the set of oriented edges obtained by giving both orientations to every undirected edge. A cavity fixed point assigns one field $h_{i\to j}$ to each $(i,j)\in\overrightarrow E$, namely the field sent from $i$ to $j$ in the graph where the edge between $i$ and $j$ has been removed. The empirical cavity-field law on this finite graph is therefore
\[
W_N^{\rm cav}(h)=\frac{1}{|\overrightarrow E|}\sum_{(i,j)\in\overrightarrow E}\delta(h-h_{i\to j})\,.
\]
When this empirical directed-edge law has a deterministic large-$N$ limit, we denote it by $W(h)$. For an Erd\H{o}s--R\'enyi graph with mean degree $c$, independent couplings with distribution $p_J(J)$, and a fixed external field $B$, this limiting cavity-field law obeys
\begin{equation}
W(h)=\sum_{\ell=0}^{\infty}e^{-c}\frac{c^\ell}{\ell!}\int\left[\prod_{r=1}^{\ell}dh_r\,W(h_r)dJ_r p_J(J_r)\right]\delta\left(h-B-\sum_{r=1}^{\ell}u(J_r,h_r)\right)\,.
\label{eq:er-cavity-field-distribution}
\end{equation}
For a general degree distribution, the Poisson law that appears in Eq.~\eqref{eq:er-cavity-field-distribution} is replaced by the appropriate excess-degree law. This distinction is crucial in sparse systems: a cavity message is attached to an oriented edge, and therefore its incoming neighborhood is distributed according to the degree distribution seen from a randomly chosen edge. This cavity-field law is distinct from the law of full site fields $h_i$, which include all neighbors of a vertex and are used to compute physical single-site observables. In random regular graphs, the cavity recursion has a fixed number $c-1$ of incoming messages; in heterogeneous networks, it involves the excess-degree distribution, equivalently the size-biased degree distribution shifted by one.

The Bethe free energy associated with a fixed point of \eqref{eq:bp-ising-message} is
\begin{equation}
-\beta F_{\rm Bethe}=\sum_{i=1}^N\log Z_i-\sum_{(i,j)\in E}\log Z_{ij}\,,
\label{eq:bethe-free-energy}
\end{equation}
where
\begin{equation}
Z_i=\sum_{\sigma_i=\pm1}e^{\beta B_i\sigma_i}\left[\prod_{k\in\partial i}\sum_{\sigma_k=\pm1}e^{\beta J_{ik}\sigma_i\sigma_k}\psi_{k\to i}(\sigma_k)\right]\,,
\label{eq:bethe-site-partition}
\end{equation}
and
\begin{equation}
Z_{ij}=\sum_{\sigma_i,\sigma_j=\pm1}e^{\beta J_{ij}\sigma_i\sigma_j}\psi_{i\to j}(\sigma_i)\psi_{j\to i}(\sigma_j)\,.
\label{eq:bethe-edge-partition}
\end{equation}
The first term in \eqref{eq:bethe-free-energy} sums the free-energy contributions of vertices, while the second subtracts the edge contributions that have been counted twice. On a tree this expression is exact. On a sparse random graph it is the finite-connectivity mean-field approximation associated with the locally tree-like structure.

The connection with random matrices is now direct. The Edwards--Jones representation rewrites the resolvent problem as a Gaussian partition function with quenched graph and weight disorder \cite{EdwardsJones1976}. For a symmetric diluted matrix $\pmb A$, the Gaussian measure associated with the spectral parameter $z=\lambda-i\epsilon$ has the form
\begin{equation}
Z_{\pmb A}(z)=\int\left[\prod_{i=1}^N\frac{dx_i}{\sqrt{2\pi}}\right]\exp\left[-\frac{i}{2}\sum_{i,j=1}^Nx_i(z\delta_{ij}-A_{ij})x_j\right]\,.
\label{eq:gaussian-spectral-partition-function}
\end{equation}
This is a continuous-spin model with complex quadratic interactions. The role played by the Ising free energy is now played by $\log Z_{\pmb A}(z)$, and the disorder average of the spectral density involves the quenched average of this logarithm. Thus the same formal obstruction appears:
\begin{equation}
\overline{\rho_{\pmb A}(\lambda)}=-\frac{2}{\pi N}\lim_{\epsilon\downarrow0}{\rm Im}\frac{\partial}{\partial\lambda}\overline{\log Z_{\pmb A}(\lambda-i\epsilon)}\,.
\label{eq:spectral-quenched-free-energy-recall}
\end{equation}
The replica method would compute this quantity by introducing $n$ copies of the Gaussian variables, averaging $Z_{\pmb{A}}^n$, and taking $n\to0$. This was the strategy initiated in the random-matrix context by Edwards and Jones and later applied to sparse random matrices by Rodgers and Bray \cite{EdwardsJones1976,RodgersBray1988}. The finite-connectivity nature of the problem again forces the order parameter to be a distribution, now of Gaussian variances or inverse variances rather than of Ising cavity fields.

The cavity method gives the same equations more directly. Because the model \eqref{eq:gaussian-spectral-partition-function} is Gaussian, all cavity marginals remain Gaussian. We may therefore write
\begin{equation}
P_{i\to j}(x_i)\propto\exp\left[-\frac{1}{2}\omega_{i\to j}x_i^2\right]\,,
\label{eq:gaussian-cavity-message-parametrization}
\end{equation}
with complex inverse variance $\omega_{i\to j}$. The proportionality constant is fixed by normalization once a branch of the complex Gaussian integral is chosen. For the derivation of the cavity recursion, only the quadratic coefficient is needed. For a locally tree-like graph, integrating over the neighboring cavity variables gives
\begin{equation}
\omega_{i\to j}=i(z-A_{ii})+\sum_{\ell\in\partial i\setminus j}\frac{A_{i\ell}^2}{\omega_{\ell\to i}}\,.
\label{eq:gaussian-cavity-recursion}
\end{equation}
The corresponding full inverse variance is
\begin{equation}
\omega_i=i(z-A_{ii})+\sum_{\ell\in\partial i}\frac{A_{i\ell}^2}{\omega_{\ell\to i}}\,,
\label{eq:gaussian-full-inverse-variance}
\end{equation}
and the diagonal resolvent is obtained as
\begin{equation}
G_{ii}(z)=\frac{i}{\omega_i}\,.
\label{eq:diagonal-resolvent-gaussian-cavity}
\end{equation}
Consequently,
\begin{equation}
\rho_{\pmb A}(\lambda)=\frac{1}{\pi N}\lim_{\epsilon\downarrow0}{\rm Im}\sum_{i=1}^N
\frac{i}{\omega_i(\lambda-i\epsilon)}\,.
\label{eq:density-gaussian-cavity}
\end{equation}
Equations \eqref{eq:gaussian-cavity-recursion}--\eqref{eq:density-gaussian-cavity} are the spectral analogue of the Ising cavity equations \eqref{eq:ising-cavity-field-recursion}--\eqref{eq:ising-full-field}. In the Ising model the message is a cavity field; in the Gaussian spectral problem the message is a complex inverse variance. The simplification is substantial: instead of an arbitrary function on a continuous variable, the Gaussian ansatz closes the cavity equations on a single complex parameter per directed edge.

\begin{examplebox}[The Gaussian cavity recursion on a three-site chain]
Consider the real symmetric matrix
\begin{equation}
\pmb A=\begin{pmatrix}
D_1 & J_{12} & 0\\
J_{12} & D_2 & J_{23}\\
0 & J_{23} & D_3
\end{pmatrix}\,,
\label{eq:rcmds-example-gaussian-chain}
\end{equation}
which corresponds to the three-site chain $1\!-\!2\!-\!3$. Let
\begin{equation}
z=\lambda-i\epsilon\,,\qquad\epsilon>0\,.
\label{eq:rcmds-example-gaussian-z}
\end{equation}
Since sites $1$ and $3$ are leaves, their cavity marginals are obtained without integrating over any neighbor:
\begin{equation}
P_{1\to 2}(x_1)\propto \exp\left[-\frac{i}{2}(z-D_1)x_1^2\right]\,,\qquad P_{3\to 2}(x_3)\propto\exp\left[-\frac{i}{2}(z-D_3)x_3^2\right]\,.
\label{eq:rcmds-example-leaf-gaussians}
\end{equation}
Thus
\begin{equation}
\omega_{1\to 2}=i(z-D_1)\,,\qquad \omega_{3\to 2}=i(z-D_3)\,.
\label{eq:rcmds-example-leaf-omegas}
\end{equation}

Now compute the full inverse variance at the middle site. By \eqref{eq:gaussian-full-inverse-variance},
\begin{equation}
\omega_2=i(z-D_2)+\frac{J_{12}^2}{\omega_{1\to 2}}+\frac{J_{23}^2}{\omega_{3\to 2}}\,.
\label{eq:rcmds-example-middle-omega}
\end{equation}
Hence the diagonal resolvent at site $2$ is
\begin{equation}
G_{22}(z)=\frac{i}{\omega_2}=\frac{i}{i(z-D_2)+\displaystyle\frac{J_{12}^2}{i(z-D_1)}+\frac{J_{23}^2}{i(z-D_3)}}\,.
\label{eq:rcmds-example-middle-green}
\end{equation}
Multiplying numerator and denominator by $-i$, this can be written as
\begin{equation}
G_{22}(z)=\frac{1}{z-D_2-\displaystyle\frac{J_{12}^2}{z-D_1}-\frac{J_{23}^2}{z-D_3}}\,.
\label{eq:rcmds-example-middle-green-schur}
\end{equation}
This is exactly the Schur-complement formula for the central diagonal entry of $(z\pmb I-\pmb A)^{-1}$.

As a simple check, take
\begin{equation}
D_1=D_2=D_3=0\,,\qquad J_{12}=J_{23}=1\,.
\label{eq:rcmds-example-simple-values}
\end{equation}
Then
\begin{equation}
G_{22}(z)=\frac{1}{z-\frac{1}{z}-\frac{1}{z}}=\frac{z}{z^2-2}\,.
\label{eq:rcmds-example-middle-green-simple}
\end{equation}
Direct inversion of
\begin{equation}
z\pmb I-\pmb A=\begin{pmatrix}
z & -1 & 0\\
-1 & z & -1\\
0 & -1 & z
\end{pmatrix}
\label{eq:rcmds-example-direct-matrix}
\end{equation}
gives the same result for the $(2,2)$ entry. This explicit example shows why the Gaussian cavity message is naturally parametrized by a single inverse variance.
\end{examplebox}

For an ensemble of sparse matrices, one can pass from the fixed-graph cavity equations to an equation for the statistics of a typical message. For a finite undirected graph $\mathcal{G}=(V,E)$, let $\overrightarrow E=\{(i,j),(j,i):\{i,j\}\in E\}$ be the set of oriented edges. If a Gaussian cavity fixed point assigns an inverse variance $\omega_{i\to j}$ to each $(i,j)\in\overrightarrow E$, its empirical directed-edge law is
\[
\pi_N^{\rm cav}(\omega)=\frac{1}{|\overrightarrow E|}\sum_{(i,j)\in\overrightarrow E}\delta(\omega-\omega_{i\to j})\,.
\]
The corresponding empirical law of full inverse variances is
\[
\pi_{N,{\rm full}}(\omega)=\frac{1}{N}\sum_{i=1}^{N}\delta(\omega-\omega_i)\,,
\]
where $\omega_i$ is obtained by including all neighbors of $i$. When these empirical laws have deterministic large-$N$ limits, we denote them by $\pi(\omega)$ and $\pi_{\rm full}(\omega)$. In the Erd\H{o}s--R\'enyi case, suppressing diagonal disorder and using independent edge weights with distribution $p_J(J)$, the limiting cavity law satisfies
\begin{equation}
\pi(\omega)=\sum_{\ell=0}^{\infty}e^{-c}\frac{c^\ell}{\ell!}\int\left[\prod_{r=1}^{\ell}d\omega_r \pi(\omega_r) dJ_r p_J(J_r)\right]\delta\left(\omega-iz-\sum_{r=1}^{\ell}\frac{J_r^2}{\omega_r}\right)\,.
\label{eq:er-gaussian-cavity-distribution}
\end{equation}
The limiting law of full inverse variances is then
\begin{equation}
\pi_{\rm full}(\omega)=\sum_{k=0}^{\infty}e^{-c}\frac{c^k}{k!}\int\left[\prod_{r=1}^{k}d\omega_r \pi(\omega_r)dJ_r p_J(J_r)\right]\delta\left(\omega-iz-\sum_{r=1}^{k}\frac{J_r^2}{\omega_r}\right)\,,
\label{eq:er-full-gaussian-cavity-distribution}
\end{equation}
and the ensemble-averaged spectral density follows from

\begin{equation}
\overline{\rho_{\pmb A}(\lambda)} =\frac{1}{\pi}\lim_{\epsilon\downarrow0}{\rm Im}\int d\omega\pi_{\rm full}(\omega)\frac{i}{\omega}\,.
\label{eq:ensemble-density-gaussian-cavity}
\end{equation}
For Erd\H{o}s--R\'enyi graphs the cavity and full degree distributions are both Poisson with mean $c$, while for a general graph ensemble the corresponding excess-degree and degree distributions differ. This replacement is one of the simplest ways in which graph topology enters the spectral density.

The cavity equations for sparse spectral problems have been developed in several equivalent forms. The replica approach to sparse random matrices was introduced by Rodgers and Bray \cite{RodgersBray1988}; later work clarified the relation with localized states and finite-connectivity effects \cite{Kuhn2008}. The Gaussian cavity method gives a direct and operational formulation for sparse symmetric matrices and sparse covariance matrices \cite{RogersTakedaPerezCastilloKuhn2008}. A detailed pedagogical account of the equivalence between the cavity and replica approaches for sparse symmetric random matrices is given in \cite{SuscaVivoKuhn2021}. The same logic extends to non-Hermitian sparse matrices after Hermitization, where each scalar Gaussian message is replaced by a small matrix-valued message \cite{RogersPerezCastillo2009,MetzNeriRogers2019}.

Several qualifications are useful before moving on. First, the cavity equations are exact on trees and become asymptotically exact on locally tree-like random graphs under the assumptions implicit in the replica-symmetric or Bethe description. When the problem has several competing pure states, or when long-range correlations are not negligible, replica-symmetry breaking or more elaborate message structures may be required. Second, the regulator $\epsilon$ in $z=\lambda-i\epsilon$ is part of the calculation. In sparse matrices, taking $\epsilon$ too small at finite $N$ reveals isolated poles and localized eigenstates; keeping $\epsilon$ finite gives a smoothed density. Third, the ensemble-level equations such as \eqref{eq:er-cavity-field-distribution} and \eqref{eq:er-gaussian-cavity-distribution} are functional fixed-point equations. Their numerical solution is usually obtained by population dynamics, while their single-instance version is ordinary belief propagation on the graph.

The main lesson of this section is that replica and cavity methods are two views of the same finite-connectivity mean-field structure. The replica method starts from the quenched logarithm, introduces copies, and identifies a functional order parameter. The cavity method starts from local tree-likeness, removes a vertex or edge, and writes recursive equations for messages on a fixed graph. In sparse random matrix theory the two viewpoints lead to the same practical outcome: fixed-instance observables are obtained from local resolvent messages, while ensemble-averaged observables are obtained from the corresponding law of a typical message. The next section develops the algorithmic form of this statement through belief propagation and population dynamics.

\begin{exerciseblock}
\exitem[Quenched versus annealed free energy]
Let $Z>0$ be any positive random variable. Use Jensen's inequality to show that
\begin{equation}
\overline{\log Z}\leq\log \overline{Z}\,.
    \label{eq:rcmds-exercise-jensen}
\end{equation}
Explain why this inequality captures the difference between quenched and annealed calculations discussed in the section.

\exitem[Replica identity]
Starting from
\begin{equation}
\frac{\mathcal Z^n-1}{n}\,,
\label{eq:rcmds-exercise-replica-identity-start}
\end{equation}
show that for $\mathcal Z>0$,
\begin{equation}
\lim_{n\to 0}\frac{\mathcal Z^n-1}{n}=\log \mathcal Z\,.
\label{eq:rcmds-exercise-replica-identity}
\end{equation}
What is the formal difficulty in using this identity in a disordered many-body problem?

\exitem[Replicated order parameter versus overlap]
For $n=2$ replicas, let
\begin{equation}
P_{++}\,,\quad P_{+-}\,,\quad P_{-+}\,,\quad P_{--}
\label{eq:rcmds-exercise-four-probabilities}
\end{equation}
be the probabilities of the four replicated local configurations. Show that
\begin{equation}
q_{12}=P_{++}+P_{--}-P_{+-}-P_{-+}\,.
\label{eq:rcmds-exercise-overlap-from-P}
\end{equation}
Construct two different distributions $P$ having the same value of $q_{12}$ but different magnetizations. Conclude that the full distribution of replicated local configurations contains more information than the overlap alone.

\exitem[Belief propagation on a three-spin chain]
Consider the three-spin chain with Hamiltonian
\begin{equation}
H(\sigma_1,\sigma_2,\sigma_3)=-J_{12}\sigma_1\sigma_2-J_{23}\sigma_2\sigma_3-B_1\sigma_1-B_2\sigma_2-B_3\sigma_3\,.
\label{eq:rcmds-exercise-chain-hamiltonian}
\end{equation}
Starting directly from the definition
\begin{equation}
\psi_{i\to j}(\sigma_i)=\frac{1}{Z_{i\to j}}e^{\beta B_i\sigma_i}\prod_{k\in\partial i\setminus j}\left[\sum_{\sigma_k=\pm1}e^{\beta J_{ik}\sigma_i\sigma_k}\psi_{k\to i}(\sigma_k)\right]\,,
\label{eq:rcmds-exercise-bp-start}
\end{equation}
derive explicitly the message $\psi_{2\to 1}(\sigma_2)$ and the full marginal of $\sigma_2$.

\exitem[Derivation of the cavity-field map]
Assume that
\begin{equation}
\psi_{k\to i}(\sigma_k)=\frac{e^{\beta h_{k\to i}\sigma_k}}{2\cosh(\beta h_{k\to i})}\,.
\label{eq:rcmds-exercise-field-parametrization}
\end{equation}
Show that
\begin{equation}
\sum_{\sigma_k=\pm1}e^{\beta J_{ik}\sigma_i\sigma_k}\psi_{k\to i}(\sigma_k)\propto e^{\beta u(J_{ik},h_{k\to i})\sigma_i}\,,
\label{eq:rcmds-exercise-u-derivation-start}
\end{equation}
with
\begin{equation}
u(J,h)=\frac{1}{\beta}\operatorname{arctanh}\left[\tanh(\beta J)\tanh(\beta h)\right]\,.
\label{eq:rcmds-exercise-u-derivation}
\end{equation}

\exitem[Distributional cavity equation]
For an Erd\H{o}s--R\'enyi graph with mean degree $c$, fixed external field $B$, and independent couplings drawn from $p_J(J)$, let $W(h)$ denote the large-$N$ law of a cavity field on a uniformly chosen directed edge. Derive the Poisson form
\begin{equation}
W(h)=\sum_{\ell=0}^{\infty}e^{-c}\frac{c^\ell}{\ell!}\int\left[\prod_{r=1}^{\ell}dh_r W(h_r)dJ_r p_J(J_r)\right]\delta\left(h-B-\sum_{r=1}^{\ell}u(J_r,h_r)\right)\,.
\label{eq:rcmds-exercise-distributional}
\end{equation}
from the local tree picture. Explain why, for Erd\H{o}s--R\'enyi graphs, the excess-degree distribution seen by a message is again Poisson with mean $c$.

\exitem[Bethe free energy on a single edge]
Take the graph consisting of two spins connected by one edge. Compute the exact partition function
\begin{equation}
\mathcal Z=\sum_{\sigma_1,\sigma_2=\pm1}e^{\beta J\sigma_1\sigma_2+\beta B_1\sigma_1+\beta B_2\sigma_2}\,.
\label{eq:rcmds-exercise-single-edge-exact}
\end{equation}
Then evaluate the Bethe expression
\begin{equation}
F_{\rm Bethe}=-\frac{1}{\beta}\left(\log Z_1+\log Z_2-\log Z_{12}\right)
\label{eq:rcmds-exercise-single-edge-bethe}
\end{equation}
using the corresponding cavity messages. Verify that it agrees exactly with the true free energy of the two-spin system.

\exitem[Gaussian cavity update]
Assume ${\rm Re}\,\omega_{\ell\to i}>0$ so that the following Gaussian integral is convergent. Starting from
\begin{equation}
P_{\ell\to i}(x_\ell)\propto e^{-\frac{1}{2}\omega_{\ell\to i}x_\ell^2}\,,
\label{eq:rcmds-exercise-gaussian-message}
\end{equation}
show by explicit Gaussian integration that
\begin{equation}
\int dx_\ell\exp\left[-\frac{1}{2}\omega_{\ell\to i}x_\ell^2+iA_{i\ell}x_ix_\ell\right]\propto\exp\left[-\frac{A_{i\ell}^2}{2\omega_{\ell\to i}}x_i^2\right]\,.
\label{eq:rcmds-exercise-gaussian-integration}
\end{equation}
Deduce the recursion
\begin{equation}
\omega_{i\to j}=i(z-A_{ii})+\sum_{\ell\in\partial i\setminus j}\frac{A_{i\ell}^2}{\omega_{\ell\to i}}\,.
\label{eq:rcmds-exercise-gaussian-recursion}
\end{equation}

\exitem[Resolvent from inverse variance]
Using the Gaussian parametrization of the full single-site marginal,
\begin{equation}
P_i(x_i)\propto\exp\left[-\frac{1}{2}\omega_i x_i^2\right]\,,
\label{eq:rcmds-exercise-full-gaussian}
\end{equation}
show that
\begin{equation}
\langle x_i^2\rangle=\frac{1}{\omega_i}\,.
\label{eq:rcmds-exercise-variance}
\end{equation}
Then use the Edwards--Jones representation to explain why
\begin{equation}
G_{ii}(z)=\frac{i}{\omega_i}\,.
\label{eq:rcmds-exercise-green-from-omega}
\end{equation}

\exitem[Programming exercise: belief propagation and Gaussian cavity on trees]
Fix a finite tree with $N$ vertices, small enough that exact summation over the $2^N$ spin configurations is feasible. For the Ising part, choose couplings $J_{ij}$ on the edges, fields $B_i$, and an inverse temperature $\beta$, and report these choices. Implement the belief-propagation update
\begin{equation}
\psi_{i\to j}(\sigma_i)=\frac{1}{Z_{i\to j}}e^{\beta B_i\sigma_i}\prod_{k\in\partial i\setminus j}\left[\sum_{\sigma_k=\pm1}e^{\beta J_{ik}\sigma_i\sigma_k}\psi_{k\to i}(\sigma_k)\right]\,,
\label{eq:rcmds-exercise-program-bp}
\end{equation}
and compare the resulting cavity magnetizations with the exact magnetizations obtained by brute-force summation of the partition function for small $N$.

For the Gaussian spectral part, choose a real symmetric matrix $\pmb A$ supported on the same tree, choose $z=\lambda-i\epsilon$ with $\epsilon>0$, and solve the Gaussian cavity equations
\begin{equation}
\omega_{i\to j}=i(z-A_{ii})+\sum_{\ell\in\partial i\setminus j}\frac{A_{i\ell}^2}{\omega_{\ell\to i}}\,.
\label{eq:rcmds-exercise-program-gaussian-cavity}
\end{equation}
Then compute the full inverse variances
\begin{equation}
\omega_i=i(z-A_{ii})+\sum_{\ell\in\partial i}\frac{A_{i\ell}^2}{\omega_{\ell\to i}}\,.
\label{eq:rcmds-exercise-program-full-omega}
\end{equation}
Compare the cavity prediction
\begin{equation}
G_{ii}(z)=\frac{i}{\omega_i}
\label{eq:rcmds-exercise-program-local-resolvent}
\end{equation}
with the corresponding diagonal entry of $(z\pmb I-\pmb A)^{-1}$ obtained by direct inversion. Report the values of $N$, $\lambda$, $\epsilon$, and a discrepancy measure such as $\max_i|G_{ii}^{\rm cav}(z)-G_{ii}^{\rm inv}(z)|$.
\end{exerciseblock}

\section{Belief propagation and population dynamics}
\label{sec:belief-propagation-population-dynamics}
The replica and cavity methods provide a conceptual route from quenched disorder to self-consistency equations. In practice, these equations still have to be solved. For sparse random matrices there are two closely related computational levels. The first is a message-passing method on a single finite graph: given one realization of the matrix, one iterates the cavity equations on its directed edges and obtains approximations to the diagonal resolvent entries. This is the belief-propagation viewpoint. In this setting, belief propagation is simply the fixed-point iteration of the cavity equations on a particular graph. The second is an ensemble-level problem: one studies the large-$N$ statistics of messages obtained by choosing a directed edge uniformly in a random sparse instance. This is the population-dynamics viewpoint. In this case the fixed-point iteration acts on the limiting message law, which is represented numerically by a large population of messages. In both cases, the procedure is an operational implementation of the same cavity principle.

Belief propagation is most naturally formulated in the context of graphical models whose probability density factorizes into local terms. For a pairwise model on a simple undirected graph $\mathcal{G}=(V,E)$, with variables $x_i$ attached to vertices, one writes
\begin{equation}
P(\pmb{x})=\frac{1}{Z}\prod_{i\in V}\phi_i(x_i)\prod_{\{i,j\}\in E}\psi_{ij}(x_i,x_j)\,,
\label{eq:pairwise-graphical-model}
\end{equation}
with $\pmb{x}=(x_1,\ldots,x_N)$ and $V=\{1,\ldots,N\}$. On a tree, marginal probabilities can be computed exactly by passing messages from leaves to the root. In the notation used here, the message from $i$ to $j$ is a cavity marginal $P_{i\to j}(x_i)$ for $x_i$ in the graph where the edge $\{i,j\}$ has been removed. It satisfies
\begin{equation}
P_{i\to j}(x_i)=\frac{\phi_i(x_i)}{Z_{i\to j}}\prod_{\ell\in\partial i\setminus j}\left[\int dx_\ell\psi_{i\ell}(x_i,x_\ell)P_{\ell\to i}(x_\ell)\right]\,,
\label{eq:general-bp-cavity-message}
\end{equation}
where $\partial i$ denotes the set of neighbors of $i$, and $Z_{i\to j}$ is a normalization constant. The corresponding single-site belief $P_i(x_i)$ is
\begin{equation}
P_{i}(x_i)=\frac{\phi_i(x_i)}{Z_i}\prod_{\ell\in\partial i}\left[\int dx_\ell\psi_{i\ell}(x_i,x_\ell)P_{\ell\to i}(x_\ell)\right]\,.
    \label{eq:general-bp-site-belief}
\end{equation}
Equations \eqref{eq:general-bp-cavity-message} and \eqref{eq:general-bp-site-belief} are exact on trees. On graphs with loops, their iterative use is called loopy belief propagation. Its fixed points are stationary points of the Bethe free energy, and this is the sense in which belief propagation is the algorithmic counterpart of the Bethe or replica-symmetric cavity approximation \cite{Bethe1935,Pearl1988,KschischangFreyLoeliger2001,YedidiaFreemanWeiss2005,MezardMontanari2009}.

For the spectral problem, the variables are continuous Gaussian variables introduced by the Edwards--Jones representation. The factorized object is now a complex Gaussian weight rather than a positive probability density, but the message-passing algebra is the same. For a real symmetric diluted matrix $\pmb A$ and spectral parameter
\begin{equation}
z=\lambda-i\epsilon\,,\qquad \epsilon>0\,,
\label{eq:bp-spectral-parameter}
\end{equation}
the Gaussian partition function may be written as
\begin{equation}
Z_{\pmb A}(z)=\int\left[\prod_{i=1}^N\frac{dx_i}{\sqrt{2\pi}}\right]\exp\left[-\frac{i}{2}\sum_{i=1}^N(z-A_{ii})x_i^2+i\sum_{\{i,j\}\in E}A_{ij}x_i x_j\right]\,.
    \label{eq:bp-gaussian-spectral-model}
\end{equation}
Here the edge sum is over unordered edges of the underlying undirected support, so each off-diagonal interaction is counted once. This has precisely the form \eqref{eq:pairwise-graphical-model}, identifying
\begin{equation}
\phi_i(x_i)=\exp\left[-\frac{i}{2}(z-A_{ii})x_i^2\right]\,,\qquad\psi_{ij}(x_i,x_j)=\exp\left[iA_{ij}x_i x_j\right]\,.
    \label{eq:bp-gaussian-factors}
\end{equation}
The choice ${\rm Im}z<0$ gives a positive damping factor in the Gaussian integral. In the convention used here, the diagonal resolvent has positive imaginary part for $\epsilon>0$.

The essential simplification is that the cavity messages remain Gaussian. We therefore parametrize them by their quadratic coefficient,
\begin{equation}
P_{i\to j}(x_i)\propto\exp\left[-\frac{1}{2}\omega_{i\to j}x_i^2\right]\,,
\label{eq:bp-gaussian-message-omega}
\end{equation}
where $\omega_{i\to j}$ is a complex inverse variance. The proportionality constant is fixed by normalization once a branch of the complex Gaussian integral is chosen; for the cavity recursion only the quadratic coefficient is needed. Substituting \eqref{eq:bp-gaussian-message-omega} into \eqref{eq:general-bp-cavity-message}, one uses the elementary Gaussian identity
\begin{equation}
\int\frac{dx_\ell}{\sqrt{2\pi}}\exp\left[-\frac{1}{2}\omega_{\ell\to i}x_\ell^2+iA_{i\ell}x_i x_\ell\right]=\omega_{\ell\to i}^{-1/2}\exp\left[-\frac{A_{i\ell}^2}{2\omega_{\ell\to i}}x_i^2\right]\,,
\label{eq:bp-gaussian-edge-integration}
\end{equation}
and obtains
\begin{equation}
\omega_{i\to j}=i(z-A_{ii})+\sum_{\ell\in\partial i\setminus j}\frac{A_{i\ell}^2}{\omega_{\ell\to i}}\,.
\label{eq:bp-omega-update}
\end{equation}
Equivalently, introducing the cavity Green function
\begin{equation}
G_{i\to j}(z)=\frac{i}{\omega_{i\to j}(z)}\,,
\label{eq:cavity-green-definition}
\end{equation}
one obtains the more transparent recursion
\begin{equation}
G_{i\to j}(z)=\frac{1}{z-A_{ii}-\sum_{\ell\in\partial i\setminus j} A_{i\ell}^2G_{\ell\to i}(z)}\,.
\label{eq:bp-cavity-green-update}
\end{equation}
Once the directed-edge messages have reached a fixed point, the full local resolvent at vertex $i$ is
\begin{equation}
G_{ii}(z)=\frac{1}{ z-A_{ii}-\sum_{\ell\in\partial i}A_{i\ell}^2G_{\ell\to i}(z)}\,.
\label{eq:bp-site-green}
\end{equation}

\begin{examplebox}[One belief-propagation sweep on a three-site chain]
Consider the symmetric matrix on the chain $1\!-\!2\!-\!3$,
\begin{equation}
\pmb A=\begin{pmatrix}
0 & J_{12} & 0\\
J_{12} & 0 & J_{23}\\
0 & J_{23} & 0
\end{pmatrix}\,,\qquad z=\lambda-i\epsilon\,,\qquad \epsilon>0\,.
\label{eq:bppd-example-chain-matrix}
\end{equation}
The directed cavity messages are
\begin{equation}
G_{1\to2}\,,\quad G_{2\to1}\,, \quad G_{2\to3}\,,\quad G_{3\to2}\,.
\label{eq:bppd-example-chain-messages}
\end{equation}
Since vertices $1$ and $3$ are leaves, the two leaf messages are
\begin{equation}
G_{1\to2}(z)=\frac{1}{z}\,,\qquad G_{3\to2}(z)=\frac{1}{z}\,.
\label{eq:bppd-example-leaf-messages}
\end{equation}
The messages sent from the middle vertex to the leaves exclude the recipient vertex. Therefore
\begin{equation}
G_{2\to1}(z)=\frac{1}{z-J_{23}^2G_{3\to2}(z)}=\frac{1}{z-\displaystyle\frac{J_{23}^2}{z}}\,,
\label{eq:bppd-example-middle-to-one}
\end{equation}
and
\begin{equation}
G_{2\to3}(z)=\frac{1}{z-J_{12}^2G_{1\to2}(z)}=\frac{1}{z-\displaystyle\frac{J_{12}^2}{z}}\,.
\label{eq:bppd-example-middle-to-three}
\end{equation}
The full local Green functions are then
\begin{equation}
G_1(z)=\frac{1}{z-J_{12}^2G_{2\to1}(z)}\,,\qquad G_3(z)=\frac{1}{z-J_{23}^2G_{2\to3}(z)}\,,
\label{eq:bppd-example-end-full-greens}
\end{equation}
and
\begin{equation}
G_2(z)=\frac{1}{z-J_{12}^2G_{1\to2}(z)-J_{23}^2G_{3\to2}(z)}=\frac{1}{z-\displaystyle\frac{J_{12}^2+J_{23}^2}{z}}\,.
\label{eq:bppd-example-middle-full-green}
\end{equation}
In this tree example, belief propagation is not an approximation: the quantities in
\eqref{eq:bppd-example-end-full-greens} and \eqref{eq:bppd-example-middle-full-green}
are exactly the diagonal entries of $(z\pmb I-\pmb A)^{-1}$. The example also shows the meaning of a cavity message: $G_{2\to1}$ is the Green function at vertex $2$ in the branch in which the edge to vertex $1$ has been removed.
\end{examplebox}

The spectral density of the instance is then estimated as
\begin{equation}
\rho_{\pmb A,\epsilon}^{\rm BP}(\lambda)=\frac{1}{\pi N}\sum_{i=1}^N{\rm Im}[G_{ii}(\lambda-i\epsilon)]\,.
\label{eq:bp-density-estimator}
\end{equation}
On a tree, \eqref{eq:bp-cavity-green-update}--\eqref{eq:bp-density-estimator} are exact. On a sparse random graph, they become asymptotically exact for locally tree-like ensembles under the same replica-symmetric assumptions that underlie the cavity method \cite{RogersTakedaPerezCastilloKuhn2008,SuscaVivoKuhn2021}.

For a fixed finite graph, belief propagation is implemented by initializing all directed-edge messages $G_{i\to j}^{(0)}$ with positive imaginary part and then iterating the map
\begin{equation}
G_{i\to j}^{(t+1)}=\mathcal{F}_{i\to j}\left(\{G_{\ell\to i}^{(t)}\}_{\ell\in\partial i\setminus j}\right)\,,
\label{eq:bp-synchronous-update}
\end{equation}
where
\begin{equation}
\mathcal{F}_{i\to j}\left(\{G_{\ell\to i}\}\right)=\frac{1}{z-A_{ii}-\sum_{\ell\in\partial i\setminus j}A_{i\ell}^2G_{\ell\to i}}\,.
\label{eq:bp-update-map}
\end{equation}
In practice, the following damped fixed-point iteration is often more stable:
\begin{equation}
G_{i\to j}^{(t+1)}=(1-\gamma)G_{i\to j}^{(t)}+\gamma\mathcal{F}_{i\to j}\left(\{G_{\ell\to i}^{(t)}\}_{\ell\in\partial i\setminus j}\right)\,,\qquad 0<\gamma\leq 1.
    \label{eq:bp-damped-update}
\end{equation}
The regulator $\epsilon$ smooths the Dirac delta functions that appear in the empirical spectral density. If $\epsilon$ is too large, the density is over-smoothed; if it is too small at finite $N$, isolated poles and localized eigenvectors may dominate the iteration. The preservation of ${\rm Im}[G_{i\to j}]>0$ for Hermitian problems with $\epsilon>0$ is a useful check on the implementation.

\begin{examplebox}[Why damping is useful near a fixed point]
For a single directed edge update, write the undamped map as
\begin{equation}
G^{(t+1)}=\mathcal F(G^{(t)})\,.
\label{eq:bppd-example-undamped-map}
\end{equation}
Suppose $G^\star$ is a fixed point:
\begin{equation}
G^\star=\mathcal F(G^\star)\,.
\label{eq:bppd-example-fixed-point}
\end{equation}
Linearizing around it with
\begin{equation}
G^{(t)}=G^\star+\delta^{(t)}\,,
\label{eq:bppd-example-linearization}
\end{equation}
one obtains
\begin{equation}
\delta^{(t+1)}=\mathcal F'(G^\star)\delta^{(t)}+O\left((\delta^{(t)})^2\right)\,.
\label{eq:bppd-example-undamped-linear}
\end{equation}
Thus the undamped update is locally stable if
\begin{equation}
|\mathcal F'(G^\star)|<1\,.
\label{eq:bppd-example-undamped-stability}
\end{equation}
The damped update is
\begin{equation}
G^{(t+1)}=(1-\gamma)G^{(t)}+\gamma \mathcal F(G^{(t)})\,,\qquad 0<\gamma\leq1\,.
\label{eq:bppd-example-damped-map}
\end{equation}
Linearizing the damped update gives
\begin{equation}
\delta^{(t+1)}=\left[1-\gamma+\gamma\mathcal F'(G^\star)\right]\delta^{(t)}+O\left((\delta^{(t)})^2\right)\,.
\label{eq:bppd-example-damped-linear}
\end{equation}
Therefore damping replaces the local multiplier $\mathcal F'(G^\star)$ by
\begin{equation}
\Lambda_\gamma=1-\gamma+\gamma\mathcal F'(G^\star)\,.
\label{eq:bppd-example-damped-multiplier}
\end{equation}
This does not guarantee convergence in every situation, but it can reduce oscillations and improve stability when the undamped fixed-point iteration is close to marginal. This is why damping is often useful when the regulator $\epsilon$ is small or when the spectral parameter is close to a region where local resolvents fluctuate strongly.
\end{examplebox} 

The same equations may be viewed not as an algorithm on one finite graph, but as a fixed-point problem for the message statistics generated by an ensemble of locally tree-like graphs. Consider a symmetric diluted matrix $\pmb A$ of the form
\begin{equation}
A_{ij}=C_{ij}J_{ij}+D_i\delta_{ij}\,,
\label{eq:bp-ensemble-sparse-matrix}
\end{equation}
where $C_{ij}$ is the adjacency matrix of a sparse graph, $J_{ij}=J_{ji}$ are independent edge weights with law $p_J$, and $D_i$ are diagonal variables with law $p_D$. Let $p_k$ be the degree distribution and
\begin{equation}
q_\ell=\frac{(\ell+1)p_{\ell+1}}{\sum_k kp_k}\,,\qquad \ell=0,1,2,\ldots\,,
\label{eq:bp-excess-degree-law}
\end{equation}
be the excess-degree distribution seen by following a uniformly chosen edge. For a finite undirected support with edge set $E$, let $\overrightarrow E=\{(i,j),(j,i):\{i,j\}\in E\}$ be the set of directed edges. A BP fixed point defines the empirical cavity-message law
\[
\mathcal P_{{\rm cav},N}(G)=\frac{1}{|\overrightarrow E|}\sum_{(i,j)\in\overrightarrow E}\delta(G-G_{i\to j})
\]
and the empirical site law
\[
\mathcal P_{{\rm site},N}(G)=\frac{1}{N}\sum_{i=1}^{N}\delta(G-G_{ii})\,.
\]
When these empirical laws converge to deterministic limits, we denote them by $\mathcal P_{\rm cav}$ and $\mathcal P_{\rm site}$. The limiting cavity-message law satisfies
\begin{equation}
\mathcal{P}_{\rm cav}(G)=\sum_{\ell=0}^{\infty}q_\ell\int dD p_D(D)\left[\prod_{r=1}^{\ell} dG_r\mathcal{P}_{\rm cav}(G_r) dJ_r p_J(J_r)\right]\delta\left(G-\frac{1}{z-D-\displaystyle\sum_{r=1}^{\ell}J_r^2G_r}\right)\,.
\label{eq:population-cavity-equation-general}
\end{equation}
The distribution of a full site Green function is
\begin{equation}
\mathcal{P}_{\rm site}(G)=\sum_{k=0}^{\infty}p_k\int dD p_D(D)\left[\prod_{r=1}^{k}
dG_r\mathcal{P}_{\rm cav}(G_r) dJ_r p_J(J_r)\right]\delta\left(G-\frac{1}{z-D-\displaystyle\sum_{r=1}^{k}J_r^2G_r}\right)\,.
    \label{eq:population-site-equation-general}
\end{equation}
The ensemble-averaged spectral density is then
\begin{equation}
\overline{\rho_{\pmb A,\epsilon}(\lambda)}=\frac{1}{\pi}{\rm Im}\int dG \mathcal{P}_{\rm site}(G)G\,,\qquad z=\lambda-i\epsilon\,,
\label{eq:population-density-general}
\end{equation}
with the limiting density obtained by taking $\epsilon\downarrow0$ after the thermodynamic limit. In numerical population dynamics one usually works at a small but finite value of $\epsilon$, rather than taking this limit explicitly. A more systematic algorithmic derivation of population dynamics, including the weighted variants used later for tilted ensembles, is given in Appendix~\ref{app:population-dynamics}.

Equations \eqref{eq:population-cavity-equation-general}--\eqref{eq:population-density-general} are solved numerically by population dynamics. The idea is to represent $\mathcal{P}_{\rm cav}$ by a large empirical population
\begin{equation}
\{G^{(1)},G^{(2)},\ldots,G^{(M)}\}\,.
\label{eq:population-representation}
\end{equation}
One update consists of drawing an excess degree $\ell$ from $q_\ell$, drawing $\ell$ members independently and uniformly, with replacement, from the population, drawing $\ell$ edge weights and one diagonal term according to their distributions $p_{J}$ and $p_D$, and replacing a randomly chosen member of the population by
\begin{equation}
G_{\rm new}=\frac{1}{z-D-\displaystyle\sum_{r=1}^{\ell}J_r^2G^{(a_r)}}\,.
\label{eq:population-update}
\end{equation}
After equilibration, observables are computed by drawing a full degree $k$ from $p_k$, drawing $k$ population members, $k$ edge weights, and one diagonal term from the same laws, and evaluating
\begin{equation}
G_{\rm site}=\frac{1}{z-D-\displaystyle\sum_{r=1}^{k}J_r^2G^{(a_r)}}\,.
\label{eq:population-site-sampling}
\end{equation}
The average of ${\rm Im}[G_{\rm site}]/\pi$ over many such samples estimates the integral appearing in Eq.~\eqref{eq:population-density-general}. Population dynamics is therefore a fixed-point iteration method in which the integrals are evaluated by a Monte Carlo procedure. Note that this is not a diagonalization of finite matrices. It solves the distributional cavity equation directly.

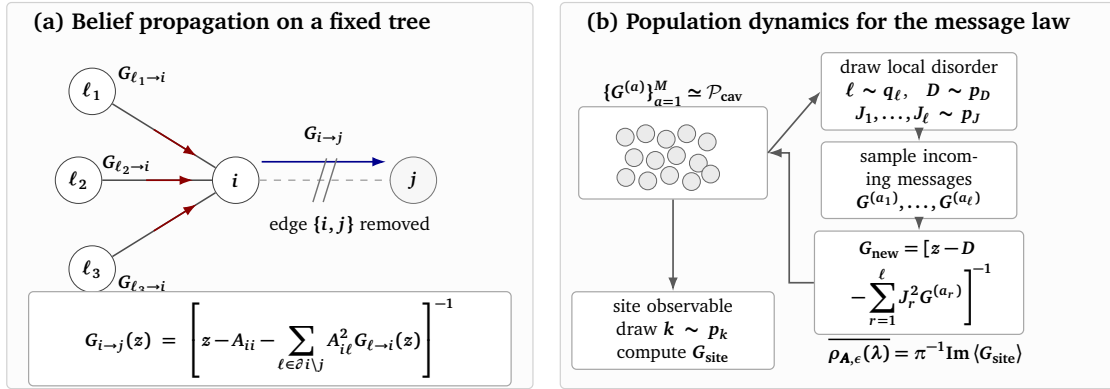
\begin{figure}[t]
\centering
\resizebox{0.98\textwidth}{!}{%
\begin{tikzpicture}[
    x=1cm,
    y=1cm,
    >=Latex,
    panel/.style={draw=black!18, fill=black!1, rounded corners=2pt, line width=0.5pt},
    vnode/.style={circle, draw=black!75, fill=white, minimum size=6.8mm, inner sep=0pt, font=\scriptsize},
    cutnode/.style={circle, draw=black!55, fill=black!3, minimum size=6.8mm, inner sep=0pt, font=\scriptsize},
    edge/.style={draw=black!70, line width=0.65pt},
    paleedge/.style={draw=black!35, line width=0.6pt, dashed},
    arrowedge/.style={draw=red!55!black, line width=0.7pt, -{Latex[length=2.0mm,width=1.4mm]}},
    bluearrow/.style={draw=blue!55!black, line width=0.7pt, -{Latex[length=2.0mm,width=1.4mm]}},
    flowarrow/.style={draw=black!65, line width=0.7pt, -{Latex[length=2.0mm,width=1.4mm]}},
    ptitle/.style={font=\bfseries\small, anchor=west},
    paneltext/.style={font=\scriptsize, align=center},
    tinytext/.style={font=\tiny, align=center},
    box/.style={draw=black!35, fill=white, rounded corners=2pt, line width=0.5pt, inner sep=3pt, font=\scriptsize, align=center},
    popdot/.style={circle, draw=black!60, fill=black!8, minimum size=2.6mm, inner sep=0pt}
]

\draw[panel] (0,0) rectangle (7.70,5.65);
\node[ptitle] at (0.25,5.35) {(a) Belief propagation on a fixed tree};

\node[vnode] (l1) at (1.25,4.35) {$\ell_1$};
\node[vnode] (l2) at (1.05,3.05) {$\ell_2$};
\node[vnode] (l3) at (1.25,1.75) {$\ell_3$};
\node[vnode] (i)  at (3.35,3.05) {$i$};
\node[cutnode] (j) at (5.95,3.05) {$j$};

\draw[edge] (l1) -- (i);
\draw[edge] (l2) -- (i);
\draw[edge] (l3) -- (i);
\draw[paleedge] (i) -- (j);

\draw[arrowedge] ($(l1)!0.43!(i)$) -- ($(l1)!0.73!(i)$);
\draw[arrowedge] ($(l2)!0.43!(i)$) -- ($(l2)!0.73!(i)$);
\draw[arrowedge] ($(l3)!0.43!(i)$) -- ($(l3)!0.73!(i)$);
\draw[bluearrow] ($(i)+(0.38,0.27)$) -- ($(j)+(-0.40,0.27)$);

\draw[black!55, line width=0.6pt] (4.48,2.73) -- (4.68,3.37);
\draw[black!55, line width=0.6pt] (4.63,2.73) -- (4.83,3.37);

\node[paneltext, anchor=east] at (2.48,4.58) {$G_{\ell_1\to i}$};
\node[paneltext, anchor=east] at (2.25,3.28) {$G_{\ell_2\to i}$};
\node[paneltext, anchor=east] at (2.48,1.52) {$G_{\ell_3\to i}$};
\node[paneltext] at (4.65,3.70) {$G_{i\to j}$};
\node[paneltext] at (5.02,2.40) {edge $\{i,j\}$\ removed};

\node[box, text width=6.85cm] at (3.85,0.73)
{$\displaystyle
G_{i\to j}(z)=\left[z-A_{ii}
-\sum_{\ell\in\partial i\setminus j}A_{i\ell}^2G_{\ell\to i}(z)\right]^{-1}$};

\draw[panel] (8.10,0) rectangle (16.25,5.65);
\node[ptitle] at (8.35,5.35) {(b) Population dynamics for the message law};

\node[box, text width=2.55cm, minimum height=1.35cm] (popbox) at (9.75,3.45) {};
\node[paneltext] at (9.75,4.32) {$\{G^{(a)}\}_{a=1}^{M}\simeq\mathcal P_{\rm cav}$};
\foreach \x/\y in {9.05/3.68,9.35/3.73,9.65/3.65,9.95/3.73,10.25/3.61,9.18/3.38,9.49/3.34,9.80/3.41,10.12/3.30,9.06/3.09,9.38/3.04,9.70/3.12,10.02/3.03,10.30/3.14}{
  \node[popdot] at (\x,\y) {};
}

\node[box, text width=2.65cm] (drawbox) at (13.35,4.35)
{draw local disorder\\$\ell\sim q_\ell$,\quad $D\sim p_D$\\$J_1,\ldots,J_\ell\sim p_J$};
\node[box, text width=2.65cm] (samplebox) at (13.35,3.05)
{sample incoming messages\\$G^{(a_1)},\ldots,G^{(a_\ell)}$};
\node[box, text width=2.85cm] (updatebox) at (13.35,1.55)
{$\displaystyle
\begin{gathered}
G_{\rm new}=\left[z-D\right.\\[-1mm]
\left.{}-\sum_{r=1}^{\ell}J_r^2G^{(a_r)}\right]^{-1}
\end{gathered}$};

\draw[flowarrow] (popbox.east) -- (drawbox.west);
\draw[flowarrow] (drawbox) -- (samplebox);
\draw[flowarrow] (samplebox) -- (updatebox);
\draw[flowarrow] (updatebox.west) -- (11.45,1.55) -- (11.45,3.45) -- (popbox.east);

\node[box, text width=2.75cm] (sitebox) at (9.75,0.85)
{site observable\\draw $k\sim p_k$\\compute $G_{\rm site}$};
\draw[flowarrow] (popbox.south) -- (sitebox.north);
\node[paneltext] at (13.45,0.55)
{$\displaystyle \overline{\rho_{\pmb A,\epsilon}(\lambda)}=\pi^{-1}{\rm Im}\,\langle G_{\rm site}\rangle$};
\end{tikzpicture}%
}
\caption{Belief propagation and population dynamics as two implementations of the same cavity recursion. On a fixed locally tree-like instance, directed-edge messages are updated by removing the recipient from the incoming neighbourhood. At the ensemble level, the message law is represented by an empirical population, and the same local recursion is sampled as a Monte Carlo fixed-point equation.}
\label{fig:bp-population-workflow}
\end{figure}

The two computational levels described above are summarized in Figure~\ref{fig:bp-population-workflow}.

\begin{examplebox}[One population-dynamics update for a Poisson graph]
Consider an Erd\H{o}s--R\'enyi sparse symmetric ensemble with zero diagonal terms and unit edge weights,
\begin{equation}
D=0\,,\qquad J=1\,.
\label{eq:bppd-example-er-simple-weights}
\end{equation}
The cavity distribution satisfies
\begin{equation}
\mathcal P_{\rm cav}(G)=\sum_{\ell=0}^{\infty}e^{-c}\frac{c^\ell}{\ell!}\int\left[\prod_{r=1}^{\ell}dG_r \mathcal P_{\rm cav}(G_r)\right]\delta\left(G-\frac{1}{z-\displaystyle\sum_{r=1}^{\ell}G_r}\right)\,.
\label{eq:bppd-example-er-population-law}
\end{equation}
A single population-dynamics update is therefore the following random experiment. First draw
\begin{equation}
\ell\sim {\rm Poisson}(c)\,.
\label{eq:bppd-example-draw-degree}
\end{equation}
Then draw $\ell$ messages from the current population,
\begin{equation}
G^{(a_1)},\ldots,G^{(a_\ell)}\,.
\label{eq:bppd-example-draw-messages}
\end{equation}
Finally compute
\begin{equation}
G_{\rm new}=\frac{1}{z-\displaystyle\sum_{r=1}^{\ell}G^{(a_r)}}\,.
\label{eq:bppd-example-er-new-message}
\end{equation}
If $\ell=0$, this gives
\begin{equation}
G_{\rm new}=\frac{1}{z}\,,
\label{eq:bppd-example-zero-neighbor}
\end{equation}
which is the cavity Green function of an isolated branch. If $\ell=1$, it gives
\begin{equation}
G_{\rm new}=\frac{1}{z-G^{(a_1)}}\,,
\label{eq:bppd-example-one-neighbor}
\end{equation}
which is the update for a branch with one incoming neighbor. Thus population dynamics is not a new approximation beyond the cavity equation: it is the Monte Carlo method for sampling the random recursion itself.
\end{examplebox} 

For Erd\H{o}s--R\'enyi graphs, the degree distribution and the excess-degree distribution are both Poisson with mean $c$,
\begin{equation}
p_k=e^{-c}\frac{c^k}{k!}\,,\qquad q_\ell=e^{-c}\frac{c^\ell}{\ell!}\,.
\label{eq:bp-er-degree-laws}
\end{equation}
For random $c$-regular graphs, they are
\begin{equation}
p_k=\delta_{k,c}\,,\qquad q_\ell=\delta_{\ell,c-1}\,.
\label{eq:bp-rrg-degree-laws}
\end{equation}
In the unweighted adjacency case, $J=1$ and $D=0$, the random-regular population collapses to a single value. The following example makes explicit how the distributional equation reduces to an ordinary algebraic self-consistency equation, and how the Kesten--McKay law is recovered from it \cite{Kesten1959,McKay1981}. Here this calculation is used only as an algorithmic benchmark; the sparse symmetric case will be treated systematically in the next section.

\begin{examplebox}[Population dynamics collapses to an algebraic equation on a regular graph]
For the unweighted adjacency matrix of a random $c$-regular graph, each cavity message has exactly $c-1$ incoming neighbors. By symmetry all cavity messages have the same value, say $G_{\rm cav}(z)$. The population equation then collapses to
\begin{equation}
G_{\rm cav}(z)=\frac{1}{z-(c-1)G_{\rm cav}(z)}\,.
\label{eq:bppd-example-rrg-cavity}
\end{equation}
Multiplying by the denominator gives
\begin{equation}
(c-1)G_{\rm cav}(z)^2-zG_{\rm cav}(z)+1=0\,.
\label{eq:bppd-example-rrg-quadratic}
\end{equation}
Thus
\begin{equation}
G_{\rm cav}(z)=\frac{z-\sqrt{z^2-4(c-1)}}{2(c-1)}\,,
\label{eq:bppd-example-rrg-cavity-solution}
\end{equation}
where the branch is chosen so that $G_{\rm cav}(z)\sim 1/z$ for large $|z|$ and ${\rm Im}\,G_{\rm cav}(\lambda-i\epsilon)>0$. The full site Green function is
\begin{equation}
G_{\rm site}(z)=\frac{1}{z-cG_{\rm cav}(z)}\,.
\label{eq:bppd-example-rrg-site}
\end{equation}
Taking the imaginary part of \eqref{eq:bppd-example-rrg-site} for $z=\lambda-i0^+$ yields the Kesten--McKay density. This example shows that population dynamics is a distributional generalization of ordinary algebraic self-consistency: in a homogeneous ensemble the population has no width, while in a heterogeneous sparse ensemble it is a genuine distribution.
\end{examplebox}

Sparse covariance and diluted Wishart matrices lead to analogous message-passing equations. As in the bipartite representation introduced earlier, it is natural to express them on the graph associated with the rectangular matrix. Let $\pmb{X}$ be an $N\times P$ sparse rectangular matrix and consider the linearized Hermitian matrix
\begin{equation}
\pmb{\mathcal L}=\frac{1}{\sqrt d}
\begin{pmatrix}
\pmb{0} &\pmb{X}\\
\pmb{X}^{\rm T} & \pmb{0}
\end{pmatrix}\,.
\label{eq:bp-bipartite-linearization}
\end{equation}
In this case, the cavity equations on the bipartite graph take the form
\begin{equation}
G_{i\to\mu}(z)=\frac{1}{z-\displaystyle\frac{1}{d}\sum_{\nu\in\partial i\setminus\mu}(X_i^\nu)^2\widehat G_{\nu\to i}(z)}\,,
\label{eq:bipartite-variable-message}
\end{equation}
and
\begin{equation}
\widehat G_{\mu\to i}(z)=\frac{1}{z-\displaystyle\frac{1}{d}\sum_{j\in\partial\mu\setminus i}(X_j^\mu)^2G_{j\to\mu}(z)}\,.
\label{eq:bipartite-factor-message}
\end{equation}

There are now two different cavity Green functions, corresponding to the two vertex classes of the bipartite graph: variable nodes and factor nodes. The corresponding full Green functions on the two sides are
\begin{equation}
\begin{split}
G_i(z)=\frac{1}{z-\displaystyle\frac{1}{d}\sum_{\nu\in\partial i}(X_i^\nu)^2\widehat G_{\nu\to i}(z)}\,,\qquad \widehat G_\mu(z)=\frac{1}{z-\displaystyle\frac{1}{d}\sum_{j\in\partial\mu}(X_j^\mu)^2G_{j\to\mu}(z)}\,.
\end{split}
\end{equation}
The covariance spectral density is the density of $\pmb W$ normalized by $N$, and it is recovered from the variable side of the linearized problem. The auxiliary matrix $\pmb{\mathcal L}$ also has its own spectral density, normalized by $N+P$; this density is obtained by tracing over both vertex classes,
\begin{eqnarray}
\rho_{\pmb{\mathcal L},\epsilon}(\lambda)=\frac{1}{\pi(N+P)}{\rm Im}\left[\sum_{i=1}^{N}G_i(\lambda-i\epsilon)+\sum_{\mu=1}^{P}\widehat G_\mu(\lambda-i\epsilon)\right]\,.   
\end{eqnarray}
This is not the covariance density itself. If $\nu_1,\ldots,\nu_r$ are the nonzero eigenvalues of $\pmb W$, then $\pmb{\mathcal L}$ has nonzero eigenvalues $\pm\sqrt{\nu_1},\ldots,\pm\sqrt{\nu_r}$, together with $N+P-2r$ zero modes. Thus the two-population formulation is useful both algorithmically and conceptually: the variable-side trace gives the covariance resolvent, while the trace over both sides gives the density of the auxiliary bipartite linearization. In a population-dynamics implementation, the variable-to-factor updates are sampled with the variable excess-degree law, and the factor-to-variable updates with the factor excess-degree law. Full site measurements use the full degree laws on the corresponding side. For the Poisson bipartite ensemble with $P=N/\alpha$, these laws are Poisson with means $d/\alpha$ on variable nodes and $d$ on factor nodes; for non-Poisson bipartite ensembles the excess and full laws must be kept distinct. This formulation is the natural algorithmic expression of sparse covariance and diluted Wishart ensembles, because the elementary disorder lives on the rectangular matrix $\pmb X$, not directly on the induced covariance matrix $\pmb X\pmb X^{\rm T}$ \cite{NagaoTanaka2007,RogersTakedaPerezCastilloKuhn2008,PerezCastilloMetz2018Wishart,PerezCastillo2022Generalized}.

\begin{examplebox}[Two populations, covariance density, and linearized density]
The two-population representation gives Green functions on the two sides of the bipartite graph. These Green functions are naturally associated with the linearized operator $\pmb{\mathcal L}$, but the covariance density is the density of $\pmb W=d^{-1}\pmb X\pmb X^{\rm T}$. The connection is most transparent at the level of the block resolvent. Let $\widetilde{\pmb W}=d^{-1}\pmb X^{\rm T}\pmb X$. The variable-variable block of $(z\pmb I_{N+P}-\pmb{\mathcal L})^{-1}$ is obtained by a Schur complement:
\begin{equation}
\left[z\pmb I_N-\frac{1}{dz}\pmb X\pmb X^{\rm T}\right]^{-1}=z\left[z^2\pmb I_N-\pmb W\right]^{-1}\,.
\label{eq:bppd-example-variable-block}
\end{equation}
Therefore
\begin{equation}
\sum_{i=1}^{N}G_i(z)=z\,{\rm Tr}_N\left(z^2\pmb I_N-\pmb W\right)^{-1}\,.
\label{eq:bppd-example-variable-trace}
\end{equation}
If $m_{\pmb W}(\zeta)=N^{-1}{\rm Tr}(\zeta\pmb I_N-\pmb W)^{-1}$, this says that
\begin{equation}
\frac{1}{N}\sum_{i=1}^{N}G_i(z)=z\,m_{\pmb W}(z^2)\,.
\label{eq:bppd-example-wishart-resolvent-from-variable-side}
\end{equation}
Thus the covariance resolvent is recovered from the variable side of the linearized problem, after the square map $\zeta=z^2$.

The factor-side trace gives the analogous expression for $\widetilde{\pmb W}$:
\begin{equation}
\sum_{\mu=1}^{P}\widehat G_\mu(z)=z\,{\rm Tr}_P\left(z^2\pmb I_P-\widetilde{\pmb W}\right)^{-1}\,.
\label{eq:bppd-example-factor-trace}
\end{equation}
Since $\pmb W$ and $\widetilde{\pmb W}$ have the same nonzero eigenvalues, the factor side carries the same nonzero singular-value information, but with the zero-mode count appropriate to the $P$-dimensional space. If $r={\rm rank}\,\pmb X$ and $\nu_1,\ldots,\nu_r$ are the nonzero eigenvalues of $\pmb W$, then
\begin{equation}
\sum_{i=1}^{N}G_i(z)=z\sum_{a=1}^{r}\frac{1}{z^2-\nu_a}+\frac{N-r}{z}\,,\qquad \sum_{\mu=1}^{P}\widehat G_\mu(z)=z\sum_{a=1}^{r}\frac{1}{z^2-\nu_a}+\frac{P-r}{z}\,.
\label{eq:bppd-example-two-traces-rank}
\end{equation}
Hence
\begin{equation}
\sum_{\mu=1}^{P}\widehat G_\mu(z)-\sum_{i=1}^{N}G_i(z)=\frac{P-N}{z}\,.
\label{eq:bppd-example-trace-difference}
\end{equation}
The trace over both vertex classes gives the density of the auxiliary linearized matrix,
\begin{equation}
\rho_{\pmb{\mathcal L},\epsilon}(\lambda)=\frac{1}{\pi(N+P)}{\rm Im}\left[\sum_{i=1}^{N}G_i(\lambda-i\epsilon)+\sum_{\mu=1}^{P}\widehat G_\mu(\lambda-i\epsilon)\right]\,.
\label{eq:bppd-example-linearized-density}
\end{equation}
This density is related to, but different from, the covariance density. Away from the zero mode,
\begin{equation}
\rho_{\pmb{\mathcal L}}(x)=\frac{2N}{N+P}|x|\,\rho_{\pmb W}(x^2)\,,\qquad x\neq0\,,
\label{eq:bppd-example-density-square-map}
\end{equation}
where $\rho_{\pmb W}$ is normalized by $N$.

Let us check these identities on the smallest nontrivial bipartite chain,
\begin{equation}
i_1-\mu-i_2\,,
\label{eq:bppd-example-bipartite-chain}
\end{equation}
with nonzero entries $X_{i_1}^{\mu}=x_1$ and $X_{i_2}^{\mu}=x_2$. The linearized matrix is
\begin{equation}
\pmb{\mathcal L}=\frac{1}{\sqrt d}\begin{pmatrix}0 & 0 & x_1\\0 & 0 & x_2\\x_1 & x_2 & 0\end{pmatrix}\,.
\label{eq:bppd-example-bipartite-linearized}
\end{equation}
Since both variable nodes are leaves, the variable-to-factor cavity messages are
\begin{equation}
G_{i_1\to\mu}(z)=\frac{1}{z}\,,\qquad G_{i_2\to\mu}(z)=\frac{1}{z}\,.
\label{eq:bppd-example-bipartite-variable-leaves}
\end{equation}
The factor-to-variable messages are
\begin{equation}
\widehat G_{\mu\to i_1}(z)=\frac{1}{z-\displaystyle\frac{x_2^2}{d}G_{i_2\to\mu}(z)}=\frac{1}{z-\displaystyle\frac{x_2^2}{dz}}\,,
\label{eq:bppd-example-factor-to-i1}
\end{equation}
and
\begin{equation}
\widehat G_{\mu\to i_2}(z)=\frac{1}{z-\displaystyle\frac{x_1^2}{d}G_{i_1\to\mu}(z)}=\frac{1}{z-\displaystyle\frac{x_1^2}{dz}}\,.
\label{eq:bppd-example-factor-to-i2}
\end{equation}
The full variable Green functions are
\begin{equation}
G_{i_1}(z)=\frac{1}{z-\displaystyle\frac{x_1^2}{d}\widehat G_{\mu\to i_1}(z)}\,,\qquad G_{i_2}(z)=\frac{1}{z-\displaystyle\frac{x_2^2}{d}\widehat G_{\mu\to i_2}(z)}\,.
\label{eq:bppd-example-bipartite-full-variables}
\end{equation}
The full factor Green function is
\begin{equation}
\widehat G_\mu(z)=\frac{1}{z-\displaystyle\frac{x_1^2}{d}G_{i_1\to\mu}(z)-\displaystyle\frac{x_2^2}{d}G_{i_2\to\mu}(z)}\,.
\label{eq:bppd-example-bipartite-full-factor}
\end{equation}
Writing
\begin{equation}
\nu=\frac{x_1^2+x_2^2}{d}\,,
\label{eq:bppd-example-one-factor-nu}
\end{equation}
one obtains
\begin{equation}
G_{i_1}(z)+G_{i_2}(z)=z\left(\frac{1}{z^2-\nu}+\frac{1}{z^2}\right)\,,\qquad \widehat G_\mu(z)=\frac{z}{z^2-\nu}\,.
\label{eq:bppd-example-small-chain-traces}
\end{equation}
This agrees with the general trace formulas: the covariance matrix has eigenvalues $\nu$ and $0$, while the factor-side matrix $\widetilde{\pmb W}$ has the single eigenvalue $\nu$.

Indeed, for this example
\begin{equation}
\pmb W=\frac{1}{d}\begin{pmatrix}x_1^2 & x_1x_2\\x_1x_2 & x_2^2\end{pmatrix}\,,
\label{eq:bppd-example-small-chain-wishart}
\end{equation}
so
\begin{equation}
\rho_{\pmb W}(\lambda)=\frac{1}{2}\delta(\lambda-\nu)+\frac{1}{2}\delta(\lambda)\,.
\label{eq:bppd-example-small-chain-wishart-density}
\end{equation}
The linearized matrix has eigenvalues $+\sqrt{\nu}$, $-\sqrt{\nu}$, and $0$, and therefore
\begin{equation}
\rho_{\pmb{\mathcal L}}(x)=\frac{1}{3}\delta(x-\sqrt{\nu})+\frac{1}{3}\delta(x+\sqrt{\nu})+\frac{1}{3}\delta(x)\,.
\label{eq:bppd-example-small-chain-linearized-density}
\end{equation}
Thus the two-population cavity equations give the Green functions of the bipartite linearized problem. The variable side recovers the covariance resolvent, while tracing both variable and factor sides gives the density of the auxiliary matrix $\pmb{\mathcal L}$.
\end{examplebox}

For non-Hermitian sparse matrices the same principle applies after Hermitization. In the simplest case, each vertex carries a doubled Gaussian variable, and the scalar cavity Green function is replaced by a small $2\times 2$ matrix-valued message. In schematic form, on the symmetrized support of the directed matrix, the update becomes
\begin{equation}
\pmb{G}_{i\to j}=\left[\pmb{Z}_i-\sum_{\ell\in\partial i\setminus j}\pmb{\mathcal A}_{i\ell}\pmb{G}_{\ell\to i}\pmb{\mathcal A}_{i\ell}^{\dagger}\right]^{-1}\,,
    \label{eq:nonhermitian-matrix-message}
\end{equation}
where $\pmb{Z}_i$ contains the complex spectral parameter and the Hermitization regulator, while the matrices $\pmb{\mathcal A}_{i\ell}$ are the Hermitized edge blocks encoding the directed couplings. The algorithmic structure is unchanged: belief propagation solves \eqref{eq:nonhermitian-matrix-message} on a single sparse instance, and population dynamics solves the corresponding ensemble-level fixed-point equation. This is the computational basis of the cavity approach to sparse non-Hermitian spectra \cite{RogersPerezCastillo2009,MetzNeriRogers2019}.

The origin of population dynamics in this context is broader than random matrix theory. Recursive distributions of Green functions appeared already in self-consistent approaches to localization on Bethe lattices \cite{AbouChacraAndersonThouless1973}. In disordered systems, population dynamics became the standard numerical method for solving finite-connectivity cavity equations, especially when the order parameter is a distribution of effective fields rather than a finite set of scalars \cite{MezardParisiVirasoro1987,MezardParisi2001,MezardMontanari2009}. In sparse random matrix theory, the same algorithm solves the distributional equations for local resolvents and therefore provides the ensemble-averaged spectral density without diagonalizing large random matrices \cite{RogersTakedaPerezCastilloKuhn2008,SuscaVivoKuhn2021}.

Several practical checks are important. The density obtained from \eqref{eq:bp-density-estimator} or \eqref{eq:population-density-general} must be nonnegative and normalized after integration over $\lambda$. The result should be stable under changes of population size, iteration time, damping, and regulator $\epsilon$. In ensembles with known limits, such as random regular graphs, dense Erd\H{o}s--R\'enyi matrices, or dense covariance matrices, the algorithm should recover the appropriate benchmark law. Finally, one should remember that finite $\epsilon$ computes a smoothed spectral density. In sparse systems this is not only a numerical convenience: the small-$\epsilon$ behavior contains information about localization, isolated spectral peaks, and the distinction between absolutely continuous and singular spectral components.

Belief propagation and population dynamics are therefore the operational form of the cavity method. Belief propagation solves the local resolvent equations on a fixed sparse instance; population dynamics solves the corresponding ensemble fixed-point problem for the statistics of a typical message. The first is useful for comparing with a given matrix realization, while the second gives direct access to disorder-averaged spectral observables. The following sections apply these tools to concrete ensembles, beginning with sparse symmetric random matrices.

\begin{exerciseblock}
\exitem[Belief propagation on a tree]
Starting from the pairwise graphical model
\begin{equation}
P(\pmb x)=\frac{1}{Z}\prod_{i\in V}\phi_i(x_i)\prod_{\{i,j\}\in E}\psi_{ij}(x_i,x_j)\,,
\label{eq:bppd-ex1-pairwise-model}
\end{equation}
on a simple undirected tree with vertex set $V=\{1,\ldots,N\}$, derive the belief-propagation recursion
\begin{equation}
P_{i\to j}(x_i)=\frac{1}{Z_{i\to j}}\phi_i(x_i)\prod_{\ell\in\partial i\setminus j}\left[\int dx_\ell \psi_{i\ell}(x_i,x_\ell)P_{\ell\to i}(x_\ell)\right]
    \label{eq:bppd-ex1-bp}
\end{equation}
by explicitly using the factorization of branches after cutting the edge $\{i,j\}$.

\exitem[Gaussian closure of the spectral messages]
Let $\pmb A$ be an $N\times N$ real symmetric matrix supported on a simple undirected graph, and let $z=\lambda-i\epsilon$ with $\epsilon>0$. Starting from
\begin{equation}
\phi_i(x_i)=\exp\left[-\frac{i}{2}(z-A_{ii})x_i^2\right]\,,\qquad\psi_{ij}(x_i,x_j)=\exp\left[
iA_{ij}x_ix_j\right]\,,
\label{eq:bppd-ex2-gaussian-factors}
\end{equation}
and assuming a Gaussian incoming message
\begin{equation}
P_{\ell\to i}(x_\ell)\propto\exp\left[-\frac{1}{2}\omega_{\ell\to i}x_\ell^2\right]\,,
\label{eq:bppd-ex2-incoming-message}
\end{equation}
perform the Gaussian integral over $x_\ell$ and derive the update
\begin{equation}
\omega_{i\to j}=i(z-A_{ii})+\sum_{\ell\in\partial i\setminus j}\frac{A_{i\ell}^2}{\omega_{\ell\to i}}\,.
\label{eq:bppd-ex2-omega-update}
\end{equation}

\exitem[From inverse variances to cavity Green functions]
Using
\begin{equation}
G_{i\to j}(z)=\frac{i}{\omega_{i\to j}(z)}\,,
\label{eq:bppd-ex3-G-omega}
\end{equation}
show that the inverse-variance recursion in Exercise 5.2 is equivalent to
\begin{equation}
G_{i\to j}(z)=\frac{1}{z-A_{ii}-\displaystyle\sum_{\ell\in\partial i\setminus j}A_{i\ell}^2G_{\ell\to i}(z)}\,.
\label{eq:bppd-ex3-green-update}
\end{equation}

\exitem[Exactness on a finite tree]
Consider the three-site chain with adjacency matrix
\begin{equation}
\pmb A=\begin{pmatrix}
0 & J_{12} & 0\\
J_{12} & 0 & J_{23}\\
0 & J_{23} & 0
\end{pmatrix}\,.
\label{eq:bppd-ex4-chain-matrix}
\end{equation}
Compute $(z\pmb I-\pmb A)^{-1}$ directly by matrix inversion and verify that the diagonal entries agree with the belief-propagation formulas derived in the worked example.

\exitem[Fast update from a full self-energy]
For a fixed finite graph, define
\begin{equation}
\Sigma_i(z)=\sum_{\ell\in\partial i} A_{i\ell}^2G_{\ell\to i}(z)\,.
\label{eq:bppd-ex5-full-self-energy}
\end{equation}
Show that
\begin{equation}
G_{i\to j}(z)=\frac{1}{z-A_{ii}-\Sigma_i(z)+A_{ij}^2G_{j\to i}(z)}\,.
\label{eq:bppd-ex5-fast-update}
\end{equation}
Explain why this identity reduces the cost of one belief-propagation sweep from a naive sum over all excluded neighborhoods to a number of operations proportional to the number of directed edges.

\exitem[Damping and local stability]
Let $G^{(t+1)}=\mathcal F(G^{(t)})$ be a scalar fixed-point iteration, and let $G^\star$ be a fixed point. Derive the linearized stability condition for the damped iteration
\begin{equation}
G^{(t+1)}=(1-\gamma)G^{(t)}+\gamma\mathcal F(G^{(t)})\,,\qquad 0<\gamma\leq1\,.
\label{eq:bppd-ex6-damped-iteration}
\end{equation}
What is the effective linear multiplier in terms of $\gamma$ and $\mathcal F'(G^\star)$?

\exitem[Population dynamics as a Monte Carlo fixed-point equation]
Let $\mathcal P_{\rm cav}(G)$ be the distribution of cavity Green functions for a locally tree-like sparse matrix ensemble with excess-degree distribution $q_\ell$, diagonal distribution $p_D(D)$, and edge-weight distribution $p_J(J)$. Starting from
\begin{equation}
\mathcal{P}_{\rm cav}(G)=\sum_{\ell=0}^{\infty}q_\ell \int dD p_D(D)\left[\prod_{r=1}^{\ell} dG_r \mathcal{P}_{\rm cav}(G_r) dJ_r p_J(J_r)\right]\delta\left(G-\frac{1}{z-D-\displaystyle\sum_{r=1}^{\ell}J_r^2G_r}\right)\,,
\label{eq:bppd-ex7-population-equation}
\end{equation}
derive the population update rule
\begin{equation}
G_{\rm new}=\frac{1}{z-D-\displaystyle\sum_{r=1}^{\ell}J_r^2G^{(a_r)}}\,.
\label{eq:bppd-ex7-population-update}
\end{equation}
Explain why the indices $a_r$ are sampled uniformly from the current population.

\exitem[Site law versus cavity law]
Explain why the cavity update uses the excess-degree distribution $q_\ell$, whereas the site observable uses the full degree distribution $p_k$. Verify explicitly that for an Erd\H{o}s--R\'enyi graph both distributions are Poisson with mean $c$, while for a random $c$-regular graph one has
\begin{equation}
p_k=\delta_{k,c}\,,\qquad q_\ell=\delta_{\ell,c-1}\,.
\label{eq:bppd-ex8-degree-laws}
\end{equation}

\exitem[Random regular graph benchmark]
For $c\geq2$, starting from
\begin{equation}
G_{\rm cav}=\frac{1}{z-(c-1)G_{\rm cav}}\,,
\label{eq:bppd-ex9-rrg-cavity}
\end{equation}
solve for $G_{\rm cav}(z)$ by choosing the branch such that $G_{\rm cav}(z)\sim 1/z$ for large $|z|$. Then compute
\begin{equation}
G_{\rm site}(z)=\frac{1}{z-cG_{\rm cav}(z)}\,.
\label{eq:bppd-ex9-rrg-site}
\end{equation}
Use the imaginary part of $G_{\rm site}(\lambda-i0^+)$ to recover the Kesten--McKay density.

\exitem[Bipartite population dynamics]
Let $\pmb X$ be an $N\times P$ sparse rectangular matrix, let $d>0$, and define the linearized sparse covariance matrix
\begin{equation}
\pmb{\mathcal L}=\frac{1}{\sqrt d}\begin{pmatrix}
\pmb 0_{N\times N} & \pmb X\\
\pmb X^{\rm T} & \pmb 0_{P\times P}
\end{pmatrix}\,.
\label{eq:bppd-ex10-linearization}
\end{equation}
Derive the two message recursions
\begin{equation}
G_{i\to\mu}(z)=\frac{1}{z-\displaystyle\frac{1}{d}\sum_{\nu\in\partial i\setminus\mu}(X_i^\nu)^2\widehat G_{\nu\to i}(z)}\,,
\label{eq:bppd-ex10-variable-message}
\end{equation}
and
\begin{equation}
\widehat G_{\mu\to i}(z)=\frac{1}{z-\displaystyle\frac{1}{d}\sum_{j\in\partial\mu\setminus i}(X_j^\mu)^2 G_{j\to\mu}(z)}\,.
\label{eq:bppd-ex10-factor-message}
\end{equation}
Why are two populations needed in the ensemble-level algorithm?

\exitem[Non-Hermitian matrix messages]
Starting from the Hermitized recursion
\begin{equation}
\pmb{G}_{i\to j}=\left[\pmb{Z}_i-\sum_{\ell\in\partial i\setminus j}\pmb{\mathcal A}_{i\ell}\pmb{G}_{\ell\to i}\pmb{\mathcal A}_{i\ell}^{\dagger}\right]^{-1}\,,
\label{eq:bppd-ex11-nh-message}
\end{equation}
explain why the message is now a matrix rather than a scalar. Identify the role of Hermitization in producing this doubled local structure.

\exitem[Programming exercise: scalar population dynamics]
Fix a mean degree $c=O(1)$, a regulator $\epsilon>0$, a grid of real values $\lambda$, a population size $M$, a burn-in time $T_{\rm burn}$, and a number $T_{\rm meas}$ of site samples used for measurement. Implement population dynamics for the unweighted Erd\H{o}s--R\'enyi adjacency ensemble using
\begin{equation}
G_{\rm new}=\frac{1}{\lambda-i\epsilon-\displaystyle\sum_{r=1}^{\ell}G^{(a_r)}}\,,\qquad\ell\sim{\rm Poisson}(c)\,.
\label{eq:bppd-ex12-program-update}
\end{equation}
After equilibration, estimate
\begin{equation}
\rho_\epsilon(\lambda)=\frac{1}{\pi}{\rm Im}\left\langle\frac{1}{\lambda-i\epsilon-\displaystyle\sum_{r=1}^{k}G^{(a_r)}}\right\rangle\,,\qquad k\sim{\rm Poisson}(c),
\label{eq:bppd-ex12-program-density}
\end{equation}
where the average is over samples from the population and over $k$. For comparison, generate $S$ independent sparse Erd\H{o}s--R\'enyi adjacency matrices of size $N_{\rm diag}$ with entries $A_{ii}=0$, $A_{ij}=A_{ji}$, and ${\rm Prob}(A_{ij}=1)=c/N_{\rm diag}$ for $i<j$. Use the same $\lambda$ grid and the same $\epsilon$ in the population-dynamics and direct-diagonalization estimates, and report $c$, $M$, $T_{\rm burn}$, $T_{\rm meas}$, $N_{\rm diag}$, $S$, $\epsilon$, the grid spacing, and a discrepancy measure.

\exitem[Programming exercise: random regular benchmark]
Using the same $\epsilon$, $\lambda$ grid, $M$, $T_{\rm burn}$, and $T_{\rm meas}$ as in Exercise 5.12, implement population dynamics for the unweighted random $c$-regular ensemble with $c\geq2$. In this case every cavity update uses $\ell=c-1$ incoming messages, and every site sample uses $k=c$ incoming messages. Verify numerically that the population collapses to a narrow distribution around the deterministic solution of \eqref{eq:bppd-ex9-rrg-cavity}. Compare the estimated density with the Kesten--McKay law broadened with the same regulator $\epsilon$ and evaluated on the same grid. Report $c$, $M$, $T_{\rm burn}$, $T_{\rm meas}$, $\epsilon$, the grid spacing, and a discrepancy measure.

\exitem[Programming exercise: two-population Wishart dynamics]
Fix an aspect ratio $\alpha>0$, a dilution $d>0$, a list of system sizes $N$ for which $P=N/\alpha$ is an integer, a regulator $\epsilon>0$, a grid of positive real values $\lambda$ for the covariance spectral parameter $z_{\rm W}=\lambda-i\epsilon$, a population size $M$, a burn-in time $T_{\rm burn}$, a number $T_{\rm meas}$ of site samples used for measurement, and a number $S$ of independent samples for direct diagonalization. Generate a sparse rectangular matrix $\pmb X$ with entries
\begin{equation}
X_i^\mu=B_i^\mu\xi_i^\mu\,, \qquad {\rm Prob}(B_i^\mu=1)=\frac{d}{N}\,, \qquad {\rm Prob}(B_i^\mu=0)=1-\frac{d}{N}\,,
\label{eq:bppd-ex14-wishart-program}
\end{equation}
where the nonzero weights $\xi_i^\mu$ may be chosen equal to one or drawn independently from a standard Gaussian distribution; report which choice is used. Implement the two-population update associated with the linearized bipartite matrix
\begin{equation}
\pmb{\mathcal L}=\frac{1}{\sqrt d}\begin{pmatrix}\pmb 0_{N\times N} & \pmb X\\ \pmb X^{\rm T} & \pmb 0_{P\times P}\end{pmatrix}\,.
\label{eq:bppd-ex14-linearized}
\end{equation}
Use the variable and factor excess-degree laws for cavity updates, and the corresponding full degree laws for site measurements. The main observable is the covariance density of
\begin{equation}
\pmb W=\frac{1}{d}\pmb X\pmb X^{\rm T}\,.
\label{eq:bppd-ex14-wishart-W}
\end{equation}
For each point of the covariance grid, choose $z_{\rm L}$ with $z_{\rm L}^2=z_{\rm W}$ and ${\rm Im}\,z_{\rm L}<0$. From the variable side of the linearized resolvent, estimate
\begin{equation}
m_{\pmb W}(z_{\rm W})=\frac{1}{Nz_{\rm L}}\sum_{i=1}^{N}G_i(z_{\rm L})\,,
\label{eq:bppd-ex14-wishart-resolvent-from-linearized}
\end{equation}
and hence
\begin{equation}
\rho_{\pmb W,\epsilon}(\lambda)=\frac{1}{\pi}{\rm Im}\,m_{\pmb W}(\lambda-i\epsilon)\,.
\label{eq:bppd-ex14-wishart-density-from-variable-side}
\end{equation}
Compare this covariance density with direct diagonalization of $\pmb W$, using the same $\epsilon$ and the same $\lambda$ grid.

As an auxiliary check, one may also compute the density of the linearized matrix itself. In that case the trace must include both vertex classes,
\begin{equation}
\rho_{\pmb{\mathcal L},\epsilon}(x)=\frac{1}{\pi(N+P)}{\rm Im}\left[\sum_{i=1}^{N}G_i(x-i\epsilon)+\sum_{\mu=1}^{P}\widehat G_\mu(x-i\epsilon)\right]\,,
\label{eq:bppd-ex14-linearized-density}
\end{equation}
and this density should be compared with direct diagonalization of $\pmb{\mathcal L}$. This is not the covariance density. Away from the zero mode, the two limiting spectral measures are related by
\begin{equation}
\rho_{\pmb{\mathcal L}}(x)=\frac{2N}{N+P}|x|\,\rho_{\pmb W}(x^2)\,,\qquad x\neq0\,.
\label{eq:bppd-ex14-density-square-map}
\end{equation}
Check separately the normalization of the covariance density, the zero-mode fraction, and the first moment
\begin{equation}
\frac{1}{N}{\rm Tr}\pmb W\,.
\label{eq:bppd-ex14-first-moment}
\end{equation}
Report $N$, $P$, $\alpha$, $d$, $M$, $T_{\rm burn}$, $T_{\rm meas}$, $S$, $\epsilon$, the grid spacing, the branch convention for $z_{\rm L}$, the nonzero-weight distribution, and the discrepancy measures used in the covariance and auxiliary linearized comparisons.
\end{exerciseblock}

\section{Spectral density of sparse symmetric random matrices}
\label{sec:sparse-symmetric-spectral-density}
We now specialize the general formalism to sparse real symmetric matrices. This is the first central example because it contains, in its simplest form, the main conceptual differences between dense random matrix theory and finite-connectivity random matrix theory. In dense Wigner-type ensembles, each row contains a diverging number of weakly correlated entries, and the limiting spectral density is controlled by self-averaging at the level of a scalar resolvent. In sparse symmetric ensembles, each row contains only a finite number of nonzero entries in the thermodynamic limit. The local environment of a vertex therefore remains random. At the ensemble level, the spectral density is obtained from the statistics of local Green functions rather than from a closed equation for a single scalar resolvent.

The ensemble considered in this section is
\begin{equation}
A_{ij}=D_i\delta_{ij}+C_{ij}J_{ij}\,,\qquad A_{ij}=A_{ji},
\label{eq:ssrm-model}
\end{equation}
where $\pmb C$ is the adjacency matrix of a sparse undirected graph, $J_{ij}=J_{ji}$ are edge weights, and $D_i$ are diagonal terms. Unless stated otherwise, the graph is locally tree-like in the thermodynamic limit, the weights on distinct edges are independent with law $p_J(J)$, and the diagonal terms are independent with law $p_D(D)$. The adjacency matrix of an unweighted graph is obtained by setting $J_{ij}=1$ and $D_i=0$, while graph Laplacians and Anderson-type operators correspond to different choices of $D_i$ and $J_{ij}$. For instance, the adjacency operator is
\begin{equation}
A_{ij}=C_{ij}\,,
\label{eq:ssrm-adjacency}
\end{equation}
whereas the combinatorial Laplacian is
\begin{equation}
L_{ij}=\left(\sum_{\ell=1}^N C_{i\ell}\right)\delta_{ij}-C_{ij}\,.
    \label{eq:ssrm-laplacian}
\end{equation}
Both are sparse symmetric matrices, but their spectra encode different graph-theoretic and physical information.

The spectral density of sparse symmetric matrices was one of the first random-matrix problems treated by statistical-mechanics methods. Edwards and Jones introduced the Gaussian integral representation of the density of states for large symmetric random matrices \cite{EdwardsJones1976}. Rodgers and Bray applied replica ideas to sparse random matrices and showed that finite connectivity leads to a functional order parameter rather than to a scalar self-consistency equation \cite{RodgersBray1988}. Later work clarified the role of localized states, finite-connectivity corrections, and approximate recursive descriptions \cite{BiroliMonasson1999,SemerjianCugliandolo2002,Kuhn2008}. The cavity formulation developed in \cite{RogersTakedaPerezCastilloKuhn2008} gives a direct message-passing derivation of the spectral density for locally tree-like sparse symmetric matrices and also connects naturally with sparse covariance ensembles. A pedagogical account of the equivalence between the replica and cavity formulations is given in \cite{SuscaVivoKuhn2021}. There is also a rigorous parallel literature based on local weak convergence and resolvent methods, which relates sparse random graphs to limiting random trees and proves refined results for random regular graphs in several regimes \cite{BordenaveLelarge2010,DumitriuPal2012,BauerschmidtHuangYau2019}.

Recall that for a fixed realization of $\pmb{A}$, the resolvent is
\begin{equation}
\pmb{G}(z)=(z\pmb{I} -\pmb{A})^{-1}\,,\qquad z=\lambda-i\epsilon\,,\qquad\epsilon>0\,,
\label{eq:ssrm-resolvent}
\end{equation}
so that the regularized spectral density is given by
\begin{equation}
\rho_{\pmb{A},\epsilon}(\lambda) =\frac{1}{\pi N}\sum_{i=1}^N{\rm Im}[\pmb{G}_{ii}(\lambda-i\epsilon)]\,.
\label{eq:ssrm-regularized-density}
\end{equation}
The aim is to compute the large-$N$ limit of \eqref{eq:ssrm-regularized-density}, and then to understand the limit $\epsilon\downarrow0$. In sparse systems these two limits are delicate. At finite $N$, the spectrum is a sum of delta peaks. At finite $\epsilon$, the resolvent smooths these peaks into Lorentzians. In the thermodynamic limit, the limiting spectral measure may contain absolutely continuous parts, isolated atoms, and singular structures associated with localized states or finite graph components. The regulator is therefore both a technical device and a diagnostic probe. In most numerical applications one therefore works at small but finite $\epsilon$, rather than attempting to take the limit explicitly.

The same cavity equations can also be derived directly from Schur complements, without starting from the probabilistic cavity interpretation. Indeed, consider first a tree, and let $G_{i\to j}(z)$ denote the diagonal resolvent at vertex $i$ in the branch obtained by deleting the edge between $i$ and $j$. On a tree, removing $i$ disconnects its neighboring branches. The matrix $z\pmb{I}-\pmb{A}$ restricted to the branch rooted at $i$ can then be written in block form as
\begin{equation}
\begin{pmatrix}
z-D_i & -\pmb{a}^{\rm T}\\
-\pmb{a} & \pmb{B}
\end{pmatrix}\,,
\label{eq:ssrm-schur-block}
\end{equation}
where the vector $\pmb{a}$ has components $A_{i\ell}=J_{i\ell}$ for $\ell\in\partial i\setminus j$, and $\pmb{B}$ is block diagonal because the branches attached to distinct neighbors $\ell\in\partial i\setminus j$ are disconnected. The Schur complement gives
\begin{equation}
G_{i\to j}(z)=\left[ z-D_i-\pmb{a}^{\rm T}\pmb{B}^{-1}\pmb{a}\right]^{-1}\,.
\label{eq:ssrm-schur-complement}
\end{equation}
Since $\pmb{B}^{-1}$ is block diagonal, only the diagonal resolvent of the neighboring branch contributes, and one obtains the exact tree recursion
\begin{equation}
G_{i\to j}(z)=\frac{1}{z-D_i -\displaystyle\sum_{\ell\in\partial i\setminus j}J_{i\ell}^2G_{\ell\to i}(z)}\,.
\label{eq:ssrm-cavity-green}
\end{equation}
The full diagonal resolvent at $i$ is obtained by including all neighbors,
\begin{equation}
G_{i}(z) =\frac{1}{z-D_i-\displaystyle\sum_{\ell\in\partial i}J_{i\ell}^2G_{\ell\to i}(z)}\,.
\label{eq:ssrm-full-green}
\end{equation}
Equations \eqref{eq:ssrm-cavity-green} and \eqref{eq:ssrm-full-green} are exact on trees. On locally tree-like random graphs they become the cavity equations for the thermodynamic spectral density.

The same equations can be written in the inverse-variance notation that follows directly from the Gaussian integral representation introduced in the previous section. Recalling that
\begin{equation}
G_{i\to j}(z)=\frac{i}{\omega_{i\to j}(z)}\,,
\label{eq:ssrm-omega-green-relation}
\end{equation}
then \eqref{eq:ssrm-cavity-green} becomes
\begin{equation}
\omega_{i\to j}(z)=i(z-D_i)+\sum_{\ell\in\partial i\setminus j}\frac{J_{i\ell}^2}{\omega_{\ell\to i}(z)}.
    \label{eq:ssrm-cavity-omega}
\end{equation}
This is the form obtained by assuming Gaussian cavity marginals in the Edwards--Jones partition function. The Green-function version \eqref{eq:ssrm-cavity-green} is often more intuitive, while the inverse-variance version \eqref{eq:ssrm-cavity-omega} makes the Gaussian nature of the cavity solution explicit.

For an uncorrelated locally tree-like graph ensemble, one describes the statistics of a cavity message selected from the directed edges of a large instance. If $E$ is the undirected edge set of a finite support, let $\overrightarrow E=\{(i,j),(j,i):\{i,j\}\in E\}$ be the set of oriented edges. A fixed point of \eqref{eq:ssrm-cavity-green} defines the empirical cavity-message law $\mathcal{P}_{{\rm cav},N}(G)=|\overrightarrow E|^{-1}\sum_{(i,j)\in\overrightarrow E}\delta(G-G_{i\to j})$. When this law has a deterministic thermodynamic limit, we denote it by $\mathcal{P}_{\rm cav}(G)$. Let $p_k$ be the asymptotic degree distribution and recall that the excess-degree distribution seen by following a uniformly chosen edge is
\begin{equation}
q_\ell=\frac{(\ell+1)p_{\ell+1}}{\sum_k kp_k}\,,\qquad\ell=0,1,2,\ldots\,.
\label{eq:ssrm-excess-degree}
\end{equation}
The limiting cavity-message law satisfies
\begin{equation}
\mathcal{P}_{\rm cav}(G)=\sum_{\ell=0}^{\infty}q_\ell\int dD p_D(D) \left[\prod_{r=1}^{\ell}dG_r \mathcal{P}_{\rm cav}(G_r)  dJ_r p_J(J_r)\right]\delta\left(G-\frac{1}{z-D-\displaystyle\sum_{r=1}^{\ell}J_r^2G_r}\right)\,.
\label{eq:ssrm-cavity-distribution}
\end{equation}
The full diagonal Green functions define the empirical site law $\mathcal{P}_{{\rm site},N}(G)=N^{-1}\sum_{i=1}^{N}\delta(G-G_i)$. Its deterministic thermodynamic limit is
\begin{equation}
\mathcal{P}_{\rm site}(G)=\sum_{k=0}^{\infty}p_k\int dD p_D(D)\left[\prod_{r=1}^{k}dG_r \mathcal{P}_{\rm cav}(G_r) dJ_r p_J(J_r)\right]\delta\left(G-\frac{1}{z-D-\displaystyle\sum_{r=1}^{k}J_r^2G_r}\right)\,.
\label{eq:ssrm-site-distribution}
\end{equation}
The ensemble-averaged spectral density is then
\begin{equation}
\overline{\rho_{\pmb A,\epsilon}(\lambda)}=\frac{1}{\pi}{\rm Im}\int dG\mathcal{P}_{\rm site}(G)G\,,\qquad z=\lambda-i\epsilon\,.
\label{eq:ssrm-density-from-site-distribution}
\end{equation}
This is the basic cavity formula for sparse symmetric random matrices. It replaces the scalar Pastur-type equation of dense random matrix theory by a distributional fixed-point equation.

For the sparse Erd\H{o}s--R\'enyi ensemble, $p_k$ is Poisson with mean $c$, and the excess-degree distribution is the same Poisson law:
\begin{equation}
p_k=e^{-c}\frac{c^k}{k!}\,,\qquad q_\ell=e^{-c}\frac{c^\ell}{\ell!}\,.
\label{eq:ssrm-er-degree-laws}
\end{equation}
For a random $c$-regular graph,
\begin{equation}
p_k=\delta_{k,c}\,,\qquad q_\ell=\delta_{\ell,c-1}\,.
\label{eq:ssrm-rrg-degree-laws}
\end{equation}
The random regular case, already used above as an algorithmic benchmark, is especially useful here because it gives the first closed-form solution of the sparse symmetric cavity equations. In the locally tree-like $c$-regular limit, the cavity law collapses to a delta peak. For the unweighted adjacency matrix with $J=1$ and $D=0$, the representative cavity Green function $G_{\rm cav}$ satisfies
\begin{equation}
G_{\rm cav}(z)=\frac{1}{z-(c-1)G_{\rm cav}(z)}\,.
\label{eq:ssrm-rrg-cavity-equation}
\end{equation}
Solving the quadratic equation gives
\begin{equation}
G_{\rm cav}(z)=\frac{z-\sqrt{z^2-4(c-1)}}{2(c-1)}\,,
\label{eq:ssrm-rrg-cavity-solution}
\end{equation}
where the branch is chosen so that $G_{\rm cav}(z)\sim 1/z$ for large $|z|$ and ${\rm Im}[G_{\rm cav}(\lambda-i\epsilon)]>0$. The full site Green function is
\begin{equation}
G_{\rm site}(z)=\frac{1}{z-cG_{\rm cav}(z)}\,,
\label{eq:ssrm-rrg-site-green}
\end{equation}
and hence
\begin{equation}
\rho_{\rm RRG}(\lambda)=\frac{c}{2\pi}\frac{\sqrt{4(c-1)-\lambda^2}}{c^2-\lambda^2}\mathbf{1}_{|\lambda|\leq 2\sqrt{c-1}}\,.
\label{eq:ssrm-kesten-mckay}
\end{equation}
This is the Kesten--McKay law \cite{Kesten1959,McKay1981}. For a connected $c$-regular graph, the Perron eigenvalue $\lambda=c$ has spectral weight $1/N$ and therefore does not contribute to the limiting empirical density, although it may be important for dynamical applications. More generally, the multiplicity of the eigenvalue $c$ is the number of connected components.

\begin{examplebox}[The $c=2$ random regular graph and the arcsine law]
The Kesten--McKay density provides a useful check in the special case $c=2$. A $2$-regular graph is a disjoint union of cycles. For large cycles, the adjacency spectrum is the same local object as the spectrum of the infinite one-dimensional chain.

Setting $c=2$ in \eqref{eq:ssrm-kesten-mckay}, we obtain
\begin{equation}
\rho_{\rm RRG}^{(c=2)}(\lambda)=\frac{2}{2\pi}\frac{\sqrt{4-\lambda^2}}{4-\lambda^2}\mathbf{1}_{|\lambda|\leq2}=\frac{1}{\pi\sqrt{4-\lambda^2}} \mathbf{1}_{|\lambda|\leq2}\,.
\label{eq:ssrm-ped-c2-arcsine}
\end{equation}
This is the arcsine law.

Let us recover the same result directly from a large cycle. The adjacency matrix of a cycle with $L$ vertices has eigenvalues
\begin{equation}
\lambda_m=2\cos\left(\frac{2\pi m}{L}\right)\,,\qquad m=0,\ldots,L-1\,.
\label{eq:ssrm-ped-cycle-eigenvalues}
\end{equation}
As $L\to\infty$, the angle
\begin{equation}
\theta=\frac{2\pi m}{L}
\label{eq:ssrm-ped-cycle-angle}
\end{equation}
becomes uniformly distributed on $[0,2\pi)$. Since
\begin{equation}
\lambda=2\cos\theta\,,
\label{eq:ssrm-ped-cycle-change-variable}
\end{equation}
we have
\begin{equation}
\left|\frac{d\lambda}{d\theta}\right|=2|\sin\theta|=\sqrt{4-\lambda^2}\,.
\label{eq:ssrm-ped-cycle-jacobian}
\end{equation}
For each $\lambda\in(-2,2)$ there are two angles in $[0,2\pi)$ with the same value of $\lambda$. Therefore
\begin{equation}
\rho(\lambda)=2\cdot\frac{1}{2\pi}\frac{1}{\sqrt{4-\lambda^2}}=\frac{1}{\pi\sqrt{4-\lambda^2}}\,,
\label{eq:ssrm-ped-cycle-density}
\end{equation}
which agrees with \eqref{eq:ssrm-ped-c2-arcsine}. This example shows how the random-regular cavity formula reduces, in the case $c=2$, to the familiar density of states of the one-dimensional tight-binding chain.
\end{examplebox} 

The Erd\H{o}s--R\'enyi adjacency matrix is less trivial. Even when $J=1$ and $D=0$, the random number of incoming messages in \eqref{eq:ssrm-cavity-distribution} prevents the distribution $\mathcal{P}_{\rm cav}$ from collapsing to a delta function. The finite-connectivity spectral density contains features that are absent in dense Wigner matrices. Isolated vertices give an atom at $\lambda=0$. Finite tree components and dangling structures generate additional singular contributions. High-degree vertices and local defects produce localized states and spectral tails. These effects were central in the early replica and effective-medium analyses of sparse spectra \cite{RodgersBray1988,BiroliMonasson1999,SemerjianCugliandolo2002,Kuhn2008}. In the cavity formulation they appear naturally through the distribution of messages: rare local neighborhoods generate rare values of the local Green function. The same low-connectivity picture appeared earlier as a quenched-versus-annealed check; here it is used instead to identify finite graph components as explicit spectral atoms.

\begin{examplebox}[The first terms of the sparse Erd\H{o}s--R\'enyi spectrum at small connectivity]
The statement that local graph structures produce singular spectral contributions can be seen already at the first order in the mean degree $c$. Consider the unweighted Erd\H{o}s--R\'enyi adjacency matrix,
\begin{equation}
A_{ij}=C_{ij}\,,\qquad {\rm Prob}(C_{ij}=1)=\frac{c}{N}\,, \qquad C_{ij}=C_{ji}\,,\qquad C_{ii}=0\,.
\label{eq:ssrm-ped-er-adjacency}
\end{equation}
For small $c$, most vertices are isolated, while the first nontrivial components are isolated edges. Components with two or more adjacent edges have probability of order $c^2$ from the viewpoint of a uniformly chosen vertex.

An isolated vertex contributes the one-dimensional adjacency matrix
\begin{equation}
\pmb A_{\rm iso}=(0),
\label{eq:ssrm-ped-isolated-vertex}
\end{equation}
and therefore contributes one eigenvalue at $\lambda=0$. The probability that a uniformly chosen vertex is isolated is
\begin{equation}
\left(1-\frac{c}{N}\right)^{N-1}=e^{-c}+o(1)=1-c+O(c^2)\,.
\label{eq:ssrm-ped-isolated-probability}
\end{equation}
Thus isolated vertices contribute, to first order in $c$,
\begin{equation}
(1-c)\delta(\lambda)\,.
\label{eq:ssrm-ped-isolated-contribution}
\end{equation}

Now consider an isolated edge. Its adjacency matrix is
\begin{equation}
\pmb A_{\rm edge} =\begin{pmatrix}
0 & 1\\
1 & 0
\end{pmatrix}\,,
\label{eq:ssrm-ped-isolated-edge-matrix}
\end{equation}
with eigenvalues
\begin{equation}
\lambda_+=1\,, \qquad \lambda_-=-1\,.
\label{eq:ssrm-ped-isolated-edge-eigenvalues}
\end{equation}
The expected number of isolated edges is
\begin{equation}
\binom{N}{2}\frac{c}{N}\left(1-\frac{c}{N}\right)^{2(N-2)}=\frac{cN}{2}e^{-2c}+o(N)\,.
\label{eq:ssrm-ped-number-isolated-edges}
\end{equation}
Therefore the expected fraction of vertices belonging to isolated edges is
\begin{equation}
\frac{2}{N}\binom{N}{2}\frac{c}{N}\left(1-\frac{c}{N}\right)^{2(N-2)}=c+O(c^2)\,.
\label{eq:ssrm-ped-fraction-isolated-edges}
\end{equation}
Since each isolated edge has two eigenvalues, one at $+1$ and one at $-1$, isolated edges contribute
\begin{equation}
\frac{c}{2}\delta(\lambda-1)+\frac{c}{2}\delta(\lambda+1)+O(c^2)\,.
\label{eq:ssrm-ped-isolated-edge-contribution}
\end{equation}
Combining the isolated vertices and isolated edges gives the first-order small-$c$ spectral density
\begin{equation}
\rho(\lambda)=(1-c)\delta(\lambda)+\frac{c}{2}\delta(\lambda-1)+\frac{c}{2}\delta(\lambda+1)+O(c^2)\,.
\label{eq:ssrm-ped-small-c-density}
\end{equation}
This calculation is elementary, but it is conceptually important. The first correction to the isolated-vertex spectrum is not a smooth perturbation of a dense random-matrix law. It is a contribution from a finite graph component. This is precisely the type of local structure that the cavity distribution of Green functions keeps track of.
\end{examplebox}

The local density of states at vertex $i$ is
\begin{equation}
\rho_i(\lambda)=\frac{1}{\pi}\lim_{\epsilon\downarrow0}{\rm Im}[G_i(\lambda-i\epsilon)]\,.
\label{eq:ssrm-local-density}
\end{equation}
The distribution of $\rho_i(\lambda)$ over vertices is therefore obtained from the distribution of the full Green function \eqref{eq:ssrm-site-distribution}. Extended states are associated with a broad set of vertices having comparable local weights, while localized states are associated with large local spectral weights concentrated on atypical neighborhoods. This connection between local resolvents, random-tree recursions, and localization is closely related to the self-consistent Green-function approach to localization on Bethe lattices \cite{AbouChacraAndersonThouless1973}. In sparse random matrices, the same recursion simultaneously gives the spectral density and probes the local structure of eigenvectors.

\begin{examplebox}[A high-degree vertex creates localized outlying modes]
A simple way to see how local structure can create large eigenvalues is to study the star graph with one central vertex connected to $K$ leaves. Its adjacency matrix has the block form
\begin{equation}
\pmb A_{\rm star}=\begin{pmatrix}
0 & \pmb 1_K^{\rm T}\\
\pmb 1_K & \pmb 0_{K\times K}
\end{pmatrix}\,,
\label{eq:ssrm-ped-star-adjacency}
\end{equation}
where $\pmb 1_K=(1,\ldots,1)^{\rm T}$. Let an eigenvector have central component $a$ and equal leaf components $b$:
\begin{equation}
\pmb u=(a,b,\ldots,b)^{\rm T}.
\label{eq:ssrm-ped-star-vector}
\end{equation}
The eigenvalue equation $\pmb A_{\rm star}\pmb u=\lambda\pmb u$ gives
\begin{equation}
Kb=\lambda a\,,\qquad a=\lambda b\,.
\label{eq:ssrm-ped-star-eigen-equations}
\end{equation}
Eliminating $a$ and $b$ gives
\begin{equation}
\lambda^2=K\,,\qquad \lambda_\pm=\pm\sqrt{K}\,.
\label{eq:ssrm-ped-star-outliers}
\end{equation}
The remaining $K-1$ eigenvalues are zero, corresponding to leaf vectors orthogonal to $\pmb 1_K$.

After normalization, the eigenvectors associated with $\lambda_\pm$ have
\begin{equation}
|a|^2=\frac{1}{2}\,,\qquad|b|^2=\frac{1}{2K}\,.
\label{eq:ssrm-ped-star-eigenvector-weights}
\end{equation}
Thus half of the eigenvector weight sits on the central vertex, while the other half is spread over the leaves. For large $K$, the eigenvalue scale $\sqrt{K}$ is much larger than the typical eigenvalue scale of a bounded-degree region. This example illustrates why high-degree vertices in sparse graphs can create spectral tails or outlying localized modes. Such effects are invisible in a scalar dense effective-medium approximation but are naturally encoded in the distribution of local Green functions.
\end{examplebox}

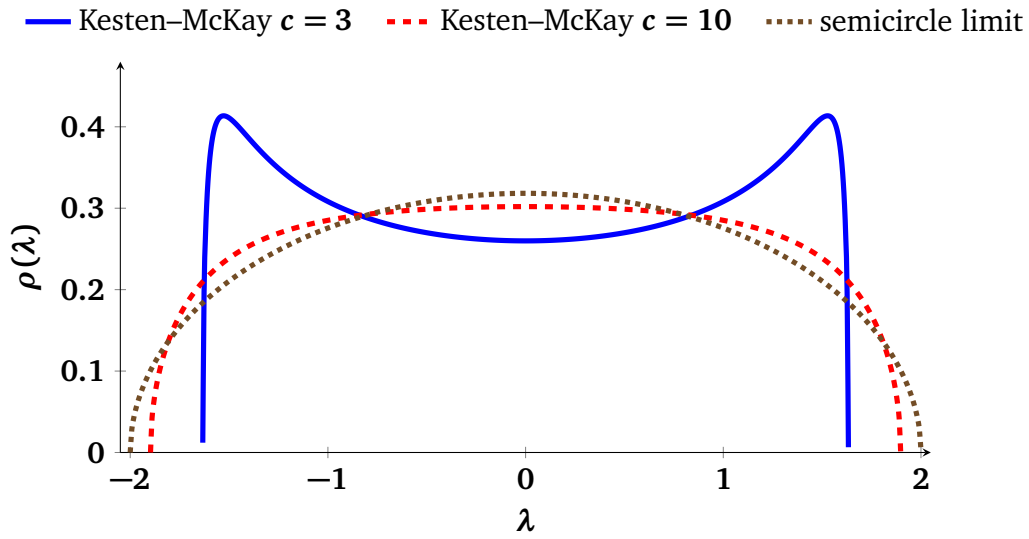
\begin{figure}[t]
\centering
\begin{tikzpicture}
\pgfmathsetmacro{\cA}{3}
\pgfmathsetmacro{\cB}{10}
\pgfmathsetmacro{\edgeA}{2*sqrt((\cA-1)/\cA)}
\pgfmathsetmacro{\edgeB}{2*sqrt((\cB-1)/\cB)}
\begin{axis}[
    width=0.82\textwidth,
    height=0.45\textwidth,
    xmin=-2.05,
    xmax=2.05,
    ymin=0,
    ymax=0.48,
    axis lines=left,
    xlabel={$\lambda$},
    ylabel={$\rho(\lambda)$},
    xlabel style={font=\large},
    ylabel style={font=\large},
    tick label style={font=\large},
    xtick={-2,-1,0,1,2},
    ytick={0,0.1,0.2,0.3,0.4},
    samples=500,
    smooth,
    clip=false,
    legend style={
        at={(0.5,1.04)},
        anchor=south,
        draw=none,
        fill=none,
        font=\large,
        legend columns=3,
        /tikz/every even column/.append style={column sep=0.35cm}
    },
    every axis plot/.append style={line width=1.95pt}
]
\addplot+[no marks, solid, domain=-\edgeA:\edgeA]
{\cA/(2*pi*(\cA-x^2))*sqrt(max(0,4*(\cA-1)/\cA - x^2))};
\addlegendentry{Kesten--McKay $c=3$}

\addplot+[no marks, dashed, domain=-\edgeB:\edgeB]
{\cB/(2*pi*(\cB-x^2))*sqrt(max(0,4*(\cB-1)/\cB - x^2))};
\addlegendentry{Kesten--McKay $c=10$}

\addplot+[no marks, dotted, domain=-2:2]
{1/(2*pi)*sqrt(max(0,4 - x^2))};
\addlegendentry{semicircle limit}
\end{axis}
\end{tikzpicture}
\caption{Kesten--McKay versus the semicircle law in the dense-connectivity scaling. The finite-$c$ curves are the analytic random-regular densities after rescaling eigenvalues by $c^{-1/2}$; the dotted curve is the Wigner semicircle law obtained when the cavity message distribution concentrates.}
\label{fig:ssrm-kesten-mckay-semicircle}
\end{figure}

It is useful to compare the cavity equation with the dense limit and check whether Wigner's semicircle law is recovered. For this check, set $D_i=0$. Suppose that the mean degree $c$ tends to infinity and the weights are scaled as
\begin{equation}
J_{ij}=\frac{\widetilde J_{ij}}{\sqrt{c}}\,,\qquad \int d\widetilde J p_{\widetilde J}(\widetilde J) \widetilde J=0\,,\qquad \int d\widetilde J p_{\widetilde J}(\widetilde J)\,\widetilde J^2=\sigma^2\,.
\label{eq:ssrm-dense-scaling}
\end{equation}
If the distribution of cavity messages concentrates in this limit, then the sum in \eqref{eq:ssrm-cavity-green} self-averages by the law of large numbers to
\begin{equation}
\sum_{r=1}^{k}J_r^2G_r \longrightarrow \sigma^2 g(z)\,,
\label{eq:ssrm-dense-self-averaging}
\end{equation}
and the limiting Green function satisfies
\begin{equation}
g(z)=\frac{1}{z-\sigma^2 g(z)}\,.
\label{eq:ssrm-wigner-equation}
\end{equation}
The dense-connectivity reduction of the sparse symmetric cavity equation is derived explicitly in Appendix~\ref{app:dense-limit-reductions}. This is the usual self-consistent equation for the Stieltjes transform of the Wigner semicircle law. Thus the cavity equations contain the dense Wigner result as a limiting case, but at finite connectivity they retain the full distributional structure of the sparse graph. Figure~\ref{fig:ssrm-kesten-mckay-semicircle} displays this finite-connectivity benchmark and its dense-connectivity semicircle limit at the level of analytic densities.

The same observation explains the role of effective-medium approximations. If one replaces $\mathcal{P}_{\rm cav}(G)$ by a delta function, or otherwise suppresses the fluctuations of the local Green function, one obtains a closed scalar approximation. Such approximations can capture broad features of the continuous part of the spectrum but miss important sparse-graph effects such as atoms, tails, and localization. The single-defect approximation improves on this by treating one atypical local environment against an effective background \cite{BiroliMonasson1999}. The full cavity or population-dynamics solution keeps the distribution of local environments and is therefore the natural finite-connectivity theory.

Several consistency checks follow directly from the equations. For an isolated vertex, $k=0$, equation \eqref{eq:ssrm-full-green} gives $G_i(z)=(z-D_i)^{-1}$, so the corresponding contribution to the density is a delta peak at $D_i$ as $\epsilon\downarrow0$. For an unweighted random regular graph, the distributional equation collapses to the scalar recursion \eqref{eq:ssrm-rrg-cavity-equation}, and the Kesten--McKay law \eqref{eq:ssrm-kesten-mckay} is recovered. In the large-connectivity scaling \eqref{eq:ssrm-dense-scaling}, the distributional recursion collapses instead to the Wigner equation \eqref{eq:ssrm-wigner-equation}. These three checks test, respectively, the local-defect limit, the homogeneous finite-connectivity limit, and the dense random-matrix limit.

From a computational point of view, equations \eqref{eq:ssrm-cavity-distribution}--\eqref{eq:ssrm-density-from-site-distribution} are solved by population dynamics. One represents $\mathcal{P}_{\rm cav}$ by a large population of complex numbers, repeatedly updates a randomly chosen member by sampling an excess degree, edge weights, diagonal disorder, and incoming cavity messages according to \eqref{eq:ssrm-cavity-distribution}, and then samples full site Green functions according to \eqref{eq:ssrm-site-distribution}. The output is the regularized density at the chosen value of $\lambda$ and $\epsilon$. Repeating the procedure over a grid of $\lambda$ gives the spectral density. The same equations can also be iterated as belief-propagation equations on a single large graph, in which case \eqref{eq:ssrm-regularized-density} gives the spectral density of that realization.

The conclusion of this section is that sparse symmetric random matrices are governed by random local resolvents. The spectral density is the average of the imaginary part of a full Green function, but that Green function is itself distributed according to the random rooted-tree environment of a typical vertex. This is the fundamental departure from dense random matrix theory. It is also the reason why the same formalism can be extended naturally to graph ensembles with degree correlations, modular structure, covariance-type bipartite ensembles, large deviations of eigenvalue counts, and non-Hermitian sparse matrices.

\begin{exerciseblock}
\exitem[Schur complement derivation of the cavity recursion]
Let $\pmb A$ be a real symmetric matrix supported on a finite tree with vertex set $V=\{1,\ldots,N\}$. Write $A_{ii}=D_i$ and $A_{i\ell}=J_{i\ell}$ when vertices $i$ and $\ell$ are connected. Isolate a vertex $i$ from the branches attached to its neighbors and write the block decomposition of $z\pmb I-\pmb A$ in which the first block corresponds to vertex $i$. Using the Schur complement, derive
\begin{equation}
G_i(z)=\frac{1}{z-D_i-\displaystyle\sum_{\ell\in\partial i}J_{i\ell}^2G_{\ell\to i}(z)}\,.
    \label{eq:ssrm-ex-schur-full}
\end{equation}
Then repeat the derivation with the edge between $i$ and $j$ removed and obtain
\begin{equation}
G_{i\to j}(z)=\frac{1}{z-D_i-\displaystyle\sum_{\ell\in\partial i\setminus j}J_{i\ell}^2G_{\ell\to i}(z)}\,.
\label{eq:ssrm-ex-schur-cavity}
\end{equation}

\exitem[Inverse-variance and Green-function parametrizations]
Starting from
\begin{equation}
\omega_{i\to j}(z)=i(z-D_i)+\sum_{\ell\in\partial i\setminus j}\frac{J_{i\ell}^2}{\omega_{\ell\to i}(z)}\,,
\label{eq:ssrm-ex-omega-recursion}
\end{equation}
and using
\begin{equation}
G_{i\to j}(z)=\frac{i}{\omega_{i\to j}(z)}\,,
\label{eq:ssrm-ex-G-omega-relation}
\end{equation}
derive the Green-function recursion \eqref{eq:ssrm-ex-schur-cavity}. Explain why the Green-function form is usually more directly connected to the spectral density.

\exitem[Isolated vertices and atoms]
Suppose a vertex has degree zero in a sparse symmetric matrix with diagonal entry $D_i$. Show that its full local Green function is
\begin{equation}
G_i(z)=\frac{1}{z-D_i}\,.
\label{eq:ssrm-ex-isolated-green}
\end{equation}
Deduce that this vertex contributes a delta peak at $\lambda=D_i$ to the local density of states. Specialize to the adjacency matrix of an Erd\H{o}s--R\'enyi graph on $N$ vertices, with $A_{ii}=0$, $A_{ij}=A_{ji}=C_{ij}$, and ${\rm Prob}(C_{ij}=1)=c/N$ for $i<j$. Compute the limiting weight of the atom at zero coming from isolated vertices.

\exitem[Isolated edges]
Consider an isolated weighted edge with matrix
\begin{equation}
\pmb A_{\rm edge}=\begin{pmatrix}
D_1 & J\\
J & D_2
\end{pmatrix}\,.
\label{eq:ssrm-ex-weighted-edge}
\end{equation}
Compute its two eigenvalues explicitly. Then compute the diagonal resolvent entries by direct inversion and compare them with the cavity formulas for a two-vertex tree.

\exitem[Small-connectivity expansion]
For the unweighted Erd\H{o}s--R\'enyi adjacency matrix with $A_{ii}=0$, $A_{ij}=A_{ji}=C_{ij}$, and ${\rm Prob}(C_{ij}=1)=c/N$ for $i<j$, use isolated vertices and isolated edges to reproduce the first-order small-$c$ density
\begin{equation}
\rho(\lambda)=(1-c)\delta(\lambda)+\frac{c}{2}\delta(\lambda-1)+\frac{c}{2}\delta(\lambda+1)+O(c^2)\,.
\label{eq:ssrm-ex-small-c-density}
\end{equation}
Which finite connected components would contribute at order $c^2$?

\exitem[Distributional cavity equation]
Starting from the local recursion
\begin{equation}
G_{i\to j}(z)=\frac{1}{z-D_i-\displaystyle\sum_{\ell\in\partial i\setminus j}J_{i\ell}^2G_{\ell\to i}(z)}\,,
\label{eq:ssrm-ex-local-recursion}
\end{equation}
derive the distributional equation
\begin{equation}
\mathcal{P}_{\rm cav}(G)=\sum_{\ell=0}^{\infty} q_\ell \int dD p_D(D)\left[\prod_{r=1}^{\ell} dG_r \mathcal{P}_{\rm cav}(G_r) dJ_r p_J(J_r)\right]\delta\left(G-\frac{1}{z-D-\displaystyle\sum_{r=1}^{\ell}J_r^2G_r}\right)\,.
\label{eq:ssrm-ex-distributional-cavity}
\end{equation}
Explain why $q_\ell$ is an excess-degree law rather than the full degree law.

\exitem[Site law versus cavity law]
Given the cavity law $\mathcal P_{\rm cav}(G)$, derive the site law
\begin{equation}
\mathcal{P}_{\rm site}(G)=\sum_{k=0}^{\infty}p_k \int dD p_D(D) \left[\prod_{r=1}^{k}  dG_r\mathcal{P}_{\rm cav}(G_r)dJ_r p_J(J_r)\right]\delta\left(G-\frac{1}{z-D-\displaystyle\sum_{r=1}^{k}J_r^2G_r}\right)\,.
\label{eq:ssrm-ex-site-law}
\end{equation}
Use it to prove
\begin{equation}
\overline{\rho_{\pmb A,\epsilon}(\lambda)}=\frac{1}{\pi}{\rm Im} \int dG \mathcal P_{\rm site}(G) G\,.
\label{eq:ssrm-ex-density-from-site-law}
\end{equation}

\exitem[Random regular graph calculation]
For the unweighted adjacency matrix in the locally tree-like $c$-regular ensemble, equivalently on the infinite $c$-regular tree, show that the cavity law collapses to a delta distribution. Derive
\begin{equation}
G_{\rm cav}=\frac{1}{z-(c-1)G_{\rm cav}}\,,
\label{eq:ssrm-ex-rrg-cavity}
\end{equation}
solve the quadratic equation with the branch satisfying $G_{\rm cav}(z)\sim 1/z$ for large $|z|$ and ${\rm Im}\,G_{\rm cav}(\lambda-i\epsilon)>0$, and use
\begin{equation}
G_{\rm site}=\frac{1}{z-cG_{\rm cav}}
\label{eq:ssrm-ex-rrg-site}
\end{equation}
to recover the Kesten--McKay density.

\exitem[The $c=2$ check]
Set $c=2$ in the Kesten--McKay density and show that
\begin{equation}
\rho(\lambda)=\frac{1}{\pi\sqrt{4-\lambda^2}}\mathbf{1}_{|\lambda|\leq2}\,.
\label{eq:ssrm-ex-c2-density}
\end{equation}
Then derive the same result from the eigenvalues
\begin{equation}
\lambda_m=2\cos\left(\frac{2\pi m}{L}\right)
\label{eq:ssrm-ex-cycle-eigenvalues}
\end{equation}
of a cycle of length $L$ in the limit $L\to\infty$.

\exitem[Dense limit and the semicircle law]
Assume $D_i=0$ and the scaling
\begin{equation}
J_{ij}=\frac{\widetilde J_{ij}}{\sqrt c}\,,\qquad\mathbb E[\widetilde J]=0\,,\qquad\mathbb E[\widetilde J^2]=\sigma^2\,,
\label{eq:ssrm-ex-dense-scaling}
\end{equation}
and let $c\to\infty$ after the thermodynamic limit. Starting from the distributional cavity equation, show that the message distribution concentrates and that the limiting Green function satisfies
\begin{equation}
g(z)=\frac{1}{z-\sigma^2g(z)}\,.
\label{eq:ssrm-ex-wigner-equation}
\end{equation}
Solve this equation and recover the Wigner semicircle law.

\exitem[Diagonal disorder]
Keep the diagonal terms $D_i$ random with distribution $p_D(D)$ while taking the dense limit in Exercise 6.10. Show that the limiting resolvent satisfies
\begin{equation}
g(z)=\int dD p_D(D) \frac{1}{z-D-\sigma^2g(z)}\,.
\label{eq:ssrm-ex-pastur-equation}
\end{equation}
Explain how this differs from the pure Wigner case.

\exitem[Star graph and localized modes]
For the star graph with one central vertex connected to $K$ leaves, prove that the adjacency spectrum is
\begin{equation}
\left\{\sqrt K,\,-\sqrt K,\,\underbrace{0,\ldots,0}_{K-1\text{ times}}\right\}\,.
    \label{eq:ssrm-ex-star-spectrum}
\end{equation}
Compute the normalized eigenvectors associated with $\pm\sqrt K$ and show that the central vertex carries half of the eigenvector weight. Explain why this is a simple model for localized high-degree effects in sparse graphs.

\exitem[Local density of states]
For a fixed sparse matrix realization, let $G_i(z)$ be the full diagonal Green function at vertex $i$. Show that the distribution over vertices of
\begin{equation}
\rho_i^{(\epsilon)}(\lambda)=\frac{1}{\pi}{\rm Im}\,G_i(\lambda-i\epsilon)
\label{eq:ssrm-ex-local-density}
\end{equation}
contains more information than the averaged density
\begin{equation}
\rho_\epsilon(\lambda)=\frac{1}{N}\sum_{i=1}^N\rho_i^{(\epsilon)}(\lambda)\,.
\label{eq:ssrm-ex-average-local-density}
\end{equation}
Discuss how a broad distribution of $\rho_i^{(\epsilon)}(\lambda)$ may indicate localized spectral weight.

\exitem[Effective-medium approximation]
For an unweighted Erd\H{o}s--R\'enyi adjacency matrix with zero diagonal, suppose one approximates
\begin{equation}
\mathcal P_{\rm cav}(G)\approx\delta(G-G_\star)\,.
    \label{eq:ssrm-ex-ema-ansatz}
\end{equation}
Using the Poisson excess-degree distribution, close the distributional equation by matching the mean cavity Green function and derive the scalar self-consistency equation
\begin{equation}
G_\star=\sum_{\ell=0}^{\infty}e^{-c}\frac{c^\ell}{\ell!}\frac{1}{z-\ell G_\star}\,.
\label{eq:ssrm-ex-ema-equation}
\end{equation}
Which features of the sparse spectrum are lost by this approximation?

\exitem[Programming exercise: population dynamics for the Erd\H{o}s--R\'enyi adjacency matrix]
Fix a mean degree $c=O(1)$, a population size $M$, a regulator $\epsilon>0$, a grid of real values of $\lambda$, a burn-in time $T_{\rm burn}$, and a number $T_{\rm meas}$ of site samples used for measurement. Implement the population-dynamics update
\begin{equation}
G_{\rm new}=\frac{1}{\lambda-i\epsilon-\displaystyle\sum_{r=1}^{\ell}G_r}\,,\qquad\ell\sim{\rm Poisson}(c)\,,
\label{eq:ssrm-ex-program-er-update}
\end{equation}
and estimate the spectral density by sampling the corresponding site Green functions with $k\sim{\rm Poisson}(c)$. For comparison, generate $S$ independent Erd\H{o}s--R\'enyi adjacency matrices of size $N_{\rm diag}$ with $A_{ii}=0$, $A_{ij}=A_{ji}$, and ${\rm Prob}(A_{ij}=1)=c/N_{\rm diag}$ for $i<j$. Use the same $\lambda$ grid and the same regulator $\epsilon$ in both estimates, and report $c$, $M$, $T_{\rm burn}$, $T_{\rm meas}$, $N_{\rm diag}$, $S$, $\epsilon$, the grid spacing, and a discrepancy measure.

\exitem[Programming exercise: random regular benchmark]
Fix an integer $c\geq2$, a list of system sizes $N$ such that $cN$ is even, a number $S$ of independent graph samples for each $N$, a regulator $\epsilon>0$, and a grid of real values of $\lambda$. Generate random $c$-regular graphs on $N$ vertices and compute their eigenvalue histograms. For the comparison with the Kesten--McKay bulk density, either remove the eigenvalue $c$ associated with each connected component before forming the histogram, or include the corresponding finite-$N$ Lorentzian contribution with the same regulator $\epsilon$. Then solve the scalar cavity equation
\begin{equation}
G_{\rm cav}=\frac{1}{\lambda-i\epsilon-(c-1)G_{\rm cav}}
\label{eq:ssrm-ex-program-rrg-cavity}
\end{equation}
numerically on the same grid and verify that it gives the same regularized density. Report $c$, $N$, $S$, $\epsilon$, the grid spacing, and a discrepancy measure.

\exitem[Programming exercise: local density statistics]
Fix a sparse graph realization with real symmetric adjacency matrix $\pmb A$, a regulator $\epsilon>0$, and several values of $\lambda$ chosen in the bulk and near a spectral tail. Compute
\begin{equation}
\rho_i^{(\epsilon)}(\lambda)=\frac{1}{\pi}{\rm Im}\left[\left((\lambda-i\epsilon)\pmb I-\pmb A\right)^{-1}\right]_{ii}
\label{eq:ssrm-ex-program-local-density}
\end{equation}
for all vertices $i$. For each value of $\lambda$, plot the empirical distribution of $\rho_i^{(\epsilon)}(\lambda)$ over vertices. Compare the behavior in the bulk of the spectrum and near a spectral tail, and check that $N^{-1}\sum_i\rho_i^{(\epsilon)}(\lambda)$ agrees with the corresponding trace-resolvent estimate. Report the graph ensemble, $N$, $c$ or the degree sequence, $\epsilon$, and the chosen values of $\lambda$.
\end{exerciseblock}

\section{Spectra of random graphs with topological constraints}
\label{sec:topologically-constrained-random-graphs}
The sparse ensembles considered so far are the simplest locally tree-like graph ensembles. In the Erd\H{o}s--R\'enyi ensemble the degrees fluctuate independently around a Poisson law, while in the random regular ensemble all vertices have the same degree. Real sparse networks, and many synthetic ensembles used to model them, contain additional topological information: heterogeneous degrees, degree--degree correlations, assortative or disassortative mixing, block structure, communities, and modular organization. From the spectral point of view these features are not perturbative decorations. They modify the statistics of the local neighborhoods seen by the resolvent, which, in turn, modify the self-consistency equations for the spectral density.

In this section we discuss how the cavity description of sparse symmetric matrices is modified when the underlying graph is sampled from an ensemble with topological constraints. The guiding example is the adjacency matrix $\pmb A$ of an undirected graph,
\begin{equation}
A_{ij}=C_{ij}\,,\qquad C_{ij}=C_{ji}\,,\qquad C_{ii}=0\,,
\label{eq:tcrg-adjacency}
\end{equation}
although the same construction applies, with minor notational changes, to weighted adjacency matrices, Laplacians, and sparse symmetric operators with diagonal disorder. The essential point is that the local recursion for a fixed tree remains unchanged whenever deleting an edge or vertex separates the adjacent branches:
\begin{equation}
G_{i\to j}(z)=\frac{1}{z-\displaystyle\sum_{\ell\in\partial i\setminus j}G_{\ell\to i}(z)}\,,\qquad G_i(z)=\frac{1}{z-\displaystyle\sum_{\ell\in\partial i}G_{\ell\to i}(z)}\,.
\label{eq:tcrg-tree-recursion}
\end{equation}
What changes is the probability law of the rooted neighborhood, and therefore the probability law of the incoming cavity messages. The problem is thus not to alter the local resolvent identity, but to compute the correct distributional recursion induced by the constrained graph ensemble.

The need for such a formulation was already apparent in early numerical studies of complex network spectra, where the spectral density of empirical or synthetic networks was found to depart strongly from the semicircle law \cite{FarkasDerenyiBarabasiVicsek2001}. Scale-free degree distributions, hubs, and local inhomogeneities generate tails and localized eigenvectors; degree correlations change the way high- and low-degree vertices are coupled; and modular structure produces isolated or nearly isolated eigenvalues carrying mesoscopic information. Analytical approaches to locally tree-like networks with correlations were developed in several forms, including recursive Green-function equations for complex networks \cite{DorogovtsevGoltsevMendesSamukhin2003}, replica calculations for constrained ensembles \cite{RogersPerezVicenteTakedaPerezCastillo2010}, and later message-passing and random-matrix treatments of networks with arbitrary degrees, expected degrees, and communities \cite{NadakuditiNewman2013,ZhangNadakuditiNewman2014,NewmanZhangNadakuditi2019}. The formulation below follows the statistical-mechanics viewpoint of \cite{RogersPerezVicenteTakedaPerezCastillo2010}: topological constraints are encoded in the ensemble measure, and the spectral density is obtained from the corresponding distribution of cavity messages. Although \cite{RogersPerezVicenteTakedaPerezCastillo2010} presents the calculation through the replica method, here we use the cavity method as the more direct pedagogical route.

The simplest topological constraint beyond the mean degree is a prescribed degree distribution. Let $p_k$ denote the asymptotic probability that a uniformly chosen vertex has degree $k$, and let
\begin{equation}
\langle k\rangle=\sum_{k\geq0}k p_k\,.
\label{eq:tcrg-mean-degree}
\end{equation}
For an uncorrelated configuration model, the degree of a vertex reached by following a uniformly chosen edge is distributed according to the size-biased law
\begin{equation}
\frac{k p_k}{\langle k\rangle}\,.
\label{eq:tcrg-size-biased-degree}
\end{equation}
Equivalently, the number of remaining neighbors of such a vertex is distributed according to the excess-degree law
\begin{equation}
q_\ell=\frac{(\ell+1)p_{\ell+1}}{\langle k\rangle}\,.
\label{eq:tcrg-excess-degree}
\end{equation}
This replacement of the Poisson excess law by \eqref{eq:tcrg-excess-degree} is already enough to change the spectral density. A broad degree distribution creates a broad distribution of local resolvents, and high-degree vertices can generate spectral tails or isolated eigenvalues. This is one of the most elementary ways in which topology enters the sparse spectral problem.

Degree--degree correlations impose a stronger constraint. Let $e_{kk'}$ be the probability that a uniformly chosen oriented edge starts at a vertex of degree $k$ and ends at a vertex of degree $k'$. This probability obviously satisfies
\begin{equation}
\sum_{k,k'}e_{kk'}=1\,,\qquad\sum_{k'}e_{kk'} = \frac{k p_k}{\langle k\rangle}.
    \label{eq:tcrg-edge-degree-consistency}
\end{equation}
For undirected graphs one has $e_{kk'}=e_{k'k}$. The corresponding conditional probability that a neighbor of a degree-$k$ vertex has degree $k'$ is
\begin{equation}
P(k'|k)=\frac{e_{kk'}}{\sum_{\ell}e_{k\ell}}=\frac{\langle k\rangle}{k p_k}e_{kk'}\,,
\qquad k\geq1\,.
\label{eq:tcrg-degree-conditional}
\end{equation}
The uncorrelated configuration model is recovered when
\begin{equation}
P(k'|k)=\frac{k'p_{k'}}{\langle k\rangle}\,,
\label{eq:tcrg-uncorrelated-neighbor-degree}
\end{equation}
independently of $k$. Deviations from \eqref{eq:tcrg-uncorrelated-neighbor-degree} encode assortative or disassortative mixing by degree. Assortative mixing means that high-degree vertices tend to connect to high-degree vertices, while disassortative mixing means that high-degree vertices preferentially connect to low-degree vertices \cite{Newman2002Assortative,Newman2003Mixing,Newman2010}.

In a locally tree-like graph ensemble in which \eqref{eq:tcrg-degree-conditional} is the nearest-neighbor degree law and distinct remaining neighbor degrees are conditionally independent at the local-tree level, the cavity message distribution can be conditioned on the degree of the source vertex. In the sums below, neighbor degrees are understood to satisfy $k_r\geq1$, since a vertex reached by following an edge cannot have degree zero; degree-zero vertices enter only through the site law. Let $\mathcal{P}_k(G)$ be the distribution of a cavity Green function sent by a degree-$k$ vertex. Then
\begin{equation}
\mathcal{P}_k(G)=\sum_{k_1,\ldots,k_{k-1}}\left[\prod_{r=1}^{k-1}P(k_r|k)\right]\int\left[\prod_{r=1}^{k-1}dG_r\mathcal{P}_{k_r}(G_r)\right]\delta\left(G-\frac{1}{z-\displaystyle\sum_{r=1}^{k-1}G_r}\right)\,,\qquad k\geq1\,.
\label{eq:tcrg-degree-correlated-cavity}
\end{equation}
The corresponding distribution of a full local Green function at a degree-$k$ vertex is
\begin{equation}
\mathcal{P}^{(\rm site)}_k(G)=\sum_{k_1,\ldots,k_k}\left[\prod_{r=1}^{k}P(k_r|k)\right]\int\left[\prod_{r=1}^{k}dG_r\mathcal{P}_{k_r}(G_r)\right]\delta\left(G-\frac{1}{z-\displaystyle\sum_{r=1}^{k}G_r}\right)\,.
\label{eq:tcrg-degree-correlated-site}
\end{equation}
The spectral density is then
\begin{equation}
\rho(\lambda)=\frac{1}{\pi}\lim_{\epsilon\downarrow0}{\rm Im}\sum_{k\geq0}p_k\int dG\mathcal{P}^{(\rm site)}_k(G)G\,,\qquad z=\lambda-i\epsilon\,.
    \label{eq:tcrg-degree-correlated-density}
\end{equation}
Equations \eqref{eq:tcrg-degree-correlated-cavity}--\eqref{eq:tcrg-degree-correlated-density} should be read as the degree-conditioned version of the sparse resolvent recursion. For $k=0$, the site law is understood as
\begin{equation}
\mathcal P^{(\rm site)}_0(G)=\delta\left(G-\frac{1}{z}\right)\,.
\end{equation}
In the uncorrelated case, \eqref{eq:tcrg-uncorrelated-neighbor-degree} removes the dependence of the neighbor degree on the source degree. Defining the mixed cavity law
\begin{equation}
\mathcal P_{\rm cav}(G)=\sum_{k\geq1}\frac{k p_k}{\langle k\rangle}\mathcal P_k(G)\,,
\end{equation}
one recovers the ordinary configuration-model cavity equation with the excess-degree distribution \eqref{eq:tcrg-excess-degree}. In the regular case, $p_k=\delta_{k,c}$, the distribution collapses to the Kesten--McKay case.

\begin{examplebox}[A two-degree correlated ensemble]
Consider an unweighted graph ensemble with only two possible degrees,
\begin{equation}
p(k)=\frac{1}{2}\delta_{k,1}+\frac{1}{2}\delta_{k,3}\,.
\label{eq:tcrg-ped-two-degree-law}
\end{equation}
The mean degree is
\begin{equation}
\langle k\rangle=\frac{1}{2}\cdot 1+\frac{1}{2}\cdot 3=2\,.
\label{eq:tcrg-ped-two-degree-mean}
\end{equation}
The degree distribution seen from a uniformly chosen oriented edge is therefore
\begin{equation}
s_k=\frac{k p(k)}{\langle k\rangle}\,,\qquad s_1=\frac{1}{4}\,,\qquad s_3=\frac{3}{4}\,.
\label{eq:tcrg-ped-edge-degree-law}
\end{equation}
To introduce degree--degree correlations, let $e_{kk'}$ be the probability that an oriented edge starts at a vertex of degree $k$ and ends at a vertex of degree $k'$. For an undirected graph we impose $e_{kk'}=e_{k'k}$ and the consistency condition
\begin{equation}
\sum_{k'}e_{kk'}=s_k\,.
\label{eq:tcrg-ped-edge-consistency}
\end{equation}
A one-parameter family satisfying these constraints is
\begin{equation}
e_{11}=a\,,\qquad e_{13}=e_{31}=\frac{1}{4}-a\,,\qquad e_{33}=\frac{1}{2}+a\,,\qquad
 0\leq a\leq\frac{1}{4}\,.
\label{eq:tcrg-ped-edge-matrix}
\end{equation}
The conditional probabilities are
\begin{equation}
P(1|1)=4a\,,\qquad P(3|1)=1-4a\,,
\label{eq:tcrg-ped-conditional-one}
\end{equation}
and
\begin{equation}
P(1|3)=\frac{1-4a}{3}\,, \qquad P(3|3)=\frac{2+4a}{3}\,.
\label{eq:tcrg-ped-conditional-three}
\end{equation}
The uncorrelated configuration model corresponds to
\begin{equation}
e_{kk'}=s_k s_{k'}\,,\qquad a=s_1^2=\frac{1}{16}\,.
\label{eq:tcrg-ped-uncorrelated-a}
\end{equation}
Larger values of $a$ make the graph more assortative by degree, because degree-one vertices connect more often to degree-one vertices and degree-three vertices connect more often to degree-three vertices.

Let $\mathcal P_1(G)$ and $\mathcal P_3(G)$ be the cavity Green-function distributions sent by degree-one and degree-three vertices, respectively. For the unweighted adjacency matrix with zero diagonal terms, a degree-one vertex has no remaining neighbors when sending a cavity message. Hence
\begin{equation}
\mathcal P_1(G)=\delta\left(G-\frac{1}{z}\right)\,.
\label{eq:tcrg-ped-P1}
\end{equation}
A degree-three vertex has two remaining neighbors in a cavity graph. Therefore
\begin{align}
\mathcal P_3(G)&=\sum_{k_1,k_2\in\{1,3\}}P(k_1|3)P(k_2|3)\int dG_1 dG_2 \mathcal P_{k_1}(G_1)\mathcal P_{k_2}(G_2)\delta\left(G-\frac{1}{z-G_1-G_2}\right)\,.
\label{eq:tcrg-ped-P3}
\end{align}
The corresponding full site distributions are
\begin{align}
\mathcal P_1^{\rm site}(G)&=\sum_{k_1\in\{1,3\}}P(k_1|1)\int dG_1 \mathcal P_{k_1}(G_1)\delta\left(G-\frac{1}{z-G_1}\right)\,,
\label{eq:tcrg-ped-P1-site}\\
\mathcal P_3^{\rm site}(G)&=\sum_{k_1,k_2,k_3\in\{1,3\}}\left[\prod_{r=1}^3 P(k_r|3)\right]\int\left[\prod_{r=1}^3 dG_r\mathcal P_{k_r}(G_r)\right]\delta\left(G-\frac{1}{z-G_1-G_2-G_3}\right)\,.
\label{eq:tcrg-ped-P3-site}
\end{align}
The averaged density is then
\begin{equation}
\rho(\lambda)=\frac{1}{\pi}\lim_{\epsilon\downarrow0}{\rm Im}\left[\frac{1}{2}\int dG\mathcal P_1^{\rm site}(G)G+\frac{1}{2}\int dG\mathcal P_3^{\rm site}(G)G\right]\,,\qquad z=\lambda-i\epsilon\,.
\label{eq:tcrg-ped-two-degree-density}
\end{equation}
This example shows explicitly how degree correlations enter the cavity equations: the local recursion is the same tree recursion, but the law of the incoming messages depends on the degree of the vertex that sends the message.
\end{examplebox}

The previous equations assume that, once the degree of a vertex is fixed, the degrees of its different neighbors are independent draws from $P(k'|k)$. More refined constrained ensembles keep track of correlations among the degrees, or generalized degrees, of different neighbors of the same vertex. A useful way to organize such constraints is through generalized degrees. Define
\begin{equation}
k_i^{(0)}=1\,,\qquad k_i^{(r)}=\sum_{j=1}^N C_{ij}k_j^{(r-1)}\,,\qquad r=1,\ldots,L\,.
\label{eq:tcrg-generalized-degree-recursion}
\end{equation}
Thus $k_i^{(1)}$ is the ordinary degree of vertex $i$, while $k_i^{(2)}$ is the sum of the degrees of the neighbors of $i$. More generally, $k_i^{(r)}$ counts the number of walks of length $r$ emanating from $i$. We denote the generalized degree vector by
\begin{equation}
\pmb{k}_i=\left(k_i^{(1)},\ldots,k_i^{(L)}\right)\,.
\label{eq:tcrg-generalized-degree-vector}
\end{equation}
Fixing the empirical law of $\pmb{k}_i$ imposes hierarchical constraints on the local topology. For $L=1$ one fixes only the degree distribution. For $L=2$ one constrains information about the degrees of neighboring vertices. Increasing $L$ imposes progressively deeper information about the rooted neighborhood. This hierarchy was used in \cite{RogersPerezVicenteTakedaPerezCastillo2010} to derive spectral-density equations for graph ensembles with generalized degree--degree correlations.

A microcanonical ensemble with prescribed generalized degrees may be written formally as
\begin{equation}
P_L(\pmb{C}|\{\pmb{k}_i\})=\frac{1}{Z_L(\{\pmb{k}_i\})}\prod_{i=1}^N\delta_{\pmb{k}_i(\pmb{C}),\pmb{k}_i}\prod_{i<j}\delta_{C_{ij},C_{ji}}\prod_{i=1}^N\delta_{C_{ii},0}\,.
\label{eq:tcrg-generalized-degree-ensemble}
\end{equation}
The local compatibility condition follows from Eqs~\eqref{eq:tcrg-generalized-degree-recursion}. If a vertex has generalized degree $\pmb{k}$ and neighbors with generalized degrees $\pmb{q}_1,\ldots,\pmb{q}_{k^{(1)}}$, then
\begin{equation}
\sum_{r=1}^{k^{(1)}} q_r^{(m-1)}=k^{(m)}\,, \qquad m=1,\ldots,L\,,
\label{eq:tcrg-generalized-degree-compatibility}
\end{equation}
where $q_r^{(0)}=1$. In a cavity message from a vertex of type $\pmb{k}$ to a neighbor of type $\pmb{q}$, the remaining neighbors must therefore satisfy
\begin{equation}
\sum_{r=1}^{k^{(1)}-1} q_r^{(m-1)}=k^{(m)}-q^{(m-1)},\qquad\,m=1,\ldots,L\,.
\label{eq:tcrg-cavity-generalized-degree-compatibility}
\end{equation}
This is the topological origin of the more complicated order parameters in the replica calculation: the message is no longer characterized only by the degree of the source vertex, but by the generalized degree of the source and, in general, by the generalized degree of the removed neighbor.

\begin{examplebox}[Generalized-degree compatibility for \(L=2\)]
For $L=2$, the generalized degree of a vertex is
\begin{equation}
\pmb k=(k^{(1)},k^{(2)})\,,
\label{eq:tcrg-ped-gdegree-L2}
\end{equation}
where $k^{(1)}$ is the ordinary degree and
\begin{equation}
k^{(2)}=\sum_{j\in\partial i} k_j^{(1)}
\label{eq:tcrg-ped-second-gdegree}
\end{equation}
is the sum of the degrees of the neighbors. Suppose a vertex has generalized degree
\begin{equation}
\pmb k=(3,7)\,.
\label{eq:tcrg-ped-vertex-type}
\end{equation}
This means that the vertex has three neighbors, and the ordinary degrees of those three neighbors must add up to seven.

Now consider a cavity message sent from this vertex to a neighbor of ordinary degree
\begin{equation}
q^{(1)}=2\,.
\label{eq:tcrg-ped-removed-neighbor-degree}
\end{equation}
After removing the edge to this neighbor, two neighbors remain in the cavity branch. The compatibility condition
\begin{equation}
\sum_{r=1}^{k^{(1)}-1}q_r^{(m-1)}=k^{(m)}-q^{(m-1)}
\label{eq:tcrg-ped-compatibility-recall}
\end{equation}
gives two constraints. For $m=1$, since $q_r^{(0)}=1$, one obtains
\begin{equation}
\sum_{r=1}^{2}q_r^{(0)}=2=k^{(1)}-q^{(0)}=3-1\,.
\label{eq:tcrg-ped-compatibility-m1}
\end{equation}
This simply says that two neighbors remain. For $m=2$, one obtains
\begin{equation}
q_1^{(1)}+q_2^{(1)}=k^{(2)}-q^{(1)}=7-2=5\,.
\label{eq:tcrg-ped-compatibility-m2}
\end{equation}
Thus the remaining two neighbors must have ordinary degrees adding up to five. The allowed ordered possibilities are, for instance,
\begin{equation}
(q_1^{(1)},q_2^{(1)})=(1,4),(2,3),(3,2),(4,1)\,,
\label{eq:tcrg-ped-compatible-neighbors}
\end{equation}
subject to whichever degree types are actually present in the ensemble.

This is the local meaning of generalized-degree constraints. They do not change the algebraic form of the resolvent recursion, but they restrict which incoming message types can appear together at a vertex. In a population-dynamics implementation, such constraints are enforced by sampling only compatible local neighborhoods.
\end{examplebox}

Let $\mathcal{P}_{\pmb{k}\to\pmb{q}}(G)$ be the distribution of a cavity Green function sent from a vertex of generalized degree $\pmb{k}$ to a vertex of generalized degree $\pmb{q}$. Let
\begin{equation}
\mathcal{W}_{\pmb{k},\pmb{q}}\left(\pmb{q}_1,\ldots,\pmb{q}_{k^{(1)}-1}\right)
\label{eq:tcrg-cavity-neighborhood-weight}
\end{equation}
denote the probability, in the constrained ensemble, that the remaining neighbors have generalized degrees $\pmb{q}_1,\ldots,\pmb{q}_{k^{(1)}-1}$, conditional on the cavity edge $\pmb{k}\to\pmb{q}$. This weight is supported only on configurations satisfying \eqref{eq:tcrg-cavity-generalized-degree-compatibility}. The ensemble-averaged cavity equations are then
\begin{equation}
\mathcal{P}_{\pmb{k}\to\pmb{q}}(G)=\sum_{\pmb{q}_1,\ldots,\pmb{q}_{k^{(1)}-1}}\mathcal{W}_{\pmb{k},\pmb{q}}\left(\pmb{q}_1,\ldots,\pmb{q}_{k^{(1)}-1}\right)\int\left[\prod_{r=1}^{k^{(1)}-1}dG_r\mathcal{P}_{\pmb{q}_r\to\pmb{k}}(G_r)\right]\delta\left(G-\frac{1}{z-\displaystyle\sum_{r=1}^{k^{(1)}-1}G_r}\right)\,.
    \label{eq:tcrg-generalized-degree-cavity}
\end{equation}
Similarly, if $\mathcal{W}_{\pmb{k}}^{(\rm site)}(\pmb{q}_1,\ldots,\pmb{q}_{k^{(1)}})$ is the corresponding full-neighborhood probability, supported on \eqref{eq:tcrg-generalized-degree-compatibility}, then
\begin{equation}
\mathcal{P}_{\pmb{k}}^{(\rm site)}(G)=\sum_{\pmb{q}_1,\ldots,\pmb{q}_{k^{(1)}}}\mathcal{W}_{\pmb{k}}^{(\rm site)}\left(
\pmb{q}_1,\ldots,\pmb{q}_{k^{(1)}}\right)\int\left[\prod_{r=1}^{k^{(1)}}dG_r\mathcal{P}_{\pmb{q}_r\to\pmb{k}}(G_r)\right]\delta\left(G-\frac{1}{z-\displaystyle\sum_{r=1}^{k^{(1)}}G_r}\right)\,.
\label{eq:tcrg-generalized-degree-site}
\end{equation}
If $p(\pmb{k})$ is the empirical distribution of generalized degrees, the spectral density is
\begin{equation}
\rho(\lambda)=\frac{1}{\pi}\lim_{\epsilon\downarrow0}\sum_{\pmb{k}}p(\pmb{k}){\rm Im}\int dG\mathcal{P}_{\pmb{k}}^{(\rm site)}(G)G\,.
\label{eq:tcrg-generalized-degree-density}
\end{equation}
Equations \eqref{eq:tcrg-generalized-degree-cavity}--\eqref{eq:tcrg-generalized-degree-density} are the cavity counterpart of the replica consistency equations for hierarchically constrained graph ensembles. They show explicitly why a scalar effective-medium approximation is insufficient: the resolvent must be averaged over the ensemble of rooted neighborhoods compatible with the imposed topology.

Another important topological constraint is community structure. In a stochastic block model, each vertex carries a group label $a\in\{1,\ldots,B\}$, with group proportions $p_a$, and edges are drawn with probabilities depending on the pair of groups \cite{HollandLaskeyLeinhardt1983,Newman2010,KarrerNewman2011}. In the sparse undirected case one may write
\begin{equation}
{\rm Prob}(C_{ij}=1|g_i=a,g_j=b)=\frac{c_{ab}}{N}\,,\qquad c_{ab}=c_{ba}\,.
\label{eq:tcrg-block-edge-probability}
\end{equation}
Up to corrections that vanish as $N\to\infty$, the expected number of neighbors in group $b$ of a vertex in group $a$ is $p_b c_{ab}$, and the expected degree of a group-$a$ vertex is
\begin{equation}
c_a=\sum_{b=1}^B p_b c_{ab}\,.
\label{eq:tcrg-block-mean-degree}
\end{equation}
For a Poisson block ensemble, the numbers of neighbors in the different groups are asymptotically independent Poisson random variables. Let $\mathcal{P}_a(G)$ be the distribution of a cavity Green function sent by a vertex in group $a$. The block-conditioned cavity equation is
\begin{equation}
\mathcal{P}_a(G)=\sum_{n_1,\ldots,n_B\geq0}\left[\prod_{b=1}^Be^{-p_b c_{ab}}\frac{(p_b c_{ab})^{n_b}}{n_b!}\right]\int\prod_{b=1}^B\prod_{r=1}^{n_b}\left[dG_{br}\mathcal{P}_b(G_{br})\right]\delta\left(G-\frac{1}{z-\displaystyle\sum_{b=1}^B\sum_{r=1}^{n_b}G_{br}}\right)\,.
\label{eq:tcrg-block-cavity}
\end{equation}
For the Poisson block model the cavity and full-neighborhood count distributions coincide in the thermodynamic limit, because conditioning on one existing edge leaves the other edge counts Poisson. Therefore the averaged spectral density is
\begin{equation}
\rho(\lambda)=\frac{1}{\pi}\lim_{\epsilon\downarrow0}\sum_{a=1}^Bp_a{\rm Im}\int dG  \mathcal{P}_a(G)G\,.
\label{eq:tcrg-block-density}
\end{equation}
If the block constraints are microcanonical, or if one fixes the number of neighbors of each block type, then the Poisson factors in \eqref{eq:tcrg-block-cavity} are replaced by the appropriate constrained neighborhood law. The structure of the recursion, however, remains the same.

\begin{examplebox}[Two-block Poisson graph and group-conditioned messages]
Consider a sparse graph with two groups, $a=1,2$, with group proportions
\begin{equation}
p_1=p\,,\qquad p_2=1-p\,.
\label{eq:tcrg-ped-two-block-proportions}
\end{equation}
Edges are drawn independently with probabilities
\begin{equation}
{\rm Prob}(C_{ij}=1|g_i=a,g_j=b)=\frac{c_{ab}}{N}\,,\qquad c_{ab}=c_{ba}\,.
\label{eq:tcrg-ped-two-block-edge-prob}
\end{equation}
A vertex in group $1$ has an asymptotically Poisson number of neighbors in group $1$ with mean $p c_{11}$, and an asymptotically Poisson number of neighbors in group $2$ with mean $(1-p)c_{12}$. Similarly, a vertex in group $2$ has mean numbers $p c_{21}$ and $(1-p)c_{22}$ of neighbors in groups $1$ and $2$.

Let $\mathcal P_1(G)$ and $\mathcal P_2(G)$ be the cavity Green-function distributions sent by vertices in groups $1$ and $2$. For the unweighted adjacency matrix with zero diagonal terms, the group-conditioned recursions are
\begin{align}
\mathcal P_1(G)&=\sum_{n_1,n_2\geq0}e^{-p c_{11}}\frac{(p c_{11})^{n_1}}{n_1!}e^{-(1-p)c_{12}}\frac{[(1-p)c_{12}]^{n_2}}{n_2!}\nonumber\\&\hspace{1cm}\times\int\left[\prod_{r=1}^{n_1}dG_r^{(1)}\mathcal P_1(G_r^{(1)})\right]\left[\prod_{s=1}^{n_2}dG_s^{(2)}\mathcal P_2(G_s^{(2)})\right]\nonumber\\
&\hspace{1cm}\times\delta\left(G-\frac{1}{z-\displaystyle\sum_{r=1}^{n_1}G_r^{(1)}-\displaystyle\sum_{s=1}^{n_2}G_s^{(2)}}\right)\,,
\label{eq:tcrg-ped-two-block-P1}
\end{align}
and
\begin{align}
\mathcal P_2(G)&=\sum_{n_1,n_2\geq0}e^{-p c_{21}}\frac{(p c_{21})^{n_1}}{n_1!}e^{-(1-p)c_{22}}\frac{[(1-p)c_{22}]^{n_2}}{n_2!}\nonumber\\
&\hspace{1cm}\times\int\left[\prod_{r=1}^{n_1}dG_r^{(1)}\,\mathcal P_1(G_r^{(1)})\right]\left[\prod_{s=1}^{n_2}dG_s^{(2)}\,\mathcal P_2(G_s^{(2)})\right]\nonumber\\
&\hspace{1cm}\times\delta\left(G-\frac{1}{z-\displaystyle\sum_{r=1}^{n_1}G_r^{(1)}-\displaystyle\sum_{s=1}^{n_2}G_s^{(2)}}\right)\,.
\label{eq:tcrg-ped-two-block-P2}
\end{align}
The corresponding density is
\begin{equation}
\rho(\lambda)=\frac{1}{\pi}\lim_{\epsilon\downarrow0}{\rm Im}\left[p\int dG \mathcal P_1(G)G+(1-p)\int dG \mathcal P_2(G)G\right]\,,\qquad z=\lambda-i\epsilon\,.
\label{eq:tcrg-ped-two-block-density}
\end{equation}
If $c_{11}=c_{12}=c_{21}=c_{22}=c$, then the group labels carry no information and the two equations collapse to the ordinary Erd\H{o}s--R\'enyi population equation with mean degree $c$. If the rows of the matrix $(c_{ab})$ are different, the two groups have different local environments, and the cavity order parameter becomes a pair of distributions rather than a single one.
\end{examplebox}

Community structure affects the spectrum in two complementary ways. First, it changes the continuous part of the density through the group-dependent local neighborhoods in \eqref{eq:tcrg-block-cavity}. Second, it can generate isolated eigenvalues associated with the low-rank block structure of the mean adjacency matrix. These isolated eigenvalues have vanishing weight in the normalized empirical density when the number of blocks is fixed, but they are crucial for spectral community detection. The transition at which informative eigenvalues merge into the bulk is the spectral manifestation of detectability loss in block-structured random graphs \cite{NadakuditiNewman2012,ZhangNadakuditiNewman2014,Peixoto2013}. Thus the spectral density and the outlier spectrum encode different aspects of the same topology: the former describes the typical local environment, while the latter can reveal mesoscopic organization.

The block model also clarifies the relation between communities and degree heterogeneity. A simple stochastic block model controls group membership but not broad degree fluctuations. Degree-corrected block models incorporate both ingredients by assigning each vertex a group label and a degree parameter \cite{KarrerNewman2011}. In cavity language this simply enlarges the type space: messages are conditioned on both the degree information and the block label. The resulting equations have the same form as \eqref{eq:tcrg-generalized-degree-cavity}, with the generalized type $\pmb k$ replaced by a composite type containing both degree and community data.

This is conceptually important because empirical networks often contain both broad degree distributions and community structure; treating only one of these constraints can give a misleading spectral null model.

All the ensembles discussed in this section remain locally tree-like. This assumption is essential for the direct cavity recursions above. Generalized degrees and block labels encode correlations in rooted neighborhoods, but they do not, by themselves, impose a finite density of short loops. Short cycles, clustering, and motifs create additional dependencies between branches and therefore require enlarged messages or different approximations. Message-passing methods for spectra of networks with short loops have been developed in other contexts \cite{Newman2019ShortLoops}, but they are not part of the locally tree-like framework used in the main line of these notes.

The limiting cases provide useful checks. If all degree correlations are removed, \eqref{eq:tcrg-degree-correlated-cavity} reduces to the configuration-model equation. If the degree distribution is $p_k=\delta_{k,c}$, it reduces to the random-regular scalar recursion and hence to the Kesten--McKay law. If the block model has a single group, \eqref{eq:tcrg-block-cavity} reduces to the Erd\H{o}s--R\'enyi population equation. If one takes a dense limit with appropriate centering and scaling, the message distribution concentrates and one recovers the corresponding dense random-matrix or finite-rank-deformation description. These checks are important because the constrained equations interpolate between ordinary sparse random matrix theory, network null models, and random graphs with mesoscopic structure.

The conclusion is that topological constraints enter sparse spectra through the law of the cavity environment. The local resolvent recursion is universal on a tree, but the distribution of its inputs is ensemble-dependent. Degree distributions, degree--degree correlations, generalized degrees, and community labels all modify the population of messages and hence the spectral density. This is the central reason why sparse random matrix theory is richer than its dense counterpart: in finite connectivity, the graph topology survives the thermodynamic limit and becomes part of the spectral order parameter.

\begin{exerciseblock}
\exitem[Edge-type consistency]
Let $p_k$ be a degree distribution with mean $\langle k\rangle$, and define
\begin{equation}
s_k=\frac{k p_k}{\langle k\rangle}\,.
\label{eq:tcrg-ex-s-edge-law}
\end{equation}
Suppose $e_{kk'}$ is the probability that a uniformly chosen oriented edge starts at a vertex of degree $k$ and ends at a vertex of degree $k'$. Show that
\begin{equation}
\sum_{k'}e_{kk'}=s_k\,.
\label{eq:tcrg-ex-edge-row-sum}
\end{equation}
For an undirected graph, explain why one also requires $e_{kk'}=e_{k'k}$.

\exitem[Conditional neighbor law]
For $k\geq1$ with $p_k>0$, starting from $e_{kk'}$, derive
\begin{equation}
P(k'|k)=\frac{e_{kk'}}{\sum_{\ell}e_{k\ell}}=\frac{\langle k\rangle}{k p_k}e_{kk'}\,.
\label{eq:tcrg-ex-conditional-neighbor-law}
\end{equation}
Then verify that
\begin{equation}
\sum_{k'}P(k'|k)=1\,.
\label{eq:tcrg-ex-conditional-normalization}
\end{equation}

\exitem[Uncorrelated configuration model]
Show that the uncorrelated configuration model corresponds to
\begin{equation}
P(k'|k)=\frac{k'p_{k'}}{\langle k\rangle}\,,
\label{eq:tcrg-ex-uncorrelated-conditional}
\end{equation}
independently of $k$. Substitute this into the degree-conditioned cavity recursion. Then define
\begin{equation}
\mathcal P_{\rm cav}(G)=\sum_{k\geq1}\frac{k p_k}{\langle k\rangle}\mathcal P_k(G)\,,
    \label{eq:tcrg-ex-uncorrelated-mixed-cavity-law}
\end{equation}
and show that the mixed law satisfies the ordinary configuration-model recursion with the excess-degree distribution.

\exitem[Assortativity in the two-degree example]
For the two-degree ensemble introduced above, compute the conditional probabilities $P(k'|k)$ from \eqref{eq:tcrg-ped-edge-matrix}. Show that $a=1/16$ gives the uncorrelated case. What happens at the two endpoints $a=0$ and $a=1/4$?

\exitem[Degree-conditioned cavity equations]
For a graph with degrees $k\in\{2,4\}$ and conditional probabilities $P(k'|k)$, take the unweighted adjacency matrix with zero diagonal terms and spectral parameter $z=\lambda-i\epsilon$, $\epsilon>0$. Write explicitly the coupled equations for $\mathcal P_2(G)$ and $\mathcal P_4(G)$. Then write the corresponding full site distributions.

\exitem[Random regular reduction]
For $c\geq2$, starting from the degree-conditioned recursion, set
\begin{equation}
p_k=\delta_{k,c}\,.
\label{eq:tcrg-ex-regular-degree-law}
\end{equation}
Show that the message distribution collapses to a single scalar recursion,
\begin{equation}
G_{\rm cav}=\frac{1}{z-(c-1)G_{\rm cav}}\,.
\label{eq:tcrg-ex-regular-recursion}
\end{equation}
Explain why this is the same recursion that leads to the Kesten--McKay law.

\exitem[Generalized-degree compatibility]
For $L=2$, consider a vertex with generalized degree
\begin{equation}
\pmb k=(4,10)\,.
\label{eq:tcrg-ex-gdegree-example}
\end{equation}
Suppose that one removed neighbor has ordinary degree $3$. Assuming that the remaining neighbor degrees are positive integers and that no further restriction on degree types is imposed, list all possible ordered triples of ordinary degrees of the remaining neighbors that satisfy the compatibility condition.

\exitem[Walk interpretation of generalized degrees]
Let $\pmb C$ be the $N\times N$ adjacency matrix of a simple undirected graph with vertex set $V=\{1,\ldots,N\}$. Using
\begin{equation}
k_i^{(r)}=\sum_{j=1}^{N}C_{ij}k_j^{(r-1)}\,,\qquad k_i^{(0)}=1\,,
\label{eq:tcrg-ex-gdegree-recursion}
\end{equation}
show that $k_i^{(r)}$ counts the number of walks of length $r$ starting at vertex $i$. Work out explicitly the cases $r=1,2,3$.

\exitem[From generalized degrees to constrained neighborhoods]
For $L=2$, show that fixing $\pmb k_i=(k_i^{(1)},k_i^{(2)})$ fixes the degree of vertex $i$ and the sum of the degrees of its neighbors, but does not fix the degrees of each neighbor individually. Explain why this leads to a constrained sum over compatible neighborhoods in the cavity equation.

\exitem[Two-block expected degrees]
Consider a sparse block model on $N$ vertices with group labels $g_i\in\{1,\ldots,B\}$. Suppose that a fraction $p_b$ of vertices belongs to group $b$, and that
\begin{equation}
{\rm Prob}(C_{ij}=1|g_i=a,g_j=b)=\frac{c_{ab}}{N}\,,
\label{eq:tcrg-ex-block-probability}
\end{equation}
with $c_{ab}=c_{ba}$ for an undirected graph. Show that, up to corrections that vanish as $N\to\infty$, the expected degree of a vertex in group $a$ is
\begin{equation}
c_a=\sum_{b=1}^{B}p_b c_{ab}\,.
\label{eq:tcrg-ex-block-expected-degree}
\end{equation}

\exitem[Two-block cavity equations]
For two groups with proportions $p$ and $1-p$, derive the equations \eqref{eq:tcrg-ped-two-block-P1} and \eqref{eq:tcrg-ped-two-block-P2}. Explain why the number of neighbors of each group is Poisson in the sparse limit.

\exitem[Collapse to Erd\H{o}s--R\'enyi]
In the two-block equations, impose
\begin{equation}
c_{11}=c_{12}=c_{21}=c_{22}=c\,.
\label{eq:tcrg-ex-block-er-limit}
\end{equation}
Show that $\mathcal P_1=\mathcal P_2$ and that the common equation is the Erd\H{o}s--R\'enyi cavity equation with mean degree $c$.

\exitem[Degree-corrected block labels]
Suppose a graph ensemble has both a degree class $k$ and a group label $a$. Define a composite type
\begin{equation}
\tau=(k,a)\,.
\label{eq:tcrg-ex-composite-type}
\end{equation}
Let $P(\tau'|\tau)$ be the conditional probability that a neighbor of a vertex of type $\tau$ has type $\tau'$. For a message sent by a vertex of type $\tau=(k,a)$ with $k\geq1$, write the type-conditioned cavity equation
\begin{equation}
\mathcal P_{\tau}(G)=\sum_{\tau_1,\ldots,\tau_{k-1}}\left[\prod_{r=1}^{k-1}P(\tau_r|\tau)\right]\int
\left[\prod_{r=1}^{k-1}dG_r\,\mathcal P_{\tau_r}(G_r)\right]\delta\left(G-\frac{1}{z-\displaystyle\sum_{r=1}^{k-1}G_r}\right)\,,
\label{eq:tcrg-ex-composite-type-cavity}
\end{equation}
where $\tau=(k,a)$. Explain why this is the correct order parameter when degree heterogeneity and community structure are both present.

\exitem[Short loops and the limit of the locally tree-like approximation]
The cavity equations in this section assume that different branches become independent after removing a vertex or an edge. Explain why a finite density of triangles violates this assumption. Give a simple example of a graph motif for which two neighbors of a vertex remain connected after the central vertex is removed.

\exitem[Programming exercise: degree correlations]
Fix a list of system sizes $N$, a number $S$ of independent samples for each $N$, a list of admissible values of the assortativity parameter $a$, and a common histogram binning or smoothing parameter for the spectral-density comparison. Generate graphs with the two-degree distribution introduced above. Take $N$ even, assign degree $1$ to $N/2$ vertices and degree $3$ to $N/2$ vertices, and choose $N$ so that the edge counts below are integers. Since $\langle k\rangle=2$, the total number of undirected edges is $M=N$. Use the oriented edge probabilities
\begin{equation}
e_{11}=a\,,\qquad e_{13}=e_{31}=\frac{1}{4}-a\,,\qquad e_{33}=\frac{1}{2}+a
\end{equation}
to prescribe the numbers of undirected edges of each type:
\begin{equation}
M_{11}=aN\,, \qquad M_{13}=2N\left(\frac{1}{4}-a\right)\,,\qquad M_{33}=\left(\frac{1}{2}+a\right)N\,.
\end{equation}
Construct the graph by pairing stubs between the corresponding degree classes. If a simple graph is required, reject a proposed pairing that creates a self-edge or multiple edge and resample that pairing, or restart the construction, while preserving the prescribed edge counts and degree sequence. For each value of $a$, compute the adjacency spectrum for the $S$ samples and compare the empirical spectral density using the same binning or smoothing convention. Also estimate the empirical conditional probabilities $P(k'|k)$ and compare them with \eqref{eq:tcrg-ped-conditional-one}--\eqref{eq:tcrg-ped-conditional-three}. Report $N$, $S$, the values of $a$, the bin width or smoothing parameter, and whether multiedges were rejected.

\exitem[Programming exercise: block structure]
Fix a list of system sizes $N$, a number $S$ of independent samples for each $N$, group proportions $p_1=p$ and $p_2=1-p$ with integer group sizes, and a common histogram binning or smoothing parameter for spectral-density comparisons. Generate simple undirected sparse two-block graphs with $C_{ii}=0$, $C_{ij}=C_{ji}$, and independent upper-triangular edges satisfying
\begin{equation}
{\rm Prob}(C_{ij}=1|g_i=a,g_j=b)=\frac{c_{ab}}{N}\,,\qquad i<j,\qquad c_{ab}=c_{ba}\,.
\end{equation}
Choose different matrices $(c_{ab})$ with the same average degree
\begin{equation}
\bar c=\sum_{a,b=1}^2 p_a p_b c_{ab}\,.
\end{equation}
Compare their empirical spectral densities using the same binning or smoothing convention. Separately record the leading Perron or mean-degree outlier and any additional isolated eigenvalues associated with the block structure. Check whether the bulk density changes when the two groups have different expected degrees
\begin{equation}
c_a=\sum_{b=1}^2 p_b c_{ab}\,.
\end{equation}
Report $N$, $S$, $p$, the matrices $(c_{ab})$, the average degree $\bar c$, the group expected degrees $c_a$, and the bin width or smoothing parameter.

\exitem[Programming exercise: typed population dynamics]
Implement population dynamics for the two-block equations \eqref{eq:tcrg-ped-two-block-P1}--\eqref{eq:tcrg-ped-two-block-P2}. Fix $p$, $c_{11}$, $c_{12}=c_{21}$, $c_{22}$, a population size $M$, a burn-in time $T_{\rm burn}$, a number $T_{\rm meas}$ of site samples used for measurement, a regulator $\epsilon>0$, and a grid of real values $\lambda$. Represent $\mathcal P_1$ and $\mathcal P_2$ by two populations of complex cavity Green functions. For a message sent by a group-$a$ vertex, draw independently
\begin{equation}
n_b\sim{\rm Poisson}(p_b c_{ab})\,,\qquad b=1,2\,,
\end{equation}
sample $n_b$ incoming messages from the population $\mathcal P_b$, and update
\begin{equation}
G_{\rm new}^{(a)}=\frac{1}{\lambda-i\epsilon-\displaystyle\sum_{b=1}^{2}\sum_{r=1}^{n_b}G_{br}}\,.
\end{equation}
After equilibration, estimate the spectral density from
\begin{equation}
\rho_\epsilon(\lambda)=\frac{1}{\pi}{\rm Im}\left[p\langle G^{(1)}_{\rm site}\rangle+(1-p)\langle G^{(2)}_{\rm site}\rangle\right]\,,
\end{equation}
where the site Green functions are sampled with the same group-dependent Poisson neighborhood laws. Compare the result with direct diagonalization of $S$ independent two-block random graphs of size $N_{\rm diag}$, using the same $\lambda$ grid and the same regulator $\epsilon$ to broaden the eigenvalue histograms. Report $p$, $(c_{ab})$, $M$, $T_{\rm burn}$, $T_{\rm meas}$, $N_{\rm diag}$, $S$, $\epsilon$, the grid spacing, and a discrepancy measure.
\end{exerciseblock}

\section{Sparse covariance and diluted Wishart matrices}
\label{sec:sparse-covariance-diluted-wishart}
We now turn from sparse symmetric matrices supported on ordinary graphs to sparse covariance matrices supported on bipartite graphs. This is the natural finite-connectivity analogue of the Wishart ensemble. In the dense setting, Wishart matrices arise as empirical covariance matrices built from rectangular data matrices \cite{Wishart1928}, and their limiting spectral density is described by the Mar\v{c}enko--Pastur law when the two matrix dimensions diverge at fixed ratio \cite{MarchenkoPastur1967,BaiSilverstein2010}. In the diluted setting, the rectangular data matrix itself is sparse. The resulting covariance matrix is still positive semidefinite, but the finite-connectivity structure of the underlying bipartite graph survives in the thermodynamic limit and changes the spectral problem qualitatively.

Let $\pmb{X}$ be an $N\times P$ random rectangular matrix. We shall take
\begin{equation}
P=\frac{N}{\alpha}\,,\qquad\alpha>0\,,
\label{eq:scdw-rectangularity}
\end{equation}
so that $\alpha=N/P$ is kept fixed as $N,P\to\infty$. The entries of $\pmb X$ are diluted according to
\begin{equation}
X_i^\mu=B_i^\mu \xi_i^\mu\,,\qquad{\rm Prob}(B_i^\mu=1)=\frac{d}{N}\,,\qquad{\rm Prob}(B_i^\mu=0)=1-\frac{d}{N}\,,
\label{eq:scdw-diluted-rectangular-matrix}
\end{equation}
where $i=1,\ldots,N$, $\mu=1,\ldots,P$, and the nonzero weights $\xi_i^\mu$ are independent random variables drawn from a distribution $p_\xi(\xi)$. The parameter $d=O(1)$ controls the dilution. A column $\mu$ has mean degree $d$, while a row $i$ has mean degree $d/\alpha$. Thus $\pmb{X}$ is naturally represented as the weighted adjacency matrix of a sparse bipartite graph, with variable nodes $i$ on one side and sample, pattern, or factor nodes $\mu$ on the other.

The diluted Wishart matrix studied in this section is
\begin{equation}
\pmb{W}=\frac{1}{d}\pmb{X}\pmb{X}^{\rm T}\,,\qquad W_{ij}=\frac{1}{d}\sum_{\mu=1}^PX_i^\mu X_j^\mu\,.
\label{eq:scdw-diluted-wishart-definition}
\end{equation}
It is an $N\times N$ symmetric positive semidefinite matrix. The normalization by $d$ is convenient in the finite-connectivity ensemble because each sample node contains $O(d)$ nonzero entries. With the convention \eqref{eq:scdw-rectangularity}, and for $\mathbb{E}[(\xi_i^\mu)^2]=1$, the typical diagonal scale of $\pmb{W}$ is $1/\alpha$. If one wants the conventional dense covariance normalization with unit diagonal scale, one may instead study $\widehat{\pmb{W}}=\alpha\pmb{W}$. The two normalizations are equivalent up to a deterministic rescaling of eigenvalues.

\begin{examplebox}[A sparse rectangular matrix and its covariance projection]
We return to the elementary rectangular example introduced earlier, now to isolate the covariance projection, positivity, and rank deficiency that enter the diluted Wishart calculation. Take
\begin{equation}
\pmb X=\begin{pmatrix}
1 & 0\\
0 & 1\\
1 & 1
\end{pmatrix}\,,\qquad N=3\,,\qquad P=2\,, \qquad d=1\,.
\label{eq:scdw-ped-example-X}
\end{equation}
The rectangular matrix $\pmb X$ is naturally represented as a bipartite graph: variable node $1$ is connected only to factor node $1$, variable node $2$ only to factor node $2$, and variable node $3$ to both factor nodes. The corresponding diluted Wishart matrix is
\begin{equation}
\pmb W=\pmb X\pmb X^{\rm T}=\begin{pmatrix}
1 & 0 & 1\\
0 & 1 & 1\\
1 & 1 & 2
\end{pmatrix}\,.
\label{eq:scdw-ped-example-W}
\end{equation}
The entry $W_{13}=1$ appears because variables $1$ and $3$ share factor node $1$, while $W_{23}=1$ appears because variables $2$ and $3$ share factor node $2$. On the other hand, $W_{12}=0$ because variables $1$ and $2$ do not share a factor node. Thus the covariance matrix is not simply the adjacency matrix of the bipartite graph; it is the one-mode projection induced by common factor neighbors.

The matrix $\pmb W$ is positive semidefinite because, for every $\pmb v\in\mathbb R^3$,
\begin{equation}
\pmb v^{\rm T}\pmb W\pmb v=\pmb v^{\rm T}\pmb X\pmb X^{\rm T}\pmb v=\left\|\pmb X^{\rm T}\pmb v\right\|^2\geq 0\,.
\label{eq:scdw-ped-example-positivity}
\end{equation}
It is also rank deficient. Since $\pmb X$ is $3\times2$,
\begin{equation}
{\rm rank}\pmb W={\rm rank}(\pmb X\pmb X^{\rm T})={\rm rank}\pmb X\leq2\,.
\label{eq:scdw-ped-example-rank-bound}
\end{equation}
Indeed, the nonzero eigenvalues of $\pmb W$ are the eigenvalues of
\begin{equation}
\pmb X^{\rm T}\pmb X=\begin{pmatrix}
2 & 1\\
1 & 2
\end{pmatrix}\,,
\label{eq:scdw-ped-example-XTX}
\end{equation}
namely $3$ and $1$. Therefore
\begin{equation}
{\rm spec}(\pmb W)=\{3,1,0\}\,.
\label{eq:scdw-ped-example-spectrum}
\end{equation}
This finite example illustrates three structural features of diluted Wishart matrices: bipartite support, positive semidefiniteness, and rank-induced zero modes.
\end{examplebox}

The entries of $\pmb{W}$ are not independent. Indeed, a sample node $\mu$ contributes the rank-one matrix
\begin{equation}
\frac{1}{d}\pmb{x}^\mu(\pmb{x}^\mu)^{\rm T}\,,\qquad \pmb{x}^\mu=(X_1^\mu,\ldots,X_N^\mu)^{\rm T}\,,
\label{eq:scdw-rank-one-contribution}
\end{equation}
and hence couples all variables connected to the same sample node. In the one-mode projection onto the variable vertices, each factor node generates a clique among its neighboring variables. This is why the bipartite graph is the correct object for the cavity calculation. Treating $\pmb{W}$ as an ordinary sparse matrix on the variable side alone would obscure the rank-one structure of the interactions.

The spectral density of $\pmb{W}$ is
\begin{equation}
\rho_{\pmb{W}}(\lambda)=\frac{1}{N}\sum_{i=1}^N\delta(\lambda-\lambda_i)\,,\qquad\lambda_i\geq0\,,
\label{eq:scdw-empirical-density}
\end{equation}
and its regularized form is obtained from the resolvent
\begin{equation}
\pmb{G}(z)=(z\pmb{I}-\pmb{W})^{-1}\,,\qquad z=\lambda-i\epsilon\,,\qquad\epsilon>0\,,
\label{eq:scdw-resolvent}
\end{equation}
as
\begin{equation}
\rho_{\pmb{W},\epsilon}(\lambda)=\frac{1}{\pi N}\sum_{i=1}^N{\rm Im} [G_{ii}(\lambda-i\epsilon)]\,.
\label{eq:scdw-regularized-density}
\end{equation}
Sparse sample covariance matrices were analyzed using statistical-mechanics methods by Nagao and Tanaka \cite{NagaoTanaka2007}. The cavity formulation of \cite{RogersTakedaPerezCastilloKuhn2008} gives a direct finite-connectivity treatment of both sparse symmetric matrices and sparse covariance matrices. Later work used related ideas to study large deviations of eigenvalue counts in diluted Wishart ensembles \cite{PerezCastilloMetz2018Wishart}, and generalized diluted Wishart ensembles provide a broader framework for sparse cross-correlation matrices \cite{PerezCastillo2022Generalized}.

Let us recall the Edwards--Jones representation in this new setting. The Gaussian structure is the same as before, but the quadratic interaction is now local on the bipartite factor graph. From
\begin{equation}
\pmb{x}^{\rm T}\pmb{W}\pmb{x}=\frac{1}{d}\sum_{\mu=1}^P\left(\sum_{i\in\partial\mu}X_i^\mu x_i\right)^2\,,
\label{eq:scdw-quadratic-form}
\end{equation}
where $\partial\mu$ denotes the set of variable nodes connected to sample node $\mu$, the Edwards--Jones partition function can be written as
\begin{equation}
Z_{\pmb{W}}(z)=\int\left[\prod_{i=1}^N\frac{dx_i}{\sqrt{2\pi}}\right]\exp\left[-\frac{i}{2}z\sum_{i=1}^N x_i^2+\frac{i}{2d}\sum_{\mu=1}^P\left(\sum_{i\in\partial\mu}X_i^\mu x_i\right)^2\right]\,.
\label{eq:scdw-gaussian-partition-function}
\end{equation}
The first term is local on variable nodes, while the second term is local on factor nodes. Thus the covariance problem is a Gaussian model on a sparse bipartite factor graph. Since the graph is locally tree-like, the cavity method can be applied directly on this bipartite representation.

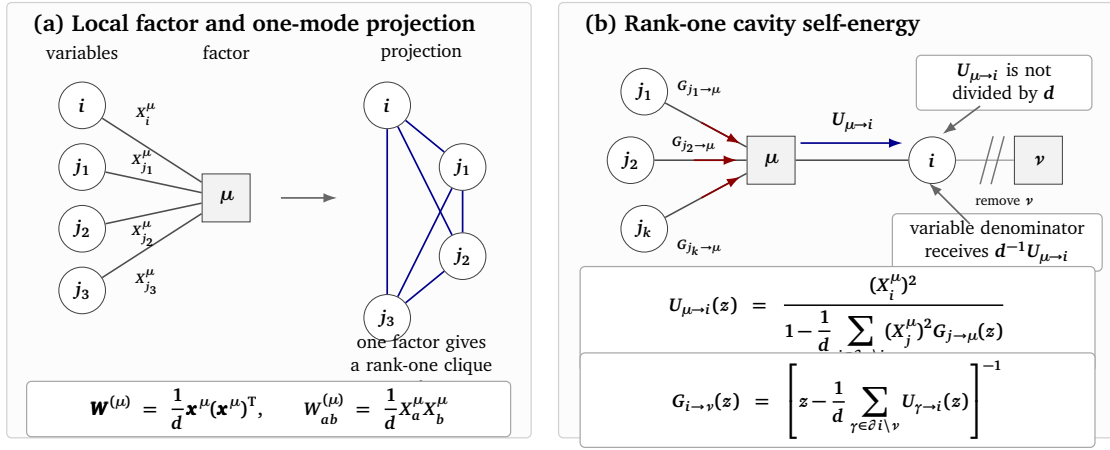
\begin{figure}[t]
\centering
\resizebox{0.98\textwidth}{!}{%
\begin{tikzpicture}[
    x=1cm,
    y=1cm,
    >=Latex,
    panel/.style={draw=black!18, fill=black!1, rounded corners=2pt, line width=0.5pt},
    vnode/.style={circle, draw=black!75, fill=white, minimum size=6.8mm, inner sep=0pt, font=\scriptsize},
    fnode/.style={rectangle, draw=black!75, fill=black!5, minimum size=7.0mm, inner sep=0pt, font=\scriptsize},
    edge/.style={draw=black!70, line width=0.65pt},
    paleedge/.style={draw=black!35, line width=0.6pt},
    projedge/.style={draw=blue!55!black, line width=0.7pt},
    arrowedge/.style={draw=red!55!black, line width=0.7pt, -{Latex[length=2.0mm,width=1.4mm]}},
    bluearrow/.style={draw=blue!55!black, line width=0.7pt, -{Latex[length=2.0mm,width=1.4mm]}},
    flowarrow/.style={draw=black!60, line width=0.7pt, -{Latex[length=2.0mm,width=1.4mm]}},
    ptitle/.style={font=\bfseries\small, anchor=west},
    paneltext/.style={font=\scriptsize, align=center},
    smalltext/.style={font=\tiny, align=center},
    box/.style={draw=black!35, fill=white, rounded corners=2pt, line width=0.5pt, inner sep=3pt, font=\scriptsize, align=center}
]
\draw[panel] (0,0) rectangle (7.65,6.35);
\node[ptitle] at (0.25,6.03) {(a) Local factor and one-mode projection};

\node[paneltext] at (1.10,5.62) {variables};
\node[paneltext] at (3.20,5.62) {factor};
\node[paneltext] at (6.05,5.62) {projection};

\node[vnode] (vi)  at (1.10,4.85) {$i$};
\node[vnode] (vj1) at (1.10,3.95) {$j_1$};
\node[vnode] (vj2) at (1.10,3.05) {$j_2$};
\node[vnode] (vj3) at (1.10,2.15) {$j_3$};
\node[fnode] (muA) at (3.20,3.50) {$\mu$};

\draw[edge] (vi) -- (muA);
\draw[edge] (vj1) -- (muA);
\draw[edge] (vj2) -- (muA);
\draw[edge] (vj3) -- (muA);
\node[smalltext] at (2.03,4.72) {$X_i^\mu$};
\node[smalltext] at (2.00,4.02) {$X_{j_1}^\mu$};
\node[smalltext] at (2.00,2.98) {$X_{j_2}^\mu$};
\node[smalltext] at (2.05,2.28) {$X_{j_3}^\mu$};

\draw[flowarrow] (4.00,3.50) -- (4.82,3.50);

\node[vnode] (pi)  at (5.55,4.85) {$i$};
\node[vnode] (pj1) at (6.65,3.95) {$j_1$};
\node[vnode] (pj2) at (6.65,2.65) {$j_2$};
\node[vnode] (pj3) at (5.55,1.75) {$j_3$};
\draw[projedge] (pi) -- (pj1);
\draw[projedge] (pi) -- (pj2);
\draw[projedge] (pi) -- (pj3);
\draw[projedge] (pj1) -- (pj2);
\draw[projedge] (pj1) -- (pj3);
\draw[projedge] (pj2) -- (pj3);

\node[paneltext, text width=2.75cm] at (6.05,1.08)
{one factor gives a rank-one clique contribution};
\node[box, text width=6.85cm] at (3.82,0.42)
{$\displaystyle \pmb{W}^{(\mu)}=\frac{1}{d}\pmb{x}^\mu(\pmb{x}^\mu)^{\rm T},\qquad
W_{ab}^{(\mu)}=\frac{1}{d}X_a^\mu X_b^\mu$};

\draw[panel] (8.05,0) rectangle (16.20,6.35);
\node[ptitle] at (8.30,6.03) {(b) Rank-one cavity self-energy};

\node[vnode] (rj1) at (9.30,5.05) {$j_1$};
\node[vnode] (rj2) at (9.10,4.05) {$j_2$};
\node[vnode] (rjk) at (9.30,3.05) {$j_k$};
\node[fnode] (muB) at (11.15,4.05) {$\mu$};
\node[vnode] (ri)  at (13.50,4.05) {$i$};
\node[fnode] (nuB) at (15.05,4.05) {$\nu$};

\draw[paleedge] (ri) -- (nuB);
\draw[edge] (rj1) -- (muB);
\draw[edge] (rj2) -- (muB);
\draw[edge] (rjk) -- (muB);
\draw[edge] (muB) -- (ri);

\draw[arrowedge] ($(rj1)!0.45!(muB)$) -- ($(rj1)!0.76!(muB)$);
\draw[arrowedge] ($(rj2)!0.45!(muB)$) -- ($(rj2)!0.76!(muB)$);
\draw[arrowedge] ($(rjk)!0.45!(muB)$) -- ($(rjk)!0.76!(muB)$);
\node[smalltext] at (10.10,5.10) {$G_{j_1\to\mu}$};
\node[smalltext] at (10.02,4.28) {$G_{j_2\to\mu}$};
\node[smalltext] at (10.10,2.78) {$G_{j_k\to\mu}$};

\draw[bluearrow] ($(muB)+(0.43,0.25)$) -- ($(ri)+(-0.43,0.25)$);
\node[paneltext] at (12.35,4.60) {$U_{\mu\to i}$};
\draw[black!55, line width=0.6pt] (14.20,3.70) -- (14.39,4.40);
\draw[black!55, line width=0.6pt] (14.36,3.70) -- (14.55,4.40);
\node[smalltext] at (14.55,3.45) {remove $\nu$};

\node[box, text width=2.35cm] at (14.55,5.18)
{$U_{\mu\to i}$ is not divided by $d$};
\draw[flowarrow] (14.20,4.84) -- (13.65,4.43);

\node[box, text width=2.85cm] at (14.45,2.85)
{variable denominator receives $d^{-1}U_{\mu\to i}$};
\draw[flowarrow] (14.05,3.12) -- (13.45,3.72);

\node[box, text width=7.30cm] at (12.12,1.73)
{$\displaystyle
U_{\mu\to i}(z)=\frac{(X_i^\mu)^2}{1-\displaystyle\frac{1}{d}\sum_{j\in\partial\mu\setminus i}(X_j^\mu)^2G_{j\to\mu}(z)}$};

\node[box, text width=7.30cm] at (12.12,0.55)
{$\displaystyle
G_{i\to\nu}(z)=\left[z-\frac{1}{d}\sum_{\gamma\in\partial i\setminus\nu}U_{\gamma\to i}(z)\right]^{-1}$};
\end{tikzpicture}%
}
\caption{Bipartite cavity structure of a diluted Wishart factor. A sample node $\mu$ contributes the rank-one matrix $d^{-1}\pmb{x}^\mu(\pmb{x}^\mu)^{\rm T}$ to the covariance matrix, producing a clique in the one-mode projection but a single local factor in the bipartite representation. The factor-to-variable message $U_{\mu\to i}$ is defined without the overall $d^{-1}$ prefactor; the contribution entering a variable Green-function denominator is $d^{-1}U_{\mu\to i}$.}
\label{fig:scdw-bipartite-wishart-factor}
\end{figure}

Let $G_{i\to\mu}(z)$ be the cavity Green function associated with variable node $i$ when factor node $\mu$ is removed. It is useful to introduce the effective self-energy--in the jargon of condensed-matter physics--sent from a factor node $\mu$ to a variable node $i$,
\begin{equation}
U_{\mu\to i}(z)=\frac{(X_i^\mu)^2}{1-\displaystyle\frac{1}{d}\sum_{j\in\partial\mu\setminus i}(X_j^\mu)^2G_{j\to\mu}(z)}\,.
\label{eq:scdw-factor-to-variable-self-energy}
\end{equation}
The rank-one Gaussian identity behind this factor update is derived in Appendix~\ref{app:gaussian-identities-resolvents}. Then the variable-to-factor cavity equation is
\begin{equation}
G_{i\to\mu}(z)=\frac{1}{z-\displaystyle\frac{1}{d}\sum_{\nu\in\partial i\setminus\mu}U_{\nu\to i}(z)}\,,
\label{eq:scdw-variable-to-factor-green}
\end{equation}
and the full diagonal resolvent at variable node $i$ is
\begin{equation}
G_i(z)=\frac{1}{z-\displaystyle\frac{1}{d}\sum_{\nu\in\partial i}U_{\nu\to i}(z)}\,.
\label{eq:scdw-full-variable-green}
\end{equation}
Equations \eqref{eq:scdw-factor-to-variable-self-energy}--\eqref{eq:scdw-full-variable-green} are the sparse-covariance analogue of the cavity equations for sparse symmetric matrices. The only difference is that an incoming factor node contributes not a single edge term $J^2G$, but the rank-one expression \eqref{eq:scdw-factor-to-variable-self-energy}, which accounts for all other variables attached to the same sample node. Figure~\ref{fig:scdw-bipartite-wishart-factor} summarizes the bipartite factor structure and the normalization convention entering the diluted-Wishart cavity equations.

\begin{examplebox}[Deriving the factor-to-variable self-energy]
We derive equation \eqref{eq:scdw-factor-to-variable-self-energy} by an explicit Schur-complement calculation. Consider a factor node $\mu$ connected to a distinguished variable $i$ and to a set $R=\partial\mu\setminus i$ of other variables. To avoid confusing the matrix entries with the Gaussian integration variables, write
\begin{equation}
\chi_i=X_i^\mu\,,\qquad \chi_j=X_j^\mu \quad (j\in R)\,.
\label{eq:scdw-ped-self-energy-weights}
\end{equation}
The factor contributes the rank-one matrix
\begin{equation}
B_{ab}^{\mu}=\frac{\chi_a \chi_b}{d}\,, \qquad a,b\in\partial\mu\,.
\label{eq:scdw-ped-factor-rank-one}
\end{equation}
Assume that the incoming cavity Green functions from the variables $j\in R$ are $G_{j\to\mu}$. Their inverse cavity propagators form the diagonal matrix
\begin{equation}
\pmb D_R={\rm diag}\left(G_{j\to\mu}^{-1}\right)_{j\in R}\,.
\label{eq:scdw-ped-DR}
\end{equation}
The block corresponding to the variables in $R$ is
\begin{equation}
\pmb D_R-\frac{1}{d}\pmb\chi_R\pmb\chi_R^{\rm T}\,,\qquad \pmb\chi_R=(\chi_j)_{j\in R}\,.
\label{eq:scdw-ped-R-block}
\end{equation}
The direct diagonal contribution of factor $\mu$ to variable $i$ is $\chi_i^2/d$, and the coupling vector from $i$ to the variables in $R$ is
\begin{equation}
\frac{\chi_i}{d}\pmb\chi_R\,.
\label{eq:scdw-ped-coupling-vector}
\end{equation}
Thus the Schur-complement contribution of factor $\mu$ to the denominator of the Green function at $i$ is
\begin{equation}
\Sigma_{\mu\to i}=\frac{\chi_i^2}{d}+\frac{\chi_i^2}{d^2}\pmb\chi_R^{\rm T}\left(\pmb D_R-\frac{1}{d}\pmb\chi_R\pmb\chi_R^{\rm T}\right)^{-1}\pmb\chi_R\,.
\label{eq:scdw-ped-self-energy-schur}
\end{equation}
To simplify this expression, define
\begin{equation}
S_R=\pmb\chi_R^{\rm T}\pmb D_R^{-1}\pmb\chi_R=\sum_{j\in R}\chi_j^2G_{j\to\mu}\,.
\label{eq:scdw-ped-SR}
\end{equation}
Using the Sherman--Morrison identity,
\begin{equation}
\left(\pmb D_R-\frac{1}{d}\pmb\chi_R\pmb\chi_R^{\rm T}\right)^{-1}=\pmb D_R^{-1}+\frac{\pmb D_R^{-1}\pmb\chi_R\pmb\chi_R^{\rm T}\pmb D_R^{-1}}{d-S_R}\,,
    \label{eq:scdw-ped-sherman-morrison}
\end{equation}
one obtains
\begin{equation}
\pmb\chi_R^{\rm T}\left(\pmb D_R-\frac{1}{d}\pmb\chi_R\pmb\chi_R^{\rm T}\right)^{-1}\pmb\chi_R=\frac{dS_R}{d-S_R}\,.
\label{eq:scdw-ped-rank-one-contraction}
\end{equation}
Substituting this into \eqref{eq:scdw-ped-self-energy-schur} gives
\begin{equation}
\Sigma_{\mu\to i}=\frac{\chi_i^2}{d}+\frac{\chi_i^2}{d^2}\frac{dS_R}{d-S_R}=\frac{\chi_i^2/d}{1-S_R/d}\,.
\label{eq:scdw-ped-self-energy-final}
\end{equation}
Since $S_R=\sum_{j\in\partial\mu\setminus i}(X_j^\mu)^2G_{j\to\mu}$, this is precisely
\begin{equation}
\Sigma_{\mu\to i}(z)=\frac{1}{d}\frac{(X_i^\mu)^2}{1-\displaystyle\frac{1}{d}\sum_{j\in\partial\mu\setminus i}(X_j^\mu)^2G_{j\to\mu}(z)}=\frac{1}{d}U_{\mu\to i}(z)\,.
\label{eq:scdw-ped-self-energy-result}
\end{equation}
The calculation shows why the rank-one structure of a Wishart factor is essential: after integrating out the neighboring variables, the factor contributes a scalar term to the denominator of the distinguished variable. With the convention used in this section, this Schur-complement contribution is $d^{-1}U_{\mu\to i}$.
\end{examplebox}

The derivation is a direct Gaussian integration expressed in Schur-complement form. Suppose the incoming cavity messages to a factor node $\mu$ are Gaussian, with diagonal Green functions $G_{j\to\mu}$. Integrating over the variables $x_j$ with $j\in\partial\mu\setminus i$ produces an effective quadratic term in $x_i$. Because the factor interaction in \eqref{eq:scdw-gaussian-partition-function} has rank one, the result is again Gaussian and is fully summarized by \eqref{eq:scdw-factor-to-variable-self-energy}. This closure is the reason why diluted Wishart matrices remain tractable by cavity methods despite the induced clique interactions in $\pmb W$.

The spectral density of a fixed realization is estimated from
\begin{equation}
\rho_{\pmb W,\epsilon}^{\rm cav}(\lambda)=\frac{1}{\pi N}\sum_{i=1}^N{\rm Im}\frac{1}{\lambda-i\epsilon-\displaystyle\frac{1}{d}\sum_{\nu\in\partial i}U_{\nu\to i}(\lambda-i\epsilon)}\,.
\label{eq:scdw-cavity-density-fixed-instance}
\end{equation}
On a tree factor graph this expression is exact. On the diluted bipartite ensembles considered here it becomes exact in the thermodynamic limit under the usual locally tree-like and replica-symmetric assumptions. Numerically, one may iterate \eqref{eq:scdw-factor-to-variable-self-energy} and \eqref{eq:scdw-variable-to-factor-green} as belief-propagation equations on a single large instance.

At the ensemble level, the same equations become distributional. Let $p^{(\rm v)}_\ell$ be the degree distribution of a variable node and $p^{(\rm f)}_k$ the degree distribution of a factor node. In the Poisson bipartite ensemble \eqref{eq:scdw-diluted-rectangular-matrix},
\begin{equation}
p^{(\rm v)}_\ell=e^{-d/\alpha}\frac{(d/\alpha)^\ell}{\ell!}\,,\qquad p^{(\rm f)}_k=e^{-d}\frac{d^k}{k!}\,.
\label{eq:scdw-bipartite-poisson-degrees}
\end{equation}
For a general bipartite ensemble, the cavity degree distributions are the corresponding excess-degree laws,
\begin{equation}
q^{(\rm v)}_\ell=\frac{(\ell+1)p^{(\rm v)}_{\ell+1}}{\sum_r r p^{(\rm v)}_r}\,,\qquad q^{(\rm f)}_k=\frac{(k+1)p^{(\rm f)}_{k+1}}{\sum_r r p^{(\rm f)}_r}\,.
\label{eq:scdw-bipartite-excess-degrees}
\end{equation}
For a bipartite ensemble with $N$ variable nodes and $P=N/\alpha$ factor nodes, the two degree laws must also satisfy the edge-count consistency
\begin{equation}
\sum_{\ell\geq0}\ell\,p^{(\rm v)}_\ell
=\frac{P}{N}\sum_{k\geq0}k\,p^{(\rm f)}_k
=\frac{1}{\alpha}\sum_{k\geq0}k\,p^{(\rm f)}_k\,.
\label{eq:scdw-bipartite-degree-consistency}
\end{equation}
Let $\mathcal{P}(G)$ denote the distribution of a variable-to-factor cavity Green function and let $\mathcal{Q}(U)$ denote the distribution of a factor-to-variable self-energy. Then
\begin{equation}
\mathcal{Q}(U)=\sum_{k=0}^\infty q^{(\rm f)}_k\int d\xi\, p_\xi(\xi)\left[\prod_{r=1}^k dG_r\,\mathcal{P}(G_r)\, d\eta_r\, p_\xi(\eta_r)\right]\delta\left(U-\frac{\xi^2}{1-\displaystyle\frac{1}{d}\sum_{r=1}^k\eta_r^2G_r}\right)\,,
\label{eq:scdw-self-energy-distribution}
\end{equation}
and
\begin{equation}
\mathcal{P}(G)=\sum_{\ell=0}^\infty q^{(\rm v)}_\ell\int\left[\prod_{r=1}^\ell dU_r\mathcal{Q}(U_r)\right]\delta\left(G-\frac{1}{z-\displaystyle\frac{1}{d}\sum_{r=1}^\ell U_r}\right)\,.
\label{eq:scdw-cavity-green-distribution}
\end{equation}
The full site distribution is
\begin{equation}
\mathcal{P}_{\rm site}(G)=\sum_{\ell=0}^\infty p^{(\rm v)}_\ell\int\left[\prod_{r=1}^\ell dU_r\mathcal{Q}(U_r)\right]\delta\left(G-\frac{1}{z-\displaystyle\frac{1}{d}\sum_{r=1}^\ell U_r}\right)\,,
\label{eq:scdw-site-green-distribution}
\end{equation}
and the averaged spectral density is
\begin{equation}
\overline{\rho_\epsilon(\lambda)}=\frac{1}{\pi}{\rm Im}\int dG\mathcal{P}_{\rm site}(G)G\,,\qquad z=\lambda-i\epsilon\,.
\label{eq:scdw-density-from-populations}
\end{equation}
For the Poisson bipartite ensemble, the excess laws coincide with the corresponding degree laws, so $q^{(\rm v)}=p^{(\rm v)}$ and $q^{(\rm f)}=p^{(\rm f)}$. For fixed-degree bipartite graphs, by contrast, the distinction between full degrees and excess degrees must be kept.

\begin{examplebox}[One population-dynamics update for the Poisson bipartite ensemble]
In the Poisson bipartite ensemble, variable degrees and factor degrees have laws
\begin{equation}
p_\ell^{(\rm v)}=e^{-d/\alpha}\frac{(d/\alpha)^\ell}{\ell!}\,,\qquad p_k^{(\rm f)}=e^{-d}\frac{d^k}{k!}\,.
\label{eq:scdw-ped-poisson-degrees}
\end{equation}
Because the Poisson distribution is invariant under size biasing followed by subtracting one, the cavity excess-degree laws are the same:
\begin{equation}
q_\ell^{(\rm v)}=p_\ell^{(\rm v)}\,,\qquad q_k^{(\rm f)}=p_k^{(\rm f)}\,.
\label{eq:scdw-ped-poisson-excess}
\end{equation}
A population-dynamics algorithm therefore uses two populations,
\begin{equation}
\{G^{(1)},\ldots,G^{(M)}\}\,,\qquad\{U^{(1)},\ldots,U^{(M)}\}\,.
\label{eq:scdw-ped-two-populations}
\end{equation}
To generate a new factor-to-variable message $U_{\rm new}$, draw
\begin{equation}
k\sim{\rm Poisson}(d)\,,
\label{eq:scdw-ped-draw-factor-degree}
\end{equation}
draw $k$ incoming Green functions $G^{(a_1)},\ldots,G^{(a_k)}$ from the $G$-population, and draw $k+1$ weights $\xi,\xi_1,\ldots,\xi_k$ from $p_\xi$. Then set
\begin{equation}
U_{\rm new}=\frac{\xi^2}{1-\displaystyle\frac{1}{d}\sum_{r=1}^{k}\xi_r^2G^{(a_r)}}\,.
\label{eq:scdw-ped-factor-update}
\end{equation}
To generate a new variable-to-factor message $G_{\rm new}$, draw
\begin{equation}
\ell\sim{\rm Poisson}(d/\alpha)\,,
\label{eq:scdw-ped-draw-variable-degree}
\end{equation}
draw $\ell$ self-energies $U^{(b_1)},\ldots,U^{(b_\ell)}$ from the $U$-population, and set
\begin{equation}
G_{\rm new}=\frac{1}{z-\displaystyle\frac{1}{d}\sum_{r=1}^{\ell}U^{(b_r)}}\,.
\label{eq:scdw-ped-variable-update}
\end{equation}
Finally, to measure the density, draw a full variable degree
\begin{equation}
\ell_{\rm site}\sim{\rm Poisson}(d/\alpha)
\label{eq:scdw-ped-site-degree}
\end{equation}
and form
\begin{equation}
G_{\rm site}=\frac{1}{z-\displaystyle\frac{1}{d}\sum_{r=1}^{\ell_{\rm site}}U^{(b_r)}}\,.
\label{eq:scdw-ped-site-sample}
\end{equation}
The regularized density is estimated by averaging
\begin{equation}
\rho_\epsilon(\lambda)=\frac{1}{\pi}{\rm Im}\langle G_{\rm site}\rangle\,,\qquad z=\lambda-i\epsilon\,.
    \label{eq:scdw-ped-pop-density}
\end{equation}
This example makes explicit why sparse covariance matrices require two coupled populations: one population lives on variable-to-factor messages, and the other on factor-to-variable self-energies.
\end{examplebox}

Equations \eqref{eq:scdw-self-energy-distribution}--\eqref{eq:scdw-density-from-populations} are solved by population dynamics. One represents $\mathcal{P}$ and $\mathcal{Q}$ by two populations of complex numbers. A new factor-to-variable self-energy is generated by drawing a factor excess degree, incoming variable messages, and weights, and then applying \eqref{eq:scdw-self-energy-distribution}. A new variable-to-factor Green function is generated by drawing a variable excess degree, incoming self-energies, and then applying \eqref{eq:scdw-cavity-green-distribution}. Once the two populations have reached stationarity, full site Green functions are sampled from \eqref{eq:scdw-site-green-distribution}. This is the direct analogue of population dynamics for sparse symmetric matrices, with the important difference that the factor-node update contains the rank-one denominator in \eqref{eq:scdw-self-energy-distribution}.

There is an alternative and often useful representation by linearization. Define the $(N+P)\times(N+P)$ sparse symmetric matrix
\begin{equation}
\pmb{\mathcal{L}}=\frac{1}{\sqrt d}
\begin{pmatrix}
\pmb{0}_{N\times N} & \pmb{X}\\
\pmb{X}^{\rm T} & \pmb{0}_{P\times P}
\end{pmatrix}\,.
\label{eq:scdw-linearization}
\end{equation}
Let $r={\rm rank} \pmb X$. If $\nu_1,\ldots,\nu_r$ are the nonzero eigenvalues of $\pmb W$, then $\pmb{\mathcal L}$ has nonzero eigenvalues $\pm\sqrt{\nu_a}$, with $a=1,\ldots,r$, together with $N+P-2r$ zero eigenvalues. In particular, there are at least $|N-P|$ zero eigenvalues when the rectangular matrix has unequal dimensions, and there may be additional zero modes if $\pmb X$ is rank deficient. The covariance spectral problem may therefore be studied through a sparse symmetric operator on the bipartite graph, provided the spectral-parameter map is kept explicit. If $w$ is the spectral parameter of $\pmb{\mathcal L}$ and $z=w^2$ is the spectral parameter of $\pmb W$, then
\begin{equation}
\left[(w\pmb I_{N+P}-\pmb{\mathcal L})^{-1}\right]_{1:N,1:N}=w (w^2\pmb I_N-\pmb W)^{-1}\,.
\label{eq:scdw-linearization-variable-block}
\end{equation}
Thus the covariance resolvent is recovered from the variable-side block of the linearized problem, while the normalized trace over all $N+P$ vertices gives the density of $\pmb{\mathcal L}$ rather than the density of $\pmb W$. Equivalently, eliminating the factor-side messages in the cavity equations for $\pmb{\mathcal L}$ gives \eqref{eq:scdw-factor-to-variable-self-energy}--\eqref{eq:scdw-variable-to-factor-green} after setting $z=w^2$ and rescaling the variable-side linearized Green functions by $w$. The linearization is especially useful for comparing sparse covariance matrices with general sparse symmetric matrices, while the direct factor form keeps the positivity and rank-one structure of $\pmb{W}$ explicit.

\begin{examplebox}[Linearization of the example matrix]
Return to the rectangular matrix in Eq.~\eqref{eq:scdw-ped-example-X},
\begin{equation}
\pmb X=\begin{pmatrix}
1 & 0\\
0 & 1\\
1 & 1
\end{pmatrix}\,,\qquad d=1\,.
\label{eq:scdw-ped-linearization-X}
\end{equation}
The bipartite linearization is
\begin{equation}
\pmb{\mathcal L}=\begin{pmatrix}
0 & 0 & 0 & 1 & 0\\
0 & 0 & 0 & 0 & 1\\
0 & 0 & 0 & 1 & 1\\
1 & 0 & 1 & 0 & 0\\
0 & 1 & 1 & 0 & 0
\end{pmatrix}\,.
\label{eq:scdw-ped-linearization-matrix}
\end{equation}
A direct multiplication gives
\begin{equation}
\pmb{\mathcal L}^2=\begin{pmatrix}
\pmb X\pmb X^{\rm T} & \pmb 0\\
\pmb 0 & \pmb X^{\rm T}\pmb X
\end{pmatrix}\,.
\label{eq:scdw-ped-linearization-square}
\end{equation}
We already computed
\begin{equation}
{\rm spec}(\pmb X\pmb X^{\rm T})=\{3,1,0\}\,,
\label{eq:scdw-ped-spec-XXT}
\end{equation}
while
\begin{equation}
\pmb X^{\rm T}\pmb X=\begin{pmatrix}
2 & 1\\
1 & 2
\end{pmatrix}
\end{equation}
has eigenvalues $3$ and $1$. Therefore
\begin{equation}
{\rm spec}(\pmb{\mathcal L})=\left\{\sqrt3,\,-\sqrt3,\,1,\,-1,\,0\right\}\,.
\label{eq:scdw-ped-linearized-spectrum}
\end{equation}
This verifies in a finite example that the nonzero eigenvalues of the bipartite linearization are $\pm\sqrt{\nu}$, where $\nu$ runs over the nonzero eigenvalues of the covariance matrix.
\end{examplebox}

The dense limit provides an important check. Let $d\to\infty$ after the thermodynamic limit, with $\mathbb{E}[\xi]=0$ and $\mathbb{E}[\xi^2]=1$. In this limit the variable Green functions concentrate, while the factor-to-variable self-energies enter the variable update only through their averaged contribution. The sums over the $O(d)$ incoming factor and variable neighborhoods self-average. Denoting the limiting site Green function by $m(z)$ for the normalization \eqref{eq:scdw-diluted-wishart-definition}, one obtains
\begin{equation}
m(z)=\frac{1}{z-\displaystyle\frac{1}{\alpha}\frac{1}{1-m(z)}}\,.
\label{eq:scdw-dense-mp-equation}
\end{equation}
This is the Mar\v{c}enko--Pastur self-consistency equation with the eigenvalue scale corresponding to $\pmb{W}=d^{-1}\pmb{X}\pmb{X}^{\rm T}$. Equivalently, for the conventionally normalized matrix $\widehat{\pmb{W}}=\alpha\pmb{W}$, the limiting density is the Mar\v{c}enko--Pastur law with aspect ratio $\alpha$,
\begin{equation}
\rho_{\rm MP}(\lambda)=\frac{1}{2\pi\alpha\lambda}\sqrt{(\lambda_+-\lambda)(\lambda-\lambda_-)}\,\mathbf{1}_{\lambda_-\leq\lambda\leq\lambda_+}\,,\qquad\lambda_\pm=(1\pm\sqrt{\alpha})^2\,,
\label{eq:scdw-mp-density}
\end{equation}
with an additional Dirac delta at the origin of weight $1-1/\alpha$ when $\alpha>1$. Thus the finite-connectivity equations reduce to the classical dense covariance result in the appropriate limit. The corresponding dense-limit reduction, in the normalization used here, is given in Appendix~\ref{app:dense-limit-reductions}.

\begin{examplebox}[The square dense limit]
Set $\alpha=1$ and assume $\mathbb E[\xi^2]=1$. In the dense limit $d\to\infty$, equation \eqref{eq:scdw-dense-mp-equation} becomes
\begin{equation}
m(z)=\frac{1}{z-\displaystyle\frac{1}{1-m(z)}}\,.
\label{eq:scdw-ped-square-dense-equation}
\end{equation}
Multiplying by the denominator gives
\begin{equation}
m(z)\left[z-\frac{1}{1-m(z)}\right]=1\,.
\label{eq:scdw-ped-square-dense-step1}
\end{equation}
Multiplying by $1-m(z)$,
\begin{equation}
z m(z)[1-m(z)]-m(z)=1-m(z).
\label{eq:scdw-ped-square-dense-step2}
\end{equation}
The terms $-m(z)$ on the two sides cancel, and we obtain
\begin{equation}
z m(z)[1-m(z)]=1\,.
\label{eq:scdw-ped-square-dense-step3}
\end{equation}
Equivalently,
\begin{equation}
z m(z)^2-z m(z)+1=0\,.
\label{eq:scdw-ped-square-dense-quadratic}
\end{equation}
The solution satisfying $m(z)\sim 1/z$ for large $|z|$ is
\begin{equation}
m(z)=\frac{1-\sqrt{1-4/z}}{2}\,.
\label{eq:scdw-ped-square-dense-solution}
\end{equation}
For $z=\lambda-i0^+$ and $0<\lambda<4$, this gives
\begin{equation}
{\rm Im}[ m(\lambda-i0^+)]=\frac{1}{2}\sqrt{\frac{4-\lambda}{\lambda}}\,.
    \label{eq:scdw-ped-square-dense-imaginary}
\end{equation}
Hence
\begin{equation}
\rho_{\rm MP}^{(\alpha=1)}(\lambda)=\frac{1}{2\pi}\sqrt{\frac{4-\lambda}{\lambda}}\mathbf 1_{0<\lambda<4}=\frac{1}{2\pi\lambda}\sqrt{(4-\lambda)\lambda}\mathbf 1_{0<\lambda<4}.
\label{eq:scdw-ped-square-mp}
\end{equation}
This is the square Mar\v{c}enko--Pastur law. The calculation shows explicitly how the diluted Wishart cavity equations reduce to the classical dense covariance result.
\end{examplebox}

At finite $d$, the spectral density differs substantially from the dense Mar\v{c}enko--Pastur law. First, the bipartite graph contains isolated variable nodes with probability
\begin{equation}
p^{(\rm v)}_0=e^{-d/\alpha},
\label{eq:scdw-isolated-variable-probability}
\end{equation}
and each such node contributes a zero eigenvalue. Additional zero modes may arise from rank deficiencies associated with finite bipartite components or with the rectangularity of $\pmb X$. Second, factor nodes of small degree generate local spectral structures: a degree-one factor contributes only to a diagonal entry, while higher-degree factors create finite rank-one couplings among small groups of variables. Third, rare high-degree nodes or atypical finite components can generate localized modes and spectral tails. These features are precisely the finite-connectivity effects that are washed out in the dense limit.

The positivity of $\pmb W$ is another useful check. Since $\pmb{W}=d^{-1}\pmb{X}\pmb{X}^{\rm T}$, the limiting spectral density must be supported on $\lambda\geq0$. At finite regulator $\epsilon$, Lorentzian broadening gives nonzero values for $\lambda<0$. If the limiting measure has a zero atom, this leakage is concentrated near the origin and need not be small in an integral over a window touching $\lambda=0$; away from the origin it must shrink as $\epsilon\downarrow0$ after the thermodynamic limit. The total mass of the limiting density must be one, and the first moment is fixed by the trace:
\begin{equation}
\frac{1}{N}{\rm Tr}\pmb{W}=\frac{1}{Nd}\sum_{\mu=1}^P\sum_{i=1}^N(X_i^\mu)^2\,.
\label{eq:scdw-first-moment}
\end{equation}
For the Poisson ensemble with $\mathbb{E}[\xi^2]=1$, this converges to $1/\alpha$. Hence
\begin{equation}
\int_0^\infty d\lambda \rho(\lambda)=1\,,\qquad \int_0^\infty d\lambda\lambda\rho(\lambda)=\frac{1}{\alpha}\,,
\label{eq:scdw-normalization-and-first-moment}
\end{equation}
for the normalization \eqref{eq:scdw-diluted-wishart-definition}. These moment checks are useful diagnostics for numerical solutions of population dynamics.

Sparse covariance matrices therefore interpolate between two familiar structures. On one side, they are covariance matrices and inherit positivity, rectangularity effects, and the dense Mar\v{c}enko--Pastur limit. On the other side, they are finite-connectivity graphical models and inherit local tree structure, message distributions, isolated components, and localization phenomena. The cavity equations above make this interpolation explicit. They also prepare the ground for the later study of diluted Wishart large deviations, where one asks not only for the typical spectral density but for the probability of atypical eigenvalue counts below a prescribed threshold.

\begin{exerciseblock}
\exitem[Mean degrees in the bipartite ensemble]
Starting from
\begin{equation}
{\rm Prob}(B_i^\mu=1)=\frac{d}{N}\,,
\label{eq:scdw-ex-bernoulli-support}
\end{equation}
show that the expected degree of a factor node $\mu$ is $d$, while the expected degree of a variable node $i$ is $d/\alpha$, with $\alpha=N/P$.

\exitem[Quadratic form and positivity]
Using
\begin{equation}
\pmb W=\frac{1}{d}\pmb X\pmb X^{\rm T}\,,
\label{eq:scdw-ex-W-definition}
\end{equation}
show that for every $\pmb v\in\mathbb R^N$,
\begin{equation}
\pmb v^{\rm T}\pmb W\pmb v=\frac{1}{d}\sum_{\mu=1}^{P}\left(\sum_{i=1}^{N}X_i^\mu v_i\right)^2\geq 0\,.
\label{eq:scdw-ex-positivity}
\end{equation}
Deduce that all eigenvalues of $\pmb W$ are nonnegative.

\exitem[Rank bound and zero modes]
Prove that
\begin{equation}
{\rm rank}\pmb W={\rm rank}(\pmb X\pmb X^{\rm T})={\rm rank}\pmb X\leq P\,.
\label{eq:scdw-ex-rank-bound}
\end{equation}
Show that if $\alpha=N/P>1$, then $\pmb W$ has at least
\begin{equation}
N-P=N\left(1-\frac{1}{\alpha}\right)
\label{eq:scdw-ex-rank-zero-modes}
\end{equation}
zero eigenvalues.

\exitem[Induced clique from one factor]
Suppose a factor node $\mu$ is connected to the variable set $\partial\mu=\{i_1,\ldots,i_k\}$ with weights $X_{i_r}^{\mu}=\xi_r$. Show that this single factor contributes
\begin{equation}
W_{i_r i_s}^{(\mu)}=\frac{1}{d}\xi_r\xi_s
\label{eq:scdw-ex-factor-contribution}
\end{equation}
to the covariance matrix. Explain why, in the one-mode projection, a factor node creates a clique among its neighboring variables.

\exitem[Deriving the Gaussian factor representation]
Starting from the Edwards--Jones representation for $\pmb W$,
\begin{equation}
Z_{\pmb W}(z)=\int\left[\prod_i\frac{dx_i}{\sqrt{2\pi}}\right]\exp\left[-\frac{i}{2}\pmb x^{\rm T}(z\pmb I-\pmb W)\pmb x\right]\,,
\label{eq:scdw-ex-EJ-W}
\end{equation}
derive
\begin{equation}
\pmb x^{\rm T}\pmb W\pmb x=\frac{1}{d}\sum_{\mu=1}^{P}\left(\sum_{i\in\partial\mu}X_i^\mu x_i\right)^2\,.
\label{eq:scdw-ex-quadratic-factor}
\end{equation}
Then recover equation \eqref{eq:scdw-gaussian-partition-function}.

\exitem[Rank-one self-energy]
Using the Schur-complement derivation in the worked example \emph{Deriving the factor-to-variable self-energy}, prove
\begin{equation}
U_{\mu\to i}(z)=\frac{(X_i^\mu)^2}{1-\displaystyle\frac{1}{d}\sum_{j\in\partial\mu\setminus i}(X_j^\mu)^2G_{j\to\mu}(z)}\,.
\label{eq:scdw-ex-rank-one-self-energy}
\end{equation}
Check explicitly that the contribution to the denominator of $G_{i\to\mu}$ is $d^{-1}U_{\mu\to i}(z)$. Then check the special case in which the factor $\mu$ has degree one.

\exitem[Variable-to-factor recursion]
Assume that factor-to-variable self-energies $U_{\nu\to i}$ have been computed. Derive
\begin{equation}
G_{i\to\mu}(z)=\frac{1}{z-\displaystyle\frac{1}{d}\sum_{\nu\in\partial i\setminus\mu}U_{\nu\to i}(z)}\,,
\label{eq:scdw-ex-variable-recursion}
\end{equation}
and the full local Green function
\begin{equation}
G_i(z)=\frac{1}{z-\displaystyle\frac{1}{d}\sum_{\nu\in\partial i}U_{\nu\to i}(z)}\,.
\label{eq:scdw-ex-full-green}
\end{equation}

\exitem[One-factor finite example]
Let
\begin{equation}
\pmb X=
\begin{pmatrix}
x_1\\
x_2
\end{pmatrix}\,,\qquad d=1\,.
\label{eq:scdw-ex-one-factor-X}
\end{equation}
Compute $\pmb W=\pmb X\pmb X^{\rm T}$ and its eigenvalues. For $z=\lambda-i\epsilon$, with $\epsilon>0$, compute the full local resolvent entries directly and compare them with the bipartite cavity equations.

\exitem[Poisson excess degrees]
For the Poisson degree laws
\begin{equation}
p_\ell^{(\rm v)}=e^{-d/\alpha}\frac{(d/\alpha)^\ell}{\ell!}\,,\qquad p_k^{(\rm f)}=e^{-d}\frac{d^k}{k!}\,,
\label{eq:scdw-ex-poisson-laws}
\end{equation}
show that the corresponding excess-degree distributions are the same laws:
\begin{equation}
q_\ell^{(\rm v)}=p_\ell^{(\rm v)}\,, \qquad q_k^{(\rm f)}=p_k^{(\rm f)}\,.
\label{eq:scdw-ex-poisson-excess}
\end{equation}

\exitem[Distributional equations]
Starting from the local recursions for $U_{\mu\to i}$ and $G_{i\to\mu}$, derive the population equations
\begin{equation}
\mathcal Q(U)=\sum_{k=0}^{\infty}q_k^{(\rm f)} \int d\xi p_\xi(\xi)\left[\prod_{r=1}^k dG_r \mathcal P(G_r)d\eta_r p_\xi(\eta_r)\right]\delta\left(U-\frac{\xi^2}{1-\displaystyle\frac{1}{d}\sum_{r=1}^k\eta_r^2 G_r}\right)
\label{eq:scdw-ex-Q-equation}
\end{equation}
and
\begin{equation}
\mathcal P(G)=\sum_{\ell=0}^{\infty}q_\ell^{(\rm v)}\int\left[\prod_{r=1}^{\ell}dU_r \mathcal Q(U_r)\right]\delta\left(G-\frac{1}{z-\displaystyle\frac{1}{d}\sum_{r=1}^{\ell}U_r}\right)\,.
\label{eq:scdw-ex-P-equation}
\end{equation}

\exitem[Site distribution and density]
Explain why the site distribution uses the full variable-degree law $p_\ell^{(\rm v)}$ rather than the excess law $q_\ell^{(\rm v)}$. Derive
\[
\mathcal P_{\rm site}(G)=\sum_{\ell=0}^{\infty}p_\ell^{(\rm v)}\int\left[\prod_{r=1}^{\ell}dU_r \mathcal Q(U_r)\right]\delta\left(G-\frac{1}{z-\displaystyle\frac{1}{d}\sum_{r=1}^{\ell}U_r}\right)
\]
and hence
\begin{equation}
\overline{\rho_{\pmb W,\epsilon}(\lambda)}=\frac{1}{\pi}{\rm Im}\int dG \mathcal P_{\rm site}(G) G\,.
\label{eq:scdw-ex-density-from-site}
\end{equation}

\exitem[Linearization]
Let
\begin{equation}
\pmb{\mathcal L}=\frac{1}{\sqrt d}\begin{pmatrix}
\pmb 0 & \pmb X\\
\pmb X^{\rm T} & \pmb 0
\end{pmatrix}\,.
\label{eq:scdw-ex-linearization}
\end{equation}
Show that
\begin{equation}
\pmb{\mathcal L}^2=\frac{1}{d}
\begin{pmatrix}
\pmb X\pmb X^{\rm T} & \pmb 0\\
\pmb 0 & \pmb X^{\rm T}\pmb X
\end{pmatrix}\,.
\label{eq:scdw-ex-linearization-square}
\end{equation}
Let $r={\rm rank}\pmb X$. Deduce that the nonzero eigenvalues of $\pmb{\mathcal L}$ are $\pm\sqrt{\nu}$, where $\nu$ runs over the nonzero eigenvalues of $\pmb W$, and that $\pmb{\mathcal L}$ has $N+P-2r$ zero eigenvalues.

\exitem[Dense limit]
Assume $\mathbb E[\xi^2]=1$ and let $d\to\infty$ after the thermodynamic limit. Starting from the population equations, show that the variable Green functions concentrate, that the sums entering the variable update self-average, and derive
\begin{equation}
m(z)=\frac{1}{z-\displaystyle\frac{1}{\alpha}\frac{1}{1-m(z)}}\,.
\label{eq:scdw-ex-dense-limit}
\end{equation}

\exitem[Square Mar\v{c}enko--Pastur law]
Set $\alpha=1$ in the dense-limit equation from the previous exercise. Choose the solution satisfying $m(z)\sim 1/z$ for large $|z|$ and ${\rm Im}\,m(\lambda-i0^+)>0$ on the support. Show that
\begin{equation}
\rho(\lambda)=\frac{1}{2\pi\lambda}\sqrt{(4-\lambda)\lambda}\mathbf 1_{0<\lambda<4}\,.
\label{eq:scdw-ex-square-mp}
\end{equation}

\exitem[First moment check]
Using
\begin{equation}
\frac{1}{N}{\rm Tr}\pmb W=\frac{1}{Nd}\sum_{\mu=1}^{P}\sum_{i=1}^{N}(X_i^\mu)^2\,,
\label{eq:scdw-ex-trace-W}
\end{equation}
show that, in the Poisson diluted ensemble,
\begin{equation}
\int_0^\infty d\lambda \lambda\rho(\lambda) = \frac{\mathbb E[\xi^2]}{\alpha}\,.
\label{eq:scdw-ex-first-moment}
\end{equation}
Specialize to $\mathbb E[\xi^2]=1$.

\exitem[Isolated variable zero modes]
In the Poisson bipartite ensemble, show that the probability that a variable node is isolated is
\begin{equation}
e^{-d/\alpha}\,.
\label{eq:scdw-ex-isolated-variable}
\end{equation}
Explain why each isolated variable contributes a zero eigenvalue to $\pmb W$.

\exitem[Programming exercise: direct simulation]
Fix a list of system sizes $N$, an aspect ratio $\alpha>0$ such that $P=N/\alpha$ is an integer for each $N$, a list of dilutions $d$, a number $S$ of independent samples for each parameter pair, a nonzero-weight distribution $p_\xi$ with zero mean and unit variance, and a common binning or Lorentzian regulator $\epsilon$ for estimating densities. Generate a sparse rectangular matrix
\begin{equation}
X_i^\mu=B_i^\mu\xi_i^\mu\,, \qquad {\rm Prob}(B_i^\mu=1)=\frac{d}{N}\,,\qquad {\rm Prob}(B_i^\mu=0)=1-\frac{d}{N}\,,
\label{eq:scdw-ex-program-support}
\end{equation}
where the nonzero weights $\xi_i^\mu$ are drawn independently from $p_\xi$. Form
\begin{equation}
\pmb W=\frac{1}{d}\pmb X\pmb X^{\rm T}.
\label{eq:scdw-ex-program-W}
\end{equation}
For each value of $N$, $d$, and $\alpha$, compute the empirical spectral density over the $S$ samples. Check positivity of the eigenvalues up to numerical tolerance, normalization of the density, and the first moment $\int d\lambda\,\lambda\rho(\lambda)=1/\alpha$. Report $N$, $P$, $d$, $\alpha$, $S$, the distribution $p_\xi$, the bin width or regulator $\epsilon$, and the numerical tolerance used to identify zero or negative eigenvalues.

\exitem[Programming exercise: two-population dynamics]
Fix $d$, $\alpha$, a regulator $\epsilon>0$, a grid of real values $\lambda$, population sizes for the $G$- and $U$-populations, a burn-in time $T_{\rm burn}$, a number $T_{\rm meas}$ of site samples used for measurement, and a nonzero-weight distribution $p_\xi$. Implement the two-population update defined by \eqref{eq:scdw-self-energy-distribution} and \eqref{eq:scdw-cavity-green-distribution}. For comparison, generate $S$ independent diluted Wishart matrices of size $N_{\rm diag}\times N_{\rm diag}$ with $P=N_{\rm diag}/\alpha$ an integer, using the same $d$, $\alpha$, and $p_\xi$. Compare the population-dynamics density with the direct-diagonalization density of $\pmb W$ using the same $\lambda$ grid and the same regulator $\epsilon$. Report $d$, $\alpha$, $p_\xi$, the population sizes, $T_{\rm burn}$, $T_{\rm meas}$, $N_{\rm diag}$, $P$, $S$, $\epsilon$, the grid spacing, and a discrepancy measure.

\exitem[Programming exercise: approach to Mar\v{c}enko--Pastur]
Fix an aspect ratio $\alpha>0$, a list of increasing dilutions $d$, a nonzero-weight distribution $p_\xi$ with zero mean and unit variance, and a common density-estimation convention. For each $d$, form the diluted Wishart matrix
\begin{equation}
\pmb W=\frac{1}{d}\pmb X\pmb X^{\rm T}
\end{equation}
and compare the empirical spectral density of the rescaled matrix $\widehat{\pmb W}=\alpha\pmb W$, equivalently the eigenvalues $\widehat\lambda=\alpha\lambda$, with the dense Mar\v{c}enko--Pastur law. If the comparison is made by direct diagonalization, report $N$, $P$, the number of samples $S$, and the bin width or regulator. If the comparison is made by population dynamics, report the population sizes, burn-in time, measurement time, regulator, and grid spacing. Track how the distribution of variable Green functions narrows as $d$ increases.
\end{exerciseblock}

\section{Products of Wishart matrices and dense consistency checks}
\label{sec:products-wishart-dense-checks}
The previous section treated a single diluted covariance matrix. A natural extension is to replace one rectangular matrix by a product of rectangular matrices. This leads to product-Wishart ensembles, or equivalently to the squared singular values of products of random rectangular matrices. Products of random matrices have a long history in probability and statistical physics, especially in problems where the number of factors grows and Lyapunov exponents become the natural observables \cite{FurstenbergKesten1960}. The setting considered here is different: the number of factors is fixed, the matrix dimensions diverge, and the observable of interest is the empirical spectral density of a positive semidefinite matrix built from the product. This places the problem in the same family as Wishart random matrix theory, but with a richer multiplicative structure.

Let
\begin{equation}
\pmb{X}_\ell\in\mathbb{R}^{N_{\ell-1}\times N_\ell}\,,\qquad \ell=1,\ldots,L\,,
\label{eq:pwdc-rectangular-factors}
\end{equation}
be independent rectangular random matrices. We define the product
\begin{equation}
\pmb{Y}_L=\pmb{X}_1\pmb{X}_2\cdots\pmb{X}_L\,,
\label{eq:pwdc-product-matrix}
\end{equation}
which is an $N_0\times N_L$ matrix, and we study the positive semidefinite matrix
\begin{equation}
\pmb{W}_L=\frac{1}{N_1N_2\cdots N_L}\pmb{Y}_L\pmb{Y}_L^{\rm T}\,.
\label{eq:pwdc-product-wishart}
\end{equation}
In the dense Gaussian consistency checks below, the entries of the rectangular factors are independent centered Gaussian variables with variance one; with this convention the prefactor in \eqref{eq:pwdc-product-wishart} keeps the eigenvalues of order one. In fact,
\begin{equation}
\frac{1}{N_0}\mathbb E\,{\rm Tr}\,\pmb W_L=1\,,
\label{eq:pwdc-first-moment-normalization}
\end{equation}
so the first moment of the limiting squared-singular-value law is one. Different entry variances would deterministically rescale the spectrum. For $L=1$, this is the standard sample covariance matrix. For $L>1$, its eigenvalues are the squared singular values of a product of rectangular random matrices. This is the appropriate Hermitian object associated with the product. The product of Wishart matrices written naively as $\pmb{W}_1\pmb{W}_2\cdots\pmb{W}_L$ is generally not symmetric unless the factors commute; the object \eqref{eq:pwdc-product-wishart} avoids this ambiguity and is the natural positive operator.

\begin{examplebox}[Dimensions, positivity, and rank in a product-Wishart matrix]
Consider two rectangular matrices
\begin{equation}
\pmb X_1\in\mathbb R^{N_0\times N_1}\,,\qquad\pmb X_2\in\mathbb R^{N_1\times N_2}\,.
\label{eq:pwdc-ped-two-factor-dimensions}
\end{equation}
Their product is
\begin{equation}
\pmb Y_2=\pmb X_1\pmb X_2\in\mathbb R^{N_0\times N_2}\,,
\label{eq:pwdc-ped-Y2}
\end{equation}
and the corresponding product-Wishart matrix is
\begin{equation}
\pmb W_2=\frac{1}{N_1N_2}\pmb Y_2\pmb Y_2^{\rm T}=\frac{1}{N_1N_2}\pmb X_1\pmb X_2\pmb X_2^{\rm T}\pmb X_1^{\rm T}\,.
\label{eq:pwdc-ped-W2}
\end{equation}
This matrix is $N_0\times N_0$ and positive semidefinite. Indeed, for any vector $\pmb v\in\mathbb R^{N_0}$,
\begin{align}
\pmb v^{\rm T}\pmb W_2\pmb v&=\frac{1}{N_1N_2}\pmb v^{\rm T}\pmb Y_2\pmb Y_2^{\rm T}\pmb v\nonumber\\
&=\frac{1}{N_1N_2}\left\|\pmb Y_2^{\rm T}\pmb v\right\|^2\geq 0\,.
\label{eq:pwdc-ped-W2-positivity}
\end{align}
Thus the eigenvalues of $\pmb W_2$ are real and nonnegative, even though the rectangular product $\pmb Y_2$ itself is not square in general.

The rank is bounded by the smallest layer dimension. Since
\begin{equation}
{\rm rank}(\pmb W_2)={\rm rank}(\pmb Y_2\pmb Y_2^{\rm T})={\rm rank}(\pmb Y_2)\,.
\label{eq:pwdc-ped-rank-step1}
\end{equation}
and
\begin{equation}
{\rm rank}(\pmb Y_2)={\rm rank}(\pmb X_1\pmb X_2)\leq\min\{N_0,N_1,N_2\}\,,
\label{eq:pwdc-ped-rank-step2}
\end{equation}
we obtain
\begin{equation}
{\rm rank}(\pmb W_2)\leq\min\{N_0,N_1,N_2\}\,.
\label{eq:pwdc-ped-W2-rank-bound}
\end{equation}
Therefore, if either $N_1$ or $N_2$ is smaller than $N_0$, the product-Wishart matrix has exact zero eigenvalues. This is the finite-dimensional origin of the zero atom discussed later in the dense limit.
\end{examplebox}

In a diluted version, each matrix $\pmb{X}_\ell$ is supported on a sparse bipartite graph connecting layer $\ell-1$ to layer $\ell$. Thus the full product is supported on a multilayer graph. A nonzero entry of $\pmb{Y}_L$ corresponds to a path crossing the layers from a vertex in layer $0$ to a vertex in layer $L$, and an entry of $\pmb{W}_L$ couples two such paths by identifying their endpoint in the last layer. This is the multilayer analogue of the bipartite structure used for sparse covariance matrices. The corresponding Edwards--Jones representation produces a Gaussian model on a multipartite factor graph, and the cavity method leads to recursive equations for messages propagating through the layers. The dense case of this problem was analyzed in detail by Dupic and the author as the first part of a program on products of diluted Wishart matrices \cite{DupicPerezCastillo2014}. In these notes we use the product ensemble mainly as a consistency check: any correct diluted theory must reduce, when the layer connectivities become large, to the known dense product-Wishart laws.

The dense product-Wishart problem is also well understood from free probability and from exact random matrix methods. In the large-dimensional limit, independent Wishart factors become asymptotically free, and the limiting spectral measure of \eqref{eq:pwdc-product-wishart} is the multiplicative free convolution of Mar\v{c}enko--Pastur laws \cite{VoiculescuDykemaNica1992,NicaSpeicher2006}. Exact finite-$N$ results and correlation kernels are known for products of complex Gaussian matrices, where the squared singular values form determinantal point processes with Meijer-$G$ kernels \cite{AkemannKieburgWei2013,AkemannIpsenKieburg2013,KuijlaarsZhang2014}. At the macroscopic level, the same limiting densities were obtained by diagrammatic and free-probability methods for products of square and rectangular Gaussian matrices \cite{BurdaJanikWaclaw2010,BurdaJaroszLivanNowakSwiech2010}. In the square case, the limiting law is the Fuss--Catalan distribution, whose density and moments can be written explicitly \cite{PensonZyczkowski2011}.

We first record the dense rectangular consistency equation. Let
\begin{equation}
\alpha_\ell=\frac{N_0}{N_\ell}\,,\qquad\ell=1,\ldots,L\,,
\label{eq:pwdc-aspect-ratios}
\end{equation}
and assume that these ratios remain finite as all dimensions diverge. Let
\begin{equation}
g_L(z)=\lim_{N_0\to\infty}\frac{1}{N_0}{\rm Tr}(z\pmb{I}-\pmb{W}_L)^{-1}
\label{eq:pwdc-stieltjes-transform}
\end{equation}
be the limiting Stieltjes transform, with $z=\lambda-i\epsilon$ in the convention used throughout these notes. Introduce
\begin{equation}
\mathfrak m_L(z)=z g_L(z)-1\,.
\label{eq:pwdc-m-transform}
\end{equation}
The dense product-Wishart law is encoded by the algebraic equation
\begin{equation}
z=\frac{1+\mathfrak m_L(z)}{\mathfrak m_L(z)}\prod_{\ell=1}^{L}\left[1+\alpha_\ell\mathfrak m_L(z)\right]\,.
\label{eq:pwdc-dense-product-equation}
\end{equation}
Equivalently,
\begin{equation}
z\mathfrak m_L(z)=\left[1+\mathfrak m_L(z)\right]\prod_{\ell=1}^{L}\left[1+\alpha_\ell\mathfrak m_L(z)\right]\,.
\label{eq:pwdc-dense-product-polynomial}
\end{equation}
The edges of the absolutely continuous part are obtained from the branch points of the map
\begin{equation}
z(\mathfrak m)=\frac{1+\mathfrak m}{\mathfrak m}\prod_{\ell=1}^{L}\left[1+\alpha_\ell\mathfrak m\right]\,,
\label{eq:pwdc-rectangular-edge-map}
\end{equation}
namely from the real critical points satisfying $dz/d\mathfrak m=0$ on the physical branch. A possible atom at $\lambda=0$ is fixed separately by the rank constraint discussed below. The physical branch is selected by the condition that $g_L(z)\sim z^{-1}$ as $|z|\to\infty$ and by the sign convention
\begin{equation}
\rho_L(\lambda)=\frac{1}{\pi}\lim_{\epsilon\downarrow0}{\rm Im}[g_L(\lambda-i\epsilon)]\,.
\label{eq:pwdc-density-from-gl}
\end{equation}
For $L=1$, equation \eqref{eq:pwdc-dense-product-equation} reduces to the usual Mar\v{c}enko--Pastur equation with aspect ratio $\alpha_1$. Thus the single covariance matrix is recovered as the first consistency check.

\begin{examplebox}[Recovering the Mar\v{c}enko--Pastur equation from the product formula]
The dense product-Wishart equation should reduce to the usual covariance equation when there is only one rectangular factor. Set $L=1$ in \eqref{eq:pwdc-dense-product-polynomial}. Then
\begin{equation}
z\mathfrak m_1(z)=\left[1+\mathfrak m_1(z)\right]\left[1+\alpha_1\mathfrak m_1(z)\right]\,.
\label{eq:pwdc-ped-L1-product}
\end{equation}
By definition,
\begin{equation}
\mathfrak m_1(z)=z g_1(z)-1\,,
\label{eq:pwdc-ped-m-g-relation}
\end{equation}
so
\begin{equation}
1+\mathfrak m_1(z)=z g_1(z)\,,\qquad 1+\alpha_1\mathfrak m_1(z)=1-\alpha_1+\alpha_1 z g_1(z)\,.
\label{eq:pwdc-ped-L1-factors}
\end{equation}
Substituting into \eqref{eq:pwdc-ped-L1-product} gives
\begin{equation}
z[z g_1(z)-1]=z g_1(z)\left[1-\alpha_1+\alpha_1 z g_1(z)\right]\,.
\label{eq:pwdc-ped-L1-substitution}
\end{equation}
Dividing by $z$ and rearranging,
\begin{equation}
\alpha_1 z g_1(z)^2+(1-\alpha_1-z)g_1(z)+1=0\,.
    \label{eq:pwdc-ped-MP-quadratic}
\end{equation}
This is the usual Mar\v{c}enko--Pastur quadratic equation for the Stieltjes transform, with aspect ratio $\alpha_1=N_0/N_1$. Thus the product-Wishart formula contains the ordinary Wishart ensemble as its first member.
\end{examplebox}

The rectangular structure also fixes the possible atom at the origin. Since
\begin{equation}
{\rm rank}\pmb{W}_L={\rm rank}(\pmb Y_L\pmb Y_L^{\rm T})={\rm rank}\pmb Y_L\leq\min\{N_0,N_1,\ldots,N_L\}\,.
\label{eq:pwdc-rank-bound}
\end{equation}
the limiting spectral measure has a zero-eigenvalue atom whenever at least one of the layers $N_1,\ldots,N_L$ has asymptotic dimension smaller than $N_0$. For full-rank generic rectangular factors, its weight is
\begin{equation}
w_0=\max\left\{0,1-\min_{1\leq \ell\leq L}\frac{N_\ell}{N_0}\right\}=\max\left\{0,1-\frac{1}{\max_{1\leq\ell\leq L}\alpha_\ell}\right\}\,.
    \label{eq:pwdc-zero-atom}
\end{equation}
For $L=1$, this is precisely the zero-mode weight of the Mar\v{c}enko--Pastur law when $N_0>N_1$. This rank check is independent of any detailed random matrix calculation and is therefore a useful diagnostic for normalizations and aspect-ratio conventions.

\begin{examplebox}[Zero modes from a bottleneck layer]
Let
\begin{equation}
N_0=1000\,,\qquad N_1=500\,,\qquad N_2=2000\,.
\label{eq:pwdc-ped-zero-mode-dimensions}
\end{equation}
Then
\begin{equation}
\pmb X_1\in\mathbb R^{1000\times500}\,,\qquad\pmb X_2\in\mathbb R^{500\times2000}\,,\qquad \pmb Y_2=\pmb X_1\pmb X_2\in\mathbb R^{1000\times2000}\,.
\label{eq:pwdc-ped-zero-mode-matrices}
\end{equation}
The rank of $\pmb Y_2$ is at most the smallest layer dimension,
\begin{equation}
{\rm rank}(\pmb Y_2)\leq\min\{1000,500,2000\}=500\,.
\label{eq:pwdc-ped-zero-mode-rank}
\end{equation}
Therefore
\begin{equation}
\pmb W_2=\frac{1}{N_1N_2}\pmb Y_2\pmb Y_2^{\rm T}
\label{eq:pwdc-ped-zero-mode-W}
\end{equation}
has at least $1000-500=500$ zero eigenvalues. The zero-mode weight is therefore at least
\begin{equation}
w_0=\frac{500}{1000}=\frac{1}{2}\,.
\label{eq:pwdc-ped-zero-mode-weight}
\end{equation}
In the notation of the text,
\begin{equation}
\alpha_1=\frac{N_0}{N_1}=2\,,\qquad\alpha_2=\frac{N_0}{N_2}=\frac{1}{2}\,.
\label{eq:pwdc-ped-zero-mode-alphas}
\end{equation}
Hence
\begin{equation}
\max_{\ell}\alpha_\ell=2\,,\qquad 1-\frac{1}{\max_\ell\alpha_\ell}=1-\frac{1}{2}=\frac{1}{2}\,,
\label{eq:pwdc-ped-zero-mode-formula-check}
\end{equation}
which agrees with the rank-counting argument. This example illustrates why rectangular product ensembles can have macroscopic atoms at the origin.
\end{examplebox}

The square case gives a particularly transparent expression. Set
\begin{equation}
N_0=N_1=\cdots=N_L=N\,, \qquad\alpha_1=\cdots=\alpha_L=1\,.
\label{eq:pwdc-square-product}
\end{equation}
Then \eqref{eq:pwdc-dense-product-polynomial} becomes
\begin{equation}
z \mathfrak m_L(z)=\left[1+\mathfrak m_L(z)\right]^{L+1}\,.
\label{eq:pwdc-fuss-catalan-equation}
\end{equation}
The corresponding moments are the Fuss--Catalan numbers of order $L$:
\begin{equation}
\int_0^\infty d\lambda\lambda^n\rho_L(\lambda)=\frac{1}{Ln+1}\binom{(L+1)n}{n}\,,\qquad n=0,1,2,\ldots\,.
\label{eq:pwdc-fuss-catalan-moments}
\end{equation}
For $L=1$, these are the Catalan moments of the square Mar\v{c}enko--Pastur law. The support is compact,
\begin{equation}
0\leq \lambda\leq\lambda_+^{(L)}\,,\qquad\lambda_+^{(L)}=\frac{(L+1)^{L+1}}{L^L}\,.
\label{eq:pwdc-fuss-catalan-support}
\end{equation}

\begin{examplebox}[The square product of two Wishart factors]
For square products with $L=2$, equation \eqref{eq:pwdc-fuss-catalan-equation} becomes
\begin{equation}
z \mathfrak m_2(z)=\left[1+\mathfrak m_2(z)\right]^3\,.
\label{eq:pwdc-ped-L2-square-equation}
\end{equation}
This cubic equation determines the limiting squared singular-value density of a product of two independent square Gaussian matrices.

The upper edge of the support can be obtained directly from the map
\begin{equation}
z(\mathfrak m)=\frac{(1+\mathfrak m)^3}{\mathfrak m}\,.
\label{eq:pwdc-ped-L2-z-map}
\end{equation}
At a spectral edge two branches of the inverse function merge, so
\begin{equation}
\frac{dz}{d\mathfrak m}=0\,.
\label{eq:pwdc-ped-L2-edge-condition}
\end{equation}
Differentiating \eqref{eq:pwdc-ped-L2-z-map},
\begin{align}
\frac{dz}{d\mathfrak m}&=\frac{3(1+\mathfrak m)^2\mathfrak m-(1+\mathfrak m)^3}{\mathfrak m^2}\nonumber\\
&=\frac{(1+\mathfrak m)^2(2\mathfrak m-1)}{\mathfrak m^2}\,.
\label{eq:pwdc-ped-L2-derivative}
\end{align}
The physical positive edge is therefore obtained from
\begin{equation}
2\mathfrak m-1=0\,,\qquad\mathfrak m=\frac{1}{2}\,.
\label{eq:pwdc-ped-L2-m-edge}
\end{equation}
Substituting into \eqref{eq:pwdc-ped-L2-z-map},
\begin{equation}
\lambda_+^{(2)}=z\left(\frac{1}{2}\right)=\frac{(3/2)^3}{1/2}=\frac{27}{4}\,.
\label{eq:pwdc-ped-L2-upper-edge}
\end{equation}
This agrees with the general Fuss--Catalan edge formula
\begin{equation}
\lambda_+^{(L)}=\frac{(L+1)^{L+1}}{L^L}
\label{eq:pwdc-ped-general-edge-check}
\end{equation}
when $L=2$.
\end{examplebox}

A useful parametrization of the density is obtained by setting
\begin{equation}
\lambda(\varphi)=\frac{\sin^{L+1}\left[(L+1)\varphi\right]}{\sin\varphi\sin^L(L\varphi)}\,, \qquad 0<\varphi<\frac{\pi}{L+1}\,.
\label{eq:pwdc-fuss-catalan-parametric-lambda}
\end{equation}
Then
\begin{equation}
\rho_L(\lambda(\varphi))=\frac{1}{\pi\lambda(\varphi)}\frac{\sin\left[(L+1)\varphi\right]\sin\varphi}{\sin(L\varphi)}\,.
\label{eq:pwdc-fuss-catalan-density}
\end{equation}
This formula reduces, for $L=1$, to the square Mar\v{c}enko--Pastur density on $[0,4]$. For general $L$, the density has an integrable singularity at the origin,
\begin{equation}
\rho_L(\lambda)\sim\lambda^{-L/(L+1)} \qquad (\lambda\downarrow0)\,,
\label{eq:pwdc-hard-edge-singularity}
\end{equation}
which becomes stronger as the number of factors increases. This hard-edge behavior is one of the characteristic signatures of products of Wishart matrices.

\begin{examplebox}[The square Mar\v{c}enko--Pastur law from the Fuss--Catalan parametrization]
Set $L=1$ in the parametrization \eqref{eq:pwdc-fuss-catalan-parametric-lambda}. Then
\begin{equation}
\lambda(\varphi)=\frac{\sin^2(2\varphi)}{\sin\varphi\,\sin\varphi}=\frac{[2\sin\varphi\cos\varphi]^2}{\sin^2\varphi}=4\cos^2\varphi\,,\qquad 0<\varphi<\frac{\pi}{2}\,.
\label{eq:pwdc-ped-L1-param-lambda}
\end{equation}
The density formula \eqref{eq:pwdc-fuss-catalan-density} gives
\begin{equation}
\rho_1(\lambda(\varphi))=\frac{1}{\pi\lambda(\varphi)}\frac{\sin(2\varphi)\sin\varphi}{\sin\varphi}
=\frac{1}{\pi\lambda(\varphi)}\sin(2\varphi)\,.
\label{eq:pwdc-ped-L1-param-density-step1}
\end{equation}
Since
\begin{equation}
\lambda=4\cos^2\varphi\,,\qquad\sin(2\varphi)=2\sin\varphi\cos\varphi=\frac{1}{2}\sqrt{\lambda(4-\lambda)}\,,
\label{eq:pwdc-ped-L1-param-identities}
\end{equation}
we find
\begin{equation}
\rho_1(\lambda)=\frac{1}{2\pi\lambda}\sqrt{\lambda(4-\lambda)}\mathbf 1_{0<\lambda<4}.
\label{eq:pwdc-ped-L1-MP-from-FC}
\end{equation}
This is precisely the square Mar\v{c}enko--Pastur law. Thus the Fuss--Catalan family reduces correctly to the ordinary covariance spectrum when the number of factors is one.
\end{examplebox}

The same result should be recovered as the dense-connectivity limit of a diluted multilayer ensemble. Suppose that the entries of each layer are diluted, but that the mean degrees of all layers are subsequently sent to infinity with the appropriate variance scaling. In the cavity equations, the incoming messages then self-average. The distributional order parameter collapses to a deterministic set of layer-dependent Green functions, and the resulting scalar equations combine into \eqref{eq:pwdc-dense-product-equation}. This is the direct analogue of the way the sparse symmetric cavity equations reduce to the Wigner semicircle equation and the diluted covariance equations reduce to the Mar\v{c}enko--Pastur equation. The product case is more algebraically involved because a perturbation entering layer $0$ is propagated through all layers before returning to layer $0$, but the final dense consistency condition is the same multiplicative free-convolution equation.

It is instructive to see how the first two checks appear explicitly. For $L=1$, equation \eqref{eq:pwdc-dense-product-polynomial} gives
\begin{equation}
z\mathfrak m_1(z)=\left[1+\mathfrak m_1(z)\right]\left[1+\alpha_1\mathfrak m_1(z)\right]\,,
\label{eq:pwdc-lone-factor-check}
\end{equation}
which is the Mar\v{c}enko--Pastur equation. For square products with $L=2$, one obtains
\begin{equation}
z\mathfrak m_2(z)=\left[1+\mathfrak m_2(z)\right]^3\,,
\label{eq:pwdc-two-factor-check}
\end{equation}
and the right edge is
\begin{equation}
\lambda_+^{(2)}=\frac{27}{4}\,.
\label{eq:pwdc-two-factor-edge}
\end{equation}
These formulas are simple but important. They fix the scale of the spectrum and therefore provide stringent checks for any future population-dynamics implementation of a diluted product ensemble.

The dense product-Wishart law should also be distinguished from the eigenvalue law of a non-Hermitian product. If the product is square, for instance when $N_0=N_L$, and one studies the normalized product
\begin{equation}
\frac{1}{\sqrt{N_1\cdots N_L}}\pmb{X}_1\pmb{X}_2\cdots\pmb{X}_L
\label{eq:pwdc-nonhermitian-product}
\end{equation}
itself, rather than its squared singular values, the eigenvalues live in the complex plane. For square independent centered Gaussian factors, the limiting eigenvalue density of this normalized product is rotationally invariant and has a power-law radial profile \cite{BurdaJanikWaclaw2010}; rigorous circular-law-type results for products of independent non-Hermitian matrices were obtained in \cite{ORourkeSoshnikov2011}. This non-Hermitian product problem is related but distinct. In the present section the object is the Hermitian positive matrix \eqref{eq:pwdc-product-wishart}, whose spectrum lies on the positive real axis and whose dense law is governed by \eqref{eq:pwdc-dense-product-equation}. This distinction prepares the next section, where non-Hermitian spectra enter through directed sparse operators and Hermitization rather than through the squared singular values of products.

For the purposes of these notes, the main conclusion is that products of Wishart matrices provide a hierarchy of dense consistency checks between ordinary covariance matrices and more elaborate diluted multilayer ensembles. The first member of the hierarchy is the Mar\v{c}enko--Pastur law. The square product hierarchy gives Fuss--Catalan laws. Rectangular products are described by the algebraic equation \eqref{eq:pwdc-dense-product-equation}, with zero modes fixed by the rank constraint \eqref{eq:pwdc-zero-atom}. In finite connectivity, one expects the same qualitative phenomena encountered for sparse covariance matrices: isolated components, zero modes, localized states, and broad distributions of local resolvents. The dense product laws reviewed here are therefore not the end of the story; they are the reference point that any statistical-mechanics theory of diluted product-Wishart matrices must reproduce.

\begin{exerciseblock}
\exitem[Dimensions of the product]
Let
\begin{equation}
\pmb X_\ell\in\mathbb R^{N_{\ell-1}\times N_\ell}\,,\qquad\ell=1,\ldots,L\,.
\label{eq:pwdc-ex-dimensions}
\end{equation}
Show that
\begin{equation}
\pmb Y_L=\pmb X_1\pmb X_2\cdots\pmb X_L
\label{eq:pwdc-ex-YL}
\end{equation}
is an $N_0\times N_L$ matrix and that
\begin{equation}
\pmb W_L=\frac{1}{N_1\cdots N_L}\pmb Y_L\pmb Y_L^{\rm T}
\label{eq:pwdc-ex-WL}
\end{equation}
is an $N_0\times N_0$ positive semidefinite matrix.

\exitem[Positive semidefiniteness]
For an arbitrary vector $\pmb v\in\mathbb R^{N_0}$, prove that
\begin{equation}
\pmb v^{\rm T}\pmb W_L\pmb v=\frac{1}{N_1\cdots N_L}\left\|\pmb Y_L^{\rm T}\pmb v\right\|^2\geq 0\,.
\label{eq:pwdc-ex-positivity}
\end{equation}
Deduce that all eigenvalues of $\pmb W_L$ are real and nonnegative.

\exitem[Rank bound]
Prove that
\begin{equation}
{\rm rank}(\pmb W_L)={\rm rank}(\pmb Y_L\pmb Y_L^{\rm T})={\rm rank}(\pmb Y_L)\leq\min\{N_0,N_1,\ldots,N_L\}\,.
\label{eq:pwdc-ex-rank-bound}
\end{equation}
Use this to show that the fraction of zero eigenvalues is at least
\begin{equation}
w_0=\max\left\{0,1-\min_{1\leq \ell\leq L}\frac{N_\ell}{N_0}\right\}\,.
\label{eq:pwdc-ex-zero-mode-weight}
\end{equation}
For independent Gaussian rectangular factors, explain why the rank bound is saturated with probability one at finite dimensions, so that this lower bound becomes the zero-mode weight in the large-dimensional limit.

\exitem[Aspect ratios and zero modes]
Using
\begin{equation}
\alpha_\ell=\frac{N_0}{N_\ell},
\label{eq:pwdc-ex-alpha}
\end{equation}
show that \eqref{eq:pwdc-ex-zero-mode-weight} can be written as
\begin{equation}
w_0=\max\left\{0,1-\frac{1}{\max_{1\leq\ell\leq L}\alpha_\ell}\right\}\,.
\label{eq:pwdc-ex-zero-mode-alpha}
\end{equation}

\exitem[Single-factor reduction]
Set $L=1$ in the product-Wishart equation
\begin{equation}
z\mathfrak m_L(z)=\left[1+\mathfrak m_L(z)\right]\prod_{\ell=1}^{L}\left[1+\alpha_\ell\mathfrak m_L(z)\right]\,.
\label{eq:pwdc-ex-product-polynomial}
\end{equation}
Using $\mathfrak m_1(z)=zg_1(z)-1$, derive the Mar\v{c}enko--Pastur quadratic equation
\begin{equation}
\alpha_1 z g_1(z)^2+(1-\alpha_1-z)g_1(z)+1=0\,.
\label{eq:pwdc-ex-MP-quadratic}
\end{equation}

\exitem[Square Fuss--Catalan equation]
For square products, $\alpha_1=\cdots=\alpha_L=1$, show that the product equation reduces to
\begin{equation}
z\mathfrak m_L(z)=\left[1+\mathfrak m_L(z)\right]^{L+1}\,.
\label{eq:pwdc-ex-FC-equation}
\end{equation}

\exitem[Upper edge of the square product law]
For the square product law, write
\begin{equation}
z(\mathfrak m)=\frac{(1+\mathfrak m)^{L+1}}{\mathfrak m}\,.
\label{eq:pwdc-ex-edge-map}
\end{equation}
Find the positive critical point of this map and show that
\begin{equation}
\lambda_+^{(L)}=\frac{(L+1)^{L+1}}{L^L}\,.
\label{eq:pwdc-ex-upper-edge}
\end{equation}

\exitem[Two-factor square product]
Set $L=2$ and derive explicitly
\begin{equation}
z\mathfrak m_2(z)=[1+\mathfrak m_2(z)]^3\,.
    \label{eq:pwdc-ex-L2-cubic}
\end{equation}
Use the map $z(\mathfrak m)=(1+\mathfrak m)^3/\mathfrak m$ and the condition $dz/d\mathfrak m=0$ to show that the support ends at
\begin{equation}
\lambda_+^{(2)}=\frac{27}{4}\,.
\label{eq:pwdc-ex-L2-edge}
\end{equation}

\exitem[Fuss--Catalan moments]
For square products, the moments are
\begin{equation}
M_n^{(L)}=\frac{1}{Ln+1}\binom{(L+1)n}{n}\,.
\label{eq:pwdc-ex-FC-moments}
\end{equation}
Compute $M_0^{(L)}$, $M_1^{(L)}$, and $M_2^{(L)}$. Then specialize to $L=1$ and verify that the moments reduce to the Catalan numbers
\begin{equation}
M_n^{(1)}=\frac{1}{n+1}\binom{2n}{n}\,.
\label{eq:pwdc-ex-catalan}
\end{equation}

\exitem[Square Mar\v{c}enko--Pastur law from the parametrization]
Using the domain $0<\varphi<\pi/(L+1)$, set $L=1$ in the parametrization
\begin{equation}
\lambda(\varphi)=\frac{\sin^{L+1}[(L+1)\varphi]}{\sin\varphi\,\sin^L(L\varphi)}\,,
\label{eq:pwdc-ex-param-lambda}
\end{equation}
and in the density formula
\begin{equation}
\rho_L(\lambda(\varphi))=\frac{1}{\pi\lambda(\varphi)}\frac{\sin[(L+1)\varphi]\sin\varphi}{\sin(L\varphi)}\,.
\label{eq:pwdc-ex-param-density}
\end{equation}
Show that one obtains
\begin{equation}
\rho_1(\lambda)=\frac{1}{2\pi\lambda}\sqrt{\lambda(4-\lambda)}\mathbf 1_{0<\lambda<4}\,.
\label{eq:pwdc-ex-square-MP}
\end{equation}

\exitem[Hard-edge singularity]
Using the parametric representation for square products, analyze the limit $\varphi\uparrow\pi/(L+1)$, equivalently $\lambda\downarrow0$, and verify the scaling
\begin{equation}
\rho_L(\lambda)\sim\lambda^{-L/(L+1)}\,.
\label{eq:pwdc-ex-hard-edge}
\end{equation}
What happens to this singularity as $L$ increases?

\exitem[Rectangular zero-mode examples]
For each of the following dimension sequences, compute the zero-mode weight using the rank formula:
\begin{equation}
(N_0,N_1,N_2)=(1000,2000,3000)\,,
\label{eq:pwdc-ex-zero-example1}
\end{equation}
\begin{equation}
(N_0,N_1,N_2)=(1000,500,3000)\,,
\label{eq:pwdc-ex-zero-example2}
\end{equation}
and
\begin{equation}
(N_0,N_1,N_2,N_3)=(1000,1200,800,2000)\,.
\label{eq:pwdc-ex-zero-example3}
\end{equation}

\exitem[Dense product versus non-Hermitian product]
Explain why
\begin{equation}
\pmb W_L=\frac{1}{N_1\cdots N_L}\pmb Y_L\pmb Y_L^{\rm T}
\label{eq:pwdc-ex-hermitian-product}
\end{equation}
has a real nonnegative spectrum, whereas
\begin{equation}
\frac{1}{\sqrt{N_1\cdots N_L}}\pmb X_1\pmb X_2\cdots\pmb X_L
\label{eq:pwdc-ex-nonhermitian-product}
\end{equation}
has, in general, complex eigenvalues when $N_0=N_L$ so that the product is square. Why are these two spectral problems different?

\exitem[Dense consistency from cavity self-averaging]
Consider a diluted multilayer ensemble whose layer connectivities are sent to infinity with the variance scaling that keeps the spectrum of $\pmb W_L$ of order one. Explain, at the level of the law of large numbers for cavity sums, why the layer-dependent message distributions are expected to concentrate around deterministic Green functions. The consistency requirement is that the deterministic dense-limit equations combine to \eqref{eq:pwdc-dense-product-equation}; verify this explicitly for the single-factor case $L=1$ and for the square two-factor case $L=2$.

\exitem[Programming exercise: square product]
Fix a list of matrix sizes $N$, a number $S$ of independent samples for each $N$, and a common density-estimation convention, such as a histogram bin width or Lorentzian regulator. For each $N$ and each sample, generate two independent $N\times N$ Gaussian matrices $\pmb X_1$ and $\pmb X_2$ with independent centered entries of variance one. Form
\begin{equation}
\pmb W_2=\frac{1}{N^2}\pmb X_1\pmb X_2\pmb X_2^{\rm T}\pmb X_1^{\rm T}\,.
\label{eq:pwdc-ex-program-W2}
\end{equation}
Compute the empirical eigenvalue density over the $S$ samples and compare the observed upper edge with $27/4$. Report $N$, $S$, the density-estimation convention, and the edge estimator used in the comparison, for example the largest eigenvalue or a high empirical quantile.

\exitem[Programming exercise: rectangular product and zero modes]
Fix $L$, prescribed dimension sequences $N_0,N_1,\ldots,N_L$, a number $S$ of independent samples for each dimension sequence, and a numerical tolerance $\tau_0$ for identifying zero eigenvalues. Generate rectangular Gaussian matrices $\pmb X_\ell\in\mathbb R^{N_{\ell-1}\times N_\ell}$ with independent centered entries of variance one. Form
\begin{equation}
\pmb W_L=\frac{1}{N_1\cdots N_L}\pmb Y_L\pmb Y_L^{\rm T}\,,\qquad\pmb Y_L=\pmb X_1\cdots\pmb X_L\,.
\label{eq:pwdc-ex-program-rectangular-WL}
\end{equation}
Measure the fraction of eigenvalues below $\tau_0$ and compare with the rank prediction \eqref{eq:pwdc-ex-zero-mode-alpha}. Report $L$, the dimension sequence, $S$, $\tau_0$, and the observed zero-eigenvalue fraction.

\exitem[Programming exercise: moments]
For square products, fix $L$, a list of matrix sizes $N$, a number $S$ of independent samples for each $N$, and a maximal moment order $n_{\max}$. Generate independent $N\times N$ Gaussian matrices $\pmb X_1,\ldots,\pmb X_L$ with independent centered entries of variance one and form
\begin{equation}
\pmb W_L=\frac{1}{N^L}\pmb X_1\cdots\pmb X_L\pmb X_L^{\rm T}\cdots\pmb X_1^{\rm T}\,.
\label{eq:pwdc-ex-program-square-WL}
\end{equation}
For $n=1,\ldots,n_{\max}$, estimate the empirical moments
\begin{equation}
\frac{1}{N}{\rm Tr}\pmb W_L^n
\label{eq:pwdc-ex-program-moments}
\end{equation}
averaged over the $S$ samples and compare them with the Fuss--Catalan moments \eqref{eq:pwdc-ex-FC-moments}. Report $L$, $N$, $S$, and $n_{\max}$.
\end{exerciseblock}

\section{Spectral density of non-Hermitian sparse matrices}
\label{sec:non-hermitian-sparse-spectral-density}
We now consider sparse random matrices without Hermitian symmetry. This class of ensembles is natural whenever the interactions encoded by the matrix are directed, asymmetric, dissipative, or generated by the linearization of non-equilibrium dynamics. In such problems the eigenvalues are generically complex, and the spectral density is a two-dimensional measure in the complex plane. This is already a substantial departure from the Hermitian case. For Hermitian matrices the resolvent is an analytic function away from the real axis, and the density is recovered from the discontinuity of its boundary values. For non-Hermitian matrices the resolvent $(z\pmb{I}-\pmb{A})^{-1}$ is not a stable object near the spectrum in the same sense, and the spectral density is obtained instead from a logarithmic potential or, equivalently, from a Hermitized problem.

Dense non-Hermitian random matrix theory begins with the Ginibre ensembles \cite{Ginibre1965}. For matrices with independent centered entries of variance $1/N$, the empirical spectral distribution converges to the uniform measure on the unit disk, the circular law \cite{Girko1984,TaoVuKrishnapur2010,BordenaveChafai2012AroundCircular}. If correlations are introduced between $A_{ij}$ and $A_{ji}$, the circular support is deformed into an ellipse; this is the elliptic law \cite{SommersCrisantiSompolinskyStein1988}. These dense results are important reference points, but they are not the main object here. In the sparse finite-connectivity regime, each row has only $O(1)$ nonzero entries. The graph structure remains visible in the thermodynamic limit, and the spectral density depends on the local directed neighborhood of a typical vertex. Sparse circular-law results exist when the mean degree diverges with $N$ \cite{Wood2012SparseCircular}, but the finite-connectivity case studied in these notes is governed by a different, graph-local self-consistency problem.

Non-Hermitian sparse matrices also appear naturally in applications. In theoretical ecology and the theory of complex systems, random interaction matrices are used to test the linear stability of large equilibria \cite{May1972,AllesinaTang2012}. In neural network theory, random connectivity matrices determine the spectrum controlling linearized dynamics and transient amplification \cite{RajanAbbott2006}. In all these settings the real part of the rightmost eigenvalue is often the stability-relevant observable, while the full complex spectral density describes the distribution of decay rates and oscillation frequencies. Non-Hermiticity also brings in eigenvector non-orthogonality: left and right eigenvectors are distinct, and their overlaps affect spectral sensitivity and transient response \cite{ChalkerMehlig1998}. The present section focuses on the spectral density itself, but these broader issues are part of the same non-Hermitian random-matrix structure.

Let $\pmb{A}$ be an $N\times N$ sparse non-Hermitian matrix. We write
\begin{equation}
A_{ij}=C_{ij}J_{ij}\,,\qquad A_{ij}\neq A_{ji}\quad\text{in general}\,,
\label{eq:nhsm-sparse-nonhermitian-matrix}
\end{equation}
where $\pmb{C}$ is a directed or partially directed adjacency matrix and $J_{ij}$ are edge weights. It is often useful to regard the underlying graph as the symmetrized support
\begin{equation}
\partial i=\left\{j\neq i: A_{ij}\neq0 \text{ or } A_{ji}\neq0\right\}\,,
\label{eq:nhsm-symmetrized-neighborhood}
\end{equation}
because Hermitization couples vertices $i$ and $j$ whenever at least one of the two directed entries is present. A convenient ensemble is obtained by drawing, for each unordered pair $\{i,j\}$, the pair of entries $(A_{ij},A_{ji})$ from
\begin{equation}
P(A_{ij},A_{ji})=\left(1-\frac{c}{N}\right)\delta(A_{ij})\delta(A_{ji})+\frac{c}{N}p_{J_1,J_2}(A_{ij},A_{ji})\,,\qquad i<j\,.
\label{eq:nhsm-pair-distribution}
\end{equation}
The joint law $p_{J_1,J_2}$ controls the reciprocal correlations. The Hermitian case corresponds to $J_2=J_1^*$. A maximally asymmetric dense-limit convention corresponds to a vanishing reciprocal moment or correlation between $J_1$ and $J_2$, while a strictly one-way sparse support is represented by a law $p_{J_1,J_2}$ supported on pairs for which one of the two entries is zero. In dense limits, the normalized reciprocal correlation is the parameter that interpolates between circular and elliptic laws. At finite connectivity, the same correlation modifies the local message recursion.

\begin{examplebox}[A single pair of vertices: reciprocal versus one-way coupling]
The role of the joint law $p_{J_1,J_2}$ in \eqref{eq:nhsm-pair-distribution} can already be seen for two vertices. Consider
\begin{equation}
\pmb A=\begin{pmatrix}
0 & J\\
K & 0
\end{pmatrix}\,,
\label{eq:nhsm-ped-two-vertex-matrix}
\end{equation}
where $J=A_{12}$ and $K=A_{21}$. The characteristic polynomial is
\begin{equation}
\det(z\pmb I-\pmb A)=\det\begin{pmatrix}
z & -J\\
-K & z
\end{pmatrix}=z^2-JK\,.
\label{eq:nhsm-ped-two-vertex-characteristic}
\end{equation}
Hence the two eigenvalues are
\begin{equation}
z_\pm=\pm\sqrt{JK}\,.
\label{eq:nhsm-ped-two-vertex-eigenvalues}
\end{equation}
Several limiting cases are instructive.

If $K=J^*$, then $JK=|J|^2$ and the eigenvalues are real:
\begin{equation}
z_\pm=\pm |J|\,.
\label{eq:nhsm-ped-hermitian-pair}
\end{equation}
This is the Hermitian reciprocal case. If $J$ and $K$ are real and have the same sign, the eigenvalues are again real. If $J$ and $K$ are real and have opposite signs, then
\begin{equation}
JK<0\,,\qquad z_\pm=\pm i\sqrt{|JK|}\,,
\label{eq:nhsm-ped-opposite-signs}
\end{equation}
so the spectrum lies on the imaginary axis. Finally, if the edge is completely one-way, say $K=0$ and $J\neq0$, then
\begin{equation}
\pmb A=\begin{pmatrix}
0 & J\\
0 & 0
\end{pmatrix}\,,\qquad\pmb A^2=\pmb 0\,,
    \label{eq:nhsm-ped-one-way-edge}
\end{equation}
and both eigenvalues are zero. Thus a directed edge without a reciprocal partner does not by itself create a nonzero eigenvalue.

This example explains why non-Hermitian sparse spectra depend sensitively on reciprocity. The product $A_{ij}A_{ji}$ appears naturally in local spectral recursions outside the support, and the joint distribution of the two directed weights on an unordered pair is therefore part of the model, not a minor detail.
\end{examplebox}

The empirical spectral density in the complex plane is
\begin{equation}
\rho_{\pmb{A}}(z)=\frac{1}{N}\sum_{i=1}^N\delta^{(2)}(z-z_i)\,,\qquad z=x+iy\,,
\label{eq:nhsm-complex-density}
\end{equation}
where $z_1,\ldots,z_N$ are the eigenvalues of $\pmb{A}$. The logarithmic potential is
\begin{equation}
\Phi_{\pmb{A},\eta}(z,z^*)=\frac{1}{N}\log\det\left[(z\pmb{I}-\pmb{A})(z^*\pmb{I}-\pmb{A}^\dagger)+\eta^2\pmb{I}\right]\,,\qquad\eta>0\,.
\label{eq:nhsm-logarithmic-potential}
\end{equation}
Using Girko's Hermitization principle, the spectral density is obtained from
\begin{equation}
\rho_{\pmb{A}}(z)=\frac{1}{\pi}\lim_{\eta\downarrow0}\partial_{z^*}\partial_z\Phi_{\pmb A,\eta}(z,z^*)\,.
    \label{eq:nhsm-density-from-potential}
\end{equation}
Equivalently, since $\partial_{z^*}\partial_z=(1/4)(\partial_x^2+\partial_y^2)$, one may write
\begin{equation}
\rho_{\pmb{A}}(z) =\frac{1}{4\pi}\lim_{\eta\downarrow0}\Delta_z\Phi_{\pmb{A},\eta}(z,z^*)\,,\qquad\Delta_z=\partial_x^2+\partial_y^2\,.
\label{eq:nhsm-density-laplacian}
\end{equation}
The regulator $\eta$ plays the role of a non-Hermitian broadening. It is analogous to the imaginary regulator $\epsilon$ in Hermitian resolvents, but here it regularizes singular values of $z\pmb I-\pmb A$ rather than distances from the real axis.

We now reuse Hermitization in its graph-local form. The determinant in \eqref{eq:nhsm-logarithmic-potential} can be written as the determinant of a doubled matrix, and this doubled structure is what turns scalar cavity messages into $2\times2$ matrix messages. Define the $2N\times2N$ Hermitized block matrix
\begin{equation}
\pmb{\mathcal B}_{\pmb{A}}(z,\eta) =\begin{pmatrix}
i\eta\pmb{I}&z\pmb{I}-\pmb{A}\\
z^*\pmb{I}-\pmb{A}^\dagger&i\eta\pmb{I}
\end{pmatrix}\,.
\label{eq:nhsm-hermitized-matrix}
\end{equation}
Up to a $z$-independent factor and a harmless phase,
\begin{equation}
\det \pmb{\mathcal B}_{\pmb{A}}(z,\eta)=\det\left[(z\pmb{I}-\pmb{A})(z^*\pmb{I}-\pmb{A}^\dagger)+\eta^2\pmb{I}\right]\,.
\label{eq:nhsm-hermitized-determinant}
\end{equation}
The determinant identity and Wirtinger conventions used in Hermitization are summarized in Appendix~\ref{app:gaussian-identities-resolvents}. Thus the non-Hermitian spectral problem is transformed into a Hermitian-type Gaussian problem with two complex fields per vertex. This is the point at which the cavity method can be applied. The price paid for Hermitization is that scalar cavity messages are replaced by $2\times2$ matrices.

\begin{examplebox}[Hermitizing the two-vertex non-Hermitian edge]
Let
\begin{equation}
\pmb A=\begin{pmatrix}
0 & J\\
K & 0
\end{pmatrix}\,.
\label{eq:nhsm-ped-herm-two-vertex-A}
\end{equation}
Then
\begin{equation}
z\pmb I-\pmb A=\begin{pmatrix}
z & -J\\
-K & z
\end{pmatrix}\,.
\label{eq:nhsm-ped-herm-zIA}
\end{equation}
The regularized logarithmic potential involves
\begin{equation}
(z\pmb I-\pmb A)(z^*\pmb I-\pmb A^\dagger)+\eta^2\pmb I\,.
\label{eq:nhsm-ped-herm-positive-matrix}
\end{equation}
For real $J,K$, a direct multiplication gives
\begin{equation}
(z\pmb I-\pmb A)(z^*\pmb I-\pmb A^{\rm T})+\eta^2\pmb I=\begin{pmatrix}
|z|^2+J^2+\eta^2&-zK-Jz^*\\
-Kz^*-zJ&|z|^2+K^2+\eta^2
\end{pmatrix}\,.
\label{eq:nhsm-ped-herm-product}
\end{equation}
The determinant is
\begin{align}
\Delta_\eta(z,z^*)&=\left(|z|^2+J^2+\eta^2\right)\left(|z|^2+K^2+\eta^2\right)-\left|zK+Jz^*\right|^2\,.
\label{eq:nhsm-ped-herm-determinant}
\end{align}
In the limit $\eta\downarrow0$, this determinant vanishes precisely when
\begin{equation}
\det(z\pmb I-\pmb A)=0\,,\qquad\text{i.e.}\qquad z^2=JK\,.
\label{eq:nhsm-ped-herm-zeros}
\end{equation}
Thus Hermitization does not change the eigenvalue problem. Instead, it replaces the non-Hermitian determinant by the determinant of a positive regularized singular-value problem for $z\pmb I-\pmb A$.

In the one-way case $K=0$, equation \eqref{eq:nhsm-ped-herm-determinant} becomes
\begin{equation}
\Delta_\eta(z,z^*)=\left(|z|^2+J^2+\eta^2\right)\left(|z|^2+\eta^2\right)-J^2|z|^2=|z|^4+\eta^2(2|z|^2+J^2)+\eta^4\,.
\label{eq:nhsm-ped-one-way-herm-det}
\end{equation}
At $\eta=0$ this vanishes only at $z=0$, consistently with the nilpotent spectrum of the one-way edge. This illustrates why directed local motifs can have very different spectral effects depending on their reciprocal structure.
\end{examplebox}

Let $\pmb{G}_{i\to j}(z,\eta)$ denote the cavity Green matrix sent from vertex $i$ to vertex $j$. Introduce the local block
\begin{equation}
\pmb{Z}_i(z,\eta)=\begin{pmatrix}
i\eta&z-A_{ii}\\
z^*-A_{ii}^*&i\eta
\end{pmatrix}\,,
\label{eq:nhsm-local-block}
\end{equation}
and the directed edge block
\begin{equation}
\pmb{\mathcal A}_{ij}=\begin{pmatrix}
0 & A_{ij}\\
A_{ji}^* & 0
\end{pmatrix}\,.
\label{eq:nhsm-edge-block}
\end{equation}
In the full Hermitized matrix \eqref{eq:nhsm-hermitized-matrix}, the off-diagonal block connecting $i$ to $j$ is $-\pmb{\mathcal A}_{ij}$, because the off-diagonal entries of $z\pmb I-\pmb A$ are $-A_{ij}$. Since
\begin{equation}
\pmb{\mathcal A}_{ji}=\pmb{\mathcal A}_{ij}^{\dagger}\,,
\end{equation}
the two minus signs cancel in the Schur complement. This is why the recursion below contains $\pmb{\mathcal A}_{i\ell}\pmb G_{\ell\to i}\pmb{\mathcal A}_{i\ell}^{\dagger}$ with no extra minus sign.

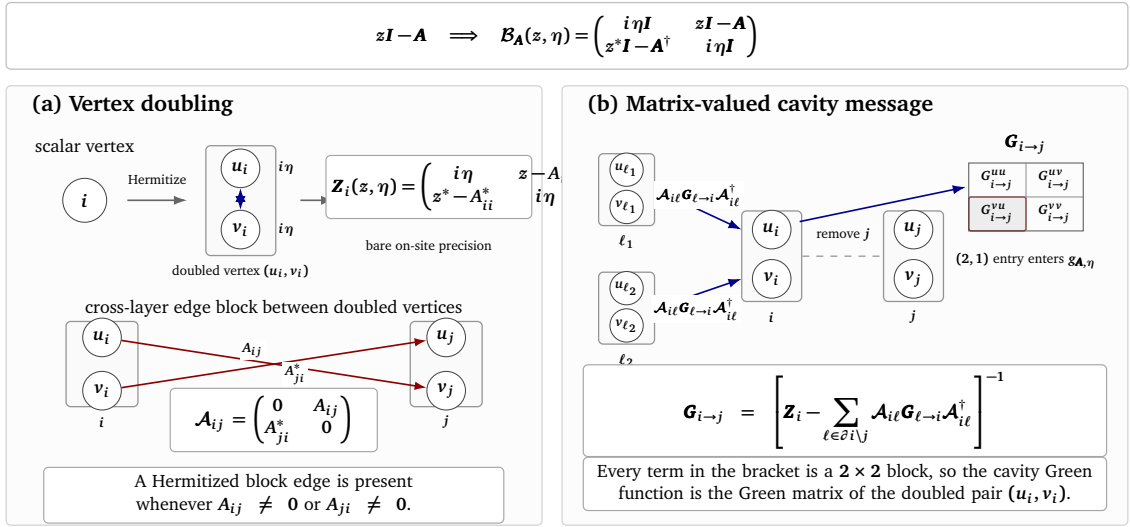
\begin{figure}[t]
\centering
\resizebox{0.99\textwidth}{!}{%
\begin{tikzpicture}[
    x=1cm,
    y=1cm,
    >=Latex,
    panel/.style={draw=black!18, fill=black!1, rounded corners=2pt, line width=0.5pt},
    header/.style={draw=black!25, fill=white, rounded corners=2pt, line width=0.55pt},
    vnode/.style={circle, draw=black!75, fill=white, minimum size=6.6mm, inner sep=0pt, font=\scriptsize},
    comp/.style={circle, draw=black!75, fill=white, minimum size=5.8mm, inner sep=0pt, font=\scriptsize},
    compSmall/.style={circle, draw=black!75, fill=white, minimum size=5.0mm, inner sep=0pt, font=\tiny},
    blockframe/.style={draw=black!42, rounded corners=2pt, line width=0.55pt, fill=black!2},
    onsite/.style={draw=blue!55!black, line width=0.65pt, -{Latex[length=1.8mm,width=1.3mm]}},
    cross/.style={draw=red!55!black, line width=0.70pt, -{Latex[length=1.8mm,width=1.3mm]}},
    flow/.style={draw=black!60, line width=0.7pt, -{Latex[length=2.0mm,width=1.4mm]}},
    dashededge/.style={draw=black!35, dashed, line width=0.65pt},
    msg/.style={draw=blue!55!black, line width=0.70pt, -{Latex[length=2.0mm,width=1.4mm]}},
    ptitle/.style={font=\bfseries\small, anchor=west},
    paneltext/.style={font=\scriptsize, align=center},
    tinytext/.style={font=\tiny, align=center},
    box/.style={draw=black!35, fill=white, rounded corners=2pt, line width=0.5pt, inner sep=3pt, font=\scriptsize, align=center}
]
\draw[header] (0,6.85) rectangle (16.85,7.85);
\node[paneltext] at (8.42,7.35)
{$\displaystyle
z\pmb{I}-\pmb{A}
\quad\Longrightarrow\quad
\pmb{\mathcal B}_{\pmb{A}}(z,\eta)=
\begin{pmatrix}
 i\eta\pmb{I}&z\pmb{I}-\pmb{A}\\
 z^*\pmb{I}-\pmb{A}^{\dagger}&i\eta\pmb{I}
\end{pmatrix}$};

\draw[panel] (0,0) rectangle (8.05,6.60);
\node[ptitle] at (0.25,6.32) {(a) Vertex doubling};

\node[paneltext] at (1.18,5.70) {scalar vertex};
\node[vnode] (orig) at (1.18,4.88) {$i$};

\draw[flow] (1.82,4.88) -- (2.72,4.88);
\node[tinytext] at (2.27,5.20) {Hermitize};

\draw[blockframe] (3.00,4.05) rectangle (4.05,5.70);
\node[comp] (ui) at (3.52,5.36) {$u_i$};
\node[comp] (vi) at (3.52,4.44) {$v_i$};
\node[tinytext, anchor=west] at (3.96,5.37) {$i\eta$};
\node[tinytext, anchor=west] at (3.96,4.44) {$i\eta$};
\draw[onsite, <->] (ui) -- (vi);
\node[tinytext] at (3.52,3.82) {doubled vertex $(u_i,v_i)$};

\draw[flow] (4.42,4.88) -- (4.93,4.88);
\node[box, text width=2.88cm] at (6.35,5.08)
{$\displaystyle
\pmb{Z}_i(z,\eta)=
\begin{pmatrix}
 i\eta&z-A_{ii}\\
 z^*-A_{ii}^*&i\eta
\end{pmatrix}$};
\node[tinytext] at (6.35,4.18) {bare on-site precision};

\node[paneltext] at (4.02,3.25) {cross-layer edge block between doubled vertices};

\draw[blockframe] (0.90,1.78) rectangle (1.98,3.06);
\node[comp] (ui2) at (1.44,2.82) {$u_i$};
\node[comp] (vi2) at (1.44,2.02) {$v_i$};
\node[tinytext] at (1.44,1.55) {$i$};

\draw[blockframe] (6.07,1.78) rectangle (7.15,3.06);
\node[comp] (uj2) at (6.61,2.82) {$u_j$};
\node[comp] (vj2) at (6.61,2.02) {$v_j$};
\node[tinytext] at (6.61,1.55) {$j$};

\draw[cross] (ui2) -- node[pos=0.43, above, tinytext, fill=white, inner sep=1pt] {$A_{ij}$} (vj2);
\draw[cross] (vi2) -- node[pos=0.57, below, tinytext, fill=white, inner sep=1pt] {$A_{ji}^*$} (uj2);

\node[box, text width=2.90cm] at (4.02,1.58)
{$\displaystyle
\pmb{\mathcal A}_{ij}=\begin{pmatrix}
0&A_{ij}\\
A_{ji}^*&0
\end{pmatrix}$};
\node[box, text width=6.70cm] at (4.02,0.45)
{A Hermitized block edge is present whenever $A_{ij}\neq0$ or $A_{ji}\neq0$.};

\draw[panel] (8.35,0) rectangle (16.85,6.60);
\node[ptitle] at (8.60,6.32) {(b) Matrix-valued cavity message};

\draw[blockframe] (8.90,4.50) rectangle (9.74,5.58);
\node[compSmall] (u1) at (9.32,5.34) {$u_{\ell_1}$};
\node[compSmall] (v1) at (9.32,4.78) {$v_{\ell_1}$};
\node[tinytext] at (9.32,4.26) {$\ell_1$};

\draw[blockframe] (8.90,2.72) rectangle (9.74,3.80);
\node[compSmall] (u2) at (9.32,3.56) {$u_{\ell_2}$};
\node[compSmall] (v2) at (9.32,3.00) {$v_{\ell_2}$};
\node[tinytext] at (9.32,2.48) {$\ell_2$};

\draw[blockframe] (11.05,3.36) rectangle (11.95,4.72);
\node[comp] (ui3) at (11.50,4.44) {$u_i$};
\node[comp] (vi3) at (11.50,3.70) {$v_i$};
\node[tinytext] at (11.50,3.13) {$i$};

\draw[blockframe] (13.18,3.36) rectangle (14.08,4.72);
\node[comp] (uj3) at (13.63,4.44) {$u_j$};
\node[comp] (vj3) at (13.63,3.70) {$v_j$};
\node[tinytext] at (13.63,3.13) {$j$};

\draw[msg] (9.74,5.05) -- node[pos=0.51, above, tinytext, fill=white, inner sep=1pt] {$\pmb{\mathcal A}_{i\ell}\pmb{G}_{\ell\to i}\pmb{\mathcal A}_{i\ell}^{\dagger}$} (11.05,4.44);
\draw[msg] (9.74,3.25) -- node[pos=0.48, below, tinytext, fill=white, inner sep=1pt] {$\pmb{\mathcal A}_{i\ell}\pmb{G}_{\ell\to i}\pmb{\mathcal A}_{i\ell}^{\dagger}$} (11.05,3.70);
\draw[dashededge] (11.96,4.04) -- (13.17,4.04);
\node[tinytext] at (12.58,4.36) {remove $j$};

\draw[msg] (11.92,4.56) -- (14.42,5.08);
\node[paneltext] at (15.32,5.72) {$\pmb{G}_{i\to j}$};
\fill[black!7] (14.46,4.42) rectangle (15.32,4.93);
\draw[red!55!black, line width=0.55pt, rounded corners=1pt] (14.46,4.42) rectangle (15.32,4.93);
\draw[black!65, line width=0.45pt] (14.46,4.42) rectangle (16.18,5.44);
\draw[black!45, line width=0.35pt] (15.32,4.42) -- (15.32,5.44);
\draw[black!45, line width=0.35pt] (14.46,4.93) -- (16.18,4.93);
\node[tinytext] at (14.89,5.19) {$G^{uu}_{i\to j}$};
\node[tinytext] at (15.75,5.19) {$G^{uv}_{i\to j}$};
\node[tinytext] at (14.89,4.68) {$G^{vu}_{i\to j}$};
\node[tinytext] at (15.75,4.68) {$G^{vv}_{i\to j}$};
\node[tinytext, text width=2.35cm] at (15.32,3.98) {$(2,1)$ entry enters $g_{\pmb{A},\eta}$};

\node[box, text width=7.65cm] at (12.60,1.72)
{$\displaystyle
\pmb{G}_{i\to j}=
\left[
\pmb{Z}_i-
\sum_{\ell\in\partial i\setminus j}
\pmb{\mathcal A}_{i\ell}\pmb{G}_{\ell\to i}\pmb{\mathcal A}_{i\ell}^{\dagger}
\right]^{-1}$};
\node[box, text width=7.65cm] at (12.60,0.62)
{Every term in the bracket is a $2\times2$ block, so the cavity Green function is the Green matrix of the doubled pair $(u_i,v_i)$.};
\end{tikzpicture}%
}
\caption{Hermitization as vertex doubling and block-valued cavity messaging. The non-Hermitian problem is lifted from one scalar field per vertex to two fields $(u_i,v_i)$ per vertex. The block $\pmb{Z}_i$ is the bare on-site precision of this doubled local pair, while $\pmb{\mathcal A}_{ij}$ gives the cross-layer coupling between doubled neighboring vertices. Consequently the cavity Green function sent from $i$ to $j$ is a $2\times2$ matrix; its lower-left entry is the component averaged to form the regularized non-Hermitian resolvent field.}
\label{fig:nhsm-hermitization-vertex-doubling}
\end{figure}

Figure~\ref{fig:nhsm-hermitization-vertex-doubling} illustrates Hermitization as a vertex-doubling construction: $\pmb{Z}_i$ is the bare on-site block of the doubled pair $(u_i,v_i)$, while $\pmb{\mathcal A}_{ij}$ gives the cross-layer coupling that makes the cavity messages $2\times2$ matrices.

On a tree, the Schur complement gives the exact recursion
\begin{equation}
\pmb{G}_{i\to j}=\left[\pmb{Z}_i-\sum_{\ell\in\partial i\setminus j}\pmb{\mathcal A}_{i\ell}\pmb{G}_{\ell\to i}\pmb{\mathcal A}_{i\ell}^{\dagger}\right]^{-1}\,.
\label{eq:nhsm-cavity-matrix-recursion}
\end{equation}

\begin{examplebox}[Matrix-valued cavity message for a leaf]
Consider a vertex $i$ that is a leaf in the symmetrized support graph, with only one neighbor $j$. The cavity message sent from $i$ to $j$ is computed in the graph where the edge to $j$ has been removed. Therefore the sum in \eqref{eq:nhsm-cavity-matrix-recursion} is empty, and
\begin{equation}
\pmb{G}_{i\to j}=\pmb{Z}_i^{-1}\,.
\label{eq:nhsm-ped-leaf-message}
\end{equation}
For zero diagonal entries this means
\begin{equation}
\pmb{Z}_i=\begin{pmatrix}
i\eta & z\\
z^* & i\eta
\end{pmatrix}\,.
\label{eq:nhsm-ped-leaf-Z}
\end{equation}
The inverse is
\begin{equation}
\pmb{G}_{i\to j}=\frac{1}{(i\eta)^2-|z|^2}\begin{pmatrix}
i\eta & -z\\
-z^* & i\eta
\end{pmatrix}=-\frac{1}{|z|^2+\eta^2}\begin{pmatrix}
i\eta & -z\\
-z^* & i\eta
\end{pmatrix}\,.
\label{eq:nhsm-ped-leaf-message-explicit}
\end{equation}
Thus
\begin{equation}
\left[\pmb{G}_{i\to j}\right]_{21}=\frac{z^*}{|z|^2+\eta^2}\,,
\label{eq:nhsm-ped-leaf-21}
\end{equation}
and outside the spectrum, in the limit $\eta\downarrow0$,
\begin{equation}
\left[\pmb{G}_{i\to j}\right]_{21}\longrightarrow\frac{1}{z}\,.
\label{eq:nhsm-ped-leaf-holomorphic}
\end{equation}
This is the holomorphic resolvent of an isolated zero eigenvalue. The example also shows explicitly why the non-Hermitian message is not a scalar: Hermitization doubles the local degrees of freedom, so even an isolated vertex carries a $2\times2$ Green matrix.
\end{examplebox}

The full local Green matrix at vertex $i$ is
\begin{equation}
\pmb{G}_{i}=\left[\pmb{Z}_i-\sum_{\ell\in\partial i}\pmb{\mathcal A}_{i\ell}\pmb{G}_{\ell\to i}\pmb{\mathcal A}_{i\ell}^{\dagger}\right]^{-1}\,.
\label{eq:nhsm-full-matrix-recursion}
\end{equation}
These equations are the non-Hermitian analogue of the scalar cavity recursions for sparse symmetric matrices. The same local tree logic is used, but each message now carries the doubled structure required by Hermitization. The cavity approach to the spectral density of sparse non-Hermitian matrices was developed in \cite{RogersPerezCastillo2009}; subsequent work derived analytical solutions and gave a systematic account of sparse non-Hermitian spectral theory \cite{NeriMetz2012,MetzNeriRogers2019}.

The scalar object entering the density is the lower-left element of the local matrix Green function. Define
\begin{equation}
g_{\pmb{A},\eta}(z,z^*)=\frac{1}{N}\sum_{i=1}^N\left[\pmb{G}_{i}(z,\eta)\right]_{21}\,.
\label{eq:nhsm-regularized-resolvent}
\end{equation}
For $\eta>0$ this is a regularized, non-holomorphic version of the non-Hermitian resolvent. In the limit $\eta\downarrow0$, it becomes holomorphic outside the spectral support and develops a nonzero $\partial_{z^*}$ derivative inside the support. The density is
\begin{equation}
\rho_{\pmb{A}}(z)=\frac{1}{\pi}\lim_{\eta\downarrow0}\partial_{z^*}g_{\pmb{A},\eta}(z,z^*)\,.
\label{eq:nhsm-density-from-regularized-resolvent}
\end{equation}
For numerical work one may evaluate the derivative by finite differences in the complex plane, or derive companion cavity equations for the response of the messages with respect to $z^*$. The important structural point is that the local matrix recursion \eqref{eq:nhsm-cavity-matrix-recursion} gives direct access to the non-holomorphic resolvent field from which the two-dimensional density follows.

The ensemble-level version of \eqref{eq:nhsm-cavity-matrix-recursion} is a distributional fixed-point equation for $2\times2$ random matrices. For the Poisson ensemble \eqref{eq:nhsm-pair-distribution}, the number of incoming neighbors in a cavity graph is Poisson with mean $c$. Let $\mathcal{P}(\pmb{G})$ be the distribution of a typical cavity message. Then
\begin{equation}
\begin{split}
\mathcal{P}(\pmb{G})&=\sum_{\ell=0}^{\infty}\frac{e^{-c}c^\ell}{\ell!}\int\left[\prod_{r=1}^{\ell}d\pmb{G}_r \mathcal{P}(\pmb{G}_r)dJ_r dK_r p_{J_1,J_2}(J_r,K_r)\right]\\
&\times\delta\left(\pmb{G}-\left[\pmb{Z}-\sum_{r=1}^{\ell}\pmb{\mathcal A}(J_r,K_r)\pmb{G}_r\pmb{\mathcal A}(J_r,K_r)^\dagger\right]^{-1}\right)\,,
\label{eq:nhsm-population-equation}
\end{split}
\end{equation}
where
\begin{equation}
\pmb{Z}=\begin{pmatrix}
i\eta & z\\
z^* & i\eta
\end{pmatrix}\,,\qquad\pmb{\mathcal A}(J,K)=\begin{pmatrix}
0 & J\\
 K^* & 0
\end{pmatrix}\,,
\label{eq:nhsm-population-blocks}
\end{equation}
for zero diagonal entries. In \eqref{eq:nhsm-population-equation}, $p_{J_1,J_2}(J,K)$ is understood as the law of the ordered pair $(A_{i\ell},A_{\ell i})$ entering the message update. If \eqref{eq:nhsm-pair-distribution} is specified only for labelled pairs $i<j$ and the pair law is not exchangeable, the ordered law must be obtained from the corresponding mixture of the two label orientations. If diagonal disorder is present, the update also draws $D=A_{ii}$ and uses
\begin{equation}
\pmb{Z}(D)=\begin{pmatrix}
i\eta & z-D\\
z^*-D^* & i\eta
\end{pmatrix}    
\end{equation}
inside the inverse before averaging. The distribution of the full local Green matrix is obtained by the same equation with the full degree distribution. For Poisson graphs this is again Poisson with mean $c$, while for a general directed or symmetrized degree distribution one must replace it by the appropriate excess-degree and full-degree laws.

In a population-dynamics implementation, $\mathcal{P}$ is represented by a large population of $2\times2$ complex matrices. One update consists of drawing a random number $\ell$ of incoming neighbors, drawing $\ell$ messages from the population, drawing $\ell$ weight pairs $(J_r,K_r)$, and replacing a population member by the matrix inverse in \eqref{eq:nhsm-population-equation}. After equilibration, full local matrices are sampled by including all neighbors, and the regularized resolvent \eqref{eq:nhsm-regularized-resolvent} is estimated by averaging the $(2,1)$ entry. This is precisely the non-Hermitian version of population dynamics for sparse symmetric matrices, with scalar Green functions replaced by matrix-valued messages.

The equations have a useful interpretation in the limit $\eta\downarrow0$. Outside the spectral support, the regularized resolvent becomes holomorphic. In the cavity equations this corresponds to a solution in which the diagonal entries of the $2\times2$ messages vanish, while the off-diagonal entries obey scalar recursions. To see this, set
\begin{equation}
\pmb{G}_{i\to j}=\begin{pmatrix}
0 & b_{i\to j}\\
a_{i\to j} & 0
\end{pmatrix}
\label{eq:nhsm-holomorphic-ansatz}
\end{equation}
at $\eta=0$. Substitution into \eqref{eq:nhsm-cavity-matrix-recursion} gives
\begin{equation}
a_{i\to j} =\frac{1}{z-A_{ii}-\displaystyle\sum_{\ell\in\partial i\setminus j}A_{i\ell}A_{\ell i}a_{\ell\to i}}\,,
\label{eq:nhsm-holomorphic-a-recursion}
\end{equation}
and the companion equation
\begin{equation}
b_{i\to j}=\frac{1}{z^*-A_{ii}^*-\displaystyle\sum_{\ell\in\partial i\setminus j}A_{i\ell}^*A_{\ell i}^* b_{\ell\to i}}\,.    
\end{equation}
In this branch, $g(z)=N^{-1}\sum_i a_i$ is analytic in $z$, and therefore $\rho(z)=0$. The spectral boundary is associated with the loss of stability of this holomorphic solution. Inside the support, the diagonal entries generated by the Hermitization regulator remain finite as $\eta\downarrow0$, the resolvent field becomes non-holomorphic, and the density is nonzero. This mechanism is the sparse analogue of the transition between the holomorphic and non-holomorphic branches in dense non-Hermitian random matrix theory.

\begin{examplebox}[The holomorphic branch for a one-way directed tree]
Consider a directed graph with no reciprocal edges:
\begin{equation}
A_{ij}A_{ji}=0\qquad\text{for every pair } i\neq j\,.
\label{eq:nhsm-ped-no-reciprocal-condition}
\end{equation}
On the holomorphic branch at $\eta=0$, the scalar recursion derived from the matrix equation is
\begin{equation}
a_{i\to j}=\frac{1}{z-A_{ii}-\displaystyle\sum_{\ell\in\partial i\setminus j}A_{i\ell}A_{\ell i}a_{\ell\to i}}\,.
\label{eq:nhsm-ped-holomorphic-recursion}
\end{equation}
For a zero-diagonal ensemble, $A_{ii}=0$. If there are no reciprocal edges, then every product $A_{i\ell}A_{\ell i}$ vanishes, and therefore
\begin{equation}
a_{i\to j}=\frac{1}{z}\,,\qquad a_i=\frac{1}{z}\,.
\label{eq:nhsm-ped-holomorphic-one-way}
\end{equation}
The corresponding holomorphic resolvent field is
\begin{equation}
g(z)=\frac{1}{z}\,.
\label{eq:nhsm-ped-holomorphic-g}
\end{equation}
Since $g(z)$ is holomorphic away from the origin,
\begin{equation}
\partial_{z^*}g(z)=0 \qquad (z\neq0)\,,
\label{eq:nhsm-ped-holomorphic-zero-density}
\end{equation}
and this branch carries no two-dimensional density.

This does not mean that a directed sparse matrix has no nonzero spectrum. It means that outside the spectral support the resolvent is described by a holomorphic solution. The spectral support is determined by the point at which this holomorphic solution becomes unstable and a genuinely non-holomorphic solution with nonzero diagonal Hermitization components takes over. This is the non-Hermitian analogue of selecting the nontrivial branch of the resolvent across the real cut in Hermitian random matrix theory.
\end{examplebox}

Several limiting cases are instructive. If $A_{ij}=A_{ji}^*$, then $\pmb A$ is Hermitian and the complex spectral density collapses onto the real axis. The matrix recursion \eqref{eq:nhsm-cavity-matrix-recursion} reduces to the scalar Hermitian cavity equation after taking the appropriate boundary value of the resolvent. If the mean degree $c$ is sent to infinity and the nonzero weights are scaled as $A_{ij}=O(c^{-1/2})$, the distribution of messages concentrates and the dense circular or elliptic laws are recovered, depending on the reciprocal correlation between $A_{ij}$ and $A_{ji}$. If the directed graph has no reciprocal edges, then the products $A_{i\ell}A_{\ell i}$ vanish on the holomorphic branch, and the outside solution reduces to $g(z)=1/z$ for centered zero-diagonal ensembles; the nontrivial spectral support is again determined by the instability of this branch rather than by the scalar recursion itself.

The finite-connectivity regime is qualitatively different from the dense one. The spectral support need not be a disk or an ellipse. Degree fluctuations, directed local motifs, reciprocal edges, dangling structures, and weight heterogeneity can deform the support and generate localized eigenvectors. The spectrum may contain isolated eigenvalues associated with hubs, source--sink structures, or low-rank mean components. Even when the bulk resembles a circular or elliptic cloud at moderate connectivities, the tails and boundary are controlled by the sparse graph structure. This is why the cavity equations retain the full distribution of local matrix messages rather than reducing immediately to a deterministic resolvent.

The non-Hermitian sparse problem also clarifies why Hermitization is indispensable. Directly iterating $(z\pmb{I}-\pmb{A})^{-1}$ would fail to capture the two-dimensional density, because outside the support the resolvent is holomorphic and inside it the limiting object is non-holomorphic. Hermitization replaces the unstable inverse of $z\pmb{I}-\pmb{A}$ by a regularized singular-value problem for $z\pmb{I}-\pmb{A}$. The cavity equations then describe the local statistics of this singular-value problem on a sparse graph. In this sense, the non-Hermitian spectral density is computed through a family of Hermitian-like problems parametrized by the complex number $z$.

\begin{examplebox}[Dense elliptic boundary from the holomorphic recursion]
The dense elliptic law can be recovered from the sparse cavity equations when the mean degree $c$ is sent to infinity and the real weight pairs are scaled as $A_{ij}=J_{ij}/\sqrt c$. Suppose
\begin{equation}
\mathbb E[J_{ij}]=0\,,\qquad\mathbb E[J_{ij}^2]=\sigma^2\,,\qquad\mathbb E[J_{ij}J_{ji}]=\tau\sigma^2\,,\qquad-1\leq \tau\leq 1\,.
\label{eq:nhsm-ped-elliptic-moments}
\end{equation}
On the holomorphic branch, the dense limit of the recursion has the form
\begin{equation}
g(z)=\frac{1}{z-\tau\sigma^2 g(z)}\,.
    \label{eq:nhsm-ped-elliptic-holomorphic}
\end{equation}
Equivalently,
\begin{equation}
z=\frac{1}{g(z)}+\tau\sigma^2 g(z)\,.
\label{eq:nhsm-ped-elliptic-z-of-g}
\end{equation}
The boundary of the spectral support is obtained when the holomorphic solution loses stability. In the dense elliptic case this occurs at
\begin{equation}
|g(z)|=\frac{1}{\sigma}\,.
\label{eq:nhsm-ped-elliptic-boundary-condition}
\end{equation}
Parametrize
\begin{equation}
g(z)=\frac{e^{-i\theta}}{\sigma}\,,\qquad0\leq\theta<2\pi\,.
\label{eq:nhsm-ped-elliptic-param-g}
\end{equation}
Substituting into \eqref{eq:nhsm-ped-elliptic-z-of-g}, we obtain
\begin{equation}
z(\theta)=\sigma e^{i\theta}+\tau\sigma e^{-i\theta}\,.
\label{eq:nhsm-ped-elliptic-param-z}
\end{equation}
Writing $z=x+iy$, this gives
\begin{equation}
x(\theta)=\sigma(1+\tau)\cos\theta\,,\qquad y(\theta)=\sigma(1-\tau)\sin\theta\,.
\label{eq:nhsm-ped-elliptic-boundary}
\end{equation}
Thus the support is an ellipse with semi-axes
\begin{equation}
\sigma(1+\tau)\qquad\text{and}\qquad\sigma(1-\tau)\,.
\label{eq:nhsm-ped-elliptic-axes}
\end{equation}
For $\tau=0$ one obtains the circular law support, while for $\tau=1$ the ellipse collapses to the Hermitian interval $[-2\sigma,2\sigma]$. This calculation shows how the dense non-Hermitian laws emerge as self-averaging limits of the sparse Hermitized cavity equations. Appendix~\ref{app:dense-limit-reductions} derives these dense circular and elliptic limits directly from the sparse Hermitized cavity equations.
\end{examplebox}

The conclusion is that the non-Hermitian extension preserves the central philosophy of the notes. Sparse spectra are computed from local graphical recursions. The Hermitian case uses scalar cavity Green functions; the non-Hermitian case uses $2\times2$ Hermitized cavity Green matrices. The ensemble-averaged density is obtained by population dynamics over these matrix-valued messages. Dense circular and elliptic laws appear as consistency checks, but the finite-connectivity theory is richer because the directed local topology remains part of the spectral order parameter.

\begin{exerciseblock}
\exitem[Two-vertex non-Hermitian spectrum]
For
\begin{equation}
\pmb A=\begin{pmatrix}
0 & J\\
K & 0
\end{pmatrix}\,,
\label{eq:nhsm-ex-two-vertex}
\end{equation}
compute the characteristic polynomial and show that the eigenvalues are $\pm\sqrt{JK}$. Discuss separately the cases $K=J^*$, $K=0$, and, for real $J$, the cases $K=J$ and $K=-J$.

\exitem[Directed nilpotent chain]
Consider the strictly directed chain
\begin{equation}
\pmb A=\begin{pmatrix}
0 & 1 & 0\\
0 & 0 & 1\\
0 & 0 & 0
\end{pmatrix}\,.
\label{eq:nhsm-ex-directed-chain}
\end{equation}
Show that $\pmb A^3=\pmb 0$ and that all eigenvalues are zero. Explain why directed acyclic structures can contribute nilpotent local pieces to a sparse non-Hermitian matrix.

\exitem[Hermitized determinant identity]
Starting from
\begin{equation}
\pmb{\mathcal B}_{\pmb A}(z,\eta)=\begin{pmatrix}
i\eta\pmb I&z\pmb I-\pmb A\\
z^*\pmb I-\pmb A^\dagger&i\eta\pmb I
\end{pmatrix}\,,
\label{eq:nhsm-ex-hermitized-block}
\end{equation}
use the block determinant formula to prove that, up to a phase independent of $z$,
\begin{equation}
\det \pmb{\mathcal B}_{\pmb A}(z,\eta)=\det\left[(z\pmb I-\pmb A)(z^*\pmb I-\pmb A^\dagger)+\eta^2\pmb I\right]\,.
\label{eq:nhsm-ex-hermitized-det}
\end{equation}

\exitem[Logarithmic potential for a scalar eigenvalue]
For the $1\times1$ matrix $\pmb A=(a)$, with $a\in\mathbb C$, show that
\begin{equation}
\Phi_\eta(z,z^*)=\log\left(|z-a|^2+\eta^2\right)
\label{eq:nhsm-ex-scalar-potential}
\end{equation}
satisfies
\begin{equation}
\frac{1}{\pi}\partial_{z^*}\partial_z\Phi_\eta(z,z^*)=\frac{1}{\pi}\frac{\eta^2}{\left(|z-a|^2+\eta^2\right)^2}\,.
\label{eq:nhsm-ex-scalar-regularized-density}
\end{equation}
Verify that this regularized density integrates to one over the complex plane and converges to $\delta^{(2)}(z-a)$ as $\eta\downarrow0$.

\exitem[Leaf message]
For a leaf vertex with $A_{ii}=0$, compute explicitly
\begin{equation}
\pmb{G}_{i\to j}=\begin{pmatrix}
i\eta & z\\
z^* & i\eta
\end{pmatrix}^{-1}\,.
\label{eq:nhsm-ex-leaf-message}
\end{equation}
Show that the $(2,1)$ entry tends to $1/z$ as $\eta\downarrow0$ away from $z=0$.

\exitem[Deriving the matrix cavity recursion]
Starting from the Schur complement of the Hermitized block matrix on a tree, derive
\begin{equation}
\pmb{G}_{i\to j}=\left[\pmb{Z}_i-\sum_{\ell\in\partial i\setminus j}\pmb{\mathcal A}_{i\ell}\pmb{G}_{\ell\to i}\pmb{\mathcal A}_{i\ell}^{\dagger}\right]^{-1}\,.
\label{eq:nhsm-ex-matrix-cavity}
\end{equation}
Make clear where the local tree assumption is used.

\exitem[Hermitian reduction]
Assume $A_{ij}=A_{ji}^*$ for all $i,j$. Explain how the non-Hermitian Hermitized recursion reduces to the scalar Hermitian cavity recursion after taking the appropriate boundary value of the resolvent on the real axis.

\exitem[Holomorphic branch]
At $\eta=0$, insert the ansatz
\begin{equation}
\pmb{G}_{i\to j}=\begin{pmatrix}
0 & b_{i\to j}\\
a_{i\to j} & 0
\end{pmatrix}
\label{eq:nhsm-ex-holomorphic-ansatz}
\end{equation}
into the matrix recursion and derive
\begin{equation}
a_{i\to j}=\frac{1}{z-A_{ii}-\displaystyle\sum_{\ell\in\partial i\setminus j}A_{i\ell}A_{\ell i}a_{\ell\to i}}\,.
\label{eq:nhsm-ex-holomorphic-recursion}
\end{equation}

\exitem[No reciprocal edges]
Use the recursion in Exercise 10.8 to show that if $A_{i\ell}A_{\ell i}=0$ for all pairs and $A_{ii}=0$, then the holomorphic branch is
\begin{equation}
a_{i\to j}=\frac{1}{z}\,,\qquad a_i=\frac{1}{z}\,.
\label{eq:nhsm-ex-no-reciprocal-holomorphic}
\end{equation}
Why does this not imply that the full non-Hermitian spectral density is zero?

\exitem[Population equation for matrix messages]
For the Poisson sparse non-Hermitian ensemble with pair weight law $p_{J_1,J_2}$, derive the distributional equation
\begin{equation}
\begin{split}
\mathcal{P}(\pmb{G})&=\sum_{\ell=0}^{\infty}e^{-c}\frac{c^\ell}{\ell!}\int\left[\prod_{r=1}^{\ell}d\pmb{G}_r \mathcal{P}(\pmb{G}_r)dJ_r dK_r p_{J_1,J_2}(J_r,K_r)\right]\\
&\times\delta\left(\pmb{G}-\left[\pmb{Z}-\sum_{r=1}^{\ell}\pmb{\mathcal A}(J_r,K_r)\pmb{G}_r\pmb{\mathcal A}(J_r,K_r)^\dagger\right]^{-1}\right)\,.
\label{eq:nhsm-ex-population-equation}
\end{split}
\end{equation}
Here $p_{J_1,J_2}$ is the law of the ordered pair $(A_{i\ell},A_{\ell i})$ entering the update, and, for zero diagonal entries,
\begin{equation}
\pmb{Z}=\begin{pmatrix}
i\eta & z\\
z^* & i\eta
\end{pmatrix}\,,\qquad
\pmb{\mathcal A}(J,K)=\begin{pmatrix}
0 & J\\
K^* & 0
\end{pmatrix}\,.
\end{equation}

\exitem[Regularized density from the non-holomorphic resolvent]
Given
\begin{equation}
g_{\eta}(z,z^*)=\frac{1}{N}\sum_{i=1}^{N}\left[\pmb{G}_i(z,\eta)\right]_{21}\,,
\label{eq:nhsm-ex-regularized-resolvent}
\end{equation}
derive
\begin{equation}
\rho(z)=\frac{1}{\pi}\lim_{\eta\downarrow0}\partial_{z^*}g_{\eta}(z,z^*)\,.
\label{eq:nhsm-ex-density-dzbar}
\end{equation}
Explain why \(g_\eta\) must be non-holomorphic inside the spectral support.

\exitem[Dense circular law boundary]
Set $\tau=0$ in the dense-limit holomorphic equation
\begin{equation}
g(z)=\frac{1}{z-\tau\sigma^2g(z)}\,.
\label{eq:nhsm-ex-dense-holomorphic}
\end{equation}
Use the boundary condition $|g(z)|=1/\sigma$ to show that the support is the circle
\begin{equation}
|z|=\sigma\,.
\label{eq:nhsm-ex-circular-boundary}
\end{equation}

\exitem[Dense elliptic law boundary]
For general $\tau$, use
\begin{equation}
z=\frac{1}{g}+\tau\sigma^2g\,,\qquad|g|=\frac{1}{\sigma}\,,
\label{eq:nhsm-ex-elliptic-boundary-start}
\end{equation}
to derive the ellipse
\begin{equation}
x=\sigma(1+\tau)\cos\theta\,,\qquad y=\sigma(1-\tau)\sin\theta\,.
\label{eq:nhsm-ex-elliptic-boundary}
\end{equation}
Check the limiting cases $\tau=0$ and $\tau=1$.

\exitem[Directed support and symmetrized neighborhood]
Give an example of a directed graph in which $A_{ij}\neq0$ but $A_{ji}=0$. Explain why the Hermitized cavity recursion still treats $i$ and $j$ as neighbors in the symmetrized support.

\exitem[Programming exercise: direct diagonalization of sparse directed matrices]
Fix a list of matrix sizes $N$, a list of mean degrees $c$, and a number $S$ of independent samples for each pair $(N,c)$. Generate sparse directed Erd\H{o}s--R\'enyi matrices with zero diagonal entries and independent off-diagonal entries
\begin{equation}
{\rm Prob}(A_{ij}=1)=\frac{c}{N}\,,\qquad{\rm Prob}(A_{ij}=0)=1-\frac{c}{N}\,,\qquad i\neq j\,.
\label{eq:nhsm-ex-program-directed-er}
\end{equation}
For each sample, plot the eigenvalues of $\pmb A/\sqrt c$ in the complex plane. Compare the bulk cloud with the unit-disk circular-law intuition as $c$ becomes large, and record separately any real outlier associated with the nonzero mean of the Bernoulli entries. Report $N$, $c$, $S$, and the plotting or density-estimation convention used.

\exitem[Programming exercise: reciprocal correlations]
Fix a list of matrix sizes $N$, a mean degree $c$, a list of correlation parameters $\tau\in[-1,1]$, and a number $S$ of independent samples for each parameter pair. For each unordered pair $\{i,j\}$, draw an edge pair with probability $c/N$. Conditional on the pair being present, draw independent standard Gaussian variables $\xi,\zeta$ and set
\[
J=\xi\,,\qquad K=\tau\xi+\sqrt{1-\tau^2}\,\zeta\,,
\]
so that
\begin{equation}
\mathbb E[J]=\mathbb E[K]=0\,,\qquad \mathbb E[J^2]=\mathbb E[K^2]=1\,,\qquad\mathbb E[JK]=\tau\,.
\label{eq:nhsm-ex-program-pair-moments}
\end{equation}
Then set
\begin{equation}
A_{ij}=\frac{J}{\sqrt c}\,,\qquad A_{ji}=\frac{K}{\sqrt c}\,.
\label{eq:nhsm-ex-program-reciprocal-entries}
\end{equation}
Vary $\tau$ and observe how the eigenvalue cloud changes shape. For large $c$, compare qualitatively with the dense elliptic boundary
\[
x=(1+\tau)\cos\theta\,,\qquad y=(1-\tau)\sin\theta\,.
\]
Report $N$, $c$, $S$, the values of $\tau$, and the plotting or density-estimation convention used.

\exitem[Programming exercise: Hermitized resolvent field]
Fix a small non-Hermitian matrix $\pmb A$, a regulator $\eta>0$, a rectangular grid of points $z=x+iy$ in the complex plane, and a finite-difference step $h$ for estimating Wirtinger derivatives. Compute
\begin{equation}
g_\eta(z,z^*)=\frac{1}{N}{\rm Tr}\left[\left(\pmb{\mathcal B}_{\pmb A}(z,\eta)^{-1}\right)_{21}\right]
\label{eq:nhsm-ex-program-hermitized-resolvent}
\end{equation}
on the grid, approximate $\partial_{z^*}g_\eta$ by finite differences, form
\begin{equation}
\rho_\eta(z)=\frac{1}{\pi}\partial_{z^*}g_\eta(z,z^*)\,,
\end{equation}
and compare this regularized density with a smoothed histogram of the eigenvalues using a smoothing scale comparable to $\eta$. Here $(\cdots)_{21}$ denotes the lower-left $N\times N$ block of the $2N\times2N$ inverse matrix. Report $\eta$, the grid spacings in $x$ and $y$, the finite-difference step $h$, and the smoothing convention used for the eigenvalue histogram.
\end{exerciseblock}

\section{Generalized diluted Wishart and cross-correlation ensembles}
\label{sec:generalized-diluted-wishart}
The diluted Wishart ensemble studied in Section~\ref{sec:sparse-covariance-diluted-wishart} is obtained from a single sparse rectangular matrix $\pmb X$ through the positive semidefinite product $d^{-1}\pmb X\pmb X^{\rm T}$. This construction is natural for covariance matrices, but it is not the most general finite-connectivity matrix built from rectangular data. In many problems one is interested in correlations between two different sets of measurements, or in the symmetric part of a cross-correlation matrix constructed from two related data arrays. In dense random matrix theory, empirical covariance and correlation matrices have long been used as null models for high-dimensional data, especially in the analysis of financial cross-correlations, where the Mar\v{c}enko--Pastur law separates a random bulk from a small number of informative modes \cite{LalouxCizeauBouchaudPotters1999,PlerouGopikrishnanRosenowAmaralStanley1999}. The ensemble introduced in this section is a sparse finite-connectivity analogue of that idea: two diluted rectangular matrices are placed on the same bipartite graph, their entries may be correlated on each occupied edge, and the object of interest is the symmetric cross-correlation matrix generated by them. Thus this section returns to Hermitian spectra, but it keeps the bipartite finite-connectivity viewpoint developed for diluted Wishart matrices and prepares the cross-correlation structure that will reappear later in non-Hermitian extensions.

Let $\pmb{X}$ and $\pmb{Y}$ be two real $N\times P$ rectangular matrices. We keep
\begin{equation}
P=\frac{N}{\alpha}\,,\qquad\alpha>0\,,
\label{eq:gdw-rectangularity}
\end{equation}
fixed in the thermodynamic limit. The generalized diluted Wishart, or symmetric diluted cross-correlation, matrix is defined by
\begin{equation}
\pmb{F}=\frac{1}{2d}\left(\pmb{X}\pmb{Y}^{\rm T}+\pmb{Y}\pmb{X}^{\rm T}\right)\,,
\label{eq:gdw-definition}
\end{equation}
or, in components,
\begin{equation}
F_{ij}=\frac{1}{2d}\sum_{\mu=1}^P\left(x_i^\mu y_j^\mu+y_i^\mu x_j^\mu\right)\,.
\label{eq:gdw-entrywise-definition}
\end{equation}

\begin{examplebox}[A two-variable cross-correlation matrix]
The generalized diluted Wishart matrix is symmetric by construction, but it is not necessarily positive semidefinite. This can be seen in the smallest nontrivial example. Take
\begin{equation}
N=2\,,\qquad P=1\,, \qquad d=1\,,
\label{eq:gdw-ped-N2P1}
\end{equation}
so that
\begin{equation}
\pmb X=\begin{pmatrix}
x_1\\
x_2
\end{pmatrix}\,,\qquad \pmb Y=\begin{pmatrix}
y_1\\
y_2
\end{pmatrix}\,.
\label{eq:gdw-ped-two-vectors}
\end{equation}
Then
\begin{equation}
\pmb F=\frac{1}{2}\left(\pmb X\pmb Y^{\rm T}+\pmb Y\pmb X^{\rm T}\right)=\begin{pmatrix}
x_1y_1&\frac{x_1y_2+y_1x_2}{2}\\
\frac{x_1y_2+y_1x_2}{2}&x_2y_2
\end{pmatrix}\,.
\label{eq:gdw-ped-two-variable-F}
\end{equation}
If $\pmb Y=\pmb X$, then
\begin{equation}
\pmb F=\pmb X\pmb X^{\rm T}\,,
\label{eq:gdw-ped-positive-limit}
\end{equation}
which is positive semidefinite. For example,
\begin{equation}
\pmb X=\begin{pmatrix}
1\\
1
\end{pmatrix}\,,\qquad
\pmb Y=\begin{pmatrix}
1\\
1
\end{pmatrix}\quad\Longrightarrow\quad\pmb F=\begin{pmatrix}
1 & 1\\
1 & 1
\end{pmatrix}\,,
\label{eq:gdw-ped-positive-example}
\end{equation}
whose eigenvalues are $2$ and $0$.

If instead
\begin{equation}
\pmb X=\begin{pmatrix}
1\\
1
\end{pmatrix}\,,\qquad\pmb Y=\begin{pmatrix}
1\\
-1
\end{pmatrix}\,,
\label{eq:gdw-ped-indefinite-choice}
\end{equation}
then
\begin{equation}
\pmb F=\frac{1}{2}\left[\begin{pmatrix}
1 & -1\\
1 & -1
\end{pmatrix}+
\begin{pmatrix}
1 & 1\\
-1 & -1
\end{pmatrix}\right]=
\begin{pmatrix}
1 & 0\\
0 & -1
\end{pmatrix}\,.
\label{eq:gdw-ped-indefinite-F}
\end{equation}
The eigenvalues are $1$ and $-1$. Thus the symmetric cross-correlation matrix is real symmetric, but it is not generically positive semidefinite. Positivity is a special feature of the perfectly correlated Wishart limit $\pmb X=\pmb Y$.
\end{examplebox}

The matrix $\pmb{F}$ is real symmetric by construction, but it is not positive semidefinite in general. Positivity is recovered only in special cases, most notably when $\pmb{X}=\pmb{Y}$, in which case
\begin{equation}
\pmb{F}=\frac{1}{d}\pmb{X}\pmb{X}^{\rm T}\,.
\label{eq:gdw-wishart-limit}
\end{equation}
Thus the ordinary diluted Wishart ensemble is contained as a limiting case. The general ensemble \eqref{eq:gdw-definition} is broader because it allows the two rectangular matrices to be partially correlated, independent, or anticorrelated.

The dilution is imposed at the level of the rectangular matrices. For each pair $(i,\mu)$, the two entries $(x_i^\mu,y_i^\mu)$ are jointly distributed as
\begin{equation}
P(x_i^\mu,y_i^\mu)=\left(1-\frac{d}{N}\right)\delta(x_i^\mu)\delta(y_i^\mu)+\frac{d}{N}
\varrho(x_i^\mu,y_i^\mu)\,,
\label{eq:gdw-entry-distribution}
\end{equation}
where $\varrho(x,y)$ denotes the law of the nonzero pair of weights; when this law is absolutely continuous we identify it with its density, while the Wishart and negative-Wishart reductions below correspond to singular laws supported on lines such as $y=x$ and $y=-x$. Equivalently, the matrices $\pmb X$ and $\pmb Y$ live on a common sparse bipartite graph connecting variable nodes $i=1,\ldots,N$ to factor nodes $\mu=1,\ldots,P$, and each occupied bipartite edge carries two weights. The average degree of a factor node is $d$, while the average degree of a variable node is $d/\alpha$. This ensemble was introduced and analyzed in \cite{PerezCastillo2022Generalized} as a generalized diluted Wishart ensemble.

The law $\varrho$ controls the correlation between the entries of $\pmb{X}$ and $\pmb{Y}$ on the same occupied edge. It is useful to introduce the moments
\begin{equation}
m_{ab}=\int dx dy  \varrho(x,y)x^a y^b\,,
\label{eq:gdw-weight-moments}
\end{equation}
with the same notation understood in the distributional sense for singular laws. When $m_{20}$ and $m_{02}$ are finite and nonzero, define
\begin{equation}
\Gamma=\frac{m_{11}}{\sqrt{m_{20}m_{02}}}\,.
\label{eq:gdw-entry-correlation}
\end{equation}
The case $\Gamma=1$ corresponds, after a possible positive rescaling, to perfectly correlated entries and hence to a positive diluted Wishart-type matrix. The case $\Gamma=-1$ corresponds, after a possible positive rescaling, to perfectly anticorrelated entries and gives a negative Wishart-type structure. The case $\Gamma=0$ corresponds to a vanishing local cross-moment; when the first moments vanish, this is local uncorrelatedness. For intermediate values, $\pmb{F}$ is a genuine symmetric cross-correlation matrix whose spectrum may have both positive and negative eigenvalues.

The trace already shows the role of the local correlation. Since
\begin{equation}
F_{ii}=\frac{1}{d}\sum_{\mu=1}^Px_i^\mu y_i^\mu\,,
\label{eq:gdw-diagonal-entry}
\end{equation}
one has
\begin{equation}
\frac{1}{N}{\rm Tr}\pmb{F}=\frac{1}{dN}\sum_{i=1}^N\sum_{\mu=1}^Px_i^\mu y_i^\mu\,.
\label{eq:gdw-trace}
\end{equation}
For the Poisson bipartite ensemble \eqref{eq:gdw-entry-distribution}, and assuming sufficient integrability for the law of large numbers, for instance $m_{22}<\infty$, this converges in probability to
\begin{equation}
\lim_{N\to\infty}\frac{1}{N}{\rm Tr}\pmb{F}=\frac{m_{11}}{\alpha}\,.
\label{eq:gdw-first-moment}
\end{equation}
Thus a zero local cross-moment $m_{11}=0$ gives a limiting density with vanishing first moment, while the Wishart case $x=y$ with $m_{20}=1$ gives the first moment $1/\alpha$ in the normalization used here.

\begin{examplebox}[Trace and local cross-correlation]
The first moment of the spectral density is fixed by the trace. Starting from
\begin{equation}
\pmb F=\frac{1}{2d}\left(\pmb X\pmb Y^{\rm T}+\pmb Y\pmb X^{\rm T}\right)\,,
\label{eq:gdw-ped-trace-start}
\end{equation}
we get
\begin{equation}
F_{ii}=\frac{1}{2d}\sum_{\mu=1}^{P}\left(x_i^\mu y_i^\mu+y_i^\mu x_i^\mu\right)=\frac{1}{d}\sum_{\mu=1}^{P}x_i^\mu y_i^\mu\,.
\label{eq:gdw-ped-diagonal-trace}
\end{equation}
Therefore
\begin{equation}
\frac{1}{N}{\rm Tr}\pmb F=\frac{1}{dN}\sum_{i=1}^{N}\sum_{\mu=1}^{P}x_i^\mu y_i^\mu\,.
\label{eq:gdw-ped-trace-average}
\end{equation}
For the diluted law
\begin{equation}
P(x_i^\mu,y_i^\mu)=\left(1-\frac{d}{N}\right)\delta(x_i^\mu)\delta(y_i^\mu)+\frac{d}{N}\varrho(x_i^\mu,y_i^\mu)\,,
\label{eq:gdw-ped-entry-law-recall}
\end{equation}
one has
\begin{equation}
\mathbb E[x_i^\mu y_i^\mu]=\frac{d}{N}\int dx dy \varrho(x,y)xy=\frac{d}{N}m_{11}\,.
\label{eq:gdw-ped-local-cross-moment}
\end{equation}
There are $NP$ possible rectangular entries and $P=N/\alpha$. Hence
\begin{equation}
\mathbb E\left[\frac{1}{N}{\rm Tr}\pmb F\right]=\frac{1}{dN}NP\frac{d}{N}m_{11}=\frac{m_{11}}{\alpha}\,.
\label{eq:gdw-ped-trace-result}
\end{equation}
This computation explains why the local cross-moment $m_{11}$ controls the center of the spectrum.
\end{examplebox}

Another structural constraint is the rank. Since
\begin{equation}
{\rm rank}\left(\pmb{X}\pmb{Y}^{\rm T}+\pmb{Y}\pmb{X}^{\rm T}\right)\leq{\rm rank}(\pmb{X}\pmb{Y}^{\rm T})+{\rm rank}(\pmb{Y}\pmb{X}^{\rm T})\leq 2P\,,
\label{eq:gdw-rank-bound}
\end{equation}
the zero-mode weight $w_0$ satisfies the lower bound
\begin{equation}
w_0\geq\max\left\{ 0,1-\frac{2}{\alpha}\right\}
\label{eq:gdw-zero-mode-bound}
\end{equation}
for any such rectangular construction. In the dense generic case this bound is saturated, before taking into account additional zero modes generated by dilution. In the Wishart limit $\pmb{X}=\pmb{Y}$, the stronger bound ${\rm rank}\pmb{F}\leq P$ gives the usual zero-mode weight $\max\{0,1-1/\alpha\}$. At finite connectivity, isolated variable nodes and finite bipartite components can generate further zero modes or singular spectral contributions. This is one of the ways in which the sparse ensemble differs from the dense cross-correlation problem.

\begin{examplebox}[Rank bounds in the generalized and Wishart limits]
The generalized matrix has a different rank bound from the ordinary Wishart matrix. Take $N=4$ and $P=1$. Then
\begin{equation}
\pmb X=\pmb x\,,\qquad \pmb Y=\pmb y\,,
\label{eq:gdw-ped-rank-vectors}
\end{equation}
with $\pmb x,\pmb y\in\mathbb R^4$, and
\begin{equation}
\pmb F=\frac{1}{2d}\left(\pmb x\pmb y^{\rm T}+\pmb y\pmb x^{\rm T}\right)\,.
\label{eq:gdw-ped-rank-one-factor}
\end{equation}
Each term $\pmb x\pmb y^{\rm T}$ and $\pmb y\pmb x^{\rm T}$ has rank at most one, so
\begin{equation}
{\rm rank}\pmb F\leq 2\,.
\label{eq:gdw-ped-rank-at-most-two}
\end{equation}
For generic linearly independent $\pmb x$ and $\pmb y$, the rank is exactly two, so $\pmb F$ has at least $N-2=2$ zero eigenvalues.

In the Wishart limit $\pmb y=\pmb x$, however,
\begin{equation}
\pmb F=\frac{1}{d}\pmb x\pmb x^{\rm T}\,,
\label{eq:gdw-ped-rank-wishart-limit}
\end{equation}
which has rank one. Therefore the Wishart limit has at least $N-1=3$ zero eigenvalues in this example. This is the finite-dimensional version of the distinction between the generic bound
\begin{equation}
{\rm rank}\pmb F\leq 2P
\label{eq:gdw-ped-rank-2P}
\end{equation}
and the stronger Wishart bound
\begin{equation}
{\rm rank}\pmb F\leq P\,.
\label{eq:gdw-ped-rank-P}
\end{equation}
\end{examplebox}

The resolvent is
\begin{equation}
\pmb{G}(z)=(z\pmb{I}-\pmb{F})^{-1}\,,\qquad z=\lambda-i\epsilon\,,\qquad\epsilon>0\,,
\label{eq:gdw-resolvent}
\end{equation}
and the regularized spectral density is
\begin{equation}
\rho_{\pmb{F},\epsilon}(\lambda)=\frac{1}{\pi N}\sum_{i=1}^N{\rm Im}[G_{ii}(\lambda-i\epsilon)]\,.
\label{eq:gdw-regularized-density}
\end{equation}
Let us repeat the Edwards--Jones construction in this generalized bipartite setting. The Gaussian structure is the same as in the diluted Wishart case, but a factor node now contributes a rank-at-most-two symmetric interaction rather than a rank-one covariance term. The quadratic form entering the Edwards--Jones representation is
\begin{equation}
\pmb{u}^{\rm T}\pmb{F}\pmb{u}=\frac{1}{d}\sum_{\mu=1}^P\left(\sum_{i\in\partial\mu}x_i^\mu u_i\right)\left(\sum_{j\in\partial\mu}y_j^\mu u_j\right)\,.
\label{eq:gdw-quadratic-form}
\end{equation}
Therefore the Gaussian partition function is
\begin{equation}
Z_{\pmb{F}}(z)=\int\left[\prod_{i=1}^N\frac{du_i}{\sqrt{2\pi}}\right]\exp\left[ -\frac{i}{2}z\sum_{i=1}^N u_i^2+\frac{i}{2d}\sum_{\mu=1}^P\left(\sum_{i\in\partial\mu}x_i^\mu u_i\right)\left(\sum_{j\in\partial\mu}y_j^\mu u_j\right)\right]\,.
\label{eq:gdw-gaussian-partition-function}
\end{equation}
This is a Gaussian model on a sparse bipartite factor graph. Each factor node contributes a rank-at-most-two symmetric quadratic interaction on its neighboring variable nodes. In the Wishart limit $x_i^\mu=y_i^\mu$, this contribution collapses to the rank-one factor already encountered in the diluted Wishart ensemble.

We now derive the cavity equations. Let $G_{i\to\mu}(z)$ denote the cavity Green function of variable node $i$ when factor node $\mu$ is removed. The contribution of factor $\mu$ to the matrix $\pmb F$ on the variables in $\partial\mu$ is
\begin{equation}
B_{ij}^{\mu}=\frac{1}{2d}\left(x_i^\mu y_j^\mu+y_i^\mu x_j^\mu\right)\,,\qquad i,j\in\partial\mu\,.
    \label{eq:gdw-factor-matrix}
\end{equation}
Fix a variable $i\in\partial\mu$ and denote by $R=\partial\mu\setminus i$ the remaining variables attached to factor $\mu$. If the incoming cavity Green functions from the variables in $R$ are $G_{j\to\mu}$, then the diagonal matrix of cavity inverse Green functions is
\begin{equation}
\pmb{D}_{\mu\to i}={\rm diag}\left(G_{j\to\mu}^{-1}\right)_{j\in R}\,.
\label{eq:gdw-cavity-inverse-matrix}
\end{equation}
By the Schur complement, the self-energy sent from factor $\mu$ to variable $i$ is
\begin{equation}
U_{\mu\to i}=B_{ii}^{\mu}+\pmb{B}_{iR}^{\mu}\left(\pmb{D}_{\mu\to i}-\pmb{B}_{RR}^{\mu}\right)^{-1}\pmb{B}_{Ri}^{\mu}\,.
\label{eq:gdw-self-energy-schur}
\end{equation}
This expression is exact on a tree factor graph. Since $\pmb B_{RR}^{\mu}$ has rank at most two, it can be reduced to a scalar formula involving only three cavity sums.

Define
\begin{equation}
S_{xx}^{\mu\to i}=\sum_{j\in\partial\mu\setminus i}(x_j^\mu)^2G_{j\to\mu}\,,\qquad S_{xy}^{\mu\to i}=\sum_{j\in\partial\mu\setminus i} x_j^\mu y_j^\mu G_{j\to\mu}\,,\qquad S_{yy}^{\mu\to i}=\sum_{j\in\partial\mu\setminus i}(y_j^\mu)^2G_{j\to\mu}\,.
\label{eq:gdw-cavity-sums}
\end{equation}
Using the rank-two form of \eqref{eq:gdw-factor-matrix}, the denominator generated by the Schur complement is
\begin{equation}
\Delta_{\mu\to i}=\left(2d-S_{xy}^{\mu\to i}\right)^2-S_{xx}^{\mu\to i}S_{yy}^{\mu\to i}\,.
\label{eq:gdw-delta}
\end{equation}
The factor-to-variable self-energy is then
\begin{equation}
U_{\mu\to i}(z)=\frac{4d x_i^\mu y_i^\mu+(y_i^\mu)^2S_{xx}^{\mu\to i}-2x_i^\mu y_i^\mu S_{xy}^{\mu\to i}+(x_i^\mu)^2S_{yy}^{\mu\to i}}{\left(2d-S_{xy}^{\mu\to i}\right)^2-S_{xx}^{\mu\to i}S_{yy}^{\mu\to i}}\,.
\label{eq:gdw-factor-to-variable-self-energy}
\end{equation}
Here $U_{\mu\to i}$ denotes the full Schur-complement contribution entering the denominator of a variable Green function. This differs from the convention of Section~\ref{sec:sparse-covariance-diluted-wishart}, where $U_{\mu\to i}^{(\rm Wishart)}$ is defined before the overall $d^{-1}$ factor and the actual denominator contribution is $d^{-1}U_{\mu\to i}^{(\rm Wishart)}$. The finite-rank Gaussian identity underlying this rank-at-most-two factor contribution is collected in Appendix~\ref{app:gaussian-identities-resolvents}.

\begin{examplebox}[Reduction of the rank-two self-energy to the Wishart self-energy]
The generalized factor-to-variable self-energy is
\begin{equation}
U_{\mu\to i}(z)=\frac{4d\,x_i^\mu y_i^\mu+(y_i^\mu)^2S_{xx}^{\mu\to i}-2x_i^\mu y_i^\mu S_{xy}^{\mu\to i}+(x_i^\mu)^2S_{yy}^{\mu\to i}}{\left(2d-S_{xy}^{\mu\to i}\right)^2-S_{xx}^{\mu\to i}S_{yy}^{\mu\to i}}\,.
\label{eq:gdw-ped-self-energy-general}
\end{equation}
Let us check the Wishart limit $y_j^\mu=x_j^\mu$ for all $j\in\partial\mu$. Then
\begin{equation}
S_{xx}^{\mu\to i}=S_{xy}^{\mu\to i}=S_{yy}^{\mu\to i}\equiv S^{\mu\to i}\,,
    \label{eq:gdw-ped-wishart-S-reduction}
\end{equation}
and $x_i^\mu y_i^\mu=(x_i^\mu)^2$. The numerator of \eqref{eq:gdw-ped-self-energy-general} becomes
\begin{align}
4d(x_i^\mu)^2+(x_i^\mu)^2S^{\mu\to i}-2(x_i^\mu)^2S^{\mu\to i}+(x_i^\mu)^2S^{\mu\to i}=4d(x_i^\mu)^2\,.
\label{eq:gdw-ped-wishart-numerator}
\end{align}
The denominator becomes
\begin{align}
\left(2d-S^{\mu\to i}\right)^2-\left(S^{\mu\to i}\right)^2=4d^2-4dS^{\mu\to i}=4d(d-S^{\mu\to i})\,.
\label{eq:gdw-ped-wishart-denominator}
\end{align}
Therefore
\begin{equation}
U_{\mu\to i}(z)=\frac{4d(x_i^\mu)^2}{4d(d-S^{\mu\to i})}=\frac{(x_i^\mu)^2/d}{1-S^{\mu\to i}/d}\,.
\label{eq:gdw-ped-wishart-self-energy}
\end{equation}
This is the full rank-one contribution to the denominator of the diluted Wishart cavity Green function, in the normalization used in this section.
\end{examplebox}

The variable-to-factor cavity Green function is
\begin{equation}
G_{i\to\mu}(z)=\frac{1}{z-\displaystyle\sum_{\nu\in\partial i\setminus\mu}U_{\nu\to i}(z)}\,,
\label{eq:gdw-variable-to-factor-green}
\end{equation}
and the full local Green function is
\begin{equation}
G_i(z)=\frac{1}{z-\displaystyle\sum_{\nu\in\partial i}U_{\nu\to i}(z)}\,.
\label{eq:gdw-full-green}
\end{equation}
The spectral density of a typical large instance is obtained from
\begin{equation}
\rho_{\pmb{F},\epsilon}^{\rm cav}(\lambda)=\frac{1}{\pi N}\sum_{i=1}^N{\rm Im}[G_i(\lambda-i\epsilon)]\,.
\label{eq:gdw-cavity-density-instance}
\end{equation}
Equations \eqref{eq:gdw-factor-to-variable-self-energy}--\eqref{eq:gdw-cavity-density-instance} are the belief-propagation form of the spectral-density calculation for a fixed diluted cross-correlation matrix.

The ensemble-level equations follow by turning the messages into random variables. Let $\mathcal{P}(G)$ be the distribution of a variable-to-factor cavity Green function and let $\mathcal{Q}(U)$ be the distribution of a factor-to-variable self-energy. In the Poisson bipartite ensemble \eqref{eq:gdw-entry-distribution}, the excess-degree distributions on both sides are again Poisson:
\begin{equation}
q^{(\rm v)}_\ell=e^{-d/\alpha}\frac{(d/\alpha)^\ell}{\ell!}\,,\qquad q^{(\rm f)}_k= e^{-d}\frac{d^k}{k!}\,.
\label{eq:gdw-poisson-excess-degrees}
\end{equation}
For a factor-to-variable message, draw the weight pair $(x,y)$ of the distinguished edge from $\varrho$, draw $k$ incoming variable messages $G_1,\ldots,G_k$ from $\mathcal{P}$, and draw $k$ incoming weight pairs $(x_r,y_r)$ from $\varrho$. Define
\begin{equation}
S_{xx}=\sum_{r=1}^k x_r^2G_r\,,\qquad S_{xy}=\sum_{r=1}^k x_ry_rG_r\,,\qquad S_{yy}=\sum_{r=1}^k y_r^2G_r\,,
\label{eq:gdw-random-cavity-sums}
\end{equation}
and
\begin{equation}
\mathcal{U}\left(x,y;\{x_r,y_r,G_r\}_{r=1}^k\right)=\frac{4dxy+y^2S_{xx}-2xyS_{xy}+x^2S_{yy}}{(2d-S_{xy})^2-S_{xx}S_{yy}}\,.
\label{eq:gdw-random-self-energy-map}
\end{equation}
Then
\begin{equation}
\begin{split}
\mathcal{Q}(U)&=\sum_{k=0}^{\infty}e^{-d}\frac{d^k}{k!}\int dxdy\varrho(x,y)\left[\prod_{r=1}^{k}dG_r \mathcal{P}(G_r) dx_r dy_r \varrho(x_r,y_r)\right]\\
&\times\delta\left[U-\mathcal{U}\left(x,y;\{x_r,y_r,G_r\}_{r=1}^k\right)\right]\,.
\label{eq:gdw-self-energy-distribution}    
\end{split}
\end{equation}
The variable-to-factor message distribution is
\begin{equation}
\mathcal{P}(G)=\sum_{\ell=0}^{\infty}\frac{e^{-d/\alpha}(d/\alpha)^\ell}{\ell!}\int\left[\prod_{r=1}^{\ell}dU_r \mathcal{Q}(U_r)\right]\delta\left(G-\frac{1}{z-\displaystyle\sum_{r=1}^{\ell}U_r}\right)\,.
\label{eq:gdw-green-distribution}    
\end{equation}
Because the variable degrees are Poisson, the full-site distribution has the same form:
\begin{equation}
\mathcal{P}_{\rm site}(G)=\sum_{\ell=0}^{\infty}e^{-d/\alpha}\frac{(d/\alpha)^\ell}{\ell!}\int\left[\prod_{r=1}^{\ell}dU_r\mathcal{Q}(U_r)\right]\delta\left(G-\frac{1}{z-\displaystyle\sum_{r=1}^{\ell}U_r}\right)\,,
\label{eq:gdw-site-green-distribution}
\end{equation}
and the ensemble-averaged regularized density is
\begin{equation}
\overline{\rho_{\pmb F,\epsilon}(\lambda)}=\frac{1}{\pi}{\rm Im}\int dG\mathcal{P}_{\rm site}(G)G\,,\qquad z=\lambda-i\epsilon\,.
\label{eq:gdw-density-from-populations}
\end{equation}
For non-Poisson bipartite ensembles, the Poisson laws in \eqref{eq:gdw-self-energy-distribution}--\eqref{eq:gdw-site-green-distribution} are replaced by the appropriate factor excess-degree law, variable excess-degree law, and full variable-degree law, exactly as in the diluted Wishart case.

\begin{examplebox}[One population-dynamics update in the generalized ensemble]
The generalized diluted Wishart ensemble is solved by two coupled populations:
\begin{equation}
\{G^{(1)},\ldots,G^{(M)}\}\,,\qquad\{U^{(1)},\ldots,U^{(M)}\}\,.
\label{eq:gdw-ped-populations}
\end{equation}
A factor-to-variable update is generated as follows. First draw
\begin{equation}
k\sim{\rm Poisson}(d)\,,
\label{eq:gdw-ped-factor-degree}
\end{equation}
then draw $k$ incoming Green functions
\begin{equation}
G^{(a_1)}\,,\ldots,G^{(a_k)}
\label{eq:gdw-ped-incoming-Gs}
\end{equation}
from the $G$-population, and draw $k+1$ weight pairs
\begin{equation}
(x,y)\,,\qquad(x_1,y_1)\,,\ldots,(x_k,y_k)
\label{eq:gdw-ped-weight-pairs}
\end{equation}
from $\varrho(x,y)$. The distinguished pair $(x,y)$ belongs to the edge receiving the outgoing self-energy; the other $k$ pairs belong to the incoming neighboring variables. Form
\begin{equation}
S_{xx}=\sum_{r=1}^{k}x_r^2G^{(a_r)}\,,\qquad S_{xy}=\sum_{r=1}^{k}x_ry_rG^{(a_r)}\,,\qquad S_{yy}=\sum_{r=1}^{k}y_r^2G^{(a_r)}\,.
\label{eq:gdw-ped-pop-sums}
\end{equation}
Then set
\begin{equation}
U_{\rm new}=\frac{4dxy+y^2S_{xx}-2xyS_{xy}+x^2S_{yy}}{(2d-S_{xy})^2-S_{xx}S_{yy}}\,.
\label{eq:gdw-ped-pop-U-new}
\end{equation}
A variable-to-factor update is generated by drawing
\begin{equation}
\ell\sim{\rm Poisson}(d/\alpha)\,,
\label{eq:gdw-ped-variable-degree}
\end{equation}
drawing $\ell$ self-energies $U^{(b_1)},\ldots,U^{(b_\ell)}$ from the $U$-population, and setting
\begin{equation}
G_{\rm new}=\frac{1}{z-\displaystyle\sum_{r=1}^{\ell}U^{(b_r)}}\,.
\label{eq:gdw-ped-pop-G-new}
\end{equation}
The ordinary diluted Wishart algorithm is recovered by imposing $y=x$ in the factor update, with $U$ interpreted as the full denominator contribution used in the present section. Thus the generalized ensemble is algorithmically close to the Wishart case, but with a rank-at-most-two rather than rank-one factor contribution.
\end{examplebox}

The reduction to ordinary diluted Wishart matrices is an important check. Set $y_i^\mu=x_i^\mu$ for all occupied edges. Then
\begin{equation}
S_{xx}^{\mu\to i}=S_{xy}^{\mu\to i}=S_{yy}^{\mu\to i}\equiv S^{\mu\to i}\,,
\label{eq:gdw-wishart-sum-reduction}
\end{equation}
and \eqref{eq:gdw-factor-to-variable-self-energy} gives
\begin{equation}
U_{\mu\to i}(z)=\frac{(x_i^\mu)^2}{d-S^{\mu\to i}}=\frac{(x_i^\mu)^2/d}{1-S^{\mu\to i}/d}\,.
\label{eq:gdw-wishart-self-energy-reduction}
\end{equation}
This is precisely the full denominator contribution of the diluted Wishart ensemble; in the notation of Section~\ref{sec:sparse-covariance-diluted-wishart}, it corresponds to $d^{-1}U_{\mu\to i}^{\rm Wishart}$. If instead $y_i^\mu=-x_i^\mu$, then $\pmb{F}=-d^{-1}\pmb{X}\pmb{X}^{\rm T}$ and
\begin{equation}
U_{\mu\to i}(z)=-\frac{(x_i^\mu)^2}{d+S^{\mu\to i}}\,,
\label{eq:gdw-negative-wishart-reduction}
\end{equation}
as expected for a negative Wishart matrix. These two limits check both the signs and the normalization in the generalized formula.

\begin{examplebox}[Positive, negative, and proportional Wishart subfamilies]
The generalized ensemble contains a one-parameter family of deterministic reductions. Suppose
\begin{equation}
\pmb Y=\gamma \pmb X\,,
\label{eq:gdw-ped-proportional-condition}
\end{equation}
where $\gamma$ is a real constant. Then
\begin{equation}
\pmb F=\frac{1}{2d}\left(\gamma\pmb X\pmb X^{\rm T}+\gamma\pmb X\pmb X^{\rm T}\right)=\frac{\gamma}{d}\pmb X\pmb X^{\rm T}\,.
\label{eq:gdw-ped-proportional-case}
\end{equation}
If $\gamma>0$, all nonzero eigenvalues of $\pmb F$ are positive. If $\gamma<0$, all nonzero eigenvalues are negative. If $\gamma=0$, the matrix vanishes.

The ordinary diluted Wishart case is $\gamma=1$:
\begin{equation}
\pmb F=\frac{1}{d}\pmb X\pmb X^{\rm T}\,.
\label{eq:gdw-ped-positive-Wishart}
\end{equation}
The negative Wishart case is $\gamma=-1$:
\begin{equation}
\pmb F=-\frac{1}{d}\pmb X\pmb X^{\rm T}.
\label{eq:gdw-ped-negative-Wishart}
\end{equation}
The generic ensemble is richer because $x$ and $y$ are not deterministically proportional; they are only drawn from a joint distribution $\varrho(x,y)$.
\end{examplebox}

The dense limit gives another consistency check. If $d\to\infty$ after the thermodynamic limit and the weight moments remain finite, the random sums in \eqref{eq:gdw-random-cavity-sums} self-average. In the perfectly correlated case $x=y$ with $m_{20}=1$, the population equations collapse to the Mar\v{c}enko--Pastur self-consistency equation already obtained for diluted Wishart matrices. More generally, the dense limit depends on the low moments of $\varrho(x,y)$, in particular on $m_{20}$, $m_{02}$, and $m_{11}$, and describes the symmetric part of a product of two correlated rectangular random matrices. The finite-connectivity equations \eqref{eq:gdw-self-energy-distribution}--\eqref{eq:gdw-density-from-populations} are therefore a genuine sparse extension of dense cross-correlation random matrix theory rather than merely a reparametrization of the ordinary Wishart ensemble.

The generalized ensemble also clarifies the role of positivity. For $x=y$, all eigenvalues of $\pmb F$ are nonnegative. Beyond positively proportional reductions such as $y_i^\mu=\gamma x_i^\mu$ with $\gamma>0$, there is no general positivity constraint. The spectral density may extend over both positive and negative values of $\lambda$, and its center is controlled at the level of the first moment by $m_{11}/\alpha$. This makes the ensemble useful as a model for symmetric interaction matrices generated by two correlated rectangular data sets. In contrast with ordinary covariance matrices, where a negative eigenvalue would signal an error in the construction, negative eigenvalues are an intrinsic feature of the generic symmetric cross-correlation ensemble.

From the point of view of sparse graphs, the ensemble has the same local phenomena as the diluted Wishart ensemble, but in a richer algebraic form. Isolated variable nodes produce zero modes. Small finite bipartite components produce isolated finite-dimensional spectra. Rare high-degree factor nodes can create outlying or localized eigenvectors. The distinction between the continuous bulk and singular contributions is again regulated by the small imaginary part $\epsilon$ in $z=\lambda-i\epsilon$. The difference is that each occupied edge now carries two coupled weights, and each factor node transmits a rank-two self-energy rather than the rank-one self-energy of an ordinary covariance matrix.

The population-dynamics implementation is correspondingly simple. One keeps two populations: a population of variable-to-factor Green functions $G$ and a population of factor-to-variable self-energies $U$. A factor update samples a Poisson number of incoming variable messages, samples the corresponding weight pairs from $\varrho(x,y)$, computes the three sums \eqref{eq:gdw-random-cavity-sums}, and applies \eqref{eq:gdw-random-self-energy-map}. A variable update samples a Poisson number of incoming self-energies and applies the scalar resolvent map in \eqref{eq:gdw-green-distribution}. The spectral density is then obtained by averaging the imaginary part of the full site Green function as in \eqref{eq:gdw-density-from-populations}. The same equations can be iterated as belief propagation on a single sparse bipartite instance.

There are several useful checks for numerical and analytical work. First, the density must be normalized:
\begin{equation}
\int d\lambda \rho(\lambda)=1\,.
\label{eq:gdw-normalization}
\end{equation}
Second, its first moment must agree with \eqref{eq:gdw-first-moment}:
\begin{equation}
\int d\lambda\lambda\rho(\lambda)=\frac{m_{11}}{\alpha}\,.
\label{eq:gdw-first-moment-check}
\end{equation}
Third, the zero-mode weight must respect the rank constraints \eqref{eq:gdw-zero-mode-bound}, with additional zero modes possible because of dilution. Fourth, the limits $x=y$ and $y=-x$ must reduce to positive and negative diluted Wishart matrices. Finally, when $d$ is increased at fixed $\alpha$, the population of messages should narrow and approach the corresponding dense cross-correlation law.

The generalized diluted Wishart ensemble is therefore a compact synthesis of the themes developed so far. It is sparse because its elementary disorder is supported on a finite-connectivity bipartite graph. It is Wishart-like because it is built from rectangular matrices. It is more general than an ordinary covariance matrix because it depends on two correlated arrays and is not necessarily positive semidefinite. The cavity method remains effective because each factor node contributes only a rank-at-most-two quadratic interaction, allowing the local Gaussian integrations to close on scalar Green functions and factor self-energies. This ensemble will also serve as a bridge to later material: the large-deviation theory for diluted Wishart matrices is recovered in the positive case, while non-Hermitian extensions of cross-correlation ensembles arise once the symmetric combination in \eqref{eq:gdw-definition} is replaced by an asymmetric one \cite{GuzmanGonzalezPerezCastillo2025}.

\begin{exerciseblock}
\exitem[Symmetry of the generalized matrix]
Starting from
\begin{equation}
\pmb F=\frac{1}{2d}\left(\pmb X\pmb Y^{\rm T}+\pmb Y\pmb X^{\rm T}\right)\,,
\label{eq:gdw-ex-F-definition}
\end{equation}
show that $\pmb F^{\rm T}=\pmb F$ for real matrices $\pmb X$ and $\pmb Y$.

\exitem[Entrywise formula]
Derive
\begin{equation}
F_{ij}=\frac{1}{2d}\sum_{\mu=1}^{P}\left(x_i^\mu y_j^\mu+y_i^\mu x_j^\mu\right)
\label{eq:gdw-ex-entrywise}
\end{equation}
from the matrix definition of $\pmb F$.

\exitem[Wishart reduction]
Show that if $\pmb X=\pmb Y$, then
\begin{equation}
\pmb F=\frac{1}{d}\pmb X\pmb X^{\rm T}.
\label{eq:gdw-ex-wishart-reduction}
\end{equation}
Explain why the spectrum is then nonnegative.

\exitem[Negative Wishart reduction]
Show that if $\pmb Y=-\pmb X$, then
\begin{equation}
\pmb F=-\frac{1}{d}\pmb X\pmb X^{\rm T}\,.
\label{eq:gdw-ex-negative-wishart}
\end{equation}
What can you say about the sign of its nonzero eigenvalues?

\exitem[Proportional weights]
Assume $y_i^\mu=\gamma x_i^\mu$ for all occupied edges, where $\gamma$ is a real constant. Show that
\begin{equation}
\pmb F=\frac{\gamma}{d}\pmb X\pmb X^{\rm T}\,.
\label{eq:gdw-ex-proportional}
\end{equation}
How does the sign of $\gamma$ affect the spectrum?

\exitem[Trace and first moment]
Starting from
\begin{equation}
F_{ii}=\frac{1}{d}\sum_{\mu=1}^{P}x_i^\mu y_i^\mu\,,
\label{eq:gdw-ex-diagonal}
\end{equation}
and using the Poisson dilution law under the same trace self-averaging assumption as in the main text, derive
\begin{equation}
\frac{1}{N}{\rm Tr}\pmb F\longrightarrow\frac{m_{11}}{\alpha}\,.
\label{eq:gdw-ex-first-moment}
\end{equation}
State clearly where the dilution probability $d/N$ and the rectangularity $P=N/\alpha$ enter.

\exitem[Correlation coefficient]
Given
\begin{equation}
\Gamma=\frac{m_{11}}{\sqrt{m_{20}m_{02}}}\,,
\label{eq:gdw-ex-Gamma}
\end{equation}
show using Cauchy--Schwarz that $|\Gamma|\leq1$, assuming $m_{20},m_{02}>0$. Interpret the cases $\Gamma=1$, $\Gamma=-1$, and $\Gamma=0$.

\exitem[Rank bound]
Prove
\begin{equation}
{\rm rank}\left(\pmb X\pmb Y^{\rm T}+\pmb Y\pmb X^{\rm T}\right)\leq2P\,.
\label{eq:gdw-ex-rank-bound}
\end{equation}
Then derive
\begin{equation}
w_0\geq\max\left\{0,1-\frac{2}{\alpha}\right\}\,.
\label{eq:gdw-ex-zero-mode-bound}
\end{equation}

\exitem[Wishart rank bound]
Explain why the Wishart limit $\pmb X=\pmb Y$ satisfies the stronger bound
\begin{equation}
{\rm rank} \pmb F\leq P\,.
\label{eq:gdw-ex-wishart-rank}
\end{equation}
Compare the corresponding zero-mode lower bound with \eqref{eq:gdw-ex-zero-mode-bound}.

\exitem[Quadratic form]
Show that
\begin{equation}
\pmb u^{\rm T}\pmb F\pmb u=\frac{1}{d}\sum_{\mu=1}^{P}\left(\sum_{i\in\partial\mu}x_i^\mu u_i\right)\left(\sum_{j\in\partial\mu}y_j^\mu u_j\right)\,.
\label{eq:gdw-ex-quadratic-form}
\end{equation}
Use this formula to explain why $\pmb F$ is not generically positive semidefinite.

\exitem[Factor matrix]
For one factor node $\mu$, show that the contribution to $\pmb F$ on the variables in $\partial\mu$ is
\begin{equation}
B_{ij}^{\mu}=\frac{x_i^\mu y_j^\mu+y_i^\mu x_j^\mu}{2d}\,.
\label{eq:gdw-ex-factor-matrix}
\end{equation}
Show that this matrix has rank at most two.

\exitem[Rank-two denominator]
Let $R=\partial\mu\setminus i$, let $\pmb D={\rm diag}(G_j^{-1})_{j\in R}$, and let the factor matrix on $R$ be
\[
\pmb B_{RR}=\frac{1}{2d}\left(\pmb x_R\pmb y_R^{\rm T}+\pmb y_R\pmb x_R^{\rm T}\right).
\]
Starting from
\begin{equation}
S_{xx}=\sum_{j\in R}x_j^2G_j\,,\qquad S_{xy}=\sum_{j\in R}x_jy_jG_j\,,\qquad S_{yy}=\sum_{j\in R}y_j^2G_j\,,
\label{eq:gdw-ex-rank-two-sums}
\end{equation}
derive
\begin{equation}
\Delta=(2d-S_{xy})^2-S_{xx}S_{yy}
\label{eq:gdw-ex-rank-two-denominator}
\end{equation}
using the matrix determinant lemma.

\exitem[Factor-to-variable self-energy]
Use the Schur complement for the rank-two factor matrix to derive
\begin{equation}
U_{\mu\to i}(z)=\frac{4d\,x_i^\mu y_i^\mu+(y_i^\mu)^2S_{xx}^{\mu\to i}-2x_i^\mu y_i^\mu S_{xy}^{\mu\to i}+(x_i^\mu)^2S_{yy}^{\mu\to i}}{(2d-S_{xy}^{\mu\to i})^2-S_{xx}^{\mu\to i}S_{yy}^{\mu\to i}}\,.
\label{eq:gdw-ex-self-energy}
\end{equation}

\exitem[Wishart self-energy check]
Set $y_j^\mu=x_j^\mu$ in \eqref{eq:gdw-ex-self-energy} and show that it reduces to
\begin{equation}
U_{\mu\to i}(z)=\frac{(x_i^\mu)^2}{d-S^{\mu\to i}}=\frac{(x_i^\mu)^2/d}{1-S^{\mu\to i}/d}\,,
\label{eq:gdw-ex-wishart-self-energy}
\end{equation}
where
\begin{equation}
S^{\mu\to i}=\sum_{j\in\partial\mu\setminus i}(x_j^\mu)^2G_{j\to\mu}(z)\,.
\label{eq:gdw-ex-wishart-S}
\end{equation}
Explain why this is the full denominator contribution in the convention of the present section.

\exitem[Negative Wishart self-energy check]
Set $y_j^\mu=-x_j^\mu$ in \eqref{eq:gdw-ex-self-energy} and derive
\begin{equation}
U_{\mu\to i}(z)=-\frac{(x_i^\mu)^2}{d+S^{\mu\to i}}\,,
\label{eq:gdw-ex-negative-self-energy}
\end{equation}
where
\begin{equation}
S^{\mu\to i}=\sum_{j\in\partial\mu\setminus i}(x_j^\mu)^2G_{j\to\mu}(z)\,.
\label{eq:gdw-ex-negative-S}
\end{equation}

\exitem[Population equations]
Starting from the local maps for $U_{\mu\to i}$ and $G_{i\to\mu}$, derive the population equations for $\mathcal Q(U)$ and $\mathcal P(G)$ in the Poisson bipartite ensemble. Use the convention of the main text, in which $U_{\mu\to i}$ is the full contribution of a factor node to the denominator of the variable Green function. Identify which Poisson law corresponds to factor updates and which corresponds to variable updates.

\exitem[Density from the site law]
Explain why the site distribution uses the full variable degree law rather than the excess-degree law. Derive
\begin{equation}
\overline{\rho_{\pmb F,\epsilon}(\lambda)}=\frac{1}{\pi}{\rm Im}\int dG \mathcal P_{\rm site}(G)G\,.
\label{eq:gdw-ex-density-site-law}
\end{equation}

\exitem[Moment checks]
Show that any numerical or analytical solution should satisfy
\begin{equation}
\int d\lambda \rho(\lambda)=1\,,\qquad\int d\lambda \lambda\rho(\lambda)=\frac{m_{11}}{\alpha}\,.
\label{eq:gdw-ex-moment-checks}
\end{equation}
Explain why the first is a normalization check and the second is a trace check.

\exitem[Finite example]
Choose $N=2$, $P=1$, $d=1$, and
\begin{equation}
\pmb X=\begin{pmatrix}
1\\
2
\end{pmatrix}\,,\qquad\pmb Y=
\begin{pmatrix}
2\\
-1
\end{pmatrix}\,.
\label{eq:gdw-ex-finite-example}
\end{equation}
Compute $\pmb F$ and its eigenvalues. Then compute $N^{-1}{\rm Tr}\pmb F$ and verify the exact finite identity
\begin{equation}
\frac{1}{N}{\rm Tr}\pmb F=\frac{1}{dN}\sum_{i=1}^{N}\sum_{\mu=1}^{P}x_i^\mu y_i^\mu\,.
\end{equation}
Explain how, in the Poisson ensemble and under the self-averaging assumption stated in the text, this identity becomes $N^{-1}{\rm Tr}\pmb F\to m_{11}/\alpha$.

\exitem[Programming exercise: varying local correlation]
Fix a list of system sizes $N$, an aspect ratio $\alpha>0$ such that $P=N/\alpha$ is an integer for each $N$, a dilution $d>0$, a number $S$ of independent samples for each parameter set, a list of correlation parameters $\Gamma\in[-1,1]$, and a common density-estimation convention, such as a histogram bin width or Lorentzian regulator. Generate sparse rectangular matrices $\pmb X$ and $\pmb Y$ on the same bipartite support with
\begin{equation}
{\rm Prob}(B_i^\mu=1)=\frac{d}{N}\,.
\end{equation}
Conditional on $B_i^\mu=1$, draw independent standard Gaussian variables $\xi_i^\mu,\zeta_i^\mu$ and set
\[
x_i^\mu=\xi_i^\mu\,,\qquad y_i^\mu=\Gamma\xi_i^\mu+\sqrt{1-\Gamma^2}\,\zeta_i^\mu\,,
\]
so that
\begin{equation}
\mathbb E[x^2]=\mathbb E[y^2]=1\,,\qquad \mathbb E[xy]=\Gamma\,.
\label{eq:gdw-ex-program-correlation}
\end{equation}
Form
\begin{equation}
\pmb F=\frac{1}{2d}\left(\pmb X\pmb Y^{\rm T}+\pmb Y\pmb X^{\rm T}\right)
\label{eq:gdw-ex-program-F}
\end{equation}
and plot the empirical spectral density for the chosen values of $\Gamma$. Track the first moment and compare it with $\Gamma/\alpha$. Report $N$, $P$, $d$, $\alpha$, $S$, the values of $\Gamma$, and the density-estimation convention.

\exitem[Programming exercise: zero modes]
Fix a list of system sizes $N$, an aspect ratio $\alpha>0$ such that $P=N/\alpha$ is an integer for each $N$, a dilution $d>0$, a number $S$ of independent samples for each parameter set, a nonzero-weight distribution $\varrho(x,y)$, and a numerical tolerance $\tau_0$ for identifying zero eigenvalues. Generate generalized diluted Wishart matrices
\begin{equation}
\pmb F=\frac{1}{2d}\left(\pmb X\pmb Y^{\rm T}+\pmb Y\pmb X^{\rm T}\right) 
\end{equation}
on a common sparse bipartite support, and estimate the fraction of eigenvalues with $|\lambda|<\tau_0$. Compare the observed zero-mode fraction with the lower bound
\begin{equation}
\max\left\{0,1-\frac{2}{\alpha}\right\}\,.
\label{eq:gdw-ex-program-zero-bound}
\end{equation}
Repeat the experiment in the Wishart limit $\pmb X=\pmb Y$ and compare with the stronger bound $\max\{0,1-1/\alpha\}$. Report $N$, $P$, $d$, $\alpha$, $S$, $\varrho$, and $\tau_0$.

\exitem[Programming exercise: population dynamics]
Fix $d$, $\alpha$, a regulator $\epsilon>0$, a grid of real values $\lambda$, population sizes for the $G$- and $U$-populations, a burn-in time $T_{\rm burn}$, a number $T_{\rm meas}$ of site samples used for measurement, and a nonzero-pair distribution $\varrho(x,y)$. Implement the two-population algorithm using the factor update
\begin{equation}
U_{\rm new}=\frac{4dxy+y^2S_{xx}-2xyS_{xy}+x^2S_{yy}}{(2d-S_{xy})^2-S_{xx}S_{yy}}\,,
\label{eq:gdw-ex-program-U-update}
\end{equation}
and the variable update
\begin{equation}
G_{\rm new}=\frac{1}{z-\displaystyle\sum_{r=1}^{\ell}U_r}\,.
\label{eq:gdw-ex-program-G-update}
\end{equation}
Estimate the density from the corresponding full site Green functions, where the full variable degree is drawn from ${\rm Poisson}(d/\alpha)$. For comparison, generate $S$ independent finite matrices of size $N_{\rm diag}\times N_{\rm diag}$ with $P=N_{\rm diag}/\alpha$ an integer, using the same $d$, $\alpha$, and $\varrho(x,y)$. Compare the population-dynamics density with direct diagonalization using the same $\lambda$ grid and the same regulator $\epsilon$. Report $d$, $\alpha$, $\varrho$, the population sizes, $T_{\rm burn}$, $T_{\rm meas}$, $N_{\rm diag}$, $P$, $S$, $\epsilon$, the grid spacing, and a discrepancy measure.
\end{exerciseblock}

\section{Index number and spectral-count large deviations}
\label{sec:index-number-large-deviations}
Up to this point the main spectral observable has been the typical density of eigenvalues. This is the first object one usually wants to know, but it is not the whole spectral problem. Once the density is known, one may ask how many eigenvalues fall in a prescribed interval and how this number fluctuates from one matrix realization to another. These questions lead to the index number and, more generally, to spectral-count statistics. They are particularly natural in sparse random matrices, where the absence of an explicit joint eigenvalue density makes the usual invariant-ensemble methods unavailable, but where determinant and cavity representations remain effective.

For a real symmetric matrix $\pmb{A}$ with eigenvalues $\lambda_1,\ldots,\lambda_N$, the index number below a threshold $\lambda$ is
\begin{equation}
\mathcal{K}_{\pmb{A}}(\lambda)=\sum_{i=1}^N\Theta(\lambda-\lambda_i)\,,
\label{eq:inld-index-number}
\end{equation}
where $\Theta$ is the Heaviside step function. In particular, with the usual convention that eigenvalues exactly at zero are treated separately, $\mathcal{K}_{\pmb A}(0)$ counts the number of negative eigenvalues and is often called the index of the matrix. More generally, for definiteness we write interval counts with the half-open convention $I=(a,b]$, with endpoint eigenvalues treated separately when needed, and set
\begin{equation}
\mathcal{N}_{\pmb{A}}(I)=\sum_{i=1}^N\mathbf{1}_{a< \lambda_i\leq b}=\mathcal{K}_{\pmb{A}}(b)-\mathcal{K}_{\pmb{A}}(a)\,.
\label{eq:inld-interval-count}
\end{equation}
The distinction between $a<\lambda_i\leq b$, $a\leq\lambda_i\leq b$, and the other endpoint conventions only affects eigenvalues exactly at the endpoints. Equivalently, for a fixed realization with no eigenvalue at the endpoints, or with the same endpoint convention understood in the distributional integral,
\begin{equation}
\frac{1}{N}\mathcal{N}_{\pmb{A}}(I)=\int_a^b d\lambda\rho_{\pmb{A}}(\lambda)\,,
\label{eq:inld-count-from-density}
\end{equation}
where $\rho_{\pmb A}$ is the empirical spectral density. At the level of typical values, the interval count is therefore obtained by integrating the spectral density. At the level of fluctuations and large deviations, however, the typical density is not enough. One must study the probability law of $\mathcal{N}_{\pmb{A}}(I)$ itself.

The index number can be written in terms of determinants. We use the same lower-half-plane convention as before,
\begin{equation}
z_\lambda=\lambda-i\epsilon\,,\qquad  \epsilon>0\,.
\label{eq:inld-lower-half-plane}
\end{equation}
For a single eigenvalue with $\lambda_i\neq\lambda$,
\begin{equation}
\lim_{\epsilon\downarrow0}{\rm Im}\log(\lambda-i\epsilon-\lambda_i)=-\pi\,\Theta(\lambda_i-\lambda)\,.
\label{eq:inld-log-step-identity}
\end{equation}
Eigenvalues exactly at the threshold are handled by the endpoint convention stated above. Therefore
\begin{equation}
\mathcal{K}_{\pmb{A}}(\lambda)=N+\frac{1}{\pi}\lim_{\epsilon\downarrow0}{\rm Im}\log\det\left[(\lambda-i\epsilon)\pmb{I}-\pmb{A}\right]\,,
\label{eq:inld-index-from-log-det}
\end{equation}
and the number of eigenvalues in $I=(a,b]$ is
\begin{equation}
\mathcal{N}_{\pmb{A}}(I)=\frac{1}{\pi}\lim_{\epsilon\downarrow0}{\rm Im}\left\{\log\det\left[(b-i\epsilon)\pmb{I}-\pmb{A}\right]-\log\det\left[(a-i\epsilon)\pmb{I}-\pmb{A}\right]\right\}\,.
\label{eq:inld-interval-count-log-det}
\end{equation}
This identity is the analogue, for spectral counting, of the resolvent formula for the spectral density. Differentiating \eqref{eq:inld-index-from-log-det} with respect to $\lambda$ gives back the trace of the resolvent and hence the density, while taking finite differences as in \eqref{eq:inld-interval-count-log-det} counts eigenvalues in an interval.

\begin{examplebox}[The index from the phase of a determinant]
Consider the $2\times2$ diagonal matrix
\begin{equation}
\pmb A=   \begin{pmatrix}
-1 & 0\\
0 & 2
\end{pmatrix}\,.
\label{eq:inld-ped-diagonal-example}
\end{equation}
The number of eigenvalues below a threshold $\lambda$ is immediate from the spectrum. For instance, at $\lambda=0$,
\begin{equation}
\mathcal K_{\pmb A}(0)=1\,.
  \label{eq:inld-ped-index-direct}
\end{equation}
Let us recover the same result from the logarithmic determinant formula. The characteristic determinant is
\begin{equation}
\det(z\pmb I-\pmb A)=(z+1)(z-2)\,.
\label{eq:inld-ped-characteristic}
\end{equation}
At the threshold $\lambda=0$, take $z=-i\epsilon$ with $\epsilon>0$. Then
\begin{equation}
\det(-i\epsilon\pmb I-\pmb A)=(1-i\epsilon)(-2-i\epsilon)\,.
\label{eq:inld-ped-det-lower}
\end{equation}
The first factor has phase tending to $0$, while the second factor has phase tending to $-\pi$ as $\epsilon\downarrow0$. Therefore
\begin{equation}
\lim_{\epsilon\downarrow0}{\rm Im}\log\det(-i\epsilon\pmb I-\pmb A)=-\pi\,.
\label{eq:inld-ped-imag-logdet}
\end{equation}
Using
\begin{equation}
\mathcal K_{\pmb A}(\lambda)=N+\frac{1}{\pi}\lim_{\epsilon\downarrow0}{\rm Im}\log\det[(\lambda-i\epsilon)\pmb I-\pmb A]\,,
\label{eq:inld-ped-index-formula}
\end{equation}
with $N=2$, we obtain
\begin{equation}
\mathcal K_{\pmb A}(0)=2+\frac{1}{\pi}(-\pi)=1\,.
\label{eq:inld-ped-index-result}
\end{equation}
Thus the index is recovered from the phase of the determinant. Eigenvalues above the threshold contribute a phase $-\pi$ in the lower-half-plane convention, while eigenvalues below the threshold contribute no such phase; the constant $N$ in \eqref{eq:inld-ped-index-formula} converts this phase count into the number of eigenvalues below $\lambda$.
\end{examplebox}

The statistical object of interest is the probability distribution of the intensive count
\begin{equation}
k_I=\frac{1}{N}\mathcal{N}_{\pmb{A}}(I)\,.
\label{eq:inld-intensive-count}
\end{equation}
For sparse random matrices one expects, and in the ensembles treated in \cite{MetzPerezCastillo2016} one obtains, a large-deviation principle with speed $N$; for admissible finite-$N$ values of $k$, or equivalently for small count windows around $k$, one writes
\begin{equation}
{\rm Prob}\left[\frac{1}{N}\mathcal{N}_{\pmb{A}}(I)=k\right]\asymp\exp\left[-N\Phi_I(k)\right]\,.
\label{eq:inld-large-deviation-principle}
\end{equation}
The rate function $\Phi_I(k)$ is nonnegative and vanishes at the typical value
\begin{equation}
k_I^\star=\int_a^b d\lambda\overline{\rho_{\pmb A}(\lambda)}\,.
\label{eq:inld-typical-count-fraction}
\end{equation}
This speed $N$ is one of the main differences between finite-connectivity ensembles and dense invariant ensembles. In Gaussian invariant ensembles, spectral-count large deviations are governed by a Coulomb gas of strongly repelling eigenvalues and have speed $N^2$ \cite{DeanMajumdar2006,MajumdarNadalScardicchioVivo2009}. For the finite-connectivity sparse ensembles considered here, the graphical disorder and local spectral fluctuations remain extensive in the number of vertices, and the corresponding large-deviation exponent is naturally of order $N$ \cite{MetzPerezCastillo2016}.

The rate function is obtained from the scaled cumulant-generating function
\begin{equation}
\psi_I(s)=\lim_{N\to\infty}\frac{1}{N}\log\overline{\exp\left[s\mathcal{N}_{\pmb{A}}(I)\right]}\,,
\label{eq:inld-scaled-cgf}
\end{equation}
where the overbar denotes the average over the random matrix ensemble. When the Legendre-Fenchel transform is regular,
\begin{equation}
\Phi_I(k)=\sup_s\left\{sk-\psi_I(s)\right\}\,.
\label{eq:inld-legendre-transform}
\end{equation}

\begin{examplebox}[A toy large-deviation calculation for independent local counts]
Before treating matrix spectra, it is useful to recall the simplest large-deviation mechanism. Suppose that a count is the sum of independent Bernoulli variables,
\begin{equation}
\mathcal N_N=\sum_{i=1}^{N}X_i\,,\qquad{\rm Prob}(X_i=1)=p\,,\qquad{\rm Prob}(X_i=0)=1-p\,.
\label{eq:inld-ped-binomial-count}
\end{equation}
The scaled cumulant-generating function is
\begin{align}
\psi(s)&=\lim_{N\to\infty}\frac{1}{N}\log\mathbb E\left[e^{s\mathcal N_N}\right]\nonumber\\
&=\lim_{N\to\infty}\frac{1}{N}\log\left[(1-p)+p e^s\right]^N\nonumber\\
&=\log\left[1-p+p e^s\right]\,.
\label{eq:inld-ped-binomial-cgf}
\end{align}
The tilted mean is
\begin{equation}
k(s)=\psi'(s)=\frac{p e^s}{1-p+p e^s}\,.
\label{eq:inld-ped-binomial-tilted-mean}
\end{equation}
Solving for $s$ gives
\begin{equation}
e^s=\frac{k(1-p)}{p(1-k)}\,.
\label{eq:inld-ped-binomial-s-of-k}
\end{equation}
The rate function is
\begin{align}
\Phi(k)&=sk-\psi(s)\nonumber\\
&=k\log\frac{k}{p}+(1-k)\log\frac{1-k}{1-p}\,.
\label{eq:inld-ped-binomial-rate}
\end{align}
This elementary example is useful because the same structure appears in spectral-count large deviations. The matrix problem is much harder because the local contributions are not independent Bernoulli variables; they are generated by a graphical resolvent problem. Nevertheless, the relation between $\psi(s)$, the tilted mean $k(s)$, and the rate function $\Phi(k)$ is the same.
\end{examplebox}

The derivatives of $\psi_I(s)$ at $s=0$ give the extensive cumulants of the count:
\begin{equation}
\lim_{N\to\infty}\frac{1}{N}\overline{\mathcal{N}_{\pmb{A}}(I)}=\psi_I'(0)\,,\qquad\lim_{N\to\infty}\frac{1}{N}{\rm Var}\mathcal{N}_{\pmb{A}}(I)=\psi_I''(0)\,,
\label{eq:inld-first-two-cumulants}
\end{equation}
and similarly for higher cumulants. Thus the same formalism gives both typical count fluctuations and rare count deviations.

It is useful to rewrite \eqref{eq:inld-scaled-cgf} directly in terms of Gaussian partition functions. Here the Edwards--Jones representation is being reused in a different role: the determinant phase now generates a spectral count rather than a density. Define
\begin{equation}
D_{\pmb{A}}(z)=\det(z\pmb{I}-\pmb{A})\,,\qquad Z_{\pmb{A}}(z)=D_{\pmb{A}}(z)^{-1/2}\,,
\label{eq:inld-det-and-partition-function}
\end{equation}
where $Z_{\pmb{A}}(z)$ is the Edwards--Jones Gaussian partition function up to a $z$-independent normalization. For real symmetric $\pmb A$, and with conjugate logarithm branches chosen consistently,
\begin{equation}
{\rm Im}\log D_{\pmb{A}}(z)=\frac{1}{2i}\left[\log D_{\pmb{A}}(z)-\log D_{\pmb{A}}(z^*)\right]\,,
\label{eq:inld-imaginary-log-det}
\end{equation}
equation \eqref{eq:inld-interval-count-log-det} gives
\begin{equation}
\exp\left[s\mathcal{N}_{\pmb{A}}(I)\right]=\lim_{\epsilon\downarrow0}\left[\frac{Z_{\pmb A}(b-i\epsilon)Z_{\pmb A}(a+i\epsilon)}{Z_{\pmb A}(a-i\epsilon)Z_{\pmb A}(b+i\epsilon)}\right]^{-\frac{s}{\pi i}}\,.
\label{eq:inld-count-generating-partitions}
\end{equation}
The complex power in \eqref{eq:inld-count-generating-partitions} is understood through the same logarithm branches used in the determinant-phase identity. This formula is the starting point of the replica calculation. The moment-generating function of the count is an average of complex powers of Gaussian partition functions evaluated at the two endpoints of the interval. In the replica language, one first introduces integer numbers of replicas associated with the four spectral parameters $a\mp i\epsilon$ and $b\mp i\epsilon$, performs the disorder average, and then analytically continues the replica numbers to values proportional to the counting field $s$. The resulting order parameter is a distribution of cavity quantities associated with the two endpoints of the interval, rather than with a single spectral parameter. The replica-symmetric saddle-point structure of these multi-sector order parameters is summarized in Appendix~\ref{app:replica-symmetric-saddle-points}.

\begin{examplebox}[From the interval count to four determinant factors]
Let $I=(a,b]$. The interval count is
\begin{equation}
\mathcal N_{\pmb A}(I)=\frac{1}{\pi}\lim_{\epsilon\downarrow0}{\rm Im}\left[\log D_{\pmb A}(b-i\epsilon)-\log D_{\pmb A}(a-i\epsilon)\right]\,,
\label{eq:inld-ped-interval-det-start}
\end{equation}
where
\begin{equation}
D_{\pmb A}(z)=\det(z\pmb I-\pmb A)\,.
\label{eq:inld-ped-D-def}
\end{equation}
Using
\begin{equation}
{\rm Im}\log D_{\pmb A}(z)=\frac{1}{2i}\left[\log D_{\pmb A}(z)-\log D_{\pmb A}(z^*)\right]\,,
\label{eq:inld-ped-imlog}
\end{equation}
we obtain
\begin{align}
\mathcal N_{\pmb A}(I)&=\frac{1}{2\pi i}\lim_{\epsilon\downarrow0}\Big[\log D_{\pmb A}(b-i\epsilon)-\log D_{\pmb A}(b+i\epsilon)\nonumber\\
&\hspace{3.4cm}-\log D_{\pmb A}(a-i\epsilon)+\log D_{\pmb A}(a+i\epsilon)\Big]\,.
\label{eq:inld-ped-four-determinants}
\end{align}
Since the Edwards--Jones partition function satisfies
\begin{equation}
Z_{\pmb A}(z)=D_{\pmb A}(z)^{-1/2}
\label{eq:inld-ped-Z-D-relation}
\end{equation}
up to a $z$-independent factor, the exponential of the count contains four partition functions, evaluated at the two endpoints and on the two sides of the real axis:
\begin{equation}
e^{s\mathcal N_{\pmb A}(I)}=\lim_{\epsilon\downarrow0}\left[\frac{Z_{\pmb A}(b-i\epsilon)Z_{\pmb A}(a+i\epsilon)}{Z_{\pmb A}(a-i\epsilon)Z_{\pmb A}(b+i\epsilon)}\right]^{-s/(\pi i)}\,.
\label{eq:inld-ped-four-Z}
\end{equation}
This is why the replica calculation of an interval count involves several replica sectors: one needs fields associated with the two endpoints $a,b$ and with the two boundary values $z=\lambda\mp i0^+$.
\end{examplebox}

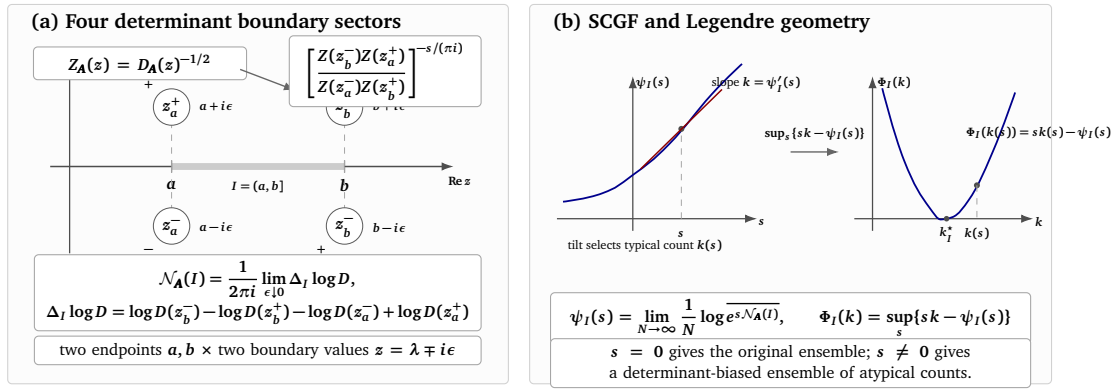
\begin{figure}[t]
\centering
\resizebox{0.98\textwidth}{!}{%
\begin{tikzpicture}[
    x=1cm,
    y=1cm,
    >=Latex,
    panel/.style={draw=black!18, fill=black!1, rounded corners=2pt, line width=0.5pt},
    axis/.style={draw=black!70, line width=0.6pt, -{Latex[length=2.0mm,width=1.4mm]}},
    realaxis/.style={draw=black!70, line width=0.7pt, -{Latex[length=2.0mm,width=1.4mm]}},
    tick/.style={draw=black!70, line width=0.55pt},
    sector/.style={circle, draw=black!65, fill=white, minimum size=6.0mm, inner sep=0pt, font=\scriptsize},
    formula/.style={draw=black!35, fill=white, rounded corners=2pt, line width=0.5pt, inner sep=3pt, font=\scriptsize, align=center},
    ptitle/.style={font=\bfseries\small, anchor=west},
    paneltext/.style={font=\scriptsize, align=center},
    tinytext/.style={font=\tiny, align=center},
    maparrow/.style={draw=black!60, line width=0.65pt, -{Latex[length=2.0mm,width=1.4mm]}},
    curve/.style={draw=blue!55!black, line width=0.85pt},
    tanline/.style={draw=red!55!black, line width=0.7pt},
    guide/.style={draw=black!35, line width=0.45pt, dashed}
]
\draw[panel] (0,0) rectangle (8.25,6.25);
\node[ptitle] at (0.25,5.95) {(a) Four determinant boundary sectors};

\draw[realaxis] (0.70,3.58) -- (7.35,3.58);
\draw[axis] (1.02,2.15) -- (1.02,5.25);
\node[tinytext] at (7.45,3.35) {${\rm Re}\,z$};
\node[tinytext] at (1.35,5.18) {${\rm Im}\,z$};

\coordinate (a0) at (2.70,3.58);
\coordinate (b0) at (5.55,3.58);
\coordinate (ap) at (2.70,4.56);
\coordinate (am) at (2.70,2.60);
\coordinate (bp) at (5.55,4.56);
\coordinate (bm) at (5.55,2.60);

\draw[black!20, line width=3.2pt] (a0) -- (b0);
\draw[tick] ($(a0)+(0,-0.10)$) -- ($(a0)+(0,0.10)$);
\draw[tick] ($(b0)+(0,-0.10)$) -- ($(b0)+(0,0.10)$);
\node[paneltext] at (2.70,3.28) {$a$};
\node[paneltext] at (5.55,3.28) {$b$};
\node[tinytext] at (4.12,3.25) {$I=(a,b]$};

\draw[guide] (a0) -- (ap);
\draw[guide] (a0) -- (am);
\draw[guide] (b0) -- (bp);
\draw[guide] (b0) -- (bm);

\node[sector] (zaP) at (ap) {$z_a^+$};
\node[sector] (zaM) at (am) {$z_a^-$};
\node[sector] (zbP) at (bp) {$z_b^+$};
\node[sector] (zbM) at (bm) {$z_b^-$};
\node[tinytext] at ($(zaP)+(-0.38,0.38)$) {$+$};
\node[tinytext] at ($(zaM)+(-0.38,-0.38)$) {$-$};
\node[tinytext] at ($(zbP)+(-0.38,0.38)$) {$-$};
\node[tinytext] at ($(zbM)+(-0.38,-0.38)$) {$+$};

\node[tinytext, anchor=west] at ($(zaP)+(0.33,0.02)$) {$a+i\epsilon$};
\node[tinytext, anchor=west] at ($(zaM)+(0.33,-0.02)$) {$a-i\epsilon$};
\node[tinytext, anchor=west] at ($(zbP)+(0.33,0.02)$) {$b+i\epsilon$};
\node[tinytext, anchor=west] at ($(zbM)+(0.33,-0.02)$) {$b-i\epsilon$};

\node[formula, text width=7.15cm] at (4.12,1.50)
{$\begin{gathered}
\mathcal N_{\pmb A}(I)=\frac{1}{2\pi i}\lim_{\epsilon\downarrow0}\Delta_I\log D,\\[-1mm]
\Delta_I\log D=\log D(z_b^-)-\log D(z_b^+)-\log D(z_a^-)+\log D(z_a^+)
\end{gathered}$};

\node[formula, text width=7.15cm] at (4.12,0.55)
{two endpoints $a,b$ $\times$ two boundary values $z=\lambda\mp i\epsilon$};

\node[formula, text width=3.30cm] at (2.18,5.28)
{$\displaystyle Z_{\pmb A}(z)=D_{\pmb A}(z)^{-1/2}$};
\draw[maparrow] (3.83,5.18) -- (4.75,4.83);
\node[formula, text width=2.95cm] at (6.22,5.15)
{$\displaystyle
\left[\frac{Z(z_b^-)Z(z_a^+)}{Z(z_a^-)Z(z_b^+)}\right]^{-s/(\pi i)}$};

\draw[panel] (8.60,0) rectangle (17.20,6.25);
\node[ptitle] at (8.85,5.95) {(b) SCGF and Legendre geometry};

\begin{scope}[shift={(9.05,2.72)}]
  \draw[axis] (0,0) -- (3.30,0);
  \draw[axis] (1.25,-0.20) -- (1.25,2.35);
  \node[tinytext] at (3.38,-0.05) {$s$};
  \node[tinytext] at (1.58,2.25) {$\psi_I(s)$};
  \draw[curve] plot[smooth, tension=0.8] coordinates {(0.10,0.28) (0.75,0.42) (1.25,0.72) (1.85,1.23) (2.55,2.05) (3.05,2.55)};
  \coordinate (spt) at (2.05,1.48);
  \fill[black!70] (spt) circle (1.4pt);
  \draw[guide] (2.05,0) -- (spt);
  \node[tinytext] at (2.05,-0.25) {$s$};
  \draw[tanline] (1.38,0.82) -- (2.72,2.13);
  \node[tinytext, anchor=west] at (2.42,2.22) {slope $k=\psi_I'(s)$};
  \node[tinytext] at (1.45,-0.45) {tilt selects typical count $k(s)$};
\end{scope}

\draw[maparrow] (12.90,3.82) -- (13.70,3.82);
\node[tinytext] at (13.30,4.12) {$\sup_s\{sk-\psi_I(s)\}$};

\begin{scope}[shift={(13.95,2.72)}]
  \draw[axis] (0,0) -- (2.95,0);
  \draw[axis] (0.30,-0.20) -- (0.30,2.35);
  \node[tinytext] at (3.05,-0.05) {$k$};
  \node[tinytext] at (0.66,2.25) {$\Phi_I(k)$};
  \draw[curve] plot[smooth, tension=0.8] coordinates {(0.46,2.15) (0.82,0.96) (1.24,0.18) (1.52,0.00) (1.88,0.30) (2.28,1.10) (2.65,2.08)};
  \coordinate (kstar) at (1.52,0.00);
  \coordinate (kpt) at (2.02,0.55);
  \fill[black!70] (kstar) circle (1.3pt);
  \fill[black!70] (kpt) circle (1.4pt);
  \draw[guide] (kpt) -- (2.02,0);
  \node[tinytext] at (1.52,-0.25) {$k_I^\star$};
  \node[tinytext] at (2.02,-0.25) {$k(s)$};
  \node[tinytext, anchor=west] at (1.72,1.42) {$\Phi_I(k(s))=s k(s)-\psi_I(s)$};
\end{scope}

\node[formula, text width=7.70cm] at (12.90,1.08)
{$\displaystyle
\psi_I(s)=\lim_{N\to\infty}\frac{1}{N}\log\overline{e^{s\mathcal N_{\pmb A}(I)}} ,\qquad
\Phi_I(k)=\sup_s\{sk-\psi_I(s)\}$};
\node[formula, text width=7.70cm] at (12.90,0.34)
{$s=0$ gives the original ensemble; $s\neq0$ gives a determinant-biased ensemble of atypical counts.};
\end{tikzpicture}%
}
\caption{Determinant sectors and Legendre geometry for spectral-count large deviations. The interval count is represented by determinant phases at the two endpoints and the two boundary values $z=\lambda\mp i0^+$, equivalently by four Edwards--Jones partition-function sectors. Exponential tilting defines the scaled cumulant-generating function $\psi_I(s)$; when the Legendre transform is regular, its slope gives the tilted count $k(s)$ and $\Phi_I(k(s))=s\,k(s)-\psi_I(s)$.}
\label{fig:inld-determinant-sectors-legendre}
\end{figure}

Figure~\ref{fig:inld-determinant-sectors-legendre} summarizes the two structures that will recur below: the four determinant boundary sectors for interval counts and the Legendre geometry relating the tilted scaled cumulant-generating function to the spectral-count rate function.

Rather than carrying out the replica saddle here, we now express the same determinant structure in Bethe/cavity language, which is the form used by the population-dynamics implementation. The same structure can be understood from the Bethe determinant factorization. On a tree, or within the Bethe approximation on a locally tree-like graph, the determinant of $z\pmb I-\pmb A$ can be written in terms of cavity Green functions. For a sparse symmetric matrix
\begin{equation}
A_{ij}=D_i\delta_{ij}+C_{ij}J_{ij}\,,
\label{eq:inld-sparse-symmetric-model}
\end{equation}
the cavity Green functions satisfy
\begin{equation}
G_{i\to j}(z)=\frac{1}{z-D_i-\displaystyle\sum_{\ell\in\partial i\setminus j}J_{i\ell}^2G_{\ell\to i}(z)}\,,
\label{eq:inld-cavity-green-recalled}
\end{equation}
and the full local Green functions are
\begin{equation}
G_i(z)=\frac{1}{z-D_i-\displaystyle\sum_{\ell\in\partial i}J_{i\ell}^2G_{\ell\to i}(z)}\,.
\label{eq:inld-full-green-recalled}
\end{equation}
For a tree one has the exact identity
\begin{equation}
\det(z\pmb I-\pmb A)=\prod_{i=1}^NG_i(z)^{-1}\prod_{\{i,j\}\in E}\left[1-J_{ij}^2G_{i\to j}(z)G_{j\to i}(z)\right]^{-1}\,.
\label{eq:inld-bethe-determinant-factorization}
\end{equation}

\begin{examplebox}[Bethe determinant factorization for a single edge]
Consider the two-vertex weighted graph
\begin{equation}
\pmb A=\begin{pmatrix}
0 & J\\
J & 0
\end{pmatrix}\,.
\label{eq:inld-ped-edge-matrix}
\end{equation}
The exact determinant is
\begin{equation}
\det(z\pmb I-\pmb A)=z^2-J^2\,.
\label{eq:inld-ped-edge-det-exact}
\end{equation}
Let us verify the Bethe factorization. Since each vertex is a leaf in the cavity graph,
\begin{equation}
G_{1\to2}(z)=\frac{1}{z}\,,\qquad G_{2\to1}(z)=\frac{1}{z}\,.
\label{eq:inld-ped-edge-cavity}
\end{equation}
The full local Green functions are
\begin{equation}
G_1(z)=\frac{1}{z-J^2G_{2\to1}(z)}=\frac{z}{z^2-J^2}\,,\qquad G_2(z)=\frac{z}{z^2-J^2}\,.
\label{eq:inld-ped-edge-full-greens}
\end{equation}
Therefore
\begin{equation}
\prod_{i=1}^{2}G_i(z)^{-1}=\left(\frac{z^2-J^2}{z}\right)^2\,.
\label{eq:inld-ped-edge-site-product}
\end{equation}
The edge correction is
\begin{equation}
\left[1-J^2G_{1\to2}(z)G_{2\to1}(z)\right]^{-1}=\left[1-\frac{J^2}{z^2}\right]^{-1}=\frac{z^2}{z^2-J^2}\,.
\label{eq:inld-ped-edge-correction}
\end{equation}
Multiplying the site product and the edge correction gives
\begin{equation}
\prod_{i=1}^{2}G_i(z)^{-1}\left[1-J^2G_{1\to2}(z)G_{2\to1}(z)\right]^{-1}=z^2-J^2\,,
\label{eq:inld-ped-edge-bethe-result}
\end{equation}
which agrees with the exact determinant. The example shows explicitly why the edge factor appears: the product of site terms counts the interaction twice, and the edge correction removes the overcounting.
\end{examplebox}

The first product contains site contributions and the second product removes the double counting of edge contributions, exactly as in the Bethe free energy. Substituting \eqref{eq:inld-bethe-determinant-factorization} into \eqref{eq:inld-interval-count-log-det} gives the additive decomposition
\begin{equation}
\mathcal{N}_{\pmb A}(I)=\sum_{i=1}^N\nu_i(I)-\sum_{\{i,j\}\in E}\nu_{ij}(I)\,,
\label{eq:inld-count-bethe-decomposition}
\end{equation}
where
\begin{equation}
\nu_i(I)=\frac{1}{\pi}\lim_{\epsilon\downarrow0}{\rm Im}\left[\log G_i(b-i\epsilon)^{-1}-\log G_i(a-i\epsilon)^{-1}\right]\,,
\label{eq:inld-site-count-contribution}
\end{equation}
and
\begin{equation}
\begin{split}
\nu_{ij}(I)=\frac{1}{\pi}\lim_{\epsilon\downarrow0}{\rm Im}\Big[&\log\left(1-J_{ij}^2G_{i\to j}(b-i\epsilon)G_{j\to i}(b-i\epsilon)\right)\\
&-\log\left(1-J_{ij}^2G_{i\to j}(a-i\epsilon)G_{j\to i}(a-i\epsilon)\right)\Big]\,.
\label{eq:inld-edge-count-contribution}
\end{split}
\end{equation}
The logarithms in \eqref{eq:inld-site-count-contribution} and \eqref{eq:inld-edge-count-contribution} must be evaluated with branches continuous along the lower-half-plane path from $a-i\epsilon$ to $b-i\epsilon$, consistently with \eqref{eq:inld-interval-count-log-det}. Otherwise the individual site and edge phases are ambiguous even though their Bethe site-minus-edge combination reproduces the determinant phase. This representation has two important consequences. First, it shows explicitly why the spectral count is an extensive observable on a locally tree-like sparse graph: it is a sum of local site and edge terms. Second, it shows why count fluctuations are controlled by the joint distribution of cavity messages at the two endpoints $a$ and $b$. The typical density requires the one-point distribution of a local resolvent at one spectral parameter. The interval-count problem requires the joint law of resolvent messages at two spectral parameters, together with the phase information of the logarithms.

At the ensemble level, the large-deviation calculation introduces a tilted measure
\begin{equation}
P_s(\pmb A)=\frac{P(\pmb A)\exp\left[s\mathcal{N}_{\pmb A}(I)\right]}{\overline{\exp\left[s \mathcal{N}_{\pmb A}(I)\right]}}\,.
\label{eq:inld-tilted-ensemble}
\end{equation}

\begin{examplebox}[Soft tilting of a two-matrix ensemble]
Consider an ensemble consisting of only two possible matrices, $\pmb A_1$ and $\pmb A_2$, each with probability $1/2$. Suppose that for a fixed interval $I$,
\begin{equation}
\mathcal N_{\pmb A_1}(I)=n_1\,,\qquad\mathcal N_{\pmb A_2}(I)=n_2\,.
\label{eq:inld-ped-two-matrix-counts}
\end{equation}
The tilted ensemble is
\begin{equation}
P_s(\pmb A)=\frac{P(\pmb A)e^{s\mathcal N_{\pmb A}(I)}}{\overline{e^{s\mathcal N_{\pmb A}(I)}}}\,.
    \label{eq:inld-ped-two-matrix-tilt}
\end{equation}
Thus
\begin{equation}
P_s(\pmb A_1)=\frac{e^{s n_1}}{e^{s n_1}+e^{s n_2}}\,,\qquad P_s(\pmb A_2)=\frac{e^{s n_2}}{e^{s n_1}+e^{s n_2}}\,.
    \label{eq:inld-ped-two-matrix-probabilities}
\end{equation}
If $s>0$, the tilted ensemble favors the matrix with the larger count. If $s<0$, it favors the matrix with the smaller count. The scaled cumulant-generating function in this finite example is
\begin{equation}
\psi_N(s)=\frac{1}{N}\log\left[\frac{1}{2}e^{s n_1}+\frac{1}{2}e^{s n_2}\right]\,.
\label{eq:inld-ped-two-matrix-cgf}
\end{equation}
This toy example illustrates the meaning of the tilted ensemble: it is not a new structural random-matrix model with modified entries, but the original ensemble biased according to a spectral observable.
\end{examplebox}

The derivative of the scaled cumulant-generating function gives the typical count in this tilted ensemble:
\begin{equation}
k_I(s)=\psi_I'(s)=\lim_{N\to\infty}\frac{1}{N}\left\langle\mathcal{N}_{\pmb A}(I)\right\rangle_s\,,
\label{eq:inld-tilted-typical-count}
\end{equation}
where $\langle\cdots\rangle_s$ denotes the average with respect to $P_s$. The rate function is then parametrically obtained as
\begin{equation}
\Phi_I(k_I(s))=s k_I(s)-\psi_I(s)\,.
\label{eq:inld-parametric-rate-function}
\end{equation}
The tilted ensemble is not obtained by simply changing the mean degree, the edge-weight distribution, or the interval endpoints. It is an ensemble of matrices conditioned, softly through the field $s$, to have an atypical number of eigenvalues in $I$. This is why the large-deviation problem is substantially richer than the computation of the typical spectral density.

For sparse Erd\H{o}s--R\'enyi graphs or random regular graphs, the replica-symmetric cavity solution of the tilted problem is implemented by population dynamics. Instead of storing a population of scalar messages $G(z)$ at one spectral parameter, one stores messages carrying the information needed at both endpoints of the interval. In a schematic notation, a cavity message is of the form
\begin{equation}
\mathcal{M}=\left(G(a-i\epsilon),G(b-i\epsilon);\text{branch and normalization data}\right)\,,
\label{eq:inld-tilted-message}
\end{equation}
where the additional data encode the local site and edge contributions entering the determinant representation of the count. The population update uses the ordinary resolvent recursion \eqref{eq:inld-cavity-green-recalled}, but the new message is accepted or reweighted according to the local contribution to \eqref{eq:inld-count-generating-partitions}, equivalently to the Bethe decomposition \eqref{eq:inld-count-bethe-decomposition}. At $s=0$ this reweighting disappears and the usual population dynamics for the spectral density is recovered. For $s\neq0$, the population is biased toward local environments that increase or decrease the number of eigenvalues in the interval, depending on the sign of $s$. This is the operational content of the method developed in \cite{MetzPerezCastillo2016}.

The first derivative at $s=0$ gives the average fraction of eigenvalues in the interval, which is a check against the integrated density:
\begin{equation}
\psi_I'(0)=\int_a^b d\lambda \overline{\rho_{\pmb A}(\lambda)}\,.
\label{eq:inld-first-derivative-check}
\end{equation}
The second derivative gives the asymptotic variance per vertex,
\begin{equation}
\psi_I''(0)=\lim_{N\to\infty}\frac{1}{N}\left[\overline{\mathcal{N}_{\pmb A}(I)^2}-\overline{\mathcal{N}_{\pmb A}(I)}^2\right]\,.
\label{eq:inld-second-derivative-check}
\end{equation}
Thus the curvature of the rate function at its minimum is
\begin{equation}
\Phi_I''(k_I^\star)=\frac{1}{\psi_I''(0)}\,,
\label{eq:inld-rate-curvature}
\end{equation}
provided the variance is nonzero and the Legendre transform is locally regular. These relations are useful because they connect the large-deviation theory with ordinary count fluctuations.

One physical application is the Anderson model on random regular graphs. In its simplest form, the operator is
\begin{equation}
H_{ij}=C_{ij}+V_i\delta_{ij}\,,
\label{eq:inld-anderson-operator}
\end{equation}
where $\pmb C$ is the adjacency matrix of a random regular graph and $V_i$ are independent random on-site potentials. The usual cavity equations describe the local density of states and localization properties of this operator. The spectral-count formalism gives additional information: it computes the variance and higher cumulants of the number of eigenvalues in an interval. This provides access to level compressibility, which is a diagnostic of spectral statistics and is sensitive to localization and delocalization properties. The application of the sparse-matrix large-deviation formalism to this problem was developed in \cite{MetzPerezCastillo2017}, where eigenvalue-count fluctuations were used to probe spectral statistics in the extended phase of the Anderson model on random regular graphs.

It is important to distinguish three levels of description. The first is the typical spectral density, obtained from the ordinary cavity distribution of local Green functions. The second is the central fluctuation regime of $\mathcal{N}_{\pmb A}(I)$ around its typical value, governed by the cumulants $\psi_I^{(n)}(0)$. The third is the large-deviation regime, where $k_I$ is constrained to differ by an amount of order one from $k_I^\star$ and the probability is controlled by the full rate function $\Phi_I(k)$. The first level is a one-point resolvent problem. The second and third levels are biased two-endpoint determinant problems. This distinction is essential: one cannot reconstruct the large-deviation function from the average density alone.

The sparse large-deviation problem also differs conceptually from the invariant Coulomb-gas problem. In invariant Gaussian ensembles, imposing an atypical index forces a collective rearrangement of a strongly interacting gas of eigenvalues. The corresponding optimal density is found by minimizing an electrostatic energy, and the large-deviation speed is $N^2$ \cite{DeanMajumdar2006,MajumdarNadalScardicchioVivo2009}. For the sparse non-invariant ensembles considered here, one does not have, in general, a closed Coulomb-gas energy for the eigenvalues. The natural variational object is instead a distribution of local cavity messages under a tilted graphical measure. This change of order parameter reflects the basic theme of these notes: in finite connectivity, the graph-local environment remains visible in spectral observables.

Several consistency checks are immediate. If the interval is the whole real line, then $\mathcal{N}_{\pmb A}(I)=N$ deterministically, so
\begin{equation}
\psi_I(s)=s\,,\qquad\Phi_I(k)=
\begin{cases}
0, & k=1\,,\\
+\infty, & k\neq1\,.
\end{cases}
\label{eq:inld-whole-line-check}
\end{equation}
If the interval is empty, then $\mathcal{N}_{\pmb A}(I)=0$ and $\psi_I(s)=0$. If $s=0$, the tilted ensemble \eqref{eq:inld-tilted-ensemble} reduces to the original random matrix ensemble, and the first derivative \eqref{eq:inld-first-derivative-check} recovers the integrated spectral density. Finally, if the graph decomposes into disconnected components, the count is the sum of the counts of the components, which is consistent with the additive Bethe representation \eqref{eq:inld-count-bethe-decomposition}.

The conclusion is that spectral-count statistics extend the cavity theory of sparse spectra from typical densities to fluctuations and rare events. The key identity is the determinant representation of the index number, \eqref{eq:inld-index-from-log-det}, and its interval version, \eqref{eq:inld-interval-count-log-det}. These identities turn the number of eigenvalues in an interval into a phase of a Gaussian partition function. The resulting large-deviation theory is naturally formulated through a replicated or tilted cavity method, whose order parameter is a distribution of two-endpoint resolvent messages. This framework will be reused in the next section for diluted Wishart matrices, where the positivity and bipartite structure of the covariance ensemble lead to a related but distinct spectral-count large-deviation problem.

\begin{exerciseblock}
\exitem[Index number from eigenvalues]
Let
\begin{equation}
\pmb A=\begin{pmatrix}
-2 & 0 & 0\\
0 & 1 & 0\\
0 & 0 & 3
\end{pmatrix}\,.
\label{eq:inld-ex-diagonal-A}
\end{equation}
Compute $\mathcal K_{\pmb A}(\lambda)$ for $\lambda=-3,0,2,4$. Draw the corresponding step function.

\exitem[Interval count as a difference of indices]
For the same matrix, compute
\begin{equation}
\mathcal N_{\pmb A}((a,b])=\mathcal K_{\pmb A}(b)-\mathcal K_{\pmb A}(a)
\label{eq:inld-ex-interval-difference}
\end{equation}
for $(a,b]=(-1,2]$ and $(a,b]=(0,4]$. Discuss how the answer changes, or does not change, under other conventions for eigenvalues exactly at the endpoints.

\exitem[Logarithmic representation of the Heaviside function]
Verify, for $x\neq0$, the identity
\begin{equation}
\Theta(x)=1+\frac{1}{2\pi i}\lim_{\epsilon\downarrow0}\left[\log(x-i\epsilon)-\log(x+i\epsilon)\right]\,.
\label{eq:inld-ex-step-log}
\end{equation}
Check separately the cases $x>0$ and $x<0$ using the principal branch of the logarithm.

\exitem[Index from the determinant]
Assuming no eigenvalue lies exactly at the threshold $\lambda$, or fixing an endpoint convention, starting from the previous exercise, derive
\begin{equation}
\mathcal K_{\pmb A}(\lambda)=N+\frac{1}{\pi}\lim_{\epsilon\downarrow0}{\rm Im}\log\det[(\lambda-i\epsilon)\pmb I-\pmb A]\,.
\label{eq:inld-ex-index-det}
\end{equation}

\exitem[Interval count from determinant phases]
Derive
\begin{equation}
\mathcal N_{\pmb A}((a,b])=\frac{1}{\pi}\lim_{\epsilon\downarrow0}{\rm Im}\left\{\log\det\left[(b-i\epsilon)\pmb I-\pmb A\right]-\log\det\left[(a-i\epsilon)\pmb I-\pmb A\right]\right\}\,.
\label{eq:inld-ex-interval-det}
\end{equation}

\exitem[Resolvent representation of the interval count]
Starting from
\begin{equation}
\frac{\partial}{\partial\lambda}\log\det[(\lambda-i\epsilon)\pmb I-\pmb A]={\rm Tr}[(\lambda-i\epsilon)\pmb I-\pmb A]^{-1}\,,
\label{eq:inld-ex-logdet-derivative}
\end{equation}
derive, with the same endpoint convention as in the previous exercise,
\begin{equation}
\mathcal N_{\pmb A}((a,b])=\frac{1}{\pi}\lim_{\epsilon\downarrow0}{\rm Im}\int_a^b d\lambda {\rm Tr}[(\lambda-i\epsilon)\pmb I-\pmb A]^{-1}\,.
\label{eq:inld-ex-resolvent-count}
\end{equation}

\exitem[Scaled cumulant-generating function]
Let
\begin{equation}
\psi_I(s)=\lim_{N\to\infty}\frac{1}{N}\log\overline{e^{s\mathcal N_{\pmb A}(I)}}\,.
\label{eq:inld-ex-cgf}
\end{equation}
Show formally that
\begin{equation}
\psi_I'(0)=\lim_{N\to\infty}\frac{1}{N}\overline{\mathcal N_{\pmb A}(I)}
\label{eq:inld-ex-first-derivative}
\end{equation}
and
\begin{equation}
\psi_I''(0)=\lim_{N\to\infty}\frac{1}{N}{\rm Var}\mathcal N_{\pmb A}(I)\,.
\label{eq:inld-ex-second-derivative}
\end{equation}

\exitem[Legendre transform in the Bernoulli model]
For the Bernoulli count model in the example \emph{A toy large-deviation calculation for independent local counts}, derive the rate function
\begin{equation}
\Phi(k)=k\log\frac{k}{p}+(1-k)\log\frac{1-k}{1-p}\,.
\label{eq:inld-ex-bernoulli-rate}
\end{equation}
Verify that $\Phi(k)$ is minimized at $k=p$.

\exitem[Curvature of the rate function]
For the Bernoulli rate function derived above, compute $\Phi''(p)$ and compare it with $1/\psi''(0)$. This verifies the general relation
\begin{equation}
\Phi''(k^\star)=\frac{1}{\psi''(0)}
\label{eq:inld-ex-curvature-relation}
\end{equation}
in a simple case.

\exitem[Edwards--Jones partition functions and determinant powers]
Using
\begin{equation}
Z_{\pmb A}(z)=D_{\pmb A}(z)^{-1/2}\,,
\label{eq:inld-ex-Z-D}
\end{equation}
up to $z$-independent factors, derive the four-partition-function representation of $e^{s\mathcal N_{\pmb A}((a,b])}$.

\exitem[Why four replica sectors appear]
Explain why the interval-count generating function involves spectral parameters
\begin{equation}
a-i\epsilon\,,\qquad a+i\epsilon\,, \qquad b-i\epsilon\,, \qquad b+i\epsilon\,.
\label{eq:inld-ex-four-sectors}
\end{equation}
What is the meaning of each boundary value?

\exitem[Bethe determinant on a two-site graph]
For
\begin{equation}
\pmb A=\begin{pmatrix}
0 & J\\
J & 0
\end{pmatrix}\,,
\label{eq:inld-ex-two-site-A}
\end{equation}
verify directly that the Bethe determinant factorization gives
\begin{equation}
\det(z\pmb I-\pmb A)=z^2-J^2\,.
\label{eq:inld-ex-two-site-det}
\end{equation}

\exitem[Site and edge index contributions]
For the same two-site graph, assume $J>0$ and choose an interval $I=(a,b]$ such that $0<a<J<b$. Use \eqref{eq:inld-site-count-contribution} and \eqref{eq:inld-edge-count-contribution} to compute the site contributions $\nu_i(I)$ and the edge contribution $\nu_{12}(I)$. Verify that
\begin{equation}
\mathcal N_{\pmb A}(I)=1\,.
\label{eq:inld-ex-two-site-count}
\end{equation}

\exitem[Whole-line and empty-interval checks]
Show that if $I=\mathbb R$, then
\begin{equation}
\mathcal N_{\pmb A}(I)=N\,,\qquad\psi_I(s)=s\,.
\label{eq:inld-ex-whole-line}
\end{equation}
If $I=\emptyset$, show that
\begin{equation}
\mathcal N_{\pmb A}(I)=0\,,\qquad \psi_I(s)=0\,.
\label{eq:inld-ex-empty-interval}
\end{equation}

\exitem[Tilted ensemble]
Given
\begin{equation}
P_s(\pmb A)=\frac{P(\pmb A)e^{s\mathcal N_{\pmb A}(I)}}{\overline{e^{s\mathcal N_{\pmb A}(I)}}}\,,
\label{eq:inld-ex-tilted-ensemble}
\end{equation}
show that
\begin{equation}
\frac{d}{ds}\psi_I(s)=\lim_{N\to\infty}\frac{1}{N}\left\langle\mathcal N_{\pmb A}(I)\right\rangle_s\,.
\label{eq:inld-ex-tilted-mean}
\end{equation}

\exitem[Why the tilted ensemble is not simply a new matrix ensemble]
Explain why $P_s(\pmb A)$ cannot generally be obtained by simply changing the mean degree or edge-weight distribution of the original sparse ensemble. What spectral information enters the definition of $P_s$?

\exitem[Anderson model index statistics]
For the Anderson-type operator
\begin{equation}
H_{ij}=C_{ij}+V_i\delta_{ij}\,,
\label{eq:inld-ex-anderson}
\end{equation}
explain what physical information is contained in the number of eigenvalues in an interval $I$. How is this related to level compressibility?

\exitem[Programming exercise: empirical cumulant-generating function]
Fix a list of system sizes $N$, a mean degree $c=O(1)$, a number $M$ of independent samples for each $N$, an interval $I=(a,b]$ with a stated endpoint convention, and a grid of real values of $s$ containing $s=0$. Generate $M$ independent sparse Erd\H{o}s--R\'enyi adjacency matrices $\pmb A^{(m)}$ with zero diagonal entries, $A_{ij}^{(m)}=A_{ji}^{(m)}$, and
\begin{equation}
{\rm Prob}(A_{ij}^{(m)}=1)=\frac{c}{N}\,, \qquad {\rm Prob}(A_{ij}^{(m)}=0)=1-\frac{c}{N}\,,
\qquad i<j\,.
\label{eq:inld-ex-program-er-law}
\end{equation}
For each sample compute the count $\mathcal N_m(I)$, and estimate
\begin{equation}
\widehat\psi_{I,N}(s)=\frac{1}{N}\log\left[\frac{1}{M}\sum_{m=1}^{M}e^{s\mathcal N_m(I)}\right]\,.
\label{eq:inld-ex-program-cgf}
\end{equation}
Compare $\widehat\psi'_{I,N}(0)$, estimated from the same $s$ grid, with the empirical average count divided by $N$. Report $N$, $c$, $M$, the interval $I$, the endpoint convention, the $s$ grid, and the numerical differentiation rule used near $s=0$.

\exitem[Programming exercise: count variance]
For the same Erd\H{o}s--R\'enyi adjacency ensemble and the same interval $I=(a,b]$, fix a list of system sizes $N$ and a number $M$ of independent samples for each $N$. Estimate
\begin{equation}
\frac{1}{N}{\rm Var}\mathcal N_{\pmb A}(I)
\label{eq:inld-ex-program-variance}
\end{equation}
from the $M$ empirical counts at each $N$. Compare this quantity with the second derivative of the empirical cumulant-generating function near $s=0$, using the same samples and the same endpoint convention as in the previous exercise. Report $N$, $M$, $c$, $I$, the endpoint convention, and the finite-difference rule used for $\widehat\psi''_{I,N}(0)$.

\exitem[Programming exercise: small-connectivity component check]
Fix a small mean degree $c$, a list of system sizes $N$, a number $M$ of independent samples for each $N$, and three intervals $I_0$, $I_+$, and $I_-$ containing respectively $0$, $1$, and $-1$, with endpoint conventions stated explicitly. Generate sparse Erd\H{o}s--R\'enyi adjacency matrices $\pmb A$ with $A_{ii}=0$, $A_{ij}=A_{ji}$, and ${\rm Prob}(A_{ij}=1)=c/N$ for $i<j$. For each sample, identify isolated vertices and isolated edges. In the component approximation, each isolated vertex contributes one eigenvalue at $0$, and each isolated edge contributes one eigenvalue at $1$ and one eigenvalue at $-1$. Use this approximation to estimate the counts in $I_0$, $I_+$, and $I_-$. Compare these approximate counts with direct diagonalization of the same samples. Report $N$, $M$, $c$, the intervals, the endpoint convention, and the discrepancy between the component approximation and direct diagonalization.
\end{exerciseblock}

\section{Large deviations for diluted Wishart matrices}
\label{sec:large-deviations-diluted-wishart}
We now apply the spectral-count formalism to diluted Wishart matrices. This section is deliberately more explicit than the preceding overview, because the diluted Wishart case is the first example in which the spectral-count large-deviation problem combines two structures at once: the determinant representation of the index number and the bipartite factor-graph structure of sparse covariance matrices. The main reference for this problem is \cite{PerezCastilloMetz2018Wishart}, which extends to diluted Wishart matrices the sparse-graph index-number formalism introduced in \cite{MetzPerezCastillo2016}. The comparison with invariant random matrices is useful but secondary: for Gaussian invariant ensembles the index large deviations are naturally formulated as a Coulomb-gas problem with speed $N^2$ \cite{DeanMajumdar2006,MajumdarNadalScardicchioVivo2009}, whereas for finite-connectivity Wishart ensembles the natural object is a distribution of local cavity messages and the large-deviation speed is $N$.

We consider again an $N\times P$ diluted rectangular matrix $\pmb X$, with
\begin{equation}
P=\frac{N}{\alpha},\qquad\alpha>0\,,
\label{eq:lddw-rectangularity}
\end{equation}
and entries
\begin{equation}
X_i^\mu=B_i^\mu \xi_i^\mu\,,\qquad{\rm Prob}(B_i^\mu=1)=\frac{d}{N}\,,\qquad{\rm Prob}(B_i^\mu=0)=1-\frac{d}{N}\,.
\label{eq:lddw-diluted-rectangular-entries}
\end{equation}
The nonzero weights $\xi_i^\mu$ are independent random variables with distribution $p_\xi(\xi)$. The associated diluted Wishart matrix is
\begin{equation}
\pmb W=\frac{1}{d}\pmb X\pmb X^{\rm T}\,,\qquad W_{ij} = \frac{1}{d} \sum_{\mu=1}^P X_i^\mu X_j^\mu\,.
\label{eq:lddw-diluted-wishart}
\end{equation}
The matrix $\pmb W$ is real symmetric and positive semidefinite. In the dense limit $d\to\infty$, with the nonzero weights normalized so that the dense covariance scaling has unit second moment, its spectral density reduces to the Mar\v{c}enko--Pastur law, after accounting for the normalization convention \cite{Wishart1928,MarchenkoPastur1967}. At finite $d$, however, the sparse bipartite graph underlying $\pmb X$ remains visible, and the spectral statistics are controlled by finite-connectivity cavity equations \cite{NagaoTanaka2007,RogersTakedaPerezCastilloKuhn2008}.

Let $\lambda_1,\ldots,\lambda_N$ be the eigenvalues of $\pmb W$. Since $\pmb W$ is positive semidefinite, $\lambda_i\geq0$. For a threshold $\lambda\in\mathbb{R}$, define the Wishart index number
\begin{equation}
\mathcal{K}_{\pmb W}(\lambda)=\sum_{i=1}^N\Theta(\lambda-\lambda_i).
\label{eq:lddw-index-number}
\end{equation}
This is the number of eigenvalues of $\pmb W$ below the threshold $\lambda$, with eigenvalues exactly at the threshold treated separately, or by a fixed endpoint convention. For $\lambda<0$ it is identically zero. For $\lambda$ larger than the top of the spectrum it is $N$. For intermediate values, $\mathcal{K}_{\pmb W}(\lambda)$ fluctuates from sample to sample. For $\lambda>0$, its typical intensive value is
\begin{equation}
k_\star(\lambda)=\lim_{N\to\infty}\frac{1}{N}\overline{\mathcal{K}_{\pmb W}(\lambda)}=\int_0^\lambda d\lambda'\overline{\rho_{\pmb W}(\lambda')}\,,
\label{eq:lddw-typical-index}
\end{equation}
where the lower limit includes the zero-mode atom according to the same endpoint convention, and the expression is understood whenever the limiting density has no atom at the threshold. The problem of this section is not merely to compute $k_\star(\lambda)$, but to compute the probability of observing an atypical value
\begin{equation}
k=\frac{1}{N}\mathcal{K}_{\pmb W}(\lambda)\,.
\label{eq:lddw-intensive-index}
\end{equation}

The large-deviation principle is written as
\begin{equation}
{\rm Prob}\left[\frac{1}{N}\mathcal{K}_{\pmb W}(\lambda)= k\right]\asymp\exp\left[-N\Phi_\lambda(k)\right]\,,
\label{eq:lddw-ldp}
\end{equation}
where $\Phi_\lambda(k)$ is the rate function. The speed is $N$ because the sparse matrix is controlled by $O(N)$ local graph variables. This is different from the $N^2$ speed of invariant dense ensembles, where changing the index requires rearranging a Coulomb gas of $N$ strongly repelling eigenvalues. The corresponding scaled cumulant-generating function is
\begin{equation}
\psi_\lambda(s)=\lim_{N\to\infty}\frac{1}{N}\log\overline{\exp\left[s\mathcal{K}_{\pmb W}(\lambda)\right]}\,.
\label{eq:lddw-scaled-cgf}
\end{equation}
When the Legendre transform is regular,
\begin{equation}
\Phi_\lambda(k)=\sup_s\left\{sk-\psi_\lambda(s)\right\}\,.
\label{eq:lddw-rate-legendre}
\end{equation}
The derivatives at $s=0$ give the cumulants per degree of freedom:
\begin{equation}
\psi_\lambda'(0)=\lim_{N\to\infty}\frac{1}{N}\overline{\mathcal{K}_{\pmb W}(\lambda)},\qquad\psi_\lambda''(0)=\lim_{N\to\infty}\frac{1}{N}{\rm Var}\mathcal{K}_{\pmb W}(\lambda)\,,
\label{eq:lddw-first-cumulants}
\end{equation}
and analogously for higher cumulants. Thus the same object $\psi_\lambda(s)$ contains both central fluctuations and large deviations.

We first derive the determinant representation of the index. Introduce
\begin{equation}
z_-=\lambda-i\epsilon\,, \qquad z_+=\lambda+i\epsilon\,, \qquad \epsilon>0\,,
\label{eq:lddw-zpm}
\end{equation}
and define
\begin{equation}
D_{\pmb W}(z)=\det(z\pmb I-\pmb W)=\prod_{i=1}^N(z-\lambda_i)\,.
\label{eq:lddw-characteristic-det}
\end{equation}
For a real number $x\neq0$, with the principal branch of the logarithm,
\begin{equation}
\Theta(x)=1+\frac{1}{2\pi i}\lim_{\epsilon\downarrow0}\left[\log(x-i\epsilon)-\log(x+i\epsilon)\right]\,.
\label{eq:lddw-step-log-identity}
\end{equation}
Indeed, if $x>0$ the two logarithms have the same limiting phase and the second term vanishes; if $x<0$ the limiting phases are $-\pi$ and $+\pi$, so the second term is $-1$. Applying \eqref{eq:lddw-step-log-identity} to $x=\lambda-\lambda_i$ and summing over $i$ gives
\begin{equation}
\mathcal{K}_{\pmb W}(\lambda)=N+\frac{1}{2\pi i}\lim_{\epsilon\downarrow0}\left[\log D_{\pmb W}(z_-)-\log D_{\pmb W}(z_+)\right]\,.
\label{eq:lddw-index-log-det}
\end{equation}
This equation is the starting point of the large-deviation calculation. It expresses a discontinuous spectral count as the phase difference of two determinants evaluated on the two sides of the real axis.

\begin{examplebox}[The Wishart index from the phase of a determinant]
Consider the one-dimensional diluted Wishart matrix
\begin{equation}
\pmb W=(w)\,,\qquad w\geq0\,.
\label{eq:lddw-ped-one-dimensional-W}
\end{equation}
Its index below a threshold $\lambda$ is simply
\begin{equation}
\mathcal K_{\pmb W}(\lambda)=\Theta(\lambda-w)\,.
\label{eq:lddw-ped-one-dimensional-index-direct}
\end{equation}
Let us recover this from the determinant representation. The characteristic determinant is
\begin{equation}
D_{\pmb W}(z)=z-w\,.
\label{eq:lddw-ped-one-dimensional-det}
\end{equation}
Equation \eqref{eq:lddw-index-log-det} gives, for $N=1$,
\begin{equation}
\mathcal K_{\pmb W}(\lambda)=1+\frac{1}{2\pi i}\lim_{\epsilon\downarrow0}\left[\log(\lambda-w-i\epsilon)-\log(\lambda-w+i\epsilon)\right]\,.
\label{eq:lddw-ped-one-dimensional-index-log}
\end{equation}
There are two cases. If $\lambda>w$, then $\lambda-w>0$, and the two logarithms have the same limiting phase. Hence
\begin{equation}
\mathcal K_{\pmb W}(\lambda)=1\,.
\label{eq:lddw-ped-one-dimensional-positive-case}
\end{equation}
If $\lambda<w$, then $\lambda-w<0$. With the principal branch,
\begin{equation}
\log(\lambda-w-i0^+)=\log|\lambda-w|-i\pi\,,\qquad\log(\lambda-w+i0^+)=\log|\lambda-w|+i\pi\,.
\label{eq:lddw-ped-one-dimensional-negative-logs}
\end{equation}
Therefore
\begin{equation}
\frac{1}{2\pi i}\left[\log(\lambda-w-i0^+)-\log(\lambda-w+i0^+)\right]=-1\,,
\label{eq:lddw-ped-one-dimensional-negative-phase}
\end{equation}
and
\begin{equation}
\mathcal K_{\pmb W}(\lambda)=0\,.
\label{eq:lddw-ped-one-dimensional-negative-case}
\end{equation}
Thus the determinant phase formula reproduces the elementary step function. This is the simplest version of the mechanism used later for large deviations: the eigenvalue count is encoded in a discontinuity of the logarithm across the real axis.
\end{examplebox}

We now reuse the Edwards--Jones representation in the spectral-count setting. The Gaussian integral is the same object used for the density, but here its logarithmic phase generates the index. Define
\begin{equation}
Z_{\pmb W}(z)=\int\left[\prod_{i=1}^N\frac{du_i}{\sqrt{2\pi}}\right]\exp\left[-\frac{i}{2}\pmb u^{\rm T}(z\pmb I-\pmb W)\pmb u\right]\,.
\label{eq:lddw-edwards-jones}
\end{equation}
For ${\rm Im} z<0$ the integral is convergent, and for the conjugate value $z_+$ it is understood by analytic continuation. Up to a $z$-independent phase,
\begin{equation}
Z_{\pmb W}(z)=\det\left[i(z\pmb I-\pmb W)\right]^{-1/2}\,.
\label{eq:lddw-z-det}
\end{equation}
The constant phase cancels in the difference between $z_-$ and $z_+$. Therefore
\begin{equation}
\mathcal{K}_{\pmb W}(\lambda)=N-\frac{1}{\pi i}\lim_{\epsilon\downarrow0}\left[\log Z_{\pmb W}(z_-)-\log Z_{\pmb W}(z_+)\right]\,.
\label{eq:lddw-index-partition-functions}
\end{equation}
Exponentiating this identity gives
\begin{equation}
\exp\left[s\mathcal{K}_{\pmb W}(\lambda)\right]=e^{sN}\lim_{\epsilon\downarrow0}\left[Z_{\pmb W}(z_-)\right]^{-s/(\pi i)}\left[Z_{\pmb W}(z_+)\right]^{s/(\pi i)}\,.
\label{eq:lddw-exp-index-partition-functions}
\end{equation}
The powers in \eqref{eq:lddw-exp-index-partition-functions} are complex when $s$ is real and must be understood through the same logarithm branches used in the determinant-phase identity. This is the origin of the replica continuation in the index problem.

The bipartite structure of the diluted Wishart ensemble enters when we substitute \eqref{eq:lddw-diluted-wishart} into the quadratic form. We have
\begin{equation}
\pmb u^{\rm T}\pmb W\pmb u=\frac{1}{d}\sum_{\mu=1}^P\left(\sum_{i=1}^NX_i^\mu u_i\right)^2=\frac{1}{d}\sum_{\mu=1}^P\left(\sum_{i\in\partial\mu}\xi_i^\mu u_i\right)^2\,,
\label{eq:lddw-wishart-quadratic-form}
\end{equation}
where $\partial\mu$ is the set of variable nodes connected to factor node $\mu$ in the bipartite graph. Hence
\begin{equation}
Z_{\pmb W}(z)=\int\left[\prod_{i=1}^N\frac{du_i}{\sqrt{2\pi}}\right]\exp\left[-\frac{i}{2}z\sum_{i=1}^N u_i^2+\frac{i}{2d}\sum_{\mu=1}^P\left(\sum_{i\in\partial\mu}\xi_i^\mu u_i\right)^2\right]\,.
\label{eq:lddw-bipartite-gaussian}
\end{equation}
This is a Gaussian factor graph. Variable nodes carry the quadratic term $-iz u_i^2/2$, and factor nodes carry rank-one interactions. The index number is therefore the phase difference of two Gaussian factor-graph partition functions, one at $z_-$ and one at $z_+$.

\begin{examplebox}[The bipartite Gaussian factor for a single sample node]
Take $N=2$, $P=1$, and $d=1$, so that
\begin{equation}
\pmb X=\begin{pmatrix}
\xi_1\\
\xi_2
\end{pmatrix},\qquad\pmb W=\pmb X\pmb X^{\rm T}\,.
\label{eq:lddw-ped-single-factor-X}
\end{equation}
Then
\begin{equation}
\pmb W=\begin{pmatrix}
\xi_1^2 & \xi_1\xi_2\\
\xi_1\xi_2 & \xi_2^2
\end{pmatrix}\,.
\label{eq:lddw-ped-single-factor-W}
\end{equation}
For a Gaussian integration variable $\pmb u=(u_1,u_2)^{\rm T}$,
\begin{align}
\pmb u^{\rm T}\pmb W\pmb u&=\xi_1^2u_1^2+2\xi_1\xi_2u_1u_2+\xi_2^2u_2^2\nonumber\\
&=\left(\xi_1u_1+\xi_2u_2\right)^2\,.
\label{eq:lddw-ped-single-factor-quadratic}
\end{align}
Therefore the Edwards--Jones partition function is
\begin{equation}
Z_{\pmb W}(z)=\int\frac{du_1du_2}{2\pi}\exp\left[-\frac{i}{2}z(u_1^2+u_2^2)+\frac{i}{2}\left(\xi_1u_1+\xi_2u_2\right)^2\right]\,.
    \label{eq:lddw-ped-single-factor-partition}
\end{equation}
This is the simplest example of the bipartite factor structure. The factor node $\mu=1$ does not create an arbitrary interaction among $u_1$ and $u_2$; it creates the square of one linear combination of the variables. This rank-one structure is what makes the diluted Wishart cavity equations close on scalar self-energies.

The eigenvalues of \eqref{eq:lddw-ped-single-factor-W} are
\begin{equation}
\lambda_1=\xi_1^2+\xi_2^2\,,\qquad\lambda_2=0\,.
\label{eq:lddw-ped-single-factor-eigenvalues}
\end{equation}
Thus the index below a positive threshold $\lambda$ has the following form:
\begin{equation}
\mathcal K_{\pmb W}(\lambda)=\begin{cases}
1, & 0<\lambda<\xi_1^2+\xi_2^2,\\
2, & \lambda>\xi_1^2+\xi_2^2,
\end{cases}
\label{eq:lddw-ped-single-factor-index}
\end{equation}
with the zero mode contributing immediately for any positive threshold. This illustrates why zero modes and rank constraints are central in the diluted Wishart index problem.
\end{examplebox}

The replica representation follows directly. For positive integers $n_-$ and $n_+$, introduce
\begin{equation}
\mathcal{Q}_N(n_-,n_+)=\overline{\left[Z_{\pmb W}(z_-)\right]^{n_-}\left[Z_{\pmb W}(z_+)\right]^{n_+}}\,.
\label{eq:lddw-replicated-generating-function}
\end{equation}
The desired scaled cumulant-generating function is obtained by the analytic continuation
\begin{equation}
n_-=-\frac{s}{\pi i}\,, \qquad n_+=\frac{s}{\pi i}\,,
\label{eq:lddw-replica-continuation}
\end{equation}
namely
\begin{equation}
\psi_\lambda(s)=s+\lim_{\epsilon\downarrow0}\lim_{N\to\infty}\frac{1}{N}\log\mathcal{Q}_N\left(-\frac{s}{\pi i},\frac{s}{\pi i}\right)\,.
\label{eq:lddw-cgf-replica}
\end{equation}

\begin{examplebox}[Why the replica numbers become complex]
The index is a phase of a determinant. This is why the replica numbers in the large-deviation calculation are analytically continued to complex values. To see this explicitly, suppress all irrelevant normalization factors and write
\begin{equation}
Z_{\pmb W}(z)=D_{\pmb W}(z)^{-1/2}\,,\qquad D_{\pmb W}(z)=\det(z\pmb I-\pmb W)\,.
\label{eq:lddw-ped-ZD-relation}
\end{equation}
From \eqref{eq:lddw-index-log-det},
\begin{equation}
\mathcal K_{\pmb W}(\lambda)=N+\frac{1}{2\pi i}\left[\log D_{\pmb W}(z_-)-\log D_{\pmb W}(z_+)\right]\,,
\label{eq:lddw-ped-index-det-recall}
\end{equation}
where $z_\pm=\lambda\pm i\epsilon$. Since
\begin{equation}
\log D_{\pmb W}(z)=-2\log Z_{\pmb W}(z)\,,
\label{eq:lddw-ped-logD-logZ}
\end{equation}
we get
\begin{equation}
\mathcal K_{\pmb W}(\lambda)=N-\frac{1}{\pi i}\left[\log Z_{\pmb W}(z_-)-\log Z_{\pmb W}(z_+)\right]\,.
\label{eq:lddw-ped-index-Z-recall}
\end{equation}
Exponentiating,
\begin{align}
e^{s\mathcal K_{\pmb W}(\lambda)}&=e^{sN}\exp\left[-\frac{s}{\pi i}\log Z_{\pmb W}(z_-)+\frac{s}{\pi i}\log Z_{\pmb W}(z_+)\right]\nonumber\\
&=e^{sN}\left[Z_{\pmb W}(z_-)\right]^{-s/(\pi i)}\left[Z_{\pmb W}(z_+)\right]^{s/(\pi i)}\,.
\label{eq:lddw-ped-exp-index-Z}
\end{align}
Therefore, if one first computes
\begin{equation}
\overline{
[Z_{\pmb W}(z_-)]^{n_-}[Z_{\pmb W}(z_+)]^{n_+}
}
\label{eq:lddw-ped-integer-replica-object}
\end{equation}
for integer $n_-$ and $n_+$, the desired index generating function is obtained by the analytic continuation
\begin{equation}
n_-=-\frac{s}{\pi i}\,,\qquad n_+=\frac{s}{\pi i}\,.
\label{eq:lddw-ped-complex-replicas}
\end{equation}
The complex replica numbers are not an arbitrary complication; they come directly from the fact that an index is a phase, not a modulus.
\end{examplebox}

For integer $n_\pm$, write the replicated variables as
\begin{equation}
\underline u_i^-=(u_i^{-,1},\ldots,u_i^{-,n_-})\,,\qquad\underline u_i^+=(u_i^{+,1},\ldots,u_i^{+,n_+})\,.
\label{eq:lddw-replicated-variables}
\end{equation}
Then
\begin{align}
[Z_{\pmb W}(z_-)]^{n_-}[Z_{\pmb W}(z_+)]^{n_+}&=\int\left[\prod_{i=1}^Nd\underline u_i^-\,d\underline u_i^+\right]\exp\Bigg[-\frac{i}{2}z_-\sum_{i=1}^N\sum_{a=1}^{n_-}(u_i^{-,a})^2\nonumber\\
&\hspace{1.0cm}-\frac{i}{2}z_+\sum_{i=1}^N\sum_{b=1}^{n_+}(u_i^{+,b})^2+\frac{i}{2d}\sum_{\mu=1}^P\sum_{a=1}^{n_-}\left(\sum_{i=1}^NX_i^\mu u_i^{-,a}\right)^2\nonumber\\
&\hspace{1.0cm}+\frac{i}{2d}\sum_{\mu=1}^P\sum_{b=1}^{n_+}\left(\sum_{i=1}^NX_i^\mu u_i^{+,b}\right)^2\Bigg]\,,
\label{eq:lddw-replicated-integral}
\end{align}
where the normalization constants in the measures are immaterial for the saddle-point equations and for the spectral-count derivatives. Since the columns of $\pmb X$ are independent, the disorder average over $\mu$ factorizes. In the sparse ensemble, a column contains a Poisson number of nonzero entries of mean $d$ in the thermodynamic limit. Thus a typical factor node selects $k$ variable nodes, each carrying the replicated variables, with probability $e^{-d}d^k/k!$, draws $k$ weights from $p_\xi$, and contributes
\begin{align}
\mathcal{Z}_{\rm f}[Q]&=\sum_{k=0}^{\infty}e^{-d}\frac{d^k}{k!}\int\left[\prod_{r=1}^{k}d\underline u_r^- d\underline u_r^+ Q(\underline u_r^-,\underline u_r^+) d\xi_r p_\xi(\xi_r)\right]\nonumber\\
&\hspace{1.0cm}\times\exp\Bigg[\frac{i}{2d}\sum_{a=1}^{n_-}\left(\sum_{r=1}^k\xi_r u_r^{-,a}\right)^2+\frac{i}{2d}\sum_{b=1}^{n_+}\left(\sum_{r=1}^k\xi_r u_r^{+,b}\right)^2\Bigg]\,,
\label{eq:lddw-factor-functional}
\end{align}
where
\begin{equation}
Q(\underline u^-,\underline u^+)=\frac{1}{N}\sum_{i=1}^N\delta(\underline u^- - \underline u_i^-)\delta(\underline u^+ - \underline u_i^+)
\label{eq:lddw-replica-order-parameter}
\end{equation}
is the empirical distribution of replicated variables. The replicated disorder average therefore takes the large-$N$ form
\begin{equation}
\mathcal{Q}_N(n_-,n_+)=\int\mathcal{D}Q\exp\left[N\mathcal{S}_{n_-,n_+}[Q]\right]\,,
\label{eq:lddw-functional-integral}
\end{equation}
with action
\begin{align}
\mathcal{S}_{n_-,n_+}[Q]&=
-\int d\underline u^- d\underline u^+ Q(\underline u^-,\underline u^+) \log Q(\underline u^-,\underline u^+)\nonumber\\
&\quad-\frac{i}{2}z_-\int d\underline u^-d\underline u^+Q(\underline u^-,\underline u^+)\sum_{a=1}^{n_-}(u^{-,a})^2\nonumber\\
&\quad-\frac{i}{2}z_+\int d\underline u^- d\underline u^+ Q(\underline u^-,\underline u^+)\sum_{b=1}^{n_+}(u^{+,b})^2+\frac{1}{\alpha}\log\mathcal{Z}_{\rm f}[Q]\,.
\label{eq:lddw-replica-action}
\end{align}
The factor $1/\alpha$ appears because there are $P=N/\alpha$ factor nodes. The saddle point of \eqref{eq:lddw-replica-action} gives the replicated free energy. This is the replica derivation of the large-deviation problem: the index number has been converted into a replicated Gaussian factor-graph theory with a functional order parameter.

One could continue from this point within the replica formalism. For the finite-connectivity diluted Wishart ensemble this is possible but notationally cumbersome, because the order parameter is a distribution over replicated Gaussian variables on a bipartite factor graph. Since the locally tree-like structure gives the same replica-symmetric fixed point more directly, we now switch to the cavity formulation. At $s=0$, or equivalently $n_-=n_+=0$, the corresponding distribution reduces to the ordinary population of cavity Green functions used for the typical diluted Wishart density. For $s\neq0$, the two spectral parameters $z_-$ and $z_+$ are coupled through the counting field, and the message must carry information about both sides of the branch cut. Thus one introduces paired cavity messages
\begin{equation}
\mathfrak{g}_{i\to\mu}=\left(G_{i\to\mu}^-,G_{i\to\mu}^+\right)\,,\qquad G_{i\to\mu}^{\pm}=G_{i\to\mu}(z_\pm)\,.
\label{eq:lddw-paired-message}
\end{equation}
For a fixed bipartite tree, these two components obey the ordinary diluted-Wishart cavity recursion at the two spectral parameters:
\begin{equation}
U_{\mu\to i}^{\pm}=\frac{(\xi_i^\mu)^2}{1-\displaystyle\frac{1}{d}\sum_{j\in\partial\mu\setminus i}(\xi_j^\mu)^2G_{j\to\mu}^{\pm}}\,,
\label{eq:lddw-factor-message-pm}
\end{equation}
and
\begin{equation}
G_{i\to\mu}^{\pm}=\frac{1}{z_\pm-\displaystyle\frac{1}{d}\sum_{\nu\in\partial i\setminus\mu}U_{\nu\to i}^{\pm}}\,.
\label{eq:lddw-variable-message-pm}
\end{equation}
The full local Green functions are
\begin{equation}
G_i^{\pm}=\frac{1}{z_\pm-\displaystyle\frac{1}{d}\sum_{\nu\in\partial i}U_{\nu\to i}^{\pm}}\,.
\label{eq:lddw-full-green-pm}
\end{equation}
Equations \eqref{eq:lddw-factor-message-pm}--\eqref{eq:lddw-full-green-pm} are not yet the large-deviation equations; they are the deterministic local recursions on which the tilted problem is built. The counting field $s$ changes the probability law of the messages by reweighting local configurations according to their contribution to the phase of the determinant.

To make this reweighting explicit, it is useful to write the determinant in Bethe form. Define the full variable inverse Green function
\begin{equation}
H_i^\pm=(G_i^\pm)^{-1}=z_\pm-\frac{1}{d}\sum_{\nu\in\partial i}U_{\nu\to i}^{\pm}\,,
\label{eq:lddw-H-site}
\end{equation}
the full factor denominator
\begin{equation}
R_\mu^\pm=1-\frac{1}{d}\sum_{i\in\partial\mu}(\xi_i^\mu)^2G_{i\to\mu}^{\pm}\,,
\label{eq:lddw-R-factor}
\end{equation}
and the edge denominator
\begin{equation}
E_{i\mu}^{\pm}=1-\frac{1}{d}G_{i\to\mu}^{\pm}U_{\mu\to i}^{\pm}\,.
\label{eq:lddw-edge-denominator}
\end{equation}
For a tree factor graph, the Gaussian Bethe factorization gives
\begin{equation}
D_{\pmb W}(z_\pm)=\det(z_\pm\pmb I-\pmb W)=\frac{\displaystyle\prod_{i=1}^N H_i^\pm\prod_{\mu=1}^P R_\mu^\pm}{\displaystyle\prod_{(i,\mu)}E_{i\mu}^{\pm}}\,.
\label{eq:lddw-bethe-determinant}
\end{equation}
Here and below, products and sums over $(i,\mu)$ run over occupied bipartite edges, equivalently over pairs with $i\in\partial\mu$.

\begin{examplebox}[Bethe determinant factorization for one factor connected to two variables]
Consider again one factor node connected to two variables, with
\begin{equation}
\pmb X=\begin{pmatrix}
\xi_1\\
\xi_2
\end{pmatrix}\,,\qquad d=1\,.
\label{eq:lddw-ped-bethe-X}
\end{equation}
Then
\begin{equation}
\pmb W=\pmb X\pmb X^{\rm T}=\begin{pmatrix}
\xi_1^2 & \xi_1\xi_2\\
\xi_1\xi_2 & \xi_2^2
\end{pmatrix}\,.
\label{eq:lddw-ped-bethe-W}
\end{equation}
The exact determinant is
\begin{equation}
D_{\pmb W}(z)=\det(z\pmb I-\pmb W)=z\left(z-\xi_1^2-\xi_2^2\right)\,.
\label{eq:lddw-ped-bethe-exact-det}
\end{equation}

Now compute the same determinant using the Bethe factorization. Since each variable is a leaf when the factor is removed,
\begin{equation}
G_{1\to\mu}(z)=\frac{1}{z}\,, \qquad G_{2\to\mu}(z)=\frac{1}{z}\,.
\label{eq:lddw-ped-bethe-variable-cavity}
\end{equation}
The full factor denominator is
\begin{equation}
R_\mu(z)=1-\xi_1^2G_{1\to\mu}(z)-\xi_2^2G_{2\to\mu}(z)=1-\frac{\xi_1^2+\xi_2^2}{z}\,.
\label{eq:lddw-ped-bethe-factor-denominator}
\end{equation}
The factor-to-variable self-energies are
\begin{equation}
U_{\mu\to1}(z)=\frac{\xi_1^2}{1-\xi_2^2/z}\,,\qquad U_{\mu\to2}(z) = \frac{\xi_2^2}{1-\xi_1^2/z}\,.
\label{eq:lddw-ped-bethe-self-energies}
\end{equation}
The full variable inverse Green functions are
\begin{equation}
H_1(z)=z-U_{\mu\to1}(z)\,, \qquad H_2(z)=z-U_{\mu\to2}(z)\,.
\label{eq:lddw-ped-bethe-H}
\end{equation}
The edge denominators are
\begin{equation}
E_{1\mu}(z)=1-G_{1\to\mu}(z)U_{\mu\to1}(z)\,, \qquad E_{2\mu}(z)=1-G_{2\to\mu}(z)U_{\mu\to2}(z)\,.
\label{eq:lddw-ped-bethe-edge-denominators}
\end{equation}
The Bethe determinant formula gives
\begin{equation}
D_{\pmb W}^{\rm Bethe}(z)=\frac{H_1(z)H_2(z)R_\mu(z)}{E_{1\mu}(z)E_{2\mu}(z)}\,.
\label{eq:lddw-ped-bethe-det-formula}
\end{equation}
Substituting \eqref{eq:lddw-ped-bethe-factor-denominator}--\eqref{eq:lddw-ped-bethe-edge-denominators} and simplifying gives
\begin{equation}
D_{\pmb W}^{\rm Bethe}(z)=z\left(z-\xi_1^2-\xi_2^2\right)\,,
\label{eq:lddw-ped-bethe-det-result}
\end{equation}
which agrees with the exact determinant \eqref{eq:lddw-ped-bethe-exact-det}. This explicit check shows the meaning of the variable, factor, and edge terms in the diluted Wishart Bethe determinant.
\end{examplebox}

Let us verify where the three factors come from. The term $H_i^\pm$ is the inverse variance of the full Gaussian marginal at variable node $i$. The term $R_\mu^\pm$ is the rank-one Gaussian integral produced by factor node $\mu$ when all incoming variable-to-factor messages are combined:
\begin{equation}
\int\prod_{i\in\partial\mu}du_i\exp\left[-\frac{i}{2}\sum_{i\in\partial\mu}(G_{i\to\mu}^{\pm})^{-1}u_i^2+\frac{i}{2d}\left(\sum_{i\in\partial\mu}\xi_i^\mu u_i\right)^2\right]\propto(R_\mu^\pm)^{-1/2}\,.
\label{eq:lddw-factor-gaussian-integral}
\end{equation}
The term $E_{i\mu}^\pm$ removes the double counting of the edge contribution between variable $i$ and factor $\mu$. For a single factor attached to one variable, $R_\mu^\pm=E_{i\mu}^\pm$, provided the edge denominator is defined with the same $d^{-1}U_{\mu\to i}$ convention as in \eqref{eq:lddw-edge-denominator}. Then \eqref{eq:lddw-bethe-determinant} reduces to the exact one-dimensional determinant $z_\pm-(\xi_i^\mu)^2/d$. This elementary check fixes the relative signs in \eqref{eq:lddw-bethe-determinant}.

Substituting \eqref{eq:lddw-bethe-determinant} into \eqref{eq:lddw-index-log-det} gives an additive decomposition of the index:
\begin{equation}
\mathcal{K}_{\pmb W}(\lambda)=N+\sum_{i=1}^N\kappa_i+\sum_{\mu=1}^P\kappa_\mu-\sum_{(i,\mu)}\kappa_{i\mu}\,,\label{eq:lddw-bethe-index-decomposition}
\end{equation}
where
\begin{eqnarray}
\kappa_i=\frac{1}{2\pi i}\lim_{\epsilon\downarrow0}\left[\log H_i^--\log H_i^+\right]\,,
\label{eq:lddw-site-index-contribution}\\
\kappa_\mu=\frac{1}{2\pi i}\lim_{\epsilon\downarrow0}\left[\log R_\mu^--\log R_\mu^+\right]\,,\label{eq:lddw-factor-index-contribution}\\
\kappa_{i\mu}=\frac{1}{2\pi i}\lim_{\epsilon\downarrow0}\left[\log E_{i\mu}^--\log E_{i\mu}^+\right]\,.
\label{eq:lddw-edge-index-contribution}
\end{eqnarray}
The logarithms in \eqref{eq:lddw-site-index-contribution}--\eqref{eq:lddw-edge-index-contribution} must be evaluated with branches consistent with the determinant phase in \eqref{eq:lddw-index-log-det}. The individual variable, factor, and edge phases are branch-dependent, while the Bethe combination in \eqref{eq:lddw-bethe-index-decomposition} reproduces the determinant phase. Equation \eqref{eq:lddw-bethe-index-decomposition} is one of the most useful formulas in this section. It shows explicitly that the index number is an extensive sum of local contributions on a sparse bipartite graph. It also explains why the large-deviation speed is $N$: the atypical index is produced by an atypical empirical distribution of local messages, not by an $N^2$-scale Coulomb-gas rearrangement.

\begin{examplebox}[Variable, factor, and edge contributions to the index]
The Bethe decomposition of the index has the form
\begin{equation}
\mathcal K_{\pmb W}(\lambda)=N+\sum_{i=1}^{N}\kappa_i+\sum_{\mu=1}^{P}\kappa_\mu-\sum_{(i,\mu)}\kappa_{i\mu}.
\label{eq:lddw-ped-index-bethe-recall}
\end{equation}
The signs are the same as in the Bethe determinant: variable and factor terms are added, while edge terms correct for overcounting.

For the one-factor, two-variable example of the previous example, the exact eigenvalues are
\begin{equation}
0\,,\qquad \xi_1^2+\xi_2^2\,.
\label{eq:lddw-ped-two-eigenvalues}
\end{equation}
Take a threshold satisfying
\begin{equation}
0<\lambda<\xi_1^2+\xi_2^2\,.
\label{eq:lddw-ped-threshold-between}
\end{equation}
Then the exact index is
\begin{equation}
\mathcal K_{\pmb W}(\lambda)=1\,.
\label{eq:lddw-ped-index-one}
\end{equation}
The determinant-phase formula gives the same result because
\begin{equation}
D_{\pmb W}(z)=z\left(z-\xi_1^2-\xi_2^2\right)\,.
\label{eq:lddw-ped-D-two-factor}
\end{equation}
At $z=\lambda-i0^+$, the factor $z$ has zero phase because $\lambda>0$, while the factor $z-\xi_1^2-\xi_2^2$ has phase $-\pi$ because it is negative and approached from below. Therefore
\begin{equation}
\mathcal K_{\pmb W}(\lambda)=2+\frac{1}{\pi}(-\pi)=1\,.
\label{eq:lddw-ped-index-phase-one}
\end{equation}
The Bethe index decomposition is a local way of producing the same total phase. The variable terms, factor term, and edge corrections distribute the phase of the determinant across the bipartite graph, but the final sum must reproduce the integer count of eigenvalues below the threshold.
\end{examplebox}

At $s=0$, the paired-message distribution is obtained by solving the ordinary population dynamics at $z_-$ and $z_+$. For the Poisson bipartite ensemble, let $\mathcal{P}(\mathfrak g)$ be the distribution of the paired variable-to-factor message and $\mathcal{Q}(\mathfrak u)$ the distribution of the paired factor-to-variable self-energy,
\begin{equation}
\mathfrak u=(U^-,U^+)\,.
\label{eq:lddw-paired-self-energy}
\end{equation}
The factor update is
\begin{equation}
\mathfrak u=\mathcal{U}\left(\xi;\{\xi_r,\mathfrak g_r\}_{r=1}^k\right)\,,
\label{eq:lddw-factor-map-symbolic}
\end{equation}
with components
\begin{equation}
U^\pm=\frac{\xi^2}{1-\displaystyle\frac{1}{d}\sum_{r=1}^k\xi_r^2G_r^\pm}\,,
\label{eq:lddw-factor-map-components}
\end{equation}
where $k$ is Poisson with mean $d$. The variable update is
\begin{equation}
\mathfrak g=\mathcal{G}\left(\{\mathfrak u_r\}_{r=1}^{\ell}\right)\,,
\label{eq:lddw-variable-map-symbolic}
\end{equation}
with components
\begin{equation}
G^\pm=\frac{1}{z_\pm-\displaystyle\frac{1}{d}\sum_{r=1}^{\ell}U_r^\pm}\,,
\label{eq:lddw-variable-map-components}
\end{equation}
where $\ell$ is Poisson with mean $d/\alpha$. These equations generate the typical population. The large-deviation calculation replaces this population by a tilted population. Formally, configurations of messages are reweighted by the factor
\begin{equation}
\exp\left[s\left(N+\sum_i\kappa_i+\sum_\mu\kappa_\mu-\sum_{(i,\mu)}\kappa_{i\mu}\right)\right]\,,
\label{eq:lddw-tilting-factor}
\end{equation}
with the local terms given by \eqref{eq:lddw-site-index-contribution}--\eqref{eq:lddw-edge-index-contribution}. In the replica derivation, this reweighting is generated automatically by the analytic continuation \eqref{eq:lddw-replica-continuation}. In a population-dynamics implementation, it appears as a biased sampling of local variable and factor neighborhoods. The scaled cumulant-generating function is the Bethe free energy of this tilted message distribution.

It is useful to write the tilted ensemble explicitly. Let
\begin{equation}
P_s(\pmb W)=\frac{P(\pmb W)\exp\left[s\mathcal{K}_{\pmb W}(\lambda)\right]}{\overline{\exp\left[s\mathcal{K}_{\pmb W}(\lambda)\right]}}\,.
\label{eq:lddw-tilted-ensemble}
\end{equation}
Then
\begin{equation}
k(s)=\psi_\lambda'(s)=\lim_{N\to\infty}\frac{1}{N}\left\langle\mathcal{K}_{\pmb W}(\lambda)\right\rangle_s\,,
\label{eq:lddw-tilted-mean-index}
\end{equation}
where $\langle\cdots\rangle_s$ denotes expectation with respect to $P_s$. The rate function is obtained parametrically as
\begin{equation}
\Phi_\lambda(k(s))=s k(s)-\psi_\lambda(s)\,.
\label{eq:lddw-parametric-rate}
\end{equation}
The tilted ensemble is not simply another diluted Wishart ensemble with a different mean degree or a different weight distribution. It is an ensemble biased by a spectral property. The cavity messages therefore acquire a biased law that depends on $\lambda$ and $s$.

The first check is obtained by setting $s=0$. Then $P_s$ is the original diluted Wishart ensemble, and
\begin{equation}
\psi_\lambda(0)=0\,.
\label{eq:lddw-psi-zero}
\end{equation}
The first derivative gives the typical index fraction:
\begin{equation}
\psi_\lambda'(0)=\lim_{N\to\infty}\frac{1}{N}\overline{\mathcal K_{\pmb W}(\lambda)}\,.
\label{eq:lddw-first-derivative-density-check}
\end{equation}
When the limiting density has no atom at the threshold, and with the endpoint convention fixed, this may be written as
\[
\psi_\lambda'(0)=\int_0^\lambda d\lambda'\overline{\rho_{\pmb W}(\lambda')}\,.
\]
This follows directly from \eqref{eq:lddw-index-number}. It also follows from the determinant representation, because differentiating the logarithmic phase recovers the imaginary part of the resolvent. The second derivative gives the asymptotic variance of the number of eigenvalues below $\lambda$,
\begin{equation}
\psi_\lambda''(0)=\lim_{N\to\infty}\frac{1}{N}\left[\overline{\mathcal{K}_{\pmb W}(\lambda)^2}-\overline{\mathcal{K}_{\pmb W}(\lambda)}^2\right]\,.
\label{eq:lddw-second-derivative-variance}
\end{equation}
When $\psi_\lambda''(0)>0$, the curvature of the rate function at its minimum is
\begin{equation}
\Phi_\lambda''(k_\star)=\frac{1}{\psi_\lambda''(0)}\,.
\label{eq:lddw-curvature-rate}
\end{equation}

\begin{examplebox}[Trivial thresholds and zero modes]
The positivity of $\pmb W$ gives two exact checks on the index large-deviation formalism. Since all eigenvalues satisfy $\lambda_a\geq0$, for any threshold $\lambda<0$,
\begin{equation}
\mathcal K_{\pmb W}(\lambda)=0
\label{eq:lddw-ped-negative-threshold-index}
\end{equation}
for every matrix in the ensemble. Therefore
\begin{equation}
\psi_\lambda(s)=0\,,\qquad \Phi_\lambda(k)=\begin{cases}
0, & k=0,\\
+\infty, & k\neq0\,.
\end{cases}
\label{eq:lddw-ped-negative-threshold-rate}
\end{equation}
Similarly, if $\lambda$ is above the largest eigenvalue for every realization under consideration, or with probability one at the large-deviation exponential scale, then
\begin{equation}
\mathcal K_{\pmb W}(\lambda)=N\,,
\label{eq:lddw-ped-large-threshold-index}
\end{equation}
and
\begin{equation}
\psi_\lambda(s)=s\,,\qquad\Phi_\lambda(k)=\begin{cases}
0, & k=1,\\
+\infty, & k\neq1.
\end{cases}
\label{eq:lddw-ped-large-threshold-rate}
\end{equation}

There is also a nontrivial zero-mode check. If $P<N$, then
\begin{equation}
{\rm rank}\pmb W\leq P\,,
\label{eq:lddw-ped-rank-check}
\end{equation}
so at least $N-P$ eigenvalues are exactly zero. With $\alpha=N/P$, this means that the zero-mode fraction is at least
\begin{equation}
1-\frac{1}{\alpha} \qquad (\alpha>1)\,.
\label{eq:lddw-ped-zero-mode-fraction}
\end{equation}
Therefore, for any positive threshold $\lambda$, the index must satisfy
\begin{equation}
\frac{1}{N}\mathcal K_{\pmb W}(\lambda)\geq  1-\frac{1}{\alpha}\,.
\label{eq:lddw-ped-positive-small-threshold}
\end{equation}
A numerical or analytical solution violating this inequality cannot be correct.
\end{examplebox}

The dense limit provides another important check. When $d\to\infty$ after the thermodynamic limit, and the weights are normalized so that the corresponding dense covariance entries have unit second moment, the diluted Wishart ensemble approaches the ordinary dense Wishart ensemble at the level of typical spectral observables. With the normalization used here, $\widehat{\pmb W}=\alpha\pmb W$ has the conventional Mar\v{c}enko--Pastur density $\rho_{\rm MP}$. Therefore, for a threshold $\lambda$ of $\pmb W$, the typical fraction of eigenvalues below $\lambda$ approaches
\begin{equation}
k_\star^{\rm MP}(\lambda)=\int_0^{\alpha\lambda} dx\,\rho_{\rm MP}(x)=\int_0^\lambda d\lambda'\,\alpha\rho_{\rm MP}(\alpha\lambda')\,.
\label{eq:lddw-mp-typical-index}
\end{equation}
At the level of large deviations, however, one must be careful. The finite-connectivity theory has speed $N$ and is controlled by local graph fluctuations. The invariant dense Wishart large-deviation problem has a Coulomb-gas structure and a different speed. Thus the dense-limit check applies directly to the typical density and typical index, while the full rate function may have a singular crossover as $d$ grows with $N$.

The practical algorithm suggested by the derivation is as follows. One fixes a threshold $\lambda$ and a regulator $\epsilon$. One stores a population of paired messages $\mathfrak g=(G^-,G^+)$. A factor update samples a Poisson number $k$ of incoming variable messages, draws the corresponding weights, and computes the paired self-energy using \eqref{eq:lddw-factor-map-components}. A variable update samples a Poisson number $\ell$ of incoming paired self-energies and computes the new paired Green function using \eqref{eq:lddw-variable-map-components}. At $s=0$ this gives the ordinary population dynamics and hence the typical index. For $s\neq0$, the same local maps are used, but the sampling is tilted by the local phase contributions \eqref{eq:lddw-site-index-contribution}--\eqref{eq:lddw-edge-index-contribution}; the normalization accumulated by this tilted dynamics gives $\psi_\lambda(s)$. The rate function then follows from \eqref{eq:lddw-parametric-rate}.

The conceptual outcome is parallel to the sparse symmetric case but with an important bipartite modification. The number of eigenvalues below $\lambda$ is a phase of a determinant. The determinant is represented by a Gaussian factor graph. Because the factor graph is locally tree-like, its determinant admits a Bethe decomposition into variable, factor, and edge contributions. Large deviations of the index are therefore described by a tilted distribution of paired cavity messages. The ordinary spectral density corresponds to the unbiased one-point message distribution; the large-deviation theory requires the biased two-point distribution associated with $z_-$ and $z_+$. This is the finite-connectivity mechanism behind the large-deviation theory for diluted Wishart matrices.

\begin{exerciseblock}
\exitem[Wishart index in a finite example]
Let
\begin{equation}
\pmb X=\begin{pmatrix}
1 & 0\\
0 & 1\\
1 & 1
\end{pmatrix}\,,\qquad d=1\,, \qquad \pmb W=\pmb X\pmb X^{\rm T}\,.
\label{eq:lddw-ex-finite-X}
\end{equation}
Compute the eigenvalues of $\pmb W$ and determine $\mathcal K_{\pmb W}(\lambda)$ for several thresholds $\lambda$. Identify the zero-mode contribution.

\exitem[Determinant representation of the index]
Assuming that the threshold is not exactly equal to an eigenvalue, or after fixing an endpoint convention, starting from the identity
\begin{equation}
\Theta(x)=1+\frac{1}{2\pi i}\lim_{\epsilon\downarrow0}\left[\log(x-i\epsilon)-\log(x+i\epsilon)\right]\,,
\label{eq:lddw-ex-theta-log}
\end{equation}
derive
\begin{equation}
\mathcal K_{\pmb W}(\lambda)=N+\frac{1}{2\pi i}\lim_{\epsilon\downarrow0}\left[\log D_{\pmb W}(\lambda-i\epsilon)-\log D_{\pmb W}(\lambda+i\epsilon)\right]\,.
\label{eq:lddw-ex-index-logdet}
\end{equation}

\exitem[Partition-function representation]
Using
\begin{equation}
Z_{\pmb W}(z)=D_{\pmb W}(z)^{-1/2}
\label{eq:lddw-ex-ZD}
\end{equation}
up to $z$-independent factors, derive
\begin{equation}
e^{s\mathcal K_{\pmb W}(\lambda)}=e^{sN}\left[Z_{\pmb W}(z_-)\right]^{-s/(\pi i)}\left[Z_{\pmb W}(z_+)\right]^{s/(\pi i)}\,.
\label{eq:lddw-ex-exp-index}
\end{equation}
Explain why this leads to the analytic continuation $n_-=-s/(\pi i)$ and $n_+=s/(\pi i)$.

\exitem[Bipartite quadratic form]
Starting from
\begin{equation}
\pmb W=\frac{1}{d}\pmb X\pmb X^{\rm T}\,,
\label{eq:lddw-ex-W}
\end{equation}
show that
\begin{equation}
\pmb u^{\rm T}\pmb W\pmb u=\frac{1}{d}\sum_{\mu=1}^{P}\left(\sum_{i\in\partial\mu}\xi_i^\mu u_i\right)^2\,.
\label{eq:lddw-ex-quadratic-form}
\end{equation}
Then derive the bipartite Gaussian representation of $Z_{\pmb W}(z)$.

\exitem[Replica sectors]
Write explicitly the replicated variables associated with $[Z_{\pmb W}(z_-)]^{n_-}$ and $[Z_{\pmb W}(z_+)]^{n_+}$. Explain why the two sectors are coupled by the same underlying disorder after averaging over $\pmb X$.

\exitem[Poisson factor average]
For one column $\mu$ of $\pmb X$, show that the number of nonzero entries converges to a Poisson random variable of mean $d$. Use this to motivate the factor functional appearing in the replicated calculation.

\exitem[Paired cavity messages]
Starting from the ordinary diluted Wishart recursions at one spectral parameter, derive the paired recursions
\begin{equation}
U_{\mu\to i}^{\pm}=\frac{(\xi_i^\mu)^2}{1-\displaystyle\frac{1}{d}\sum_{j\in\partial\mu\setminus i}(\xi_j^\mu)^2G_{j\to\mu}^{\pm}}\,,
\label{eq:lddw-ex-factor-pm}
\end{equation}
and
\begin{equation}
G_{i\to\mu}^{\pm}=\frac{1}{z_\pm-\displaystyle\frac{1}{d}\sum_{\nu\in\partial i\setminus\mu}U_{\nu\to i}^{\pm}}\,.
\label{eq:lddw-ex-variable-pm}
\end{equation}
Why are the two components statistically coupled even though the maps are componentwise?

\exitem[Rank-one factor denominator]
For a factor node $\mu$, prove that
\begin{equation}
R_\mu^\pm=1-\frac{1}{d}\sum_{i\in\partial\mu}(\xi_i^\mu)^2G_{i\to\mu}^{\pm}
\label{eq:lddw-ex-R-factor}
\end{equation}
is the determinant correction generated by the rank-one Gaussian integration at the factor node.

\exitem[Edge denominator]
Using the variable-to-factor Green function $G_{i\to\mu}^{\pm}$ and the factor-to-variable self-energy $U_{\mu\to i}^{\pm}$, derive the edge denominator
\begin{equation}
E_{i\mu}^{\pm}=1-\frac{1}{d}G_{i\to\mu}^{\pm}U_{\mu\to i}^{\pm}\,.
\label{eq:lddw-ex-edge-denominator}
\end{equation}
Explain its role as an overcounting correction in the Bethe determinant.

\exitem[Bethe determinant for one factor]
For the matrix
\begin{equation}
\pmb X=\begin{pmatrix}
\xi_1\\
\xi_2\\
\xi_3
\end{pmatrix},\qquad d=1\,,
\label{eq:lddw-ex-one-factor-X}
\end{equation}
compute $\det(z\pmb I-\pmb X\pmb X^{\rm T})$ directly and show that it is
\begin{equation}
z^2\left(z-\xi_1^2-\xi_2^2-\xi_3^2\right)\,.
\end{equation}
Then verify the Bethe determinant factorization using the variable, factor, and edge terms.

\exitem[Bethe index decomposition]
Starting from the Bethe determinant formula, derive
\begin{equation}
\mathcal K_{\pmb W}(\lambda)=N+\sum_i\kappa_i+\sum_\mu\kappa_\mu-\sum_{(i,\mu)}\kappa_{i\mu}\,.
\label{eq:lddw-ex-bethe-index}
\end{equation}
Make explicit which logarithmic phase produces each contribution.

\exitem[Trivial threshold checks]
Prove that for $\lambda<0$,
\begin{equation}
\mathcal K_{\pmb W}(\lambda)=0
\label{eq:lddw-ex-negative-threshold}
\end{equation}
for every realization. Then show that if $\lambda$ is above the whole spectrum,
\begin{equation}
\mathcal K_{\pmb W}(\lambda)=N\,.
\label{eq:lddw-ex-large-threshold}
\end{equation}
Derive the corresponding forms of $\psi_\lambda(s)$ and $\Phi_\lambda(k)$.

\exitem[Zero-mode lower bound]
Use
\begin{equation}
{\rm rank}\pmb W\leq P
\label{eq:lddw-ex-rank-bound}
\end{equation}
to show that for $\alpha=N/P>1$ the fraction of zero eigenvalues is at least
\begin{equation}
1-\frac{1}{\alpha}\,.
\label{eq:lddw-ex-zero-bound}
\end{equation}
Explain how this constrains $\mathcal K_{\pmb W}(\lambda)/N$ for every positive threshold $\lambda$.

\exitem[First derivative of the cumulant-generating function]
Starting from
\begin{equation}
\psi_\lambda(s)=\lim_{N\to\infty}\frac{1}{N}\log\overline{e^{s\mathcal K_{\pmb W}(\lambda)}}\,,
\label{eq:lddw-ex-cgf}
\end{equation}
show formally that
\begin{equation}
\psi_\lambda'(0)=\lim_{N\to\infty}\frac{1}{N}\overline{\mathcal K_{\pmb W}(\lambda)}.
\label{eq:lddw-ex-first-derivative}
\end{equation}
Relate this to the integrated spectral density, stating explicitly the convention used if the limiting density has an atom at the threshold.

\exitem[Second derivative and variance]
Show that
\begin{equation}
\psi_\lambda''(0)=\lim_{N\to\infty}\frac{1}{N}{\rm Var}\mathcal K_{\pmb W}(\lambda)\,.
\label{eq:lddw-ex-second-derivative}
\end{equation}
Explain why this quantity measures the extensive part of index fluctuations.

\exitem[Legendre relation]
Assuming differentiability of $\psi_\lambda(s)$, derive the parametric relation
\begin{equation}
k(s)=\psi_\lambda'(s)\,,\qquad \Phi_\lambda(k(s))=s k(s)-\psi_\lambda(s)\,.
\label{eq:lddw-ex-legendre-parametric}
\end{equation}

\exitem[Dense-limit typical index]
In the dense limit, the conventionally normalized matrix $\widehat{\pmb W}=\alpha\pmb W$ has the Mar\v{c}enko--Pastur density. For a threshold $\lambda$ of $\pmb W$, write the typical index as
\begin{equation}
k_\star^{\rm MP}(\lambda)=\int_0^{\alpha\lambda} dx\,\rho_{\rm MP}(x)\,.
\label{eq:lddw-ex-MP-index}
\end{equation}
For $\alpha=1$, evaluate this integral explicitly or numerically using the square Mar\v{c}enko--Pastur density
\[
\rho_{\rm MP}(\lambda)=\frac{1}{2\pi}\sqrt{\frac{4-\lambda}{\lambda}}\mathbf 1_{0<\lambda<4}\,.
\]

\exitem[Programming exercise: empirical index distribution]
Fix a list of system sizes $N$, an aspect ratio $\alpha>0$ such that $P=N/\alpha$ is an integer for each $N$, a dilution $d>0$, a number $M$ of independent samples for each parameter set, a nonzero-weight distribution $p_\xi$, and a threshold $\lambda$ with an explicitly stated convention for eigenvalues exactly at the threshold. Generate diluted Wishart matrices from sparse rectangular matrices
\begin{equation}
X_i^\mu=B_i^\mu\xi_i^\mu\, \qquad {\rm Prob}(B_i^\mu=1)=\frac{d}{N}\,, \qquad {\rm Prob}(B_i^\mu=0)=1-\frac{d}{N}\,,
\end{equation}
with nonzero weights $\xi_i^\mu$ drawn independently from $p_\xi$, and form
\begin{equation}
\pmb W=\frac{1}{d}\pmb X\pmb X^{\rm T}.
\label{eq:lddw-ex-program-W}
\end{equation}
For each value of $N$, compute the empirical distribution of $\mathcal K_{\pmb W}(\lambda)$ over the $M$ samples. Report $N$, $P$, $\alpha$, $d$, $M$, $p_\xi$, $\lambda$, and the threshold convention.

\exitem[Programming exercise: empirical cumulant-generating function]
Using the same ensemble and the same $M$ samples as in the previous exercise, fix a grid of real values of $s$ containing $s=0$ and estimate
\begin{equation}
\widehat\psi_{\lambda,N}(s)=\frac{1}{N}\log\left[\frac{1}{M}\sum_{m=1}^{M}e^{s\mathcal K_m(\lambda)}\right]\,,
\label{eq:lddw-ex-program-cgf}
\end{equation}
where $\mathcal K_m(\lambda)$ is the index of the $m$th sampled matrix. Estimate $\widehat\psi'_{\lambda,N}(0)$ from $N^{-1}$ times the sample mean of $\mathcal K_m(\lambda)$, and estimate $\widehat\psi''_{\lambda,N}(0)$ from $N^{-1}$ times the sample variance. Compare these estimates with finite-difference derivatives of $\widehat\psi_{\lambda,N}(s)$ near $s=0$. Report the $s$ grid and the finite-difference rule.

\exitem[Programming exercise: zero modes]
Fix a list of system sizes $N$, several values of $\alpha>1$ such that $P=N/\alpha$ is an integer, a dilution $d>0$, a number $M$ of independent samples for each parameter set, a nonzero-weight distribution $p_\xi$, and a numerical tolerance $\tau_0$ for identifying zero eigenvalues. Generate diluted Wishart matrices
\begin{equation}
\pmb W=\frac{1}{d}\pmb X\pmb X^{\rm T}
\end{equation}
from the same sparse rectangular ensemble used above, and measure the fraction of eigenvalues with $|\lambda|<\tau_0$. Compare the observed fraction with the rank lower bound $1-1/\alpha$ and discuss additional zero modes caused by dilution. Report $N$, $P$, $\alpha$, $d$, $M$, $p_\xi$, and $\tau_0$.
\end{exerciseblock}

\section{Conditioned spectral density of non-invariant random matrices}
\label{sec:conditioned-spectral-density}
The large-deviation formalism of the previous sections answers the following question: what is the probability that a random matrix has an atypical number of eigenvalues in a prescribed region? There is a second, more refined, question. If the matrix is forced to have such an atypical number of eigenvalues, what does its spectrum look like? This is the question of the conditioned spectral density. It is a natural continuation of the index-number problem because the index is only one scalar constraint, while the conditioned spectral density describes the whole empirical measure under that constraint.

For invariant random matrix ensembles this problem can be formulated in terms of a constrained Coulomb gas. The eigenvalues have an explicit joint probability density, and conditioning the index amounts to minimizing the Coulomb-gas energy under an additional constraint on the mass placed below a threshold \cite{DeanMajumdar2006,MajumdarNadalScardicchioVivo2009,Forrester2010}. For the sparse and diluted ensembles emphasized in these notes, no such eigenvalue gas is available. The matrices are non-invariant: their probability law is not invariant under orthogonal conjugation, and eigenvalues are not the only natural variables. The correct order parameter is instead graphical. The conditioned spectral density must be computed from the distribution of local resolvent messages in a tilted ensemble. This is the point of the theory developed in \cite{PerezCastilloMetz2018Conditioned}.

We consider first a general real symmetric random matrix $\pmb A$ with eigenvalues $\lambda_1,\ldots,\lambda_N$. Let $x$ be the threshold used for the conditioning. The index number below $x$ is
\begin{equation}
\mathcal{K}_{\pmb A}(x)=\sum_{i=1}^N\Theta(x-\lambda_i)\,.
\label{eq:csd-index}
\end{equation}
The empirical spectral density is
\begin{equation}
\rho_{\pmb A}(\lambda)=\frac{1}{N}\sum_{i=1}^N\delta(\lambda-\lambda_i)\,.
\label{eq:csd-empirical-density}
\end{equation}
The conditioned spectral density at fixed intensive index $k$ is defined by
\begin{equation}
\rho_x(\lambda|k)=\lim_{N\to\infty}\frac{\overline{\rho_{\pmb A}(\lambda)\delta\left(\mathcal{K}_{\pmb A}(x)-Nk\right)}}{\overline{\delta\left(\mathcal{K}_{\pmb A}(x)-Nk\right)}}\,.
\label{eq:csd-definition}
\end{equation}
Here the overbar denotes the average over the matrix ensemble. At finite $N$, the symbol $\delta$ in \eqref{eq:csd-definition} should be read as a Kronecker delta on the integer-valued count, or equivalently as a narrow count window in a large-deviation formulation; $Nk$ is understood to be compatible with the chosen finite-$N$ sequence. Eigenvalues exactly at the threshold $x$ are treated separately, or by a fixed endpoint convention used consistently in all integrals below $x$. The threshold is denoted by $x$, while $\lambda$ denotes the spectral variable at which the density is measured. This separation of notation is useful because the conditioned density is a function of the whole spectral variable $\lambda$, not only of the threshold $x$.

\begin{examplebox}[A finite conditioned spectral density]
Consider an ensemble with only two possible $2\times2$ diagonal matrices,
\begin{equation}
\pmb A_1=\begin{pmatrix}
-1 & 0\\
0 & 2
\end{pmatrix}\,,\qquad\pmb A_2=\begin{pmatrix}
-3 & 0\\
0 & -2
\end{pmatrix}\,,
\label{eq:csd-ped-two-matrix-ensemble}
\end{equation}
each chosen with probability $1/2$. Let the conditioning threshold be
\begin{equation}
x=0\,.
\label{eq:csd-ped-threshold-zero}
\end{equation}
The index numbers are
\begin{equation}
\mathcal K_{\pmb A_1}(0)=1\,,\qquad\mathcal K_{\pmb A_2}(0)=2\,.
\label{eq:csd-ped-two-indices}
\end{equation}
The empirical spectral densities are
\begin{equation}
\rho_{\pmb A_1}(\lambda)=\frac{1}{2}\delta(\lambda+1)+\frac{1}{2}\delta(\lambda-2)\,,
\label{eq:csd-ped-density-A1}
\end{equation}
and
\begin{equation}
\rho_{\pmb A_2}(\lambda)=\frac{1}{2}\delta(\lambda+3)+\frac{1}{2}\delta(\lambda+2).
\label{eq:csd-ped-density-A2}
\end{equation}
Conditioning on the event
\begin{equation}
\frac{1}{2}\mathcal K_{\pmb A}(0)=\frac{1}{2}
\label{eq:csd-ped-condition-k-half}
\end{equation}
selects only $\pmb A_1$, so the conditioned density is
\begin{equation}
\rho_0\left(\lambda\middle|\frac{1}{2}\right)=\rho_{\pmb A_1}(\lambda)=\frac{1}{2}\delta(\lambda+1)+\frac{1}{2}\delta(\lambda-2)\,.
\label{eq:csd-ped-conditioned-half}
\end{equation}
Conditioning on
\begin{equation}
\frac{1}{2}\mathcal K_{\pmb A}(0)=1
\label{eq:csd-ped-condition-k-one}
\end{equation}
selects only $\pmb A_2$, and therefore
\begin{equation}
\rho_0(\lambda|1)=\rho_{\pmb A_2}(\lambda)=\frac{1}{2}\delta(\lambda+3)+\frac{1}{2}\delta(\lambda+2)\,.
\label{eq:csd-ped-conditioned-one}
\end{equation}
In both cases the conditioned density satisfies the defining constraint:
\begin{equation}
\int_{-\infty}^{0}d\lambda\rho_0\left(\lambda\middle|\frac{1}{2}\right)=\frac{1}{2}\,,\qquad\int_{-\infty}^{0}d\lambda\rho_0(\lambda|1)=1\,.
\label{eq:csd-ped-conditioned-check}
\end{equation}
This finite example is deliberately simple: it shows that the conditioned spectral density is not the density of a new matrix formula, but the ordinary empirical density averaged over those samples that satisfy a spectral constraint.
\end{examplebox}

The conditioning imposes the constraint
\begin{equation}
\int_{-\infty}^{x}d\lambda\rho_x(\lambda|k)=k\,,
\label{eq:csd-index-constraint}
\end{equation}
and the density is normalized:
\begin{equation}
\int_{-\infty}^{\infty}d\lambda\rho_x(\lambda|k)=1\,.
\label{eq:csd-normalization}
\end{equation}
These two equations are not assumptions; they follow directly from \eqref{eq:csd-definition}. They are the first checks that any calculation of a conditioned density must satisfy. If $k$ equals its typical value,
\begin{equation}
k_\star(x)=\int_{-\infty}^{x}d\lambda\overline{\rho_{\pmb A}(\lambda)}\,,
\label{eq:csd-typical-index}
\end{equation}
then the conditioned density must reduce to the ordinary spectral density:
\begin{equation}
\rho_x(\lambda|k_\star)=\overline{\rho_{\pmb A}(\lambda)}\,.
\label{eq:csd-typical-density-check}
\end{equation}

To compute \eqref{eq:csd-definition}, we first replace the hard conditioning by a tilted ensemble. Define the scaled cumulant-generating function of the index number,
\begin{equation}
\psi_x(s)=\lim_{N\to\infty}\frac{1}{N}\log\overline{\exp\left[s\mathcal{K}_{\pmb A}(x)\right]}\,.
\label{eq:csd-index-cgf}
\end{equation}
When the corresponding rate function is obtained by a regular Legendre--Fenchel transform, $\Phi_x(k)=\sup_s\{sk-\psi_x(s)\}$. The tilted ensemble is
\begin{equation}
P_s(\pmb A)=\frac{P(\pmb A)\exp\left[s\mathcal{K}_{\pmb A}(x)\right]}{\overline{\exp\left[s\mathcal{K}_{\pmb A}(x)\right]}}\,.
    \label{eq:csd-tilted-ensemble}
\end{equation}
Let
\begin{equation}
\langle O\rangle_s=\frac{\overline{O(\pmb A)\exp\left[s\mathcal{K}_{\pmb A}(x)\right]}}{\overline{\exp\left[s\mathcal{K}_{\pmb A}(x)\right]}}
\label{eq:csd-tilted-average}
\end{equation}
denote expectation in this tilted ensemble. If $\psi_x(s)$ is differentiable, the typical index in the tilted ensemble is
\begin{equation}
k(s)=\psi_x'(s)=\lim_{N\to\infty}\frac{1}{N}\left\langle\mathcal{K}_{\pmb A}(x)\right\rangle_s\,.
\label{eq:csd-tilted-index}
\end{equation}
The hard-conditioned density at $k=k(s)$ is then equivalent, at the level of thermodynamic local observables, to the tilted density
\begin{equation}
\rho_x(\lambda|k(s))=\lim_{N\to\infty}\left\langle\rho_{\pmb A}(\lambda)\right\rangle_s\,.
    \label{eq:csd-equivalence-conditioned-tilted}
\end{equation}

\begin{examplebox}[Soft tilting versus hard conditioning in a toy ensemble]
The equivalence between hard conditioning and exponential tilting becomes transparent in the two-matrix ensemble of the previous example. The tilted probability of a matrix is
\begin{equation}
P_s(\pmb A)=\frac{P(\pmb A)e^{s\mathcal K_{\pmb A}(0)}}{\overline{e^{s\mathcal K_{\pmb A}(0)}}}\,.
\label{eq:csd-ped-toy-tilted-probability}
\end{equation}
Since
\begin{equation}
\mathcal K_{\pmb A_1}(0)=1\,,\qquad \mathcal K_{\pmb A_2}(0)=2\,,
\label{eq:csd-ped-toy-indices-recall}
\end{equation}
we obtain
\begin{equation}
P_s(\pmb A_1)=\frac{e^s}{e^s+e^{2s}}\,,\qquad P_s(\pmb A_2)=\frac{e^{2s}}{e^s+e^{2s}}\,.
\label{eq:csd-ped-toy-tilted-weights}
\end{equation}
The tilted mean intensive index is
\begin{equation}
k(s)=\frac{1}{2}\left[P_s(\pmb A_1)\cdot 1+P_s(\pmb A_2)\cdot 2\right]=\frac{1}{2}\frac{e^s+2e^{2s}}{e^s+e^{2s}}\,.
\label{eq:csd-ped-toy-tilted-index}
\end{equation}
As $s\to-\infty$,
\begin{equation}
P_s(\pmb A_1)\to1\,,\qquad k(s)\to\frac{1}{2}\,,
\label{eq:csd-ped-toy-negative-s}
\end{equation}
while as $s\to+\infty$,
\begin{equation}
P_s(\pmb A_2)\to1\,, \qquad k(s)\to1\,.
\label{eq:csd-ped-toy-positive-s}
\end{equation}
The tilted density is
\begin{equation}
\rho_s(\lambda)=P_s(\pmb A_1)\rho_{\pmb A_1}(\lambda)+P_s(\pmb A_2)\rho_{\pmb A_2}(\lambda)\,.
\label{eq:csd-ped-toy-tilted-density}
\end{equation}
Thus the hard-conditioned densities at $k=1/2$ and $k=1$ are recovered as limiting tilted densities. In large random-matrix ensembles the same principle holds at the thermodynamic level: for differentiable $\psi_x(s)$, the ensemble tilted by $s$ realizes the conditioned density at $k=\psi_x'(s)$.
\end{examplebox}

Let us derive this equivalence explicitly at the large-deviation level. Using a formal inverse-Laplace representation of the count constraint,
\begin{equation}
\delta\left(\mathcal{K}_{\pmb A}(x)-Nk\right)=\int_{\gamma-i\infty}^{\gamma+i\infty}\frac{ds}{2\pi i}\exp\left[s\mathcal{K}_{\pmb A}(x)-sNk\right]\,,
    \label{eq:csd-delta-laplace}
\end{equation}
the denominator of \eqref{eq:csd-definition} becomes, at exponential order,
\begin{equation}
\overline{\delta\left(\mathcal{K}_{\pmb A}(x)-Nk\right)}\asymp\int ds\exp\left[N\{\psi_x(s)-sk\}\right]\,.
\label{eq:csd-denominator-saddle}
\end{equation}
The saddle point satisfies
\begin{equation}
\psi_x'(s)=k\,.
\label{eq:csd-saddle-condition}
\end{equation}
The numerator has the same exponential factor, but with the observable $\rho_{\pmb A}(\lambda)$ evaluated in the tilted measure at the saddle. Hence \eqref{eq:csd-equivalence-conditioned-tilted} follows. When $\psi_x$ is not differentiable, the equivalence between the hard-conditioned and tilted ensembles may fail or may require a convexified description; in these notes we stay within the differentiable regime.

We now connect the tilted density with determinants and resolvents. Introduce
\begin{equation}
z_x^- = x-i\epsilon\,, \qquad z_x^+ = x+i\epsilon\,, \qquad \epsilon>0\,,
\label{eq:csd-threshold-parameters}
\end{equation}
and
\begin{equation}
\Delta_{\pmb A}(z)=\det(z\pmb I-\pmb A)\,.
\label{eq:csd-characteristic-determinant}
\end{equation}
For a real number $u\neq0$,
\begin{equation}
\Theta(u)=1+\frac{1}{2\pi i}\lim_{\epsilon\downarrow0}\left[\log(u-i\epsilon)-\log(u+i\epsilon)\right]\,.
\label{eq:csd-step-log-identity}
\end{equation}
If $u>0$, the logarithms have the same phase and the second term vanishes. If $u<0$, the two phases differ by $-2\pi$, and the second term is $-1$. Applying this identity to $u=x-\lambda_i$ and summing over eigenvalues gives
\begin{equation}
\mathcal{K}_{\pmb A}(x)=N+\frac{1}{2\pi i}\lim_{\epsilon\downarrow0}\left[\log \Delta_{\pmb A}(z_x^-)-\log \Delta_{\pmb A}(z_x^+)\right]\,.
\label{eq:csd-index-determinant}
\end{equation}
The logarithms in \eqref{eq:csd-index-determinant} are understood with the same branch convention as in \eqref{eq:csd-step-log-identity}. Thus the tilted ensemble \eqref{eq:csd-tilted-ensemble} is a determinant-biased ensemble.

\begin{examplebox}[The determinant phase fixes the constraint]
Let
\begin{equation}
\pmb A=\begin{pmatrix}
-1 & 0\\
0 & 2
\end{pmatrix}\,, \qquad x=0\,.
\label{eq:csd-ped-phase-example-matrix}
\end{equation}
The characteristic determinant is
\begin{equation}
\Delta_{\pmb A}(z)=(z+1)(z-2)\,.
\label{eq:csd-ped-phase-characteristic}
\end{equation}
At $z_x^-= -i\epsilon$, with $\epsilon>0$,
\begin{equation}
\Delta_{\pmb A}(-i\epsilon)=(1-i\epsilon)(-2-i\epsilon)\,.
\label{eq:csd-ped-phase-lower}
\end{equation}
As $\epsilon\downarrow0$, the first factor has phase $0$ and the second factor has phase $-\pi$. Thus
\begin{equation}
\lim_{\epsilon\downarrow0} {\rm Im}\log\Delta_{\pmb A}(-i\epsilon) = -\pi\,.
\label{eq:csd-ped-phase-imlog-lower}
\end{equation}
Similarly,
\begin{equation}
\lim_{\epsilon\downarrow0}{\rm Im}\log\Delta_{\pmb A}(+i\epsilon)=+\pi\,.
\label{eq:csd-ped-phase-imlog-upper}
\end{equation}
Therefore
\begin{align}
\mathcal K_{\pmb A}(0) &= 2+\frac{1}{2\pi i} \left[ \log\Delta_{\pmb A}(-i0^+) - \log\Delta_{\pmb A}(+i0^+) \right]\nonumber\\
&=2+\frac{1}{2\pi i}(-2\pi i)=1\,.
\label{eq:csd-ped-phase-index}
\end{align}
Thus the determinant phase is exactly the scalar constraint that defines the tilted ensemble. The conditioned spectral density is obtained by measuring a separate probe resolvent in the ensemble biased by this phase.
\end{examplebox}

The spectral density at the point $\lambda$ is obtained from a separate resolvent. Let
\begin{equation}
z_\lambda=\lambda-i\eta\,, \qquad\eta>0\,.
\label{eq:csd-probe-parameter}
\end{equation}
The regularized density is
\begin{equation}
\rho_{\pmb A,\eta}(\lambda)=\frac{1}{\pi N}{\rm Im}{\rm Tr}(z_\lambda\pmb I-\pmb A)^{-1}\,.
\label{eq:csd-regularized-density}
\end{equation}
Equivalently, using the Edwards--Jones Gaussian partition function
\begin{equation}
Z_{\pmb A}(z)=\int\prod_{i=1}^N\frac{du_i}{\sqrt{2\pi}}\exp\left[-\frac{i}{2}\pmb u^{\rm T}(z\pmb I-\pmb A)\pmb u\right]\,,
\label{eq:csd-edwards-jones}
\end{equation}
one has
\begin{equation}
\rho_{\pmb A,\eta}(\lambda)=-\frac{2}{\pi N}{\rm Im}\frac{\partial}{\partial\lambda}\log Z_{\pmb A}(z_\lambda)\,.
\label{eq:csd-density-free-energy}
\end{equation}
The tilted conditioned density can therefore be computed from the tilted quenched free energy
\begin{equation}
\mathcal{F}_x^{(s)}(z_\lambda)=\lim_{N\to\infty}\frac{1}{N}\left\langle\log Z_{\pmb A}(z_\lambda)\right\rangle_s\,,
\label{eq:csd-tilted-free-energy}
\end{equation}
through
\begin{equation}
\rho_x(\lambda|k(s))=-\frac{2}{\pi}\lim_{\eta\downarrow0}{\rm Im}\frac{\partial}{\partial\lambda}\mathcal{F}_x^{(s)}(\lambda-i\eta).
\label{eq:csd-density-from-tilted-free-energy}
\end{equation}

The replica representation of \eqref{eq:csd-tilted-free-energy} is obtained by adding a spectator replica at the probe spectral parameter. Define
\begin{equation}
\varphi_x(s,n;z_\lambda)=\lim_{N\to\infty}\frac{1}{N}\log\overline{\exp\left[s\mathcal{K}_{\pmb A}(x)\right][Z_{\pmb A}(z_\lambda)]^n}\,.
\label{eq:csd-joint-generating-function}
\end{equation}
Then
\begin{equation}
\mathcal{F}_x^{(s)}(z_\lambda)=\left.\frac{\partial}{\partial n}\varphi_x(s,n;z_\lambda)\right|_{n=0}\,.
\label{eq:csd-free-energy-replica-derivative}
\end{equation}
The field $s$ is associated with the determinant ratio at $z_x^-$ and $z_x^+$, while the auxiliary replica number $n$ probes the density at $z_\lambda$. The limit $n\to0$ ensures that the probe does not change the tilted ensemble; it only measures a local resolvent inside it. One could continue from this point within the replica formalism. For finite-connectivity ensembles this is possible but notationally cumbersome, because the order parameter must keep track of the replicated fields associated with the two threshold parameters and with the probe resolvent. The corresponding three-sector replica-symmetric saddle is described in Appendix~\ref{app:replica-symmetric-saddle-points}. Since the locally tree-like structure gives the same replica-symmetric fixed point more directly, we now switch to the cavity formulation.

We now specialize these formulas to sparse symmetric matrices, because in that case the graphical meaning is transparent. Let
\begin{equation}
A_{ij}=D_i\delta_{ij}+C_{ij}J_{ij}\,,\qquad A_{ij}=A_{ji}\,.
\label{eq:csd-sparse-symmetric-model}
\end{equation}
For every directed edge $i\to j$ we introduce three cavity Green functions:
\begin{equation}
\pmb{G}_{i\to j}=\left(G_{i\to j}^-,G_{i\to j}^+,G_{i\to j}^0\right)\,,
\label{eq:csd-message-triple}
\end{equation}
where
\begin{equation}
G_{i\to j}^- = G_{i\to j}(z_x^-)\,, \qquad G_{i\to j}^+ = G_{i\to j}(z_x^+)\,, \qquad G_{i\to j}^0 = G_{i\to j}(z_\lambda)\,.
\label{eq:csd-message-components}
\end{equation}
On a tree, or within the locally tree-like cavity approximation, these three components obey the same Schur-complement recursion at their respective spectral parameters:
\begin{equation}
G_{i\to j}^{a}= \frac{1}{z_a-D_i-\displaystyle\sum_{\ell\in\partial i\setminus j}J_{i\ell}^2G_{\ell\to i}^{a}}\,,\qquad a\in\{-,+,0\}\,,
\label{eq:csd-triple-cavity-recursion}
\end{equation}
with
\begin{equation}
z_- = z_x^-\,, \qquad z_+ = z_x^+\,,\qquad z_0 = z_\lambda\,.
\label{eq:csd-z-triple}
\end{equation}
The corresponding full local Green functions are
\begin{equation}
G_i^{a}=\frac{1}{z_a-D_i-\displaystyle\sum_{\ell\in\partial i}J_{i\ell}^2G_{\ell\to i}^{a}}\,,\qquad a\in\{-,+,0\}\,.
\label{eq:csd-triple-full-green}
\end{equation}
The two components $G^-$ and $G^+$ encode the index constraint, while $G^0$ probes the density at $\lambda$.

\begin{examplebox}[Why three Green functions appear in the conditioned density]
In the conditioned-density problem one has two different spectral tasks. The first task is to impose the index constraint below the threshold $x$. This requires the two boundary values
\begin{equation}
z_- = x-i\epsilon\,, \qquad z_+ = x+i\epsilon\,.
\label{eq:csd-ped-three-green-zpm}
\end{equation}
The second task is to measure the density at a possibly different spectral point $\lambda$. This requires the probe parameter
\begin{equation}
z_0=\lambda-i\eta\,.
\label{eq:csd-ped-three-green-z0}
\end{equation}
Therefore a cavity message must carry the three components
\begin{equation}
\pmb{G}_{i\to j}=\left(G_{i\to j}^-,G_{i\to j}^+,G_{i\to j}^0\right)\,.
\label{eq:csd-ped-triple-message}
\end{equation}

For example, consider the two-site graph
\begin{equation}
\pmb A=\begin{pmatrix}
0 & J\\
J & 0
\end{pmatrix}\,.
\label{eq:csd-ped-two-site-triple}
\end{equation}
Since each vertex is a leaf in the cavity graph,
\begin{equation}
G_{1\to2}^{a}= \frac{1}{z_a}\,, \qquad G_{2\to1}^{a} = \frac{1}{z_a}\,, \qquad a\in\{-,+,0\}\,.
\label{eq:csd-ped-two-site-cavity-triples}
\end{equation}
The full Green functions are
\begin{equation}
G_1^{a}=G_2^{a}=\frac{1}{z_a-J^2/z_a}=\frac{z_a}{z_a^2-J^2}\,,\qquad a\in\{-,+,0\}\,.
    \label{eq:csd-ped-two-site-full-triples}
\end{equation}
The pair $(G^-,G^+)$ determines the phase of the determinant at the threshold $x$, while $G^0$ gives the local contribution to the density at $\lambda$. Thus the triple message is not an arbitrary complication: it is the minimal object that simultaneously enforces the index constraint and measures the conditioned density.
\end{examplebox}

The local decomposition of the determinant makes the tilted measure explicit. On a tree one has
\begin{equation}
\Delta_{\pmb A}(z)=\prod_{i=1}^N G_i(z)^{-1}\prod_{\{i,j\}\in E}\left[1-J_{ij}^2G_{i\to j}(z)G_{j\to i}(z)\right]^{-1}\,.
\label{eq:csd-bethe-determinant}
\end{equation}

\begin{examplebox}[Bethe determinant identity for a two-site graph]
For the weighted two-site graph
\begin{equation}
\pmb A=\begin{pmatrix}
0 & J\\
J & 0
\end{pmatrix}\,,
\label{eq:csd-ped-bethe-two-site}
\end{equation}
the exact determinant is
\begin{equation}
\Delta_{\pmb A}(z)=z^2-J^2\,.
\label{eq:csd-ped-bethe-exact}
\end{equation}
The cavity Green functions are
\begin{equation}
G_{1\to2}(z)=\frac{1}{z}\,,\qquad G_{2\to1}(z)=\frac{1}{z}\,.
\label{eq:csd-ped-bethe-cavity}
\end{equation}
The full local Green functions are
\begin{equation}
G_1(z)=G_2(z)=\frac{1}{z-J^2/z}=\frac{z}{z^2-J^2}\,.
\label{eq:csd-ped-bethe-full}
\end{equation}
Therefore the product of site contributions is
\begin{equation}
\prod_{i=1}^{2}G_i(z)^{-1}=\left(\frac{z^2-J^2}{z}\right)^2\,.
\label{eq:csd-ped-bethe-site-product}
\end{equation}
The edge correction is
\begin{equation}
\left[1-J^2G_{1\to2}(z)G_{2\to1}(z)\right]^{-1}=\left[1-\frac{J^2}{z^2}\right]^{-1}=\frac{z^2}{z^2-J^2}\,.
\label{eq:csd-ped-bethe-edge-correction}
\end{equation}
Multiplying site and edge terms gives
\begin{equation}
\left(\frac{z^2-J^2}{z}\right)^2\frac{z^2}{z^2-J^2}=z^2-J^2\,,
\label{eq:csd-ped-bethe-result}
\end{equation}
which is the exact determinant. This example explains why the index decomposition has site terms minus edge terms: the determinant itself factorizes into site contributions corrected by edge overcounting factors.
\end{examplebox}

Let us indicate the derivation. For a branch rooted at $i$ with the edge to $j$ removed, the Schur complement gives
\begin{equation}
G_{i\to j}(z)^{-1}=z-D_i-\sum_{\ell\in\partial i\setminus j}J_{i\ell}^2G_{\ell\to i}(z).
\label{eq:csd-branch-schur}
\end{equation}
The determinant of the same branch is the product of the determinants of the neighboring branches times the Schur complement \eqref{eq:csd-branch-schur}. Iterating this identity over a tree gives \eqref{eq:csd-bethe-determinant}. The edge factors remove the overcounting produced by multiplying all full site contributions.

Substituting \eqref{eq:csd-bethe-determinant} into \eqref{eq:csd-index-determinant} gives
\begin{equation}
\mathcal{K}_{\pmb A}(x)=N+\sum_{i=1}^N\kappa_i-\sum_{\{i,j\}\in E}\kappa_{ij}\,,
\label{eq:csd-index-bethe}
\end{equation}
with site contribution
\begin{equation}
\kappa_i=\frac{1}{2\pi i}\lim_{\epsilon\downarrow0}\left[\log (G_i^-)^{-1}-\log (G_i^+)^{-1}\right]\,,
\label{eq:csd-site-index-contribution}
\end{equation}
and edge contribution
\begin{equation}
\kappa_{ij}=\frac{1}{2\pi i}\lim_{\epsilon\downarrow0}\left[\log\left(1-J_{ij}^2G_{i\to j}^-G_{j\to i}^-\right)-\log\left(1-J_{ij}^2G_{i\to j}^+G_{j\to i}^+\right)\right]\,.
\label{eq:csd-edge-index-contribution}
\end{equation}
The logarithms in \eqref{eq:csd-site-index-contribution} and \eqref{eq:csd-edge-index-contribution} must be evaluated with branches consistent with the determinant phase in \eqref{eq:csd-index-determinant}. The individual site and edge phases are branch-dependent, while the Bethe combination in \eqref{eq:csd-index-bethe} is the integer index fixed by the determinant. Equivalently,
\begin{equation}
\exp\left[s\mathcal{K}_{\pmb A}(x)\right]=e^{sN}\prod_{i=1}^Ne^{s\kappa_i}\prod_{\{i,j\}\in E}e^{-s\kappa_{ij}}\,.
\label{eq:csd-local-tilt-factorization}
\end{equation}
This formula is important. It says that the matrix ensemble conditioned on an atypical index becomes, at the Bethe level, a graphical model whose local weights are phases of cavity determinants at $z_x^-$ and $z_x^+$. The probe resolvent $G^0$ is then propagated through this biased graphical model.

\begin{figure}[t]
\centering
\resizebox{0.98\textwidth}{!}{%
\begin{tikzpicture}[
    x=1cm,
    y=1cm,
    >=Latex,
    panel/.style={draw=black!18, fill=black!1, rounded corners=2pt, line width=0.5pt},
    ptitle/.style={font=\bfseries\small, anchor=west},
    paneltext/.style={font=\scriptsize, align=center},
    tinytext/.style={font=\tiny, align=center},
    axisline/.style={draw=black!70, line width=0.65pt, -{Latex[length=2.0mm,width=1.4mm]}},
    guide/.style={draw=black!35, dashed, line width=0.55pt},
    threshold/.style={draw=red!55!black, line width=0.80pt},
    probe/.style={draw=blue!55!black, line width=0.80pt},
    flow/.style={draw=black!60, line width=0.70pt, -{Latex[length=2.0mm,width=1.4mm]}},
    redflow/.style={draw=red!55!black, line width=0.70pt, -{Latex[length=2.0mm,width=1.4mm]}},
    blueflow/.style={draw=blue!55!black, line width=0.70pt, -{Latex[length=2.0mm,width=1.4mm]}},
    box/.style={draw=black!35, fill=white, rounded corners=2pt, line width=0.5pt, inner sep=3pt, font=\scriptsize, align=center},
    redbox/.style={draw=red!45!black, fill=red!2, rounded corners=2pt, line width=0.55pt, inner sep=3pt, font=\scriptsize, align=center},
    bluebox/.style={draw=blue!45!black, fill=blue!2, rounded corners=2pt, line width=0.55pt, inner sep=3pt, font=\scriptsize, align=center},
    sector/.style={draw=black!45, fill=white, rounded corners=2pt, line width=0.5pt, minimum height=8.0mm, font=\scriptsize, align=center},
    vnode/.style={circle, draw=black!75, fill=white, minimum size=6.8mm, inner sep=0pt, font=\scriptsize},
    edge/.style={draw=black!65, line width=0.65pt},
    msg/.style={draw=blue!55!black, line width=0.70pt, -{Latex[length=2.0mm,width=1.4mm]}}
]
\draw[panel] (0,5.80) rectangle (16.60,11.60);
\node[ptitle] at (0.35,11.25) {(a) Threshold $x$ fixes the conditioning; $\lambda$ is a separate probe};

\draw[axisline] (1.00,8.80) -- (15.35,8.80);
\node[tinytext] at (15.48,8.57) {spectral axis};
\draw[black!18, line width=5.0pt] (1.12,8.80) -- (6.70,8.80);
\node[tinytext] at (3.95,8.43) {eigenvalues counted below $x$};
\foreach \xx in {1.45,2.05,3.10,4.25,5.50,7.60,8.85,10.50,12.20,13.85}{
  \fill[black!55] (\xx,8.80) circle (1.2pt);
}

\coordinate (xpos) at (6.70,8.80);
\draw[threshold] ($(xpos)+(0,-1.35)$) -- ($(xpos)+(0,1.70)$);
\node[paneltext, text=red!55!black] at (6.70,7.27) {threshold $x$};
\node[redbox, text width=2.18cm] (zxplus) at (6.70,10.45) {$z_x^+=x+i\epsilon$};
\node[redbox, text width=2.18cm] (zxminus) at (6.70,6.35) {$z_x^-=x-i\epsilon$};
\draw[redflow] (zxplus.south) -- (6.70,9.20);
\draw[redflow] (zxminus.north) -- (6.70,8.40);
\node[redbox, text width=3.75cm] at (3.15,10.25)
{$G^+,G^-$ are the determinant sectors that encode $\mathcal K_{\pmb A}(x)$};

\coordinate (lpos) at (11.50,8.80);
\draw[probe] ($(lpos)+(0,-1.00)$) -- ($(lpos)+(0,1.00)$);
\node[paneltext, text=blue!55!black] at (11.50,10.07) {probe $\lambda$};
\node[bluebox, text width=2.35cm] (zl) at (11.50,6.98) {$z_\lambda=\lambda-i\eta$};
\draw[blueflow] (zl.north) -- (11.50,8.28);
\draw[probe, <->] (9.35,9.55) -- (13.70,9.55);
\node[tinytext, text=blue!55!black] at (11.50,9.80) {$\lambda$ scans the density};
\node[bluebox, text width=3.40cm] at (13.72,10.25)
{$G^0$ is the probe resolvent sector; it does not set the index constraint};

\node[box, text width=4.05cm] (hard) at (3.50,6.65)
{hard conditioning\\[-1mm]
$\displaystyle \mathcal K_{\pmb A}(x)/N=k$};
\node[box, text width=4.70cm] (soft) at (11.40,6.23)
{soft tilted ensemble\\[-1mm]
$\displaystyle P_s(\pmb A)\propto P(\pmb A)e^{s\mathcal K_{\pmb A}(x)}$,\quad $\displaystyle \psi_x'(s)=k$};
\draw[flow] (hard.east) to[bend left=4] node[pos=0.50, above, tinytext] {equivalent for local observables at the saddle} (soft.west);

\draw[panel] (0,0) rectangle (16.60,5.45);
\node[ptitle] at (0.35,5.10) {(b) Three-sector message and tilted Bethe sampling law};

\node[vnode] (lone) at (1.05,3.70) {$\ell_1$};
\node[vnode] (ltwo) at (1.05,2.40) {$\ell_2$};
\node[vnode] (i) at (3.10,3.05) {$i$};
\node[vnode] (j) at (4.92,3.05) {$j$};
\draw[edge] (lone) -- (i);
\draw[edge] (ltwo) -- (i);
\draw[edge, black!35, dashed] (i) -- (j);
\draw[msg] ($(lone)!0.45!(i)$) -- ($(lone)!0.78!(i)$);
\draw[msg] ($(ltwo)!0.45!(i)$) -- ($(ltwo)!0.78!(i)$);
\draw[msg] ($(i)+(0.37,0.22)$) -- ($(j)+(-0.37,0.22)$);
\node[tinytext] at (2.18,3.93) {$\pmb G_{\ell_1\to i}$};
\node[tinytext] at (2.16,2.14) {$\pmb G_{\ell_2\to i}$};
\node[tinytext] at (4.00,3.52) {$\pmb G_{i\to j}$};
\node[tinytext] at (4.00,2.62) {remove $j$};

\node[sector, text width=1.50cm] (gm) at (6.55,3.75) {$G^-_{i\to j}$\\[-1mm]$z_x^-$};
\node[sector, text width=1.50cm] (gp) at (8.15,3.75) {$G^+_{i\to j}$\\[-1mm]$z_x^+$};
\node[sector, text width=1.50cm] (g0) at (9.75,3.75) {$G^0_{i\to j}$\\[-1mm]$z_\lambda$};
\node[paneltext] at (8.15,4.55) {$\pmb G_{i\to j}=(G^-_{i\to j},G^+_{i\to j},G^0_{i\to j})$};

\node[redbox, text width=4.20cm] (bias) at (7.35,2.25)
{tilted Bethe weights use the determinant sectors\\[-1mm]
$e^{s\kappa_i}$ and $e^{-s\kappa_{ij}}$ depend on $G^-,G^+$};
\node[bluebox, text width=4.20cm] (probeBox) at (12.50,2.25)
{the probe sector is averaged in the same tilted law\\[-1mm]
$\displaystyle \rho_x(\lambda|k(s))=\frac{1}{\pi}{\rm Im}\langle G^0_{\rm site}\rangle_{s}$};
\draw[redflow] ($(gm.south)+(0,-0.03)$) -- ($(gm.south)+(0,-0.35)$) -- ($(bias.north)+(-0.65,0)$);
\draw[redflow] ($(gp.south)+(0,-0.03)$) -- ($(gp.south)+(0,-0.35)$) -- ($(bias.north)+(0.65,0)$);
\draw[blueflow] (g0.east) -- (probeBox.north west);
\draw[flow] (bias.east) -- node[pos=0.50, above, tinytext] {same tilted population} (probeBox.west);

\node[box, text width=14.70cm] at (8.30,0.82)
{$\displaystyle
G^a_{i\to j}=\left[z_a-D_i-\sum_{\ell\in\partial i\setminus j}J_{i\ell}^2G^a_{\ell\to i}\right]^{-1},\qquad
z_- = z_x^- ,\quad z_+=z_x^+ ,\quad z_0=z_\lambda ,\quad a\in\{-,+,0\}$};
\node[box, text width=14.70cm] at (8.30,0.25)
{The update is componentwise, but the three components are sampled jointly because the Bethe tilt changes the distribution of local environments.};
\end{tikzpicture}%
}
\caption{Threshold/probe separation and three-sector messages for the conditioned spectral density. The threshold $x$ fixes the index constraint through the two determinant boundary values $z_x^-$ and $z_x^+$, while the spectral parameter $\lambda$ is a separate probe used to measure the density. In the cavity formulation each directed edge therefore carries the triple $(G^-,G^+,G^0)$: the first two sectors define the tilted Bethe measure, and the third sector is averaged in that tilted measure to obtain the conditioned density.}
\label{fig:csd-threshold-probe-three-sector}
\end{figure}
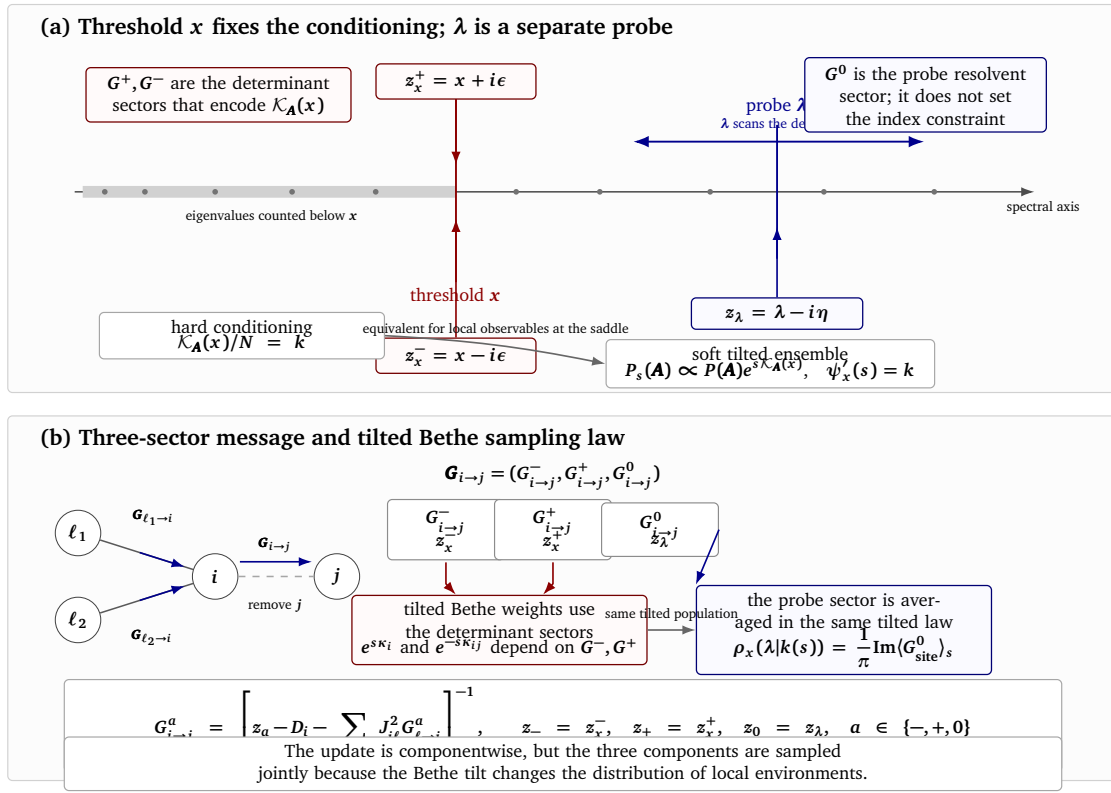

Figure~\ref{fig:csd-threshold-probe-three-sector} summarizes the separation between the conditioning threshold, the probe spectral parameter, and the three-sector cavity message used to compute the conditioned density.

Let $\mathbb{P}_s$ denote the replica-symmetric tilted law of the message triples \eqref{eq:csd-message-triple}. This law is obtained from the usual population dynamics, but with the site and edge weights in \eqref{eq:csd-local-tilt-factorization} included in the Bethe measure. The conditioned density is
\begin{equation}
\rho_x(\lambda|k(s))=\frac{1}{\pi}\lim_{\eta\downarrow0}{\rm Im}\int dG^- dG^+ dG^0 \mathcal{P}_{\rm site}^{(s)}(G^-,G^+,G^0)G^0\,,
\label{eq:csd-cavity-conditioned-density}
\end{equation}
where $\mathcal{P}_{\rm site}^{(s)}$ is the tilted distribution of full site Green-function triples. More explicitly, for a vertex of degree $m$, a full site triple is generated from incoming cavity triples by
\begin{equation}
G^{a}=\frac{1}{z_a-D-\displaystyle\sum_{r=1}^{m}J_r^2G_r^{a}}\,,\qquad a\in\{-,+,0\}\,,
\label{eq:csd-site-triple-map}
\end{equation}
and the tilted law weights the generated configuration by the index contribution
\begin{equation}
e^{s\kappa_{\rm site}}=\exp\left\{\frac{s}{2\pi i}\left[\log (G^-)^{-1}-\log (G^+)^{-1}\right]\right\}\,,
\label{eq:csd-site-tilt-weight}
\end{equation}
together with the corresponding edge corrections
\begin{equation}
e^{-s\kappa_{\rm edge}}=\exp\left\{-\frac{s}{2\pi i}\left[\log\left(1-J^2G_{i\to j}^-G_{j\to i}^-\right)-\log\left(1-J^2G_{i\to j}^+G_{j\to i}^+\right)\right]\right\}\,.
\label{eq:csd-edge-tilt-weight}
\end{equation}
Equations \eqref{eq:csd-triple-cavity-recursion}--\eqref{eq:csd-cavity-conditioned-density} are the practical cavity formulation of the conditioned spectral density. At $s=0$, the weights \eqref{eq:csd-site-tilt-weight} and \eqref{eq:csd-edge-tilt-weight} are equal to one, and the ordinary spectral-density population dynamics is recovered. For $s\neq0$, neighborhoods that increase or decrease the number of eigenvalues below $x$ are statistically favored according to the sign and magnitude of $s$.

The same construction applies to diluted Wishart matrices, with the ordinary graph replaced by the bipartite factor graph. Let
\begin{equation}
\pmb W=\frac{1}{d}\pmb X\pmb X^{\rm T}\,.
\label{eq:csd-wishart-matrix}
\end{equation}
For every directed variable-to-factor edge $i\to\mu$ we introduce
\begin{equation}
\pmb{G}_{i\to\mu}=\left(G_{i\to\mu}^-,G_{i\to\mu}^+,G_{i\to\mu}^0\right)\,,
\label{eq:csd-wishart-message-triple}
\end{equation}
and for every factor-to-variable edge $\mu\to i$ we introduce the corresponding self-energy triple
\begin{equation}
\pmb{U}_{\mu\to i}=\left(U_{\mu\to i}^-,U_{\mu\to i}^+,U_{\mu\to i}^0\right)\,.
\label{eq:csd-wishart-self-energy-triple}
\end{equation}
The local recursions are, for $a\in\{-,+,0\}$,
\begin{equation}
U_{\mu\to i}^{a}=\frac{(\xi_i^\mu)^2}{1-\displaystyle\frac{1}{d}\sum_{j\in\partial\mu\setminus i}(\xi_j^\mu)^2G_{j\to\mu}^{a}}\,,
\label{eq:csd-wishart-factor-recursion}
\end{equation}
and
\begin{equation}
G_{i\to\mu}^{a}=\frac{1}{z_a-\displaystyle\frac{1}{d}\sum_{\nu\in\partial i\setminus\mu}U_{\nu\to i}^{a}}\,.
\label{eq:csd-wishart-variable-recursion}
\end{equation}
The full local Green functions are
\begin{equation}
G_i^{a}=\frac{1}{z_a-\displaystyle\frac{1}{d}\sum_{\nu\in\partial i}U_{\nu\to i}^{a}}\,.
    \label{eq:csd-wishart-full-green}
\end{equation}
The index decomposition now contains variable, factor, and edge terms:
\begin{equation}
\mathcal{K}_{\pmb W}(x)=N+\sum_{i=1}^N\kappa_i+\sum_{\mu=1}^P\kappa_\mu-\sum_{(i,\mu)}\kappa_{i\mu}.
\label{eq:csd-wishart-index-bethe}
\end{equation}

\begin{examplebox}[Conditioning in a one-factor Wishart example]
Let
\begin{equation}
\pmb X=\begin{pmatrix}
\xi_1\\
\xi_2
\end{pmatrix}\,,\qquad d=1\,,\qquad\pmb W=\pmb X\pmb X^{\rm T}\,.
\label{eq:csd-ped-wishart-one-factor}
\end{equation}
The eigenvalues are
\begin{equation}
0\,,\qquad \xi_1^2+\xi_2^2\,.
\label{eq:csd-ped-wishart-one-factor-eigs}
\end{equation}
If the conditioning threshold satisfies
\begin{equation}
0<x<\xi_1^2+\xi_2^2\,,
\label{eq:csd-ped-wishart-threshold}
\end{equation}
then
\begin{equation}
\mathcal K_{\pmb W}(x)=1\,.
\label{eq:csd-ped-wishart-index-one}
\end{equation}
The empirical density is
\begin{equation}
\rho_{\pmb W}(\lambda)=\frac{1}{2}\delta(\lambda)+\frac{1}{2}\delta\left(\lambda-\xi_1^2-\xi_2^2\right)\,.
\label{eq:csd-ped-wishart-density}
\end{equation}
Therefore, with the convention that the zero mode is included in the count below any positive threshold,
\begin{equation}
\int_{-\infty}^x d\lambda\rho_{\pmb W}(\lambda)=\frac{1}{2}\,,
\label{eq:csd-ped-wishart-density-constraint}
\end{equation}
which agrees with $\mathcal K_{\pmb W}(x)/N=1/2$. In larger diluted Wishart ensembles, the same statement becomes nontrivial: the tilted cavity equations bias the bipartite graph so that the resulting local density satisfies the prescribed index constraint.
\end{examplebox}

The variable contribution is
\begin{equation}
\kappa_i=\frac{1}{2\pi i}\lim_{\epsilon\downarrow0}\left[\log (G_i^-)^{-1}-\log (G_i^+)^{-1}\right]\,,
\label{eq:csd-wishart-variable-contribution}
\end{equation}
the factor contribution is
\begin{equation}
\kappa_\mu=\frac{1}{2\pi i}\lim_{\epsilon\downarrow0}\left[\log R_\mu^--\log R_\mu^+\right]\,,
\label{eq:csd-wishart-factor-contribution}
\end{equation}
with
\begin{equation}
R_\mu^{a}=1-\frac{1}{d}\sum_{i\in\partial\mu}(\xi_i^\mu)^2G_{i\to\mu}^{a}\,,\qquad a\in\{-,+\}\,,
\label{eq:csd-wishart-factor-denominator}
\end{equation}
and the edge contribution is
\begin{equation}
\kappa_{i\mu}=\frac{1}{2\pi i}\lim_{\epsilon\downarrow0}\left[\log E_{i\mu}^--\log E_{i\mu}^+\right]\,,
\label{eq:csd-wishart-edge-contribution}
\end{equation}
with
\begin{equation}
E_{i\mu}^{a}=1-\frac{1}{d}G_{i\to\mu}^{a}U_{\mu\to i}^{a}\,,\qquad a\in\{-,+\}\,.
\label{eq:csd-wishart-edge-denominator}
\end{equation}
The logarithms in \eqref{eq:csd-wishart-variable-contribution}--\eqref{eq:csd-wishart-edge-contribution} must be evaluated with branches consistent with the determinant phase. The individual variable, factor, and edge phases are branch-dependent, while the Bethe combination in \eqref{eq:csd-wishart-index-bethe} gives the index. The conditioned Wishart density is then
\begin{equation}
\rho_x^{(\rm W)}(\lambda|k(s))=\frac{1}{\pi}\lim_{\eta\downarrow0}{\rm Im}\int dG^-dG^+dG^0\mathcal{P}_{{\rm site},\,{\rm W}}^{(s)}(G^-,G^+,G^0)G^0\,.
\label{eq:csd-wishart-conditioned-density}
\end{equation}
The meaning of \eqref{eq:csd-wishart-conditioned-density} is exactly the same as in the sparse symmetric case: the ordinary local density is averaged in the ensemble tilted by the index below $x$. The only difference is that the local graph is bipartite and the Bethe determinant contains variable, factor, and edge contributions.

We can now summarize the calculation in a way that is useful for implementation. One first fixes the conditioning threshold $x$ and the desired fraction $k$. Instead of imposing $k$ directly, one solves the tilted large-deviation problem for a value of $s$ such that
\begin{equation}
\psi_x'(s)=k\,.
\label{eq:csd-find-s}
\end{equation}
One then propagates, in the corresponding tilted population, an additional probe resolvent at $z_\lambda=\lambda-i\eta$. The conditioned density is obtained from the imaginary part of the full probe Green function. In equations, for sparse symmetric matrices,
\begin{equation}
\rho_x(\lambda|k)=\frac{1}{\pi}\lim_{\eta\downarrow0}{\rm Im}\left\langle G^0_{\rm site}\right\rangle_{s(k)}\,.
\label{eq:csd-final-sparse-formula}
\end{equation}
For diluted Wishart matrices the same formula holds with the bipartite site distribution:
\begin{equation}
\rho_x^{(\rm W)}(\lambda|k)=\frac{1}{\pi}\lim_{\eta\downarrow0}{\rm Im}\left\langle G^0_{\rm variable}\right\rangle_{s(k)}\,.
\label{eq:csd-final-wishart-formula}
\end{equation}
The notation $\langle\cdot\rangle_{s(k)}$ means that the average is not taken in the original matrix ensemble, but in the tilted ensemble whose typical index is $k$.

Several checks follow immediately. First, setting $s=0$ gives $k=k_\star(x)$ and the tilted measure becomes the original measure. Thus \eqref{eq:csd-final-sparse-formula} reduces to the ordinary spectral density. Second, integrating the conditioned density below the threshold gives
\begin{equation}
\int_{-\infty}^{x}d\lambda \rho_x(\lambda|k(s))=\lim_{N\to\infty}\frac{1}{N}\left\langle\mathcal{K}_{\pmb A}(x)\right\rangle_s=\psi_x'(s)=k(s)\,,
\label{eq:csd-integral-check}
\end{equation}
which verifies the conditioning. Third, integrating over all $\lambda$ gives one, because every matrix in the tilted ensemble has a normalized empirical spectral measure. Fourth, if the threshold is below the entire spectrum or above the entire spectrum, the tilted ensemble is trivial: it conditions on an event that is already deterministic, and the conditioned density is again the ordinary density.

The physical interpretation is different from the Coulomb-gas picture. In an invariant ensemble, imposing an atypical index forces a collective rearrangement of a gas of strongly repelling eigenvalues. In a sparse non-invariant ensemble, imposing an atypical index biases the distribution of local graph environments and local resolvent messages. The conditioned density is therefore obtained by averaging local spectral weights in a biased graphical ensemble. This is why the conditioned density of sparse matrices can behave very differently from the conditioned density of invariant random matrices: the constraint is not implemented by moving a continuum of repelling charges, but by changing the relative statistical weight of local structures that contribute eigenvalues below the threshold.

The main lesson of this section is that the conditioned spectral density is not an independent object requiring a completely new method. It is the ordinary resolvent density evaluated in the large-deviation ensemble of the index number. The determinant identity \eqref{eq:csd-index-determinant} creates the tilted ensemble; the Edwards--Jones identity \eqref{eq:csd-density-free-energy} inserts a probe resolvent; and the cavity method computes the resulting local Green function in the tilted graphical measure. This is the natural extension of the spectral-density and index-number calculations to the full spectrum conditioned on an atypical spectral count.

\begin{exerciseblock}
\exitem[Definition of the conditioned density]
Starting from
\begin{equation}
\rho_x(\lambda|k)=\lim_{N\to\infty}\frac{\overline{\rho_{\pmb A}(\lambda)\delta(\mathcal K_{\pmb A}(x)-Nk)}}{\overline{\delta(\mathcal K_{\pmb A}(x)-Nk)}}\,,
\label{eq:csd-ex-definition}
\end{equation}
show that
\begin{equation}
\int d\lambda \rho_x(\lambda|k)=1
\label{eq:csd-ex-normalization}
\end{equation}
provided each empirical density is normalized.

\exitem[Index constraint]
Using the same definition, prove that
\begin{equation}
\int_{-\infty}^{x}d\lambda \rho_x(\lambda|k)=k\,.
\label{eq:csd-ex-index-constraint}
\end{equation}
What assumptions about eigenvalues exactly at the threshold are being made?

\exitem[Typical density as a special case]
Let
\begin{equation}
k_\star(x)=\int_{-\infty}^{x}d\lambda \overline{\rho_{\pmb A}(\lambda)}\,.
\label{eq:csd-ex-kstar}
\end{equation}
Explain why one expects
\begin{equation}
\rho_x(\lambda|k_\star)=\overline{\rho_{\pmb A}(\lambda)}
\label{eq:csd-ex-typical-density}
\end{equation}
when the conditioning imposes the typical value of the index, assuming the same convention for possible eigenvalues at the threshold $x$ on both sides.

\exitem[Formal Laplace representation of the constraint]
Using the formal large-deviation representation
\begin{equation}
\delta(\mathcal K_{\pmb A}(x)-Nk)=\int_{\gamma-i\infty}^{\gamma+i\infty}\frac{ds}{2\pi i}e^{s\mathcal K_{\pmb A}(x)-sNk}\,,
\label{eq:csd-ex-laplace}
\end{equation}
derive the saddle-point condition
\begin{equation}
\psi_x'(s)=k\,.
\label{eq:csd-ex-saddle-condition}
\end{equation}

\exitem[Equivalence between hard conditioning and tilting]
Using the saddle-point calculation, show that
\begin{equation}
\rho_x(\lambda|k(s))=\lim_{N\to\infty}\left\langle\rho_{\pmb A}(\lambda)\right\rangle_s\,,
\label{eq:csd-ex-hard-soft-equivalence}
\end{equation}
where $\langle\cdots\rangle_s$ denotes expectation in the tilted ensemble.

\exitem[Determinant representation of the index]
Use, for $u\neq0$ and with eigenvalues at the threshold treated by the fixed endpoint convention,
\begin{equation}
\Theta(u)=1+\frac{1}{2\pi i}\lim_{\epsilon\downarrow0}\left[\log(u-i\epsilon)-\log(u+i\epsilon)\right]
\label{eq:csd-ex-theta-log}
\end{equation}
to derive
\begin{equation}
\mathcal K_{\pmb A}(x)=N+\frac{1}{2\pi i}\lim_{\epsilon\downarrow0}\left[\log\Delta_{\pmb A}(x-i\epsilon)-\log\Delta_{\pmb A}(x+i\epsilon)\right]\,.
\label{eq:csd-ex-index-det}
\end{equation}

\exitem[Probe resolvent]
Show that
\begin{equation}
\rho_{\pmb A,\eta}(\lambda)=-\frac{2}{\pi N}{\rm Im}\frac{\partial}{\partial\lambda}\log Z_{\pmb A}(\lambda-i\eta)\,.
\label{eq:csd-ex-probe-resolvent}
\end{equation}
Explain why the conditioned density calculation needs both the threshold parameters $x\mp i\epsilon$ and the probe parameter $\lambda-i\eta$.

\exitem[Three-sector replica structure]
Explain why the joint generating function
\begin{equation}
\overline{e^{s\mathcal K_{\pmb A}(x)}[Z_{\pmb A}(z_\lambda)]^n}
\label{eq:csd-ex-three-sector-generating}
\end{equation}
contains two sectors associated with the index constraint and one additional sector associated with the density probe.

\exitem[Triple cavity recursion]
Starting from the ordinary sparse symmetric cavity recursion, derive
\begin{equation}
G_{i\to j}^{a}=\frac{1}{z_a-D_i-\displaystyle\sum_{\ell\in\partial i\setminus j}J_{i\ell}^2G_{\ell\to i}^{a}}\,,\qquad a\in\{-,+,0\}\,.
\label{eq:csd-ex-triple-recursion}
\end{equation}
Explain why the three components are statistically coupled even though the recursion is componentwise.

\exitem[Bethe determinant on a two-site graph]
For
\begin{equation}
\pmb A=\begin{pmatrix}
0 & J\\
J & 0
\end{pmatrix}\,,
\label{eq:csd-ex-two-site}
\end{equation}
verify the Bethe determinant identity
\begin{equation}
\Delta_{\pmb A}(z)=\prod_i G_i(z)^{-1}\prod_{\{i,j\}}\left[1-J^2G_{i\to j}(z)G_{j\to i}(z)\right]^{-1}\,.
\label{eq:csd-ex-bethe-two-site}
\end{equation}

\exitem[Index decomposition]
Starting from the Bethe determinant, derive
\begin{equation}
\mathcal K_{\pmb A}(x)=N+\sum_i\kappa_i-\sum_{\{i,j\}\in E}\kappa_{ij}\,.
\label{eq:csd-ex-index-decomposition}
\end{equation}
Give explicit formulas for $\kappa_i$ and $\kappa_{ij}$.

\exitem[Local tilt factorization]
Show that
\begin{equation}
e^{s\mathcal K_{\pmb A}(x)}=e^{sN}\prod_i e^{s\kappa_i}\prod_{\{i,j\}}e^{-s\kappa_{ij}}\,.
\label{eq:csd-ex-local-tilt}
\end{equation}
Explain why this is the graphical origin of the tilted cavity measure.

\exitem[Conditioned density formula]
Using the tilted site distribution, derive
\begin{equation}
\rho_x(\lambda|k(s))=\frac{1}{\pi}\lim_{\eta\downarrow0}{\rm Im} \int dG^- dG^+ dG^0\mathcal P_{\rm site}^{(s)}(G^-,G^+,G^0)G^0\,.
\label{eq:csd-ex-conditioned-density-formula}
\end{equation}

\exitem[Wishart triple messages]
For the diluted Wishart case, derive the triple recursions
\begin{equation}
U_{\mu\to i}^{a}=\frac{(\xi_i^\mu)^2}{1-\displaystyle\frac{1}{d}\sum_{j\in\partial\mu\setminus i}(\xi_j^\mu)^2G_{j\to\mu}^{a}}\,,
\label{eq:csd-ex-wishart-factor-triple}
\end{equation}
and
\begin{equation}
G_{i\to\mu}^{a}=\frac{1}{z_a-\displaystyle\frac{1}{d}\sum_{\nu\in\partial i\setminus\mu}U_{\nu\to i}^{a}}\,,\qquad a\in\{-,+,0\}\,.
\label{eq:csd-ex-wishart-variable-triple}
\end{equation}

\exitem[Wishart Bethe index decomposition]
Derive the bipartite decomposition
\begin{equation}
\mathcal K_{\pmb W}(x)=N+\sum_i\kappa_i+\sum_\mu\kappa_\mu-\sum_{(i,\mu)}\kappa_{i\mu}\,.
\label{eq:csd-ex-wishart-index-decomposition}
\end{equation}
Explain why factor terms appear in addition to variable and edge terms.

\exitem[Conditioned Wishart density]
Show that the conditioned Wishart density can be written as
\begin{equation}
\rho_x^{(\rm W)}(\lambda|k(s))=\frac{1}{\pi}\lim_{\eta\downarrow0}{\rm Im}\left\langle G_{\rm variable}^0 \right\rangle_s\,.
\label{eq:csd-ex-wishart-conditioned-density}
\end{equation}

\exitem[Trivial conditioning]
Show that if $x$ is below the entire spectrum or above the entire spectrum, then the index is deterministic and the conditioned density is the ordinary density.

\exitem[Invariant versus sparse mechanism]
Explain in words why, in an invariant ensemble, conditioning the index suggests a Coulomb-gas constrained minimization, while in a sparse non-invariant ensemble it suggests a tilted distribution of local graph environments.

\exitem[Programming exercise: rejection conditioning]
Fix a list of system sizes $N$, a mean degree $c=O(1)$, a number $M$ of independent samples for each $N$, a threshold $x$, an endpoint convention for eigenvalues exactly at $x$, and a common density-estimation convention, such as a histogram bin width or a Lorentzian regulator. Generate sparse Erd\H{o}s--R\'enyi adjacency matrices $\pmb A$ with zero diagonal entries, $A_{ij}=A_{ji}$, and
\begin{equation}
{\rm Prob}(A_{ij}=1)=\frac{c}{N}\,,\qquad{\rm Prob}(A_{ij}=0)=1-\frac{c}{N}\,,\qquad i<j\,.
\label{eq:csd-ex-program-er-law}
\end{equation}
Compute their spectra. For each value of $N$, choose an integer target count $K_{\rm target}$, define $k_N=K_{\rm target}/N$, and retain only samples satisfying
\begin{equation}
\mathcal K_{\pmb A}(x)=K_{\rm target}\,.
\label{eq:csd-ex-program-rejection}
\end{equation}
Estimate the empirical conditioned density from the retained samples and compare it with the unconditioned density using the same density-estimation convention. Report $N$, $c$, $M$, $x$, the endpoint convention, $K_{\rm target}$, the number of retained samples, and the bin width or regulator.

\exitem[Programming exercise: soft tilting]
Using the same samples and the same density-estimation convention as in the previous exercise, let $\rho_m(\lambda)$ be the empirical spectral density of sample $m$ and let $\mathcal K_m(x)$ be its index below $x$. Fix a grid of real values of $s$ and estimate the tilted density
\begin{equation}
\rho_s(\lambda)=\frac{\sum_m e^{s\mathcal K_m(x)}\rho_m(\lambda)}{\sum_m e^{s\mathcal K_m(x)}}\,.
\label{eq:csd-ex-program-soft-tilt}
\end{equation}
For each $s$, estimate the tilted mean intensive index
\[
k(s)=\frac{1}{N}\frac{\sum_m e^{s\mathcal K_m(x)}\mathcal K_m(x)}{\sum_m e^{s\mathcal K_m(x)}}\,.
\]
Compare $\rho_s(\lambda)$ with the hard-conditioned density at nearby values of $k_N$ when enough retained samples are available. Report the $s$ grid and an effective sample-size diagnostic for the tilted weights.

\exitem[Programming exercise: conditioned Wishart spectra]
Fix a list of system sizes $N$, an aspect ratio $\alpha>0$ such that $P=N/\alpha$ is an integer for each $N$, a dilution $d>0$, a number $M$ of independent samples for each parameter set, a nonzero-weight distribution $p_\xi$ with zero mean and unit variance, a positive threshold $x$, an endpoint convention for eigenvalues exactly at $x$, and a common density-estimation convention. Generate diluted Wishart matrices from sparse rectangular matrices
\begin{equation}
X_i^\mu=B_i^\mu\xi_i^\mu\,,\qquad {\rm Prob}(B_i^\mu=1)=\frac{d}{N}\,, \qquad {\rm Prob}(B_i^\mu=0)=1-\frac{d}{N}\,,
\label{eq:csd-ex-program-wishart-support}
\end{equation}
with independent nonzero weights $\xi_i^\mu$ drawn from $p_\xi$, and form
\begin{equation}
\pmb W=\frac{1}{d}\pmb X\pmb X^{\rm T}\,.
\label{eq:csd-ex-program-wishart-W}
\end{equation}
For each sample, compute $\mathcal K_{\pmb W}(x)$ and the empirical spectral density. Compare conditioned densities for samples with unusually many and unusually few eigenvalues below $x$, using either explicit count windows or selected quantiles of the empirical distribution of $\mathcal K_{\pmb W}(x)$. Report $N$, $P$, $\alpha$, $d$, $M$, $p_\xi$, $x$, the endpoint convention, the count windows or quantiles used for conditioning, and the density-estimation convention.
\end{exerciseblock}

\section{Number statistics of non-Hermitian random matrices}
\label{sec:number-statistics-non-hermitian}
The Hermitian index number counts eigenvalues on one side of a point on the real axis, or inside an interval. For non-Hermitian matrices the eigenvalues are complex, and the analogous observable is the number of eigenvalues inside a two-dimensional domain of the complex plane. This section develops the corresponding formalism explicitly. The main point is that the number of eigenvalues inside a domain bounded by a contour can be written either as a winding number of the characteristic polynomial or as a boundary integral of the Hermitized logarithmic potential. These two representations are equivalent. The first one is the direct non-Hermitian analogue of the argument-principle formula in complex analysis; the second one is the form naturally adapted to Gaussian integrals, Hermitization, and cavity methods. The statistical-mechanics approach to this problem was developed in \cite{RamosSanchezGuzmanGonzalezPerezCastilloMetz2021}, building on the Hermitization ideas used in non-Hermitian random matrix theory \cite{Girko1984,FeinbergZee1997,TaoVuKrishnapur2010} and on the cavity treatment of sparse non-Hermitian matrices \cite{RogersPerezCastillo2009,MetzNeriRogers2019}.

Let $\pmb A$ be a general $N\times N$ real or complex non-Hermitian matrix, with eigenvalues
\begin{equation}
z_1,\ldots,z_N\in\mathbb{C}\,.
\label{eq:nhns-eigenvalues}
\end{equation}
The empirical spectral density in the complex plane is
\begin{equation}
\rho_{\pmb A}(z)=\frac{1}{N}\sum_{i=1}^{N}\delta^{(2)}(z-z_i)\,,\qquad z=x+iy\,.
\label{eq:nhns-empirical-density}
\end{equation}
Let $\mathcal{D}\subset\mathbb{C}$ be a bounded domain with positively oriented boundary $\partial\mathcal{D}=\Gamma$, and assume for the moment that no eigenvalue lies exactly on $\Gamma$. The number of eigenvalues in $\mathcal{D}$ is
\begin{equation}
\mathcal{N}_{\pmb A}(\mathcal{D})=\sum_{i=1}^{N}\mathbf{1}_{z_i\in\mathcal{D}}=N\int_{\mathcal{D}}d^2z\rho_{\pmb A}(z)\,.
\label{eq:nhns-number-definition}
\end{equation}

\begin{examplebox}[Counting eigenvalues in a domain]
Consider the diagonal non-Hermitian matrix
\begin{equation}
\pmb A=\begin{pmatrix}
1+i & 0 & 0\\
0 & -1+\frac{i}{2} & 0\\
0 & 0 & 2-2i
\end{pmatrix}\,.
\label{eq:nhns-ped-diagonal-example}
\end{equation}
Its eigenvalues are
\begin{equation}
z_1=1+i\,,\qquad z_2=-1+\frac{i}{2}\,,\qquad z_3=2-2i\,.
\label{eq:nhns-ped-diagonal-eigenvalues}
\end{equation}
Let $\mathcal D_R$ be the disk of radius $R$ centered at the origin,
\begin{equation}
\mathcal D_R=\{z\in\mathbb C: |z|<R\}\,.
\label{eq:nhns-ped-disk}
\end{equation}
The moduli of the eigenvalues are
\begin{equation}
|z_1|=\sqrt2\,, \qquad |z_2|=\sqrt{1+\frac14}=\frac{\sqrt5}{2}\,, \qquad |z_3|=\sqrt8=2\sqrt2\,.
\label{eq:nhns-ped-eigenvalue-moduli}
\end{equation}
Therefore, if
\begin{equation}
\frac{\sqrt5}{2}<R<\sqrt2\,,
\label{eq:nhns-ped-first-radius-window}
\end{equation}
only $z_2$ lies in the disk and
\begin{equation}
\mathcal N_{\pmb A}(\mathcal D_R)=1\,.
\label{eq:nhns-ped-count-one}
\end{equation}
If
\begin{equation}
\sqrt2<R<2\sqrt2\,,
\label{eq:nhns-ped-second-radius-window}
\end{equation}
then $z_1$ and $z_2$ lie in the disk, and
\begin{equation}
\mathcal N_{\pmb A}(\mathcal D_R)=2\,.
\label{eq:nhns-ped-count-two}
\end{equation}
Finally, if
\begin{equation}
R>2\sqrt2,
\label{eq:nhns-ped-third-radius-window}
\end{equation}
all three eigenvalues lie in the disk, so
\begin{equation}
\mathcal N_{\pmb A}(\mathcal D_R)=3\,.
\label{eq:nhns-ped-count-three}
\end{equation}
In terms of the empirical density,
\begin{equation}
\rho_{\pmb A}(z)=\frac{1}{3}\sum_{i=1}^{3}\delta^{(2)}(z-z_i)\,,
\label{eq:nhns-ped-diagonal-density}
\end{equation}
the intensive count is
\begin{equation}
\frac{1}{3}\mathcal N_{\pmb A}(\mathcal D_R)=\int_{\mathcal D_R}d^2z\rho_{\pmb A}(z)\,.
\label{eq:nhns-ped-count-density-relation}
\end{equation}
This finite example fixes the meaning of the observable: in the non-Hermitian problem the spectral count is a two-dimensional count of eigenvalues inside a domain, not a one-dimensional count across an interval.
\end{examplebox}

The intensive number is
\begin{equation}
k_{\mathcal{D}}=\frac{1}{N}\mathcal{N}_{\pmb A}(\mathcal{D})\,.
\label{eq:nhns-intensive-number}
\end{equation}
For a fixed domain, the first task is to compute the distribution of $\mathcal{N}_{\pmb A}(\mathcal{D})$ over the random matrix ensemble. The mean count is $N\int_{\mathcal D}d^2z\,\overline{\rho_{\pmb A}(z)}$, or equivalently the mean intensive count is obtained by integrating the average density. The variance, higher cumulants, and large deviations require a genuine number-statistics calculation.

The most elementary representation follows from the argument principle. Define the characteristic polynomial
\begin{equation}
p_{\pmb A}(z)=\det(z\pmb I-\pmb A)=\prod_{i=1}^{N}(z-z_i)\,.
\label{eq:nhns-characteristic-polynomial}
\end{equation}
For $z\notin\{z_1,\ldots,z_N\}$,
\begin{equation}
\partial_z\log p_{\pmb A}(z)={\rm Tr} (z\pmb I-\pmb A)^{-1}=\sum_{i=1}^{N}\frac{1}{z-z_i}\,.
\label{eq:nhns-log-derivative}
\end{equation}
Therefore
\begin{equation}
\frac{1}{2\pi i}\oint_{\Gamma}dz \partial_z\log p_{\pmb A}(z)=\frac{1}{2\pi i}\sum_{i=1}^{N}\oint_{\Gamma}\frac{dz}{z-z_i}=\sum_{i=1}^{N}\mathbf{1}_{z_i\in\mathcal{D}}=\mathcal{N}_{\pmb A}(\mathcal{D})\,.
\label{eq:nhns-argument-principle}
\end{equation}
Equivalently,
\begin{equation}
\mathcal{N}_{\pmb A}(\mathcal{D})=\frac{1}{2\pi i}\oint_{\Gamma}dz {\rm Tr}(z\pmb I-\pmb A)^{-1}\,.
\label{eq:nhns-number-resolvent-contour}
\end{equation}
This is the non-Hermitian counterpart of the Hermitian formula expressing the number of eigenvalues in an interval as an integral of the resolvent discontinuity. The difference is geometric: the Hermitian count uses the boundary values across a segment of the real axis, while the non-Hermitian count uses a contour integral in the complex plane.

\begin{examplebox}[The argument principle for two eigenvalues]
Let
\begin{equation}
p(z)=(z-a)(z-b)\,,
\label{eq:nhns-ped-two-zero-polynomial}
\end{equation}
and let $\Gamma$ be a positively oriented simple closed contour. Assume that $a$ lies inside $\Gamma$ and $b$ lies outside $\Gamma$. Then
\begin{equation}
\frac{d}{dz}\log p(z)=\frac{1}{z-a}+\frac{1}{z-b}\,.
\label{eq:nhns-ped-log-derivative-two-zero}
\end{equation}
The argument-principle integral is
\begin{align}
\frac{1}{2\pi i}\oint_{\Gamma}dz\frac{d}{dz}\log p(z)&=\frac{1}{2\pi i}\oint_{\Gamma}\frac{dz}{z-a}+\frac{1}{2\pi i}\oint_{\Gamma}\frac{dz}{z-b}\nonumber\\
&=1+0=1\,.
\label{eq:nhns-ped-two-zero-argument-principle}
\end{align}
Thus the contour integral counts exactly the zeros of $p$ inside the contour.

For a non-Hermitian matrix, the characteristic polynomial is
\begin{equation}
p_{\pmb A}(z)=\det(z\pmb I-\pmb A)\,,
\label{eq:nhns-ped-characteristic-polynomial-recall}
\end{equation}
and its zeros are the eigenvalues of $\pmb A$. Therefore
\begin{equation}
\frac{1}{2\pi i}\oint_{\Gamma}dz\partial_z\log p_{\pmb A}(z)
\label{eq:nhns-ped-matrix-argument-principle}
\end{equation}
counts the eigenvalues inside $\Gamma$. This is the direct non-Hermitian analogue of counting Hermitian eigenvalues in an interval by integrating the resolvent discontinuity.
\end{examplebox}

Let us make the winding interpretation explicit. Parametrize the contour by
\begin{equation}
z=z(\theta)\,,\qquad 0\leq\theta\leq2\pi\,, \qquad z(2\pi)=z(0)\,,
\label{eq:nhns-contour-parametrization}
\end{equation}
with positive orientation. Then
\begin{equation}
\mathcal{N}_{\pmb A}(\mathcal{D})=\frac{1}{2\pi i}\int_0^{2\pi}d\theta \frac{d}{d\theta}\log p_{\pmb A}(z(\theta))\,.
\label{eq:nhns-winding-log}
\end{equation}
Although $p_{\pmb A}(z(2\pi))=p_{\pmb A}(z(0))$, the logarithm need not return to the same branch. The change in the continuous argument of $p_{\pmb A}(z(\theta))$ is $2\pi$ times the number of zeros enclosed by the contour:
\begin{equation}
\mathcal{N}_{\pmb A}(\mathcal{D})=\frac{1}{2\pi}\Delta_\Gamma\arg p_{\pmb A}(z)\,.
\label{eq:nhns-winding-argument}
\end{equation}
This is why a naive product of determinant ratios around a closed discrete contour would appear to cancel if one ignored the branch of the logarithm. The observable is not the final value of the determinant, but the winding of its phase along the contour.

\begin{examplebox}[The winding number for a single eigenvalue]
Let
\begin{equation}
p(z)=z-a
\label{eq:nhns-ped-single-zero-polynomial}
\end{equation}
and let $\Gamma$ be the circle
\begin{equation}
z(\theta)=a+Re^{i\theta}\,, \qquad 0\leq\theta\leq2\pi\,.
\label{eq:nhns-ped-circle-around-a}
\end{equation}
Then
\begin{equation}
p(z(\theta))=Re^{i\theta}\,.
\label{eq:nhns-ped-p-on-circle}
\end{equation}
As $\theta$ goes from $0$ to $2\pi$, the phase of $p(z(\theta))$ increases by $2\pi$. Therefore
\begin{equation}
\frac{1}{2\pi}\Delta_\Gamma\arg p(z)=1\,.
\label{eq:nhns-ped-winding-one}
\end{equation}
The contour winds once around the zero.

Now take instead a circle not enclosing $a$, for example
\begin{equation}
z(\theta)=Re^{i\theta}\,,\qquad R<|a|\,.
\label{eq:nhns-ped-circle-not-around-a}
\end{equation}
Then the curve $p(z(\theta))=Re^{i\theta}-a$ is a circle centered at $-a$ with radius $R$, and it does not enclose the origin. Its net change of argument is zero:
\begin{equation}
\Delta_\Gamma\arg p(z)=0\,.
\label{eq:nhns-ped-winding-zero}
\end{equation}
This is the geometric content of the argument principle. The number statistic counts how many times the characteristic polynomial winds around the origin as the spectral parameter moves along the boundary of the domain.
\end{examplebox}

The generating function of the number statistic is
\begin{equation}
\mathcal{G}_{\mathcal{D},N}(s)=\overline{\exp\left[s \mathcal{N}_{\pmb A}(\mathcal{D})\right]}\,,
\label{eq:nhns-generating-function-finite}
\end{equation}
and the scaled cumulant-generating function, in the finite-connectivity setting emphasized in these notes, is
\begin{equation}
\psi_{\mathcal{D}}(s)=\lim_{N\to\infty}\frac{1}{N}\log\mathcal{G}_{\mathcal{D},N}(s)\,.
\label{eq:nhns-scaled-cgf}
\end{equation}
When a large-deviation principle with speed $N$ holds, for admissible values of $k$ along a finite-$N$ sequence, or equivalently for small count windows around $k$, one writes
\begin{equation}
{\rm Prob}\left[\frac{1}{N}\mathcal{N}_{\pmb A}(\mathcal{D})= k\right]\asymp\exp\left[-N\Phi_{\mathcal{D}}(k)\right]\,,
\label{eq:nhns-ldp}
\end{equation}
with
\begin{equation}
\Phi_{\mathcal{D}}(k)=\sup_s\left\{sk-\psi_{\mathcal{D}}(s)\right\}
\label{eq:nhns-rate-function}
\end{equation}
whenever the Legendre-Fenchel transform is regular. The derivatives of $\psi_{\mathcal{D}}$ at the origin give the cumulants per degree of freedom:
\begin{equation}
\psi_{\mathcal{D}}'(0)=\lim_{N\to\infty}\frac{1}{N}\overline{\mathcal{N}_{\pmb A}(\mathcal{D})}\,,\qquad \psi_{\mathcal{D}}''(0)=\lim_{N\to\infty}\frac{1}{N}{\rm Var}\mathcal{N}_{\pmb A}(\mathcal{D})\,.
\label{eq:nhns-first-cumulants}
\end{equation}
For dense invariant non-Hermitian ensembles, macroscopic number large deviations may have a different speed because of Coulomb-gas eigenvalue interactions, as happens in Hermitian invariant ensembles \cite{DeanMajumdar2006,MajumdarNadalScardicchioVivo2009}. In sparse and other non-invariant finite-connectivity ensembles, the natural statistical-mechanics object is instead a tilted graphical measure, and the speed $N$ is the analogue of the Hermitian sparse index problem.

Although the argument-principle formula is conceptually simple, it is not the most convenient starting point for Gaussian integral methods. We now reuse Hermitization in a different role: the same doubled determinant used earlier for non-Hermitian densities is integrated through its normal derivative along the counting boundary. The reason is that the non-Hermitian resolvent is unstable near the spectrum, and ordinary bosonic Gaussian integrals involving $z\pmb I-\pmb A$ are not convergent in general. Hermitization cures this. Define, for $\eta>0$,
\begin{equation}
\pmb H_{\pmb A}(z,\eta)=(z\pmb I-\pmb A)(z^*\pmb I-\pmb A^\dagger)+\eta^2\pmb I\,.
\label{eq:nhns-positive-hermitized-matrix}
\end{equation}
Then
\begin{equation}
\log\det\pmb H_{\pmb A}(z,\eta)=\sum_{i=1}^{N}\log\left(|z-z_i|^2+\eta^2\right)
\end{equation}
only when $\pmb A$ is normal. For a general non-normal matrix, the singular values of $z\pmb I-\pmb A$ replace the eigenvalue distances. Nevertheless, Girko's Hermitization principle states that the regularized logarithmic potential
\begin{equation}
\Phi_{\pmb A,\eta}(z,z^*)=\log\det\left[(z\pmb I-\pmb A)(z^*\pmb I-\pmb A^\dagger)+\eta^2\pmb I\right]
\label{eq:nhns-log-potential}
\end{equation}
recovers the eigenvalue density through
\begin{equation}
\rho_{\pmb A}(z)=\frac{1}{\pi N}\lim_{\eta\downarrow0}\partial_{z^*}\partial_z\Phi_{\pmb A,\eta}(z,z^*)\,.
\label{eq:nhns-density-hermitization}
\end{equation}
Since
\begin{equation}
\Delta_z=\partial_x^2+\partial_y^2=4\partial_{z^*}\partial_z\,,
\label{eq:nhns-laplacian-convention}
\end{equation}
this can also be written as
\begin{equation}
\rho_{\pmb A}(z)=\frac{1}{4\pi N}\lim_{\eta\downarrow0}\Delta_z\Phi_{\pmb A,\eta}(z,z^*)\,.
\label{eq:nhns-density-laplacian}
\end{equation}

Integrating \eqref{eq:nhns-density-laplacian} over $\mathcal{D}$ gives another representation of the number:
\begin{equation}
\mathcal{N}_{\pmb A}(\mathcal{D})=\frac{1}{4\pi}\lim_{\eta\downarrow0}\int_{\mathcal{D}}d^2z\,\Delta_z\Phi_{\pmb A,\eta}(z,z^*)\,.
\label{eq:nhns-number-bulk-laplacian}
\end{equation}
Using Green's theorem,
\begin{equation}
\int_{\mathcal{D}}d^2z\Delta_z f(z,z^*)=\oint_{\Gamma}ds \partial_n f(z,z^*)\,,
\label{eq:nhns-green-theorem}
\end{equation}
where $\partial_n$ denotes the outward normal derivative on the boundary, we obtain the boundary formula
\begin{equation}
\mathcal{N}_{\pmb A}(\mathcal{D})=\frac{1}{4\pi}\lim_{\eta\downarrow0}\oint_{\Gamma}ds\partial_n\Phi_{\pmb A,\eta}(z,z^*)\,.
    \label{eq:nhns-number-boundary-hermitized}
\end{equation}
This equation is the Hermitized version of the argument principle. It is especially useful for statistical mechanics because $\Phi_{\pmb A,\eta}$ is the logarithm of a positive regularized determinant.

\begin{examplebox}[Boundary formula for a single eigenvalue in a disk]
Consider the $1\times1$ matrix $\pmb A=(a)$ and the disk
\begin{equation}
\mathcal D_R=\{z\in\mathbb C: |z-a|<R\}\,.
\label{eq:nhns-ped-single-disk}
\end{equation}
The answer should be
\begin{equation}
\mathcal N_{\pmb A}(\mathcal D_R)=1\,.
\label{eq:nhns-ped-single-disk-count}
\end{equation}
Let us verify this from the Hermitized boundary formula. For one eigenvalue,
\begin{equation}
\Phi_\eta(z,z^*)=\log\left(|z-a|^2+\eta^2\right)\,.
\label{eq:nhns-ped-single-potential}
\end{equation}
On the boundary of the disk, write
\begin{equation}
z=a+Re^{i\theta}\,,\qquad ds=R d\theta\,.
\label{eq:nhns-ped-single-boundary-param}
\end{equation}
The outward normal derivative is the radial derivative with respect to $R$:
\begin{equation}
\partial_n\Phi_\eta=\frac{\partial}{\partial R}\log(R^2+\eta^2)=\frac{2R}{R^2+\eta^2}\,.
\label{eq:nhns-ped-single-normal-derivative}
\end{equation}
Therefore
\begin{equation}
\frac{1}{4\pi}\oint_{\partial\mathcal D_R}ds\,\partial_n\Phi_\eta=\frac{1}{4\pi}\int_0^{2\pi}R d\theta\frac{2R}{R^2+\eta^2}=\frac{R^2}{R^2+\eta^2}\,.
\label{eq:nhns-ped-single-boundary-count}
\end{equation}
Taking $\eta\downarrow0$ gives
\begin{equation}
\lim_{\eta\downarrow0}\frac{R^2}{R^2+\eta^2}=1\,,
    \label{eq:nhns-ped-single-boundary-limit}
\end{equation}
as expected.

If instead the disk does not enclose $a$, the logarithmic potential is harmonic inside the disk in the limit $\eta\downarrow0$, and the boundary integral gives zero. Thus the Hermitized boundary formula reproduces the same count as the argument principle.
\end{examplebox}

To express this determinant as a Gaussian theory, introduce the $2N\times2N$ Hermitized block matrix
\begin{equation}
\pmb{\mathcal B}_{\pmb A}(z,\eta)=\begin{pmatrix}
i\eta\pmb I&z\pmb I-\pmb A\\
z^*\pmb I-\pmb A^\dagger&i\eta\pmb I
\end{pmatrix}\,.
\label{eq:nhns-hermitized-block-matrix}
\end{equation}
Up to a $z$-independent phase,
\begin{equation}
\det\pmb{\mathcal B}_{\pmb A}(z,\eta)=\det\left[(z\pmb I-\pmb A)(z^*\pmb I-\pmb A^\dagger)+\eta^2\pmb I\right]\,.
\label{eq:nhns-block-determinant}
\end{equation}
Thus
\begin{equation}
\Phi_{\pmb A,\eta}(z,z^*)=\log\det\pmb{\mathcal B}_{\pmb A}(z,\eta)+\text{constant}\,,
\label{eq:nhns-potential-block}
\end{equation}
where the constant is independent of $z$ and does not contribute to normal derivatives. Whenever $\det\pmb{\mathcal B}_{\pmb A}$ is used below in place of the positive Hermitized determinant, this $z$-independent phase is understood to be fixed consistently. The logarithm of the determinant is generated by replicas through
\begin{equation}
\log\det\pmb{\mathcal B}=\lim_{n\to0}\frac{[\det(\pmb{\mathcal B})]^n-1}{n}\,,
\label{eq:nhns-replica-log-det}
\end{equation}
or, equivalently, by Gaussian and Grassmann integral representations depending on the sign of the determinant power. The important point for the present lecture notes is that the number statistic is a functional of a family of Hermitized Gaussian theories indexed by points on the boundary $\Gamma$.

The boundary representation can be discretized in a way that makes the replica structure explicit. Let $z_m$, $m=1,\ldots,M$, be points along $\Gamma$, with arc-length increments $\Delta s_m$, and let $n_m$ be the outward unit normal. For a small normal displacement $\delta>0$, define
\begin{equation}
z_m^{\rm out}=z_m+\frac{\delta}{2}n_m\,,\qquad z_m^{\rm in}=z_m-\frac{\delta}{2}n_m\,.
\label{eq:nhns-normal-displacements}
\end{equation}
Then
\begin{equation}
\partial_n\Phi_{\pmb A,\eta}(z_m,z_m^*)=\lim_{\delta\downarrow0}\frac{\Phi_{\pmb A,\eta}(z_m^{\rm out},(z_m^{\rm out})^*)-\Phi_{\pmb A,\eta}(z_m^{\rm in},(z_m^{\rm in})^*)}{\delta}\,.
\label{eq:nhns-normal-derivative-discrete}
\end{equation}
Substituting this into \eqref{eq:nhns-number-boundary-hermitized} gives
\begin{equation}
\mathcal{N}_{\pmb A}(\mathcal{D})=\lim_{\eta\downarrow0}\lim_{\delta\downarrow0}\lim_{M\to\infty}\sum_{m=1}^{M}\frac{\Delta s_m}{4\pi\delta}\Big[\log\det\pmb{\mathcal B}_{\pmb A}(z_m^{\rm out},\eta)-\log\det\pmb{\mathcal B}_{\pmb A}(z_m^{\rm in},\eta)\Big]\,.
\label{eq:nhns-discrete-number-determinants}
\end{equation}

\begin{examplebox}[Discretizing a circular boundary]
Let $\mathcal D$ be a disk of radius $R$ centered at the origin. A convenient boundary discretization is
\begin{equation}
z_m=Re^{2\pi i m/M}\,,\qquad m=0,\ldots,M-1\,.
\label{eq:nhns-ped-discrete-circle}
\end{equation}
The outward unit normal is
\begin{equation}
n_m=e^{2\pi i m/M}\,,
\label{eq:nhns-ped-discrete-normal}
\end{equation}
and the arc-length increment is
\begin{equation}
\Delta s_m=\frac{2\pi R}{M}\,.
\label{eq:nhns-ped-discrete-arclength}
\end{equation}
For a small normal displacement $\delta>0$, the outer and inner points are
\begin{equation}
z_m^{\rm out}=\left(R+\frac{\delta}{2}\right)e^{2\pi i m/M}\,,\qquad z_m^{\rm in}=\left(R-\frac{\delta}{2}\right)e^{2\pi i m/M}\,.
\label{eq:nhns-ped-discrete-in-out}
\end{equation}
The boundary approximation to the number statistic is
\begin{align}
\mathcal N_{\pmb A}(\mathcal D)&\simeq\frac{1}{4\pi}\sum_{m=0}^{M-1}\frac{\Delta s_m}{\delta}\left[\Phi_{\pmb A,\eta}(z_m^{\rm out},(z_m^{\rm out})^*)-\Phi_{\pmb A,\eta}(z_m^{\rm in},(z_m^{\rm in})^*)\right]\,.
    \label{eq:nhns-ped-discrete-boundary-count}
\end{align}
Substituting $\Phi_{\pmb A,\eta}=\log\det\pmb{\mathcal B}_{\pmb A}$ gives precisely the product of outer determinant powers divided by inner determinant powers that appears in the replicated boundary representation. Thus the many-replica object in the non-Hermitian number-statistics calculation is just the discretized version of a normal derivative along the counting contour.
\end{examplebox}

Exponentiating, one obtains a product of powers of Hermitized determinants:
\begin{align}
\exp\left[s\mathcal{N}_{\pmb A}(\mathcal{D})\right]&=\lim_{\eta\downarrow0}\lim_{\delta\downarrow0}\lim_{M\to\infty}\prod_{m=1}^{M}\left[\det\pmb{\mathcal B}_{\pmb A}(z_m^{\rm out},\eta)\right]^{s\Delta s_m/(4\pi\delta)}\nonumber\\
&\hspace{1.0cm}\times\prod_{m=1}^{M}\left[\det\pmb{\mathcal B}_{\pmb A}(z_m^{\rm in},\eta)\right]^{-s\Delta s_m/(4\pi\delta)}\,.
    \label{eq:nhns-replicated-boundary-product}
\end{align}
This is the non-Hermitian analogue of the determinant powers used in the Hermitian index problem. The replicas are now attached to boundary points of a two-dimensional domain rather than to two endpoints of an interval. Formula \eqref{eq:nhns-replicated-boundary-product} is formal until the determinant powers are first taken as integers and then analytically continued, but it makes the structure of the calculation clear.

Rather than carrying out the boundary-replica saddle explicitly, we now translate the same determinant structure into cavity language for sparse non-Hermitian matrices. The local recursion is the same Hermitized recursion used for the non-Hermitian spectral density, but the message now has to be evaluated along the counting boundary. Consider a sparse matrix $\pmb A$ with pairwise support
\begin{equation}
A_{ij}=C_{ij}J_{ij}\,, \qquad A_{ij}\neq A_{ji} \quad\text{in general}\,.
\label{eq:nhns-sparse-nonhermitian-model}
\end{equation}
Here the set of edges \(E\) denotes the symmetrized support: an unordered pair \(\{i,j\}\) belongs to \(E\) whenever at least one of \(A_{ij}\) or \(A_{ji}\) is nonzero. For each $z$ and $\eta$, the Hermitized cavity message on a directed edge $i\to j$ is a $2\times2$ matrix
\begin{equation}
\pmb{G}_{i\to j}(z,\eta)\,.
\label{eq:nhns-cavity-message}
\end{equation}
With
\begin{equation}
\pmb{Z}_i(z,\eta)=\begin{pmatrix}
i\eta&z-A_{ii}\\
z^*-A_{ii}^*&i\eta
\end{pmatrix}\,,
    \label{eq:nhns-local-block}
\end{equation}
and
\begin{equation}
\pmb{\mathcal A}_{ij}=\begin{pmatrix}
0 & A_{ij}\\
A_{ji}^* & 0
\end{pmatrix}\,,
\label{eq:nhns-edge-block}
\end{equation}
the message recursion on a tree is
\begin{equation}
\pmb{G}_{i\to j}(z,\eta)=\left[\pmb{Z}_i(z,\eta)-\sum_{\ell\in\partial i\setminus j}\pmb{\mathcal A}_{i\ell}\pmb{G}_{\ell\to i}(z,\eta)\pmb{\mathcal A}_{i\ell}^{\dagger}\right]^{-1}\,.
\label{eq:nhns-cavity-recursion}
\end{equation}
With this definition, $\pmb{\mathcal A}_{ji}=\pmb{\mathcal A}_{ij}^{\dagger}$; the two orientations of an unordered edge are therefore adjoint block couplings. The full local message is
\begin{equation}
\pmb{G}_{i}(z,\eta)=\left[\pmb{Z}_i(z,\eta)-\sum_{\ell\in\partial i}\pmb{\mathcal A}_{i\ell}\pmb{G}_{\ell\to i}(z,\eta)\pmb{\mathcal A}_{i\ell}^{\dagger}\right]^{-1}\,.
\label{eq:nhns-full-local-message}
\end{equation}
These are the same equations used for the non-Hermitian spectral density, but now they must be solved for all points on the boundary of $\mathcal{D}$, or for a discretization of that boundary.

The local determinant also has a Bethe factorization. On a tree, the determinant of the Hermitized block matrix can be written as
\begin{equation}
\det\pmb{\mathcal B}_{\pmb A}(z,\eta)=\prod_{i=1}^{N}\det\left[\pmb{G}_i(z,\eta)^{-1}\right]\prod_{\{i,j\}\in E}\det\left[\pmb I_2-\pmb{G}_{i\to j}(z,\eta)\pmb{\mathcal A}_{ij}\pmb{G}_{j\to i}(z,\eta)\pmb{\mathcal A}_{ji}\right]^{-1}\,.
\label{eq:nhns-bethe-determinant}
\end{equation}
This is the matrix-valued analogue of the scalar Bethe determinant identity used for Hermitian sparse matrices. To verify the structure, consider a two-vertex tree. The cavity messages are the inverses of the two local blocks. Multiplying the two full site Schur complements gives two copies of the edge determinant; the edge factor in \eqref{eq:nhns-bethe-determinant} removes one copy and leaves the exact determinant of the $4\times4$ block matrix.

Taking the logarithm of \eqref{eq:nhns-bethe-determinant} and substituting into \eqref{eq:nhns-number-boundary-hermitized}, the number of eigenvalues in $\mathcal{D}$ becomes a sum of local boundary contributions:
\begin{equation}
\mathcal{N}_{\pmb A}(\mathcal{D})=\sum_{i=1}^{N}\nu_i(\mathcal{D})-\sum_{\{i,j\}\in E}\nu_{ij}(\mathcal{D})\,,
\label{eq:nhns-bethe-number-decomposition}
\end{equation}
where
\begin{equation}
\nu_i(\mathcal{D})=\frac{1}{4\pi}\lim_{\eta\downarrow0}\oint_{\Gamma}ds \partial_n \log\det\left[\pmb{G}_i(z,\eta)^{-1}\right]\,,
\label{eq:nhns-site-number-contribution}
\end{equation}
and
\begin{equation}
\nu_{ij}(\mathcal{D})=\frac{1}{4\pi}\lim_{\eta\downarrow0}\oint_{\Gamma}ds\partial_n\log\det\left[\pmb I_2-\pmb{G}_{i\to j}(z,\eta)\pmb{\mathcal A}_{ij}\pmb{G}_{j\to i}(z,\eta)\pmb{\mathcal A}_{ji}\right]\,.
\label{eq:nhns-edge-number-contribution}
\end{equation}
This is the central local formula for non-Hermitian number statistics on sparse graphs. It says that the two-dimensional eigenvalue count is an extensive observable built from site and edge contributions, each of which is obtained by integrating the normal derivative of a Hermitized local determinant along the boundary of the domain. Note that the logarithms in \eqref{eq:nhns-site-number-contribution} and \eqref{eq:nhns-edge-number-contribution} must be evaluated with branches, and with the $z$-independent phases in \eqref{eq:nhns-block-determinant}, chosen consistently with the global Hermitized determinant in \eqref{eq:nhns-potential-block}. The individual site and edge boundary contributions are convention-dependent, while the Bethe combination in \eqref{eq:nhns-bethe-number-decomposition} reproduces the integer count. 

\begin{examplebox}[Local number decomposition on a two-site tree]
Consider again a two-vertex sparse non-Hermitian matrix
\begin{equation}
\pmb A=\begin{pmatrix}
0 & J\\
K & 0
\end{pmatrix}\,.
\label{eq:nhns-ped-two-site-nh}
\end{equation}
The Hermitized determinant factorization on a tree has the structure
\begin{equation}
\det\pmb{\mathcal B}_{\pmb A}(z,\eta)=\prod_{i=1}^{2}\det\left[\pmb{G}_i(z,\eta)^{-1}\right]\det\left[\pmb I_2-\pmb G_{1\to2}\pmb{\mathcal A}_{12}\pmb{G}_{2\to1}\pmb{\mathcal A}_{21}\right]^{-1}\,.
\label{eq:nhns-ped-two-site-bethe}
\end{equation}
Taking the logarithm and applying the boundary operator
\begin{equation}
\frac{1}{4\pi}\lim_{\eta\downarrow0}\oint_{\Gamma}ds\partial_n
\label{eq:nhns-ped-boundary-operator}
\end{equation}
gives
\begin{equation}
\mathcal N_{\pmb A}(\mathcal D)=\nu_1(\mathcal D)+\nu_2(\mathcal D)-\nu_{12}(\mathcal D)\,.
\label{eq:nhns-ped-two-site-local-count}
\end{equation}
The site terms count the spectral contribution suggested by the two full local Hermitized Green matrices, while the edge term removes the double counting of the interaction between the two vertices. This is the non-Hermitian analogue of the Bethe determinant decompositions used earlier for Hermitian interval counts.
\end{examplebox}

The corresponding tilted ensemble is
\begin{equation}
P_s(\pmb A)=\frac{P(\pmb A)\exp\left[s\mathcal{N}_{\pmb A}(\mathcal{D})\right]}{\overline{\exp\left[s\mathcal{N}_{\pmb A}(\mathcal{D})\right]}}\,.
\label{eq:nhns-tilted-ensemble}
\end{equation}
Using \eqref{eq:nhns-bethe-number-decomposition}, the tilt factor becomes
\begin{equation}
\exp\left[s\mathcal{N}_{\pmb A}(\mathcal{D})\right]=\prod_{i=1}^{N}e^{s\nu_i(\mathcal{D})}\prod_{\{i,j\}\in E}e^{-s\nu_{ij}(\mathcal{D})}.
\label{eq:nhns-local-tilt}
\end{equation}
Thus the number-statistics problem is a biased graphical model whose messages are not single matrices at one spectral parameter, but functions of the boundary point $z\in\Gamma$. In a discretized implementation, a message is the collection
\begin{equation}
\pmb{\mathfrak G}_{i\to j}=\left\{\pmb{G}_{i\to j}(z_m^{\rm out},\eta),\pmb{G}_{i\to j}(z_m^{\rm in},\eta)\right\}_{m=1}^{M}\,.
\label{eq:nhns-boundary-message}
\end{equation}
The unbiased update is obtained by applying \eqref{eq:nhns-cavity-recursion} independently at every boundary point. The tilted update uses the same local maps but samples neighborhoods with weights built from the local contributions \eqref{eq:nhns-site-number-contribution} and \eqref{eq:nhns-edge-number-contribution}, equivalently from the factorization \eqref{eq:nhns-local-tilt}. At $s=0$ the tilt disappears, and one recovers the ordinary unconditioned Hermitized population dynamics evaluated at the boundary points; the mean count is then obtained by the boundary formula, while the pointwise density is obtained by the usual Wirtinger derivative.

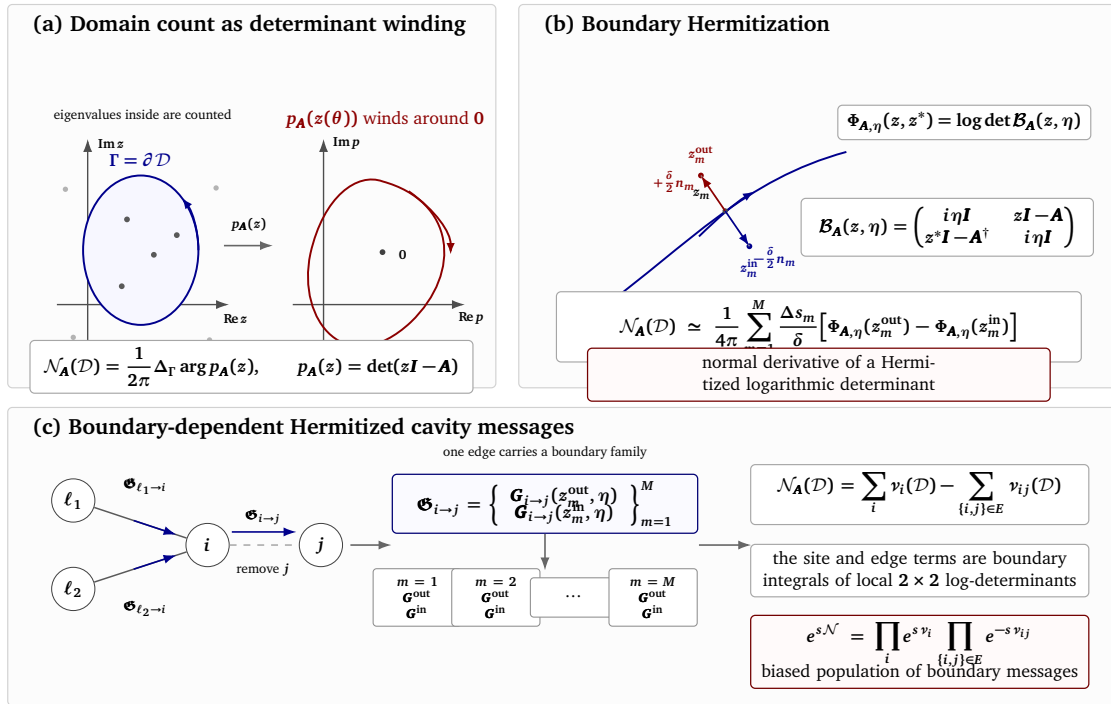
\begin{figure}[t]
\centering
\resizebox{0.98\textwidth}{!}{%
\begin{tikzpicture}[
    x=1cm,
    y=1cm,
    >=Latex,
    panel/.style={draw=black!18, fill=black!1, rounded corners=2pt, line width=0.5pt},
    ptitle/.style={font=\bfseries\small, anchor=west},
    paneltext/.style={font=\scriptsize, align=center},
    tinytext/.style={font=\tiny, align=center},
    axisline/.style={draw=black!70, line width=0.6pt, -{Latex[length=2.0mm,width=1.4mm]}},
    contour/.style={draw=blue!55!black, line width=0.85pt},
    contourarrow/.style={draw=blue!55!black, line width=0.85pt, -{Latex[length=1.8mm,width=1.3mm]}},
    domainfill/.style={fill=blue!3},
    winding/.style={draw=red!55!black, line width=0.85pt},
    windingarrow/.style={draw=red!55!black, line width=0.85pt, -{Latex[length=1.8mm,width=1.3mm]}},
    normal/.style={draw=red!55!black, line width=0.75pt, -{Latex[length=1.9mm,width=1.3mm]}},
    guide/.style={draw=black!35, dashed, line width=0.55pt},
    flow/.style={draw=black!60, line width=0.70pt, -{Latex[length=2.0mm,width=1.4mm]}},
    msg/.style={draw=blue!55!black, line width=0.70pt, -{Latex[length=2.0mm,width=1.4mm]}},
    box/.style={draw=black!35, fill=white, rounded corners=2pt, line width=0.5pt, inner sep=3pt, font=\scriptsize, align=center},
    bluebox/.style={draw=blue!45!black, fill=blue!2, rounded corners=2pt, line width=0.55pt, inner sep=3pt, font=\scriptsize, align=center},
    redbox/.style={draw=red!45!black, fill=red!2, rounded corners=2pt, line width=0.55pt, inner sep=3pt, font=\scriptsize, align=center},
    vnode/.style={circle, draw=black!75, fill=white, minimum size=6.6mm, inner sep=0pt, font=\scriptsize},
    edge/.style={draw=black!65, line width=0.65pt},
    matbox/.style={draw=black!45, fill=white, rounded corners=1.5pt, line width=0.45pt, minimum height=6.0mm, font=\tiny, align=center}
]
\draw[panel] (0,4.85) rectangle (7.45,10.70);
\node[ptitle] at (0.25,10.38) {(a) Domain count as determinant winding};

\draw[axisline] (0.75,6.12) -- (3.38,6.12);
\draw[axisline] (1.22,5.25) -- (1.22,8.65);
\node[tinytext] at (3.48,5.91) {${\rm Re}\,z$};
\node[tinytext] at (1.58,8.58) {${\rm Im}\,z$};
\fill[domainfill] (2.03,6.95) ellipse [x radius=0.88, y radius=1.14];
\draw[contour] (2.03,6.95) ellipse [x radius=0.88, y radius=1.14];
\draw[contourarrow] (2.91,6.95) arc [start angle=0, end angle=45, x radius=0.88, y radius=1.14];
\node[paneltext, text=blue!55!black] at (2.02,8.34) {$\Gamma=\partial\mathcal D$};
\node[tinytext, text=blue!55!black] at (2.02,5.45) {$\mathcal D$};
\foreach \x/\y in {1.82/7.42,2.23/6.88,1.74/6.40,2.58/7.18}{\fill[black!65] (\x,\y) circle (1.25pt);}
\foreach \x/\y in {0.88/7.88,3.18/7.90,3.08/5.58,1.00/5.62}{\fill[black!30] (\x,\y) circle (1.15pt);}
\node[tinytext] at (2.06,9.00) {eigenvalues inside are counted};

\draw[flow] (3.30,7.02) -- (4.05,7.02);
\node[tinytext] at (3.68,7.35) {$p_{\pmb A}(z)$};

\draw[axisline] (4.35,6.12) -- (6.98,6.12);
\draw[axisline] (4.82,5.25) -- (4.82,8.65);
\node[tinytext] at (7.08,5.91) {${\rm Re}\,p$};
\node[tinytext] at (5.18,8.58) {${\rm Im}\,p$};
\fill[black!75] (5.72,6.92) circle (1.25pt);
\node[tinytext, anchor=west] at (5.84,6.88) {$0$};
\draw[winding]
  (5.72,8.00) .. controls (6.82,7.70) and (6.95,6.62) .. (6.15,5.85)
  .. controls (5.18,4.95) and (4.40,5.86) .. (4.62,6.90)
  .. controls (4.82,7.78) and (5.34,8.14) .. (5.72,8.00);
\draw[windingarrow] (6.12,7.85) .. controls (6.54,7.55) and (6.76,7.22) .. (6.76,6.88);
\node[paneltext, text=red!55!black] at (5.76,8.95) {$p_{\pmb A}(z(\theta))$ winds around $0$};

\node[box, text width=6.55cm] at (3.72,5.17)
{$\displaystyle \mathcal N_{\pmb A}(\mathcal D)=\frac{1}{2\pi}\Delta_{\Gamma}\arg p_{\pmb A}(z),\qquad
p_{\pmb A}(z)=\det(z\pmb I-\pmb A)$};

\draw[panel] (7.80,4.85) rectangle (16.95,10.70);
\node[ptitle] at (8.05,10.38) {(b) Boundary Hermitization};

\draw[contour] (9.05,6.05) .. controls (10.40,7.05) and (11.25,8.03) .. (12.80,8.45);
\draw[contourarrow] (10.55,7.16) .. controls (10.85,7.45) and (11.10,7.67) .. (11.40,7.85);
\coordinate (zm) at (10.95,7.55);
\coordinate (zout) at (10.58,8.09);
\coordinate (zin) at (11.32,7.01);
\fill[black!70] (zm) circle (1.15pt);
\fill[red!55!black] (zout) circle (1.25pt);
\fill[blue!55!black] (zin) circle (1.25pt);
\draw[guide] (zin) -- (zout);
\draw[normal] (zm) -- (zout);
\draw[normal, blue!55!black] (zm) -- (zin);
\node[tinytext, anchor=south east] at ($(zm)+(-0.08,0.05)$) {$z_m$};
\node[tinytext, text=red!55!black, anchor=south] at ($(zout)+(0,0.06)$) {$z_m^{\rm out}$};
\node[tinytext, text=blue!55!black, anchor=north] at ($(zin)+(0,-0.06)$) {$z_m^{\rm in}$};
\node[tinytext, text=red!55!black] at (10.18,8.00) {$+\frac{\delta}{2}n_m$};
\node[tinytext, text=blue!55!black] at (11.72,6.84) {$-\frac{\delta}{2}n_m$};
\node[paneltext, text=blue!55!black] at (12.88,8.77) {$\Gamma$};

\node[box, text width=3.65cm] at (14.60,8.90)
{$\displaystyle \Phi_{\pmb A,\eta}(z,z^*)=\log\det\pmb{\mathcal B}_{\pmb A}(z,\eta)$};
\node[box, text width=4.25cm] at (14.34,7.30)
{$\displaystyle
\pmb{\mathcal B}_{\pmb A}(z,\eta)=
\begin{pmatrix}
i\eta\pmb I&z\pmb I-\pmb A\\
z^*\pmb I-\pmb A^\dagger&i\eta\pmb I
\end{pmatrix}$};
\node[box, text width=7.80cm] at (12.38,5.78)
{$\displaystyle
\mathcal N_{\pmb A}(\mathcal D)\simeq
\frac{1}{4\pi}\sum_{m=1}^{M}\frac{\Delta s_m}{\delta}
\Big[\Phi_{\pmb A,\eta}(z_m^{\rm out})-\Phi_{\pmb A,\eta}(z_m^{\rm in})\Big]$};
\node[redbox, text width=6.85cm] at (12.38,5.05)
{normal derivative of a Hermitized logarithmic determinant};

\draw[panel] (0,0) rectangle (16.95,4.55);
\node[ptitle] at (0.25,4.23) {(c) Boundary-dependent Hermitized cavity messages};

\node[vnode] (l1) at (1.00,3.13) {$\ell_1$};
\node[vnode] (l2) at (1.00,1.78) {$\ell_2$};
\node[vnode] (i)  at (3.05,2.45) {$i$};
\node[vnode] (j)  at (4.78,2.45) {$j$};
\draw[edge] (l1) -- (i);
\draw[edge] (l2) -- (i);
\draw[edge, black!35, dashed] (i) -- (j);
\draw[msg] ($(l1)!0.45!(i)$) -- ($(l1)!0.77!(i)$);
\draw[msg] ($(l2)!0.45!(i)$) -- ($(l2)!0.77!(i)$);
\draw[msg] ($(i)+(0.37,0.20)$) -- ($(j)+(-0.37,0.20)$);
\node[tinytext] at (2.10,3.35) {$\pmb{\mathfrak G}_{\ell_1\to i}$};
\node[tinytext] at (2.10,1.48) {$\pmb{\mathfrak G}_{\ell_2\to i}$};
\node[tinytext] at (3.90,2.88) {$\pmb{\mathfrak G}_{i\to j}$};
\node[tinytext] at (3.90,2.08) {remove $j$};

\draw[flow] (5.23,2.45) -- (5.86,2.45);

\node[bluebox, text width=4.45cm] (msgbox) at (8.20,3.08)
{$\displaystyle
\pmb{\mathfrak G}_{i\to j}=
\left\{\begin{array}{c}
\pmb G_{i\to j}(z_m^{\rm out},\eta)\\[-1mm]
\pmb G_{i\to j}(z_m^{\rm in},\eta)
\end{array}\right\}_{m=1}^{M}$};
\node[tinytext] at (8.20,3.86) {one edge carries a boundary family};

\node[matbox, text width=1.10cm] at (6.25,1.65) {$m=1$\\$\pmb G^{\rm out}$\\$\pmb G^{\rm in}$};
\node[matbox, text width=1.10cm] at (7.45,1.65) {$m=2$\\$\pmb G^{\rm out}$\\$\pmb G^{\rm in}$};
\node[matbox, text width=1.10cm] at (8.65,1.65) {$\cdots$};
\node[matbox, text width=1.10cm] at (9.85,1.65) {$m=M$\\$\pmb G^{\rm out}$\\$\pmb G^{\rm in}$};
\draw[flow] (8.20,2.62) -- (8.20,2.06);

\node[box, text width=4.95cm] at (13.92,3.22)
{$\displaystyle
\mathcal N_{\pmb A}(\mathcal D)=
\sum_i\nu_i(\mathcal D)-\sum_{\{i,j\}\in E}\nu_{ij}(\mathcal D)$};
\node[box, text width=4.95cm] at (13.92,2.06)
{the site and edge terms are boundary integrals of local $2\times2$ log-determinants};
\node[redbox, text width=4.95cm] at (13.92,0.82)
{$\displaystyle e^{s\mathcal N}=\prod_i e^{s\nu_i}
\prod_{\{i,j\}\in E}e^{-s\nu_{ij}}$\\[-1mm]
biased population of boundary messages};
\draw[flow] (10.55,2.45) -- (11.35,2.45);
\end{tikzpicture}%
}
\caption{Non-Hermitian number statistics as a boundary and Hermitized-cavity problem. Counting eigenvalues in a complex domain can be written as the winding of $p_{\pmb A}(z)$ along $\Gamma=\partial\mathcal D$. Hermitization converts the same count into a normal derivative of the logarithmic determinant $\Phi_{\pmb A,\eta}$, evaluated by inner--outer boundary displacements. On a sparse graph, the Bethe representation turns this boundary integral into site and edge contributions depending on boundary families of $2\times2$ Hermitized cavity messages.}
\label{fig:nhns-domain-winding-boundary-messages}
\end{figure}

Figure~\ref{fig:nhns-domain-winding-boundary-messages} summarizes the non-Hermitian number-statistics construction: the domain count is a winding number, Hermitization turns it into an inner--outer boundary determinant problem, and the sparse cavity formulation promotes each edge message to a boundary-indexed family of $2\times2$ Hermitized Green matrices.

The typical number in the tilted ensemble is
\begin{equation}
k_{\mathcal{D}}(s)=\psi_{\mathcal{D}}'(s)=\lim_{N\to\infty}\frac{1}{N}\left\langle\mathcal{N}_{\pmb A}(\mathcal{D})\right\rangle_s\,,
    \label{eq:nhns-tilted-typical-number}
\end{equation}
where $\langle\cdot\rangle_s$ denotes the average with respect to \eqref{eq:nhns-tilted-ensemble}. The rate function is obtained parametrically as
\begin{equation}
\Phi_{\mathcal{D}}(k_{\mathcal{D}}(s))=s k_{\mathcal{D}}(s)-\psi_{\mathcal{D}}(s)\,.
\label{eq:nhns-parametric-rate-function}
\end{equation}
The variance of the count in the original ensemble is
\begin{equation}
{\rm Var}\mathcal{N}_{\pmb A}(\mathcal{D})=N\psi_{\mathcal{D}}''(0)+o(N)
\label{eq:nhns-number-variance}
\end{equation}
in the finite-connectivity scaling. Higher derivatives of $\psi_{\mathcal{D}}$ give higher cumulants.

It is useful to compare this with a direct density integration. The average intensive number of eigenvalues in $\mathcal{D}$ is
\begin{equation}
\psi_{\mathcal{D}}'(0)=\lim_{N\to\infty}\frac{1}{N}\overline{\mathcal{N}_{\pmb A}(\mathcal{D})}=\int_{\mathcal{D}}d^2z\,\overline{\rho_{\pmb A}(z)}\,.
\label{eq:nhns-mean-number-density-check}
\end{equation}
This equation is only the first derivative at $s=0$. It does not determine the variance or the rate function. The full number-statistics problem requires correlations of the spectral density over the domain, or equivalently the tilted boundary-message distribution described above. This is the same lesson as in the Hermitian index problem: integrating the typical density gives the typical count, but not the probability of atypical counts.

Several elementary checks are immediate. If $\mathcal{D}$ is empty, then
\begin{equation}
\mathcal{N}_{\pmb A}(\mathcal{D})=0\,,\qquad\psi_{\mathcal{D}}(s)=0\,.
\label{eq:nhns-empty-domain-check}
\end{equation}
If $\mathcal{D}$ contains the entire spectrum for every realization under consideration, or with probability one at the large-deviation exponential scale, then
\begin{equation}
\mathcal{N}_{\pmb A}(\mathcal{D})=N\,,\qquad\psi_{\mathcal{D}}(s)=s\,.
\label{eq:nhns-full-domain-check}
\end{equation}
If $\mathcal{D}_1$ and $\mathcal{D}_2$ are disjoint and separated by contours that do not cross eigenvalues, then
\begin{equation}
\mathcal{N}_{\pmb A}(\mathcal{D}_1\cup\mathcal{D}_2)=\mathcal{N}_{\pmb A}(\mathcal{D}_1)+\mathcal{N}_{\pmb A}(\mathcal{D}_2)\,.
\label{eq:nhns-additivity}
\end{equation}
At the level of the argument principle this is just additivity of winding numbers. At the level of the Hermitized representation it follows from additivity of the boundary integral over domains.

There is also an important Hermitian limit. Suppose that $\pmb A$ is Hermitian, so all eigenvalues lie on the real axis, and assume that no eigenvalue lies exactly at $a$ or $b$, or that the half-open convention used in Section~\ref{sec:index-number-large-deviations} is adopted. Let $\mathcal{D}$ be a thin strip around the interval $I=(a,b]$ on the real axis. As the strip thickness goes to zero, the contour integral \eqref{eq:nhns-number-resolvent-contour} collapses to the difference of boundary values of the Hermitian resolvent above and below the real axis. One recovers
\begin{equation}
\mathcal{N}_{\pmb A}((a,b])=\frac{1}{\pi}\lim_{\epsilon\downarrow0}{\rm Im}\int_a^b d\lambda{\rm Tr}[(\lambda-i\epsilon)\pmb I-\pmb A]^{-1},
    \label{eq:nhns-hermitian-limit}
\end{equation}
which is the interval-count formula used in the Hermitian sections. Thus the non-Hermitian number statistic is a genuine extension of the Hermitian index statistic, not a separate construction.

For dense Ginibre-type matrices, the eigenvalues form a strongly correlated two-dimensional gas, and exact determinantal methods are available in special cases \cite{Ginibre1965,Forrester2010}. In such ensembles, number statistics in disks or other domains are related to two-dimensional Coulomb-gas fluctuations. The approach described here has a different purpose. It is designed for ensembles where one does not want to start from an eigenvalue joint density, either because the matrix distribution is non-invariant or because the sparse graph structure is the natural disorder variable. The price is that the order parameter is no longer an eigenvalue density alone; it is a distribution of Hermitized cavity messages along the boundary of the counting domain.

The final picture is therefore parallel to the Hermitian large-deviation theory. The Hermitian index below a threshold is the phase jump of a determinant across the real axis. The non-Hermitian number inside a domain is the winding number of the determinant along the boundary of that domain. Hermitization turns this winding problem into a boundary integral of a regularized logarithmic determinant. On a sparse graph, the Hermitized determinant has a Bethe decomposition into local site and edge terms. Consequently, number fluctuations and large deviations are described by a tilted population of boundary-dependent cavity messages. At zero tilt one recovers the ordinary unconditioned Hermitized cavity population; the full number-statistics problem is obtained by keeping the tilt and computing the Bethe free energy of the biased message ensemble.

\begin{exerciseblock}
\exitem[Counting eigenvalues in a disk]
Let $\pmb A$ be diagonal with eigenvalues
\begin{equation}
z_1=0\,,\qquad z_2=1+i\,,\qquad z_3=-2\,.
\label{eq:nhns-ex-diagonal-eigs}
\end{equation}
Compute $\mathcal N_{\pmb A}(\mathcal D_R)$ for disks $\mathcal D_R=\{z:|z|<R\}$ with $R=1/2$, $R=3/2$, and $R=3$.

\exitem[Argument principle]
Let
\begin{equation}
p(z)=\prod_{i=1}^{N}(z-z_i)\,.
\label{eq:nhns-ex-polynomial}
\end{equation}
Assume that $\Gamma$ is positively oriented and that no zero lies on $\Gamma$. Prove that
\begin{equation}
\frac{1}{2\pi i}\oint_{\Gamma}dz \partial_z\log p(z)
\label{eq:nhns-ex-argument-principle}
\end{equation}
equals the number of zeros of $p$ inside $\Gamma$.

\exitem[Resolvent contour formula]
Assume that no eigenvalue of $\pmb A$ lies on $\partial\mathcal D$ and that $\partial\mathcal D$ is positively oriented. Use
\begin{equation}
\partial_z\log\det(z\pmb I-\pmb A)={\rm Tr}(z\pmb I-\pmb A)^{-1}
\label{eq:nhns-ex-logdet-resolvent}
\end{equation}
to derive
\begin{equation}
\mathcal N_{\pmb A}(\mathcal D) =\frac{1}{2\pi i}\oint_{\partial\mathcal D}dz{\rm Tr}(z\pmb I-\pmb A)^{-1}\,.
\label{eq:nhns-ex-contour-resolvent}
\end{equation}

\exitem[Winding number]
For $p(z)=z-a$, compute explicitly
\begin{equation}
\Delta_\Gamma\arg p(z)
\label{eq:nhns-ex-winding}
\end{equation}
when $\Gamma$ is a circle centered at $a$ and when $\Gamma$ is a circle that does not enclose $a$.

\exitem[Scaled cumulant-generating function]
Starting from
\begin{equation}
\psi_{\mathcal D}(s)=\lim_{N\to\infty}\frac{1}{N}\log\overline{e^{s\mathcal N_{\pmb A}(\mathcal D)}
}\,,
\label{eq:nhns-ex-cgf}
\end{equation}
show formally that
\begin{equation}
\psi_{\mathcal D}'(0)=\lim_{N\to\infty}\frac{1}{N}\overline{\mathcal N_{\pmb A}(\mathcal D)}
\label{eq:nhns-ex-first-derivative}
\end{equation}
and
\begin{equation}
\psi_{\mathcal D}''(0)=\lim_{N\to\infty}\frac{1}{N}{\rm Var}\mathcal N_{\pmb A}(\mathcal D)\,.
\label{eq:nhns-ex-second-derivative}
\end{equation}

\exitem[Hermitized logarithmic potential]
For a $1\times1$ matrix $\pmb A=(a)$, with $a\in\mathbb C$, derive
\begin{equation}
\Phi_\eta(z,z^*)=\log\left(|z-a|^2+\eta^2\right)\,.
\label{eq:nhns-ex-single-potential}
\end{equation}
Then compute
\begin{equation}
\frac{1}{\pi}\partial_{z^*}\partial_z\Phi_\eta(z,z^*)=\frac{1}{\pi}\frac{\eta^2}{(|z-a|^2+\eta^2)^2}\,.
\label{eq:nhns-ex-single-density}
\end{equation}

\exitem[Normalization of the regularized scalar density]
Verify that
\begin{equation}
\int_{\mathbb C}d^2z\frac{1}{\pi}\frac{\eta^2}{(|z-a|^2+\eta^2)^2}=1\,.
\label{eq:nhns-ex-single-normalization}
\end{equation}

\exitem[Green's theorem and the boundary formula]
Starting from
\begin{equation}
\rho_{\pmb A}(z)=\frac{1}{4\pi N}\lim_{\eta\downarrow0}\Delta_z\Phi_{\pmb A,\eta}(z,z^*)\,,
\label{eq:nhns-ex-density-laplacian}
\end{equation}
multiply by $N$, integrate over a domain $\mathcal D$, and use Green's theorem to derive
\begin{equation}
\mathcal N_{\pmb A}(\mathcal D)=\frac{1}{4\pi}\lim_{\eta\downarrow0}\oint_{\partial\mathcal D}ds \partial_n\Phi_{\pmb A,\eta}(z,z^*)\,.
\label{eq:nhns-ex-boundary-formula}
\end{equation}

\exitem[Boundary formula for a disk]
For the scalar matrix $\pmb A=(a)$ and the disk centered at $a$ with radius $R$, compute the boundary integral explicitly and show that it tends to one as $\eta\downarrow0$.

\exitem[Hermitized block determinant]
Starting from
\begin{equation}
\pmb{\mathcal B}_{\pmb A}(z,\eta)=\begin{pmatrix}
i\eta\pmb I&z\pmb I-\pmb A\\
z^*\pmb I-\pmb A^\dagger&i\eta\pmb I
\end{pmatrix}\,,
\label{eq:nhns-ex-hermitized-block}
\end{equation}
derive
\begin{equation}
\det\pmb{\mathcal B}_{\pmb A}(z,\eta)=\det\left[(z\pmb I-\pmb A)(z^*\pmb I-\pmb A^\dagger)+\eta^2\pmb I\right]
    \label{eq:nhns-ex-hermitized-det}
\end{equation}
up to a phase independent of $z$.

\exitem[Discrete boundary approximation]
For a circular domain of radius $R$, use the discretization
\begin{equation}
z_m=Re^{2\pi i m/M}\,,\qquad m=0,\ldots,M-1\,,\qquad \Delta s_m=\frac{2\pi R}{M}
\label{eq:nhns-ex-discrete-circle}
\end{equation}
with
\[
z_m^{\rm out}=\left(R+\frac{\delta}{2}\right)e^{2\pi i m/M}\,,\qquad
z_m^{\rm in}=\left(R-\frac{\delta}{2}\right)e^{2\pi i m/M}\,,
\]
to derive the finite-difference approximation of the normal derivative used in
\begin{equation}
\mathcal N_{\pmb A}(\mathcal D)\simeq\frac{1}{4\pi}\sum_m\frac{\Delta s_m}{\delta}\left[\Phi_{\pmb A,\eta}(z_m^{\rm out})-\Phi_{\pmb A,\eta}(z_m^{\rm in})\right]\,.
\label{eq:nhns-ex-discrete-boundary}
\end{equation}

\exitem[Replica powers on the boundary]
Starting from the discretized boundary formula, show why
\begin{equation}
e^{s\mathcal N_{\pmb A}(\mathcal D)}
\label{eq:nhns-ex-exp-number}
\end{equation}
can be represented as a product of powers of Hermitized determinants at outer and inner boundary points. Explain why the replica construction now involves boundary points rather than interval endpoints.

\exitem[Matrix-valued cavity recursion]
Use the Schur complement on the Hermitized block matrix, with local block
\[
\pmb Z_i(z,\eta)=
\begin{pmatrix}
i\eta&z-A_{ii}\\
z^*-A_{ii}^*&i\eta
\end{pmatrix}
\]
and edge block
\[
\pmb{\mathcal A}_{ij}=
\begin{pmatrix}
0&A_{ij}\\
A_{ji}^*&0
\end{pmatrix},
\]
to derive
\begin{equation}
\pmb{G}_{i\to j}(z,\eta)=\left[\pmb{Z}_i(z,\eta)-\sum_{\ell\in\partial i\setminus j}\pmb{\mathcal A}_{i\ell}\pmb{G}_{\ell\to i}(z,\eta)\pmb{\mathcal A}_{i\ell}^{\dagger}\right]^{-1}\,.
\label{eq:nhns-ex-cavity-recursion}
\end{equation}
Explain why the messages are $2\times2$ matrices.

\exitem[Bethe determinant factorization]
For a tree, derive the factorized form
\begin{equation}
\det\pmb{\mathcal B}_{\pmb A}(z,\eta)=\prod_i\det\left[\pmb{G}_i(z,\eta)^{-1}\right]\prod_{\{i,j\}}\det\left[\pmb I_2-\pmb{G}_{i\to j}\pmb{\mathcal A}_{ij}\pmb{G}_{j\to i}\pmb{\mathcal A}_{ji}\right]^{-1}\,.
\label{eq:nhns-ex-bethe-det}
\end{equation}

\exitem[Local number decomposition]
Use the Bethe determinant factorization to derive
\begin{equation}
\mathcal N_{\pmb A}(\mathcal D)=\sum_i\nu_i(\mathcal D)-\sum_{\{i,j\}\in E}\nu_{ij}(\mathcal D)\,.
\label{eq:nhns-ex-local-decomposition}
\end{equation}
Write explicit expressions for $\nu_i$ and $\nu_{ij}$.

\exitem[Tilted ensemble]
Show that the tilted ensemble
\begin{equation}
P_s(\pmb A)=\frac{P(\pmb A)e^{s\mathcal N_{\pmb A}(\mathcal D)}}{\overline{e^{s\mathcal N_{\pmb A}(\mathcal D)}}}
\label{eq:nhns-ex-tilted-ensemble}
\end{equation}
satisfies
\begin{equation}
\psi_{\mathcal D}'(s)=\lim_{N\to\infty}\frac{1}{N}\left\langle\mathcal N_{\pmb A}(\mathcal D)\right\rangle_s\,.
\label{eq:nhns-ex-tilted-mean}
\end{equation}

\exitem[Additivity of disjoint domains]
Suppose $\mathcal D_1$ and $\mathcal D_2$ are disjoint domains whose boundaries do not cross eigenvalues. Prove
\begin{equation}
\mathcal N_{\pmb A}(\mathcal D_1\cup\mathcal D_2)=\mathcal N_{\pmb A}(\mathcal D_1)+\mathcal N_{\pmb A}(\mathcal D_2)\,.
\label{eq:nhns-ex-additivity}
\end{equation}
Interpret this both from the eigenvalue definition and from the winding-number representation.

\exitem[Hermitian thin-strip limit]
Assume $\pmb A$ is Hermitian, assume that no eigenvalue lies exactly at $a$ or $b$, and let $\mathcal D$ be a thin strip around the interval $[a,b]$ on the real axis. Explain how the contour formula reduces to
\begin{equation}
\mathcal N_{\pmb A}([a,b])=\frac{1}{\pi}\lim_{\epsilon\downarrow0}{\rm Im}\int_a^b d\lambda{\rm Tr}\left[(\lambda-i\epsilon)\pmb I-\pmb A\right]^{-1}\,.
\label{eq:nhns-ex-hermitian-limit}
\end{equation}

\exitem[Programming exercise: winding number]
Fix a small matrix size $N$, for example $N=20$, a disk radius $R$, a boundary discretization size $M$, and a number $S$ of independent samples. For each sample, generate a real non-Hermitian matrix $\pmb A$ with independent Gaussian entries of mean zero and variance $1/N$. Choose the circular contour
\begin{equation}
z_m=Re^{2\pi i m/M}\,,\qquad m=0,\ldots,M-1\,.
\label{eq:nhns-ex-program-contour}
\end{equation}
Evaluate numerically
\begin{equation}
\frac{1}{2\pi}\Delta_\Gamma\arg\det(z\pmb I-\pmb A)
\label{eq:nhns-ex-program-winding}
\end{equation}
by unwrapping the phase of $\det(z_m\pmb I-\pmb A)$ along the contour. Compare the result, sample by sample, with the direct count of eigenvalues of $\pmb A$ inside the disk of radius $R$. Report $N$, $R$, $M$, $S$, and the discrepancy between the winding estimate and direct eigenvalue counting.

\exitem[Programming exercise: number in a disk]
Fix a list of matrix sizes $N$, a mean degree $c=O(1)$, a list of disk radii $R$, and a number $M_{\rm samp}$ of independent samples for each $N$. Generate sparse directed Erd\H{o}s--R\'enyi matrices $\pmb A$ with zero diagonal entries and independent off-diagonal entries
\begin{equation}
{\rm Prob}(A_{ij}=1)=\frac{c}{N}\,,\qquad {\rm Prob}(A_{ij}=0)=1-\frac{c}{N}\,,\qquad i\neq j\,.
\label{eq:nhns-ex-program-directed-er}
\end{equation}
For each sample and each radius, compute the eigenvalues and count
\begin{equation}
\mathcal N_{\pmb A}(\mathcal D_R)\,,\qquad\mathcal D_R=\{z\in\mathbb C: |z|<R\}\,.
\label{eq:nhns-ex-program-disk-count}
\end{equation}
Estimate the empirical mean and variance of this count as functions of $R$. Report $N$, $c$, $M_{\rm samp}$, the list of radii, and whether any eigenvalues lie numerically close to the boundary $|z|=R$.

\exitem[Programming exercise: Hermitized boundary estimate]
Fix a small non-Hermitian matrix $\pmb A$, a circular domain $\mathcal D_R$, a regulator $\eta>0$, a boundary discretization size $M$, and a normal displacement $\delta>0$. For boundary points
\begin{equation}
z_m=Re^{2\pi i m/M}\,,\qquad m=0,\ldots,M-1\,,
\label{eq:nhns-ex-program-boundary-points}
\end{equation}
define inner and outer points
\begin{equation}
z_m^{\rm out}=\left(R+\frac{\delta}{2}\right)e^{2\pi i m/M}\,,\qquad z_m^{\rm in}=\left(R-\frac{\delta}{2}\right)e^{2\pi i m/M}\,.
    \label{eq:nhns-ex-program-inner-outer}
\end{equation}
Compute
\begin{equation}
\Phi_{\pmb A,\eta}(z,z^*)=\log\det\left[(z\pmb I-\pmb A)(z^*\pmb I-\pmb A^\dagger)+\eta^2\pmb I\right]\,,
\label{eq:nhns-ex-program-potential}
\end{equation}
and estimate
\begin{equation}
\mathcal N_{\pmb A}(\mathcal D_R)\simeq\frac{1}{4\pi}\sum_{m=0}^{M-1}\frac{2\pi R/M}{\delta}\left[\Phi_{\pmb A,\eta}(z_m^{\rm out},(z_m^{\rm out})^*)-\Phi_{\pmb A,\eta}(z_m^{\rm in},(z_m^{\rm in})^*)\right]\,.
    \label{eq:nhns-ex-program-boundary-estimator}
\end{equation}
Compare this estimate with direct eigenvalue counting inside $\mathcal D_R$ as $M$ is increased and as $\eta$ and $\delta$ are decreased. Report $N$, $R$, $M$, $\eta$, $\delta$, the direct count, the boundary estimate, and the absolute discrepancy.
\end{exerciseblock}

\section{Algorithms, checks, and numerical validation}
\label{sec:algorithms-checks-validation}
The previous sections developed the cavity and replica equations for several sparse-matrix ensembles. Those equations are only useful if one can solve them reliably and check that the numerical solution is consistent with the underlying spectral problem. The purpose of this section is therefore practical. We explain how the cavity equations become numerical algorithms, how one reduces their computational cost, which observables should be measured, and which checks should be performed before any numerical result is trusted. Since these notes are meant to be used pedagogically, we work explicitly through the algebra that turns the formal cavity recursions into efficient update rules. The two basic computational viewpoints are the same throughout the subject: one may solve the belief-propagation equations on a single large instance, or one may solve the distributional fixed-point equations directly by population dynamics \cite{Pearl1988,KschischangFreyLoeliger2001,YedidiaFreemanWeiss2005,MezardMontanari2009,RogersTakedaPerezCastilloKuhn2008,SuscaVivoKuhn2021}. The first route is useful for comparison with direct diagonalization of a fixed matrix, while the second route gives the thermodynamic-limit disorder average more directly. Appendix~\ref{app:population-dynamics} gives the generic population-dynamics construction; the present section focuses on implementation checks and validation diagnostics.

The formulas below deliberately repeat cavity equations derived in earlier sections, but their role is different here. They are no longer being introduced as formal self-consistency relations; they are being reorganized as update rules, cost reductions, diagnostics, and validation checks.

We start with sparse symmetric matrices. Consider
\begin{equation}
A_{ij}=D_i\delta_{ij}+C_{ij}J_{ij}\,,\qquad A_{ij}=A_{ji}\,,
\label{eq:acv-symmetric-matrix}
\end{equation}
and the spectral parameter
\begin{equation}
z=\lambda-i\epsilon\,,\qquad\epsilon>0\,.
\label{eq:acv-spectral-parameter}
\end{equation}
The cavity equations on a fixed graph are
\begin{equation}
G_{i\to j}(z)=\frac{1}{z-D_i-\displaystyle\sum_{\ell\in\partial i\setminus j}J_{i\ell}^2G_{\ell\to i}(z)}\,,
\label{eq:acv-symmetric-cavity}
\end{equation}
and
\begin{equation}
G_i(z)=\frac{1}{z-D_i-\displaystyle\sum_{\ell\in\partial i}J_{i\ell}^2G_{\ell\to i}(z)}\,.
\label{eq:acv-symmetric-full}
\end{equation}
At first sight, \eqref{eq:acv-symmetric-cavity} seems to require, for every directed edge $i\to j$, a sum over all neighbors of $i$ except $j$. If one implements this naively, one sweep of updates costs
\begin{equation}
\sum_{i=1}^N\sum_{j\in\partial i}O(k_i)=O\left(\sum_{i=1}^N k_i^2\right)\,,
\label{eq:acv-naive-cost}
\end{equation}
where $k_i=|\partial i|$ is the degree of vertex $i$. On graphs with broad degree fluctuations this may be much larger than the number of edges. The first useful derivation is therefore an algebraic reduction of the update cost.

Define the full incoming self-energy at vertex $i$,
\begin{equation}
\Sigma_i(z)=\sum_{\ell\in\partial i}J_{i\ell}^2G_{\ell\to i}(z)\,.
\label{eq:acv-symmetric-self-energy}
\end{equation}
Then \eqref{eq:acv-symmetric-full} becomes
\begin{equation}
G_i(z)=\frac{1}{z-D_i-\Sigma_i(z)}.
\label{eq:acv-symmetric-full-self-energy}
\end{equation}
The cavity sum in \eqref{eq:acv-symmetric-cavity} is the same quantity with the contribution of neighbor $j$ removed:
\begin{equation}
\sum_{\ell\in\partial i\setminus j}J_{i\ell}^2G_{\ell\to i}(z)=\Sigma_i(z)-J_{ij}^2G_{j\to i}(z)\,.
\label{eq:acv-symmetric-cavity-subtraction}
\end{equation}
Hence
\begin{equation}
G_{i\to j}(z)=\frac{1}{z-D_i-\Sigma_i(z)+J_{ij}^2G_{j\to i}(z)}\,.
\label{eq:acv-symmetric-fast-update}
\end{equation}
This formula is the one that should be implemented. In one sweep, one first computes all $\Sigma_i(z)$ from the incoming messages and then updates all directed-edge messages with \eqref{eq:acv-symmetric-fast-update}. Since each undirected edge contributes to exactly two values of $\Sigma_i$, the cost of computing all $\Sigma_i$ is $O(|E|)$, where $|E|$ is the number of undirected edges. The cost of the subsequent directed-edge update is also $O(|E|)$. Thus one full sweep costs
\begin{equation}
O(|E|)\,,
\label{eq:acv-linear-edge-cost}
\end{equation}
not $O(\sum_i k_i^2)$. This simple subtraction identity is essential in practice.

\begin{examplebox}[Why the fast update matters on a high-degree vertex]
The subtraction identity behind \eqref{eq:acv-symmetric-fast-update} is not only a notational convenience; it changes the computational scaling of the belief-propagation sweep. Consider a star graph with one central vertex $0$ and $K$ leaves $1,\ldots,K$. Let all weights be equal to one and all diagonal terms vanish:
\begin{equation}
J_{0\ell}=1\,,\qquad D_i=0\,.
\label{eq:acv-ped-star-settings}
\end{equation}
The cavity message from the central vertex to leaf $j$ is
\begin{equation}
G_{0\to j}(z)=\frac{1}{z-\displaystyle\sum_{\ell=1,\,\ell\neq j}^{K}G_{\ell\to0}(z)}\,.
\label{eq:acv-ped-star-naive-update}
\end{equation}
If this is computed naively for all leaves $j=1,\ldots,K$, one recomputes a sum of $K-1$ terms for each outgoing message. The cost at the central vertex is therefore $O(K^2)$.

Instead define the full incoming self-energy at the central vertex,
\begin{equation}
\Sigma_0(z)=\sum_{\ell=1}^{K}G_{\ell\to0}(z)\,.
\label{eq:acv-ped-star-full-sigma}
\end{equation}
Then
\begin{equation}
\sum_{\ell=1,\,\ell\neq j}^{K}G_{\ell\to0}(z)=\Sigma_0(z)-G_{j\to0}(z)\,,
\label{eq:acv-ped-star-subtraction}
\end{equation}
and hence
\begin{equation}
G_{0\to j}(z)=\frac{1}{z-\Sigma_0(z)+G_{j\to0}(z)}\,.
    \label{eq:acv-ped-star-fast-update}
\end{equation}
Now $\Sigma_0(z)$ is computed once at cost $O(K)$, and all $K$ outgoing messages are updated at total cost $O(K)$. Thus the fast update reduces the central-vertex cost from $O(K^2)$ to $O(K)$. On sparse graphs with bounded mean degree this distinction may look mild, but on heterogeneous graphs with hubs it is essential.
\end{examplebox}

A damped synchronous iteration is then
\begin{equation}
G_{i\to j}^{(t+1)}(z)=(1-\gamma)G_{i\to j}^{(t)}(z)+\gamma\frac{1}{z-D_i-\Sigma_i^{(t)}(z)+J_{ij}^2G_{j\to i}^{(t)}(z)}\,,\qquad 0<\gamma\leq1\,,
\label{eq:acv-symmetric-damped-update}
\end{equation}
with
\begin{equation}
\Sigma_i^{(t)}(z)=\sum_{\ell\in\partial i}J_{i\ell}^2G_{\ell\to i}^{(t)}(z)\,.
\label{eq:acv-symmetric-sigma-iteration}
\end{equation}
The damping factor $\gamma$ is often needed when $\epsilon$ is small or when the graph contains vertices of very large degree. A natural stopping criterion is the mean change per directed edge,
\begin{equation}
\Delta^{(t)}(z)=\frac{1}{2|E|}\sum_{i=1}^{N}\sum_{j\in\partial i}\left|G_{i\to j}^{(t+1)}(z)-G_{i\to j}^{(t)}(z)\right|\,.
\label{eq:acv-symmetric-convergence}
\end{equation}
One stops when $\Delta^{(t)}(z)$ falls below a prescribed tolerance. Once the fixed point has been reached, one computes the local Green functions from
\begin{equation}
G_i(z)=\frac{1}{z-D_i-\Sigma_i(z)}\,,
\label{eq:acv-symmetric-local-green}
\end{equation}
and the regularized density of the instance from
\begin{equation}
\rho_{\epsilon}^{\rm BP}(\lambda)=\frac{1}{\pi N}\sum_{i=1}^{N}{\rm Im} G_i(\lambda-i\epsilon).
\label{eq:acv-symmetric-density-estimator}
\end{equation}
This is the single-instance belief-propagation algorithm for sparse symmetric matrices.

It is useful to understand why this fixed point may fail to converge. Linearizing \eqref{eq:acv-symmetric-cavity} around a fixed point $G_{i\to j}^\star$ gives
\begin{equation}
\delta G_{i\to j}^{(t+1)}=\sum_{\ell\in\partial i\setminus j}J_{i\ell}^2\left(G_{i\to j}^{\star}\right)^2\delta G_{\ell\to i}^{(t)}\,.
\label{eq:acv-symmetric-linearization}
\end{equation}
This follows from the elementary derivative
\begin{equation}
\frac{\partial}{\partial G_{\ell\to i}}\frac{1}{z-D_i-\sum_{r\in\partial i\setminus j}J_{ir}^2G_{r\to i}}=J_{i\ell}^2\left(G_{i\to j}\right)^2\,.
    \label{eq:acv-symmetric-map-derivative}
\end{equation}
Hence the Jacobian of the update map is explicitly known. If the spectral radius of the linear operator in \eqref{eq:acv-symmetric-linearization} is smaller than one, the fixed point is locally stable. In numerical work one rarely computes this spectral radius directly, but \eqref{eq:acv-symmetric-linearization} explains three basic empirical facts: convergence becomes slower when $\epsilon$ is small, convergence becomes slower near spectral edges or localization transitions, and damping improves stability by reducing the effective Jacobian.

The ensemble-level alternative to belief propagation is population dynamics. Suppose the graph ensemble has degree distribution $p_k$ and excess-degree distribution
\begin{equation}
q_\ell=\frac{(\ell+1)p_{\ell+1}}{\sum_{r}rp_r}.
\label{eq:acv-excess-degree}
\end{equation}
The distributional cavity equation is
\begin{equation}
\mathcal{P}_{\rm cav}(G)=\sum_{\ell=0}^{\infty}q_\ell\int dD p_D(D)\left[\prod_{r=1}^{\ell}dG_r \mathcal{P}_{\rm cav}(G_r) dJ_r p_J(J_r)\right]\delta\left(G-\frac{1}{z-D-\displaystyle\sum_{r=1}^{\ell}J_r^2G_r}\right)\,.
    \label{eq:acv-population-equation}
\end{equation}
To turn \eqref{eq:acv-population-equation} into an algorithm, we represent the unknown law by an empirical population
\begin{equation}
\left\{G^{(1)},G^{(2)},\ldots,G^{(M)}\right\}\,,
\label{eq:acv-population}
\end{equation}
with large $M$. One update consists of drawing an excess degree $\ell$ from $q_\ell$, drawing $\ell$ indices $a_1,\ldots,a_\ell$ uniformly from $\{1,\ldots,M\}$, drawing $D$ from $p_D$ and $J_1,\ldots,J_\ell$ from $p_J$, and computing
\begin{equation}
G_{\rm new}=\frac{1}{z-D-\displaystyle\sum_{r=1}^{\ell}J_r^2G^{(a_r)}}\,.
\label{eq:acv-population-update}
\end{equation}
A randomly selected population member is then replaced by $G_{\rm new}$. Repeating this procedure many times produces an empirical approximation of the fixed-point distribution. To measure observables, one draws a full degree $k$ from $p_k$, generates
\begin{equation}
G_{\rm site}=\frac{1}{z-D-\displaystyle\sum_{r=1}^{k}J_r^2G^{(a_r)}}\,,
\label{eq:acv-site-sampling}
\end{equation}
and averages
\begin{equation}
\rho_\epsilon(\lambda)\approx\frac{1}{\pi n_{\rm obs}}\sum_{m=1}^{n_{\rm obs}}{\rm Im}G_{{\rm site},m}(\lambda-i\epsilon)\,.
\label{eq:acv-population-density-estimator}
\end{equation}
The statistical error of \eqref{eq:acv-population-density-estimator} has the usual Monte Carlo form and decreases as $n_{\rm obs}^{-1/2}$ at fixed population, while the error due to the finite population size decreases as $M^{-1/2}$ when the fixed point is sufficiently regular. In practice, one must check both sources of error.

The same logic applies to sparse covariance and diluted Wishart matrices, but now the graph is bipartite. We keep the diluted Wishart convention used earlier: $U_{\mu\to i}$ is not divided by $d$, and the actual contribution to the variable denominator is $d^{-1}U_{\mu\to i}$. The cavity equations are
\begin{equation}
U_{\mu\to i}(z)=\frac{(\xi_i^\mu)^2}{1-\displaystyle\frac{1}{d}\sum_{j\in\partial\mu\setminus i}(\xi_j^\mu)^2G_{j\to\mu}(z)}\,,
\label{eq:acv-wishart-factor-update}
\end{equation}
and
\begin{equation}
G_{i\to\mu}(z)=\frac{1}{z-\displaystyle\frac{1}{d}\sum_{\nu\in\partial i\setminus\mu}U_{\nu\to i}(z)}\,.
\label{eq:acv-wishart-variable-update}
\end{equation}
Again, the naïve implementation repeats the same partial sums many times. The efficient form is obtained by introducing
\begin{equation}
T_\mu(z)=\frac{1}{d}\sum_{j\in\partial\mu}(\xi_j^\mu)^2G_{j\to\mu}(z)\,,
\label{eq:acv-wishart-factor-sum}
\end{equation}
and
\begin{equation}
S_i(z)=\frac{1}{d}\sum_{\nu\in\partial i}U_{\nu\to i}(z)\,.
\label{eq:acv-wishart-variable-sum}
\end{equation}
Then
\begin{equation}
U_{\mu\to i}(z)=\frac{(\xi_i^\mu)^2}{1-T_\mu(z)+(\xi_i^\mu)^2G_{i\to\mu}(z)/d}\,,
\label{eq:acv-wishart-fast-factor}
\end{equation}
because the exclusion of the edge $(i,\mu)$ is implemented by subtracting the corresponding contribution from $T_\mu$. Similarly,
\begin{equation}
G_{i\to\mu}(z)=\frac{1}{z-S_i(z)+\frac{1}{d}U_{\mu\to i}(z)}\,,
\label{eq:acv-wishart-fast-variable}
\end{equation}
and
\begin{equation}
G_i(z)=\frac{1}{z-S_i(z)}\,.
\label{eq:acv-wishart-full-green}
\end{equation}
Thus one sweep costs $O(|E_{\rm bip}|)$, where $|E_{\rm bip}|$ is the number of occupied bipartite edges. The regularized density is
\begin{equation}
\rho_\epsilon^{\rm BP}(\lambda)=\frac{1}{\pi N}\sum_{i=1}^{N}{\rm Im}G_i(\lambda-i\epsilon)\,.
\label{eq:acv-wishart-density-estimator}
\end{equation}

The population-dynamics version of the diluted Wishart problem uses two populations, one for variable-to-factor messages and one for factor-to-variable self-energies. Let
\begin{equation}
\left\{G^{(1)},\ldots,G^{(M)}\right\}\,,\qquad\left\{U^{(1)},\ldots,U^{(M)}\right\}\,,
\label{eq:acv-wishart-two-populations}
\end{equation}
be the two empirical populations. A factor update draws a factor excess degree $k$ from the corresponding law, draws $k$ incoming variable messages and $k+1$ weights, and computes
\begin{equation}
U_{\rm new}=\frac{\xi^2}{1-\displaystyle\frac{1}{d}\sum_{r=1}^{k}\xi_r^2G^{(a_r)}}\,.
\label{eq:acv-wishart-pop-factor}
\end{equation}
A variable update draws a variable excess degree $\ell$ and computes
\begin{equation}
G_{\rm new}=\frac{1}{z-\displaystyle\frac{1}{d}\sum_{r=1}^{\ell}U^{(b_r)}}\,.
    \label{eq:acv-wishart-pop-variable}
\end{equation}
The corresponding site measurement draws a full variable degree $k_{\rm site}$ and computes
\begin{equation}
G_{\rm site}=\frac{1}{z-\displaystyle\frac{1}{d}\sum_{r=1}^{k_{\rm site}}U^{(b_r)}}\,.
\label{eq:acv-wishart-pop-site}
\end{equation}
Measurement of the density then averages ${\rm Im}\,G_{\rm site}/\pi$ over many such site samples, exactly as in the sparse symmetric case. The important point is that the factor and variable populations must be updated alternately, because the two message types live on different sides of the bipartite graph \cite{NagaoTanaka2007,RogersTakedaPerezCastilloKuhn2008}.

For the generalized diluted Wishart ensemble, the same strategy works, but the factor update depends on three partial sums rather than one. If the occupied edge $(i,\mu)$ carries the pair of weights $(x_i^\mu,y_i^\mu)$, the factor-to-variable self-energy is
\begin{equation}
U_{\mu\to i}(z)=\frac{4d\,x_i^\mu y_i^\mu+(y_i^\mu)^2S_{xx}^{\mu\to i}-2x_i^\mu y_i^\mu S_{xy}^{\mu\to i}+(x_i^\mu)^2S_{yy}^{\mu\to i}}{\left(2d-S_{xy}^{\mu\to i}\right)^2-S_{xx}^{\mu\to i}S_{yy}^{\mu\to i}}\,,
\label{eq:acv-gdw-factor-update}
\end{equation}
with
\begin{equation}
S_{xx}^{\mu\to i}=\sum_{j\in\partial\mu\setminus i}(x_j^\mu)^2G_{j\to\mu}\,,\qquad
S_{xy}^{\mu\to i}=\sum_{j\in\partial\mu\setminus i}x_j^\mu y_j^\mu G_{j\to\mu}\,,\qquad S_{yy}^{\mu\to i}=\sum_{j\in\partial\mu\setminus i}(y_j^\mu)^2G_{j\to\mu}\,.
\label{eq:acv-gdw-cavity-sums}
\end{equation}
The efficient implementation introduces the full factor sums
\begin{equation}
T_{xx,\mu}=\sum_{j\in\partial\mu}(x_j^\mu)^2G_{j\to\mu}\,,\qquad T_{xy,\mu}=\sum_{j\in\partial\mu}x_j^\mu y_j^\mu G_{j\to\mu}\,,\qquad T_{yy,\mu}=\sum_{j\in\partial\mu}(y_j^\mu)^2G_{j\to\mu}\,,
    \label{eq:acv-gdw-full-sums}
\end{equation}
and then subtracts the contribution of the distinguished variable:
\begin{equation}
S_{xx}^{\mu\to i}=T_{xx,\mu}-(x_i^\mu)^2G_{i\to\mu}\,,\quad S_{xy}^{\mu\to i} = T_{xy,\mu}-x_i^\mu y_i^\mu G_{i\to\mu}\,, \quad S_{yy}^{\mu\to i} = T_{yy,\mu}-(y_i^\mu)^2G_{i\to\mu}\,.
\label{eq:acv-gdw-subtracted-sums}
\end{equation}
The variable update remains scalar:
\begin{equation}
G_{i\to\mu}(z)=\frac{1}{z-\displaystyle\sum_{\nu\in\partial i\setminus\mu}U_{\nu\to i}(z)}\,.
\label{eq:acv-gdw-variable-update}
\end{equation}
For the same linear-cost reason as in the ordinary diluted Wishart case, one should also introduce the full variable sum
\begin{equation}
S_i(z)=\sum_{\nu\in\partial i}U_{\nu\to i}(z)\,.
\label{eq:acv-gdw-variable-sum}
\end{equation}
Then
\begin{equation}
G_{i\to\mu}(z)=\frac{1}{z-S_i(z)+U_{\mu\to i}(z)}\,,\qquad G_i(z)=\frac{1}{z-S_i(z)}\,.
\label{eq:acv-gdw-fast-variable}
\end{equation}
Thus the generalized ensemble remains computationally manageable because both the rank-two factor sums and the variable sums are updated by subtraction. In practice, the numerical cost differs from ordinary diluted Wishart matrices only by a larger constant factor.

For sparse non-Hermitian matrices the messages are $2\times2$ matrices rather than complex scalars. The cavity recursion is
\begin{equation}
\pmb{G}_{i\to j}(z,\eta)=\left[\pmb{Z}_i(z,\eta)-\sum_{\ell\in\partial i\setminus j}\pmb{\mathcal A}_{i\ell}\pmb{G}_{\ell\to i}(z,\eta)\pmb{\mathcal A}_{i\ell}^{\dagger}\right]^{-1}\,,
\label{eq:acv-nh-cavity}
\end{equation}
where
\begin{equation}
\pmb{Z}_i(z,\eta)=\begin{pmatrix}
i\eta & z-A_{ii}\\
z^*-A_{ii}^* & i\eta
\end{pmatrix}\,,\qquad
\pmb{\mathcal A}_{ij}=\begin{pmatrix}
0 & A_{ij}\\
A_{ji}^* & 0
\end{pmatrix}\,.
\label{eq:acv-nh-local-objects}
\end{equation}
The efficient form is again obtained by introducing the full matrix self-energy
\begin{equation}
\pmb{\Sigma}_i(z,\eta)=\sum_{\ell\in\partial i}\pmb{\mathcal A}_{i\ell}\pmb{G}_{\ell\to i}(z,\eta)\pmb{\mathcal A}_{i\ell}^{\dagger}\,.
\label{eq:acv-nh-self-energy}
\end{equation}
Then
\begin{equation}
\pmb{G}_{i\to j}(z,\eta)=\left[\pmb{Z}_i(z,\eta)-\pmb{\Sigma}_i(z,\eta)+\pmb{\mathcal A}_{ij}\pmb{G}_{j\to i}(z,\eta)\pmb{\mathcal A}_{ij}^{\dagger}\right]^{-1}\,,
    \label{eq:acv-nh-fast-update}
\end{equation}
and
\begin{equation}
\pmb{G}_{i}(z,\eta)=\left[\pmb{Z}_i(z,\eta)-\pmb{\Sigma}_i(z,\eta)\right]^{-1}\,.
\label{eq:acv-nh-full-green}
\end{equation}
The regularized non-Hermitian resolvent field is
\begin{equation}
g_\eta(z,z^*)=\frac{1}{N}\sum_{i=1}^{N}\left[\pmb{G}_{i}(z,\eta)\right]_{21}\,,
\label{eq:acv-nh-regularized-resolvent}
\end{equation}
and the density is
\begin{equation}
\rho(z)=\frac{1}{\pi}\lim_{\eta\downarrow0}\partial_{z^*}g_\eta(z,z^*).
\label{eq:acv-nh-density}
\end{equation}
Numerically, one evaluates $g_\eta$ on a grid of points $z=x+iy$ and approximates the Wirtinger derivative by finite differences,
\begin{equation}
\partial_{z^*}g=\frac{1}{2}\left(\partial_x+i\partial_y\right)g\,.
\label{eq:acv-wirtinger-derivative}
\end{equation}
Thus, if $\Delta x=\Delta y=h$,
\begin{equation}
\partial_{z^*}g(x,y)\approx\frac{1}{4h}\left[g(x+h,y)-g(x-h,y)\right]+\frac{i}{4h}\left[g(x,y+h)-g(x,y-h)\right]\,.
    \label{eq:acv-finite-difference-dzbar}
\end{equation}
The corresponding population dynamics uses a population of $2\times2$ matrices rather than complex numbers \cite{RogersPerezCastillo2009,MetzNeriRogers2019}. Apart from this, the algorithmic structure is unchanged.

The large-deviation and conditioned-density calculations require enlarged messages rather than a new numerical philosophy. In the Hermitian index problem the message carries the cavity Green functions at the two threshold values
\begin{equation}
z_x^- = x-i\epsilon\,,\qquad z_x^+ = x+i\epsilon\,,
\label{eq:acv-threshold-pair}
\end{equation}
and, for conditioned densities, also at the probe point $z_\lambda=\lambda-i\eta$. Thus the message is
\begin{equation}
\mathfrak m=\left(G^-,G^+,G^0\right)\,,
\label{eq:acv-enlarged-message}
\end{equation}
or the analogous pair of self-energies in a bipartite problem. The local update is the ordinary cavity map applied componentwise, while the tilted problem introduces a local weight $w_s(\mathfrak m,\{\mathfrak m_r\})$ derived from the determinant phase or Bethe free-energy contribution of the observable of interest \cite{MetzPerezCastillo2016,PerezCastilloMetz2018Wishart,PerezCastilloMetz2018Conditioned,RamosSanchezGuzmanGonzalezPerezCastilloMetz2021}. The generic structure of the distributional equation is therefore
\begin{equation}
\mathcal{P}_s(\mathfrak m)=\frac{1}{\mathcal C_s}\int\left[\prod_{r}d\mathfrak m_r \mathcal P_s(\mathfrak m_r)\right]d\omega \Pi(\omega) w_s(\omega,\{\mathfrak m_r\})\delta\left(\mathfrak m-\mathcal F(\omega,\{\mathfrak m_r\})\right)\,,
    \label{eq:acv-generic-tilted-population}
\end{equation}
where $\omega$ denotes the local disorder variables and $\mathcal C_s$ is the normalization required to keep $\mathcal P_s$ normalized. Equation \eqref{eq:acv-generic-tilted-population} is the algorithmic heart of the large-deviation and conditioning problems: the map $\mathcal F$ is the same one used in the unbiased spectral-density calculation, but the empirical population is now reweighted by the local tilting factor.

The validation step is as important as the algorithm itself. The first family of checks consists of exact sum rules. For a symmetric matrix, the density obtained from either belief propagation or population dynamics must satisfy
\begin{equation}
\int_{-\infty}^{\infty}d\lambda\rho(\lambda)=1\,.
\label{eq:acv-normalization-check}
\end{equation}
The first moment is
\begin{equation}
\int_{-\infty}^{\infty}d\lambda \lambda\rho(\lambda)=\lim_{N\to\infty}\frac{1}{N}{\rm Tr} \pmb A.
    \label{eq:acv-first-moment-check-general}
\end{equation}
For Eq. \eqref{eq:acv-symmetric-matrix},
\begin{equation}
\frac{1}{N}{\rm Tr}\pmb A=\frac{1}{N}\sum_{i=1}^{N}D_i\,,
\label{eq:acv-trace-symmetric}
\end{equation}
so the ensemble average is
\begin{equation}
\int d\lambda\lambda\overline{\rho_{\pmb A}(\lambda)}=\int dD p_D(D) D\,.
\label{eq:acv-first-moment-ensemble}
\end{equation}
The second moment is also explicit:
\begin{equation}
\frac{1}{N}{\rm Tr}\pmb A^2=\frac{1}{N}\sum_{i=1}^{N}D_i^2+\frac{2}{N}\sum_{\{i,j\}\in E}J_{ij}^2\,.
\label{eq:acv-second-moment-symmetric}
\end{equation}
Hence for a graph of mean degree $c$,
\begin{equation}
\int d\lambda \lambda^2\overline{\rho_{\pmb A}(\lambda)}=\int dD p_D(D) D^2+ c\int dJ p_J(J) J^2\,.
\label{eq:acv-second-moment-ensemble}
\end{equation}
These moment checks are elementary but indispensable. Strictly, they are checks on the limiting spectral measure, on direct eigenvalue sums, or on the high-$|z|$ expansion of the resolvent. At finite $\epsilon$, Lorentzian regularization has long tails, so moment integrals over a finite grid should be interpreted with the chosen window and regulator kept explicit. When the corresponding controlled checks fail, the numerical density should not be trusted.

\begin{examplebox}[Moment checks on a weighted three-site chain]
Moment checks are among the simplest ways of detecting an implementation error. Consider
\begin{equation}
\pmb A=\begin{pmatrix}
D_1 & J_{12} & 0\\
J_{12} & D_2 & J_{23}\\
0 & J_{23} & D_3
\end{pmatrix}\,.
\label{eq:acv-ped-three-chain}
\end{equation}
The first spectral moment is
\begin{equation}
\frac{1}{3}{\rm Tr}\pmb A=\frac{D_1+D_2+D_3}{3}\,.
\label{eq:acv-ped-first-moment-chain}
\end{equation}
The second moment is
\begin{equation}
\frac{1}{3}{\rm Tr}\pmb A^2\,.
\label{eq:acv-ped-second-moment-chain-start}
\end{equation}
Let us compute it explicitly. The diagonal entries of $\pmb A^2$ are
\begin{align}
(\pmb A^2)_{11}&=D_1^2+J_{12}^2\,,\label{eq:acv-ped-A2-11}\\
(\pmb A^2)_{22}&=D_2^2+J_{12}^2+J_{23}^2\,,\label{eq:acv-ped-A2-22}\\
(\pmb A^2)_{33}&=D_3^2+J_{23}^2\,.\label{eq:acv-ped-A2-33}
\end{align}
Therefore
\begin{equation}
\frac{1}{3}{\rm Tr}\pmb A^2=\frac{1}{3}\left[D_1^2+D_2^2+D_3^2+2J_{12}^2+2J_{23}^2\right]\,.
\label{eq:acv-ped-second-moment-chain}
\end{equation}
This is exactly the finite-size version of the general sparse-matrix identity
\begin{equation}
\frac{1}{N}{\rm Tr}\pmb A^2=\frac{1}{N}\sum_{i=1}^{N}D_i^2+\frac{2}{N}\sum_{\{i,j\}\in E}J_{ij}^2\,.
\label{eq:acv-ped-second-moment-general-check}
\end{equation}
A numerical density obtained from belief propagation, population dynamics, or direct diagonalization should reproduce these moments after integration over $\lambda$, up to the effects of finite integration range, finite grid spacing, and finite regulator.
\end{examplebox}

For diluted Wishart matrices,
\begin{equation}
\pmb W=\frac{1}{d}\pmb X\pmb X^{\rm T}\,,
\label{eq:acv-wishart-matrix}
\end{equation}
the spectrum must be nonnegative. Therefore
\begin{equation}
\rho(\lambda)=0\qquad \text{for } \lambda<0\,,
\label{eq:acv-wishart-positivity}
\end{equation}
in the limit $\epsilon\downarrow0$. At finite $\epsilon$ the numerical estimator has Lorentzian leakage into $\lambda<0$. Away from the origin this leakage must shrink as $\epsilon\downarrow0$, but if the limiting measure has a zero atom, the leakage near $\lambda=0$ should be treated as a broadening artifact rather than as negative spectral support. The first moment is
\begin{equation}
\frac{1}{N}{\rm Tr}\pmb W=\frac{1}{Nd}\sum_{i=1}^{N}\sum_{\mu=1}^{P}(X_i^\mu)^2\,,
\label{eq:acv-wishart-trace}
\end{equation}
and in the Poisson diluted ensemble it converges to
\begin{equation}
\int_0^\infty d\lambda \lambda\overline{\rho_{\pmb W}(\lambda)}=\frac{\langle \xi^2\rangle}{\alpha}.
    \label{eq:acv-wishart-first-moment}
\end{equation}

\begin{examplebox}[A finite Wishart moment and positivity check]
Let
\begin{equation}
\pmb X=\begin{pmatrix}
1 & 0\\
0 & 1\\
1 & 1
\end{pmatrix}\,, \qquad d=1\,,\qquad\pmb W=\pmb X\pmb X^{\rm T}\,.
\label{eq:acv-ped-X-Wishart-check}
\end{equation}
Then
\begin{equation}
\pmb W=\begin{pmatrix}
1 & 0 & 1\\
0 & 1 & 1\\
1 & 1 & 2
\end{pmatrix}\,.
\label{eq:acv-ped-W-Wishart-check}
\end{equation}
The trace is
\begin{equation}
\frac{1}{3}{\rm Tr}\pmb W=\frac{1+1+2}{3}=\frac{4}{3}\,.
\label{eq:acv-ped-Wishart-trace-check}
\end{equation}
Using the rectangular matrix directly,
\begin{equation}
\frac{1}{Nd}\sum_{i=1}^{N}\sum_{\mu=1}^{P}(X_i^\mu)^2=\frac{1}{3}(1+1+1+1)=\frac{4}{3}\,,
\label{eq:acv-ped-Wishart-rectangular-trace-check}
\end{equation}
which agrees with the trace of $\pmb W$.

The eigenvalues are
\begin{equation}
{\rm spec}(\pmb W)=\{3,1,0\}\,.
\label{eq:acv-ped-Wishart-spectrum-check}
\end{equation}
Hence the spectrum is nonnegative, as it must be for a matrix of the form $\pmb X\pmb X^{\rm T}$. A numerical density for a Wishart ensemble should therefore pass three elementary checks: nonnegative support in the limit $\epsilon\downarrow0$, normalization to one, and first moment equal to $N^{-1}{\rm Tr}\pmb W$ or its ensemble average.
\end{examplebox}

For the generalized diluted Wishart ensemble,
\begin{equation}
\pmb F=\frac{1}{2d}\left(\pmb X\pmb Y^{\rm T}+\pmb Y\pmb X^{\rm T}\right)\,,
\label{eq:acv-gdw-matrix}
\end{equation}
the first moment becomes
\begin{equation}
\int d\lambda \lambda\overline{\rho_{\pmb F}(\lambda)}=\frac{m_{11}}{\alpha}\,,
\label{eq:acv-gdw-first-moment}
\end{equation}
with
\begin{equation}
m_{11}=\int dx dy \varrho(x,y) xy\,.
\label{eq:acv-gdw-local-correlation}
\end{equation}
These formulas are immediate consequences of the trace identities and should always be checked numerically.

The second family of checks uses exactly solvable limiting laws. For the adjacency matrix of a random $c$-regular graph, the cavity equations must recover the Kesten--McKay density \cite{Kesten1959,McKay1981}. For sparse symmetric matrices in the large-connectivity scaling, they must reduce to the Wigner self-consistency equation. For diluted Wishart matrices, the dense limit must reproduce the Mar\v{c}enko--Pastur law \cite{MarchenkoPastur1967}. For dense products of Wishart matrices, the numerical solution must satisfy the polynomial equation derived from the product-Wishart law \cite{DupicPerezCastillo2014,BurdaJaroszLivanNowakSwiech2010}. For sparse non-Hermitian matrices, the large-connectivity limit must approach the circular or elliptic laws \cite{Ginibre1965,SommersCrisantiSompolinskyStein1988}. These checks are valuable because they test not only the implementation but also the normalization and scaling conventions.

The third family of checks compares cavity results with direct diagonalization of finite instances. For a Hermitian matrix with eigenvalues $\lambda_1,\ldots,\lambda_N$, the exact Lorentzian-broadened density is
\begin{equation}
\rho_{\epsilon,N}^{\rm diag}(\lambda)=\frac{1}{\pi N}\sum_{i=1}^{N}\frac{\epsilon}{(\lambda-\lambda_i)^2+\epsilon^2}\,.
\label{eq:acv-diag-hermitian-density}
\end{equation}

\begin{examplebox}[Why one must compare at the same regulator]
Consider a matrix with two eigenvalues,
\begin{equation}
\lambda_1=-1\,, \qquad \lambda_2=1\,.
\label{eq:acv-ped-two-eigenvalues}
\end{equation}
The exact empirical density is
\begin{equation}
\rho_N(\lambda)=\frac{1}{2}\delta(\lambda+1)+\frac{1}{2}\delta(\lambda-1)\,.
\label{eq:acv-ped-exact-two-density}
\end{equation}
At finite regulator $\epsilon>0$, the object computed from the resolvent is not \eqref{eq:acv-ped-exact-two-density}, but the Lorentzian-broadened density
\begin{equation}
\rho_{\epsilon,N}(\lambda)=\frac{1}{2\pi}\frac{\epsilon}{(\lambda+1)^2+\epsilon^2}+\frac{1}{2\pi}\frac{\epsilon}{(\lambda-1)^2+\epsilon^2}\,.
\label{eq:acv-ped-two-lorentzians}
\end{equation}
Thus, if a cavity or population-dynamics calculation is performed at finite $\epsilon$, the correct diagonalization benchmark is \eqref{eq:acv-ped-two-lorentzians}, not a raw histogram of delta peaks.

This also explains a common numerical pitfall. If $\epsilon$ is chosen much smaller than the mean eigenvalue spacing of a finite matrix, then direct diagonalization displays very sharp sample-dependent peaks. A population-dynamics calculation, on the other hand, describes the thermodynamic smoothed density at the same regulator. The correct comparison is therefore
\begin{equation}
N\to\infty\quad\text{at fixed }\epsilon>0\,,\qquad\text{then}\qquad\epsilon\downarrow0\,.
\label{eq:acv-ped-order-of-limits-reminder}
\end{equation}
This is why the regulator is a validation parameter, not just a formal device.
\end{examplebox}

Accordingly, a convenient error measure is
\begin{equation}
\mathcal E_1(\epsilon)=\int d\lambda\left|\rho_{\epsilon,N}^{\rm BP}(\lambda)-\rho_{\epsilon,N}^{\rm diag}(\lambda)\right|\,,
\label{eq:acv-hermitian-L1-error}
\end{equation}
or the corresponding ensemble-averaged version obtained by averaging both densities over disorder realizations.

For non-Hermitian matrices, direct diagonalization can be compared with the Hermitized cavity estimate only after the regularization and grid have been matched. One convenient diagonalization benchmark is a smoothed empirical density,
\begin{equation}
\rho_{h,N}^{\rm diag}(z)=\frac{1}{\pi h^2N}\sum_{i=1}^{N}\exp\left[-\frac{|z-z_i|^2}{h^2}\right]\,,
\label{eq:acv-diag-nonhermitian-density}
\end{equation}
or alternatively a binning procedure on a fine grid in the complex plane. The precise mollifier is less important than consistency: the smoothing scale $h$, the Hermitization regulator $\eta$, the finite-difference grid, and the computational box must all be stated. A practical validation is then obtained by comparing a cavity density $\rho_{\eta,h,N}^{\rm cav}$, evaluated on the same grid and with the stated regulator, with the smoothed diagonalization density:
\begin{equation}
\mathcal E_2(\eta,h)=\int_{\mathcal B}d^2z\left|\rho_{\eta,h,N}^{\rm cav}(z)-\rho_{h,N}^{\rm diag}(z)\right|
\label{eq:acv-nonhermitian-L1-error}
\end{equation}
over a box $\mathcal B$ that contains the numerical support.

The fourth family of checks concerns the regulator. In the Hermitian problem one solves the cavity equations at $\epsilon>0$, while in the non-Hermitian problem one solves the Hermitized equations at $\eta>0$. The solution should vary smoothly when the regulator is decreased. A robust numerical strategy is continuation. One starts at a value of $\epsilon$ or $\eta$ for which the fixed-point iteration is strongly contractive, converges the messages, and then uses this fixed point as the initial condition for a smaller regulator. If
\begin{equation}
\epsilon_0>\epsilon_1>\cdots>\epsilon_n
\label{eq:acv-epsilon-schedule}
\end{equation}
is a decreasing schedule, then the fixed point at $\epsilon_{m-1}$ is used to initialize the iteration at $\epsilon_m$. This procedure is especially effective near spectral edges and in large-deviation or conditioned calculations, where the enlarged message space can make the map less contractive.

\begin{examplebox}[Regulator continuation as a numerical protocol]
Suppose one wants the spectral density near a point $\lambda$ where the cavity iteration is unstable for very small $\epsilon$. Instead of starting directly at the desired regulator, one may use a continuation schedule
\begin{equation}
\epsilon_0>\epsilon_1>\cdots>\epsilon_n\,.
\label{eq:acv-ped-epsilon-schedule}
\end{equation}
A typical choice is geometric:
\begin{equation}
\epsilon_m=\epsilon_0 r^m\,,\qquad 0<r<1\,.
\label{eq:acv-ped-geometric-epsilon}
\end{equation}
At the first value $\epsilon_0$, one initializes the messages by a simple stable guess, for example
\begin{equation}
G_{i\to j}^{(0)}=\frac{1}{\lambda-i\epsilon_0-D_i}\,.
\label{eq:acv-ped-initial-guess}
\end{equation}
After convergence at $\epsilon_0$, the resulting fixed point is used as the initial condition for $\epsilon_1$. Repeating this procedure gives a sequence of fixed points
\begin{equation}
\{G_{i\to j}^{(\epsilon_0)}\}\longrightarrow\{G_{i\to j}^{(\epsilon_1)}\}\longrightarrow\cdots\longrightarrow\{G_{i\to j}^{(\epsilon_n)}\}\,.
\label{eq:acv-ped-continuation-flow}
\end{equation}
The validation check is that observables such as
\begin{equation}
\rho_{\epsilon_m}(\lambda)=\frac{1}{\pi N}\sum_i{\rm Im}G_i(\lambda-i\epsilon_m)
\label{eq:acv-ped-continuation-density}
\end{equation}
vary smoothly with $\epsilon_m$, except where the underlying spectral measure contains atoms or very sharp localized structures. Sudden jumps, loss of the sign condition ${\rm Im}\,G_i>0$, or dependence on the initialization indicate that the fixed-point iteration has not been validated.
\end{examplebox}

The fifth family of checks concerns self-averaging. Population dynamics computes the ensemble law directly, whereas belief propagation computes the density of one large instance. For a self-averaging observable, one expects
\begin{equation}
\rho_{\epsilon,N}^{\rm BP}(\lambda)\approx\overline{\rho_{\epsilon,N}^{\rm diag}(\lambda)}\approx\rho_{\epsilon}^{\rm pop}(\lambda)
\label{eq:acv-self-averaging}
\end{equation}
for large $N$. The first approximation compares one large instance with the disorder average; the second compares the finite-$N$ average with the thermodynamic-limit population dynamics. When these comparisons fail badly, there are only a few possible explanations: the system size is too small, the regulator is too small, the observable is not self-averaging in the regime under study, or the numerical fixed point is unstable or inaccurate.

\begin{examplebox}[A minimal validation workflow]
A useful validation workflow for a sparse symmetric ensemble is the following. Choose a benchmark ensemble, for instance random $c$-regular graphs with unweighted adjacency matrices.

First, compute the exact benchmark density. For random $c$-regular graphs this is the Kesten--McKay law,
\begin{equation}
\rho_{\rm KM}(\lambda)=\frac{c}{2\pi}\frac{\sqrt{4(c-1)-\lambda^2}}{c^2-\lambda^2}\mathbf 1_{|\lambda|\leq2\sqrt{c-1}}\,.
\label{eq:acv-ped-KM-benchmark}
\end{equation}
Second, generate finite random regular graphs and compute the broadened diagonalization density
\begin{equation}
\rho_{\epsilon,N}^{\rm diag}(\lambda)=\frac{1}{\pi N}\sum_{\alpha=1}^{N}\frac{\epsilon}{(\lambda-\lambda_\alpha)^2+\epsilon^2}\,.
\label{eq:acv-ped-diag-broadened-workflow}
\end{equation}
For a connected $c$-regular graph, this density contains the trivial Perron eigenvalue $c$ with weight $1/N$. In a comparison with the limiting Kesten--McKay bulk, one should either remove this contribution from the diagonalization density or add its finite-$N$ Lorentzian contribution explicitly to the benchmark.

Third, solve the cavity equation
\begin{equation}
G_{\rm cav}=\frac{1}{z-(c-1)G_{\rm cav}}\,,\qquad G_{\rm site}=\frac{1}{z-cG_{\rm cav}}\,,
\label{eq:acv-ped-cavity-workflow}
\end{equation}
and compute
\begin{equation}
\rho_{\epsilon}^{\rm cav}(\lambda)=\frac{1}{\pi}{\rm Im} G_{\rm site}(\lambda-i\epsilon)\,.
\label{eq:acv-ped-cavity-density-workflow}
\end{equation}
Fourth, compare the three objects at the same regulator:
\begin{equation}
\rho_{\epsilon,N}^{\rm diag}(\lambda)\,,\qquad\rho_{\epsilon}^{\rm cav}(\lambda)\,,\qquad\rho_{\rm KM}(\lambda)\text{ broadened at the same }\epsilon\,.
\label{eq:acv-ped-three-way-comparison}
\end{equation}
Finally, check the moments
\begin{equation}
\int d\lambda \rho(\lambda)=1\,,\qquad \int d\lambda \lambda \rho(\lambda)=0\,,\qquad \int d\lambda \lambda^2\rho(\lambda)=c\,.
\label{eq:acv-ped-rrg-moment-checks}
\end{equation}
Only after these checks pass should the same implementation be trusted for ensembles without a closed-form benchmark.
\end{examplebox}

The large-deviation and conditioned-density algorithms come with additional exact checks. For the Hermitian index problem,
\begin{equation}
\psi_x'(0)=\int_{-\infty}^{x}d\lambda\overline{\rho_{\pmb A}(\lambda)}\,,
\label{eq:acv-index-first-derivative-check}
\end{equation}
and
\begin{equation}
\psi_x''(0)=\lim_{N\to\infty}\frac{1}{N}{\rm Var}\mathcal{K}_{\pmb A}(x)\,.
\label{eq:acv-index-second-derivative-check}
\end{equation}
These identities can be checked by direct diagonalization on moderate sizes, using
\begin{equation}
\widehat{\psi}_{x,N}(s)=\frac{1}{N}\log\left[\frac{1}{M}\sum_{m=1}^{M}e^{s\mathcal K_m(x)}\right]\,,
\label{eq:acv-index-sample-cgf}
\end{equation}
where $\mathcal K_m(x)$ is the index of the $m$th sample. Likewise, for conditioned densities one must check
\begin{equation}
\int_{-\infty}^{x}d\lambda\rho_x(\lambda|k)=k\,,
\label{eq:acv-conditioned-density-check}
\end{equation}
and
\begin{equation}
\int_{-\infty}^{\infty}d\lambda\rho_x(\lambda|k)=1\,.
\label{eq:acv-conditioned-density-normalization}
\end{equation}
For non-Hermitian number statistics in a domain $\mathcal D$,
\begin{equation}
\psi_{\mathcal D}'(0)=\int_{\mathcal D}d^2z\,\overline{\rho_{\pmb A}(z)}
\label{eq:acv-nh-number-mean-check}
\end{equation}
which again provides a direct consistency test between the ordinary density and the number-statistics formalism.

The role of direct diagonalization in these checks should be emphasized. For sparse Hermitian matrices of the sizes typically used in validation, full dense diagonalization is often unnecessary; one only needs the spectrum or the spectral measure on a grid. For the lecture-note level discussion, however, the conceptual point is simpler than the implementation detail: whatever exact or numerically exact spectral data are available for finite instances should be compared with the cavity predictions after using the same smoothing and the same observable. Comparing a raw histogram with a finite-$\epsilon$ cavity density is not meaningful. Comparing a complex-plane density to a real-axis marginal is not meaningful. One must compare like with like.

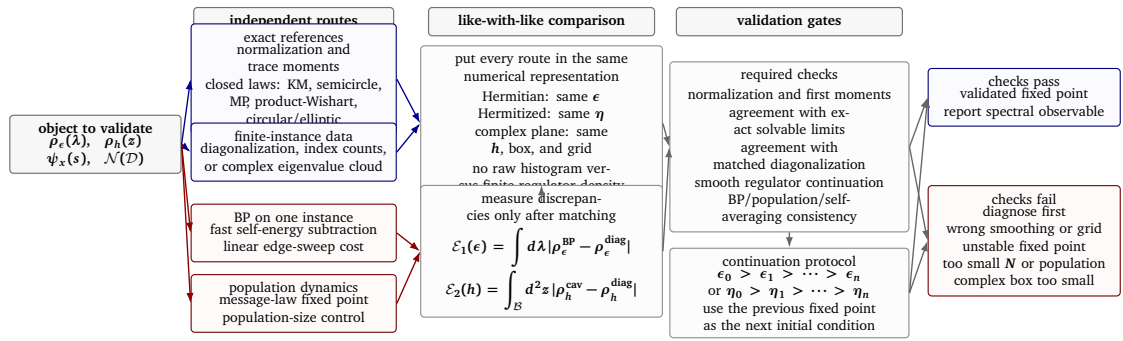
\begin{figure}[t]
\centering
\resizebox{0.98\textwidth}{!}{%
\begin{tikzpicture}[
    x=1cm,
    y=1cm,
    >=Latex,
    head/.style={draw=black!55, fill=black!3, rounded corners=2pt, line width=0.65pt, inner sep=4pt, font=\scriptsize\bfseries, align=center},
    ref/.style={draw=blue!45!black, fill=blue!2, rounded corners=2pt, line width=0.60pt, inner sep=4pt, font=\scriptsize, align=center},
    cav/.style={draw=red!45!black, fill=red!2, rounded corners=2pt, line width=0.60pt, inner sep=4pt, font=\scriptsize, align=center},
    gate/.style={draw=black!45, fill=black!1, rounded corners=2pt, line width=0.60pt, inner sep=4pt, font=\scriptsize, align=center},
    pass/.style={draw=blue!45!black, fill=blue!2, rounded corners=2pt, line width=0.60pt, inner sep=4pt, font=\scriptsize, align=center},
    fail/.style={draw=red!45!black, fill=red!2, rounded corners=2pt, line width=0.60pt, inner sep=4pt, font=\scriptsize, align=center},
    arrow/.style={draw=black!60, line width=0.75pt, -{Latex[length=2.0mm,width=1.4mm]}},
    bluearrow/.style={draw=blue!55!black, line width=0.75pt, -{Latex[length=2.0mm,width=1.4mm]}},
    redarrow/.style={draw=red!55!black, line width=0.75pt, -{Latex[length=2.0mm,width=1.4mm]}}
]
\node[head, text width=3.00cm] (target) at (1.75,4.90)
{object to validate\\[-1mm]
$\rho_\epsilon(\lambda)$,\quad $\rho_h(z)$\\
$\psi_x(s)$,\quad $\mathcal N(\mathcal D)$};

\node[head, text width=3.55cm] (routeHead) at (5.55,7.25)
{independent routes};

\node[ref, text width=3.65cm] (exact) at (5.55,6.15)
{exact references\\[-1mm]
normalization and trace moments\\
closed laws: KM, semicircle,\\
MP, product-Wishart,\\
circular/elliptic};

\node[ref, text width=3.65cm] (diag) at (5.55,4.75)
{finite-instance data\\[-1mm]
diagonalization, index counts,\\
or complex eigenvalue cloud};

\node[cav, text width=3.65cm] (bp) at (5.55,3.20)
{BP on one instance\\[-1mm]
fast self-energy subtraction\\
linear edge-sweep cost};

\node[cav, text width=3.65cm] (pd) at (5.55,1.85)
{population dynamics\\[-1mm]
message-law fixed point\\
population-size control};

\node[head, text width=4.05cm] (matchHead) at (10.30,7.25)
{like-with-like comparison};

\node[gate, text width=4.35cm] (match) at (10.30,5.30)
{put every route in the same numerical representation\\[0.5mm]
Hermitian: same $\epsilon$\\
Hermitized: same $\eta$\\
complex plane: same $h$, box, and grid\\[0.5mm]
no raw histogram versus finite-regulator density};

\node[gate, text width=4.35cm] (error) at (10.30,2.85)
{measure discrepancies only after matching\\[0.5mm]
$\displaystyle
\mathcal E_1(\epsilon)=\int d\lambda\,
|\rho_{\epsilon}^{\rm BP}-\rho_{\epsilon}^{\rm diag}|$\\[0.5mm]
$\displaystyle
\mathcal E_2(h)=\int_{\mathcal B}d^2z\,
|\rho_{h}^{\rm cav}-\rho_{h}^{\rm diag}|$};

\node[head, text width=4.05cm] (gateHead) at (15.05,7.25)
{validation gates};

\node[gate, text width=4.30cm] (checks) at (15.05,4.85)
{required checks\\[0.5mm]
normalization and first moments\\
agreement with exact solvable limits\\
agreement with matched diagonalization\\
smooth regulator continuation\\
BP/population/self-averaging consistency};

\node[gate, text width=4.30cm] (cont) at (15.05,2.05)
{continuation protocol\\[-1mm]
$\epsilon_0>\epsilon_1>\cdots>\epsilon_n$\\
or $\eta_0>\eta_1>\cdots>\eta_n$\\
use the previous fixed point as the next initial condition};

\node[pass, text width=3.40cm] (ok) at (19.55,5.80)
{checks pass\\[-1mm]
validated fixed point\\
report spectral observable};

\node[fail, text width=3.40cm] (bad) at (19.55,3.05)
{checks fail\\[-1mm]
diagnose first\\
wrong smoothing or grid\\
unstable fixed point\\
too small $N$ or population\\
complex box too small};

\draw[bluearrow] (target.east) -- (exact.west);
\draw[bluearrow] (target.east) -- (diag.west);
\draw[redarrow] (target.east) -- (bp.west);
\draw[redarrow] (target.east) -- (pd.west);

\draw[bluearrow] (exact.east) -- (match.west);
\draw[bluearrow] (diag.east) -- (match.west);
\draw[redarrow] (bp.east) -- (error.west);
\draw[redarrow] (pd.east) -- (error.west);

\draw[arrow] (match.south) -- (error.north);
\draw[arrow] (match.east) -- (checks.west);
\draw[arrow] (error.east) -- (checks.west);
\draw[arrow] (checks.south) -- (cont.north);
\draw[arrow] (checks.east) -- (ok.west);
\draw[arrow] (checks.east) -- (bad.west);
\draw[arrow] (cont.east) -- (ok.west);
\draw[arrow] (cont.east) -- (bad.west);
\end{tikzpicture}%
}
\caption{Validation workflow for numerical cavity calculations. A cavity fixed point is trusted only after it is compared with independent information expressed as the same smoothed observable at the same regulator or grid resolution. The validation gates combine normalization and moment checks, exactly solvable limits, matched direct diagonalization, regulator continuation, and consistency between single-instance BP and population dynamics. Failed checks are diagnostic information: they should be traced to smoothing, grids, instability, finite-size effects, population size, or an insufficient computational box before any spectral conclusion is drawn.}
\label{fig:acv-validation-workflow}
\end{figure}

Finally, it is useful to record the most common failure modes. If the cavity sweep converges but the density is not normalized, the error is usually a discretization or implementation mistake in the measurement stage, not in the fixed point itself. If the sweep oscillates or diverges as $\epsilon\downarrow0$, the fixed point may be unstable and damping or continuation is required. If population dynamics and belief propagation disagree while direct diagonalization agrees with one of them, the disagreement usually traces back to insufficient population size, insufficiently large single-instance size, or failure of self-averaging in the chosen regime. If a Wishart density shows substantial weight for $\lambda<0$ at very small $\epsilon$, one should first check whether that weight is merely the Lorentzian broadening of a large atom at zero before concluding that the cavity solver is wrong. If the non-Hermitian density does not integrate to one over the computational box, the box is either too small to contain the support or the numerical $\partial_{z^*}$ derivative is too coarse. Figure~\ref{fig:acv-validation-workflow} summarizes the validation logic advocated in this section: cavity fixed points should be compared with independent information only after the same observable, regulator, smoothing, and computational domain have been fixed.

The practical moral of this section is simple. The cavity equations are not only formal fixed-point relations; they are the basis of efficient numerical solvers whose cost scales linearly with the number of edges per sweep for sparse instances and linearly with the population size per sweep for ensemble calculations. The necessary algebraic reductions are the subtraction identities that convert cavity sums into full sums plus one correction term. The necessary validation steps are normalization, moment checks, exact solvable limits, comparison with direct diagonalization at fixed smoothing, and regulator continuation. Without these checks one has only a fixed point of an iterative map. With them, one has numerical evidence that the fixed point is the spectral object predicted by the theory.

\begin{exerciseblock}
\exitem[Fast update for sparse symmetric matrices]
Starting from
\begin{equation}
G_{i\to j}(z)=\frac{1}{z-D_i-\displaystyle\sum_{\ell\in\partial i\setminus j}J_{i\ell}^2G_{\ell\to i}(z)}\,,
\label{eq:acv-ex-fast-start}
\end{equation}
define
\begin{equation}
\Sigma_i(z)=\sum_{\ell\in\partial i}J_{i\ell}^2G_{\ell\to i}(z)\,.
\label{eq:acv-ex-fast-sigma}
\end{equation}
Derive
\begin{equation}
G_{i\to j}(z)=\frac{1}{z-D_i-\Sigma_i(z)+J_{ij}^2G_{j\to i}(z)}\,.
\label{eq:acv-ex-fast-update}
\end{equation}
Explain why this reduces the computational cost of one sweep.

\exitem[Cost comparison on a star graph]
For a star graph with $K$ leaves, compute the number of additions required to update all messages from the central vertex using the naive formula and using the fast self-energy formula. Verify the $O(K^2)$ versus $O(K)$ scaling.

\exitem[Damped fixed-point iteration]
Let
\begin{equation}
G^{(t+1)}=(1-\gamma)G^{(t)}+\gamma\mathcal F(G^{(t)})\,.
\label{eq:acv-ex-damped-map}
\end{equation}
Linearize around a fixed point $G^\star=\mathcal F(G^\star)$ and derive the effective multiplier
\begin{equation}
\Lambda_\gamma=1-\gamma+\gamma\mathcal F'(G^\star)\,.
\label{eq:acv-ex-damped-multiplier}
\end{equation}
Discuss how damping can reduce oscillations.

\exitem[Jacobian of the Gaussian BP map]
Starting from the sparse symmetric cavity recursion, derive the linearized perturbation equation
\begin{equation}
\delta G_{i\to j}^{(t+1)}=\sum_{\ell\in\partial i\setminus j}J_{i\ell}^2\left(G_{i\to j}^{\star}\right)^2\delta G_{\ell\to i}^{(t)}\,.
\label{eq:acv-ex-linearized-BP}
\end{equation}
What does this tell you about convergence near spectral edges?

\exitem[Population-dynamics estimator]
Starting from the distributional equation for $\mathcal P_{\rm cav}(G)$, derive the empirical population update
\begin{equation}
G_{\rm new}=\frac{1}{z-D-\displaystyle\sum_{r=1}^{\ell}J_r^2G^{(a_r)}}\,.
\label{eq:acv-ex-pop-update}
\end{equation}
Then derive the site estimator
\begin{equation}
\rho_\epsilon(\lambda)\approx\frac{1}{\pi n_{\rm obs}}\sum_{m=1}^{n_{\rm obs}}{\rm Im}G_{{\rm site},m}\,.
\label{eq:acv-ex-pop-estimator}
\end{equation}

\exitem[Wishart fast update]
Starting from the diluted Wishart factor update, define
\begin{equation}
T_\mu(z)=\frac{1}{d}\sum_{j\in\partial\mu}(X_j^\mu)^2G_{j\to\mu}(z)\,.
\label{eq:acv-ex-wishart-T}
\end{equation}
Show that
\begin{equation}
U_{\mu\to i}(z)=\frac{(X_i^\mu)^2}{1-T_\mu(z)+(X_i^\mu)^2G_{i\to\mu}(z)/d}\,.
\label{eq:acv-ex-wishart-fast-factor}
\end{equation}

\exitem[Wishart variable fast update]
Define
\begin{equation}
S_i(z)=\frac{1}{d}\sum_{\nu\in\partial i}U_{\nu\to i}(z)\,.
\label{eq:acv-ex-wishart-S}
\end{equation}
Derive
\begin{equation}
G_{i\to\mu}(z)=\frac{1}{z-S_i(z)+\frac{1}{d}U_{\mu\to i}(z)}\,,\qquad G_i(z)=\frac{1}{z-S_i(z)}\,.
\label{eq:acv-ex-wishart-fast-variable}
\end{equation}

\exitem[Generalized Wishart fast sums]
For the generalized diluted Wishart ensemble, define the full factor sums
\begin{equation}
T_{xx,\mu}=\sum_{j\in\partial\mu}(x_j^\mu)^2G_{j\to\mu}\,,\qquad T_{xy,\mu}=\sum_{j\in\partial\mu}x_j^\mu y_j^\mu G_{j\to\mu}\,,\qquad T_{yy,\mu} =\sum_{j\in\partial\mu}(y_j^\mu)^2G_{j\to\mu}\,.
\label{eq:acv-ex-gdw-full-sums}
\end{equation}
Derive the corresponding subtracted cavity sums $S_{xx}^{\mu\to i}$, $S_{xy}^{\mu\to i}$, and $S_{yy}^{\mu\to i}$. Then define $S_i(z)=\sum_{\nu\in\partial i}U_{\nu\to i}(z)$ and derive
\begin{equation}
G_{i\to\mu}(z)=\frac{1}{z-S_i(z)+U_{\mu\to i}(z)}\,,\qquad
G_i(z)=\frac{1}{z-S_i(z)}\,.
\end{equation}
Explain why the cost remains linear in the number of occupied bipartite edges.

\exitem[Non-Hermitian fast update]
Starting from the matrix-valued cavity recursion, define
\begin{equation}
\pmb{\Sigma}_i(z,\eta)=\sum_{\ell\in\partial i}\pmb{\mathcal A}_{i\ell}\pmb{G}_{\ell\to i}\pmb{\mathcal A}_{i\ell}^{\dagger}\,.
\label{eq:acv-ex-nh-sigma}
\end{equation}
Derive
\begin{equation}
\pmb{G}_{i\to j}=\left[\pmb{Z}_i-\pmb{\Sigma}_i+\pmb{\mathcal A}_{ij}\pmb{G}_{j\to i}\pmb{\mathcal A}_{ij}^{\dagger}\right]^{-1}\,.
\label{eq:acv-ex-nh-fast-update}
\end{equation}

\exitem[Wirtinger derivative by finite differences]
Starting from
\begin{equation}
\partial_{z^*}=\frac{1}{2}\left(\partial_x+i\partial_y\right)\,,
    \label{eq:acv-ex-wirtinger}
\end{equation}
derive the centered finite-difference approximation
\begin{equation}
\partial_{z^*}g(x,y)\approx\frac{g(x+h,y)-g(x-h,y)}{4h}+\frac{i[g(x,y+h)-g(x,y-h)]}{4h}\,.
\label{eq:acv-ex-finite-difference}
\end{equation}

\exitem[Normalization check]
Explain why every Hermitian spectral density must satisfy
\begin{equation}
\int d\lambda\rho(\lambda)=1\,.
\label{eq:acv-ex-normalization}
\end{equation}
In a numerical calculation on a finite grid, list at least three reasons why the numerical integral might deviate from one.

\exitem[First and second moment checks]
For
\begin{equation}
A_{ij}=D_i\delta_{ij}+C_{ij}J_{ij}\,,
\label{eq:acv-ex-symmetric-model}
\end{equation}
derive
\begin{equation}
\frac{1}{N}{\rm Tr}\pmb A=\frac{1}{N}\sum_iD_i
\label{eq:acv-ex-first-moment}
\end{equation}
and
\begin{equation}
\frac{1}{N}{\rm Tr}\pmb A^2=\frac{1}{N}\sum_iD_i^2+\frac{2}{N}\sum_{\{i,j\}\in E}J_{ij}^2\,.
\label{eq:acv-ex-second-moment}
\end{equation}

\exitem[Wishart first moment]
For
\begin{equation}
\pmb W=\frac{1}{d}\pmb X\pmb X^{\rm T}\,,
\label{eq:acv-ex-Wishart-W}
\end{equation}
show that
\begin{equation}
\frac{1}{N}{\rm Tr}\pmb W=\frac{1}{Nd}\sum_{i,\mu}(X_i^\mu)^2\,.
\label{eq:acv-ex-Wishart-first-moment}
\end{equation}
Derive the ensemble limit $\langle \xi^2\rangle/\alpha$ in the Poisson diluted ensemble.

\exitem[Generalized Wishart first moment]
For
\begin{equation}
\pmb F=\frac{1}{2d}\left(\pmb X\pmb Y^{\rm T}+\pmb Y\pmb X^{\rm T}\right)\,,
\label{eq:acv-ex-gdw-F}
\end{equation}
derive
\begin{equation}
\frac{1}{N}{\rm Tr} \pmb F\longrightarrow\frac{m_{11}}{\alpha}\,.
\label{eq:acv-ex-gdw-first-moment}
\end{equation}

\exitem[Exact benchmark laws]
List the exact benchmark laws discussed in the section and state which numerical solver each one validates: Kesten--McKay, Wigner semicircle, Mar\v{c}enko--Pastur, dense product-Wishart laws, and circular or elliptic laws.

\exitem[Comparison at fixed smoothing]
Explain why a cavity density computed at regulator $\epsilon$ should be compared with
\begin{equation}
\rho_{\epsilon,N}^{\rm diag}(\lambda)=\frac{1}{\pi N}\sum_{i=1}^{N}\frac{\epsilon}{(\lambda-\lambda_i)^2+\epsilon^2}
\label{eq:acv-ex-fixed-smoothing}
\end{equation}
rather than with an unsmoothed histogram.

\exitem[Order of limits]
Discuss the order of limits
\begin{equation}
N\to\infty\quad\text{at fixed }\epsilon>0,\qquad\epsilon\downarrow0\quad\text{afterwards}\,.
\label{eq:acv-ex-order-limits}
\end{equation}
Give an example of what can go wrong if $\epsilon$ is made too small at fixed $N$.

\exitem[Regulator continuation]
Design a regulator schedule
\begin{equation}
\epsilon_0>\epsilon_1>\cdots>\epsilon_n
\label{eq:acv-ex-regulator-schedule}
\end{equation}
and explain how the fixed point at $\epsilon_m$ should be used to initialize the iteration at $\epsilon_{m+1}$.

\exitem[Self-averaging test]
For an ensemble of sparse graphs, define a procedure to compare
\begin{equation}
\rho_{\epsilon,N}^{\rm BP}(\lambda)\,,\qquad\overline{\rho_{\epsilon,N}^{\rm diag}(\lambda)}\,,\qquad\rho_\epsilon^{\rm pop}(\lambda)\,.
\label{eq:acv-ex-self-averaging-comparison}
\end{equation}
What would constitute evidence of self-averaging?

\exitem[Large-deviation derivative check]
For an index observable $\mathcal K_{\pmb A}(x)$, and with a fixed convention for eigenvalues exactly at the threshold $x$, verify formally that
\begin{equation}
\psi_x'(0)=\int_{-\infty}^{x}d\lambda \overline{\rho_{\pmb A}(\lambda)}\,.
\label{eq:acv-ex-index-derivative}
\end{equation}
Explain how this identity validates a tilted population-dynamics calculation.

\exitem[Conditioned-density check]
For a conditioned density $\rho_x(\lambda|k)$, explain why one must verify both
\begin{equation}
\int d\lambda \rho_x(\lambda|k)=1
\label{eq:acv-ex-conditioned-normalization}
\end{equation}
and
\begin{equation}
\int_{-\infty}^{x}d\lambda \rho_x(\lambda|k)=k\,.
\label{eq:acv-ex-conditioned-index}
\end{equation}

\exitem[Non-Hermitian number check]
For a non-Hermitian domain $\mathcal D$, show that the first derivative of the number-statistics generating function satisfies
\begin{equation}
\psi_{\mathcal D}'(0)=\int_{\mathcal D}d^2z \overline{\rho_{\pmb A}(z)}\,.
\label{eq:acv-ex-nh-number-check}
\end{equation}

\exitem[Programming exercise: validation workflow for random regular graphs]
Fix an integer $c\geq2$, a list of system sizes $N$ such that $cN$ is even, a number $S$ of independent graph samples for each $N$, a regulator $\epsilon>0$, and a grid of real values of $\lambda$. Generate random $c$-regular graphs, compute the broadened diagonalization density, solve the scalar cavity equation, and compare both with the Kesten--McKay law broadened at the same regulator. Check normalization, first moment, and second moment using the same grid and density-estimation convention. Report $c$, $N$, $S$, $\epsilon$, the grid spacing, and a discrepancy measure.

\exitem[Programming exercise: regulator dependence]
Fix a sparse graph realization, its edge weights and diagonal terms, a grid of real values of $\lambda$, a damping parameter $\gamma$, a convergence tolerance, and a decreasing regulator schedule
\[
\epsilon_0>\epsilon_1>\cdots>\epsilon_n>0\,.
\]
For each $\epsilon_m$, compute the BP density, using the fixed point at $\epsilon_{m-1}$ as the initialization for $\epsilon_m$ when $m\geq1$. Track the number of sweeps to convergence, the smoothness of the density, the final value of the convergence criterion, and preservation of the sign condition ${\rm Im}\,G_i>0$. Report the graph ensemble or fixed graph, $N$, the regulator schedule, $\gamma$, the tolerance, and any initialization dependence.

\exitem[Programming exercise: Wishart validation]
Fix a list of system sizes $N$, an aspect ratio $\alpha>0$ such that $P=N/\alpha$ is an integer for each $N$, a dilution $d>0$, a number $S$ of independent samples for each parameter set, a nonzero-weight distribution $p_\xi$ with zero mean and unit variance, a regulator $\epsilon>0$, and a grid of real values of $\lambda$. Generate diluted Wishart matrices from sparse rectangular matrices
\begin{equation}
X_i^\mu=B_i^\mu\xi_i^\mu\,,\qquad{\rm Prob}(B_i^\mu=1)=\frac{d}{N}\,,\qquad{\rm Prob}(B_i^\mu=0)=1-\frac{d}{N}\,,
\label{eq:acv-ex-program-wishart-support}
\end{equation}
with independent nonzero weights drawn from $p_\xi$, and form
\begin{equation}
\pmb W=\frac{1}{d}\pmb X\pmb X^{\rm T}\,.
\label{eq:acv-ex-program-wishart-W}
\end{equation}
Compare direct diagonalization with the two-population cavity algorithm using the Wishart convention of this section, where $U$ has numerator $\xi^2$ and the variable update contains the factor $1/d$. Use the same $\epsilon$ and $\lambda$ grid in both estimates. Check positivity up to Lorentzian broadening, normalization, first moment $\langle\xi^2\rangle/\alpha$, and the fraction of zero modes using a stated numerical tolerance. Report $N$, $P$, $\alpha$, $d$, $S$, $p_\xi$, $\epsilon$, the grid spacing, the zero-mode tolerance, and a discrepancy measure.

\exitem[Programming exercise: non-Hermitian density validation]
Fix a list of matrix sizes $N$, a mean degree $c=O(1)$, a number $S$ of independent samples for each $N$, a Hermitization regulator $\eta>0$, a rectangular grid in the complex plane, and a smoothing convention for the empirical eigenvalue density. Generate sparse directed matrices $\pmb A$ with zero diagonal entries and independent off-diagonal entries
\begin{equation}
{\rm Prob}(A_{ij}=1)=\frac{c}{N}\,,\qquad{\rm Prob}(A_{ij}=0)=1-\frac{c}{N}\,,\qquad i\neq j\,.
    \label{eq:acv-ex-program-nh-er}
\end{equation}
Compute the eigenvalue cloud and compare a smoothed empirical complex-plane density with the Hermitized cavity estimate on the same grid. Use the same computational box and smoothing scale when comparing the two densities, and record any eigenvalues outside the box. Vary the grid spacing and the Hermitization regulator $\eta$, and check whether the density integrates to one over the chosen computational box. Report $N$, $c$, $S$, $\eta$, the grid spacings, the box, the smoothing convention, and an integral or $L^1$ discrepancy.

\exitem[Programming exercise: failure-mode diagnosis]
Construct a numerical experiment in which a cavity iteration appears to converge but the resulting density fails a validation check. For example, use too small a population size in population dynamics, too small a regulator in BP, or a grid that does not cover the full non-Hermitian spectral support. For the chosen failure mode, report the ensemble, matrix size or population size, regulator, convergence tolerance, grid or computational box, and the validation check that fails. Then repeat the computation after changing one diagnostic parameter, such as increasing the population size, increasing the computational box, increasing the regulator, or refining the grid. Identify whether the original error came from the fixed point, the measurement stage, the integration range, insufficient population size, finite-size effects, or the smoothing procedure.
\end{exerciseblock}

\section{Applications and physical interpretation}
\label{sec:applications-physical-interpretation}
At this stage it is useful to pause and ask what the spectral observables computed in the previous sections actually mean in physical terms. The answer is simple but important: a random matrix is rarely interesting by itself. It becomes interesting when it represents a linear operator acting on a system. The spectrum then tells us how perturbations propagate, how correlations are organized, how fast relaxation takes place, how many unstable directions are present, and how rare atypical samples are. The purpose of this section is to make these connections explicit.

The formulas below deliberately recall objects introduced earlier, but they are not rederived. Each paragraph identifies the matrix, the spectral observable, and the physical meaning carried by that observable.

The most basic interpretation comes from linear evolution. Suppose that a vector of amplitudes $\pmb u(t)\in\mathbb{C}^N$ evolves according to
\begin{equation}
\frac{d\pmb u}{dt}=\pmb M\pmb u\,.
\label{eq:api-linear-evolution}
\end{equation}
If $\pmb M$ is diagonalizable, with right eigenvectors $\pmb r_\alpha$ and eigenvalues $\zeta_\alpha$, then
\begin{equation}
\pmb u(t)=\sum_{\alpha=1}^{N}c_\alpha e^{\zeta_\alpha t}\pmb r_\alpha\,.
\label{eq:api-mode-expansion}
\end{equation}
Thus the spectrum of $\pmb M$ is the spectrum of growth rates and frequencies: the real parts ${\rm Re} \zeta_\alpha$ determine exponential growth or decay, while the imaginary parts ${\rm Im} \zeta_\alpha$ determine oscillations. This elementary observation already explains why the spectral density matters. In a Hermitian problem it gives the density of relaxation rates or excitation energies. In a non-Hermitian problem it gives the density of complex growth exponents. The resolvent appears when one studies forced response. If
\begin{equation}
\frac{d\pmb u}{dt}=-\mu \pmb u+\pmb A\pmb u+\pmb h\,e^{-i\omega t}\,,
\label{eq:api-forced-dynamics}
\end{equation}
and one looks for a stationary oscillatory response $\pmb u(t)=\pmb v\,e^{-i\omega t}$, then
\begin{equation}
\left[(\mu-i\omega)\pmb I-\pmb A\right]\pmb v=\pmb h\,,
\label{eq:api-forced-response-equation}
\end{equation}
so that
\begin{equation}
\pmb v=\left[(\mu-i\omega)\pmb I-\pmb A\right]^{-1}\pmb h\,.
\label{eq:api-forced-response-resolvent}
\end{equation}
The resolvent is therefore a response function. This is the first general interpretation to keep in mind: the spectral methods developed in these notes are also methods for computing the linear response of sparse disordered systems.

For undirected graphs, the adjacency spectrum has an immediate combinatorial meaning. Let $\pmb C$ be the adjacency matrix of a simple graph. Its empirical spectral density is
\begin{equation}
\rho_{\pmb C}(\lambda)=\frac{1}{N}\sum_{i=1}^{N}\delta(\lambda-\lambda_i)\,.
\label{eq:api-adjacency-density}
\end{equation}
The $n$th moment is
\begin{equation}
\mu_n=\int d\lambda \lambda^n\rho_{\pmb C}(\lambda)=\frac{1}{N}{\rm Tr}\pmb C^n\,.
\label{eq:api-adjacency-moments}
\end{equation}
Now
\begin{equation}
(\pmb C^n)_{ii}=\sum_{i_1,\ldots,i_{n-1}}C_{ii_1}C_{i_1i_2}\cdots C_{i_{n-1}i}\,,
\label{eq:api-closed-walk-count}
\end{equation}
and since $C_{ab}\in\{0,1\}$ for an unweighted graph, each term in \eqref{eq:api-closed-walk-count} is equal to one if and only if the sequence
\begin{equation}
i\to i_1\to i_2\to\cdots\to i_{n-1}\to i
\label{eq:api-closed-walk}
\end{equation}
is a closed walk of length $n$. Therefore ${\rm Tr}\pmb C^n$ counts closed walks of length $n$ in the graph. The spectral density is thus not merely an abstract measure on the eigenvalues; it is a compact encoding of walk statistics, loops, and local topology \cite{FarkasDerenyiBarabasiVicsek2001,Newman2010}. This explains physically why degree heterogeneity, assortative mixing, and community structure change the spectral density: they change the statistics of the local walks seen by a typical vertex.

The graph Laplacian has an equally direct dynamical interpretation. For
\begin{equation}
L_{ij}=k_i\delta_{ij}-C_{ij}\,,\qquad k_i=\sum_{j=1}^{N}C_{ij}\,,
\label{eq:api-laplacian}
\end{equation}
consider the diffusion equation
\begin{equation}
\frac{d\pmb p}{dt}=-\pmb L\pmb p\,.
\label{eq:api-diffusion}
\end{equation}
Its solution is
\begin{equation}
\pmb p(t)=e^{-t\pmb L}\pmb p(0).
\label{eq:api-diffusion-solution}
\end{equation}
A useful global observable is the average return probability
\begin{equation}
P_{\rm ret}(t)=\frac{1}{N}{\rm Tr} e^{-t\pmb L}.
\label{eq:api-return-probability}
\end{equation}
Using the spectral decomposition of $\pmb L$,
\begin{equation}
P_{\rm ret}(t)=\frac{1}{N}\sum_{i=1}^{N}e^{-t\lambda_i}=\int_{0}^{\infty}d\lambda e^{-t\lambda}\rho_{\pmb L}(\lambda)\,.
\label{eq:api-return-probability-density}
\end{equation}
Thus the small-$\lambda$ edge of the Laplacian density controls long relaxation times. Exact zero modes correspond to disconnected components or conservation laws. Very small but nonzero Laplacian eigenvalues correspond to slow inter-community exchange or bottlenecks in the network. Likewise, isolated eigenvalues of adjacency or modularity-type matrices are often signatures of mesoscale organization such as block structure or communities \cite{NadakuditiNewman2012,ZhangNadakuditiNewman2014,Peixoto2013}. The graph applications of spectral observables are therefore not restricted to combinatorics: Laplacian spectra directly govern diffusion and relaxation, while adjacency or modularity spectra help diagnose detectability.

A second class of applications concerns disordered transport and localization. Consider the Anderson-type operator
\begin{equation}
H_{ij}=C_{ij}+V_i\delta_{ij}\,,
\label{eq:api-anderson-operator}
\end{equation}
where $\pmb C$ is the adjacency matrix of a sparse graph and the on-site potentials $V_i$ are random \cite{Anderson1958}. The local density of states is
\begin{equation}
\rho_j(\lambda)=\sum_{i=1}^{N}|u_j^{(i)}|^2\delta(\lambda-\lambda_i)=\frac{1}{\pi}\lim_{\epsilon\downarrow0}{\rm Im}G_{jj}(\lambda-i\epsilon),
\label{eq:api-local-density-of-states}
\end{equation}
where $u_j^{(i)}$ is the $j$th component of the eigenvector associated with $\lambda_i$. This formula makes the physical meaning of the local Green function transparent: it measures how strongly a given energy is represented on a given site. If an eigenstate is extended, many sites contribute comparable weights and the distribution of $\rho_j(\lambda)$ over sites is relatively narrow. If an eigenstate is localized, the weight is concentrated on a few sites and the distribution of $\rho_j(\lambda)$ becomes broad \cite{AbouChacraAndersonThouless1973}. In this way the cavity distribution of local Green functions carries much more information than the average density alone.

The spectral-count formalism adds a further physical interpretation. Let $I=[a,b]$ be an interval and let
\begin{equation}
\mathcal{N}_H(I)=\sum_{i=1}^{N}\mathbf{1}_{\lambda_i\in I}
\label{eq:api-number-in-interval}
\end{equation}
be the number of eigenvalues in that interval. A natural normalized fluctuation measure is the level compressibility
\begin{equation}
\chi_I=\frac{\displaystyle\lim_{N\to\infty}\frac{1}{N}{\rm Var}\mathcal{N}_H(I)}{\displaystyle\lim_{N\to\infty}\frac{1}{N}\overline{\mathcal{N}_H(I)}}=\frac{\psi_I''(0)}{\psi_I'(0)}\,,
\label{eq:api-level-compressibility}
\end{equation}
provided $\psi_I'(0)\neq0$. For count statistics close to a Poisson process on the scale of $I$, one expects $\chi_I$ to be close to one, while in a more rigid spectrum it is smaller. On random regular graphs, the large-deviation formalism for interval counts therefore gives direct information about localization properties in the extended and localized regimes \cite{MetzPerezCastillo2017}. Here again the physical interpretation is immediate: number fluctuations measure spectral rigidity, and spectral rigidity is tied to wave-function structure.

Sparse covariance matrices have an equally clear meaning. Let the columns of the $N\times P$ matrix $\pmb X$ be sample vectors,
\begin{equation}
\pmb x^{\,\mu}=(X_1^\mu,\ldots,X_N^\mu)^{\rm T}\,, \qquad\mu=1,\ldots,P\,.
\label{eq:api-sample-vectors}
\end{equation}
The diluted Wishart matrix is
\begin{equation}
\pmb W=\frac{1}{d}\pmb X\pmb X^{\rm T}\,.
\label{eq:api-wishart-matrix}
\end{equation}
For any vector $\pmb v\in\mathbb{R}^N$,
\begin{equation}
\pmb v^{\rm T}\pmb W\pmb v=\frac{1}{d}\sum_{\mu=1}^{P}\left(\pmb v\cdot\pmb x^{\,\mu}\right)^2\geq0\,.
\label{eq:api-wishart-quadratic-form}
\end{equation}
This identity is the most direct interpretation of the eigenvalues. If $\pmb v$ is a normalized eigenvector of $\pmb W$ with eigenvalue $\lambda$, then
\begin{equation}
\lambda=\frac{1}{d}\sum_{\mu=1}^{P}\left(\pmb v\cdot\pmb x^{\,\mu}\right)^2\,.
\label{eq:api-wishart-eigenvalue-variance}
\end{equation}
Hence large eigenvalues correspond to directions of large empirical variance in the data, while small eigenvalues correspond to directions with little observed variance. This is the sparse finite-connectivity analogue of the usual principal-component interpretation of covariance spectra \cite{Wishart1928,MarchenkoPastur1967}. When the data matrix itself is sparse, however, the finite-connectivity corrections are not a nuisance but part of the physics: missing observations, sparse incidence structure, and local bipartite fluctuations produce zero modes, localized modes, and deviations from the dense Mar\v{c}enko--Pastur law \cite{NagaoTanaka2007,RogersTakedaPerezCastilloKuhn2008}.

The zero modes have a simple origin. Since
\begin{equation}
{\rm rank}\pmb W\leq{\rm rank}\pmb X\leq P\,,
\label{eq:api-wishart-rank-bound}
\end{equation}
one has at least
\begin{equation}
N-P=N\left(1-\frac{1}{\alpha}\right)
\label{eq:api-wishart-rank-deficiency}
\end{equation}
exact zero modes whenever $\alpha>1$. In the sparse setting, additional zero modes may come from isolated variables or finite disconnected bipartite components. This illustrates a general lesson of the lecture notes: finite-connectivity spectra retain local structural information that is completely washed out in the dense limit.

The generalized diluted Wishart ensemble gives a similarly explicit interpretation for cross-correlations. Recall that
\begin{equation}
\pmb F=\frac{1}{2d}\left(\pmb X\pmb Y^{\rm T}+\pmb Y\pmb X^{\rm T}\right)\,.
\label{eq:api-generalized-wishart}
\end{equation}
Here $\pmb x^{\,\mu}$ and $\pmb y^{\,\mu}$ denote the $\mu$th columns of $\pmb X$ and $\pmb Y$. For any vector $\pmb v$,
\begin{equation}
\pmb v^{\rm T}\pmb F\pmb v=\frac{1}{d}\sum_{\mu=1}^{P}\left(\pmb v\cdot\pmb x^{\,\mu}\right)\left(\pmb v\cdot\pmb y^{\,\mu}\right)\,.
\label{eq:api-generalized-wishart-quadratic-form}
\end{equation}
This equation is the correct physical interpretation of the generalized ensemble. If $\pmb v$ is a normalized eigenvector with eigenvalue $\lambda$, then
\begin{equation}
\lambda=\frac{1}{d}\sum_{\mu=1}^{P}\left(\pmb v\cdot\pmb x^{\,\mu}\right)\left(\pmb v\cdot\pmb y^{\,\mu}\right)\,.
\label{eq:api-generalized-wishart-eigenvalue}
\end{equation}
A positive eigenvalue means that the two projected data arrays have positive empirical cross-correlation along $\pmb v$, while a negative eigenvalue means that the projected arrays are anticorrelated along that direction. The first moment of the density follows from the trace:
\begin{equation}
\frac{1}{N}{\rm Tr}\pmb F=\frac{1}{dN}\sum_{i=1}^{N}\sum_{\mu=1}^{P}x_i^\mu y_i^\mu\longrightarrow\frac{m_{11}}{\alpha}\,,
\label{eq:api-generalized-wishart-trace}
\end{equation}
where $m_{11}$ is the local cross moment of the nonzero weights, and the arrow is understood under the self-averaging assumptions used in Section~\ref{sec:generalized-diluted-wishart}. Thus the generalized diluted Wishart ensemble is a natural spectral model for sparse cross-correlation problems, paired measurements, or multimodal data in which two sparse observation matrices live on the same bipartite support \cite{PerezCastillo2022Generalized}. When $\pmb X=\pmb Y$, \eqref{eq:api-generalized-wishart-quadratic-form} reduces to the positive quadratic form \eqref{eq:api-wishart-quadratic-form}, and one recovers the ordinary diluted Wishart ensemble.

The non-Hermitian case enters as soon as the underlying interactions are directed or asymmetric. Consider the linearized dynamics
\begin{equation}
\frac{d\pmb u}{dt}=-\mu\pmb u+\pmb A\pmb u\,,
\label{eq:api-nonhermitian-dynamics}
\end{equation}
where $\pmb A$ is a sparse non-Hermitian random matrix. If $\pmb A$ is diagonalizable, the contribution of an eigenmode with eigenvalue $\zeta_\alpha$ evolves as
\begin{equation}
e^{(\zeta_\alpha-\mu)t}\,.
\label{eq:api-nonhermitian-modes}
\end{equation}
A necessary spectral stability condition is therefore
\begin{equation}
\max_\alpha {\rm Re} \zeta_\alpha<\mu\,.
\label{eq:api-spectral-stability}
\end{equation}
In dense random-matrix models this idea is the basis of May's stability criterion for large complex systems \cite{May1972}. In sparse systems the same spectral criterion remains useful, but the spectral support is no longer a simple disk or ellipse. Degree fluctuations, reciprocal correlations, motifs, and local heterogeneity change the support, and the non-Hermitian cavity equations provide a way to compute the resulting stability boundary \cite{AllesinaTang2012,RogersPerezCastillo2009,MetzNeriRogers2019}. The imaginary parts ${\rm Im}\,\zeta_\alpha$ give the frequencies of oscillatory modes, so the two-dimensional spectral density is the density of decay rates and frequencies.

The resolvent interpretation \eqref{eq:api-forced-response-resolvent} makes this even more concrete. In the forced system
\begin{equation}
\frac{d\pmb u}{dt}=-\mu\pmb u+\pmb A\pmb u+\pmb h e^{-i\omega t}\,,
\label{eq:api-forced-nonhermitian}
\end{equation}
the response amplitude is
\begin{equation}
\pmb v(\omega)=\left[(\mu-i\omega)\pmb I-\pmb A\right]^{-1}\pmb h\,.
\label{eq:api-frequency-response}
\end{equation}
Thus the non-Hermitian resolvent at
\begin{equation}
z=\mu-i\omega
\label{eq:api-response-spectral-parameter}
\end{equation}
is the frequency-response operator. This gives a direct physical meaning to the complex spectral parameter used in the cavity equations. In ecology, the rightmost edge of the spectrum is associated with the onset of instability \cite{May1972,AllesinaTang2012}. In random neural-network models, the support of the spectrum sets the distribution of relaxation rates and oscillation frequencies, and reciprocal correlations deform the support in a way that changes the collective dynamics \cite{RajanAbbott2006}. Since sparse non-Hermitian matrices are generally non-normal, the full transient response also depends on eigenvector overlaps, not only on eigenvalues \cite{ChalkerMehlig1998}. Still, the spectral density is the first quantity one must know before moving on to those genuinely non-normal effects.

The number-statistics formalism has an equally transparent physical meaning in the non-Hermitian setting. For a finite matrix, the unstable modes are the eigenvalues in the half-plane ${\rm Re}\,z>\mu$. To keep the contour-counting formulation in the bounded-domain setting used earlier, one may take
\begin{equation}
\mathcal{D}_R=\left\{z\in\mathbb{C}:{\rm Re} z>\mu\,,\ |z|<R\right\}\,,
\label{eq:api-unstable-domain}
\end{equation}
with $R$ chosen large enough to contain the relevant spectrum. Then
\begin{equation}
\mathcal{N}_{\pmb A}(\mathcal{D}_R)=\sum_{i=1}^{N}\mathbf{1}_{\zeta_i\in\mathcal{D}_R}
\label{eq:api-unstable-count}
\end{equation}
counts unstable modes. More generally, if $\mathcal{D}$ is a disk centered at $i\omega_0$, then $\mathcal{N}_{\pmb A}(\mathcal{D})$ counts modes with frequency near $\omega_0$ and with growth rates in the chosen band. Thus the non-Hermitian number statistics derived earlier are not merely formal contour-counting tools. They count dynamically relevant subsets of modes.

The Hermitian index number also has a direct physical meaning. Suppose that $V(\pmb q)$ is a potential or energy function and $\pmb q_\star$ is a stationary point. Expanding around $\pmb q_\star$,
\begin{equation}
V(\pmb q_\star+\delta\pmb q)=V(\pmb q_\star)+\frac{1}{2}\delta\pmb q^{\rm T}\pmb H\delta\pmb q+O(\|\delta\pmb q\|^3)\,,
\label{eq:api-hessian-expansion}
\end{equation}
where $\pmb H$ is the Hessian at $\pmb q_\star$. The number of unstable directions of the stationary point is the number of negative eigenvalues of $\pmb H$, namely
\begin{equation}
\mathcal{K}_{\pmb H}(0)\,.
\label{eq:api-morse-index}
\end{equation}
This is the Morse index of the stationary point. Large deviations of the index therefore quantify the rarity of stationary points with an atypical fraction of unstable directions, and the conditioned spectral density describes how the full Hessian spectrum rearranges under such a constraint. Even when one does not have an explicit energy landscape in mind, the same interpretation applies to a symmetric stability operator once the sign convention has been fixed: the relevant index counts the unstable directions.

This interpretation clarifies why the conditioned density is a useful object. Knowing the typical density tells us what a typical sample looks like. Knowing the rate function of the index tells us how rare an atypical sample is. But if one wants to understand \emph{how} the atypical sample differs spectrally from the typical one, one needs the conditioned density
\begin{equation}
\rho_x(\lambda|k)\,,
\label{eq:api-conditioned-density-symbol}
\end{equation}
that is, the spectral density under the constraint that a fraction $k$ of eigenvalues lies below the threshold $x$. In invariant ensembles, this conditioning is often visualized as a rigid rearrangement of a Coulomb gas. In the sparse non-invariant ensembles of these notes, the mechanism is different: the conditioning changes the distribution of local graph environments and therefore the distribution of local cavity Green functions. This is the physical reason why conditioned spectra of sparse ensembles can differ qualitatively from their invariant counterparts \cite{PerezCastilloMetz2018Conditioned}.

The same comment applies to large deviations in the Wishart setting. For an ordinary diluted Wishart matrix, the quantity
\begin{equation}
\mathcal{K}_{\pmb W}(x)=\sum_{i=1}^{N}\Theta(x-\lambda_i)
    \label{eq:api-wishart-index}
\end{equation}
counts the number of empirical variance modes below the threshold $x$. In a data-analysis language, small eigenvalues correspond to poorly sampled directions or to directions with very weak variance. Large deviations of $\mathcal{K}_{\pmb W}(x)$ therefore describe rare samples with an atypically large or small number of such directions. In the sparse setting, this may happen because of rare bipartite neighborhoods, isolated variables, or atypical local concentrations of observations. The large-deviation theory is thus a way of quantifying rare but structurally meaningful sampling configurations.

Taken together, these examples explain the philosophy of the lecture notes. The spectral density tells us how modes are distributed. The local density of states tells us where those modes live. The non-Hermitian support tells us which modes grow, decay, or oscillate. The index tells us how many stable or unstable directions are present. The conditioned density tells us what an atypical sample must look like in order to realize a prescribed spectral count. In sparse and diluted systems, all these objects are sensitive to local graph structure, because finite connectivity prevents the graph from disappearing in the thermodynamic limit. This is why cavity and replica methods are so natural here: they compute spectral observables directly from the statistics of local environments rather than from an eigenvalue gas.

\section{Outlook: rate-functional theory and non-Hermitian diluted Wishart extensions}
\label{sec:outlook-rate-functional-non-hermitian-wishart}
The previous sections developed several concrete pieces of the finite-connectivity spectral theory: the typical spectral density, the large deviations of a scalar spectral count, the conditioned density under a scalar constraint, and the non-Hermitian density of sparse matrices. There are many possible directions beyond these notes. Rather than attempting a survey, we mention two natural continuations suggested by the methods developed here. The first possible direction is conceptual: instead of constraining one scalar observable, such as the number of eigenvalues below a threshold or inside a domain, one would like to describe large deviations of the \emph{whole} empirical spectral measure. The second possible direction is structural: one would like to merge the bipartite finite-connectivity geometry of diluted Wishart ensembles with the complex spectral geometry of non-Hermitian matrices. The purpose of this final technical section is not to present these directions as closed theories, but to explain how they arise from the material already developed in the notes.

We start with the first point. In the Hermitian sections we studied observables such as
\begin{equation}
\mathcal{K}_{\pmb A}(x)=\sum_{i=1}^{N}\Theta(x-\lambda_i)=N\int_{-\infty}^{x}d\lambda\rho_{\pmb A}(\lambda)\,,
\label{eq:outlook-index-as-linear-statistic}
\end{equation}
and in the non-Hermitian sections we studied
\begin{equation}
\mathcal{N}_{\pmb A}(\mathcal D)=N\int_{\mathcal D}d^2z\rho_{\pmb A}(z)\,,
\label{eq:outlook-domain-count-as-linear-statistic}
\end{equation}
where $\rho_{\pmb A}$ is the empirical spectral density. Both are examples of linear statistics of the empirical measure. This suggests introducing the general Hermitian linear statistic
\begin{equation}
L_f(\pmb A)=N\int_{-\infty}^{\infty}d\lambda f(\lambda)\rho_{\pmb A}(\lambda)=\sum_{i=1}^{N}f(\lambda_i),
\label{eq:outlook-hermitian-linear-statistic}
\end{equation}
for a real function $f$, and the corresponding functional cumulant-generating object
\begin{equation}
\Psi[f]=\lim_{N\to\infty}\frac{1}{N}\log\overline{e^{L_f(\pmb A)}}\,.
\label{eq:outlook-hermitian-functional-cgf}
\end{equation}
This is the natural generalization of the scalar cumulant-generating functions introduced earlier. Indeed, if
\begin{equation}
f(\lambda)=s \Theta(x-\lambda)\,,
\label{eq:outlook-step-test-function}
\end{equation}
then
\begin{equation}
L_f(\pmb A)=s\mathcal{K}_{\pmb A}(x)\,,
\label{eq:outlook-linear-statistic-index}
\end{equation}
and therefore
\begin{equation}
\Psi[f]=\psi_x(s)\,,
\label{eq:outlook-functional-cgf-reduces-index}
\end{equation}
the scalar cumulant-generating function of the index number. Thus the previously studied large deviations of the index are already contained in \eqref{eq:outlook-hermitian-functional-cgf}; they correspond to a very special choice of the field $f$.

The benefit of \eqref{eq:outlook-hermitian-functional-cgf} is that it keeps the full spectral density as the variable rather than collapsing everything to one number. To see this, differentiate \eqref{eq:outlook-hermitian-functional-cgf} formally with respect to the field $f$. Since
\begin{equation}
\frac{\delta L_f(\pmb A)}{\delta f(\lambda)}=N\rho_{\pmb A}(\lambda)\,,
\label{eq:outlook-functional-derivative-linear-statistic}
\end{equation}
we obtain
\begin{equation}
\frac{\delta \Psi[f]}{\delta f(\lambda)}=\lim_{N\to\infty}\frac{\overline{\rho_{\pmb A}(\lambda)e^{L_f(\pmb A)}}}{\overline{e^{L_f(\pmb A)}}}=\rho_f^\star(\lambda)\,,
    \label{eq:outlook-tilted-density-as-functional-derivative}
\end{equation}
provided the thermodynamic limit and the functional derivative commute and the tilted measure has a unique typical density. Equation \eqref{eq:outlook-tilted-density-as-functional-derivative} is the functional version of the relation between the scalar cumulant-generating function and the conditioned spectral density. It says that once $\Psi[f]$ is known, the spectral density in the ensemble tilted by the field $f$ is obtained by a functional derivative.

The Legendre structure becomes equally transparent. Suppose the empirical measure satisfies a large-deviation principle with speed $N$,
\begin{equation}
{\rm Prob}\left[\rho_{\pmb A}\approx \rho\right]\asymp\exp\left[-N\mathcal I[\rho]\right]\,.
\label{eq:outlook-rate-functional-ldp}
\end{equation}
Then the Laplace principle gives
\begin{equation}
\Psi[f]=\sup_{\rho}\left\{\int d\lambda f(\lambda)\rho(\lambda)-\mathcal I[\rho]\right\}\,,
\label{eq:outlook-laplace-principle}
\end{equation}
and, conversely,
\begin{equation}
\mathcal I[\rho]=\sup_f\left\{\int d\lambda f(\lambda)\rho(\lambda)-\Psi[f]\right\}\,.
\label{eq:outlook-rate-functional-legendre}
\end{equation}
This is the precise sense in which a rate-functional theory would generalize the scalar large-deviation functions derived earlier. If $f$ is a step function, \eqref{eq:outlook-laplace-principle} reduces to the scalar cumulant-generating function of the index. If $f$ is arbitrary, it biases the whole spectral density.

The scalar rate function is recovered by contraction. Let
\begin{equation}
k=\int_{-\infty}^{x}d\lambda \rho(\lambda)\,.
\label{eq:outlook-scalar-contraction-constraint}
\end{equation}
Then
\begin{equation}
\Phi_x(k)=\inf_{\rho:\int_{-\infty}^{x}d\lambda\rho(\lambda)= k}\mathcal I[\rho]\,.
\label{eq:outlook-contraction-index}
\end{equation}
This equation contains, in one line, the relation between the scalar rate function and the yet-to-be-constructed rate functional. It also explains the conditioned density. The density conditioned on the event that a fraction $k$ of eigenvalues lies below $x$ should be the minimizer of \eqref{eq:outlook-contraction-index}:
\begin{equation}
\rho_x^\star(\lambda|k)=\arg\min_{\rho}\left\{\mathcal I[\rho]:\int d\lambda\rho(\lambda)=1,\ \int_{-\infty}^{x}d\lambda\rho(\lambda)=k\right\}\,.
\label{eq:outlook-conditioned-density-minimizer}
\end{equation}
In the earlier sections we obtained the conditioned density by tilting with a scalar field $s$. Equation \eqref{eq:outlook-conditioned-density-minimizer} shows how that construction would sit inside a full rate-functional theory. The scalar field $s$ is simply the Lagrange multiplier enforcing the constraint \eqref{eq:outlook-scalar-contraction-constraint}.

An analogous construction can be written for the non-Hermitian case. Let
\begin{equation}
L_f(\pmb A)=N\int_{\mathbb C}d^2z\, f(z,z^*)\rho_{\pmb A}(z)=\sum_{i=1}^{N}f(z_i,z_i^*),
\label{eq:outlook-nonhermitian-linear-statistic}
\end{equation}
and define
\begin{equation}
\Psi[f]=\lim_{N\to\infty}\frac{1}{N}\log\overline{e^{L_f(\pmb A)}}\,.
\label{eq:outlook-nonhermitian-functional-cgf}
\end{equation}
If
\begin{equation}
f(z,z^*)=s \mathbf 1_{\mathcal D}(z)\,,
\label{eq:outlook-domain-test-function}
\end{equation}
then
\begin{equation}
L_f(\pmb A)=s \mathcal N_{\pmb A}(\mathcal D)\,,
\label{eq:outlook-number-statistic-special-case}
\end{equation}
and \eqref{eq:outlook-nonhermitian-functional-cgf} reduces to the scalar cumulant-generating function of the number of eigenvalues in the domain $\mathcal D$. If a non-Hermitian rate functional $\mathcal I[\rho]$ exists, then
\begin{equation}
\Phi_{\mathcal D}(k)=\inf_{\rho:\,\int_{\mathcal D}d^2z\,\rho(z)=k}\mathcal I[\rho]\,,
\label{eq:outlook-domain-contraction}
\end{equation}
and the conditioned density is the corresponding constrained minimizer. Thus, within such a functional formulation, the Hermitian index problem and the non-Hermitian number-statistics problem would be two scalar contractions of analogous functional large-deviation objects.

At this point one should ask which functional is really fundamental for sparse, non-invariant ensembles. For invariant ensembles the natural candidate is indeed $\mathcal I[\rho]$, because the eigenvalue density is already the main order parameter. For sparse matrices, however, the cavity construction suggests a different answer. On a fixed locally tree-like instance, the local spectral density is computed from cavity messages; at the ensemble level, their empirical law becomes the object from which the spectral density is obtained. This means that the primary rate functional is likely not a functional of $\rho$ itself, but a functional of the empirical distribution of messages.

To see this explicitly, consider sparse symmetric matrices. For a fixed spectral parameter $z=\lambda-i\epsilon$, the message on a directed edge is a complex number $G_{i\to j}(z)$. If we want the whole spectral density, not just its value at one $\lambda$, then the natural message is the entire profile
\begin{equation}
\mathfrak G_{i\to j}:\lambda\mapsto G_{i\to j}(\lambda-i0^+)\,.
\label{eq:outlook-message-profile}
\end{equation}
The empirical distribution of messages on a graph with $2|E|$ directed edges is then
\begin{equation}
\Pi_N[\mathfrak G]=\frac{1}{2|E|}\sum_{i=1}^{N}\sum_{j\in\partial i}\delta\left[\mathfrak G-\mathfrak G_{i\to j}\right]\,.
\label{eq:outlook-empirical-message-law}
\end{equation}
If a large-deviation principle of the form
\begin{equation}
{\rm Prob}\left[\Pi_N\approx \Pi\right]\asymp\exp\left[-N\mathcal J[\Pi]\right]
\label{eq:outlook-message-rate-functional}
\end{equation}
holds, then the spectral rate functional is obtained by contraction through the cavity map.

Let us write that map explicitly. Given a message law $\Pi$, the corresponding site density at spectral value $\lambda$ is
\begin{align}
\mathcal R[\Pi](\lambda)&=\frac{1}{\pi}\lim_{\epsilon\downarrow0}{\rm Im}\sum_{k=0}^{\infty}p_k\int dD p_D(D)\left[\prod_{r=1}^{k}d\Pi(\mathfrak G_r) dJ_r p_J(J_r)\right]\nonumber\\
&\hspace{1cm}\times\frac{1}{\lambda-i\epsilon-D-\displaystyle\sum_{r=1}^{k}J_r^2 \mathfrak G_r(\lambda)}\,.
\label{eq:outlook-density-from-message-law}
\end{align}
Then
\begin{equation}
\mathcal I[\rho]=\inf_{\Pi:\,\mathcal R[\Pi]=\rho}\mathcal J[\Pi]\,.
    \label{eq:outlook-density-contraction-from-messages}
\end{equation}
Equation \eqref{eq:outlook-density-contraction-from-messages} is the main conceptual point of the rate-functional outlook. In a sparse non-invariant ensemble, the full spectral large-deviation problem would naturally be formulated in terms of a rate functional on the cavity-message population, and the rate functional of the spectral density would be a contraction of that more microscopic object. The scalar rate functions studied earlier are then obtained by one more contraction, as in \eqref{eq:outlook-contraction-index} and \eqref{eq:outlook-domain-contraction}. In other words, the natural hierarchy of order parameters is
\begin{equation}
\text{message distribution}\longrightarrow\text{spectral density}\longrightarrow\text{scalar counts}\,.
    \label{eq:outlook-order-parameter-hierarchy}
\end{equation}
This is the hierarchy suggested by the cavity derivations carried out throughout the notes.

We now turn to the second possible direction, namely the non-Hermitian extension of diluted Wishart and cross-correlation ensembles. In the generalized diluted Wishart section we considered the symmetric matrix
\begin{equation}
\pmb F=\frac{1}{2d}\left(\pmb X\pmb Y^{\rm T}+\pmb Y\pmb X^{\rm T}\right)\,,
\label{eq:outlook-symmetric-cross-correlation}
\end{equation}
with sparse rectangular matrices $\pmb X$ and $\pmb Y$. One natural non-Hermitian extension is obtained by dropping the symmetrization and defining
\begin{equation}
\pmb M=\frac{1}{d}\pmb X\pmb Y^{\rm T},\qquad M_{ij}=\frac{1}{d}\sum_{\mu=1}^{P}x_i^\mu y_j^\mu\,.
    \label{eq:outlook-nonhermitian-diluted-wishart}
\end{equation}
We assume, as before, that
\begin{equation}
P=\frac{N}{\alpha}\,,\qquad P(x_i^\mu,y_i^\mu)=\left(1-\frac{d}{N}\right)\delta(x_i^\mu)\delta(y_i^\mu)+\frac{d}{N}\varrho(x_i^\mu,y_i^\mu)\,,
\label{eq:outlook-nonhermitian-entry-law}
\end{equation}
so that $\pmb X$ and $\pmb Y$ live on a common sparse bipartite graph. The symmetric matrix \eqref{eq:outlook-symmetric-cross-correlation} is simply
\begin{equation}
\pmb F=\frac{1}{2}\left(\pmb M+\pmb M^{\rm T}\right)\,.
\label{eq:outlook-symmetric-part}
\end{equation}
If $x_i^\mu=y_i^\mu$ on every occupied edge, then
\begin{equation}
\pmb M=\frac{1}{d}\pmb X\pmb X^{\rm T}
\label{eq:outlook-wishart-reduction}
\end{equation}
and the matrix becomes Hermitian positive semidefinite. Thus \eqref{eq:outlook-nonhermitian-diluted-wishart} really interpolates between the ordinary diluted Wishart ensemble and a genuinely non-Hermitian cross-correlation ensemble. This extension was analyzed in \cite{GuzmanGonzalezPerezCastillo2025}.

A first exact property is the rank bound
\begin{equation}
{\rm rank} \pmb M\leq\min\{{\rm rank} \pmb X,{\rm rank}\pmb Y\}\leq P\,.
\label{eq:outlook-nonhermitian-rank-bound}
\end{equation}
Hence, if $\alpha>1$, the matrix $\pmb M$ has at least
\begin{equation}
N-P=N\left(1-\frac{1}{\alpha}\right)
\label{eq:outlook-nonhermitian-zero-modes}
\end{equation}
zero eigenvalues. This is the same algebraic origin of zero modes as in the Hermitian Wishart case. The difference is that the nonzero eigenvalues are now generally complex; for real weights they occur in complex-conjugate pairs, while for genuinely complex weights no such conjugation symmetry is implied.

A useful derivation begins with a linearization. Define the $(N+P)\times(N+P)$ block matrix
\begin{equation}
\pmb{\mathcal L}=\frac{1}{\sqrt d}\begin{pmatrix}
\pmb 0_{N\times N} & \pmb X\\
\pmb Y^{\rm T} & \pmb 0_{P\times P}
\end{pmatrix}\,.
\label{eq:outlook-linearized-nonhermitian-wishart}
\end{equation}
Then
\begin{equation}
\pmb{\mathcal L}^2=\frac{1}{d}\begin{pmatrix}
\pmb X\pmb Y^{\rm T} & \pmb 0\\
\pmb 0 & \pmb Y^{\rm T}\pmb X
\end{pmatrix}=
\begin{pmatrix}
\pmb M & \pmb 0\\
\pmb 0 & \frac{1}{d}\pmb Y^{\rm T}\pmb X
\end{pmatrix}\,.
\label{eq:outlook-linearization-square}
\end{equation}
This identity is worth checking directly. Multiplying the blocks gives
\begin{equation}
\begin{pmatrix}
\pmb 0 & \pmb X\\
\pmb Y^{\rm T} & \pmb 0
\end{pmatrix}
\begin{pmatrix}
\pmb 0 & \pmb X\\
\pmb Y^{\rm T} & \pmb 0
\end{pmatrix}=
\begin{pmatrix}
\pmb X\pmb Y^{\rm T} & \pmb 0\\
\pmb 0 & \pmb Y^{\rm T}\pmb X
\end{pmatrix}\,,
\label{eq:outlook-linearization-square-check}
\end{equation}
and the factor $1/d$ comes from the two factors of $1/\sqrt d$. The spectral consequence is immediate. If
\begin{equation}
\pmb{\mathcal L}
\begin{pmatrix}
\pmb u\\
\pmb v
\end{pmatrix}=\zeta
\begin{pmatrix}
\pmb u\\
\pmb v
\end{pmatrix}\,,
\label{eq:outlook-linearized-eigenproblem}
\end{equation}
then
\begin{equation}
\frac{1}{\sqrt d}\pmb X\pmb v=\zeta\pmb u\,,\qquad\frac{1}{\sqrt d}\pmb Y^{\rm T}\pmb u=\zeta\pmb v\,.
\label{eq:outlook-linearized-eigenproblem-components}
\end{equation}
Eliminating $\pmb v$ gives
\begin{equation}
\frac{1}{d}\pmb X\pmb Y^{\rm T}\pmb u=\zeta^2\pmb u\,,
\label{eq:outlook-square-map}
\end{equation}
that is,
\begin{equation}
\pmb M\pmb u=\zeta^2\pmb u\,.
\label{eq:outlook-square-map-eigenvalue}
\end{equation}
Thus every nonzero eigenvalue $\lambda$ of $\pmb M$ is the square of an eigenvalue $\zeta$ of the sparse non-Hermitian bipartite matrix $\pmb{\mathcal L}$. Conversely, every nonzero eigenvalue of $\pmb{\mathcal L}$ produces an eigenvalue of $\pmb M$ by squaring. This is the key observation: the non-Hermitian diluted Wishart problem can be reduced to the spectral problem of a sparse non-Hermitian matrix on a bipartite graph.

The linearized matrix has an additional symmetry. If
\begin{equation}
\pmb{\mathcal L}
\begin{pmatrix}
\pmb u\\
\pmb v
\end{pmatrix}=
\zeta
\begin{pmatrix}
\pmb u\\
\pmb v
\end{pmatrix},
\label{eq:outlook-chiral-pairing-1}
\end{equation}
then
\begin{equation}
\pmb{\mathcal L}
\begin{pmatrix}
\pmb u\\
-\pmb v
\end{pmatrix}=
-\zeta
\begin{pmatrix}
\pmb u\\
-\pmb v
\end{pmatrix}\,.
\label{eq:outlook-chiral-pairing-2}
\end{equation}
Hence the nonzero spectrum of $\pmb{\mathcal L}$ is symmetric under $\zeta\mapsto -\zeta$. This pairing is exactly what one expects from the square relation \eqref{eq:outlook-square-map-eigenvalue}. It also means that the spectral density of $\pmb M$ is the push-forward of the nonzero spectral density of $\pmb{\mathcal L}$ under the map $\lambda=\zeta^2$. In weak form, for any test function $\phi$,
\begin{equation}
\frac{1}{N}\sum_{\lambda_i\neq0}\phi(\lambda_i)=\frac{N+P}{2N}\int_{\mathbb C}d^2\zeta \rho_{\mathcal L}^{\rm nz}(\zeta)\phi(\zeta^2)\,,
\label{eq:outlook-pushforward-measure}
\end{equation}
where $\rho_{\mathcal L}^{\rm nz}$ denotes the nonzero part of the normalized spectral density of $\pmb{\mathcal L}$, retaining its original spectral weight. Away from $\lambda=0$, this gives the density of $\pmb M$ by a change of variables.

The advantage of the linearization is that the cavity method for sparse non-Hermitian matrices applies immediately. The graph of $\pmb{\mathcal L}$ is bipartite: variable nodes $i=1,\ldots,N$ are connected to factor nodes $\mu=1,\ldots,P$. In the formulas below the edge weights are written in the real case; for complex weights the same Hermitized recursion applies with the conjugations fixed by the definition of the adjoint block. The local Hermitized block is
\begin{equation}
\pmb{Z}(w,\eta)=
\begin{pmatrix}
i\eta & w\\
w^* & i\eta
\end{pmatrix}\,,
\label{eq:outlook-hermitized-local-block}
\end{equation}
where $w$ is now the spectral parameter in the $\zeta$-plane of $\pmb{\mathcal L}$. The edge block associated with the occupied bipartite edge $(i,\mu)$ is
\begin{equation}
\pmb{\mathcal A}_{i\mu}=
\begin{pmatrix}
0 & x_i^\mu/\sqrt d\\
y_i^\mu/\sqrt d & 0
\end{pmatrix}\,.
\label{eq:outlook-edge-block-bipartite}
\end{equation}
If $\pmb{G}_{i\to\mu}(w,\eta)$ is the variable-to-factor cavity message and $\widehat{\pmb{G}}_{\mu\to i}(w,\eta)$ the factor-to-variable message, the locally tree-like Schur-complement recursions are
\begin{equation}
\pmb{G}_{i\to\mu}=\left[\pmb{Z} -\sum_{\nu\in\partial i\setminus\mu}\pmb{\mathcal A}_{i\nu}\widehat{\pmb{G}}_{\nu\to i}\pmb{\mathcal A}_{i\nu}^{\dagger}\right]^{-1}\,,
\label{eq:outlook-variable-to-factor-nh-recursion}
\end{equation}
and
\begin{equation}
\widehat{\pmb{G}}_{\mu\to i}=\left[\pmb{Z}-\sum_{j\in\partial\mu\setminus i}\pmb{\mathcal A}_{j\mu}^{\dagger}\pmb{G}_{j\to\mu}\pmb{\mathcal A}_{j\mu}\right]^{-1}\,.
\label{eq:outlook-factor-to-variable-nh-recursion}
\end{equation}
These equations are just the bipartite version of the non-Hermitian sparse-matrix cavity equations derived earlier. The only novelty is that the upper-right and lower-left edge weights are now generated by the two rectangular matrices $\pmb X$ and $\pmb Y$.

Once the full local Hermitized Green matrices are obtained,
\begin{equation}
\pmb{G}_i=\left[\pmb{Z}-\sum_{\nu\in\partial i}\pmb{\mathcal A}_{i\nu}\widehat{\pmb{G}}_{\nu\to i}\pmb{\mathcal A}_{i\nu}^{\dagger}\right]^{-1}\,,
\label{eq:outlook-full-variable-green-bipartite}
\end{equation}
\begin{equation}
\widehat{\pmb{G}}_\mu=\left[\pmb{Z}-\sum_{j\in\partial\mu}\pmb{\mathcal A}_{j\mu}^{\dagger}\pmb{G}_{j\to\mu}\pmb{\mathcal A}_{j\mu}\right]^{-1}\,,
\label{eq:outlook-full-factor-green-bipartite}
\end{equation}
the regularized non-Hermitian resolvent field of $\pmb{\mathcal L}$ is
\begin{equation}
g_{\mathcal L,\eta}(w,w^*)=\frac{1}{N+P}\left[\sum_{i=1}^{N}\left(\pmb{G}_i\right)_{21}+\sum_{\mu=1}^{P}\left(\widehat{\pmb{G}}_\mu\right)_{21}\right]\,,
\label{eq:outlook-linearized-resolvent-field}
\end{equation}
and the density follows from
\begin{equation}
\rho_{\mathcal L}(w)=\frac{1}{\pi}\lim_{\eta\downarrow0}\partial_{w^*}g_{\mathcal L,\eta}(w,w^*)\,.
\label{eq:outlook-linearized-density}
\end{equation}
Through the square map \eqref{eq:outlook-square-map-eigenvalue}, this gives the spectrum of the non-Hermitian diluted Wishart matrix $\pmb M$.

This construction indicates how the two themes of the outlook section could be combined. On the one hand, a rate-functional theory would seek to go beyond scalar observables and describe large deviations of the whole spectral density; for sparse non-invariant ensembles, the natural microscopic object suggested by the preceding cavity derivations is the empirical distribution of cavity messages. On the other hand, the non-Hermitian diluted Wishart extension indicates that Hermitized cavity methods can be adapted to asymmetric sparse cross-correlations once one uses the appropriate bipartite linearization. A possible combined program would be a non-Hermitian rate-functional theory for diluted Wishart-type matrices, formulated as a large-deviation principle for boundary-dependent Hermitized message populations on the bipartite graph, together with contractions that produce domain counts, conditioned spectra, and the full spectral measure.

This is why we place these topics in the outlook rather than in the main body of the lecture notes. The main sections developed the finite-connectivity machinery for concrete observables. The outlook shows how that machinery points toward two broader directions: a possible functional large-deviation theory whose natural order parameter would be a cavity-message distribution, and a non-Hermitian bipartite extension that merges sparse random matrix theory with diluted cross-correlation ensembles. In both cases the essential lesson remains the same as throughout these notes: in sparse systems the graph does not disappear in the thermodynamic limit, and the correct spectral order parameter is therefore local and graphical rather than purely eigenvalue-based \cite{MetzPerezCastillo2016,PerezCastilloMetz2018Wishart,PerezCastilloMetz2018Conditioned,RamosSanchezGuzmanGonzalezPerezCastilloMetz2021,PerezCastillo2022Generalized,GuzmanGonzalezPerezCastillo2025,RogersPerezCastillo2009,MetzNeriRogers2019}.

\section{Conclusion}
\label{sec:conclusion}
The main objective of these lecture notes has been to show that sparse and diluted random matrices can be studied with the same conceptual tools that are used in the statistical mechanics of disordered systems. The central difficulty of the subject is easy to state. In dense invariant ensembles one can often work directly with the eigenvalues, because the joint law of eigenvalues is known, or one can exploit a scalar self-consistency equation for the global resolvent. In sparse ensembles neither simplification is generally available. The matrix keeps the memory of the graph on which it is supported: different vertices see different local neighborhoods, and the diagonal resolvent entries are therefore described through local cavity equations. On a fixed instance these equations involve edge- or site-dependent messages; for an ensemble of locally tree-like graphs, the same local recursion leads to a self-consistency problem for the statistics of a typical message. The whole logic of the notes follows from this observation.

For Hermitian sparse matrices the starting point is the familiar relation between the spectral density and the resolvent,
\begin{equation}
\rho_{\pmb A}(\lambda)=\frac{1}{\pi}\lim_{\epsilon\downarrow0}{\rm Im}\,\frac{1}{N}{\rm Tr}\left[(\lambda-i\epsilon)\pmb I-\pmb A\right]^{-1}\,.
\label{eq:conc-resolvent-density}
\end{equation}
The point of the Edwards--Jones construction is that the trace of the resolvent is obtained by differentiating the logarithm of a Gaussian partition function,
\begin{equation}
Z_{\pmb A}(z)=\int\left[\prod_{i=1}^{N}\frac{du_i}{\sqrt{2\pi}}\right]\exp\left[-\frac{i}{2}\pmb u^{\rm T}(z\pmb I-\pmb A)\pmb u\right].
\label{eq:conc-gaussian-partition}
\end{equation}
This step is fundamental because it translates the spectral problem into a problem about a disordered Gaussian model on a sparse graph. Once this is done, the cavity method becomes natural rather than ad hoc. On a locally tree-like graph the removal of one edge disconnects the neighboring branches, and the cavity marginals remain Gaussian. The message sent from $i$ to $j$ is therefore summarized by a single complex number,
\begin{equation}
G_{i\to j}(z)=\frac{1}{z-D_i-\displaystyle\sum_{\ell\in\partial i\setminus j}J_{i\ell}^2G_{\ell\to i}(z)}\,.
\label{eq:conc-hermitian-cavity}
\end{equation}
The full local Green function is obtained by restoring the missing neighbor,
\begin{equation}
G_i(z)=\frac{1}{z-D_i-\displaystyle\sum_{\ell\in\partial i}J_{i\ell}^2G_{\ell\to i}(z)}\,,
\label{eq:conc-hermitian-local-green}
\end{equation}
and the spectral density follows by averaging ${\rm Im}\,G_i(z)/\pi$. This is the first basic lesson of the notes: for sparse matrices the spectrum is computed from local messages, not from a closed equation for a global scalar resolvent \cite{EdwardsJones1976,RodgersBray1988,RogersTakedaPerezCastilloKuhn2008,SuscaVivoKuhn2021}.

Passing from a single realization to an ensemble, the same local recursions induce a stochastic fixed-point problem for the probability law of a typical message. If $\mathcal P_{\rm cav}(G)$ denotes the distribution of a typical cavity Green function, then the self-consistency equation takes the form
\begin{equation}
\mathcal P_{\rm cav}(G)=\sum_{\ell=0}^{\infty}q_\ell\int dD p_D(D)\left[\prod_{r=1}^{\ell}dG_r \mathcal P_{\rm cav}(G_r) dJ_r p_J(J_r)\right]\delta\left(G-\frac{1}{z-D-\displaystyle\sum_{r=1}^{\ell}J_r^2G_r}\right)\,,
    \label{eq:conc-message-distribution}
\end{equation}
where $q_\ell$ is the excess-degree distribution. Population dynamics is a numerical method for solving this stochastic fixed-point problem. In this way the random-matrix problem is reduced from a global eigenvalue calculation to a recursive description of local messages. This is the second basic lesson of the notes: in the thermodynamic limit, sparse spectral theory is organized by random recursive local messages.

The same logic survives when one changes the ensemble. For random graphs with topological constraints, the local recursion \eqref{eq:conc-hermitian-cavity} does not change, but the law of the incoming messages does. Degree correlations, generalized degrees, and community labels alter the rooted neighborhood seen by a typical vertex, and therefore alter the message distribution. This is why sparse spectra are so sensitive to topology. At finite connectivity the graph is not a microscopic detail that disappears in the large-$N$ limit; it is part of the order parameter. One may summarize this by saying that the dense theory is largely controlled by entry statistics, whereas the sparse theory is controlled by the joint statistics of entries and local graph environments.

The bipartite covariance and diluted Wishart problems fit the same framework once one writes them on the correct graph. For
\begin{equation}
\pmb W=\frac{1}{d}\pmb X\pmb X^{\rm T},
\label{eq:conc-wishart-def}
\end{equation}
the correct geometry is not the projected covariance matrix alone, but the sparse bipartite graph of the rectangular matrix $\pmb X$. This leads to two types of messages: variable-to-factor Green functions and factor-to-variable self-energies. The latter have the explicit rank-one form
\begin{equation}
U_{\mu\to i}(z)=\frac{(\xi_i^\mu)^2}{1-\displaystyle\frac{1}{d}\sum_{j\in\partial\mu\setminus i}(\xi_j^\mu)^2G_{j\to\mu}(z)}\,,
\label{eq:conc-wishart-factor-message}
\end{equation}
and the variable message is
\begin{equation}
G_{i\to\mu}(z)=\frac{1}{z-\displaystyle\frac{1}{d}\sum_{\nu\in\partial i\setminus\mu}U_{\nu\to i}(z)}\,.
\label{eq:conc-wishart-variable-message}
\end{equation}
With this convention, the actual contribution of factor $\mu$ to the denominator of the variable Green function is $d^{-1}U_{\mu\to i}$. This pair of equations explains why the diluted Wishart problem is tractable. The factor node contributes only a rank-one interaction, so the Gaussian cavity marginals still close on scalar messages. The same closure persists for generalized diluted Wishart and cross-correlation ensembles because the factor-node interaction remains of finite rank, although with a more complicated algebraic form \cite{NagaoTanaka2007,RogersTakedaPerezCastilloKuhn2008,PerezCastillo2022Generalized}.

The non-Hermitian case requires one extra step, namely Hermitization. One does not try to work directly with $(z\pmb I-\pmb A)^{-1}$ near the spectrum. Instead one introduces the regularized block matrix
\begin{equation}
\pmb{\mathcal B}_{\pmb A}(z,\eta)=\begin{pmatrix}
i\eta\pmb I & z\pmb I-\pmb A\\
z^*\pmb I-\pmb A^\dagger & i\eta\pmb I
\end{pmatrix}\,,
\label{eq:conc-hermitization}
\end{equation}
and computes the non-Hermitian density from the corresponding Hermitized resolvent field. The cavity messages are now $2\times2$ matrices,
\begin{equation}
\pmb{G}_{i\to j}(z,\eta)=\left[\pmb{Z}_i(z,\eta)-\sum_{\ell\in\partial i\setminus j}\pmb{\mathcal A}_{i\ell}\pmb{G}_{\ell\to i}(z,\eta)\pmb{\mathcal A}_{i\ell}^{\dagger}\right]^{-1}\,.
    \label{eq:conc-nonhermitian-cavity}
\end{equation}
This is the non-Hermitian analogue of \eqref{eq:conc-hermitian-cavity}. The algebra becomes matrix-valued, but the conceptual structure is unchanged: the spectral problem is still solved by local graphical recursions \cite{RogersPerezCastillo2009,MetzNeriRogers2019}.

A central theme of the lecture notes has been that one can go beyond the typical spectral density without abandoning the same framework. With a fixed convention for eigenvalues exactly at the threshold, the number of eigenvalues below $x$,
\begin{equation}
\mathcal K_{\pmb A}(x)=\sum_{i=1}^{N}\Theta(x-\lambda_i)\,,
\label{eq:conc-index}
\end{equation}
is a phase of the characteristic determinant,
\begin{equation}
\mathcal K_{\pmb A}(x)=N+\frac{1}{2\pi i}\lim_{\epsilon\downarrow0}\left[\log\det\left((x-i\epsilon)\pmb I-\pmb A\right)-\log\det\left((x+i\epsilon)\pmb I-\pmb A\right)\right]\,.
\label{eq:conc-index-determinant}
\end{equation}
This identity turns large deviations of the index into a tilted determinant problem. The same structure appears for the number of eigenvalues inside a domain in the non-Hermitian case: the winding-number representation can be rewritten as a boundary integral of a Hermitized logarithmic potential. Thus the large-deviation and conditioned-density calculations are not conceptually separate from the cavity construction. They use the same local messages, but in tilted or constrained ensembles. This is the third basic lesson of the notes: typical densities, count fluctuations, large deviations, and conditioned densities all belong to one common statistical-mechanics framework \cite{MetzPerezCastillo2016,PerezCastilloMetz2018Wishart,PerezCastilloMetz2018Conditioned,RamosSanchezGuzmanGonzalezPerezCastilloMetz2021}.

Another theme of the notes has been physical interpretation. The moments of the adjacency spectrum count closed walks, and the adjacency spectrum itself controls linear propagation on networks. The Laplacian spectrum governs diffusion and relaxation. The sparse covariance and diluted Wishart spectra describe the distribution of empirical variance directions when the data matrix itself is sparse. The generalized diluted Wishart spectrum describes sparse cross-correlations between two sets of measurements. The non-Hermitian spectrum encodes modal growth rates and frequencies in directed linear dynamics. The Hermitian index counts unstable directions of a Hessian when the threshold is zero, while non-Hermitian number statistics count modes of a Jacobian in a chosen complex-domain stability region. The conditioned spectral density describes what a rare sample with an atypical spectral count must look like. This is why the theory is worth developing in the first place: spectral observables are not merely formal quantities; they encode physically meaningful information about disordered systems, networks, stability, transport, and inference problems.

It is also useful to state clearly what has \emph{not} been done. The methods in these notes rely heavily on local tree-likeness and on replica-symmetric or Bethe-type assumptions. Short loops, strong clustering, and situations in which several pure-state structures compete may require more elaborate message spaces or a more refined replica analysis. Likewise, while the cavity approach gives access to local densities of states and count statistics, a full theory of eigenvector correlations, non-normal amplification, or dynamical observables beyond the resolvent may require additional work. These are not defects of the method; they are the natural boundaries of the level of approximation that has been systematically developed here.

The reader should nevertheless come away with a clear chain of ideas. One begins with a spectral observable. One rewrites it in terms of a determinant or resolvent. One represents the determinant structure by a Gaussian integral. One uses the sparse graph structure to derive cavity recursions for local messages. One solves the resulting fixed-point problem by belief propagation on a single instance or by population dynamics at the ensemble level. Finally, one interprets the result physically. In symbols, the route is
\begin{equation}
\begin{split}
   \text{spectral observable}
&   \longrightarrow
    \text{resolvent, determinant, or Hermitized determinant}\\
&   \longrightarrow
    \text{Gaussian model on a sparse graph}\\
&    \longrightarrow
    \text{cavity messages}\\
&   \longrightarrow
    \text{densities, rate functions, and conditioned densities}.
    \label{eq:conc-logic-chain}
\end{split}
\end{equation}
This chain is the real conclusion of the notes. It is the unifying mechanism behind all the examples we have discussed.

For this reason, perhaps the most important conceptual message is not any individual formula, but a change of viewpoint. Sparse random matrices should not be thought of as dense random matrices with many zeros inserted. They should be thought of as graphical disordered systems whose spectral observables are local statistical-mechanics quantities. Once this viewpoint is adopted, the appearance of Gaussian integrals, replicas, cavity fields, message passing, and population dynamics is no longer surprising. It becomes the natural language of the problem.

In that sense, the subject of these lecture notes can be summarized in one sentence: the statistical mechanics of random matrices at finite connectivity is the statistical mechanics of local resolvents on random graphs. The rest of the theory---typical densities, covariance spectra, non-Hermitian spectra, index large deviations, conditioned densities, and the outlook toward rate-functional formulations---is the systematic unfolding of this idea.

\section*{Acknowledgements}
I would like to thank my colleagues at the Instituto de Ciencias F\'{\i}sicas (UNAM, Cuernavaca) for their hospitality and support during a difficult period. In particular, I thank Juan Carlos, Antonio, Luis, and Thomas. This work was initiated with support from Christof Jung Kohl's CONAHCYT project (No.\ 425854). Last but not least, I thank Matteo Marsili, who is always there not only as a colleague but also as a friend when life takes you onto unexpected paths.

I dedicate this set of lecture notes to my recently born baby daughter, Emilia Themis. You are now the light that guides my path in life; wherever it takes me, I will gladly and proudly walk it with you.

\begin{appendix}
\numberwithin{equation}{section}

\section{Gaussian integral identities and resolvent conventions}
\label{app:gaussian-identities-resolvents}
In the main text we repeatedly used a small number of Gaussian identities and resolvent conventions. Since the same formulas appear in the Hermitian, Wishart, and non-Hermitian problems, it is convenient to collect them in one place and derive them explicitly. The purpose of this appendix is therefore not to introduce new ideas, but to make precise the algebra that was used throughout the notes.

We begin with the Hermitian resolvent convention. Let $\pmb A$ be a real symmetric or Hermitian $N\times N$ matrix with eigenvalues $\lambda_1,\ldots,\lambda_N$. We define the resolvent and its normalized trace by
\begin{equation}
\pmb G(z)=(z\pmb I-\pmb A)^{-1}\,,\qquad g(z)=\frac{1}{N}{\rm Tr}\,\pmb G(z)\,.
\label{eq:app-resolvent-definition}
\end{equation}
Throughout these notes we use the lower-half-plane convention
\begin{equation}
z=\lambda-i\epsilon\,,\qquad\epsilon>0\,.
\label{eq:app-lower-half-plane}
\end{equation}
This choice is convenient because the Gaussian integrals used later become convergent when ${\rm Im}\,z<0$. To see how the density is recovered, diagonalize $\pmb A$ as
\begin{equation}
\pmb A=\pmb U\,{\rm diag}(\lambda_1,\ldots,\lambda_N)\,\pmb U^{-1}\,.
\label{eq:app-spectral-decomposition}
\end{equation}
Then
\begin{equation}
g(z)=\frac{1}{N}\sum_{i=1}^{N}\frac{1}{z-\lambda_i}\,.
\label{eq:app-resolvent-spectral-sum}
\end{equation}
Substituting $z=\lambda-i\epsilon$ gives
\begin{equation}
\frac{1}{\lambda-i\epsilon-\lambda_i}=\frac{\lambda-\lambda_i+i\epsilon}{(\lambda-\lambda_i)^2+\epsilon^2}\,,
\label{eq:app-lorentzian-identity}
\end{equation}
and therefore
\begin{equation}
{\rm Im}\frac{1}{\lambda-i\epsilon-\lambda_i}=\frac{\epsilon}{(\lambda-\lambda_i)^2+\epsilon^2}\,.
\label{eq:app-imaginary-part-lorentzian}
\end{equation}
Since
\begin{equation}
\frac{1}{\pi}\lim_{\epsilon\downarrow0}\frac{\epsilon}{x^2+\epsilon^2}=\delta(x)
\label{eq:app-delta-lorentzian}
\end{equation}
in the sense of distributions, we obtain
\begin{equation}
\rho(\lambda)=\frac{1}{\pi}\lim_{\epsilon\downarrow0}{\rm Im} g(\lambda-i\epsilon)=\frac{1}{N}\sum_{i=1}^{N}\delta(\lambda-\lambda_i)\,.
\label{eq:app-density-from-resolvent}
\end{equation}
This is the convention used everywhere in the main text. If one chooses instead the upper-half-plane convention $z=\lambda+i\epsilon$, then the sign in front of the imaginary part changes. The two conventions are equivalent as long as one is consistent.

The same computation gives the local density of states. Let $u_i^{(j)}$ be the $i$th component of the normalized eigenvector associated with $\lambda_j$. Then
\begin{equation}
G_{ii}(z)=\sum_{j=1}^{N}\frac{|u_i^{(j)}|^2}{z-\lambda_j}\,,
\label{eq:app-local-spectral-representation}
\end{equation}
so that
\begin{equation}
\rho_i(\lambda)=\frac{1}{\pi}\lim_{\epsilon\downarrow0}{\rm Im}G_{ii}(\lambda-i\epsilon)
=\sum_{j=1}^{N}|u_i^{(j)}|^2\delta(\lambda-\lambda_j)\,.
\label{eq:app-local-density-of-states}
\end{equation}
Averaging over $i$ gives the global density:
\begin{equation}
\rho(\lambda)=\frac{1}{N}\sum_{i=1}^{N}\rho_i(\lambda)\,.
\label{eq:app-global-density-from-local}
\end{equation}

We next derive the Gaussian integral identity that underlies the Edwards--Jones representation. We start in one dimension. If $a\in\mathbb C$ has positive real part, ${\rm Re}\,a>0$, then
\begin{equation}
\int_{-\infty}^{\infty}\frac{du}{\sqrt{2\pi}} e^{-\frac{1}{2}au^2}=a^{-1/2}\,,
\label{eq:app-one-dimensional-gaussian}
\end{equation}
where the square root is taken on the principal branch. The condition ${\rm Re}\,a>0$ guarantees convergence. For a complex symmetric matrix $\pmb M$ with positive-definite real part, the multidimensional version is
\begin{equation}
\int_{\mathbb R^N}\left[\prod_{i=1}^{N}\frac{du_i}{\sqrt{2\pi}}\right]\exp\left[-\frac{1}{2}\pmb u^{\rm T}\pmb M\pmb u\right]=\det(\pmb M)^{-1/2}\,.
\label{eq:app-multidimensional-gaussian}
\end{equation}
To prove \eqref{eq:app-multidimensional-gaussian} in the real symmetric positive-definite case, diagonalize
\begin{equation}
\pmb M=\pmb O^{\rm T}{\rm diag}(m_1,\ldots,m_N)\pmb O\,,
\label{eq:app-matrix-diagonalization}
\end{equation}
with $\pmb O$ orthogonal and $m_i>0$. Since $d\pmb u=d\pmb v$ under the rotation $\pmb v=\pmb O\pmb u$, the integral factorizes:
\begin{equation}
\int\left[\prod_{i=1}^{N}\frac{dv_i}{\sqrt{2\pi}}\right] e^{-\frac{1}{2}\sum_{i=1}^{N}m_i v_i^2}=\prod_{i=1}^{N}m_i^{-1/2}=\det(\pmb M)^{-1/2}\,.
\label{eq:app-factorized-gaussian}
\end{equation}
The complex Gaussian identities used in the Edwards--Jones representation are then obtained by analytic continuation to complex symmetric matrices whose real part is positive definite.

Now set
\begin{equation}
\pmb M=i(z\pmb I-\pmb A)\,,\qquad z=\lambda-i\epsilon\,,\qquad\epsilon>0\,.
\label{eq:app-spectral-gaussian-matrix}
\end{equation}
If $\pmb A$ is real symmetric, the eigenvalues of $i(z\pmb I-\pmb A)$ are
\begin{equation}
i(\lambda-\lambda_i)+\epsilon\,,\qquad i=1,\ldots,N\,,
\label{eq:app-gaussian-eigenvalues}
\end{equation}
whose real part is $\epsilon>0$. Thus the Gaussian integral converges and gives
\begin{equation}
Z_{\pmb A}(z)=\int_{\mathbb R^N}\left[\prod_{i=1}^{N}\frac{du_i}{\sqrt{2\pi}}\right]\exp\left[-\frac{i}{2}\pmb u^{\rm T}(z\pmb I-\pmb A)\pmb u\right]=\det\left[i(z\pmb I-\pmb A)\right]^{-1/2}\,.
\label{eq:app-edwards-jones-identity}
\end{equation}
This is the precise real-Gaussian form of the Edwards--Jones representation used throughout the lecture notes for real symmetric sparse matrices.

The connection with the resolvent follows by differentiating the logarithm. Since
\begin{equation}
\log Z_{\pmb A}(z)=-\frac{1}{2}\log\det\left[i(z\pmb I-\pmb A)\right]\,,
\label{eq:app-log-partition}
\end{equation}
we obtain
\begin{equation}
\frac{\partial}{\partial z}\log Z_{\pmb A}(z)=-\frac{1}{2}{\rm Tr}(z\pmb I-\pmb A)^{-1}\,,
\label{eq:app-log-derivative-partition}
\end{equation}
because for any invertible matrix $\pmb M(z)$,
\begin{equation}
\frac{\partial}{\partial z}\log\det \pmb M(z)={\rm Tr}\left[\pmb M(z)^{-1}\frac{\partial \pmb M(z)}{\partial z}\right]\,.
\label{eq:app-log-det-derivative}
\end{equation}
Applying \eqref{eq:app-log-derivative-partition} to $z=\lambda-i\epsilon$ gives
\begin{equation}
g(\lambda-i\epsilon)=-\frac{2}{N}\frac{\partial}{\partial\lambda}\log Z_{\pmb A}(\lambda-i\epsilon)\,,
\label{eq:app-resolvent-from-partition}
\end{equation}
and therefore
\begin{equation}
\rho(\lambda)=-\frac{2}{\pi N}\lim_{\epsilon\downarrow0}{\rm Im}\frac{\partial}{\partial\lambda}\log Z_{\pmb A}(\lambda-i\epsilon)\,.
\label{eq:app-density-from-partition}
\end{equation}
This is the Edwards--Jones representation in the conventions of the main text.

A second identity used repeatedly is the Schur-complement formula. Let
\begin{equation}
\pmb M=\begin{pmatrix}
a & \pmb b^{\rm T}\\
\pmb b & \pmb B
\end{pmatrix}\,,
\label{eq:app-block-matrix}
\end{equation}
where $a$ is a scalar, $\pmb b$ is a column vector, and $\pmb B$ is an invertible $(N-1)\times(N-1)$ matrix. Then
\begin{equation}
(\pmb M^{-1})_{11}=\left(a-\pmb b^{\rm T}\pmb B^{-1}\pmb b\right)^{-1}\,.
\label{eq:app-schur-complement}
\end{equation}
To see this, write the inverse in block form,
\begin{equation}
\pmb M^{-1}=\begin{pmatrix}
\alpha & \pmb \beta^{\rm T}\\
\pmb \gamma & \pmb \Delta
\end{pmatrix}\,,
\label{eq:app-block-inverse}
\end{equation}
and impose $\pmb M\pmb M^{-1}=\pmb I$. The $(1,1)$ block gives
\begin{equation}
a\alpha+\pmb b^{\rm T}\pmb\gamma=1\,,
\label{eq:app-schur-step1}
\end{equation}
while the lower-left block gives
\begin{equation}
\pmb b\,\alpha+\pmb B\pmb\gamma=\pmb 0\,.
\label{eq:app-schur-step2}
\end{equation}
Solving \eqref{eq:app-schur-step2} for $\pmb\gamma$,
\begin{equation}
\pmb\gamma=-\pmb B^{-1}\pmb b\,\alpha\,,
\label{eq:app-schur-step3}
\end{equation}
and substituting into \eqref{eq:app-schur-step1}, we get
\begin{equation}
\left(a-\pmb b^{\rm T}\pmb B^{-1}\pmb b\right)\alpha=1\,,
\label{eq:app-schur-step4}
\end{equation}
which proves \eqref{eq:app-schur-complement}.

The same identity is recovered directly from the Gaussian integral by completing the square. Consider
\begin{equation}
I(u_1)=\int_{\mathbb R^{N-1}}\left[\prod_{i=2}^{N}\frac{du_i}{\sqrt{2\pi}}\right]\exp\left[-\frac{1}{2}\begin{pmatrix}
u_1\\
\pmb u
\end{pmatrix}^{\rm T}
\pmb M
\begin{pmatrix}
u_1\\
\pmb u
\end{pmatrix}
\right]\,,
\label{eq:app-gaussian-marginal}
\end{equation}
where $\pmb u=(u_2,\ldots,u_N)^{\rm T}$. Expanding the exponent,
\begin{equation}
-\frac{1}{2}\left[a u_1^2+2u_1\pmb b^{\rm T}\pmb u+\pmb u^{\rm T}\pmb B\pmb u\right]\,.
\label{eq:app-expanded-exponent}
\end{equation}
Now complete the square:
\begin{equation}
\pmb u^{\rm T}\pmb B\pmb u+2u_1\pmb b^{\rm T}\pmb u=\left(\pmb u+u_1\pmb B^{-1}\pmb b\right)^{\rm T}\pmb B\left(\pmb u+u_1\pmb B^{-1}\pmb b\right)-u_1^2\pmb b^{\rm T}\pmb B^{-1}\pmb b\,.
\label{eq:app-completing-square}
\end{equation}
Therefore
\begin{equation}
I(u_1)\propto\exp\left[-\frac{1}{2}\left(a-\pmb b^{\rm T}\pmb B^{-1}\pmb b\right)u_1^2\right]\,.
\label{eq:app-effective-variance}
\end{equation}
Thus the effective inverse variance of $u_1$ is exactly the Schur complement. This is the elementary mechanism behind the cavity recursions in the main text.

Two further Gaussian identities are especially useful for bipartite and generalized diluted Wishart ensembles. The first is the matrix determinant lemma. If $\pmb D$ is invertible and $\pmb U$ and $\pmb V$ are matrices of size $N\times r$, then
\begin{equation}
\det\left(\pmb D+\pmb U\pmb V^{\rm T}\right)=\det(\pmb D)\det\left(\pmb I_r+\pmb V^{\rm T}\pmb D^{-1}\pmb U\right)\,.
\label{eq:app-matrix-determinant-lemma}
\end{equation}
The proof is a one-line application of
\begin{equation}
\det(\pmb I+\pmb A\pmb B)=\det(\pmb I+\pmb B\pmb A)
\label{eq:app-det-identity-ABBA}
\end{equation}
to
\begin{equation}
\det\left(\pmb D+\pmb U\pmb V^{\rm T}\right)=\det(\pmb D)\det\left(\pmb I+\pmb D^{-1}\pmb U\pmb V^{\rm T}\right)\,.
\label{eq:app-lemma-proof-step}
\end{equation}

For the diluted Wishart factor node we need only the rank-one case. Let
\begin{equation}
\pmb D={\rm diag}(a_1,\ldots,a_k)\,,\qquad\pmb \xi=(\xi_1,\ldots,\xi_k)^{\rm T}\,,
\label{eq:app-rank-one-data}
\end{equation}
and consider
\begin{equation}
I_{\rm rank\mbox{-}1}=\int\left[\prod_{r=1}^{k}\frac{du_r}{\sqrt{2\pi}}\right]\exp\left[-\frac{1}{2}\sum_{r=1}^{k}a_r u_r^2+\frac{c}{2}\left(\sum_{r=1}^{k}\xi_r u_r\right)^2\right]\,.
    \label{eq:app-rank-one-integral}
\end{equation}
In vector notation,
\begin{equation}
I_{\rm rank\mbox{-}1}=\int\left[\prod_{r=1}^{k}\frac{du_r}{\sqrt{2\pi}}\right]\exp\left[-\frac{1}{2}\pmb u^{\rm T}\left(\pmb D-c\,\pmb \xi\pmb \xi^{\rm T}\right)\pmb u\right]\,.
    \label{eq:app-rank-one-integral-matrix}
\end{equation}
Applying \eqref{eq:app-multidimensional-gaussian} and then \eqref{eq:app-matrix-determinant-lemma},
\begin{align}
I_{\rm rank\mbox{-}1}&=\det\left(\pmb D-c\,\pmb \xi\pmb \xi^{\rm T}\right)^{-1/2}\nonumber\\
&=\det(\pmb D)^{-1/2}\left(1-c\,\pmb \xi^{\rm T}\pmb D^{-1}\pmb \xi\right)^{-1/2}\nonumber\\
&=\left(\prod_{r=1}^{k}a_r^{-1/2}\right)\left(1-c\sum_{r=1}^{k}\frac{\xi_r^2}{a_r}\right)^{-1/2}\,.
\label{eq:app-rank-one-result}
\end{align}
This is exactly the identity used in the diluted Wishart factor update.

The generalized diluted Wishart ensemble requires the rank-two version. Let
\begin{equation}
\pmb x=(x_1,\ldots,x_k)^{\rm T}\,,\qquad\pmb y=(y_1,\ldots,y_k)^{\rm T}\,,
\label{eq:app-rank-two-vectors}
\end{equation}
and consider
\begin{equation}
I_{\rm rank\mbox{-}2}=\int\left[\prod_{r=1}^{k}\frac{du_r}{\sqrt{2\pi}}\right]\exp\left[-\frac{1}{2}\pmb u^{\rm T}\left(\pmb D-\frac{1}{2d}(\pmb x\pmb y^{\rm T}+\pmb y\pmb x^{\rm T})\right)\pmb u\right]\,.
\label{eq:app-rank-two-integral}
\end{equation}
Write
\begin{equation}
\pmb U=\begin{pmatrix}
\pmb x & \pmb y
\end{pmatrix},\qquad\pmb V=\frac{1}{2d}\begin{pmatrix}
\pmb y & \pmb x
\end{pmatrix}\,,
\label{eq:app-rank-two-UV}
\end{equation}
so that
\begin{equation}
\pmb U\pmb V^{\rm T}=\frac{1}{2d}(\pmb x\pmb y^{\rm T}+\pmb y\pmb x^{\rm T})\,.
\label{eq:app-rank-two-factorization}
\end{equation}
Then
\begin{equation}
\det\left(\pmb D-\frac{1}{2d}(\pmb x\pmb y^{\rm T}+\pmb y\pmb x^{\rm T})\right)=\det(\pmb D)\det\left(\pmb I_2-\pmb V^{\rm T}\pmb D^{-1}\pmb U\right)\,.
\label{eq:app-rank-two-det-reduction}
\end{equation}
Define
\begin{equation}
S_{xx}=\pmb x^{\rm T}\pmb D^{-1}\pmb x\,,\qquad S_{xy}=\pmb x^{\rm T}\pmb D^{-1}\pmb y\,,\qquad S_{yy}=\pmb y^{\rm T}\pmb D^{-1}\pmb y\,.
\label{eq:app-rank-two-sums}
\end{equation}
Then
\begin{equation}
\pmb V^{\rm T}\pmb D^{-1}\pmb U=\frac{1}{2d}\begin{pmatrix}
S_{xy} & S_{yy}\\
S_{xx} & S_{xy}
\end{pmatrix}\,,
\label{eq:app-rank-two-reduced-matrix}
\end{equation}
and therefore
\begin{align}
\det\left(\pmb I_2-\pmb V^{\rm T}\pmb D^{-1}\pmb U\right)&=\left(1-\frac{S_{xy}}{2d}\right)^2-\frac{S_{xx}S_{yy}}{4d^2}\nonumber\\
&=\frac{(2d-S_{xy})^2-S_{xx}S_{yy}}{4d^2}\,.
\label{eq:app-rank-two-denominator}
\end{align}
Hence
\begin{equation}
I_{\rm rank\mbox{-}2}=\det(\pmb D)^{-1/2}\left[\frac{4d^2}{(2d-S_{xy})^2-S_{xx}S_{yy}}\right]^{1/2}\,.
\label{eq:app-rank-two-result}
\end{equation}
This is the finite-rank Gaussian identity behind the generalized diluted Wishart factor contribution.

We now collect the non-Hermitian conventions. Write
\begin{equation}
z=x+iy\,,\qquad d^2z=dx\,dy\,,\qquad \delta^{(2)}(z-z_0)=\delta(x-x_0)\delta(y-y_0)\,.
\label{eq:app-complex-measure}
\end{equation}
We use the Wirtinger derivatives
\begin{equation}
\partial_z=\frac{1}{2}\left(\partial_x-i\partial_y\right)\,,\qquad\partial_{z^*}=\frac{1}{2}\left(\partial_x+i\partial_y\right)\,,
\label{eq:app-wirtinger-derivatives}
\end{equation}
so that
\begin{equation}
\Delta=\partial_x^2+\partial_y^2=4\partial_z\partial_{z^*}\,.
\label{eq:app-laplacian-wirtinger}
\end{equation}
The basic distributional identity is
\begin{equation}
\partial_{z^*}\frac{1}{z-z_0}=\pi\delta^{(2)}(z-z_0)\,.
\label{eq:app-dzbar-cauchy-kernel}
\end{equation}
One way to understand \eqref{eq:app-dzbar-cauchy-kernel} is to test it against a smooth compactly supported function $\varphi$ and integrate by parts:
\begin{equation}
\int d^2z\frac{\partial_{z^*}\varphi(z,z^*)}{z-z_0}=-\pi\varphi(z_0,z_0^*)\,,
\label{eq:app-test-function-identity}
\end{equation}
which is the Cauchy--Pompeiu formula in distributional form.

For a fixed finite matrix $\pmb A$, the trace resolvent satisfies the meromorphic identity
\begin{equation}
g(z)=\frac{1}{N}{\rm Tr}(z\pmb I-\pmb A)^{-1}=\frac{1}{N}\sum_{\alpha=1}^{N}\frac{1}{z-z_\alpha}\,,
\label{eq:app-normal-resolvent}
\end{equation}
where the eigenvalues are counted with algebraic multiplicity. Therefore, at finite $N$ in the distributional sense,
\begin{equation}
\rho_{\pmb A}(z)=\frac{1}{\pi}\partial_{z^*}g(z)\,.
\label{eq:app-normal-density-dzbar}
\end{equation}
For non-normal random matrices, however, this identity is not a stable starting point for large-$N$ spectral-density calculations. Near the spectrum, the relevant regularization is controlled by the small singular values of $z\pmb I-\pmb A$, and the limiting non-holomorphic field is obtained by Hermitization.

A convenient Hermitized block matrix is
\begin{equation}
\pmb{\mathcal B}_{\pmb A}(z,\eta)=\begin{pmatrix}
\eta\pmb I & i(z\pmb I-\pmb A)\\
i(z^*\pmb I-\pmb A^\dagger) & \eta\pmb I
\end{pmatrix}\,,\qquad \eta>0\,.
\label{eq:app-hermitized-block}
\end{equation}
At the level of determinants, this convention is equivalent, up to a $z$-independent phase, to the form with diagonal blocks $i\eta\pmb I$ and off-diagonal blocks $z\pmb I-\pmb A$ and $z^*\pmb I-\pmb A^\dagger$ used in the non-Hermitian sections. Such $z$-independent factors do not affect the logarithmic-potential formulas for the density.
Let
\begin{equation}
\pmb C=z\pmb I-\pmb A\,.
\label{eq:app-C-definition}
\end{equation}
Then
\begin{equation}
\pmb{\mathcal B}_{\pmb A}(z,\eta)=\begin{pmatrix}
\eta\pmb I & i\pmb C\\
i\pmb C^\dagger & \eta\pmb I
\end{pmatrix}\,.
\label{eq:app-hermitized-block-C}
\end{equation}
Using the block determinant formula with the upper-left block $\eta\pmb I$,
\begin{align}
\det \pmb{\mathcal B}_{\pmb A}(z,\eta)&=\det(\eta\pmb I)\det\left(\eta\pmb I-i\pmb C^\dagger(\eta\pmb I)^{-1}i\pmb C\right)\nonumber\\
&=\eta^{N}\det\left(\eta\pmb I+\eta^{-1}\pmb C^\dagger\pmb C\right)\nonumber\\
&=\det\left(\eta^2\pmb I+\pmb C^\dagger\pmb C\right)\,.
    \label{eq:app-hermitized-determinant-proof}
\end{align}
Since $\pmb C^\dagger\pmb C$ and $\pmb C\pmb C^\dagger$ have the same nonzero eigenvalues,
\begin{equation}
\det\left(\eta^2\pmb I+\pmb C^\dagger\pmb C\right)=\det\left(\eta^2\pmb I+\pmb C\pmb C^\dagger\right)\,,
    \label{eq:app-singular-value-equality}
\end{equation}
and therefore
\begin{equation}
\det \pmb{\mathcal B}_{\pmb A}(z,\eta)=\det\left[(z\pmb I-\pmb A)(z^*\pmb I-\pmb A^\dagger)+\eta^2\pmb I\right]\,.
    \label{eq:app-hermitized-determinant}
\end{equation}
This is the determinant identity used throughout the non-Hermitian sections.

The regularized logarithmic potential is then
\begin{equation}
\Phi_{\pmb A,\eta}(z,z^*)=\frac{1}{N}\log\det\left[(z\pmb I-\pmb A)(z^*\pmb I-\pmb A^\dagger)+\eta^2\pmb I\right]\,.
    \label{eq:app-logarithmic-potential}
\end{equation}
The non-Hermitian spectral density is recovered through
\begin{equation}
\rho_{\pmb A}(z)=\frac{1}{\pi}\lim_{\eta\downarrow0}\partial_{z^*}\partial_z\Phi_{\pmb A,\eta}(z,z^*)\,,
\label{eq:app-nonhermitian-density-potential}
\end{equation}
or, equivalently,
\begin{equation}
\rho_{\pmb A}(z)=\frac{1}{4\pi}\lim_{\eta\downarrow0}\Delta\Phi_{\pmb A,\eta}(z,z^*)\,.
    \label{eq:app-nonhermitian-density-laplacian}
\end{equation}
Equations \eqref{eq:app-hermitized-block}--\eqref{eq:app-nonhermitian-density-laplacian} fix the non-Hermitian determinant and density conventions used in the main text.

To summarize, the identities collected in this appendix are the minimal algebraic toolbox behind the whole lecture-note construction. Equations \eqref{eq:app-density-from-resolvent} and \eqref{eq:app-density-from-partition} explain how Hermitian spectral densities are obtained from resolvents and Gaussian partition functions. Equations \eqref{eq:app-schur-complement} and \eqref{eq:app-effective-variance} explain why local Gaussian marginals close under cavity recursion. Equations \eqref{eq:app-rank-one-result} and \eqref{eq:app-rank-two-result} are the finite-rank Gaussian formulas needed in diluted Wishart and generalized diluted Wishart ensembles. Finally, equations \eqref{eq:app-dzbar-cauchy-kernel} and \eqref{eq:app-hermitized-determinant}--\eqref{eq:app-nonhermitian-density-laplacian} fix the non-Hermitian conventions used in the Hermitization approach. Once these identities are in place, all the cavity and replica constructions in the main text follow from them by repeated application on sparse graphical structures.

\section{Dense-limit reductions}
\label{app:dense-limit-reductions}
In the main text we used the sparse cavity equations as the primary description of the spectrum, and we repeatedly stated that the classical dense laws are recovered when the mean connectivity becomes large. The purpose of this appendix is to show this explicitly. The derivations are simple once one identifies the correct scaling. The general mechanism is always the same: each cavity sum contains many terms of size $O(c^{-1})$ or $O(d^{-1})$, the fluctuations of the sum are therefore of order $c^{-1/2}$ or $d^{-1/2}$, and the sums entering the denominators self-average. In homogeneous cases this collapse is literally a collapse of the message law to a deterministic fixed point; with diagonal disorder or distinguished edge weights, the remaining local variables are averaged only after the dense self-energy has been formed. In that limit the full distributional cavity equation reduces to an ordinary deterministic self-consistency equation.

The order of limits matters. Throughout this appendix one should understand that the thermodynamic limit is taken first, and only afterwards the mean connectivity is sent to infinity. If one mixes the two limits carelessly, one may confuse a genuinely dense law with a finite-connectivity law evaluated at large but finite degree.

We begin with sparse symmetric matrices. Consider
\begin{equation}
A_{ij}= D_i\delta_{ij}+C_{ij}J_{ij}\,,\qquad A_{ij}=A_{ji}\,,
\label{eq:app-dense-sparse-symmetric-model}
\end{equation}
on a sparse graph of mean degree $c$, and impose the dense scaling
\begin{equation}
J_{ij}=\frac{\widetilde J_{ij}}{\sqrt c}\,,\qquad \int d\widetilde J p_{\widetilde J}(\widetilde J) \widetilde J=0\,,\qquad \int d\widetilde J p_{\widetilde J}(\widetilde J) \widetilde J^2=\sigma^2\,.
\label{eq:app-dense-symmetric-scaling}
\end{equation}
The cavity equation is
\begin{equation}
G_{i\to j}(z)=\frac{1}{z-D_i-\displaystyle\frac{1}{c}\sum_{\ell\in\partial i\setminus j}\widetilde J_{i\ell}^2G_{\ell\to i}(z)}\,.
\label{eq:app-dense-symmetric-cavity}
\end{equation}
We now analyze the sum in the denominator. Since the degree of a typical vertex is $k_i=c+O(\sqrt c)$, the number of terms is $O(c)$. Each term is $O(c^{-1})$, because $\widetilde J_{i\ell}^2=O(1)$ and $G_{\ell\to i}(z)=O(1)$ away from poles. Hence the whole sum is $O(1)$. More precisely, if the message distribution has a finite first moment, then by the law of large numbers
\begin{equation}
\frac{1}{c}\sum_{\ell\in\partial i\setminus j}\widetilde J_{i\ell}^2G_{\ell\to i}(z)\longrightarrow\sigma^2 g(z)\,,
\label{eq:app-dense-self-averaging-symmetric}
\end{equation}
where
\begin{equation}
g(z)=\lim_{N\to\infty}\frac{1}{N}{\rm Tr}(z\pmb I-\pmb A)^{-1}
\label{eq:app-dense-global-resolvent}
\end{equation}
is the deterministic limiting Stieltjes transform. The exclusion of one neighbor becomes irrelevant in the dense limit, because removing one term changes the sum by $O(c^{-1})$. Therefore the cavity and full Green functions coincide at leading order:
\begin{equation}
G_{i\to j}(z)=G_i(z)+O(c^{-1})\,.
\label{eq:app-dense-cavity-full-coincide}
\end{equation}
Substituting \eqref{eq:app-dense-self-averaging-symmetric} into \eqref{eq:app-dense-symmetric-cavity} gives
\begin{equation}
G_i(z)\longrightarrow\frac{1}{z-D_i-\sigma^2 g(z)}\,.
\label{eq:app-dense-site-green-with-diagonal}
\end{equation}
Averaging over the diagonal disorder $D_i$ with law $p_D(D)$, we find
\begin{equation}
g(z)=\int dD p_D(D) \frac{1}{z-D-\sigma^2 g(z)}\,.
\label{eq:app-pastur-equation}
\end{equation}
Equation \eqref{eq:app-pastur-equation} is the Pastur self-consistency equation for a Wigner-type ensemble with diagonal disorder \cite{PasturShcherbina2011}. It is the dense reduction of the sparse distributional cavity equation.

The most familiar special case is $D_i=0$. Then \eqref{eq:app-pastur-equation} becomes
\begin{equation}
g(z)=\frac{1}{z-\sigma^2 g(z)}\,.
\label{eq:app-wigner-self-consistency}
\end{equation}
Multiplying both sides by the denominator gives
\begin{equation}
\sigma^2 g(z)^2-zg(z)+1=0\,.
\label{eq:app-wigner-quadratic}
\end{equation}
Solving the quadratic,
\begin{equation}
g(z)=\frac{z\pm\sqrt{z^2-4\sigma^2}}{2\sigma^2}\,.
\label{eq:app-wigner-quadratic-solution}
\end{equation}
To select the correct branch, we impose two conditions. First, for large $|z|$ one must have
\begin{equation}
g(z)\sim \frac{1}{z}\,,
\label{eq:app-wigner-large-z}
\end{equation}
because
\begin{equation}
(z\pmb I-\pmb A)^{-1}=\frac{1}{z}\pmb I+O(z^{-2})\,.
\label{eq:app-resolvent-large-z}
\end{equation}
Second, with the convention $z=\lambda-i\epsilon$, $\epsilon>0$, one must have ${\rm Im}\,g(z)>0$. These two conditions give
\begin{equation}
g(z)=\frac{z-\sqrt{z^2-4\sigma^2}}{2\sigma^2}\,,
\label{eq:app-wigner-physical-branch}
\end{equation}
where the square root is chosen so that $\sqrt{z^2-4\sigma^2}\sim z$ for large $|z|$. For $\lambda\in[-2\sigma,2\sigma]$ and $z=\lambda-i0^+$,
\begin{equation}
\sqrt{\lambda^2-4\sigma^2}=-i\sqrt{4\sigma^2-\lambda^2}\,,
\label{eq:app-square-root-on-cut}
\end{equation}
and therefore
\begin{equation}
{\rm Im} g(\lambda-i0^+)=\frac{1}{2\sigma^2}\sqrt{4\sigma^2-\lambda^2}\,.
\label{eq:app-wigner-imaginary-part}
\end{equation}
Using the convention
\begin{equation}
\rho(\lambda)=\frac{1}{\pi}{\rm Im}g(\lambda-i0^+)\,,
\label{eq:app-density-convention-reminder}
\end{equation}
we obtain the semicircle law
\begin{equation}
\rho_{\rm sc}(\lambda)=\frac{1}{2\pi\sigma^2}\sqrt{4\sigma^2-\lambda^2}\mathbf 1_{|\lambda|\leq2\sigma}\,.
\label{eq:app-semicircle-law}
\end{equation}
Thus the sparse symmetric cavity equations reduce, in the large-connectivity scaling, to the classical Wigner law \cite{Wigner1955,Mehta2004}.

The same idea applies to diluted Wishart matrices. Recall that
\begin{equation}
\pmb W=\frac{1}{d}\pmb X\pmb X^{\rm T}\,,\qquad X_i^\mu=B_i^\mu \xi_i^\mu\,,\qquad
{\rm Prob}(B_i^\mu=1)=\frac{d}{N}\,,
\label{eq:app-dense-wishart-model}
\end{equation}
with
\begin{equation}
P=\frac{N}{\alpha}\,.
\label{eq:app-dense-wishart-rectangularity}
\end{equation}
The cavity equations are
\begin{equation}
U_{\mu\to i}(z)=\frac{(\xi_i^\mu)^2}{1-\displaystyle\frac{1}{d}\sum_{j\in\partial\mu\setminus i}(\xi_j^\mu)^2G_{j\to\mu}(z)}\,,
    \label{eq:app-dense-wishart-factor}
\end{equation}
and
\begin{equation}
G_{i\to\mu}(z)=\frac{1}{z-\displaystyle\frac{1}{d}\sum_{\nu\in\partial i\setminus\mu}U_{\nu\to i}(z)}\,.
\label{eq:app-dense-wishart-variable}
\end{equation}
We assume
\begin{equation}
\int d\xi p_\xi(\xi) \xi^2=1
\label{eq:app-dense-wishart-second-moment}
\end{equation}
for convenience. The factor degree is $d+O(\sqrt d)$, and the variable degree is $d/\alpha+O(\sqrt d)$. As before, the cavity/full distinction disappears at leading order:
\begin{equation}
G_{i\to\mu}(z)=G_i(z)+O(d^{-1})\,,\qquad U_{\mu\to i}(z)=U_{\mu\to i}^{\rm full}(z)+O(d^{-1})\,.
\label{eq:app-dense-wishart-cavity-full}
\end{equation}
Now consider the denominator of \eqref{eq:app-dense-wishart-factor}. It contains $O(d)$ terms, each of order $d^{-1}$, so by the law of large numbers
\begin{equation}
\frac{1}{d}\sum_{j\in\partial\mu\setminus i}(\xi_j^\mu)^2G_{j\to\mu}(z)\longrightarrow G(z)\,,
\label{eq:app-dense-wishart-factor-self-averaging}
\end{equation}
where $G(z)$ is now the deterministic limiting diagonal Green function on the variable side. Substituting \eqref{eq:app-dense-wishart-factor-self-averaging} into \eqref{eq:app-dense-wishart-factor} gives
\begin{equation}
U_{\mu\to i}(z)\longrightarrow\frac{(\xi_i^\mu)^2}{1-G(z)}\,.
\label{eq:app-dense-wishart-factor-limit}
\end{equation}
The variable node receives $d/\alpha+O(\sqrt d)$ such terms, but each enters the denominator through the contribution $d^{-1}U_{\nu\to i}$. Averaging over $\xi_i^\mu$ and summing, we find the deterministic self-energy
\begin{equation}
\Sigma(z)=\frac{1}{d}\sum_{\nu\in\partial i}U_{\nu\to i}(z)\longrightarrow\frac{1}{\alpha[1-G(z)]}\,.
\label{eq:app-dense-wishart-self-energy}
\end{equation}
Hence
\begin{equation}
G(z)=\frac{1}{z-\displaystyle\frac{1}{\alpha[1-G(z)]}}\,.
\label{eq:app-dense-wishart-self-consistency}
\end{equation}
This is the dense-limit equation in the normalization and message convention used in the main text, namely $\pmb W=d^{-1}\pmb X\pmb X^{\rm T}$ with $U_{\mu\to i}$ not divided by $d$.

To compare with the conventional Mar\v{c}enko--Pastur law, it is better to rescale the matrix as
\begin{equation}
\widehat{\pmb W}=\alpha \pmb W\,.
\label{eq:app-standard-wishart-rescaling}
\end{equation}
If $\widehat g(z)$ denotes the Stieltjes transform of $\widehat{\pmb W}$, then
\begin{equation}
G(z)=\alpha\widehat g(\alpha z)\,,\qquad\widehat g(z)=\frac{1}{\alpha}G\!\left(\frac{z}{\alpha}\right)\,.
\label{eq:app-resolvent-rescaling}
\end{equation}
Substituting \eqref{eq:app-resolvent-rescaling} into \eqref{eq:app-dense-wishart-self-consistency}, with $z$ replaced by $z/\alpha$, gives
\begin{equation}
\widehat g(z)=\frac{1}{z-\displaystyle\frac{1}{1-\alpha \widehat g(z)}}\,.
\label{eq:app-marcenko-pastur-self-consistency}
\end{equation}
Multiplying both sides by the denominator yields
\begin{equation}
\alpha z \widehat g(z)^2+(1-\alpha-z)\widehat g(z)+1=0\,.
\label{eq:app-marcenko-pastur-quadratic}
\end{equation}
Solving the quadratic,
\begin{equation}
\widehat g(z)=\frac{z+\alpha-1-\sqrt{(z-\lambda_-)(z-\lambda_+)}}{2\alpha z}\,,\qquad\lambda_\pm=(1\pm\sqrt\alpha)^2\,,
\label{eq:app-marcenko-pastur-solution}
\end{equation}
where the branch is fixed by $\widehat g(z)\sim z^{-1}$ for large $|z|$. Therefore
\begin{equation}
\rho_{\rm MP}(\lambda)=\frac{1}{2\pi\alpha\lambda}\sqrt{(\lambda_+-\lambda)(\lambda-\lambda_-)}\mathbf 1_{\lambda_-\leq\lambda\leq\lambda_+}
\label{eq:app-marcenko-pastur-density}
\end{equation}
for the rescaled matrix $\widehat{\pmb W}$, together with an atom of weight $1-1/\alpha$ at the origin when $\alpha>1$ \cite{MarchenkoPastur1967,BaiSilverstein2010}. Returning to the normalization used in the main text, the absolutely continuous support is divided by $\alpha$, and the zero atom, when present, keeps the same weight:
\begin{equation}
\lambda_\pm^{(W)}=\frac{(1\pm\sqrt\alpha)^2}{\alpha}\,,\qquad\rho_{W}(\lambda)=\alpha\rho_{\rm MP}(\alpha\lambda)\,.
\label{eq:app-main-text-wishart-density}
\end{equation}
Thus the diluted Wishart cavity equations reduce to the Mar\v{c}enko--Pastur law in the dense limit.

The same reduction can be carried out for the generalized diluted Wishart ensemble. Recall that
\begin{equation}
\pmb F=\frac{1}{2d}\left(\pmb X\pmb Y^{\rm T}+\pmb Y\pmb X^{\rm T}\right)\,,
\label{eq:app-generalized-wishart}
\end{equation}
with nonzero edge weights $(x_i^\mu,y_i^\mu)$ drawn from the joint law $\varrho(x,y)$. Define the moments
\begin{equation}
m_{20}=\int dx dy\varrho(x,y)x^2\,,\qquad m_{11}=\int dxdy\varrho(x,y)xy\,, \qquad m_{02}=\int dxdy\varrho(x,y)y^2\,,
\label{eq:app-generalized-wishart-moments}
\end{equation}
and
\begin{equation}
\Delta=m_{20}m_{02}-m_{11}^2\,.
\label{eq:app-generalized-wishart-delta}
\end{equation}
In the finite-connectivity cavity equation, the factor-to-variable self-energy depends on the three sums
\begin{equation}
S_{xx}^{\mu\to i}=\sum_{j\in\partial\mu\setminus i}(x_j^\mu)^2G_{j\to\mu}\,,\qquad S_{xy}^{\mu\to i}=\sum_{j\in\partial\mu\setminus i}x_j^\mu y_j^\mu G_{j\to\mu}\,,\qquad S_{yy}^{\mu\to i}=\sum_{j\in\partial\mu\setminus i}(y_j^\mu)^2G_{j\to\mu}\,.
\label{eq:app-generalized-cavity-sums}
\end{equation}
When $d\to\infty$, these sums self-average:
\begin{equation}
\frac{S_{xx}^{\mu\to i}}{d}\to m_{20}G(z)\,,\qquad\frac{S_{xy}^{\mu\to i}}{d}\to m_{11}G(z)\,,\qquad\frac{S_{yy}^{\mu\to i}}{d}\to m_{02}G(z)\,.
\label{eq:app-generalized-self-averaging}
\end{equation}
Substituting into the finite-connectivity factor formula, one finds that a single factor message is of order $d^{-1}$:
\begin{equation}
U_{\mu\to i}(z)\longrightarrow\frac{1}{d}\frac{4xy+y^2m_{20}G(z)-2xym_{11}G(z)+x^2m_{02}G(z)}{\left(2-m_{11}G(z)\right)^2-m_{20}m_{02}G(z)^2}\,.
\label{eq:app-generalized-factor-limit}
\end{equation}
A variable vertex receives $d/\alpha$ such contributions. Averaging over $(x,y)$ and summing, the self-energy becomes
\begin{equation}
\Sigma(z)=\frac{1}{\alpha}\frac{4m_{11}+2\Delta G(z)}{4-4m_{11}G(z)-\Delta G(z)^2}\,.
\label{eq:app-generalized-dense-self-energy}
\end{equation}
Hence the dense-limit Green function satisfies
\begin{equation}
G(z)=\frac{1}{z-\Sigma(z)}=\frac{1}{z-\displaystyle\frac{1}{\alpha}\frac{4m_{11}+2\Delta G(z)}{4-4m_{11}G(z)-\Delta G(z)^2}}\,.
    \label{eq:app-generalized-dense-equation}
\end{equation}
Multiplying through gives the cubic equation
\begin{equation}
\alpha\Delta z G(z)^3+\left[4\alpha m_{11}z+(2-\alpha)\Delta
\right]G(z)^2+4\left[m_{11}(1-\alpha)-\alpha z\right]G(z)+4\alpha=0\,.
\label{eq:app-generalized-dense-cubic}
\end{equation}
This is the dense reduction of the generalized diluted Wishart cavity equation. Two checks are immediate. If $x=y$, then
\begin{equation}
m_{20}=m_{11}=m_{02}=1\,,\qquad\Delta=0\,,
\label{eq:app-generalized-wishart-check1}
\end{equation}
and \eqref{eq:app-generalized-dense-equation} reduces to
\begin{equation}
G(z)=\frac{1}{z-\displaystyle\frac{1}{\alpha[1-G(z)]}}\,,
\label{eq:app-generalized-to-wishart}
\end{equation}
which is exactly the dense diluted-Wishart equation. If $y=-x$, then $m_{11}=-1$ and $\Delta=0$, so one obtains the same equation with the opposite sign in the self-energy, corresponding to a negative Wishart matrix. Thus the dense limit of the generalized ensemble interpolates between positive and negative covariance-type laws, as it should \cite{PerezCastillo2022Generalized}.

Finally, we consider the dense limit of sparse non-Hermitian matrices. For simplicity we take real entries, zero diagonal, and the paired-support convention used in the sparse non-Hermitian ensemble: for each unordered pair ${i,j}$, the support variable satisfies $C_{ij}=C_{ji}$, the support has mean degree $c\to\infty$, and
\begin{equation}
A_{ij}=\frac{C_{ij}}{\sqrt c} J_{ij}\,,\qquad A_{ji}=\frac{C_{ij}}{\sqrt c} J_{ji}\,.
\label{eq:app-dense-nonhermitian-model}
\end{equation}
The pair $(J_{ij},J_{ji})$ has moments
\begin{equation}
\mathbb E[J_{ij}]=0\,,\qquad\mathbb E[J_{ij}^2]=\sigma^2\,,\qquad\mathbb E[J_{ij}J_{ji}]=\tau\sigma^2\,,\qquad-1\leq\tau\leq1\,.
\label{eq:app-dense-nonhermitian-moments}
\end{equation}
The parameter $\tau$ measures the reciprocal correlation between opposite edges. The Hermitized cavity recursion is
\begin{equation}
\pmb{G}_{i\to j}(z,\eta)=\left[\pmb{Z}(z,\eta)-\sum_{\ell\in\partial i\setminus j}\pmb{\mathcal A}_{i\ell}\pmb{G}_{\ell\to i}(z,\eta)\pmb{\mathcal A}_{i\ell}^{\dagger}\right]^{-1}\,,
\label{eq:app-dense-nonhermitian-cavity}
\end{equation}
where
\begin{equation}
\pmb{Z}(z,\eta)=\begin{pmatrix}
i\eta & z\\
z^* & i\eta
\end{pmatrix}\,,\qquad\pmb{\mathcal A}_{i\ell}=\frac{1}{\sqrt c}\begin{pmatrix}
0 & J_{i\ell}\\
J_{\ell i} & 0
\end{pmatrix}\,.
\label{eq:app-dense-nonhermitian-objects}
\end{equation}
Assume, as before, that the message distribution collapses onto a deterministic matrix
\begin{equation}
\pmb{G}(z,\eta)=\begin{pmatrix}
a & b\\
c & d
   \end{pmatrix}\,.
\label{eq:app-dense-nonhermitian-message}
\end{equation}
We now compute the average contribution of one edge explicitly:
\begin{equation}
\pmb{\mathcal A}_{i\ell}\pmb{G}\pmb{\mathcal A}_{i\ell}^{\dagger}=\frac{1}{c}\begin{pmatrix}
J_{i\ell}^2 d&J_{i\ell}J_{\ell i} c\\
J_{i\ell}J_{\ell i} b&J_{\ell i}^2 a
\end{pmatrix}\,.
\label{eq:app-dense-edge-block-product}
\end{equation}
Averaging over the pair $(J_{i\ell},J_{\ell i})$ and summing over $c$ neighbors, we obtain the deterministic self-energy
\begin{equation}
\sum_{\ell\in\partial i}\pmb{\mathcal A}_{i\ell}\pmb{G}\pmb{\mathcal A}_{i\ell}^{\dagger}\longrightarrow\sigma^2\begin{pmatrix}
d & \tau c\\
\tau b & a
\end{pmatrix}\,.
\label{eq:app-dense-nonhermitian-self-energy}
\end{equation}
Thus the dense Hermitized fixed-point equation is
\begin{equation}
\pmb{G}(z,\eta)=\left[\begin{pmatrix}
i\eta & z\\
z^* & i\eta
\end{pmatrix}-\sigma^2\begin{pmatrix}
d & \tau c\\
\tau b & a
\end{pmatrix}
\right]^{-1}\,.
    \label{eq:app-dense-nonhermitian-matrix-equation}
\end{equation}

Outside the spectral support the solution is holomorphic. In that regime the diagonal components vanish as $\eta\downarrow0$, so we set
\begin{equation}
a=d=0\,, \qquad c=g(z)\,, \qquad b=g(z)^*\,.
\label{eq:app-holomorphic-ansatz}
\end{equation}
Substituting into \eqref{eq:app-dense-nonhermitian-matrix-equation} and setting $\eta=0$, we must invert
\begin{equation}
\begin{pmatrix}
0 & z-\tau\sigma^2 g(z)\\
z^*-\tau\sigma^2 g(z)^* & 0
\end{pmatrix}\,.
\label{eq:app-holomorphic-matrix}
\end{equation}
The inverse of
\begin{equation}
\begin{pmatrix}
0 & B\\
C & 0
\end{pmatrix}
\label{eq:app-off-diagonal-matrix}
\end{equation}
is
\begin{equation}
\begin{pmatrix}
0 & C^{-1}\\
B^{-1} & 0
\end{pmatrix}\,,
\label{eq:app-off-diagonal-inverse}
\end{equation}
so we obtain
\begin{equation}
g(z)=\frac{1}{z-\tau\sigma^2 g(z)}\,.
\label{eq:app-elliptic-holomorphic-branch}
\end{equation}
This is the dense non-Hermitian resolvent equation outside the support. For $\tau=0$ it reduces to $g(z)=1/z$, the holomorphic branch of the circular law. For $\tau=1$ it becomes the Hermitian Wigner equation.

To locate the spectral boundary we linearize around the holomorphic branch. Let $a$ and $d$ be infinitesimal. Write
\begin{equation}
\pmb M=\begin{pmatrix}-\sigma^2 d & z-\tau\sigma^2 g\\
z^*-\tau\sigma^2 g^* & -\sigma^2 a
\end{pmatrix}\,,
\label{eq:app-linearized-hermitized-matrix}
\end{equation}
so that
\begin{equation}
\pmb{G}=\pmb M^{-1}\,.
\label{eq:app-linearized-inverse}
\end{equation}
The determinant of $\pmb M$ is
\begin{equation}
\Delta=\sigma^4 ad-|z-\tau\sigma^2 g|^2\,.
\label{eq:app-linearized-determinant}
\end{equation}
Using the formula for the inverse of a $2\times2$ matrix,
\begin{equation}
a=\frac{-\sigma^2 a}{\Delta}\,,\qquad d=\frac{-\sigma^2 d}{\Delta}\,.
\label{eq:app-linearized-diagonal-equations}
\end{equation}
A nonzero infinitesimal solution requires
\begin{equation}
\Delta=-\sigma^2\,.
\label{eq:app-boundary-condition-delta}
\end{equation}
On the holomorphic branch $a=d=0$, so \eqref{eq:app-linearized-determinant} reduces to
\begin{equation}
\Delta=-|z-\tau\sigma^2 g|^2\,.
\label{eq:app-delta-holomorphic}
\end{equation}
Hence the boundary of the spectrum is determined by
\begin{equation}|z-\tau\sigma^2 g(z)|^2=\sigma^2\,.
\label{eq:app-elliptic-boundary-implicit}
\end{equation}
Using \eqref{eq:app-elliptic-holomorphic-branch}, we have
\begin{equation}
z-\tau\sigma^2 g(z)=\frac{1}{g(z)}\,,
\label{eq:app-rewrite-boundary}
\end{equation}
so \eqref{eq:app-elliptic-boundary-implicit} becomes
\begin{equation}
|g(z)|=\frac{1}{\sigma}\,.
\label{eq:app-g-modulus-boundary}
\end{equation}
We may therefore parametrize
\begin{equation}
g(z)=\frac{e^{-i\theta}}{\sigma}\,,\qquad 0\leq\theta<2\pi\,.
\label{eq:app-g-parametrization}
\end{equation}
Substituting into
\begin{equation}
z=\frac{1}{g(z)}+\tau\sigma^2 g(z)\,,
\label{eq:app-z-from-g}
\end{equation}
we find
\begin{equation}
z(\theta)=\sigma e^{i\theta}+\tau\sigma e^{-i\theta}\,.
\label{eq:app-ellipse-parametrization}
\end{equation}
Separating real and imaginary parts,
\begin{equation}
x(\theta)=\sigma(1+\tau)\cos\theta\,,\qquad y(\theta)= \sigma(1-\tau)\sin\theta\,.
    \label{eq:app-ellipse-axes}
\end{equation}
Thus the dense limit of the sparse non-Hermitian cavity equations recovers the support of the elliptic law \cite{SommersCrisantiSompolinskyStein1988}. The non-holomorphic branch gives the corresponding bulk density; the calculation above fixes the boundary. The special case $\tau=0$ gives the circle
\begin{equation}
x(\theta)=\sigma\cos\theta\,, \qquad y(\theta)=\sigma\sin\theta\,,
\label{eq:app-circular-law-boundary}
\end{equation}
that is, the circular law \cite{Ginibre1965,Girko1984,TaoVuKrishnapur2010}. The special case $\tau=1$ gives the interval
\begin{equation}
x(\theta)=2\sigma\cos\theta\,, \qquad y(\theta)=0\,,
\label{eq:app-hermitian-limit-ellipse}
\end{equation}
which is exactly the support of the Hermitian semicircle law.

The general lesson is now clear. In every ensemble treated in the main text, the sparse cavity equations contain a sum of many weak contributions. When the connectivity becomes large and the edge weights are scaled appropriately, those sums self-average. In homogeneous cases the message distribution collapses to a delta function; with residual diagonal or edge-weight randomness, the dense reduction is instead a deterministic self-consistency equation for the averaged Green function or self-energy. For sparse symmetric matrices this gives the Pastur equation and the semicircle law. For diluted Wishart matrices it gives the Mar\v{c}enko--Pastur equation. For the generalized diluted Wishart ensemble it gives a deterministic cubic equation interpolating between covariance and cross-correlation limits. For sparse non-Hermitian matrices it gives the circular and elliptic dense limits. These reductions provide one of the most important checks on the finite-connectivity formalism developed in the main text.

\section{Population-dynamics algorithms}
\label{app:population-dynamics}
Population dynamics is the numerical realization of a distributional self-consistency equation. Since this algorithm is used repeatedly in these notes, it is worth deriving it carefully once and for all. We will do this in a way that makes the logic transparent. First we derive the generic algorithm from a fixed-point equation for a probability law. Then we specialize the construction to the scalar messages of sparse symmetric matrices, the two-population structure of diluted Wishart ensembles, the finite-rank factor updates of generalized diluted Wishart matrices, the matrix-valued messages of sparse non-Hermitian ensembles, and finally the tilted populations that appear in large-deviation and conditioned-density problems. The algorithm itself is standard in the statistical mechanics of disordered systems \cite{MezardParisiVirasoro1987,MezardParisi2001,MezardMontanari2009}; what matters here is how it arises from the specific cavity equations of sparse random matrices \cite{NagaoTanaka2007,RogersTakedaPerezCastilloKuhn2008,RogersPerezCastillo2009,SuscaVivoKuhn2021,MetzNeriRogers2019}.

We begin with a generic distributional fixed-point equation. Let $h$ denote a cavity message. Depending on the problem, $h$ may be a complex number, a tuple of complex numbers, or a small matrix. Suppose the cavity law $\mathcal P(h)$ satisfies
\begin{equation}
\mathcal P(h)=\sum_{\ell=0}^{\infty}q_\ell\int d\omega\Pi_\ell(\omega)\left[\prod_{r=1}^{\ell} dh_r \mathcal P(h_r)\right]\delta\left(h-\mathcal F_\ell(\omega;h_1,\ldots,h_\ell)\right)\,,
\label{eq:app-pd-generic-fixed-point}
\end{equation}
where $q_\ell$ is the cavity branching law, $\omega$ denotes the local disorder variables, $\Pi_\ell(\omega)$ is their distribution for a neighborhood with $\ell$ incoming messages, and $\mathcal F_\ell$ is the local cavity map. This equation says the following: to generate one sample from $\mathcal P$, one first draws the number $\ell$ of incoming neighbors, then draws the local disorder, then draws $\ell$ incoming messages independently from $\mathcal P$, and finally applies the deterministic map $\mathcal F_\ell$.

The idea of population dynamics is to replace the unknown law $\mathcal P$ by an empirical measure supported on a large population of messages. Let
\begin{equation}
\widehat{\mathcal P}_M(h)=\frac{1}{M}\sum_{\alpha=1}^{M}\delta\left(h-h^{(\alpha)}\right)\,,
\label{eq:app-pd-empirical-measure}
\end{equation}
where
\begin{equation}
\left\{h^{(1)},h^{(2)},\ldots,h^{(M)}\right\}
\label{eq:app-pd-population}
\end{equation}
is the current population. To understand the algorithm, we do not start from a recipe but from the action of the fixed-point map on test functions. Let $\varphi(h)$ be a bounded test function. Integrating \eqref{eq:app-pd-generic-fixed-point} against $\varphi$ gives
\begin{align}
\int dh \varphi(h)\mathcal P(h)&=\sum_{\ell=0}^{\infty}q_\ell\int d\omega \Pi_\ell(\omega)\left[\prod_{r=1}^{\ell}dh_r \mathcal P(h_r)\right]\varphi\left(\mathcal F_\ell(\omega;h_1,\ldots,h_\ell)\right)\,.
\label{eq:app-pd-test-function-fixed-point}
\end{align}
Now replace $\mathcal P$ by the empirical measure \eqref{eq:app-pd-empirical-measure}. Then
\begin{align}
\int dh\varphi(h)\widehat{\mathcal P}_M(h)=\frac{1}{M}\sum_{\alpha=1}^{M}\varphi\left(h^{(\alpha)}\right)\,,
\label{eq:app-pd-test-function-empirical}
\end{align}
and the right-hand side of \eqref{eq:app-pd-test-function-fixed-point} becomes
\begin{align}
&\sum_{\ell=0}^{\infty}q_\ell\int d\omega \Pi_\ell(\omega)\left[\prod_{r=1}^{\ell}dh_r \widehat{\mathcal P}_M(h_r)\right]\varphi\left(\mathcal F_\ell(\omega;h_1,\ldots,h_\ell)\right)\nonumber\\
&=\sum_{\ell=0}^{\infty}q_\ell\int d\omega\,\Pi_\ell(\omega)\frac{1}{M^\ell}\sum_{\alpha_1,\ldots,\alpha_\ell=1}^{M}\varphi\left(\mathcal F_\ell(\omega;h^{(\alpha_1)},\ldots,h^{(\alpha_\ell)})\right)\,.
\label{eq:app-pd-test-function-discrete-map}
\end{align}
Equation \eqref{eq:app-pd-test-function-discrete-map} is simply an expectation over random draws. It says that if one draws $\ell$, then draws $\omega$, then draws $\ell$ indices $\alpha_1,\ldots,\alpha_\ell$ independently and uniformly from $\{1,\ldots,M\}$, and finally defines
\begin{equation}
h_{\rm new}=\mathcal F_\ell(\omega;h^{(\alpha_1)},\ldots,h^{(\alpha_\ell)})\,,
\label{eq:app-pd-new-sample}
\end{equation}
then
\begin{equation}
\mathbb E\left[\varphi(h_{\rm new})\middle|\widehat{\mathcal P}_M\right]=\int dh \varphi(h)\mathcal T[\widehat{\mathcal P}_M](h)\,,
\label{eq:app-pd-expectation-map}
\end{equation}
where $\mathcal T$ is the fixed-point map defined by the right-hand side of \eqref{eq:app-pd-generic-fixed-point}. This is the derivation of the elementary population update.

There are two standard ways to use \eqref{eq:app-pd-new-sample}. In the synchronous, or batch, version one generates a whole new population
\begin{equation}
h_{\rm new}^{(1)},\ldots,h_{\rm new}^{(M)}
\label{eq:app-pd-batch-population}
\end{equation}
from the current one and then replaces the entire population at once. In the asynchronous version one picks a random index $\beta\in\{1,\ldots,M\}$ and performs the replacement
\begin{equation}
h^{(\beta)}\leftarrow h_{\rm new}\,.
\label{eq:app-pd-asynchronous-replacement}
\end{equation}
The asynchronous version is the one most often used in practice. Let us derive its mean action on test functions. Define
\begin{equation}
\widehat{\mathcal P}_M'(h)=\widehat{\mathcal P}_M(h)+\frac{1}{M}\left[\delta(h-h_{\rm new})-\delta(h-h^{(\beta)})\right]\,.
\label{eq:app-pd-one-step-empirical}
\end{equation}
Then
\begin{align}
\mathbb E\left[\int dh \varphi(h)\widehat{\mathcal P}_M'(h) \middle| \widehat{\mathcal P}_M\right]&=\left(1-\frac{1}{M}\right)\int dh\varphi(h)\widehat{\mathcal P}_M(h)\nonumber\\
&\quad+\frac{1}{M}\int dh\varphi(h)\mathcal T[\widehat{\mathcal P}_M](h)\,.
\label{eq:app-pd-asynchronous-mean}
\end{align}
Thus one asynchronous replacement is a stochastic Euler step toward the fixed-point map, and one sweep of $M$ such replacements approximates one application of $\mathcal T$ to the population. This is the mathematical content of population dynamics.

The next step is to explain how observables are measured once the cavity law has been approximated. Suppose the corresponding site law is
\begin{equation}
\mathcal P_{\rm site}(g)=\sum_{k=0}^{\infty}p_k\int d\omega \Pi^{\rm site}_k(\omega)\left[\prod_{r=1}^{k} dh_r\mathcal P(h_r)\right]\delta\left(g-\mathcal G_k(\omega;h_1,\ldots,h_k)\right)\,,
\label{eq:app-pd-site-law}
\end{equation}
where $p_k$ is the full degree distribution and $\mathcal G_k$ is the site map. If $\mathcal O(g)$ is a site observable, then
\begin{equation}
\langle \mathcal O\rangle_{\rm site}=\int dg\mathcal P_{\rm site}(g)\mathcal O(g)\,.
\label{eq:app-pd-site-observable}
\end{equation}
The empirical estimator is obtained by drawing $n_{\rm obs}$ independent site samples from \eqref{eq:app-pd-site-law} with the current population and setting
\begin{equation}
\widehat{\mathcal O}=\frac{1}{n_{\rm obs}}\sum_{m=1}^{n_{\rm obs}}\mathcal O(g_m)\,.
\label{eq:app-pd-site-estimator}
\end{equation}
Thus the population itself approximates the cavity law, while the site observable is measured by a second Monte Carlo average on top of the population.

We now specialize the generic construction to the main ensembles of these notes. For sparse symmetric matrices the message is a complex scalar Green function $G$, and the fixed-point equation is
\begin{equation}
\mathcal P_{\rm cav}(G)=\sum_{\ell=0}^{\infty}q_\ell\int dD p_D(D)\left[\prod_{r=1}^{\ell}dG_r \mathcal P_{\rm cav}(G_r)\,dJ_r\,p_J(J_r)\right]\delta\left(G-\frac{1}{z-D-\displaystyle\sum_{r=1}^{\ell}J_r^2G_r}\right)\,,
\label{eq:app-pd-symmetric-fixed-point}
\end{equation}
with
\begin{equation}
z=\lambda-i\epsilon\,,\qquad\epsilon>0\,.
\label{eq:app-pd-hermitian-z}
\end{equation}
The elementary population update follows directly from \eqref{eq:app-pd-new-sample}:
\begin{equation}
G_{\rm new}=\frac{1}{z-D-\displaystyle\sum_{r=1}^{\ell}J_r^2G^{(\alpha_r)}}\,.
\label{eq:app-pd-symmetric-update}
\end{equation}
The corresponding site sample is
\begin{equation}
G_{\rm site}=\frac{1}{z-D-\displaystyle\sum_{r=1}^{k}J_r^2G^{(\alpha_r)}}\,,
\label{eq:app-pd-symmetric-site-sample}
\end{equation}
where now $k$ is drawn from $p_k$ rather than $q_\ell$. The regularized spectral density is therefore estimated by
\begin{equation}
\widehat\rho_\epsilon(\lambda)=\frac{1}{\pi n_{\rm obs}}\sum_{m=1}^{n_{\rm obs}}{\rm Im} G_{{\rm site},m}\,.
\label{eq:app-pd-symmetric-density-estimator}
\end{equation}
There are three elementary consistency checks for the scalar population. First, all messages should satisfy
\begin{equation}
{\rm Im} G^{(\alpha)}>0
\label{eq:app-pd-hermitian-imaginary-sign}
\end{equation}
with the convention $z=\lambda-i\epsilon$, $\epsilon>0$. Second, the density estimator \eqref{eq:app-pd-symmetric-density-estimator} should be nonnegative. Third, numerical integration over $\lambda$ should give approximately one when the density is measured over a sufficiently large interval and the regularization is small.

The diluted Wishart ensemble requires two message laws. The variable-to-factor message is a complex scalar $G$, while the factor-to-variable message is a self-energy $U$. We use the convention of the main text: $U$ is not divided by $d$, and the actual contribution to the variable denominator is $d^{-1}U$. The distributional equations are
\begin{equation}
\mathcal Q(U)=\sum_{k=0}^{\infty}q_k^{(\rm f)}\int d\xi p_\xi(\xi)\prod_{r=1}^{k}\left[dG_r \mathcal P(G_r) d\xi_r p_\xi(\xi_r)\right]\delta\left(U-\frac{\xi^2}{1-\displaystyle\frac{1}{d}\sum_{r=1}^{k}\xi_r^2G_r}\right)\,,
\label{eq:app-pd-wishart-factor-law}
\end{equation}
and
\begin{equation}
\mathcal P(G)=\sum_{\ell=0}^{\infty}q_\ell^{(\rm v)}\int\prod_{r=1}^{\ell}\left[dU_r \mathcal Q(U_r)\right]\delta\left(G-\frac{1}{z-\displaystyle\frac{1}{d}\sum_{r=1}^{\ell}U_r}\right)\,.
\label{eq:app-pd-wishart-variable-law}
\end{equation}
The derivation of the algorithm is now a direct application of the generic one, but with two populations,
\begin{equation}
\left\{G^{(1)},\ldots,G^{(M_G)}\right\}\,,\qquad\left\{U^{(1)},\ldots,U^{(M_U)}\right\}\,.
\label{eq:app-pd-two-populations}
\end{equation}
A factor update is
\begin{equation}
U_{\rm new}=\frac{\xi^2}{1-\displaystyle\frac{1}{d}\sum_{r=1}^{k}\xi_r^2G^{(\alpha_r)}}\,,
\label{eq:app-pd-wishart-factor-update}
\end{equation}
where $k\sim q_k^{(\rm f)}$, while a variable update is
\begin{equation}
G_{\rm new}=\frac{1}{z-\displaystyle\frac{1}{d}\sum_{r=1}^{\ell}U^{(\beta_r)}}\,,
\label{eq:app-pd-wishart-variable-update}
\end{equation}
where $\ell\sim q_\ell^{(\rm v)}$. The site sample is
\begin{equation}
G_{\rm site}=\frac{1}{z-\displaystyle\frac{1}{d}\sum_{r=1}^{k}U^{(\beta_r)}}\,,\qquad k\sim p_k^{(\rm v)}\,,
\label{eq:app-pd-wishart-site-sample}
\end{equation}
and the density estimator is again
\begin{equation}
\widehat\rho_\epsilon(\lambda)=\frac{1}{\pi n_{\rm obs}}\sum_{m=1}^{n_{\rm obs}}{\rm Im}G_{{\rm site},m}\,.
\label{eq:app-pd-wishart-density-estimator}
\end{equation}
The positivity of the Wishart spectrum implies a further check away from the origin:
\begin{equation}
\widehat\rho_\epsilon(\lambda)\approx0\qquad\text{for }\lambda<0
\label{eq:app-pd-wishart-positivity-check}
\end{equation}
for negative $\lambda$ on scales not touching a possible zero atom, up to the Lorentzian broadening produced by finite $\epsilon$.

The generalized diluted Wishart and sparse cross-correlation ensemble is only slightly more complicated at the algorithmic level, because the factor contribution remains finite rank. The self-energy distribution is determined by the map
\begin{equation}
U=\frac{4dxy+y^2S_{xx}-2xyS_{xy}+x^2S_{yy}}{(2d-S_{xy})^2-S_{xx}S_{yy}}\,,
\label{eq:app-pd-generalized-factor-map}
\end{equation}
where
\begin{equation}
S_{xx}=\sum_{r=1}^{k}x_r^2G_r\,,\qquad S_{xy}=\sum_{r=1}^{k}x_ry_rG_r\,,\qquad S_{yy}=\sum_{r=1}^{k}y_r^2G_r\,.
    \label{eq:app-pd-generalized-sums}
\end{equation}
Thus a factor update consists of drawing $k\sim q_k^{(\rm f)}$, drawing $k$ incoming Green functions from the $G$ population, drawing the corresponding pairs $(x_r,y_r)$ from the local weight law $\varrho(x,y)$, drawing an additional distinguished pair $(x,y)$, forming the sums \eqref{eq:app-pd-generalized-sums}, and substituting them into \eqref{eq:app-pd-generalized-factor-map}. In the generalized diluted Wishart convention used here, $U$ is already the full contribution to the variable denominator, so the variable update is
\begin{equation}
G_{\rm new}=\frac{1}{z-\displaystyle\sum_{r=1}^{\ell}U^{(\beta_r)}}\,,
\label{eq:app-pd-generalized-variable-update}
\end{equation}
where $\ell$ is drawn from the variable excess-degree law $q_\ell^{(\rm v)}$. The corresponding site sample is
\begin{equation}
G_{\rm site}=\frac{1}{z-\displaystyle\sum_{r=1}^{k_{\rm site}}U^{(\beta_r)}}\,,
\label{eq:app-pd-generalized-site-sample}
\end{equation}
where $k_{\rm site}$ is drawn from the full variable-degree law $p_k^{(\rm v)}$. The density estimator is
\begin{equation}
\widehat\rho_\epsilon(\lambda)=\frac{1}{\pi n_{\rm obs}}\sum_{m=1}^{n_{\rm obs}}{\rm Im}G_{{\rm site},m}\,.
\label{eq:app-pd-generalized-density-estimator}
\end{equation}
Thus the only extra numerical cost is the evaluation of the three sums and the rational map \eqref{eq:app-pd-generalized-factor-map}. This is one of the reasons why generalized diluted Wishart ensembles remain computationally accessible despite their richer algebraic structure \cite{PerezCastillo2022Generalized}.

For sparse non-Hermitian matrices the message becomes a $2\times2$ matrix. Denote one such message by
\begin{equation}
\pmb{G}=\begin{pmatrix}
a & b\\
c & d
\end{pmatrix}\,.
\label{eq:app-pd-matrix-message}
\end{equation}
The distributional fixed-point equation is
\begin{equation}
\mathcal P(\pmb{G})=\sum_{\ell=0}^{\infty}q_\ell\int\left[\prod_{r=1}^{\ell}d\pmb{G}_r \mathcal P(\pmb{G}_r) dJ_r dK_r p_{J_1,J_2}(J_r,K_r)\right]\delta\left(\pmb{G}-\left[\pmb{Z}-\sum_{r=1}^{\ell}\pmb{\mathcal A}(J_r,K_r)\,\pmb{G}_r \pmb{\mathcal A}(J_r,K_r)^\dagger\right]^{-1}\right)\,,
\label{eq:app-pd-nonhermitian-law}
\end{equation}
with
\begin{equation}
\pmb{Z}=\begin{pmatrix}
i\eta & z\\
z^* & i\eta
\end{pmatrix},\qquad
\pmb{\mathcal A}(J,K)=
\begin{pmatrix}
0 & J\\
K^* & 0
\end{pmatrix}\,.
\label{eq:app-pd-nonhermitian-building-blocks}
\end{equation}
The empirical population is now a set of matrices
\begin{equation}
\left\{\pmb{G}^{(1)},\ldots,\pmb{G}^{(M)}\right\},
\label{eq:app-pd-matrix-population}
\end{equation}
and the elementary update is simply
\begin{equation}
\pmb{G}_{\rm new}=\left[\pmb{Z}-\sum_{r=1}^{\ell}\pmb{\mathcal A}(J_r,K_r)\pmb{G}^{(\alpha_r)}\pmb{\mathcal A}(J_r,K_r)^\dagger\right]^{-1}\,.
\label{eq:app-pd-nonhermitian-update}
\end{equation}
The site sample is generated by the same formula with the full degree law $p_k$. The regularized non-Hermitian resolvent field is estimated by
\begin{equation}
\widehat g_\eta(z,z^*)=\frac{1}{n_{\rm obs}}\sum_{m=1}^{n_{\rm obs}}\left[\pmb{G}_{{\rm site},m}\right]_{21}\,,
\label{eq:app-pd-nonhermitian-resolvent-estimator}
\end{equation}
and the regularized density by
\begin{equation}
\widehat\rho_\eta(z)=\frac{1}{\pi}\partial_{z^*}\widehat g_\eta(z,z^*)\,.
\label{eq:app-pd-nonhermitian-density-estimator}
\end{equation}
In practice, one either approximates the Wirtinger derivative by finite differences on a grid in the complex plane or introduces companion populations for the response of the messages with respect to $z^*$. The finite-difference route is simpler and is usually sufficient for plotting the density.

The large-deviation and conditioned-density problems require one more ingredient, namely weighted populations. The formal structure is again completely transparent. Suppose the tilted cavity law satisfies
\begin{equation}
\mathcal P_s(h)=\frac{1}{\mathcal C_s}\sum_{\ell=0}^{\infty}q_\ell\int d\omega \Pi_\ell(\omega)\left[\prod_{r=1}^{\ell}dh_r \mathcal P_s(h_r)\right]w_s(\omega;h_1,\ldots,h_\ell)\delta\left(h-\mathcal F_\ell(\omega;h_1,\ldots,h_\ell)\right)\,,
\label{eq:app-pd-weighted-fixed-point}
\end{equation}
where $\mathcal C_s$ is the normalization
\begin{equation}
\mathcal C_s=\sum_{\ell=0}^{\infty}q_\ell\int d\omega \Pi_\ell(\omega)\left[\prod_{r=1}^{\ell}dh_r \mathcal P_s(h_r)\right]w_s(\omega;h_1,\ldots,h_\ell)\,.
\label{eq:app-pd-weighted-normalization}
\end{equation}
This is the generic form of the tilted distributions that appear in the index problem, in conditioned spectral densities, and in the non-Hermitian number-statistics problem. The local map $\mathcal F_\ell$ is the same cavity map as before. The new ingredient is only the weight $w_s$, which is built from the local contribution of the observable used in the tilt \cite{MetzPerezCastillo2016,PerezCastilloMetz2018Wishart,PerezCastilloMetz2018Conditioned,RamosSanchezGuzmanGonzalezPerezCastilloMetz2021}. The normalization $\mathcal C_s$ in \eqref{eq:app-pd-weighted-normalization} is the normalization of a tilted message update. In concrete spectral-count calculations, the scaled cumulant-generating function is obtained from the corresponding Bethe combination of local normalizations, with the appropriate site, factor, and edge terms for the ensemble under consideration.

To derive the weighted algorithm, integrate \eqref{eq:app-pd-weighted-fixed-point} against a test function $\varphi$:
\begin{align}
\int dh \varphi(h)\mathcal P_s(h)=\frac{\mathbb E\left[w_s\varphi(h_{\rm new})\right]}{\mathbb E\left[w_s\right]}\,,
\label{eq:app-pd-weighted-test-function}
\end{align}
where the expectation is taken with respect to the sampling of $\ell$, $\omega$, and the incoming messages from the current population, and where $h_{\rm new}$ is still defined by \eqref{eq:app-pd-new-sample}. Equation \eqref{eq:app-pd-weighted-test-function} immediately suggests the batch algorithm. One generates $M$ candidates
\begin{equation}
\widetilde h^{(1)},\ldots,\widetilde h^{(M)}
\label{eq:app-pd-weighted-candidates}
\end{equation}
and their associated weights
\begin{equation}
w^{(1)},\ldots,w^{(M)}.
\label{eq:app-pd-weighted-weights}
\end{equation}
When the weights are nonnegative, the new population is obtained by resampling the candidates with probabilities
\begin{equation}
\pi_m=\frac{w^{(m)}}{\sum_{n=1}^{M}w^{(n)}}\,.
\label{eq:app-pd-resampling-probabilities}
\end{equation}
This is importance sampling in message space. If a complex or sign-changing replica representation is used directly, one must keep explicit signed or complex weights rather than interpreting \eqref{eq:app-pd-resampling-probabilities} as probabilities. The empirical average of $\varphi$ in the resampled population converges to the right-hand side of \eqref{eq:app-pd-weighted-test-function} whenever the probability-resampling interpretation is valid. In lecture-note language, the real-tilt large-deviation algorithms used here are ordinary population dynamics plus importance-resampling.

For the Hermitian index problem, the message is not one Green function but a tuple,
\begin{equation}
h=\left(G^-,G^+\right)
\label{eq:app-pd-index-message}
\end{equation}
or, for conditioned densities,
\begin{equation}
h=\left(G^-,G^+,G^0\right)\,,
\label{eq:app-pd-conditioned-message}
\end{equation}
with
\begin{equation}
G^-=G(x-i\epsilon),,\qquad G^+=G(x+i\epsilon)\,,\qquad G^0=G(\lambda-i\eta)\,.
\label{eq:app-pd-multi-spectral-message}
\end{equation}
The map $\mathcal F_\ell$ is simply the scalar cavity map applied to each component, while the weight $w_s$ is built from the relevant Bethe phase contribution to the determinant. For sparse symmetric matrices these are site and edge terms; for diluted Wishart matrices they are variable, factor, and edge terms, and the message tuple contains both variable and factor components. Thus the formal structure of the weighted algorithm does not change from one problem to another.

We conclude with a few remarks on convergence and numerical practice. Since the appendix is about algorithms, it is useful to state explicitly what one should monitor. First, the observables of interest should stabilize as the population size $M$ is increased. Second, at fixed $M$, the running average of the measured observable should become stationary after a burn-in time. Third, when the messages are complex Green functions with the lower-half-plane convention $z=\lambda-i\epsilon$, one should monitor that
\begin{equation}
{\rm Im} G^{(\alpha)}>0
\label{eq:app-pd-imaginary-check}
\end{equation}
for all population members, up to numerical roundoff. Fourth, independent runs with different random seeds should agree within Monte Carlo errors. Finally, the observables obtained from population dynamics should agree with the corresponding sum rules and dense-limit checks derived elsewhere in the notes.

The main lesson of this appendix is therefore straightforward. Population dynamics is not an ad hoc numerical recipe. It is the Monte Carlo implementation of a distributional fixed-point equation. The empirical population \eqref{eq:app-pd-empirical-measure} approximates the unknown cavity law, the local map \eqref{eq:app-pd-new-sample} generates one draw from the fixed-point operator, and the weighted extension \eqref{eq:app-pd-weighted-fixed-point}--\eqref{eq:app-pd-resampling-probabilities} implements the large-deviation and conditioning problems. Once this generic structure is understood, all the specific algorithms of these lecture notes are just concrete realizations of the same idea.

\section{Replica-symmetric saddle-point derivations}
\label{app:replica-symmetric-saddle-points}
In the main text we emphasized that the cavity equations and the replica equations are two descriptions of the same finite-connectivity mean-field structure. The purpose of this appendix is to show this explicitly at the level of the saddle point. We will derive the replica-symmetric saddle-point equations for the simplest sparse symmetric ensemble, and then show how the same derivation extends to the replicated generating functions used for spectral-count large deviations and conditioned spectral densities. The derivation follows the general strategy introduced in the early replica treatments of sparse random matrices \cite{EdwardsJones1976,RodgersBray1988,BrayRodgers1988}, but is written here in the notation of these lecture notes and in a way that makes the connection with the cavity distributions transparent \cite{Kuhn2008,SuscaVivoKuhn2021}.

To keep the algebra readable, we first consider a Poissonian sparse symmetric ensemble. Let
\begin{equation}
A_{ij}=D_i\delta_{ij}+C_{ij}J_{ij}\,,\qquad A_{ij}=A_{ji}\,,\qquad C_{ii}=0\,,
\label{eq:app-rs-model}
\end{equation}
where, for $i<j$,
\begin{equation}
{\rm Prob}(C_{ij}=1)=\frac{c}{N}\,,\qquad{\rm Prob}(C_{ij}=0)=1-\frac{c}{N}\,,
\label{eq:app-rs-bernoulli-graph}
\end{equation}
the edge weights $J_{ij}$ are independent with distribution $p_J(J)$, and the diagonal terms $D_i$ are independent with distribution $p_D(D)$. The spectral density is obtained from the Edwards--Jones partition function
\begin{equation}
Z_{\pmb A}(z)=\int\left[\prod_{i=1}^{N}\frac{du_i}{\sqrt{2\pi}}\right]\exp\left[-\frac{i}{2}\pmb u^{\rm T}(z\pmb I-\pmb A)\pmb u\right]\,,\qquad z=\lambda-i\epsilon\,,\qquad\epsilon>0\,,
\label{eq:app-rs-edwards-jones}
\end{equation}
through
\begin{equation}
\overline{\rho_{\pmb A}(\lambda)}=-\frac{2}{\pi N}\lim_{\epsilon\downarrow0}{\rm Im}\frac{\partial}{\partial \lambda}\overline{\log Z_{\pmb A}(\lambda-i\epsilon)}\,.
\label{eq:app-rs-density-from-logZ}
\end{equation}
The quenched logarithm is handled formally through the thermodynamic replica relation
\begin{equation}
\lim_{N\to\infty}\frac{1}{N}\overline{\log Z_{\pmb A}(z)}=\lim_{n\to0}\lim_{N\to\infty}\frac{1}{nN}\log\overline{[Z_{\pmb A}(z)]^n}\,.
\label{eq:app-rs-replica-identity}
\end{equation}

We therefore begin with integer $n\geq1$ and compute $\overline{[Z_{\pmb A}(z)]^n}$. Introducing replicated variables
\begin{equation}
\underline u_i=(u_i^1,\ldots,u_i^n)\,,\qquad\underline u_i^{\,2}=\sum_{a=1}^{n}(u_i^a)^2\,,\qquad\underline u_i\cdot\underline u_j=\sum_{a=1}^{n}u_i^a u_j^a\,,
\label{eq:app-rs-replicated-variables}
\end{equation}
we may write
\begin{align}
[Z_{\pmb A}(z)]^n&=\int\left[\prod_{i=1}^{N}\prod_{a=1}^{n}\frac{du_i^a}{\sqrt{2\pi}}\right]\exp\Bigg[-\frac{i}{2}\sum_{i=1}^{N}(z-D_i)\underline u_i^{2}+i\sum_{i<j}C_{ij}J_{ij}\underline u_i\cdot\underline u_j\Bigg]\,.
\label{eq:app-rs-replicated-Z}
\end{align}
The disorder average over the graph and the weights factorizes over pairs $i<j$. Since
\begin{equation}
\overline{e^{iC_{ij}J_{ij}\underline u_i\cdot \underline u_j}}=1-\frac{c}{N}+\frac{c}{N}\int dJ p_J(J)e^{iJ \underline u_i\cdot\underline u_j}\,,
\label{eq:app-rs-edge-average}
\end{equation}
we obtain
\begin{align}
\overline{[Z_{\pmb A}(z)]^n}&=\int\left[\prod_{i=1}^{N}\prod_{a=1}^{n}\frac{du_i^a}{\sqrt{2\pi}}\right]\prod_{i=1}^{N}\left[\int dD_i p_D(D_i) e^{-\frac{i}{2}(z-D_i)\underline u_i^{2}}\right]\nonumber\\
&\hspace{1cm}\times\prod_{i<j}\left[1-\frac{c}{N}+\frac{c}{N}\int dJ p_J(J) e^{ iJ \underline u_i\cdot\underline u_j}\right]\,.
\label{eq:app-rs-average-before-expansion}
\end{align}
The product over pairs is of the standard sparse form $1+O(N^{-1})$. Writing
\begin{equation}
F(\underline u_i,\underline u_j)=\int dJ p_J(J)e^{iJ\underline u_i\cdot\underline u_j}\,,
\label{eq:app-rs-F-definition}
\end{equation}
we have
\begin{equation}
\prod_{i<j}\left[1+\frac{c}{N}\left(F(\underline u_i,\underline u_j)-1\right)\right]=\exp\left[\frac{c}{2N}\sum_{i,j=1}^{N}\left(F(\underline u_i,\underline u_j)-1\right)+O(1)\right]\,.
\label{eq:app-rs-pair-product}
\end{equation}
The $O(1)$ term is negligible in the large-$N$ saddle-point analysis, since the dominant contribution to $\log \overline{Z^n}$ is of order $N$.

At this point the interaction between different sites enters only through the empirical distribution of replicated vectors. We therefore introduce the order parameter
\begin{equation}
\varrho(\underline u)=\frac{1}{N}\sum_{i=1}^{N}\delta(\underline u-\underline u_i)\,,\qquad \int d\underline u \varrho(\underline u)=1\,,
\label{eq:app-rs-order-parameter}
\end{equation}
where
\begin{equation}
d\underline u=\prod_{a=1}^{n}du^a\,.
\label{eq:app-rs-measure-replicas}
\end{equation}
To enforce \eqref{eq:app-rs-order-parameter}, we insert the functional identity
\begin{equation}
1=\int \mathcal D\varrho \mathcal D\hat\varrho\exp\left[-iN\int d\underline u \hat\varrho(\underline u)\varrho(\underline u)+i\sum_{i=1}^{N}\hat\varrho(\underline u_i)\right]\,.
\label{eq:app-rs-functional-identity}
\end{equation}
Substituting \eqref{eq:app-rs-functional-identity} into \eqref{eq:app-rs-average-before-expansion}, the integrals over the replicated site variables factorize. After straightforward rearrangement, we obtain
\begin{equation}
\overline{[Z_{\pmb A}(z)]^n}=\int\mathcal D\varrho\mathcal D\hat\varrho\exp\left[NS_n[\varrho,\hat\varrho]\right]\,,
\label{eq:app-rs-functional-integral}
\end{equation}
with action
\begin{align}
S_n[\varrho,\hat\varrho]&=-i\int d\underline u\hat\varrho(\underline u)\varrho(\underline u)+\frac{c}{2}\int d\underline u d\underline v\varrho(\underline u)\varrho(\underline v)\left[F(\underline u,\underline v)-1\right]\nonumber\\
&\quad+\log\int dD p_D(D)\int d\underline u\exp\left[-\frac{i}{2}(z-D)\underline u^{2}+i\hat\varrho(\underline u)\right]\,.
\label{eq:app-rs-action}
\end{align}
The large-$N$ limit is now obtained by a saddle point. Varying with respect to $\hat\varrho(\underline u)$ gives
\begin{equation}
\varrho(\underline u)=\frac{\int dD p_D(D)\exp\left[-\frac{i}{2}(z-D)\underline u^{2}+i\hat\varrho(\underline u)\right]}{\int dD p_D(D)\int d\underline w\exp\left[-\frac{i}{2}(z-D)\underline w^{2}+i\hat\varrho(\underline w)\right]}\,.
\label{eq:app-rs-saddle-1}
\end{equation}
Varying with respect to $\varrho(\underline u)$ gives
\begin{equation}
i\hat\varrho(\underline u)=c\int d\underline v\varrho(\underline v)\left[F(\underline u,\underline v)-1\right]\,.
\label{eq:app-rs-saddle-2}
\end{equation}
Substituting \eqref{eq:app-rs-saddle-2} into \eqref{eq:app-rs-saddle-1}, we obtain a single functional saddle-point equation:
\begin{equation}
\varrho(\underline u)=\frac{1}{\mathcal N_n}\int dD p_D(D)\exp\left[-\frac{i}{2}(z-D)\underline u^{2}+c\int d\underline v\varrho(\underline v)\left(F(\underline u,\underline v)-1\right)\right]\,,
\label{eq:app-rs-single-functional-saddle}
\end{equation}
where $\mathcal N_n$ is the normalization that enforces $\int d\underline u\,\varrho(\underline u)=1$.

Equation \eqref{eq:app-rs-single-functional-saddle} is exact at the replica-symmetric saddle-point level, but it is still an equation for a function of an $n$-component replica vector. The simplification comes from the Gaussian structure of the original problem. Since the replicated variables entered through a Gaussian integral, and since the cavity solution of the main text uses Gaussian messages, we look for a replica-symmetric superposition of identical Gaussians in replica space. We therefore introduce the normalized Gaussian
\begin{equation}
\phi(\underline u|\omega)=\prod_{a=1}^{n}\left(\frac{\omega}{2\pi}\right)^{1/2}\exp\left[-\frac{\omega}{2}(u^a)^2\right]\,,\qquad{\rm Re}\omega>0\,,
\label{eq:app-rs-gaussian-component}
\end{equation}
and make the ansatz
\begin{equation}
\varrho(\underline u)=\int d\omega \pi_n(\omega)\phi(\underline u|\omega)\,,
\label{eq:app-rs-gaussian-mixture}
\end{equation}
with
\begin{equation}
\int d\omega\,\pi_n(\omega)=1\,.
\label{eq:app-rs-mixture-normalization}
\end{equation}
Replica symmetry means that all replicas are statistically equivalent, so the Gaussian factor \eqref{eq:app-rs-gaussian-component} depends only on the same inverse variance $\omega$ in each replica direction.

We now evaluate the interaction term in \eqref{eq:app-rs-single-functional-saddle}. Using \eqref{eq:app-rs-F-definition} and \eqref{eq:app-rs-gaussian-mixture},
\begin{align}
\int d\underline v\varrho(\underline v)F(\underline u,\underline v)&=\int dJ p_J(J)\int d\omega \pi_n(\omega)\int d\underline v\phi(\underline v|\omega) e^{iJ\underline u\cdot\underline v}\,.
\label{eq:app-rs-interaction-start}
\end{align}
The integral over $\underline v$ factorizes over replicas. For one replica,
\begin{equation}
\int_{-\infty}^{\infty}dv\left(\frac{\omega}{2\pi}\right)^{1/2}\exp\left[-\frac{\omega}{2}v^2+iJuv\right]=\exp\left[-\frac{J^2}{2\omega}u^2\right]\,.
    \label{eq:app-rs-one-replica-gaussian-transform}
\end{equation}
Hence
\begin{equation}
\int d\underline v\phi(\underline v|\omega)e^{iJ\underline u\cdot\underline v}=\exp\left[-\frac{J^2}{2\omega}\underline u^{2}\right]\,,
\label{eq:app-rs-all-replicas-gaussian-transform}
\end{equation}
and therefore
\begin{equation}
\int d\underline v\varrho(\underline v)\left[F(\underline u,\underline v)-1\right]=\int d\omega \pi_n(\omega)\int dJ p_J(J) \left[e^{-J^2\underline u^{2}/(2\omega)}-1\right]\,.
\label{eq:app-rs-interaction-result}
\end{equation}
Substituting \eqref{eq:app-rs-interaction-result} into \eqref{eq:app-rs-single-functional-saddle}, we obtain
\begin{align}
\varrho(\underline u)&=\frac{1}{\mathcal N_n}\int dD p_D(D)\exp\Bigg[-\frac{i}{2}(z-D)\underline u^{2}\nonumber\\
&\hspace{1.7cm}+c\int d\omega \pi_n(\omega)\int dJ p_J(J)\left(e^{-J^2\underline u^{\,2}/(2\omega)}-1\right)\Bigg]\,.
\label{eq:app-rs-before-poisson-expansion}
\end{align}
At first sight \eqref{eq:app-rs-before-poisson-expansion} does not look Gaussian. The key step is to expand the exponential in Poisson form:
\begin{align}
&\exp\left[c\int d\omega \pi_n(\omega)\int dJ p_J(J)\left(e^{-J^2\underline u^{2}/(2\omega)}-1\right)\right]\nonumber\\
&=e^{-c}\sum_{k=0}^{\infty}\frac{c^k}{k!}\left[\int d\omega \pi_n(\omega)\int dJ p_J(J) e^{-J^2\underline u^{\,2}/(2\omega)}\right]^k\nonumber\\
&=\sum_{k=0}^{\infty}e^{-c}\frac{c^k}{k!}\int\left[\prod_{r=1}^{k}d\omega_r \pi_n(\omega_r) dJ_r p_J(J_r)\right]\exp\left[-\frac{1}{2}\underline u^{\,2}\sum_{r=1}^{k}\frac{J_r^2}{\omega_r}\right]\,.
\label{eq:app-rs-poisson-expansion}
\end{align}
Substituting \eqref{eq:app-rs-poisson-expansion} into \eqref{eq:app-rs-before-poisson-expansion}, we find
\begin{align}
\varrho(\underline u)&=\frac{1}{\mathcal N_n}\sum_{k=0}^{\infty}e^{-c}\frac{c^k}{k!}\int dD p_D(D)\left[\prod_{r=1}^{k}d\omega_r \pi_n(\omega_r) dJ_r p_J(J_r)\right]\exp\left[-\frac{\Omega}{2}\underline u^{\,2}\right]\,,
\label{eq:app-rs-gaussian-restored}
\end{align}
where
\begin{equation}
\Omega=i(z-D)+\sum_{r=1}^{k}\frac{J_r^2}{\omega_r}\,.
\label{eq:app-rs-new-inverse-variance}
\end{equation}
Equation \eqref{eq:app-rs-gaussian-restored} is the crucial step. The order parameter is again a superposition of Gaussians, so the Gaussian-mixture ansatz is self-consistent.

To read off the induced distribution of $\Omega$, note that
\begin{equation}
e^{-\frac{\Omega}{2}\underline u^{2}}=\left(\frac{2\pi}{\Omega}\right)^{n/2}\phi(\underline u|\Omega)\,.
\label{eq:app-rs-rewrite-unnormalized-gaussian}
\end{equation}
Hence \eqref{eq:app-rs-gaussian-restored} has the same form as \eqref{eq:app-rs-gaussian-mixture}, with the induced mixture
\begin{align}
\pi_n(\Omega)&=\frac{1}{\mathcal Z_n}\sum_{k=0}^{\infty}e^{-c}\frac{c^k}{k!}\int dD p_D(D)\left[\prod_{r=1}^{k}d\omega_r \pi_n(\omega_r) dJ_r p_J(J_r)\right]\left(\frac{2\pi}{\Omega}\right)^{n/2}\nonumber\\
&\hspace{4cm}\times\delta\left(\Omega-i(z-D)-\sum_{r=1}^{k}\frac{J_r^2}{\omega_r}\right)\,,
\label{eq:app-rs-pi-n-recursion}
\end{align}
where $\mathcal Z_n$ normalizes $\pi_n$. In the replica limit $n\to0$,
\begin{equation}
\left(\frac{2\pi}{\Omega}\right)^{n/2}=1+O(n)\,,
\label{eq:app-rs-prefactor-nzero}
\end{equation}
so the saddle-point equation for the cavity inverse-variance distribution becomes
\begin{align}
\pi(\Omega)&=\sum_{k=0}^{\infty}e^{-c}\frac{c^k}{k!}\int dD p_D(D)\left[\prod_{r=1}^{k}d\omega_r \pi(\omega_r) dJ_r p_J(J_r)\right]\delta\!\left(\Omega-i(z-D)-\sum_{r=1}^{k}\frac{J_r^2}{\omega_r}\right)\,.
\label{eq:app-rs-final-omega-recursion}
\end{align}
This is exactly the distributional cavity equation written in the inverse-variance variable
\begin{equation}
\omega=iG^{-1}\,.
\label{eq:app-rs-omega-g-relation}
\end{equation}
Indeed, since the local Gaussian covariance of one replica is $\omega^{-1}$, while the Gaussian partition function \eqref{eq:app-rs-edwards-jones} has inverse covariance $i(z\pmb I-\pmb A)$, we have
\begin{equation}
\langle u_i^2\rangle=\frac{1}{\omega_i}=-iG_{ii}(z)\,,\qquad G_{ii}(z)=\frac{i}{\omega_i}\,.
\label{eq:app-rs-green-from-omega}
\end{equation}
For a Poisson ensemble the cavity and site degree laws coincide, so the full-site inverse-variance distribution is the same as the cavity one. More generally, if $p_k$ is the full degree law and $q_k$ the excess-degree law, the message distribution $\pi$ is generated by the analogue of \eqref{eq:app-rs-final-omega-recursion} with $e^{-c}c^k/k!$ replaced by $q_k$, while the site distribution uses the full degree law:
\begin{align}
\pi_{\rm site}(\Omega)&=\sum_{k=0}^{\infty}p_k\int dD p_D(D)\left[\prod_{r=1}^{k}d\omega_r \pi(\omega_r) dJ_r p_J(J_r)\right]\delta\left(\Omega-i(z-D)-\sum_{r=1}^{k}\frac{J_r^2}{\omega_r}\right)\,.
\label{eq:app-rs-site-distribution}
\end{align}
Therefore the spectral density is
\begin{equation}
\overline{\rho_{\pmb A}(\lambda)}=\frac{1}{\pi}\lim_{\epsilon\downarrow0}{\rm Im}\int d\Omega\pi_{\rm site}(\Omega)\frac{i}{\Omega}\,.
\label{eq:app-rs-density-from-site-law}
\end{equation}
Equation \eqref{eq:app-rs-density-from-site-law} is exactly the cavity formula derived in the main text. This completes the explicit replica-symmetric derivation of the standard sparse-symmetric cavity equations.

It is useful to make the logic of the derivation explicit. The replica trick turned the quenched logarithm into an annealed replicated partition function. The functional order parameter \eqref{eq:app-rs-order-parameter} encoded the empirical law of replicated local variables. The Gaussian mixture \eqref{eq:app-rs-gaussian-mixture} was the replica-symmetric ansatz appropriate to a Gaussian problem. The Poisson expansion \eqref{eq:app-rs-poisson-expansion} restored Gaussianity after disorder averaging. Finally, the inverse-variance distribution \eqref{eq:app-rs-final-omega-recursion} coincided with the cavity-message distribution. This is why the replica and cavity methods agree in this setting.

The same derivation extends to tilted problems. We first treat the large deviations of the number of eigenvalues below a threshold $x$. Define
\begin{equation}
z_- = x-i\epsilon\,,\qquad z_+ = x+i\epsilon\,, \qquad\epsilon>0\,,
\label{eq:app-rs-zpm}
\end{equation}
and the replicated generating function
\begin{equation}
\mathcal Q_N(n_-,n_+)=\overline{[Z_{\pmb A}(z_-)]^{n_-}[Z_{\pmb A}(z_+)]^{n_+}}\,.
\label{eq:app-rs-two-sector-generating}
\end{equation}
For integer $n_-,n_+\geq1$, introduce two replica sectors
\begin{equation}
\underline u_i^-=(u_i^{-,1},\ldots,u_i^{-,n_-})\,,\qquad\underline u_i^+=(u_i^{+,1},\ldots,u_i^{+,n_+})\,,
\label{eq:app-rs-two-sector-variables}
\end{equation}
and define the joint order parameter
\begin{equation}
\varrho(\underline u^-,\underline u^+)=\frac{1}{N}\sum_{i=1}^{N}\delta(\underline u^--\underline u_i^-)\delta(\underline u^+-\underline u_i^+)\,.
\label{eq:app-rs-two-sector-order-parameter}
\end{equation}
Repeating the same steps that led to \eqref{eq:app-rs-action}, we obtain the two-sector saddle-point equation
\begin{align}
\varrho(\underline u^-,\underline u^+)&=\frac{1}{\mathcal N_{n_-,n_+}}\int dD p_D(D) \exp\Bigg[-\frac{i}{2}(z_--D)(\underline u^-)^2-\frac{i}{2}(z_+-D)(\underline u^+)^2\nonumber\\
&\hspace{2cm}+c\int d\underline v^- d\underline v^+ \varrho(\underline v^-,\underline v^+)\left(e^{iJ(\underline u^-\cdot\underline v^-+\underline u^+\cdot\underline v^+)}-1\right)_{J\text{-avg}}\Bigg]\,,
\label{eq:app-rs-two-sector-saddle}
\end{align}
where
\begin{equation}
\left(e^{iJ(\underline u^-\cdot\underline v^-+\underline u^+\cdot\underline v^+)}\right)_{J\text{-avg}}\equiv\int dJ p_J(J)e^{ iJ(\underline u^-\cdot\underline v^-+\underline u^+\cdot\underline v^+)}\,.
\label{eq:app-rs-two-sector-javg}
\end{equation}
Replica symmetry now means permutation symmetry within each sector separately. The Gaussian factors in the $z_+$ sector are understood by analytic continuation of the same formulas, since the Edwards--Jones integral is directly convergent only for the lower-half-plane boundary value. We therefore use the product-Gaussian ansatz
\begin{equation}
\varrho(\underline u^-,\underline u^+)=\int d\omega^- d\omega^+ \pi_{n_-,n_+}(\omega^-,\omega^+)\phi(\underline u^-|\omega^-)\phi(\underline u^+|\omega^+)\,.
\label{eq:app-rs-two-sector-ansatz}
\end{equation}
The Gaussian integral \eqref{eq:app-rs-one-replica-gaussian-transform} factorizes independently in the two sectors, so
\begin{align}
&\int d\underline v^- d\underline v^+\phi(\underline v^-|\omega^-)\phi(\underline v^+|\omega^+)e^{ iJ(\underline u^-\cdot\underline v^-+\underline u^+\cdot\underline v^+)}\nonumber\\
&=\exp\left[-\frac{J^2}{2\omega^-}(\underline u^-)^2-\frac{J^2}{2\omega^+}(\underline u^+)^2\right]\,.
\label{eq:app-rs-two-sector-transform}
\end{align}
The same Poisson expansion as before then yields a Gaussian mixture again. Keeping the finite replica numbers $n_-$ and $n_+$, the joint recursion is
\begin{align}
\pi_{n_-,n_+}(\Omega^-,\Omega^+)
&=\frac{1}{\mathcal Z_{n_-,n_+}}
\sum_{k=0}^{\infty}e^{-c}\frac{c^k}{k!}\int dD p_D(D)
\left[\prod_{r=1}^{k}d\omega_r^- d\omega_r^+ \pi_{n_-,n_+}(\omega_r^-,\omega_r^+) dJ_r p_J(J_r)\right]\nonumber\\
&\hspace{1cm}\times
\left(\frac{2\pi}{\Omega^-}\right)^{n_-/2}
\left(\frac{2\pi}{\Omega^+}\right)^{n_+/2}
\delta\left(\Omega^--i(z_--D)-\sum_{r=1}^{k}\frac{J_r^2}{\omega_r^-}\right)\nonumber\\
&\hspace{1cm}\times\delta\left(\Omega^+-i(z_+-D)-\sum_{r=1}^{k}\frac{J_r^2}{\omega_r^+}\right)\,.
\label{eq:app-rs-two-sector-final}
\end{align}
In the strict $n_-\to0$, $n_+\to0$ limit, the prefactor in \eqref{eq:app-rs-two-sector-final} tends to one and the recursion reduces to the unweighted joint cavity law for the two boundary values $z_-$ and $z_+$. For the finite-$s$ index problem, however, the same prefactor is retained after the analytic continuation \eqref{eq:app-rs-index-continuation}; it is the replica origin of the tilted or weighted population laws used in the main text.
This equation deserves emphasis. The two components are not independent even though the map factorizes componentwise. They are correlated because the same local disorder variables $D$ and $J_r$ appear in both delta constraints. This is exactly what is needed in the large-deviation problem: the distribution of the two resolvents at $z_-$ and $z_+$ is coupled through the same local graph environment.

The corresponding site law is
\begin{align}
\pi_{{\rm site},n_-,n_+}(\Omega^-,\Omega^+)
&=\frac{1}{\mathcal Z^{\rm site}_{n_-,n_+}}
\sum_{k=0}^{\infty}p_k\int dD p_D(D)
\left[\prod_{r=1}^{k}d\omega_r^- d\omega_r^+ \pi_{n_-,n_+}(\omega_r^-,\omega_r^+) dJ_r p_J(J_r)\right]\nonumber\\
&\hspace{1cm}\times\left(\frac{2\pi}{\Omega^-}\right)^{n_-/2}
\left(\frac{2\pi}{\Omega^+}\right)^{n_+/2}
\delta\left(\Omega^--i(z_--D)-\sum_{r=1}^{k}\frac{J_r^2}{\omega_r^-}\right)\nonumber\\
&\hspace{1cm}\times\delta\left(\Omega^+-i(z_+-D)-\sum_{r=1}^{k}\frac{J_r^2}{\omega_r^+}\right)\,.
\label{eq:app-rs-two-sector-site-law}
\end{align}
After the analytic continuation
\begin{equation}
n_-=-\frac{s}{\pi i}\,,\qquad n_+=\frac{s}{\pi i}\,,
\label{eq:app-rs-index-continuation}
\end{equation}
the large-deviation theory of the index is recovered from the corresponding replica free energy, with the additive constant in the determinant-phase identity for $\mathcal K_{\pmb A}(x)$ contributing the separate term $s$ to the scaled cumulant-generating function. At the level of the saddle-point equation, however, the essential content is already visible in \eqref{eq:app-rs-two-sector-final}: the scalar replica-symmetric order parameter becomes a \emph{joint} distribution of two inverse variances.

The conditioned spectral density requires one more sector. Introduce the probe spectral parameter
\begin{equation}
z_0=\lambda-i\eta\,,\qquad\eta>0\,,
\label{eq:app-rs-probe-parameter}
\end{equation}
and consider
\begin{equation}
\mathcal Q_N(n_-,n_+,n_0)=\overline{[Z_{\pmb A}(z_-)]^{n_-}[Z_{\pmb A}(z_+)]^{n_+}[Z_{\pmb A}(z_0)]^{n_0}}\,.
\label{eq:app-rs-three-sector-generating}
\end{equation}
The order parameter is now
\begin{equation}
\varrho(\underline u^-,\underline u^+,\underline u^0)\,,
\label{eq:app-rs-three-sector-order-parameter}
\end{equation}
and the replica-symmetric ansatz is
\begin{equation}
\varrho(\underline u^-,\underline u^+,\underline u^0)=\int d\omega^- d\omega^+ d\omega^0\pi_{n_-,n_+,n_0}(\omega^-,\omega^+,\omega^0)\phi(\underline u^-|\omega^-)\phi(\underline u^+|\omega^+)\phi(\underline u^0|\omega^0)\,.
\label{eq:app-rs-three-sector-ansatz}
\end{equation}
Carrying through exactly the same steps, one obtains
\begin{align}
\pi_{n_-,n_+,n_0}(\Omega^-,\Omega^+,\Omega^0)
&=\frac{1}{\mathcal Z_{n_-,n_+,n_0}}
\sum_{k=0}^{\infty}e^{-c}\frac{c^k}{k!}\int dD p_D(D)\nonumber\\
&\hspace{1cm}\times\left[\prod_{r=1}^{k}d\omega_r^- d\omega_r^+ d\omega_r^0 \pi_{n_-,n_+,n_0}(\omega_r^-,\omega_r^+,\omega_r^0) dJ_r p_J(J_r)\right]\nonumber\\
&\hspace{1cm}\times\left(\frac{2\pi}{\Omega^-}\right)^{n_-/2}
\left(\frac{2\pi}{\Omega^+}\right)^{n_+/2}
\left(\frac{2\pi}{\Omega^0}\right)^{n_0/2}
\nonumber\\
&\hspace{1cm}\times
\delta\left(\Omega^--i(z_--D)-\sum_{r=1}^{k}\frac{J_r^2}{\omega_r^-}\right)
\delta\left(\Omega^+-i(z_+-D)-\sum_{r=1}^{k}\frac{J_r^2}{\omega_r^+}\right)\nonumber\\
&\hspace{1cm}\times
\delta\left(\Omega^0-i(z_0-D)-\sum_{r=1}^{k}\frac{J_r^2}{\omega_r^0}\right)\,.
\label{eq:app-rs-three-sector-final}
\end{align}
In the conditioned-density calculation one eventually takes \(n_0\to0\), so the probe sector measures the local Green function without changing the tilted ensemble. The two threshold sectors, however, retain the finite analytic continuation associated with the conditioning field. The conditioned density is then obtained from the site law associated with \eqref{eq:app-rs-three-sector-final} by taking the imaginary part of the probe component:
\begin{equation}
\rho_x(\lambda|k(s))=\frac{1}{\pi}\lim_{\eta\downarrow0}{\rm Im}\int d\Omega^-d\Omega^+ d\Omega^0 \pi_{\rm site}^{(s)}(\Omega^-,\Omega^+,\Omega^0)\frac{i}{\Omega^0}\,.
\label{eq:app-rs-conditioned-density}
\end{equation}
Once again, the structure matches the cavity picture exactly. The replica-symmetric saddle point is a joint law of the inverse variances at all spectral parameters involved in the observable.

Two final remarks are worth making. First, the derivation above was presented for the Poissonian sparse symmetric ensemble because it makes the algebra as transparent as possible. For arbitrary degree distributions or ensembles with topological constraints, one introduces degree- or type-conditioned order parameters. The Gaussian replica-symmetric ansatz then yields the corresponding degree- or type-conditioned cavity laws discussed in the main text. Second, for bipartite ensembles such as diluted Wishart matrices, one needs two replicated order parameters, one on the variable side and one on the factor side, but the same logic applies: the Gaussian replica-symmetric ansatz turns the replicated saddle-point problem into the same recursive distributional equations that follow from the cavity method.

The main point of this appendix can therefore be summarized succinctly. The replica-symmetric saddle-point derivation does not produce a different theory from the cavity method. It produces the same message variables and distributional structure, but from the replicated disorder average rather than from local tree recursions. In the simple spectral-density problem one obtains the single distribution \eqref{eq:app-rs-final-omega-recursion}; in the index problem one obtains the two-sector law \eqref{eq:app-rs-two-sector-final}; and in the conditioned-density problem one obtains the three-sector law \eqref{eq:app-rs-three-sector-final}. The finite-replica normalization factors are the replica origin of the local tilted weights used in the large-deviation and conditioned-density algorithms; the corresponding scaled cumulant-generating functions are obtained from the associated Bethe free-energy combinations. This is the precise technical sense in which the replica and cavity approaches coincide in the finite-connectivity problems studied in these notes.

\end{appendix}

\bibliography{biblio}

\end{document}